\documentclass[a4paper,twoside,english]{book}

\usepackage[T1]{fontenc}
\usepackage{amsthm}                  
\usepackage{newpxtext}
\usepackage{newpxmath}
\usepackage[utf8]{inputenc}

\usepackage{algorithm}
\usepackage{algpseudocode}
\usepackage{array}
\usepackage{adjustbox}
\usepackage{booktabs} 
\usepackage{longtable} 
\usepackage{arydshln}

\usepackage[square,numbers,sort]{natbib} 
\usepackage{doi}

\usepackage{babel}
\usepackage{balance}

\usepackage{blkarray} 
\usepackage{breqn} 
\usepackage{caption}
\usepackage{changepage}
\usepackage{color, colortbl}
\usepackage{makecell}
\usepackage{multirow} 
\usepackage{tabularx}

\usepackage{comment}
\usepackage{csquotes}
\usepackage{enumitem}
\usepackage{epigraph}
\usepackage{etoolbox}
\usepackage{eurosym}
\usepackage{fancyhdr}
\usepackage{fancyvrb}
\usepackage{float}
\usepackage[nomessages]{fp}
\usepackage{framed}
\usepackage{graphbox}
\usepackage{graphicx}
\usepackage{hyphenat}
\usepackage{hyperref}
\usepackage{lineno} 
\usepackage{lscape}
\usepackage{mathtools}
\usepackage{makecell}
\usepackage[font=itshape]{quoting}
\usepackage[counterclockwise]{rotating}
\usepackage{setspace}
\usepackage{soul}

\usepackage{siunitx} 
\usepackage{subcaption}
\usepackage{spverbatim}

\usepackage{imakeidx}
\usepackage{indentfirst}
\usepackage{pdflscape}
\usepackage{pdfpages}
\usepackage{placeins}

\usepackage{listings}
\usepackage{tcolorbox}

\usepackage[flushleft]{threeparttable}
\usepackage[nottoc]{tocbibind}
\usepackage{tocloft}

\usepackage{xcolor}
\usepackage{xparse}
\usepackage{xspace}
\usepackage{xstring}
\usepackage{uniudtesi}
\usepackage{url}
\usepackage{pageslts}
\usepackage[Lenny]{fncychap}

\usepackage{mathrsfs}

\let\oldtextsc\textsc
\renewcommand{\textsc}[1]{\oldtextsc{\scalefont{1.1}#1}}

\cftsetindents{figure}{0em}{3em}
\cftsetindents{table}{0em}{3em}

\graphicspath{{./figures/}}

\soulregister\cite7
\soulregister\ref7
\soulregister\pageref7
\setstcolor{red}
\setul{}{.2ex}

\tcbuselibrary{breakable}

\expandafter\def\expandafter\UrlBreaks\expandafter{\UrlBreaks
\do\a\do\b\do\c\do\d\do\e\do\f\do\g\do\h\do\i\do\j
\do\k\do\l\do\m\do\n\do\o\do\p\do\q\do\r\do\s\do\t
\do\u\do\v\do\w\do\x\do\y\do\z\do\A\do\B\do\C\do\D
\do\E\do\F\do\G\do\H\do\I\do\J\do\K\do\L\do\M\do\N
\do\O\do\P\do\Q\do\R\do\S\do\T\do\U\do\V\do\W\do\X
\do\Y\do\Z}

\makeatletter 
 
\@addtoreset{algorithm}{chapter} 
\makeatother

\makeatletter
\lst@InstallKeywords k{attributes}{attributestyle}\slshape{attributestyle}{}ld
\makeatother

\definecolor{background}{HTML}{EEEEEE}
\colorlet{punct}{red!60!black}
\definecolor{delim}{RGB}{20,105,176}
\colorlet{numb}{magenta!60!black}

\definecolor{editorLightGray}{cmyk}{0.05, 0.05, 0.05, 0.1}
\definecolor{editorGray}{cmyk}{0.6, 0.55, 0.55, 0.2}
\definecolor{editorPurple}{cmyk}{0.5, 1, 0, 0}
\definecolor{editorWhite}{cmyk}{0, 0, 0, 0}
\definecolor{editorBlack}{cmyk}{1, 1, 1, 1}
\definecolor{editorOrange}{cmyk}{0, 0.8, 1, 0}
\definecolor{editorBlue}{cmyk}{1, 0.6, 0, 0}
\definecolor{editorPink}{cmyk}{0, 1, 0, 0}
\definecolor{editorGreen}{rgb}{0, 0.5, 0} 

\titolo{In Crowd Veritas}
\sottotitolo{\textsc{Leveraging Human Intelligence} \\ \textsc{To Fight Misinformation}}
\candidate{\textsc{Michael Soprano}}
\annoaccademico{2023}
\dipartimento{Mathematics, Computer Science and Physics}
\corsodilaurea{Computer Science, Mathematics and Physics}
\supervisor[Prof.]{Stefano Mizzaro}
\cosupervisor[Ph.D.]{Kevin Roitero}

\hypersetup{
  bookmarksopen,
  bookmarksopenlevel=4,
  pdftitle={In Crowd Veritas: Leveraging Human Intelligence To Fight Misinformation},
  pdfsubject={},
  pdfauthor={Michael Soprano},
  pdfkeywords={crowdsourcing, misinformation, fact-checking, cognitive biases, truthfulness assessment, human computation},
  colorlinks=true,
  urlcolor=black,
  linkcolor=black,
  citecolor=black
}

\VerbatimFootnotes

\fvset{showspaces=false}

\newcolumntype{L}[1]{>{\raggedright\let\newline\\\arraybackslash\hspace{0pt}}m{#1}}
\newcolumntype{C}[1]{>{\centering\let\newline\\\arraybackslash\hspace{0pt}}m{#1}}
\newcolumntype{R}[1]{>{\raggedleft\let\newline\\\arraybackslash\hspace{0pt}}m{#1}}
\newcolumntype{P}[1]{>{\raggedright\arraybackslash}p{#1}}

\definecolor{editorBackground}{rgb}{0.96,0.96,0.96}

\lstdefinelanguage{CustomHTML}{
  language=HTML,
  sensitive=true,
  alsoletter={<>=-},
  morestring=[b]",
  morestring=[s]{`}{`},
  morecomment=[s]{<!--}{-->},
  commentstyle=\color{gray}\ttfamily,
  stringstyle=\color{teal},
  keywordstyle=\color{blue}\bfseries,
  tag=[s],
  morekeywords={
    html, head, body, div, span, p, a, ul, ol, li,
    h1, h2, h3, form, input, label, section, article,
    img, script, link, meta, crowd-form, crowd-instructions, short-summary, detailed-instructions,
    positive-example, negative-example, crowd-input,
    crowd-checkbox, crowd-slider
  }
}

\lstdefinestyle{htmlstyle}{
  language=CustomHTML,
  basicstyle=\ttfamily\small,
  backgroundcolor=\color{editorBackground},
  frame=leftline,
  rulecolor=\color{black},
  framesep=2mm,
  framerule=1pt,
  xleftmargin=1.5em,
  numbers=left,
  numberstyle=\tiny\color{gray},
  numbersep=10pt,
  breaklines=true,
  breakatwhitespace=false,
  columns=fullflexible,
  keepspaces=true,
  showstringspaces=false
}

\lstdefinelanguage{json}{
  basicstyle=\ttfamily\small,
  showstringspaces=false,
  morestring=[b]",
  morestring=[s]{'}{'},
  stringstyle=\color{teal},
  literate=
   *{0}{{{\color{black}0}}}{1}
    {1}{{{\color{black}1}}}{1}
    {2}{{{\color{black}2}}}{1}
    {3}{{{\color{black}3}}}{1}
    {4}{{{\color{black}4}}}{1}
    {5}{{{\color{black}5}}}{1}
    {6}{{{\color{black}6}}}{1}
    {7}{{{\color{black}7}}}{1}
    {8}{{{\color{black}8}}}{1}
    {9}{{{\color{black}9}}}{1}
    {:}{{{\color{black}:}}}{1}
    {,}{{{\color{black},}}}{1}
    {\{}{{{\color{black}\{}}}{1}
    {\}}{{{\color{black}\}}}}{1}
    {[}{{{\color{black}[}}}{1}
    {]}{{{\color{black}]}}}{1},
}

\lstdefinestyle{jsonstyle}{
  language=json,
  basicstyle=\ttfamily\small,
  backgroundcolor=\color{editorBackground},
  frame=leftline,
  framerule=1pt,
  framesep=2mm,
  xleftmargin=1.5em,
  numbers=left,
  numberstyle=\tiny\color{gray},
  numbersep=10pt,
  breaklines=true,
  breakatwhitespace=false,
  columns=fullflexible,
  keepspaces=true,
  showstringspaces=false
}

\lstdefinestyle{curlstyle}{
  basicstyle=\ttfamily\small,
  backgroundcolor=\color{editorBackground},
  frame=leftline,
  framerule=1pt,
  xleftmargin=1.5em,
  numbers=left,
  numberstyle=\tiny\color{gray},
  numbersep=10pt,
  breaklines=true,
  breakatwhitespace=false,
  columns=fullflexible,
  keepspaces=true,
  showstringspaces=false
}

\lstdefinestyle{textstyle}{
  backgroundcolor=\color{editorBackground},
  frame=leftline,
  rulecolor=\color{black},
  framesep=2mm,
  framerule=1pt,
  xleftmargin=1.5em,
  numbers=left,
  numberstyle=\tiny\color{gray},
  numbersep=10pt,
  basicstyle=\ttfamily\small,
  columns=fullflexible,
  keepspaces=true,
  breaklines=true,
  breakatwhitespace=false,
  showstringspaces=false
}

\lstdefinestyle{csvstyle}{
  basicstyle=\ttfamily\small,
  backgroundcolor=\color{editorBackground},
  frame=leftline,
  rulecolor=\color{black},
  framesep=2mm,
  framerule=1pt,
  xleftmargin=1.5em,
  numbers=left,
  numberstyle=\tiny\color{gray},
  numbersep=10pt,
  breaklines=true,
  breakatwhitespace=false,
  columns=fullflexible,
  keepspaces=true,
  showstringspaces=false,
  morecomment=[l]{\#}
}

\lstdefinelanguage{JavaScript}{
  keywords={typeof, new, true, false, catch, function, return, null, switch, var, if, in, while, do, else, case, break, let, const},
  keywordstyle=\color{blue}\bfseries,
  ndkeywords={class, export, boolean, throw, implements, import, this},
  ndkeywordstyle=\color{purple},
  identifierstyle=\color{black},
  sensitive=false,
  comment=[l]{//},
  morecomment=[s]{/*}{*/},
  commentstyle=\color{gray}\ttfamily,
  stringstyle=\color{teal}\ttfamily,
  morestring=[b]',
  morestring=[b]"
}

\lstdefinestyle{customjs}{
  language=JavaScript,
  basicstyle=\ttfamily\small,
  backgroundcolor=\color{gray!10},
  frame=leftline,
  framerule=1pt,
  xleftmargin=1.5em,
  numbers=left,
  numberstyle=\tiny\color{gray},
  numbersep=10pt,
  breaklines=true,
  breakatwhitespace=false,
  columns=fullflexible,
  keepspaces=true,
  showstringspaces=false
}

\newcommand*{\myquotingsource}{}

\SaveVerb{credentials.json}|credentials.json|
\SaveVerb{AdministratorAccess}|AdministratorAccess|
\SaveVerb{build/environments}|build/environments/|
\SaveVerb{build/task}|build/task/|
\SaveVerb{build/config}|build/config/|
\SaveVerb{build/skeleton}|build/skeleton/|
\SaveVerb{build/mturk}|build/mturk/|
\SaveVerb{build/toloka}|build/toloka/|
\SaveVerb{build/deploy}|build/deploy/|
\SaveVerb{result/Data}|result/Data/|
\SaveVerb{result/Dataframe}|result/Dataframe/|
\SaveVerb{result/Crawling}|result/Crawling/|
\SaveVerb{result/Resources}|result/Resources/|
\SaveVerb{result/Task}|result/Task/|
\SaveVerb{workers-acl}|workers_acl|
\SaveVerb{workers-answers}|workers_answers|

\newcommand\BeraMonottfamily{%
  \def\fvm@Scale{0.4} 
  \fontfamily{fvm}\selectfont 
}

\algnewcommand\algorithmicforeach{\textbf{forEach}}
\algdef{S}[FOR]{ForEach}[1]{\algorithmicforeach\ #1\ \algorithmicdo}

\algnewcommand\algorithmicinput{\textbf{await:}}
\algnewcommand\Await{\item[\algorithmicinput]}

\makeatletter\@addtoreset{chapter}{part}\makeatother

\setlist[itemize]{noitemsep, topsep=5pt}

\makeatletter
\newlength\epitextskip
\pretocmd{\@epitext}{\em}{}{}
\apptocmd{\@epitext}{\em}{}{}
\patchcmd{\epigraph}{\@epitext{#1}\\}{\@epitext{#1}\\[\epitextskip]}{}{}

\makeatother

\newcommand{\myparagraph}[1]{\vspace{0.5\baselineskip}\noindent{\textbf{#1}}~}

\setstcolor{red}
\setul{}{.2ex}

\newcommand{\listequationsname}{\normalfont{List of Equations}}
\newlistof{myequations}{equ}{\listequationsname}
\newcommand{\myequations}[1]{%
\addcontentsline{equ}{myequations}{\protect\numberline{\theequation}#1}\par}
\addto\captionsenglish{%
}

\newenvironment{dedication}[1][]{
\begin{titlepage}
\vspace*{70pt}
\vskip 0pt plus.3fil
\begin{flushright}
}{
\end{flushright}
\vskip 0pt plus.6fil
\end{titlepage}

}

\newcommand{\mrqone}{\ref{cap:research-questions-meta:identify-misinfo}: Judging Information Truthfulness\xspace}
\newcommand{\mrqtwo}{\ref{cap:research-questions-meta:impact-bias}: Cognitive Biases in Truthfulness Judgment\xspace}
\newcommand{\mrqthree}{\ref{cap:research-questions-meta:predict-explain}: Predicting and Explaining Truthfulness\xspace}

\newcommand{\N}{\mathbb{N}\index{\mathbb{N}}}

\newcommand{\crowdframe}{Crowd\_Frame\index{Crowd\_Frame}\xspace}
\newcommand{\generator}{\textsf{Generator}\index{Generator}\xspace}
\newcommand{\skeleton}{\textsf{Skeleton}\index{Skeleton}\xspace}
\newcommand{\logger}{\textsf{Logger}\index{Logger}\xspace}
\newcommand{\searchengine}{\textsf{Search Engine}\index{Search Engine}\xspace}

\newcommand{\mturk}{Amazon Mechanical Turk\index{Amazon!Mechanical Turk}\xspace}
\newcommand{\prolific}{Prolific\index{Prolific}\xspace}
\newcommand{\toloka}{Toloka\index{Toloka}\xspace}

\newcommand{\factcheckorg}{FactCheck.org\xspace}

\newcommand{\politifact}{PolitiFact\index{PolitiFact}\xspace}
\newcommand{\politifactzero}{\texttt{Pants-On-Fire}\index{Pants-On-Fire}\xspace}
\newcommand{\politifactpantsfire}{\texttt{Pants-On-Fire}\index{Pants-On-Fire}\xspace}
\newcommand{\politifactlie}{\texttt{Lie}\index{Lie}\xspace}
\newcommand{\politifactone}{\texttt{False}\index{False}\xspace}
\newcommand{\politifactfalse}{\texttt{False}\index{False}\xspace}
\newcommand{\politifactmostlyfalse}{\texttt{Mostly-False}\index{Mostly-False}\xspace}
\newcommand{\politifacttwo}{\texttt{Mostly-False}\index{Mostly-False}\xspace}
\newcommand{\politifactthree}{\texttt{Half-True}\index{Half-True}\xspace}
\newcommand{\politifacthalftrue}{\texttt{Half-True}\index{Half-True}\xspace}
\newcommand{\politifactfour}{\texttt{Mostly-True}\index{Mostly-True}\xspace}
\newcommand{\politifactmostlytrue}{\texttt{Mostly-True}\index{Mostly-True}\xspace}
\newcommand{\politifactfive}{\texttt{True}\index{True}\xspace}
\newcommand{\politifacttrue}{\texttt{True}\index{True}\xspace}

\newcommand{\politifactthreebinszero}{\texttt{01}\index{01}\xspace}
\newcommand{\politifactthreebinsone}{\texttt{23}\index{23}\xspace}
\newcommand{\politifactthreebinstwo}{\texttt{45}\index{45}\xspace}
\newcommand{\politifacttwoebinszero}{\texttt{012}\index{012}\xspace}
\newcommand{\politifacttwobinsone}{\texttt{234}\index{234}\xspace}

\newcommand{\generictrue}{\texttt{True}\index{True}\xspace}
\newcommand{\generichalftrue}{\texttt{Half-True}\index{Half-True}\xspace}
\newcommand{\genericfalse}{\texttt{False}\index{False}\xspace}
\newcommand{\genericmixed}{\texttt{Mixed}\index{Mixed}\xspace}
\newcommand{\genericother}{\texttt{Other}\index{Other}\xspace}

\newcommand{\abc}{RMIT ABC Fact Check\index{RMIT ABC Fact Check}\xspace}
\newcommand{\abczero}{\texttt{False}\index{False}\xspace}
\newcommand{\abcnegative}{\texttt{False}\index{False}\xspace}
\newcommand{\abcone}{\texttt{In-Between}\index{In-Between}\xspace}
\newcommand{\abcinbetween}{\texttt{In-Between}\index{In-Between}\xspace}
\newcommand{\abctwo}{\texttt{True}\index{True}\xspace}
\newcommand{\abcpositive}{\texttt{True}\index{True}\xspace}
\newcommand{\abccorrect}{\texttt{Correct}\index{Correct}\xspace}
\newcommand{\abcchecksout}{\texttt{Checks Out}\index{Checks Out}\xspace}
\newcommand{\abcmisleading}{\texttt{Misleading}\index{Misleading}\xspace}
\newcommand{\abcnotthefullstory}{\texttt{Not The Full Story}\index{Not The Full Story}\xspace}
\newcommand{\abcoverstated}{\texttt{Overstated}\index{Overstated}\xspace}
\newcommand{\abcwrong}{\texttt{Wrong}\index{Wrong}\xspace}

\newcommand{\high}{\texttt{High}\index{Low}\xspace}
\newcommand{\low}{\texttt{Low}\index{High}\xspace}

\newcommand{\three}{S$_{3}$\index{S!$_{3}$}\xspace}
\newcommand{\six}{S$_{6}$\index{S!$_{6}$}\xspace}
\newcommand{\onehundred}{S$_{100}$\index{S!$_{100}$}\xspace}

\newcommand{\experttwo}{E$_{2}$\index{E!$_{2}$}\xspace}
\newcommand{\expertthree}{E$_{3}$\index{E!$_{3}$}\xspace}
\newcommand{\expertsix}{E$_{6}$\index{E!$_{6}$}\xspace}
\newcommand{\crowdsix}{C$_{6}$\index{C!$_{6}$}\xspace}
\newcommand{\crowdthree}{C$_{3}$\index{C!$_{3}$}\xspace}
\newcommand{\crowdtwo}{C$_{2}$\index{C!$_{2}$}\xspace}

\newcommand{\completelydisagree}{\texttt{Completely Disagree (-2)}\index{Completely!Disagree}\xspace}

\newcommand{\disagree}{\texttt{Disagree (-1)}\index{Disagree}\xspace}

\newcommand{\neitheraord}{\texttt{Neither Agree Nor Disagree (0)}\index{Neither Agree Nor Disagree}\xspace}

\newcommand{\agree}{\texttt{Agree (+1)}\index{Agree}\xspace}

\newcommand{\completelyagree}{\texttt{Completely Agree (+2)}\index{Completely!Agree}\xspace}

\newcommand{\batchone}{\texttt{Batch1}\index{Batch!1}\xspace}
\newcommand{\batchtwo}{\texttt{Batch2}\index{Batch!2}\xspace}
\newcommand{\batchthree}{\texttt{Batch3}\index{Batch!3}\xspace}
\newcommand{\batchfour}{\texttt{Batch4}\index{Batch!4}\xspace}
\newcommand{\batchtwofromone}{\texttt{Batch2$_{\mbox{\scriptsize{from1}}}$}\index{Batch!2!$_{\mbox{\scriptsize{from1}}}$}\xspace}
\newcommand{\batchthreefromone}{\texttt{Batch3$_{\mbox{\scriptsize{from1}}}$}\index{Batch!3!$_{\mbox{\scriptsize{from1}}}$}\xspace}
\newcommand{\batchthreefromtwo}{\texttt{Batch3$_{\mbox{\scriptsize{from2}}}$}\index{Batch!3!$_{\mbox{\scriptsize{from2}}}$}\xspace}

\newcommand{\batchthreefromoneortwo}{\texttt{Batch3$_{\mbox{\scriptsize{from1or2}}}$}\index{Batch!3!$_{\mbox{\scriptsize{from1or2}}}$}\xspace}
\newcommand{\batchfourfromone}{\texttt{Batch4$_{\mbox{\scriptsize{from1}}}$}\xspace\index{Batch!4!$_{\mbox{\scriptsize{from1}}}$}}
\newcommand{\batchfourfromtwo}{\texttt{Batch4$_{\mbox{\scriptsize{from2}}}$}\xspace\index{Batch!4!$_{\mbox{\scriptsize{from2}}}$}}
\newcommand{\batchfourfromthree}{\texttt{Batch4$_{\mbox{\scriptsize{from3}}}$}\xspace\index{Batch!4!$_{\mbox{\scriptsize{from3}}}$}}
\newcommand{\batchfourfromoneortwoorthree}{\texttt{Batch4$_{\mbox{\scriptsize{from1or2or3}}}$}\xspace\index{Batch!4!$_{\mbox{\scriptsize{from1or2or3}}}$}}
\newcommand{\batchall}{\texttt{Batch$_{\mbox{\scriptsize{all}}}$}\xspace\index{Batch!$_{\mbox{\scriptsize{all}}}$}}
\newcommand{\statement}[1]{S$_{#1}$\index{S!$_{#1}$}}

\newcommand{\cem}{\texttt{CEM}$^{\mbox{\tiny{ORD}}}$\index{CEM$^{\mbox{\tiny{ORD}}}$}\xspace}

\newcommand{\m}{\texttt{m}\index{m}\xspace}
\newcommand{\meanerr}{\texttt{MAE}\index{MAE}\xspace}
\newcommand{\exterr}{\texttt{eE}\index{eE}\xspace}
\newcommand{\extmeanerr}{\texttt{eME}\index{eME}\xspace}
\newcommand{\extabserr}{\texttt{eAE}\index{eAE}\xspace}
\newcommand{\extmeanabserr}{\texttt{eMAE}\index{eMAE}\xspace}
\newcommand{\interr}{\texttt{iE}\index{iE}\xspace}
\newcommand{\intmeanerr}{\texttt{iME}\index{iMe}\xspace}

\newcommand{\negativegt}{\texttt{Neg\-a\-tive}\index{Negative}\xspace}
\newcommand{\neutralgt}{\texttt{Neu\-tral}\index{Neutral}\xspace}
\newcommand{\positivegt}{\texttt{Pos\-i\-tive}\index{Positive}\xspace}
\newcommand{\veryconservative}{\texttt{Ve\-ry Con\-ser\-va\-ti\-ve}\index{Very!Conservative}\xspace}
\newcommand{\conservative}{\texttt{Con\-ser\-va\-ti\-ve}\index{Conservative}\xspace}
\newcommand{\republican}{\texttt{Re\-pub\-li\-can}\index{Republican}\xspace}
\newcommand{\democratic}{\texttt{Dem\-o\-crat}\index{Democrat}\xspace}
\newcommand{\liberal}{\texttt{Lib\-er\-al}\index{Liberal}\xspace}
\newcommand{\veryliberal}{\texttt{Ve\-ry Lib\-er\-al}\index{Very!Liberal}\xspace}
\newcommand{\moderate}{\texttt{Mod\-er\-ate}\index{Moderate}\xspace}
\newcommand{\labor}{\texttt{La\-bor}\index{Labor}\xspace}
\newcommand{\independent}{\texttt{In\-de\-pend\-ent}\index{Independent}\xspace}

\newcommand{\hyphoonea}{Hy\-poth\-e\-sis 1a (\texttt{H1a})\index{Hypothesis!1a}\xspace}
\newcommand{\hyphooneb}{Hy\-poth\-e\-sis 1b (\texttt{H1b})\index{Hypothesis!1b}\xspace}
\newcommand{\hyphoonec}{Hy\-poth\-e\-sis 1c (\texttt{H1c})\index{Hypothesis!1c}\xspace}
\newcommand{\hyphotwoa}{Hy\-poth\-e\-sis 2a (\texttt{H2a})\index{Hypothesis!2a}\xspace}
\newcommand{\hyphotwob}{Hy\-poth\-e\-sis 2b (\texttt{H2b})\index{Hypothesis!2b}\xspace}
\newcommand{\hyphotwoc}{Hy\-poth\-e\-sis 2c (\texttt{H2c})\index{Hypothesis!2c}\xspace}
\newcommand{\hyphotwod}{Hy\-poth\-e\-sis 2d (\texttt{H2d})\index{Hypothesis!2d}\xspace}

\newcommand{\hyphooneashort}{\texttt{H1a}\index{Hypothesis!1a}\xspace}
\newcommand{\hyphoonebshort}{\texttt{H1b}\index{Hypothesis!1b}\xspace}
\newcommand{\hyphoonecshort}{\texttt{H1c}\index{Hypothesis!1c}\xspace}
\newcommand{\hyphotwoashort}{\texttt{H2a}\index{Hypothesis!2a}\xspace}
\newcommand{\hyphotwobshort}{\texttt{H2b}\index{Hypothesis!2b}\xspace}
\newcommand{\hyphotwocshort}{\texttt{H2c}\index{Hypothesis!2c}\xspace}
\newcommand{\hyphotwodshort}{\texttt{H2d}\index{Hypothesis!2d}\xspace}

\newcommand{\beliefscience}{\texttt{Be\-lief In Sci\-ence}\index{Belief!In Science}\xspace}
\newcommand{\cognitivereflection}{\texttt{Cog\-ni\-ti\-ve Rea\-son\-ing A\-bi\-li\-ti\-es}\index{Cognitive!Reasoning Abilities}\xspace}
\newcommand{\meanconfidence}{\texttt{Mean Con\-fi\-dence}\index{Mean Confidence}\xspace}
\newcommand{\trustpolitics}{\texttt{Trust In Pol\-i\-tics}\index{Trust In Politics}\xspace}
\newcommand{\affectspeaker}{\texttt{Af\-fect For State\-ment Speak\-er}\index{Affect For Statement Speaker}\xspace}
\newcommand{\politicalpartyaffiliation}{\texttt{Po\-li\-ti\-cal Par\-ty Af\-fil\-i\-a\-tion}\index{Political Party Affiliation}\xspace}
\newcommand{\statementsupport}{\texttt{State\-ment Sup\-port}\index{Statement Support}\xspace}

\newcommand{\covid}{COVID-19\index{COVID-19}\xspace}

\newcommand{\correctness}{\texttt{Cor\-rect\-ness}\index{Correctness}\xspace}
\newcommand{\neutrality}{\texttt{Neu\-tra\-lit\-y}\index{Neutrality}\xspace}
\newcommand{\comprehensibility}{\texttt{Com\-pre\-hen\-si\-bi\-li\-ty}\index{Comprehensibility}\xspace}
\newcommand{\precision}{\texttt{Pre\-ci\-si\-on}\index{Precision}\xspace}
\newcommand{\completeness}{\texttt{Com\-ple\-ten\-ess}\index{Completeness}\xspace}
\newcommand{\speakertrustworthiness}{\texttt{Speak\-er’s \-Trust\-wor\-thi\-ness}\index{Speaker's Trustworthiness}\xspace}
\newcommand{\informativeness}{\texttt{In\-form\-a\-tive\-ness}\index{Informativeness}\xspace}
\newcommand{\overalltruthfulness}{\texttt{O\-ver\-all Truth\-ful\-ness}\index{Overall Truthfulness}\xspace}
\newcommand{\confidence}{\texttt{Con\-fi\-den\-ce}\index{Confidence}\xspace}
\newcommand{\polarity}{\texttt{Po\-la\-ri\-ty}\index{Polarity}\xspace}

\newcommand{\prisma}{\texttt{PRISMA}\index{PRISMA}\xspace}
\newcommand{\numbias}{221\xspace}
\newcommand{\numbiasused}{39\xspace}
\newcommand{\numbiaswithscenario}{23\xspace}
\newcommand{\numbiaswithnoscenario}{16\xspace}
\newcommand{\numcountermeasures}{11\xspace}
\newcommand{\mybiasname}[2]{\label{#1} \textsf{#2}\index{#2}}
\newcommand{\mybiasnamecustomindex}[3]{\label{#1} \textsf{#2}\index{#3}}
\newcommand{\mycitebias}[2]{\textsf{#2}\,\textsf{#1}\index{#1}}

\newcommand{\mycitebiashyphennoindex}[4]{\textsf{#3}\,\textsf{#1-#2}\index{#4}}
\newcommand{\mycitebiasnoindex}[3]{\textsf{#2}\,\textsf{#1}\index{#3}}
\newcommand{\mycountermeasurename}[2]{\label{#1} \textit{#2}}
\newcommand{\mycitecountermeasure}[2]{\textit{#2}\,\textit{#1}}

\newcommand{\transformer}{\texttt{Transformer}\index{Transformer}\xspace}
\newcommand{\bert}{\texttt{BERT}\index{BERT}\xspace}
\newcommand{\bart}{\texttt{BART}\index{BART}\xspace}
\newcommand{\bartlarge}{\texttt{BART-Large}\index{BART!Large}\xspace}
\newcommand{\bartseparate}{\texttt{Separate-Bart}\index{BART!Separate}\xspace}
\newcommand{\ebart}{\texttt{E-BART}\index{E-BART}\xspace}
\newcommand{\ebartsmall}{\texttt{E-BARTSmall}\index{E-BART!Small}\xspace}
\newcommand{\ebartfull}{\texttt{E-BARTFull}\index{E-BART!Full}\xspace}
\newcommand{\efeversmall}{\texttt{e-FEVERSmall}\index{e-FEVER!Small}\xspace}
\newcommand{\efeverfull}{\texttt{e-FEVERFull}\index{e-FEVER!Full}\xspace}
\newcommand{\fever}{\texttt{FEVER}\index{FEVER}\xspace}
\newcommand{\efever}{\texttt{e-FEVER}\index{e-FEVER}\xspace}
\newcommand{\snli}{\texttt{SNLI}\index{SNLI}\xspace}
\newcommand{\esnli}{\texttt{e-SNLI}\index{e-SNLI}\xspace}
\newcommand{\domlin}{\texttt{DOMLIN}\index{DOMLIN}\xspace}
\newcommand{\bertbased}{\texttt{BERT-BASED}\index{BERT!BASED}\xspace}
\newcommand{\uclmr}{\texttt{UCL MR}\index{UCL MR}\xspace}
\newcommand{\unc}{\texttt{UNC}\index{UNC}\xspace}
\newcommand{\ukpathene}{\texttt{UKP-Athene}\index{UKP-Athene}\xspace}
\newcommand{\camtl}{\texttt{CA-MTL}\index{CA-MTL}\xspace}
\newcommand{\sembert}{\texttt{SemBERT}\index{BERT!Sem}\xspace}
\newcommand{\mtdnn}{\texttt{MT-DNN}\index{MT-DNN}\xspace}
\newcommand{\sjrc}{\texttt{SJRC}\index{SJRC}\xspace}
\newcommand{\dcrcoan}{\texttt{D-CRCo-AN}\index{D-CRCo-AN}\xspace}
\newcommand{\lmtrans}{\texttt{LMTransformer}\index{LMTransformer}\xspace}
\newcommand{\nulltext}{\texttt{null}\index{null}\xspace}
\newcommand{\notenoughinfo}{\texttt{NOT ENOUGH INFO}\index{NOT ENOUGH INFO}\xspace}
\newcommand{\jointpredictionhead}{\texttt{Joint Prediction Head}\index{Joint Prediction Head}\xspace}
\newcommand{\supports}{\texttt{SUPPORTS}\index{SUPPORTS}\xspace}
\newcommand{\refutes}{\texttt{REFUTES}\index{REFUTES}\xspace}
\newcommand{\bleu}{\texttt{BLEU}\index{BLEU}\xspace}
\newcommand{\rouge}{\texttt{ROUGE}\index{Rouge}\xspace}
\newcommand{\entailment}{\texttt{Entailment}\index{Entailment}\xspace}
\newcommand{\neutral}{\texttt{Neutral}\index{Neutral}\xspace}
\newcommand{\barttrue}{\texttt{True}\index{True}\xspace}
\newcommand{\bartfalse}{\texttt{False}\index{True}\xspace}
\newcommand{\taskone}{\texttt{Task 1}\index{Task!1}\xspace}
\newcommand{\tasktwo}{\texttt{Task 2}\index{Task!2}\xspace}
\newcommand{\taskthree}{\texttt{Task 3}\index{Task!3}\xspace}
\newcommand{\taskfour}{\texttt{Task 4}\index{Task!4}\xspace}
\newcommand{\esnlineutral}{\texttt{neutral}\index{Neutral}\xspace}
\newcommand{\esnlientail}{\texttt{entail}\index{Entailment}\xspace}
\newcommand{\esnlicontradiction}{\texttt{contradiction}\index{Contradiction}\xspace}

\newcommand{\workermotivation}{\texttt{worker\_mo\-ti\-va\-tion}\index{worker!\_motivation}\xspace}
\newcommand{\taskeasiness}{\texttt{task\_eas\-i\-ness}\index{task!\_easiness}\xspace}
\newcommand{\taskpayment}{\texttt{task\_pay\-ment}\index{task!\_payment}\xspace}
\newcommand{\taskfeatures}{\texttt{task\_feat\-ures}\index{task!\_features}\xspace}
\newcommand{\taskinterest}{\texttt{task\_inte\-re\-st}\index{task!\_interest}\xspace}
\newcommand{\workerfeatures}{\texttt{worker\_feat\-ures}\index{worker!\_features}\xspace}
\newcommand{\requesterfeatures}{\texttt{requester\_feat\-ures}\index{requester!\_features}\xspace}
\newcommand{\requesterreliability}{\texttt{requester\_re\-lia\-bi\-li\-ty}\index{requester!\_reliability}\xspace}
\newcommand{\lsfeatures}{\texttt{ls\_feat\-ures}\index{ls!\_features}\xspace}
\newcommand{\lsprogress}{\texttt{ls\_pro\-gress}\index{ls!\_progress}\xspace}
\newcommand{\platformfeatures}{\texttt{platform\_feat\-ures}\index{platform!\_features}\xspace}
\newcommand{\platformadequacy}{\texttt{platform\_a\-de\-qua\-cy}\index{platform!\_adequacy}\xspace}
\newcommand{\answeruseless}{\texttt{answer\_usel\-ess}\index{answer\_useless}\xspace}
\newcommand{\nosuggestion}{\texttt{no\_sug\-ges\-ti\-ons}\index{no\_suggestions}\xspace}

\newcommand{\numworkers}{300\index{300}\xspace}
\newcommand{\numexperiences}{547\index{547}\xspace}
\newcommand{\numpractices}{5\index{5}\xspace}
\newcommand{\numrecommendations}{8\index{8}\xspace}

\newcommand{\pone}{\texttt{P1}\index{P!1}\xspace}
\newcommand{\ptwo}{\texttt{P2}\index{P!2}\xspace}

\setlistdepth{9}
\newlist{myEnumerate}{enumerate}{9}

\makeindex[intoc, title=Analytical Index]

\begin{document}

\frontmatter

\maketitle

\begin{dedication}
		We can only see a short distance ahead,\\
		but we can see plenty there that needs to be done. \\
		\bigskip
		\hphantom{ }
		\emph{ -- Alan Turing}
\end{dedication}

\chapter{Abstract}

The spread of online misinformation poses serious threats to the stability of democratic societies, as the information people consume on a daily basis shapes their decision-making processes. In this context, the ability to determine which information can be trusted is critical. False or misleading content is frequently disseminated to manipulate public opinion and serve specific political agendas. Assessing the truthfulness of such information is a complex task that has traditionally been carried out by expert fact-checkers, such as journalists. These professionals investigate the sources of information, search for supporting evidence behind claims made in public discourse, and ultimately issue a judgment regarding their truthfulness. Fact-checking often involves collaboration among multiple professionals, who revise their assessments until a consensus is reached. However, the enormous volume of digital content available on the web and social media, combined with the immediacy of access and sharing, has made it increasingly difficult to verify information at scale. This challenge is further amplified by the growing popularity of social media platforms, which accelerate the dissemination of misinformation.

An alternative approach to tackling the spread of misinformation is to rely on the vast number of individuals who consume information daily. In this model, a crowd of non-expert judges takes on fact-checking, replacing the traditional role of expert professionals. While this approach opens up new opportunities for scalability and diversity of perspectives, it also introduces significant challenges. There is no assurance that non-experts can reliably identify and objectively assess online misinformation. The concept of truthfulness itself is nuanced and difficult to capture in simple terms. Statements may be imprecise, inaccurate, or outright false, and these distinctions are not easily captured by a single-dimensional truth scale. Moreover, additional factors can influence a person’s judgment of truthfulness. For example, a statement may touch on personal or sensitive topics, or it may be very recent at the time of judgment. In both cases, perception can be affected.

Judges, being human, are also prone to errors that can affect their ability to assess the truthfulness of an information item. Some of these errors stem from the limitations of human cognition and are known as cognitive biases. These biases are both frequent and consequential, shaping how people interpret misinformation as well as verified content. While they can reduce the cognitive cost of decision-making, they also interfere with the ability to make objective judgments, particularly in fact-checking scenarios. Various debiasing strategies that take cognitive factors into account have been proposed, such as supporting people's memory for misinformation. However, a systematic characterization of the specific biases that undermine fact-checking remains lacking. Identifying and describing these biases is a necessary first step, but it is not sufficient on its own. Additional effort is required to uncover their influence in practice, which would in turn support the development of effective bias mitigation techniques for existing datasets.

Machine learning approaches offer a promising solution to the scalability challenges inherent in relying only on human experts for fact-checking. Over the past years, several automated fact-checking systems have been developed to combat the spread of misinformation. Automated fact-checking systems typically use natural language processing to assess whether a statement is true given a set of evidence. This task is challenging for humans and even more so for machines. Although such systems can partially replicate evidence retrieval and synthesis, they have yet to meaningfully support or replace traditional fact-checkers due to limitations in their design. Their complexity often renders them opaque to end users, which in turn undermines trust in the system. Furthermore, cognitive biases may be embedded in the datasets used to train these models, occasionally leading to serious errors. Identifying and addressing such biases would contribute to building more reliable training resources.

This thesis addresses the challenge of misinformation spread by leveraging human intelligence across three main research directions: \emph{misinformation assessment}, \emph{cognitive biases}, and \emph{automated fact-checking systems}. To explore these themes, a series of large-scale crowdsourcing studies is conducted. Thousands of truthfulness judgments are gathered from diverse groups of non-expert individuals through micro-task platforms. The primary goal is to examine how effectively non-experts can identify and categorize (mis)information, and what factors influence their performance. While prior research has addressed human judgment, bias, and machine learning in truthfulness assessment, the role of cognitive biases remains underexplored. Understanding and mitigating such biases is important not only for improving fact-checking systems but also for broader applications involving user-generated data. To guide this investigation, three meta-research questions (MRQs) are posed, each corresponding to a thematic area, and further expanded into thirty-two specific research questions (RQs).

The findings show that non-expert workers can produce truthfulness judgments that often align with those of experts. Judgment quality is influenced by factors like timing and worker experience, with longitudinal analyses showing improvement over time. A multidimensional approach enhances interpretability and predictive modeling by capturing different facets of truthfulness. Cognitive biases are identified and empirically tested through controlled experiments. A survey of longitudinal practices offers design guidelines for sustained participation. Finally, the proposed model jointly predicts truthfulness and generates coherent, human-readable explanations, supporting the development of more transparent and trustworthy fact-checking systems.

Ultimately, this thesis contributes to the design and development of systems to combat the spread of misinformation. The research not only advances the three main topics but also provides opportunities for future exploration and expansion. This work aims to support a better understanding of how to leverage crowd-powered human intelligence to build robust, trustworthy, explainable, and transparent systems for countering misinformation, in line with the principles that fact-checking organizations strive to uphold. The full set of data collected and analyzed is publicly released to the research community at: \url{https://doi.org/10.17605/OSF.IO/JR6VC}.

\chapter{Acknowledgments}

Finding the right words to express gratitude toward the people who have allowed me to begin, pursue, and complete the PhD program in Computer Science, Mathematics and Physics at the University of Udine is both the hardest and the most rewarding task.

I would like to start by thanking my supervisor, Stefano Mizzaro, for believing in me since 2017, when I was just a master's student in Computer Science looking for a thesis project. He convinced me to embark on this journey, which began in 2019 and has proven to be immensely fruitful. His guidance has been fundamental in steering my research in the right direction.
I would also like to express my gratitude to my co-supervisor, Kevin Roitero. I am certain that without his relentless drive to pursue every goal with determination, while also motivating me and many others, I would have achieved far less. Thank you, Kevin, for being such a great friend, researcher, collaborator, and, most importantly, mentor.

I am grateful to my lab mates, Marc Donada and Mihai Horia Popescu, with whom I have shared countless coffees, work hours, and laughs. They are probably the ones who heard most of my complaints when research did not go as planned. Last but not least, thanks to David La Barbera for being one of my closest friends for 17 years and counting.

I also wish to thank the researchers I had the honor of working with over the past three and a half years. First, Gianluca Demartini from the University of Queensland and Damiano Spina from RMIT University. Although they are based in Australia and we only met in person in December 2022, we collaborated successfully, even during a global pandemic. I am also thankful to Davide Ceolin, not only for our joint work but also for giving me the opportunity to spend three months at the Centrum Wiskunde \& Informatica in Amsterdam. While I was there, I finally had the chance to meet Tim Draws from Delft University of Technology. Living in such an international and vibrant city has been a remarkable experience for both my research and future career.

\vspace{\baselineskip}
\noindent
\begin{minipage}[l]{6cm}
\flushleft
\emph{Michael Soprano}
\end{minipage}%
\hfill
\begin{minipage}[t]{6cm}
\flushright
\emph{Udine, 24 April 2023}
\end{minipage}

\newpage

\chapter*{Author Note}
\addcontentsline{toc}{chapter}{Author Note}

On 22 May 2023 I defended my PhD thesis. Because three of the included articles were still under review, the copy submitted for the defence contained pre-publication versions of those works, a compromise that never completely satisfied me. All three articles have since appeared in their final, peer-reviewed form, and, as soon as I found a bit of time, I replaced the draft texts with their camera-ready versions. I also took the opportunity to refine and update the surrounding chapters. This document is therefore the definitive version of my doctoral work.

In recognition of its contribution, this thesis was selected as the recipient of the \emph{PhD Award 2024} for the best doctoral dissertation in the scientific area at the University of Udine.

Readers who wish to jump straight to particular themes will find a suggested \emph{Reading Order} on the next page. If you build on specific results, please also cite the corresponding journal, conference, or workshop publication. Section~\ref{cap:intro-sec:publications} lists the formal citations for each chapter. All datasets and supplementary material are made available to the research community at

\begin{center}
  \url{https://osf.io/jr6vc/}
\end{center}

For the sake of transparency, the version submitted for my defence, which still contains the pre-publication articles, is available at

\begin{center}
  \url{https://air.uniud.it/handle/11390/1252385}
\end{center}

Over the past two years, I have continued my research at the University of Udine as a Postdoctoral Researcher and have published new articles, some of which already follow the future directions outlined in these pages. I would like to close by expressing my sincere gratitude to my supervisors and co-authors, whose support has been invaluable to both the research presented here and my development as a scholar.

I hope you find this work relevant and useful in your own research.

\vspace{\baselineskip}
\noindent
\begin{minipage}[l]{6cm}
\flushleft
\emph{Michael Soprano}
\end{minipage}%
\hfill
\begin{minipage}[t]{6cm}
\flushright
\emph{Udine, 11 June 2025}
\end{minipage}

\newpage

\chapter{Reading Order}

This thesis investigates how human intelligence can be leveraged in the fight against misinformation, with a particular focus on supporting fact-checking activities. It spans multiple themes, including human judgment, cognitive bias, and machine learning, all framed around the central question of how to improve the reliability, transparency, and scalability of truthfulness assessment. While the chapters are organized in a logical progression, readers interested in specific aspects of the topic may follow one of the four focused reading paths described below.

\begin{itemize}
  \item \mrqone
  \begin{itemize}
    \item Chapters~\ref{cap:paper_sigir2020}, \ref{cap:paper_pauc2021}, \ref{cap:paper_tsc2024}, and \ref{cap:paper_ipm2021} explore whether non-expert crowd workers can reliably identify and categorize true and false information. These chapters address challenges such as using effective judgment scales, assessing the influence of time and experience in longitudinal studies, and developing multidimensional judgment approaches for truthfulness assessment.
  \end{itemize}

  \item \mrqtwo
  \begin{itemize}
	\item Chapters~\ref{cap:paper_ipm2023_bias} and \ref{cap:paper_facct2022} focus on the role of cognitive biases in shaping truthfulness judgments. They present a systematic review of relevant cognitive biases, introduce a categorization framework, propose countermeasures, and analyze empirical data to detect patterns of biased behavior among crowd workers.
  \end{itemize}

  \item \mrqthree
  \begin{itemize}
    \item Chapter~\ref{cap:paper_jdiq2022} introduces a machine learning-based architecture that jointly predicts the truthfulness of a statement and generates a natural language explanation. The model is evaluated using both automatic and human-centered methods, with particular attention to explainability and transparency.
  \end{itemize}

  \item Crowdsourcing Infrastructure
  \begin{itemize}
	\item Appendix~\ref{cap:paper_wsdm2022} describes \texttt{\crowdframe}, a platform-independent system used throughout the thesis to deploy and manage crowdsourcing experiments. The system supports flexible task design and can be used by researchers for a wide range of human computation studies.
  \end{itemize}
\end{itemize}

Each of these paths offers a self-contained exploration of a major research theme. Readers may choose to follow one or more paths depending on their interests or revisit earlier chapters for complementary context.

\newpage

\tableofcontents

\newpage

\listoffigures

\newpage

\listoftables

\newpage 

\listofmyequations
\addcontentsline{toc}{chapter}{List Of Equations}

\newpage

\lstlistoflistings
\addcontentsline{toc}{chapter}{List Of Listings}

\newpage

\listofalgorithms
\addcontentsline{toc}{chapter}{List Of Algorithms}

\newpage

\mainmatter

\chapter{Introduction}

\label{cap:intro}

\section{The Rise Of Misinformation}

The rise of (online) misinformation is a problem that harms society, and the information we consume every day influences our decision-making process~\cite{visser2020reason}. Thus, understanding what information should be trusted and which should not is crucial for democratic processes to function as supposed to, since it is often done with the intended mean of deceiving people towards a certain political agenda. The sheer size of digital content on the web and social media and the ability to immediately access and share it has made it difficult to perform timely fact-checking at scale~\cite{verge-snopes}. The rate at which such a problem propagates continues to increase, largely aided by the increasing popularity of social media platforms~\cite{pennycook2021shifting}. 

The task of checking the truthfulness of published information has been traditionally performed by expert fact checkers, that is, journalists who perform the task by verifying information sources and searching for evidence that supports the claims made by the document or statement they are verifying. Indeed, it is infeasible for journalists to provide fact-checking results for all news which are being continuously published. Also, relying on fact-checking results requires trusting those who performed the fact-checking job. This is something the average web user may not be willing to accept. Even worse, fact-checking might actually decrease trust in news outlets~\cite{doi:10.1080/21670811.2022.2031240}. Significant efforts have been made by different research communities on developing techniques and datasets to automatize fact-checking, also defined as the information credibility assessment task~\cite{atanasova2019automatic, conroy2015automatic, elsayed2019overview, hansen2019neural, wang2017liar}. Key approaches to automatically differentiate between false and valid statements also include neural models~\cite{ruchansky2017csi, singhania20173han, wang2018eann}. 

The spread of misinformation is further exacerbated by events such as the \covid pandemic. The problem is (and was) so serious that the World Health Organization (WHO) used the neologism \lq\lq infodemic\rq\rq{} \index{Infodemic}to refer to the problem of misinformation, during the peak of the \covid pandemic~\cite{alam2020fighting}.
\begin{quoting}
\lq\lq We're concerned about the levels of rumours and misinformation that are hampering the response. [...] we're not just fighting an epidemic; we're fighting an infodemic. Fake news spreads faster and more easily than this virus, and is just as dangerous. That's why we're also working with search and media companies like Facebook, Google, Pinterest, Tencent, Twitter, TikTok, YouTube and others to counter the spread of rumours and misinformation. We call on all governments, companies and news organizations to work with us to sound the appropriate level of alarm, without fanning the flames of hysteria.\rq\rq{}
\end{quoting}
These are the alarming words used by Dr.~Tedros Adhanom Ghebreyesus, the WHO (World Health Organization) Director General during his speech at the Munich Security Conference on 15 February 2020.\footnote{\url{https://www.who.int/dg/speeches/detail/munich-security-conference}} Such words tell us that the WHO Director General chooses to target explicitly misinformation-related problems. Indeed, all of us have experienced mis- and dis-information during the \covid health emergency. The research community has focused on several \covid related issues \cite{bullock2020mapping}, ranging from machine learning systems aiming to classify statements and claims based on their truthfulness~\cite{wani2021evaluating}, search engines tailored to the \covid related literature, as in the TREC-COVID Challenge~\cite{TREC-COVID:JAMIA:2020}, topic-specific workshops like the NLP for COVID-19 workshop at ACL 2020~\cite{nlp-covid19-2020-nlp} and evaluation initiatives like the TREC Health Misinformation Track~\cite{DBLP:conf/trec/ClarkeRSMZ20}.\footnote{\url{https://trec-health-misinfo.github.io/}} Besides the academic research community, commercial social media platforms also have looked at this issue.

These considerations show that it is still necessary to involve humans in the fact-checking process. A more scalable and decentralized approach that relies on a (large) crowd of non-expert would allow fact-checking to be more widely available. As it is well known, \emph{crowdsourcing} means to outsource a task -- which is usually performed by a limited number of experts -- to a large mass (the \lq\lq crowd\rq\rq{}) of unknown people (the \lq\lq crowd workers\rq\rq{}), using an open call. The idea that the crowd can identify misinformation might sound implausible at first -- isn't the crowd the very means by which misinformation is spread? -- However, recent research has shown that people can reliably perform fact-checking using crowdsourcing-based approaches~\cite{la2020crowdsourcing, RSDM:2018} and assess information quality across multiple truthfulness dimensions or quality aspects~\cite{INRA:2018, tschiatschek2017detecting}, provided that adequate countermeasures and quality assurance techniques are employed. The recent works mentioned specifically crowdsource the task of misinformation identification, or rather the judgment of the truthfulness of statements made by public figures (e.g., politicians), usually on political, economical, and societal issues. Even though experts are still considered the most reliable when it comes to truthfulness judgments, leveraging them to judge and render a verdict on the truthfulness of news becomes too expensive and impractical if performed at scale. Crowdsourced fact-checking is indeed widely used in academic research~\cite{pennycook2019fighting, pinto2019towards, sethi2017crowdsourcing, shabani2018hybrid, wang2017liar} and has already found applications in industry~\cite{allen2021scaling, prollochs2021community}.

\section{The Process Of Fact-Checking}

\label{cap:intro-sec:fact-checking}

Fact-checking is a complex process that involves several activities \cite{mena2019principles, vlachos2014fact}. An abstract and general pipeline for fact-checking might include the following steps (not necessarily in this order): check-worthiness (i.e., ensure that a statement is of great interest for a possibly large audience), evidence retrieval (i.e., retrieve the evidence needed to fact-check the statement), veracity classification or truthfulness assessment, discussion among the assessors to reach a consensus, and assignment and publication of the final truthfulness score for the information item inspected. Thus, it is interesting to examine briefly the fact-checking processes adopted in practice by three famous organizations, namely \texttt{\politifact}, \texttt{\abc}, and \texttt{\factcheckorg} -- verified signatories to the International Fact-Checking Network (IFCN, \url{https://www.poynter.org/ifcn/}) -- given that they set a de-facto standard for the pipeline required to perform fact-checking at scale.

\politifact\footnote{\url{https://politifact.com/}} fact-checks information items by US Politicians. \citet{politifact-principles} details the process as follows. The reporter in charge of running the fact-checking proposes, to perform the truthfulness assessment step, a rating using a six-level ordinal scale (\politifactpantsfire, \politifactfalse, \politifactmostlyfalse, \politifacthalftrue, \politifactmostlytrue and \politifacttrue). Such assessment is reported to an editor. The reporter and the editor work together to reach a consensus on the rating proposed by adding clarifications and details if needed. Then, the statement is shown to two additional editors, which review the work of the editor and the reporter by providing answer to a set of four questions~\cite[Section \lq\lq How We Determine Truth-O-Meter Ratings\rq\rq]{politifact-principles}. The questions are:
	\begin{enumerate}
		\item Is the statement literally true?
		\item Is there another way to read the statement?
		\item Did the speaker provide evidence? Did the speaker prove the statement to be true?
		\item How have we handled similar statements in the past? What is \politifact’s jurisprudence?
	\end{enumerate}
	Then, the definitive rating of the statement is decided upon using the majority vote of the score submitted by the editors, final edits are made to ensure consistency, and the report is finally published.

\abc,\footnote{\url{https://www.abc.net.au/news/factcheck}} on the other hand, focuses on information items made by Australian public figures, advocacy groups, and institutions. Their process works as follows \cite{rmit-abc-fact-check-principles}. The statement to be checked needs to be approved by the director which assesses its check-worthiness. Then, one of the researchers at \abc contacts experts in the field and occasionally the speaker to retrieve evidence and get back data which can be helpful for fact-checking. The researcher writes the data and the information. An expert fact-checker inspects and reviews them. In this stage, the expert fact-checker identifies possible problems and questions the researcher on anything that they might have missed (e.g., missing or not exhaustive evidence retrieved). The expert fact-checker and the researcher revise the draft until the fact-checker is satisfied with the outcome; then, the whole team discusses the final verdict for the statement. The final verdict of the statement is expressed on a fine-grained categorical scale, which is used in their publications. For documentation purposes, the verdict is further refined into a three-level ordinal scale defining its truthfulness value: \abczero, \abcone, \abctwo. This choice is based on previous work demonstrating that a three-level scale may be the most suitable.

\factcheckorg\footnote{\url{https://www.factcheck.org/}} fact-checks information items dealing with US politicians. Their process works as follows \cite{factcheckorg-principles}. As for the check-worthiness step, they select statements made by the president of the United States and important politicians, focusing on those made by presidential candidates during public appearances, top senate races, and congress actions. To perform evidence retrieval, they seek through video transcripts or articles to identify possible misleading or false statement and ask the organization or the person making the statement itself to prove its truthfulness by providing supporting documentation. If no evidence is provided, \factcheckorg searches trusted sources for evidence confirming or refusing the item. Finally, the verdict about the information item is published, without assigning a fine-grained truthfulness label. At \factcheckorg, each item is revised in most cases by four people~\cite[Section \lq\lq Editing\rq\rq]{factcheckorg-principles}: a line editor (reviewing content), a copy editor (reviewing style and grammar), a fact-checker (in charge of the fact-checking process), and the director of the Annenberg Public Policy Center.\footnote{\url{https://www.annenbergpublicpolicycenter.org/}}

In summary, the fact-checking processes of the three organizations share similarities and differences, described in Table~\ref{cap:intro-sec:fact-checking-tab:fact-checking-summary}. The table reports, for each fact-checking organization considered, the information items provenance, together with the speakers considered, the truthfulness scale used, and the number of expert fact-checkers involved in the process. 

All three organizations are committed to upholding the principles of the International Fact-Checking Network (IFCN) and focus on checking statements made by politicians and public figures. However, they differ in the specific process followed for evidence retrieval, truthfulness assessment, and rating of the statements. \politifact focuses on US politicians, using a six-level rating scale and a consensus-based process among editors and reporters to determine the final rating. \abc targets Australian public figures, engaging field experts and a collaborative review process where fact-checking is performed by three experts, thus having the whole team decide the final verdict. \factcheckorg also concentrates on US politicians, seeking evidence from speakers and trusted sources, and has a four-person team to review each statement. Despite these differences, all three organizations demonstrate a strong commitment to accuracy, transparency, and thoroughness in their fact-checking processes, providing valuable resources for the public to access reliable information on political statements. However, the process of fact-checking as implemented by the organizations has the clear limitation of being not scalable: it requires experts and, being rather time-consuming, is clearly not capable of coping with the huge amount of information shared online every day \cite{das2023state, demartini2020human, spina2023human}. 

Moreover, it is worth mentioning that since all three organizations considered rely exclusively on human judgment for their evaluations, their processes are potentially susceptible to the systematic errors due to the limits of human cognition, known as cognitive biases. 

\begin{table}[tpb]
\centering
\caption{Summary statistics on the fact-checking process across selected organizations.}
\label{cap:intro-sec:fact-checking-tab:fact-checking-summary}
\begin{tabular}{p{2.6cm}lp{3.2cm}p{3.1cm}C{1.5cm}}
\toprule
\textbf{Organization} & \textbf{Country} & \textbf{Speakers} & \textbf{Truthfulness Scale} & \textbf{Experts Involved}  \\
\midrule
\politifact & USA & Politicians & \politifactzero, \politifactone, \politifacttwo, \politifactthree, \politifactfour, \politifactfive & 4 \\
\midrule
\abc & Australia & Public Figures, Advocacy Groups, Institutions & \abczero, \abcone, \abctwo & 3 \\
\midrule
\factcheckorg & USA & Politicians & / & 4 \\
\bottomrule
\end{tabular}
\end{table}

\section{The Impact Of Biases}

During an experiment that took place in 1974, \citet{Tversky:1974:Science:17835457} showed to a group of people brief personality descriptions of several individuals, allegedly sampled at random from a group of 100 professional engineers and lawyers. The subjects of the experiment were asked to assess, for each description, whether it referred to an engineer or a lawyer. The odds of any particular description belonging to an engineer rather than to a lawyer were roughly the same for each of the two groups. Subjects were split into two experimental groups; the former was told that the group of professionals from which the descriptions were drawn consisted of 70 engineers and 30 lawyers, while the latter was told the opposite (i.e., those descriptions were drawn from a group of 30 engineers and 70 lawyers). According to Bayes probabilities, the two groups should have reported unbalanced annotations while, in reality, the subjects in the two conditions reported roughly the same probability judgments, ignoring the prior probabilities of the two categories and relying only on the degree to which the description was representative of the two stereotypes. 

\citeauthor{Tversky:1974:Science:17835457} use their example to illustrate that people often rely on a limited number of heuristic principles in their cognitive processes, such as the \lq\lq judgment by representativeness\rq\rq{} detailed in the example. These heuristics, despite being useful, sometimes lead to severe and systematic errors and they are usually known as \lq\lq biases\rq\rq{}. A general definition of bias is, according to the Oxford Dictionary: 
\begin{quoting}
A strong feeling in favour of or against one group of people, or one side in an argument, is often not based on fair judgment.
\end{quoting}
People such as fact-checkers, being they experts \cite{politifact-principles, factcheckorg-principles, rmit-abc-fact-check-principles} or crowd workers \cite{demartini2020human, roitero2020crowd, roitero2020covid, SOPRANO2021102710}, can thus be subject to errors that can harm the information assessment process. Indeed, crowdsourcing often relies on contributions from large groups of laypeople with different backgrounds, expertise, and skills. Systematic errors among those workers may reduce the quality of their annotations~\cite{draws2021ChecklistCombatCognitive, eickhoff2018cognitive, hube2019understanding}. In fact-checking tasks, factors such as workers' political affiliation or their general trust in politics may affect their ability to correctly identify misinformation.

Systematic errors due to the limits of human cognition are called \lq\lq cognitive biases\rq\rq{}. There exist different types of biases: cognitive biases, conflicts of interest, statistical biases, and prejudices. Cognitive biases must be focused on because they are systematic biases due to limits in human cognition that can unintentionally affect the effectiveness of fact-checking processes. From a psychological point of view, evolutionary studies suggest that humans developed behavioral biases to minimize the cost of making mistakes in the long period, as they can improve decision-making processes \cite{johnson2013evolution}. In more detail, cognitive biases are defined by error management studies that aim at explaining human processes in decision-making \cite{johnson2013evolution} as: 
\begin{quoting}
Biases that skew our assessments away from an objective perception of information.
\end{quoting}
Such biases have been favored by nature in order to minimize whichever error caused a great cost \cite{haselton2005paranoid, nesse2005natural, nesse2001natural}. In fact, decision-making processes are often complex, and we are not always capable of keeping up to date -- and statistically correct -- the estimations of the error probabilities involved in such processes; thus, natural selection might have favored cognitive biases to simplify the overall decision process \cite{cosmides1994better, todd2000precis}. To summarize, cognitive biases evolved because of the intrinsic limitations of humans when making a decision. Cognitive biases play a major role in the way (mis)information and verified content are consumed, and different debiasing strategies have been proposed in relation to cognitive factors such as people's memory for misinformation \cite{lewandowsky2012misinformation}.

It is important to remark that biases can have far-reaching consequences. Keeping the focus on fact-checking, machine learning-based approaches are an interesting potential solution to address the obvious scalability issues of the approach based on human experts \cite{ciampaglia2015computational, 10.1145/3386253, wang2017liar, weiss2010structured}. In this respect, biases not only interfere with the human fact-checking activity in practice, but they also create issues for automatic approaches as they creep into the datasets that are then used to train the machine learning systems, in some cases contributing to blatant errors, such as the famous \lq\lq Gorilla Case\rq\rq{} \cite{ai-gorillaz}, where an image recognition algorithm misclassified black people as \lq\lq gorillas\rq\rq{}. Moreover, such biases might affect the accuracy (or even question the feasibility) of human-in-the-loop hybrid systems that try to identify misinformation at scale by combining experts, crowd, and automatic machine learning systems~\cite{demartini2020human}. Since biases introduce errors stemming from systematic limitations in human cognition that may be shared among multiple individuals, it is fundamental to focus on bias prevention, management, and control. Additionally, these efforts should enable the implementation of bias mitigation methods for existing datasets.

\section{Automated Fact-Checking}

\label{cap:intro-sec:automated-fact-checking}


Automated fact-checking (AFC) uses natural language processing (NLP) techniques to determine the truthfulness of a claim. The problem is defined in the following way: given a statement (claim) and some evidence, determine whether the statement is true with respect to the evidence \cite{stammbach2020fever}. The automated approaches previously hinted at can fall under this category. This is a challenging task for a human, let alone an autonomous system \cite{graves2018understanding}. However, AFC systems can approximate this process of evidence retrieval and synthesis with some degree of success \cite{stammbach2020fever, vlachos2014fact}. The benefits and applications of an AFC system are numerous. Indeed, they are starting to become a critical tool in combating the sheer quantity of claims that need to be verified.

AFC systems have been unable to supplement traditional fact-checkers due to a limitation in their design, even though they are accurate \cite{portelli2020distilling, stammbach2020fever}. A user may not accept to believe in a statement without first understanding the concepts and facts underpinning that statement. Such justifications are often expected when reading journalistic fact-checking outcomes such as on \politifact. The fact-check outcome is accompanied by an explanation informing the reader of how the decision was reached. Without providing users with an explanation, the decision provided by an automated system is far less likely to be trusted \cite{toreini2020relationship}, especially as it is not generated by humans. Automated systems have recently been developed to this effect, and have demonstrated promising initial results \cite{graves2018understanding}. While these initial results are unquestionably impressive, critical evaluation of the work reveals that many of these systems use separate models for veracity prediction and explanation generation. It can be argued that systems such as these are not describing their own actions and decision processes and that the truthfulness prediction model is not made any more transparent.

\section{The Crowdsourcing Activity Workflow}

\label{cap:intro-sec:crowdsourcing-activity}


In recent years, micro-task crowdsourcing has become a popular method for collecting human labels on a large scale. Typically, platforms host the tasks to be performed, which are then allocated to a crowd of workers using a first-come, first-served approach. Several crowdsourcing platforms, such as \mturk, have emerged to support the ever-increasing demand for crowd-powered data gathered through outsourced tasks. These platforms aim to help task requesters outsource their work to a diverse, distributed, and large workforce capable of performing them. While an increasing number of published studies show the use of \mturk \cite{10.3389/fpsyg.2017.01359}, alternatives are also in use \cite{PEER2017153}. Popular alternatives include \prolific, a platform dedicated to the scientific community where crowd workers are explicitly recruited for research tasks \cite{PALAN201822}, and \toloka, a platform primarily focused on data labeling tasks.

A task requester is an individual who wants to deploy a crowdsourcing task on a chosen platform. The workflow involves several phases. Initially, the requester sets parameters such as the number of crowd workers required, the time allowed for each worker to perform the task, and so on. Then, they design the task layout. Usually, a markup language is provided to help build the user interface. The requester must also write the logic to handle the task using client-side programming. Once the project is finalized, the task can be deployed an arbitrary number of times using a support file to vary the input data. Each instance of a task assigned to a worker is typically called a \index{HIT} (Human Intelligence Task). \citet{paolacci2010running} define a set of HITs as a batch. When a worker completes the assigned HITs, the requester can approve the submission and pay the worker or reject it. This workflow presents several difficulties across all platforms: the requester must have advanced programming skills; the user interface is built by mixing presentation and business logic; the input data-passing mechanism is often cumbersome; and the responsibility for storing the data produced by each worker lies with the requester in non-trivial experiments.

Requesters sometimes need to run studies that require a specific worker to perform new tasks over multiple days, weeks, or months, known as longitudinal studies (LS). These studies consist of a series of self-contained work units assigned to the same worker by the same requester, published regularly over time, and requiring ongoing participation. Each longitudinal study comprises a collection of sessions, each separated by a temporal delay, with each session involving a set of virtual work units allocated to workers. Running longitudinal studies on crowdsourcing platforms has gained popularity, as shown by \citet{Litman2017}, who introduced a tool for longitudinal study functions on \mturk. This trend is largely driven by the convenience and accessibility that crowdsourcing platforms provide for recruiting study participants. However, despite the growing popularity of crowdsourcing-based research over traditional lab studies~\cite{gadiraju2017crowdsourcing}, there is limited understanding of how workers perceive longitudinal studies. Factors such as what motivates or deters worker participation, why workers drop out, and how insights from worker experiences can improve these studies remain largely unexplored. Additionally, there is a need for a better understanding of how platforms can support longitudinal research to enhance worker engagement and the overall effectiveness of the studies.

\section{Meta-Research Questions}

\label{cap:research-questions}

Several researchers have previously investigated human-powered judgments, cognitive biases, and machine learning models for truthfulness. While significant efforts have been made to automate fact-checking, the involvement of humans remains essential. The influence of cognitive biases is often underestimated, yet maintaining a focus on bias in user-generated data can be valuable across multiple domains.

This thesis addresses the problem of misinformation spread by leveraging human intelligence along three main research directions: \emph{misinformation assessment}, \emph{cognitive biases}, and \emph{automated fact-checking systems}. To this end, a series of extensive crowdsourcing-based studies is conducted. Thousands of truthfulness judgments are collected by recruiting diverse crowds of non-expert judges (workers) through multiple experiments run on micro-task crowdsourcing platforms. The overarching goal is to assess the extent to which non-experts can objectively identify and categorize (mis)information, as well as the factors that influence their performance. Specifically, three meta-research questions (MRQs), aligned with the three thematic areas, are proposed and further elaborated into thirty-two detailed research questions (RQs).

\begin{enumerate}[label = \textsf{MRQ\arabic*}, leftmargin=3.2em]
    \item \label{cap:research-questions-meta:identify-misinfo}
    Are human assessors able to detect and objectively categorize online (mis)\-in\-for\-ma\-tion? Can their judgments be compared to and aligned with those of experts? What is the optimal environment for obtaining reliable results in the assessment of information truthfulness? Can a multidimensional notion of truthfulness be defined?

    \item \label{cap:research-questions-meta:impact-bias}
    What is the impact of cognitive biases on human assessors when judging information truthfulness? Can such biases be systematically detected? Are there countermeasures to mitigate their effects? Can a bias-aware judgment pipeline for fact-checking be defined?

    \item \label{cap:research-questions-meta:predict-explain}
    Can the collected truthfulness judgments be leveraged using machine learning-based approaches? Is it possible to develop a model capable of predicting the truthfulness of information while generating a natural language explanation to support the prediction? Are machine-generated explanations helpful for human assessors in improving their truthfulness evaluations?
\end{enumerate}

\section{Terminology}
\label{cap:intro-sec:terminology}

A specific set of nouns and technical terms from the field of crowdsourcing is used throughout the work. For the reader's convenience, a glossary of these terms is provided below. Some definitions are based on earlier work by \citet{howe2006rise} and \citet{paolacci2010running}, and have been further refined here.

\begin{itemize}[label=--]
    \item \textit{Crowdsourcing}: the act of an organization outsourcing a function previously performed by employees to an undefined (and generally large) group of people via an open call.
    \item \textit{Platform}: commercial micro-task marketplaces that allow individuals and businesses to outsource processes and jobs to a distributed workforce capable of completing tasks virtually.
    \item \textit{Human Intelligence Task (HIT)}: a single, self-contained virtual task assigned to and completed by an individual worker.
    \item \textit{Element}: an item that a worker evaluates, uses, or addresses within a HIT. A HIT is composed of a set of elements.
    \item \textit{Batch}: a set of multiple HITs published by a single individual (usually called a \emph{requester}).
    \item \textit{Requester}: an individual or organization that recruits workers from a platform to complete HITs in exchange for a wage (typically referred to as a \emph{reward}).
    \item \textit{Worker}: an individual who joins a crowdsourcing platform to perform and complete HITs published by requesters.
    \item \textit{Session}: the full set of HITs made available to and completed by the same group of workers within a specific time window.
   \item \textit{Interval Between Sessions}: the time that elapses between the completion of one session and the start of the next.
    \item \textit{Session Duration}: the time taken by a worker to complete a session.
    \item \textit{Longitudinal Study (LS)}: a sequence of HITs from the same requester, published periodically over time, and requiring repeated participation by the same workers. A longitudinal study consists of a series of sessions with temporal delays between them. Two related terms specific to LS are defined as follows:
    \begin{itemize}[label=--]
        \item \textit{Duration (of the LS)}: the total length of time required to complete a longitudinal study, from the beginning of the first session to the end of the last, including all intervals.
        \item \textit{Frequency (of the LS)}: the number of sessions a longitudinal study requires a worker to complete within a given timespan.
    \end{itemize}
\end{itemize}

\section{Synopsis}

\label{cap:intro-sec:synopsis}

This thesis features eleven (11) chapters and eight (8) appendices. Chapter~\ref{cap:related_work} reviews several studies related to the research questions addressed in the remainder of the work. The topics can be broadly grouped into three main categories: fact-checking and information truthfulness judgments, the impact of cognitive biases, and automated approaches for fact-checking. Crowdsourcing plays a role to varying extents in many of the reviewed studies. Chapter~\ref{cap:dataset} describes the five (5) data sources used in the experiments presented in the following chapters.

Chapter~\ref{cap:paper_sigir2020} focuses on collecting truthfulness judgments for publicly available fact-checked statements across two datasets, using three different judgment scales with varying levels of granularity. It also measures the political bias and cognitive background of the workers to evaluate their influence on the reliability of the collected data. Chapter~\ref{cap:paper_pauc2021} addresses recent (mis)information related to the \covid pandemic. A crowd of workers evaluates fact-checked statements using a six-level truthfulness scale, and the experiment is repeated over time with both novice and experienced workers. This longitudinal design provides insights into how judgment behavior and quality evolve. Chapter~\ref{cap:paper_tsc2024} investigates the challenges of running longitudinal studies on crowdsourcing platforms through a large-scale survey across several commercial platforms. It includes a detailed quantitative and qualitative analysis, offers a set of recommendations for researchers and practitioners, and outlines best practices for platform designers. Chapter~\ref{cap:paper_ipm2021} introduces a multidimensional notion of truthfulness. Workers assess fact-checked statements across seven dimensions selected from the literature.

Chapter~\ref{cap:paper_ipm2023_bias} presents a review of cognitive biases that may arise during fact-checking, conducted following the \prisma 2020 guidelines. A list of countermeasures is compiled to mitigate these biases, and a bias-aware fact-checking pipeline is introduced. Chapter~\ref{cap:paper_facct2022} explores which systematic biases may degrade the quality of crowdsourced truthfulness judgments. An exploratory analysis of the previously collected datasets is conducted, and specific hypotheses are tested through a new crowdsourcing experiment.

Chapter~\ref{cap:paper_jdiq2022} proposes a machine learning-based architecture capable of predicting the truthfulness of a statement and jointly generating a human-readable explanation. The model performs competitively with state-of-the-art methods. Calibration and validation are conducted, and a human evaluation assesses the impact of the generated explanations. Chapter~\ref{cap:conclusions} summarizes the main contributions, discusses practical implications, outlines future work, and closes the work.

Appendix~\ref{cap:paper_wsdm2022} provides a detailed description of \crowdframe, the software system used to design and run the crowdsourcing experiments.  
Appendix~\ref{cap:paper_sigir2020-appendix:quest-stat} reports the demographic questionnaire and the \index{Cognitive!Reflection Test}CRT tests used in several crowdsourcing experiments.  
Appendix~\ref{cap:paper_pauc2021:-appendix:statements} lists the statements used in the experiments and the longitudinal study.  
Appendix~\ref{cap:paper_tsc2024-appendix:survey-questions} provides the survey instrument used to study barriers in longitudinal crowdsourcing tasks.  
Appendix~\ref{cap:paper_ipm2021:-appendix:instructions} presents the instructions for the experiment on multidimensional truthfulness.  
Appendix~\ref{cap:paper_ipm2023_bias-appendix:prisma} provides the \prisma checklists used to characterize cognitive biases relevant to fact-checking.  
Appendix~\ref{cap:paper_ipm2023_bias-appendix:biases} contains the full list of cognitive biases identified in the literature.  
Appendix~\ref{cap:paper_facct2022-appendix:quest-crt} presents additional questionnaires used to evaluate the impact of a subset of cognitive biases.

\section{Publications}

\label{cap:intro-sec:publications}

The content of this thesis is based on thirteen published articles. Each article forms the foundation of a chapter, section, or appendix. The following list presents the articles in chronological order of publication.

\begin{enumerate}[leftmargin=*, label=\arabic*.]
    \item Chapter~\ref{cap:paper_sigir2020}: Kevin Roitero, \textbf{Michael Soprano}, Shaoyang Fan, Damiano Spina, Stefano Mizzaro, and Gianluca Demartini (2020). \textsf{Can The Crowd Identify Misinformation Objectively? The Effects of Judgment Scale and Assessor’s Background}. In: \emph{Proceedings of the 43st International ACM SIGIR Conference on Research and Development in Information Retrieval} (SIGIR 2020). Pages: 439–448. Xi’an, China (Virtual Event). July 25-30, 2020. Conference Ranks: GGS A++, Core A*. DOI: \href{https://doi.org/10.1145/3397271.3401112}{10.1145/3397271.3401112}. Reference Number: \cite{roitero2020crowd}. 
    
    \item Chapter~\ref{cap:paper_pauc2021}: Kevin Roitero, \textbf{Michael Soprano}, Beatrice Portelli, Massimiliano De Luise, Damiano Spina, Vincenzo Della Mea, Giuseppe Serra, Stefano Mizzaro, and Gianluca Demartini (2021). \textsf{Can The Crowd Judge Truthfulness? A Longitudinal Study On Recent Misinformation About COVID-19}. In: \emph{Personal and Ubiquitous Computing}. ISSN: 1617-4917. Journal Ranks: Journal Citation Reports (JCR) Q2 (2020), Scimago (SJR) Q1 (2021). DOI: \href{https://doi.org/10.1007/s00779-021-01604-6}{10.1007/s00779-021-01604-6}. Reference Number: \cite{roitero2021crowd}.
    \begin{itemize}
        \item[--] Journal extension of: Kevin Roitero, \textbf{Michael Soprano}, Beatrice Portelli, Massimiliano De Luise, Damiano Spina, Vincenzo Della Mea, Giuseppe Serra, Stefano Mizzaro, and Gianluca Demartini (2020). \textsf{The COVID-19 Infodemic: Can the Crowd Judge Recent Misinformation Objectively?} In \emph{Proceedings of the 29th ACM International Conference on Information and Knowledge Management} (CIKM2020). Pages: 1305–1314. Galway, Ireland (Virtual Event). October 19-23, 2020. Conference Ranks: GGS A+, Core A. DOI: \href{https://doi.org/10.1145/3340531.3412048}{10.1145/3340531.3412048}.  Reference Number: \cite{roitero2020covid}.
    \end{itemize}
    
    \item Chapter~\ref{cap:paper_ipm2021}: \textbf{Michael Soprano}, Kevin Roitero, David La Barbera, Davide Ceolin, Damiano Spina, Stefano Mizzaro, and Gianluca Demartini (2021). \textsf{The Many Dimensions of Truthfulness: Crowdsourcing Misinformation Assessments on a Multidimensional Scale}. In: \emph{Information Processing \& Management}, 58(6). Journal Ranks: Journal Citation Reports (JCR) Q1 (2021), Scimago (SJR) Q1 (2021). DOI: \href{https://doi.org/10.1016/j.ipm.20-21.102710}{10.1016/j.ipm.2021.102710}. Reference Number: \cite{SOPRANO2021102710}.
    
    \item Appendix~\ref{cap:paper_wsdm2022}: \textbf{Michael Soprano}, Kevin Roitero, Francesco Bombassei De Bona, and Stefano Mizzaro (2022). \textsf{Crowd\_Frame: A Simple and Complete Framework to Deploy Complex Crowdsourcing Tasks Off-the-Shelf}. In: \emph{Proceedings of the Fifteenth ACM International Conference on Web Search and Data Mining} (WSDM ’22). Pages: 1605–1608. Virtual Event, Arizona, USA. Conference Ranks: GGS A+, Core A*. DOI:\-\href{https://doi.org/10.1145/3488560.3502182}{10.1145/3488560.3502182}. Reference Number: \cite{10.1145/3488560.3502182}.
    
    \item Chapter~\ref{cap:paper_facct2022}: Tim Draws, David La Barbera, \textbf{Michael Soprano}, Kevin Roitero, Davide Ceolin, Checco Alessandro, and Stefano Mizzaro (2022). \textsf{The Effects of Crowd Worker Biases in Fact-Checking Tasks}. In \emph{Proceedings of the 2022 ACM Conference on Fairness, Accountability, and Transparency} (FAccT 2022). Pages: 2114–2124. DOI: \href{https://doi.org/10.1145/3531146.3534629}{10.1145/3531146.3534629}. Conference Ranks: Not Available. Reference Number: \cite{draws2022bias}.
    
    \item Chapter~\ref{cap:paper_jdiq2022}: Erik Brand, Kevin Roitero, \textbf{Michael Soprano}, Afshin Rahimi, and Gianluca Demartini (2022). \textsf{A Neural Model to Jointly Predict and Explain Truthfulness of Statements}. \emph{ACM Journal of Data and Information Quality}, 58(6). Journal Ranks: Journal Citation Reports (JCR) Q3 (2021), Scimago (SJR) Q2 (2021). DOI: \href{https://doi.org/10.1145/3546917}{10.1145/3546917}. Reference Number: \cite{brand2022neural}.
        \begin{itemize}
        \item[--] Journal extension of: Erik Brand, Kevin Roitero, \textbf{Michael Soprano}, Afshin Rahimi, and Gianluca Demartini (2021). \textsf{E-BART: Jointly Predicting and Explaining Truthfulness}. In: \emph{Proceedings of the 2021 Truth and Trust Online Conference} (TTO 2021). Editors: Isabelle Augenstein, Paolo Papotti, and Dustin Wright. Virtual Event. October 7-8, 2021 (pp. 18–27). Url: \url{https://truthandtrustonline.com/wp-content/uploads/2021/10/TTO2021_paper_16-1.pdf}. Reference Number: \cite{brand2021jointly}.
    	\end{itemize}
    	
    \item Section~\ref{cap:paper_is2022}: Davide Ceolin, Giuseppe Primiero, \textbf{Michael Soprano}, and Jan Wielemaker (2022). \textsf{Transparent Assessment of Information Quality of Online Reviews Using Formal Argumentation Theory}. In: \emph{Information Systems}. ISSN: 0306-4379. Journal Ranks: Journal Citation Reports (JCR) Q2 (2021), Scimago (SJR) Q1 (2021). DOI: \href{https://doi.org/10.1016/j.is.2022.102107}{10.1016/j.is.2022.102107}. Reference Number: \cite{CEOLIN2022102107}.
    \begin{itemize}
        \item[--] Journal extension of: Davide Ceolin, Giuseppe Primiero, \textbf{Michael Soprano}, and Jan Wielemaker (2021). \textsf{Assessing the Quality of Online Reviews Using Formal Argumentation Theory}. In \emph{Proceedings of the 20th International Conference on Web Engineering}. Editors: Marco Brambilla, Richard Chbeir, Flavius Frasincar, and Ioana Manolescu. Pages: 71–87. Springer International Publishing. Conference Ranks: GGS B-, Core B. DOI: \href{https://doi.org/10.1007/978-3-030-74296-6_6}{10.1007/978-3-030-74296-6\_6}. Reference Number: \cite{10.1007/978-3-030-74296-6_6}.
    \end{itemize}
    
    \item Section~\ref{cap:paper_tois2023}: Kevin Roitero, David La Barbera, \textbf{Michael Soprano}, Gianluca Demartini, Stefano Mizzaro, and Tetsuya Sakai (2023). \textsf{How Many Assessors Do I Need? On Statistical Power When Crowdsourcing Relevance Judgments}. In: \emph{ACM Transactions on Information Systems}. Journal Ranks: Journal Citation Reports (JCR) Q1 (2021), Scimago (SJR) Q1 (2021). DOI: \href{https://doi.org/10.1145/3546917}{10.1145/3546917}. Reference Number: \cite{roitero2023setsize}.
    
    \item Chapter~\ref{cap:paper_ipm2023_bias}: \textbf{Michael Soprano}, Kevin Roitero, Davide Ceolin, David La Barbera, Damiano Spina, Gianluca Demartini, and Stefano Mizzaro (2024). \textsf{Cognitive Biases in Fact-Checking and Their Countermeasures: A Review}. In: \emph{Information Processing \& Management}. Journal Ranks: Journal Citation Reports (JCR) Q1 (2023), Scimago (SJR) Q1 (2023). Reference Number: \cite{SOPRANO2024103672}. DOI: \href{https://doi.org/10.1016/j.ipm.2024.103672}{10.1016/j.ipm.2024.103672}.
    
    \item Chapter~\ref{cap:paper_tsc2024}: \textbf{Michael Soprano}, Kevin Roitero, Ujwal Gadiraju, Eddy Maddalena, and Gianluca Demartini (2024). \textsf{Longitudinal Loyalty: Understanding the Barriers to Running Longitudinal Studies on Crowdsourcing Platforms}. In: \emph{ACM Transactions on Social Computing}. Journal Ranks: Scimago (SJR) Q2 (2023). DOI: \href{https://doi.org/10.1145/367488}{10.1145/367488}. Reference Number: \cite{soprano2023loyalty}. 
\end{enumerate}

\section{Contributions}
\label{cap:intro-sec:contributions}

This section summarizes the main contributions of the thesis, presented in two parts. Section~\ref{cap:intro-sec:contributions-general} introduces the three main research directions, each aligned with a meta-research question (MRQ). Section~\ref{cap:intro-sec:contributions-detailed} details the findings associated with 32 specific research questions (RQs).

\subsection{General Summary}

\label{cap:intro-sec:contributions-general}

To understand if and how people can judge (mis)information~(\ref{cap:research-questions-meta:identify-misinfo}, \ref{cap:paper_sigir2020-sec:research-questions_1}--\ref{cap:paper_ipm2021-sec:research-questions_5}), judgments provided by expert fact-checkers are compared with those of multiple crowds of non-expert individuals recruited through commercial micro-task crowdsourcing platforms. Initially, judgments are collected using three different truthfulness scales, each with a different level of granularity. The political bias and cognitive background of the workers are also measured to assess their influence on the reliability of the collected data.

In a subsequent experiment, workers are asked to assess the truthfulness of statements related to the \covid pandemic, in order to analyze the effects of information that is both recent and sensitive. The task is repeated several times, involving both novice and experienced workers. This longitudinal design provides insights into how workers’ behavior and the quality of their judgments evolve over time. Although longitudinal studies are not uncommon in the literature, limited understanding exists about the factors that affect worker participation across different micro-task crowdsourcing platforms. To address this gap, a large-scale survey is conducted to investigate how such studies are currently implemented and perceived across multiple platforms.

To better capture the multidimensional nature of truthfulness, a further experiment involves a crowd of workers who assess publicly available information across seven dimensions drawn from the literature: \correctness, \neutrality, \comprehensibility, \precision, \completeness, \speakertrustworthiness, and \informativeness.

Concerning the impact of cognitive biases, this thesis systematically reviews those that affect human reasoning from a psychological perspective and identifies those that may arise in the fact-checking process. The review is conducted following the \prisma guidelines~\cite{moher2009preferred, Pagen71}. The previously collected datasets are then analyzed in an exploratory study to identify which biases occur systematically in crowdsourced truthfulness judgments. Based on the findings, specific hypotheses are formulated about the role of information- and judge-related characteristics in shaping the quality of assessments. An additional crowdsourcing experiment is carried out to test these hypotheses.

Turning to automated fact-checking systems, this thesis introduces a new model, \ebart, designed to jointly predict the truthfulness of an information item and generate a human-readable explanation. The goal is to increase trust in automated systems by integrating explanation as an inherent component rather than an after-the-fact addition. Unlike post-hoc explainability methods, this approach ensures that explanations are aligned with the model’s actual predictions.

\subsection{Detailed Findings}

\label{cap:intro-sec:contributions-detailed}

The detailed findings can be summarized as follows. Starting with human judgments of (mis)information~(\ref{cap:research-questions-meta:identify-misinfo}, \ref{cap:paper_sigir2020-sec:research-questions_1}--\ref{cap:paper_ipm2021-sec:research-questions_5}), the studies show that non-expert crowds can objectively assess and categorize the truthfulness of publicly available information. Their judgments often align with those of expert fact-checkers~(\ref{cap:paper_sigir2020-sec:research-questions_1}), including for recent or sensitive statements~(\ref{cap:paper_pauc2021-sec:research-questions_1}). Aggregating and transforming these judgments further improves quality~(\ref{cap:paper_sigir2020-sec:research-questions_2}, \ref{cap:paper_pauc2021-sec:research-questions_2}). Workers actively search for supporting evidence, consult multiple sources, and provide textual justifications whose quality correlates with their judgment accuracy~(\ref{cap:paper_sigir2020-sec:research-questions_3}, \ref{cap:paper_pauc2021-sec:research-questions_5}). Background factors such as political bias and cognitive ability influence, but do not always determine, the quality of their outputs~(\ref{cap:paper_sigir2020-sec:research-questions_4}, \ref{cap:paper_pauc2021-sec:research-questions_3}, \ref{cap:paper_pauc2021-sec:research-questions_4}, \ref{cap:paper_ipm2021-sec:research-questions_3}).

In longitudinal studies, the timing of judgment collection has a strong impact on quality. Workers recruited soon after an item is published produce more consistent judgments. Novice workers yield more stable assessments~(\ref{cap:paper_pauc2021-sec:research-questions_6}), while returning workers tend to spend more time and deliver higher-quality judgments~(\ref{cap:paper_pauc2021-sec:research-questions_7}). A detailed failure analysis identifies four common judgment errors and shows that the time elapsed since publication strongly influences quality for both novice and experienced workers~(\ref{cap:paper_pauc2021-sec:research-questions_8}). 

The multidimensional assessment confirms that crowd workers produce high-quality judgments when compared to expert standards~(\ref{cap:paper_ipm2021-sec:research-questions_1}). The selected dimensions are non-redundant and capture distinct aspects of truthfulness~(\ref{cap:paper_ipm2021-sec:research-questions_2}). These dimensions help interpret worker decisions and cannot be easily replicated through automatic generation~(\ref{cap:paper_ipm2021-sec:research-questions_4}). Moreover, signals from worker behavior can be leveraged to predict expert judgments~(\ref{cap:paper_ipm2021-sec:research-questions_5}).

Regarding longitudinal studies more broadly, a 22-question survey was conducted: 11 questions explored current worker perceptions, and 11 addressed factors influencing future participation~(\ref{cap:paper_tsc2024-sec:research-questions_1}). The findings indicate that some platforms host more experienced workers and support longer-term studies, while workers also report cross-platform participation. Session durations typically range from one day to one month, with up to two hours per session. Payment is typically partial and session-based. Workers consider monetary rewards both an incentive and a limiting factor. Despite this, most are open to continued participation and report high completion rates. They prefer daily or weekly sessions lasting about 90 minutes and are willing to allocate up to 180 minutes per day. The main benefits are improved productivity and reduced job search time, while downsides include inflexibility and long-term commitment. Workers emphasized the importance of clear communication, fair reward structures, and appropriate dropout handling mechanisms~(\ref{cap:paper_tsc2024-sec:research-questions_2}). Based on these insights, this thesis proposes \numrecommendations\ recommendations for task requesters and \numpractices\ best practices for platform designers~(\ref{cap:paper_tsc2024-sec:research-questions_3}).

Focusing again on cognitive biases~(\ref{cap:research-questions-meta:impact-bias}, \ref{cap:paper_ipm2023_bias-sec:research-questions_1}--\ref{cap:paper_facct2022-sec:research-questions_3}), the systematic review offers a comprehensive mapping of biases relevant to fact-checking. Out of \numbias\ known psychological biases, \numbiasused\ are found to be likely to manifest in fact-checking~(\ref{cap:paper_ipm2023_bias-sec:research-questions_1}) and are categorized according to an established classification scheme~(\ref{cap:paper_ipm2023_bias-sec:research-questions_2}). A total of \numcountermeasures\ countermeasures are identified to mitigate the effects of these biases, which are then mapped to a bias-aware assessment pipeline~(\ref{cap:paper_ipm2023_bias-sec:research-questions_4}). The follow-up study on bias manifestation links certain worker traits to biased behavior. For example, a stronger reported belief in science is paradoxically associated with lower accuracy~(\ref{cap:paper_facct2022-sec:research-questions_1}). The study confirms that specific biases manifest and affect different dimensions of truthfulness in distinct ways; for example, biased workers tend to overrate \neutrality\ and underrate \comprehensibility~(\ref{cap:paper_facct2022-sec:research-questions_2}, \ref{cap:paper_facct2022-sec:research-questions_3}).

Finally, in the area of explainable automated fact-checking~(\ref{cap:research-questions-meta:predict-explain}, \ref{cap:paper_jdiq2022-sec:research-questions_1}--\ref{cap:paper_jdiq2022-sec:research-questions_4}), the proposed \ebart\ model builds a \jointpredictionhead\ on top of \bart\ to jointly predict truthfulness and generate explanations~(\ref{cap:paper_jdiq2022-sec:research-questions_1}). Evaluation shows that it is competitive with state-of-the-art systems on \efever~\cite{stammbach2020fever} and \esnli~\cite{camburu2018snli}, with no significant degradation in performance due to explanation generation. Jointly generated explanations are more coherent with predictions than separate ones~(\ref{cap:paper_jdiq2022-sec:research-questions_2}). Human evaluation reveals that \ebart\ explanations increase skepticism and improve human ability to detect misinformation. They are also competitive with expert-curated explanations~(\ref{cap:paper_jdiq2022-sec:research-questions_3}). Model calibration using temperature scaling~\cite{10.5555/3305381.3305518} enhances the reliability of confidence scores~(\ref{cap:paper_jdiq2022-sec:research-questions_4}).

All crowdsourcing experiments in this thesis are implemented using a custom software system, \crowdframe~\cite{10.1145/3488560.3502182}. This system allows researchers to easily design and deploy diverse types of crowdsourcing tasks. It is freely available to the research community at: \url{https://github.com/Miccighel/Crowd_Frame}.

In summary, this thesis shows that non-expert judges can reliably assess online (mis)\-in\-formation, even across multiple dimensions of truthfulness~(\ref{cap:research-questions-meta:identify-misinfo}). It identifies \numbiasused\ cognitive biases relevant to the fact-checking process and provides preliminary empirical evidence of their impact~(\ref{cap:research-questions-meta:impact-bias}). It also introduces the \ebart\ model, which jointly predicts and explains truthfulness assessments, contributing to the development of more transparent, explainable, and trustworthy fact-checking systems~(\ref{cap:research-questions-meta:predict-explain}).

\chapter{Related Work}

\label{cap:related_work}

\section{Fact-Checking Using Crowdsourcing-Based Approaches}

\label{cap:related_work-sec:crowdsourcing-truthfulness}

%
The research community has been actively exploring automatic check-worthiness predictions. \citet{gencheva2017context} create a corpus of political debates containing fact-checked claims to train machine learning models for predicting which claims should be prioritized for fact-checking. In a related vein, \citet{vasileva2019takes} introduce a deep learning-based approach for estimating check-worthiness. \citet{atanasova2019automatic} propose a different model to detect check-worthy claims and employe a neural network to fact-check them.

Other researchers focus on describing the truthfulness of information items. \citet{kim2019homogeneity} model true and false news spread in social networks by considering their topic. \citet{vo2018rise} develop a recommender system that allows users to correct misinformation by referring to fact-checking URLs. Later, they propose a machine learning model to perform a fact-checking URL recommendation task \citet{you2019attributed}. Other researchers focus on the credibility and trust of sources of information. \citet{epstein2020will} investigated, in their survey experiment, the perceived trust of approximately 1,000 Americans in various news sites. Their results show that participants tend to place more trust in mainstream sources compared to hyper-partisan or fake news sources. \citet{bhuiyan2020investigating} collect credibility annotations related to climate change from both crowd workers and students enrolled in journalism or media programs. The study involved a comparison of these annotations with expert-provided ones, revealing discrepancies in the performance of the two groups of workers. 

Researchers also look at how to use crowdsourcing to collect reliable truthfulness labels in order to scale up and help study the manual fact-checking effort \cite{demartini2017introduction, demartini2020human,  pinto2019towards, visser2020reason}. For example, \citet{Kriplean:2014:IOF:2531602.2531677} analyze volunteer crowdsourcing when applied to fact-checking. \citet{zubiaga2014tweet} investigate using crowdsourcing the reliability of tweets in the setting of disaster management. Their results show that it is difficult for crowd workers to properly assess information truthfulness, but also that the source reliability is a good indicator of trustworthy information. Related to this, the CLEF initiative develop a Fact-Checking Lab \cite{barron2020overview, elsayed2019overview, clef2018checkthat, 10.1007/978-3-031-13643-6_29, 10.1007/978-3-030-72240-1_75} to address the issue of ranking sentences according to some fact-checking property. The SemEval-2019 Task 8 \cite{mihaylova2019semeval} requires providing truthfulness labels for factual information needs. \citet{INRA:2018} look at assessing news quality along eight different quality dimensions using crowdsourcing. \citet{pennycook2019fighting} crowdsourced news source quality labels. \citet{giachanou2020battle} introduce a tutorial on online harmful information that includes social media and fake news. \citet{ghenai2017catching} use crowdsourcing and machine learning to track misinformation on Twitter.  

Fact-checking websites collect a large number of high-quality labels generated by experts. However, each fact-checking site and dataset defines its own labels and rating system used to describe the truthfulness of the content. Therefore converging to a common rating scale becomes very important to integrate multiple datasets. \citet{vlachos2014fact} align labels from Channel 4 and \politifact to a five-level scale: \politifactfalse, \politifactmostlyfalse, \politifacthalftrue \politifactmostlytrue, and \politifacttrue. \citet{clef2018checkthat} retrieve evaluations of different articles at \url{factcheck.org} to assess claims made in American political debates. Then, they generate labels on a three-level scale: \genericfalse, \generichalftrue, and \generictrue. \citet{vosoughi2018spread} check the consistency between multiple fact-checking websites on three levels: \genericfalse, \genericmixed, and \generictrue. \citet{tchechmedjiev2019claimskg} look at rating distributions over different datasets and define a standardized rating scheme using four basic categories: \genericfalse, \genericmixed, \generictrue, and \genericother. 

Samples of statements from the \politifact dataset, originally published by \citet{wang2017liar}, are used to analyze the agreement of workers with labels provided by experts in the dataset itself. Workers are asked to provide the truthfulness of the selected statements using different fine-grained rating scales. \citet{RSDM:2018} compare two fine-grained scales: one in the $[0,100]$ range and one in the $(0,+\infty)$ range, on the basis of Magnitude Estimation \cite{moskowitz1977magnitude}. They find that both scales allow the collection of reliable truthfulness judgments that are in agreement with the ground truth. Furthermore, they show that the scale with one hundred levels leads to slightly higher agreement levels with the expert judgments. \citet{la2020crowdsourcing} ask workers to use the original scale proposed by the \politifact experts and the scale in the $[0,100]$ range on a larger sample of \politifact statements. They find that aggregated judgments have a high level of agreement with expert judgments. They also find evidence of differences in the way workers provide judgments, influenced by the sources they examine. In more detail, \citet{la2020crowdsourcing} find that the majority of workers use indeed the website to provide judgments. These works allow concluding that different datasets use different scales and that meta-analyses try to merge scales and aggregate ratings together. While no clear preferred scale has yet emerged, there seems to be a preference towards coarse-grained scales with just a few (e.g., from three to six) levels as they may be more user-friendly when labels need to be interpreted by information consumers.

Compared to previous work, this thesis provides a more detailed analysis of the factors influencing the quality of truthfulness assessments. Chapter~\ref{cap:paper_sigir2020} investigates the impact of assessors' background and the choice of judgment scale—specifically, a six-level scale, a coarser three-level scale, and a fine-grained hundred-level scale. Chapter~\ref{cap:paper_pauc2021} explores the role of information recency, focusing on recent \covid-related information items. Finally, Chapter~\ref{cap:paper_ipm2021} examines whether prompting crowd workers to evaluate truthfulness along multiple dimensions affects the quality of their judgments.

\section{The Effect of Information Recency}

\label{cap:related_work-sec:info-recency}

The number of initiatives that apply for Information Access and, more generally, Artificial Intelligence techniques to combat the \covid \index{Infodemic}infodemic has been rapidly increasing (see \citet[p.~16]{bullock2020mapping} for a survey). There is a significant effort by researchers like \citet{cinelli2020covid19}, \citet{gallotti2020assessing} and \citet{yang2020prevalence} on analyzing \covid information on social media and linking to data from external fact-checking organizations to quantify the spread of misinformation. \citet{mejova2020advertisers} analyze Facebook advertisements related to \covid, and find that around 5\% of them contain errors or misinformation. \citet{desai2020crowdsourcing} use a crowdsourcing-based methodology to collect and analyze data from patients with cancer who are affected by the \covid pandemic. \citet{10.48550/arxiv.2212.09683} propose a system based on NLP methods for human-in-the-loop fact-checking of tweets in the domain of COVID-19 treatments. The 2021 edition of the CheckThat! laboratory \cite{10.1007/978-3-030-72240-1_75} at the CLEF initiative focused on \covid related statements.

Compared to previous work, this thesis examines the effect of recent \covid-related information associated with the \index{Infodemic}infodemic on truthfulness judgments collected via crowdsourcing (Chapter~\ref{cap:paper_pauc2021}). It investigates whether the health domain influences crowd workers' ability to identify and accurately classify (mis)information. The study employs a single truthfulness scale, motivated by prior findings that the choice of scale does not significantly affect judgment quality (Chapter~\ref{cap:paper_sigir2020}). Additionally, workers are asked to provide a textual justification for their judgments. These justifications are analyzed to gain insight into workers' reasoning processes and to assess their potential for extracting useful information. Finally, a longitudinal study consisting of three crowdsourcing experiments is conducted over a four-month period, enabling the collection of further data and evidence, including responses from both returning and new workers.

\section{Crowdsourcing-Based Longitudinal Studies}

\label{cap:related_work-sec:longitudinal-studies}

The research community has looked extensively at workers' needs and experience on crowdsourcing platforms wherein workers receive monetary compensation for successfully completing a micro-task \cite{7156008}. 

\citet{wu2017confusing} examined the impact of task design choices on worker experience and performance, while \citet{10.1145/3494522} studied task assignment methods that address plurality problems. \citet{nouri2021unclear, nouri2023supporting} highlighted the importance of clear instructions and proposed computational tools to assist task requesters in designing clear tasks. \citet{irani2013turkopticon} and \citet{williams2019perpetual} investigated the impact of using tools to support crowd work, demonstrating how they introduce task switching and multitasking while improving productivity. Another approach to enhancing crowd work experience is through coaching by fellow workers, as described by \citet{chiang2018crowd}. Previous studies have suggested the concept of conversational crowdsourcing, using worker avatars and metaphors intelligently to enhance worker engagement and improve their overall experience \cite{jung2022great, qiu2021using, qiu2020improving, de2024we}.

There have been several efforts to empower crowd workers and support their work experiences to overcome challenges related to fair wages, power asymmetry, and unfair rejections that have plagued different crowdsourcing marketplaces~\cite{gadiraju2019understanding, edixhoven2021improving}. Reputation systems have been proposed to help propagate high-quality work and safeguard worker interests \cite{gaikwad2016boomerang}. Self-organization has been suggested to help crowd workers obtain stronger negotiation power with platforms and requesters~\cite{salehi2015we}. Related to their experience and earnings, \citet{hara2018data} adopted a quantitative lens to analyze earnings on crowdsourcing platforms, showing how workers are underpaid on average.  \citet{10.1093/icc/dtab022} explored the differences between the earnings of crowd workers based in Europe and the United States. \citet{whiting2019fair} proposed a method to ensure fair pay for workers on \mturk. \citet{fan2020crowdco} proposed a reward mechanism that allows workers to share these risks and rewards and achieve a standardized hourly wage equally split for all participating workers within cooperatives. \citet{10.1145/3491102.3501834} discussed the difficulties faced by low-income Indian women through a qualitative study. \citet{toxtli2021quantifying} analyzed the time spent by workers on non-rewarded activities, which further decrease hourly wages. \citet{doi:10.1080/07421222.2019.1705506} addressed both the nature of the task performed and the financial compensation from the worker's perspective. 

Other individual and social factors influence workers' attitudes and behavior. \citet{abbas2022goal} explored the goal-setting practices of crowd workers on \mturk and \prolific and highlighted the challenges that workers face. \citet{fulker2023cooperation} focused on exploring factors that lead crowd workers to cooperative efforts towards completing the task, while \citet{doi:10.1177/1035304620959750} studied justice expectations of workers involved in different types of crowdsourcing platforms, showing that they perceive injustices in four areas: planning insecurity, lack of transparency in performance evaluation, lack of clarity in task instructions, and low remuneration. 

The original definition of longitudinal study \cite{JTD5822} has been proposed in the past by researchers in the fields of psychology and medicine. \citet{10.1002/ir.102} described various types of longitudinal designs along with practical considerations on how to conduct them. \citet{Ployhart2011} proposed and answered a list of 12 questions that typically researchers must address when designing and conducting longitudinal studies. More recently, researchers ran longitudinal studies on crowdsourcing platforms, within different fields of study. 

The research community has focused from a longitudinal perspective, for instance, on (mis)information assessment. \citet{10.1145/3563359.3597396} propose a tool for the longitudinal assessment of the misinformation shared by Twitter accounts. \citet{fan2020crowdco} repeated a crowdsourcing task multiple times inviting the same group of participating workers each day for 20 days observing a sharp decline in return rates over time.

Longitudinal studies often address health-related issues and challenges. \citet{strickland2018feasibility} conducted a study on alcohol use, using a weekly survey over 18 weeks. The study involved an initial task that took 21 minutes to complete, followed by regular 2-minute follow-up tasks. High response rates (64.1\%-86.8\%) were observed across the 18 weeks. Active participation was incentivized through entry into a raffle for one of five \$50 bonuses if participants completed 14 or more weekly surveys. \citet{doi:10.1080/14459795.2017.1284250} described a study aimed at gathering data on gambling-related behaviors, tendencies, and traits. They conducted three crowdsourcing experiments and a fourth two-wave longitudinal study, which included 13.5\% of Study 1 participants and 14.8\% of Study 2 participants. This longitudinal study demonstrated acceptable test-retest reliability for the identified problem. Similarly, \citet{BROOKS2023107605} conducted a longitudinal study involving 636 young adults to investigate the gambling-related issue of loot boxes in video games. \citet{strickland2019use} provided an overview of using \mturk to conduct longitudinal studies for addiction science. They show a fourfold increase in the number of papers utilizing this platform for participant recruitment from 2014 to 2017.
\citet{GOODWIN2023209011}, on the other hand, examined the potential of Reddit as a recruitment strategy for addiction science research, arguing that it could be useful for conducting longitudinal follow-up surveys. \citet{10.1371/journal.pone.0284101} explored the relationship between domestic pets and their owners during the COVID-19 pandemic through a four-staged longitudinal study involving 4,237 workers. In a related context, \citet{DAYTON202214} investigated testing hesitancy and disclosure stigma in a four-wave study with 355 workers, while \citet{info:doi/10.2196/37004} studied COVID-19's progression characteristics and recovery patterns by collecting audio samples from 212 individuals. \citet{MUN20221234} conducted a two-year longitudinal study on 1453 adults with chronic pain, surveying them three times to explore pain severity, interference, emotional distress, and opioid misuse during the pandemic. Additionally, \citet{MUN2023106} investigated the impact of insomnia severity and evening chronotype on chronic pain in 884 adults over 21 months. They found that insomnia may be a stronger predictor of changes in pain and emotional distress. The literature review by \citet{doi:10.1080/17469899.2023.2200935} examined crowdsourcing-based approaches in ophthalmology, analyzing 17 longitudinal studies.
\citet{SCHOBER2022100718} investigated pollen allergies through a longitudinal study, analyzing approximately 25,000 crowdsourced search queries from citizens spanning 2017 to 2020.
\citet{Rajamani2023} used a longitudinal crowdsourcing approach to gather ideas and feedback for enhancing electronic health record systems, collecting 294 responses between 2019 and 2022.

Other researchers address human-related aspects while employing longitudinal-based crowdsourcing approaches. \citet{daly2015swapping} conducted three studies. The first focused on a two-month re-response rate among a US \mturk sample (n = 752; 75\%). The second study (n = 373) explored four- and eight-month re-response rates among US immigrants (56\% and 38\%, respectively). The third study examined a thirteen-month re-response rate (47\%), all involving a 23-minute task.
\citet{qiu2020towards} explored human memorability in the context of information retrieval on the web in a longitudinal study spanning 2 sessions across 7 days with at least a 3 day gap between the two sessions. The authors recruited participants from \mturk, and measured knowledge gain and long-term memorability of participants in their study.
\citet{tolmeijer2021second} investigated trust development in a house recommendation system through a \prolific study spanning three sessions within a week. Initially, 255 workers participated, with 83\% returning for the second session two days later. Of those, 96\% completed the third session, resulting in 203 participants who finished all three sessions, representing a nearly 80\% retention rate throughout the study.
\citet{Li2022lb} conducted a large-scale longitudinal study about recruitment and retention in remote research. They recruit 10,000 workers across two phases, gathering 12 weeks of daily surveys and passive smartphone data, resulting in 330,000 days (equivalent to 900 years) of observation.
\citet{wang2020understanding} introduced a two-week game with a purpose. Through longitudinal studies, they examine individuals' experiences with hedonic and social factors in early stages and expand to include hedonic, social, and usability-related factors in later stages.
\citet{leung2021crowd} surveyed 1000 \mturk workers to uncover factors influencing continued participation. Their findings highlight two main triggers: external regulation, such as monetary rewards, and workers' intrinsic motivation.
\citet{GRANT2023378} explored fairness in crowdsourcing through two theoretical lenses: organizational justice and institutional logic. They conduct a longitudinal netnographic study to understand workers' perceptions of fairness.
\citet{10.1007/978-3-031-05563-8_20} presented initial findings from recruiting qualified yet anonymous workers for hacking experiments involving defensive cyber deception. These experiments are part of a longitudinal study examining malicious cybersecurity experiments on crowdsourcing platforms~\cite{10.1145/3468920.3468942}.
\citet{10.1145/3573051.3593390} designed a crowdsourcing platform for a longitudinal study analyzing incorrect answers from 2015-2020 academic years across two mathematics courses, aiming to understand how to enhance student learning through remediation.

Sometimes, the specific (micro-task) commercial crowdsourcing platform chosen can hamper the overall worker experience. \citet{PEER2017153} showed that \mturk shows a lower population replenishment rate and tends to have more dishonest workers compared to platforms like \prolific. In a subsequent study, \citet{Peer2022-ki} highlighted \prolific's data quality across various measures relevant to behavioral research. Given the relevance of longitudinal studies to behavioral research \cite{doi:10.1080/14459795.2017.1284250, strickland2018feasibility, strickland2019use}, platform choice becomes a crucial consideration. \citet{hata2017glimpse} analyzed longitudinal crowdsourcing platform data and found that work quality remains stable over time for the same worker, suggesting that long-term work quality can be predicted after the first five tasks.
Additionally, \citet{su12083091} explored the motivations behind continued participation of crowd workers in crowd logistics platforms, confirming the importance of monetary incentives as well as workers' trust and cooperation.

Retention rates of workers vary significantly across longitudinal studies and decrease as time passes \cite{holden2013assessing, buhrmester2016amazon, shapiro2013using, MUN20221234, lanaj2014beginning}, starting from the 80\% obtained by \citet{shapiro2013using} after a week to the 56\% over an year obtained by \citet{MUN20221234}. Various studies used different reward schemes and incentives to increase retention rates, with strategies predominantly revolving around payment schemes. A common approach involves incentivizing worker retention through supplementary payments. \citet{difallah2014scaling} showed that offering a bonus to achieve a milestone is the most effective method for retaining workers up to a predefined milestone within a continuous series of tasks with no interruptions. \citet{auer2021pay} compared traditional work to crowd work in longitudinal studies regarding performance payment effects. They found no difference in performance but emphasized the importance of ethically rewarding workers due to their limited bargaining power. Pay significantly affects attrition (i.e., single task abandonment) but not retention in the second wave of longitudinal studies. 
\citet{Benbunan-Fich2023} investigated the question of whether workers who quit a study before its completion should receive monetary compensation. They propose that determining an appropriate partial payment, especially for longitudinal studies, involves complex considerations beyond simple monetary compensation.

Compared to previous work, this thesis addresses the experience of crowd workers in longitudinal studies, which require sustained engagement beyond standard micro-tasks (Chapter~\ref{cap:paper_tsc2024}). The findings complement earlier studies by providing guidelines and recommendations for task designers and requesters on how to structure tasks and effectively engage workers in longitudinal crowdsourcing settings.

\section{The Multidimensionality Of Truthfulness}

\label{cap:related_work-sec:multidimensionality}

The research community looks at how assessors perform judgments when using multiple dimensions and at comparing experts and non-experts. Multidimensional scales proved to be effective in the setting of information retrieval when dealing with relevance. \citet{BARRY1998219} and \citet{10.5555/1133031.1133039} list the different relevance criteria used to perform relevance evaluation. \citet{10.1145/2600428.2609577} extend the psychometric framework for multidimensional relevance proposed by \citet{10.1007/s10791-012-9206-z} by using crowdsourcing, detailing its limitations, and describing various quality control methods derived from psychometrics which can be applied to the information retrieval context. \citet{jiang2017comparing} collect multidimensional relevance along with contextual feedback from users and correlate their judgments with user metrics. Furthermore, they investigate two variants of TREC-style relevance judgments used in information retrieval and study contextual judgments by collecting multidimensional judgments using four different dimensions. \citet{uprety2020quantum} define multidimensional relevance using a quantum-inspired structure.  

Given the amount of research done and the demonstrated effectiveness of multidimensional relevance judgments, it seems natural to try and apply the same approach to truthfulness judgments. There is indeed some preliminary work in this direction. \citet{ceolin2016capturing} collect multidimensional truthfulness judgments on web documents dealing with vaccines, where few experts provided the assessments. Their results showed that experts manifest a high level of agreement, but also that the task is very demanding, and that the availability of experts online is rather limited. \citet{INRA:2018} extend the work by \citet{ceolin2016capturing} by comparing crowd and expert truthfulness assessment for a small dataset of 20 selected documents dealing with vaccines. Results show that experts are inclined to use lower values than crowd workers (i.e., they are more critical) and that the agreement between crowd and experts is high, but not total.

This thesis, as compared to previous work, describes the collection of a large number of truthfulness judgments using a multidimensional scale inspired by the literature, thus making it available to the research community (Chapter~\ref{cap:paper_ipm2021}).

\section{Cognitive Biases, Echo Chambers, And Filter Bubbles In User Generated Data}

\label{cap:related_work-sec:worker-bias}

The activities of the fact-checking process are driven by humans, either through direct human assessments in the truthfulness assessment activity or by using human-labeled data to train machine learning models. Thus, fact-checking is prone to various biases, including cognitive biases. According to the literature, more than 180 cognitive biases exist \cite{caverni1990cognitive,  haselton2015evolution, hilbert2012toward, kahneman2002representativeness}. Even if a standard conceptualization or classification of such biases is a debated problem \cite{gigerenzer2008bounded, hilbert2012toward}, many researchers confirm their presence in many domains using reproducible studies \cite{thomas2018two}, for example in information seeking and retrieval contexts \cite{azzopardi2021cognitive}. Furthermore, biases are often classified by their generative mechanism \cite{hilbert2012toward}. \citet{doi:10.1177/17456916221148147}, for instance, argue that multiple biases can be generated by a given fundamental belief. Research also agreed that multiple biases can occur at the same time \cite{maccoun1998biases, nickerson1998confirmation}.

Particularly, the effect of cognitive biases has been extensively studied across multiple disciplines. For instance, \citet{ehrlinger2016decision} and several other researchers \cite{barnes1984cognitive, das1999cognitive, hilbert2012toward, swets2000psychological} study their effect in decision processes and planning, while \citet{fisher2000cognitive} focus on market forecasting. \citet{doi:10.1177/23197145211057343} examine the influence of financial literacy, gender, annual family income, and the personality trait of neuroticism on the probability of investors experiencing selected cognitive biases. Furthermore, \citeauthor{baeza2018bias} presents surveys on potential effects derived from search and recommendation systems \cite{baeza2020bias} and biases on the web in general~\cite{baeza2018bias}. \citet{chou2020we} study the role of cognitive biases in social media platforms. \citet{otterbacher2017competent} show that human bias and stereotypes are reflected in search engine results, while \citet{yue2010beyond} investigate presentation bias in click-through data generated by a search engine. \citet{10.1145/3469595.3469615} study biases related to presentation format using conversational interfaces in the context of systems for argument search. Several researches \cite{draws2021ordered, epstein2015search, pogacar2017positive, rieger2021item, white2014belief} investigate the role of commonly occurring cognitive biases in a web search on debated topics. 

Many researchers focus specifically on issues related to bias management in user-ge\-ne\-ra\-ted and crowdsourced data.  \citet{love1981comparison} studies different user biases in peer assessment methods. \citet{chandar2018estimating} estimate click-through bias in cascade models for information retrieval.  \citet{eickhoff2018cognitive} shows, in the context of crowdsourced relevance judgments, how common types of bias can impact the judgments collected and the results of information retrieval evaluation initiatives. \citet{yildirim2013user} and \citet{lee2012s} study bias in user-generated data dealing with news media, while  \citet{muchnik2013social} focus on social influence bias. \citet{chou2020we} study the role of cognitive biases in social media platforms. \citet{hube2019understanding} analyze the effect of workers' opinions in subjective tasks. There is also evidence of differences in the way workers provide judgments, influenced by the impact of worker bias. \citet{la2020crowdsourcing} find that political background has an impact on how workers provide truthfulness judgments. In more detail, workers are more tolerant and moderate when judging statements from their very own political party.  \citet{draws2021ChecklistCombatCognitive} create a checklist to cope with common cognitive biases.  Pennycook et al.  \cite{pennycook2019fighting, pennycook2019lazy, pennycook2018falls} evaluate the ability of humans in identifying true and false news and find a positive with cognitive skills usually measured using cognitive reflection tests \cite{Frederick2005}.  \citet{limetal2020annotating} propose a news bias dataset to facilitate the development and evaluation of approaches for debiasing news articles. The biases present in datasets made using user-generated data may also impact machine learning models \cite{doi:10.1126/science.aal4230}.

Other researchers study the role of specific cognitive biases in relation to the misinformation topic. \citet{Park_Park_Kang_2021} discuss the fact-checking activity, asserting that its effectiveness varies due to multiple factors. They illustrate how statements that are neither entirely false nor true, often resulting in borderline judgments, can manifest unexpected cognitive biases in human perception. \citet{10.1145/3395046} survey and evaluate methods used to detect misinformation from four perspectives. They argue that people's trust in fake news can be built when the fake news confirms one’s preexisting political biases (i.e., particular cognitive biases), thus providing resources to evaluate the partisanship of news publishers. 
\citet{Mastroianni2023} show, in a recent study, how biased exposure to information and biased memory for information makes people believe that morality is declining for decades. \citet{zollo2019dealing} studies how information spreads across communities on Facebook, focusing on echo chambers and confirmation bias. They provide empirical evidence of echo chambers and filter bubbles, showing that confirmation bias plays a crucial role in content selection and diffusion \cite{CINELLI2022113819, doi:10.1073/pnas.2023301118, cinelli2020covid, doi:10.1073/pnas.1517441113, zollo2018misinformation}.
\citet{wesslen2019investigating} explore the role of visual anchors in the decision-making process related to Twitter misinformation. They find that these visual anchors significantly impact users in terms of activity, speed, confidence, and accuracy.
\citet{karduni2018can} focus on uncertainty in truthfulness assessment when using visual analysis, while \citet{acerbi2019cognitive} analyzes a cognitive attraction phenomenon in online misinformation, identifying a set of cognitive features that contribute to the spread of misinformation. \citet{doi:10.1177/1090198120980675} investigate factors such as biases driving misinformation sharing and acceptance in the context of \covid. 
\citet{TRABERG2022111269} study perceived source credibility to mitigate the effect of political bias. \citet{zhou2022confirmation} consider confirmation bias on misinformation related to the topic of climate change. \citet{ceci2020psychology} propose \lq\lq adversarial fact-checking\rq\rq, wherein fact-checkers from different sociopolitical backgrounds are paired as a mechanism to address biases that may arise when verifying political statements. 

While the previously cited works focus on specific aspects, there are literature reviews that relate to the issues of cognitive biases and the overall problem of misinformation spreading.  \citet{RUFFO2023100531} address the most important psychological effects that provide provisional explanations for reported empirical observations regarding the mechanisms behind the spread of misinformation on social networks. \citet{WANG2019112552} reviews works specifically focused on the spread of health-related misinformation, explicitly choosing to avoid the extensive literature related to cognitive biases. \citet{tucker2018social} reviews research findings related to cognitive biases in political discourse, with a focus on the detection of computational propaganda. More generally, the literature includes recent reviews that address cognitive biases in various fields of study unrelated to misinformation and fact-checking. \citet{10.1093/bjs/znad004} specifically explores biases that may impact surgical events and discusses mitigation strategies used to reduce their effects. Similarly, \citet{doi:10.1080/09638237.2020.1766000} investigates biases affecting military personnel. \citet{Eberhard2023} reviews strategies to mitigate the effects of cognitive biases resulting from visualization strategies on judgment and decision-making. Additionally, \citet{doi:10.1177/01655515211001777} collect data on cognitive biases in information retrieval.

The spread of information through social media and the Web, in general, has been widely studied, leading to the discovery of numerous phenomena that were not as evident in the pre-Web world, differently from human biases. Among those, echo chambers and epistemic bubbles seem to be central concepts \cite{nguyen_2020, pariser2011filter}. \citet{doi:10.1177/2158244019832705} investigate the extent of ideological echo chambers on social media using well-known media organizations and political actors as anchors. \citet{flaxman2016filter} examine the browsing history of US-based users who read news articles. They find that both search engines and social networks increase the ideological distance between individuals and that they increase the exposure of the user to the material of opposing political views.  These effects can be exploited to spread misinformation. \citet{pmid30235239} models how echo chambers contribute to the virality of misinformation, by providing an initial environment1 in which misinformation is propagated up to some level that makes it easier to expand outside the echo chamber. This helps to explain why clusters, usually known to restrain the diffusion of information, become central enablers of spread. 

On the other side, acting against misinformation seems not to be an easy task, at least due to the backfire effect. It is the effect for which someone's belief hardens when confronted with evidence opposite to its opinion. \citet{8456379} study the backfire effect and presented a collaborative framework aimed at fighting it by making the user understand her/his emotions and biases. However, the paper does not discuss the ways techniques for recognizing misinformation can be effectively translated into actions for fighting it in practice.

Compared to previous work, this thesis investigates the role of cognitive biases in crowdsourced truthfulness judgments. Chapter~\ref{cap:paper_sigir2020} describes the collection of crowd assessors’ background information and cognitive bias indicators, with the aim of identifying patterns in their judgment behavior. Chapter~\ref{cap:paper_ipm2023_bias} presents a systematic review of cognitive biases that may manifest during fact-checking tasks. Chapter~\ref{cap:paper_facct2022} analyzes which systematic biases can negatively affect the quality of truthfulness assessments produced via crowdsourcing.

\section{Argument Mining For Fact-Checking}

\label{cap:related_work-sec:argument-mining}

\citet{DUNG1995321} abstract argumentation framework emerged as a central formalism in formal argumentation. Throughout the years, it has been extended by the research community and different families of argumentations frameworks exist \cite{baroni_caminada_giacomin_2011}. Such families include Preferential Argumentation Frameworks~\cite{amgoud02, amgoud11, modgil09} and Value-based Argumentation Frameworks~\cite{bench03, bench02}. \citet{dunne11} propose a specific approach represented by systems defining preferences based on weighted attacks, establishing that some inconsistencies are tolerated in the set of arguments, provided that the sum of the weights of attacks does not exceed a given value. Weights can be used to provide a total order of attacks \cite{martinez08}. This approach can be generalized in several ways. \citeauthor{coste12} \cite{coste12, marquis12} present a different approach for relaxing the admissibility condition and strengthening the notion of defence. Furthermore, they propose different selections of extensions based on the order of weights.

Truthfulness classification and the fact-checking activity are strongly related to the scrutiny of factual information extensively studied in argumentation theory \cite{Atkinson_Baroni_Giacomin_Hunter_Prakken_Reed_Simari_Thimm_Villata_2017, lawrence-reed-2019-argument, sethi2017spotting, snaith2020modular, toulmin1958uses, visser2020reason}. Argument mining, which is the automatic identification and extraction of the structure of inference and reasoning expressed as arguments presented in natural language, is also related. \citet{lawrence-reed-2019-argument} survey the techniques used for argument mining and detail how crowdsourcing-based approaches can be used to overcome the limitations of manual analysis. \citet{sethi2017spotting} proposes a prototype social argumentation framework to curb the propagation of fake news where the argumentation structure is crowdsourced and reviewed/moderated by a set of experts in a virtual community.  \citet{sethi2019fact} develop a recommender system that makes use of argumentation and pedagogical agents to fight misinformation. \citet{snaith2020modular} presents a platform based on a modular architecture and distributed open source for argumentation and dialogue. \citet{visser2020reason} shows how to use argument mining to increase the skills of workers that assess media reports.

Compared to previous work, this thesis examines the use of an argumentation framework to enhance the quality of crowdsourced truthfulness judgments (Chapter~\ref{cap:paper_ipm2021}). In particular, it explores whether providing crowd workers with tools to analyze the argument structure of statements can improve their assessment quality.

\section{Automated Fact-Checking Using Machine Learning Techniques}

\label{cap:related_work-sec:afc-models}

The research community investigate the usage of machine learning techniques to cope with disinformation besides human-powered systems \cite{hassan2015quest, thorne2018automated}.  These techniques rely on training a machine learning algorithm on a labelled dataset which is usually built using human assessors. \citet{vlachos2014fact} define the setting and the challenges needed to create a benchmark dataset for fact-checking. \citet{ferreira2016emergent} describe a dataset for stance classification. \citet{wang2017liar} creates the \texttt{LIAR}\index{\texttt{LIAR}} dataset which contains a large collection of fact-checked statements. Several researchers focus on the algorithms which can be employed to build a fully automatic methodology to fact-check information: \citet{weiss2010structured} develop a method based on adversarial networks, while \citet{alhindi2018your} leverage justification modeling. Furthermore, \citet{ciampaglia2015computational} use an approach based on knowledge networks. \citet{reis2019explainable} and \citet{wu2020evidence} discuss explainable machine learning algorithms that can be employed for fake news detection. \citet{oeldorf2020posted} and \citet{evans2020news} consider information sources and their metadata.

Automated fact-checking aims at replacing experts, i.e., usually journalists, in performing the fact-checking process.  As an example of such methods, \citet{10.1145/3386253} propose a deep neural network model to detect misinformation statements. Their model is based on a feature extractor which works both at the textual and at the user level, an attention layer used to detect important and specific user responses, and a pooling algorithm to do feature aggregation. Their results on two datasets show that the developed model reaches an accuracy level higher than $0.9$ within 5 minutes of the spread of the misinformation statement. \citet{limetal2020annotating} use crowdsourcing to gather bias labels on news articles and propose an automatic approach for analyzing and detecting them. \citet{li2020misinformation} propose to identify possible misinformation on Twitter by learning a topic-based model from expert-provided assessment. However, fact-checking still requires manual effort, as evidenced by the approaches that exploit machine learning to build completely automatic classifiers. Such an effort is usually performed by expert fact-checkers to generate labels that can eventually lead to the training of supervised methods like the ones described. 

The research community propose various techniques for generating explanations to accompany fact-checking decisions. Saliency-based methods, such as those proposed by \citet{shu2019defend, wu2020dtca}, use attention mechanisms to highlight the input that is most useful in determining the veracity prediction and present this information to the end user as a form of explanation. Logic-based approaches make use of graphs \cite{denaux2020linked}, rule mining, and probabilistic answer set programming \cite{ahmadi2019explainable} to output a series of logical rules that result in a veracity prediction. Summarisation techniques provide an explanation by summarising the evidence retrieved. \citet{atanasova2020generating} propose a system that uses DistilBERT \cite{sanh2019distilbert} to pass contextual representations of the claim and evidence to two task-specific feed-forward networks which produce a classification and an extractive summary. 

\citet{stammbach2020fever} proposes a framework that also produces abstractive explanations but places a higher emphasis on the evidence retrieval process. The framework consists of two components. These components are an evidence retrieval and veracity prediction module, and an explanation generation module.  The first component is an enhanced version of the DOMLIN system \cite{stammbach2019team}, which uses separate BERT-based models for evidence retrieval and veracity prediction. GPT-3 \cite{brown2020language}, a large pertained multi-purpose NLP model based on the Transformer, is used in \lq\lq few-shots\rq\rq{} mode to generate a summary of the evidence with respect to the claim provided as an explanation.

BART \cite{lewis2019bart} is a transformer \cite{vaswani2017attention}  model that aims to generalise the capabilities of both BERT \cite{devlin2018bert} and GPT-style models. It consists of a bi-directional encoder, similar to BERT, as well as an auto-regressive decoder, similar to GPT. BART is pre-trained on a de-noising task whereby input text is corrupted and the model aims to reconstruct the original document, minimising the reconstruction loss. In contrast to existing de-noising models, BART is more flexible in that it is not trained to rectify a specific type of input corruption, but rather any arbitrarily corrupted document. The pre-trained BART model can be fine-tuned for a number of downstream tasks. \citet{lewis2019bart} note that BART performs comparably to other models, such as RoBERTa \cite{liu2019roberta}, on natural language inference tasks. They also note that BART outperforms current state-of-the-art models on natural language generation tasks, such as summarisation \cite{lewis2019bart, shleifer2020pre}. Its ability to perform well on these two contrasting tasks made it an attractive choice as the base model for a system that can jointly predict the truthfulness of a claim (an inference task) and provide an explanation (a generative task).

Compared to previous work, this thesis describes in Chapter~\ref{cap:paper_sigir2020} how assessors’ background information and bias indicators are collected to identify patterns in their assessment behavior. The work presented in Chapter~\ref{cap:paper_ipm2021} complements approaches that rely on manual annotations by expert fact-checkers to generate training labels for supervised methods in automated fact-checking. Chapter~\ref{cap:paper_jdiq2022} introduces an approach aimed at supporting expert fact-checkers in producing labels suitable for machine learning-based fact-checking. In contrast to prior work, which typically employs separate models for truthfulness prediction and explanation generation, this approach uses a single model to jointly produce both a truthfulness label and an abstractive justification.

\section{Supporting Crowdsourcing-Based Approaches}

\label{cap:related_work-sec:crowdsourcing-tools}

Individuals and organizations who need to gather data of some kind using crowd\-sour\-cing-based approaches may rely on different crowdsourcing platforms. These platforms help task requesters access the global human workforce available. \mturk is one of the most well-known platforms. \citet{paolacci2010running} presented, in the past, demographic data about the Mechanical Turk worker population, reviewing the strengths of Mechanical Turk relative to other online and offline methods of recruiting subjects, and comparing the magnitude of effects obtained using Mechanical Turk and traditional worker pools. More recently, \citet{10.1111/add.15032} reviews different research conducted using Mechanical Turk, provides examples and discusses the limitations and best practices of the platform. 

In the last years, researchers are arguing that the quality of data collected by recruiting workers from Mechanical Turk is declining. According to \citet{PEER2017153}, already in 2017, Mechanical Turk workers were becoming less naive. In more detail, they define workers as \lq\lq professional survey-takers\rq\rq{}. Related to this, \citet{doi:10.1177/1948550619875149} conducted before, during and after the summer of 2018 an experiment related to psychological research, finding empirical evidence of a substantial decrease in data quality. \citet{kennedy2020turkquality} show the presence of many fraudulent responders that provide low-quality data. \citet{doi:10.1177/1948550617698203} explain that a substantial number of participants misrepresent relevant characteristics to meet the eligibility criteria expected in the studies. \citet{Webb2022} explain that the data collected for their research were valid for the 2.6\% of humans recruited and claim that a call for caution is needed while using Mechanical Turk. However, other crowdsourcing platforms seem to be a viable alternative. \citet{Peer2022} compare 5 crowdsourcing platforms and panels by examining aspects of data quality for online behavioral research. They conduct two studies and only the \index{Prolific} platform \cite{PALAN201822} provides high data quality on all measures for both studies, while CloudResearch (formerly known as TurkPrime \cite{Litman2017}) only for the second study. \citet{Litman2022} analyze the claims of  \citet{Peer2022} and point out the presence of methodological decisions undisclosed in their work that limit the inference that can be drawn from the data collected. In more detail, they assert that  \citet{Peer2022} chose to turn off the recommended data quality filters while using the CloudResearch platform. They thus replicate the studies with CloudResearch using the recommended options, finding that the final data are of better quality. \citet{Litman2022} point out in their work that  \citet{Peer2022} are members of \prolific \index{Prolific}. However, \citet{Litman2022} clarify that they are part of the CloudResearch team. While other researchers suggest that using \prolific \index{Prolific} leads to some extent to results of better quality \cite{arxiv.2202.14036},  \citet{Litman2021conducting} advocates that the platform should be selected by matching the study’s goals and the platform’s strengths and weaknesses.

Several tools that aid requesters during the whole crowdsourcing activity exist. \citet{vukovic2009crowdsourcing} proposes a taxonomy for the categorization of crowdsourcing platforms and evaluates a set of systems with respect to such taxonomy. \citet{erickson2011some} proposes a conceptual framework for the design of systems to support crowdsourcing and human computation. \citet{liu2014crisis} proposes a set of best practices to develop crowdsourcing systems designed to support the articulation work needed to facilitate spontaneous volunteer effort during emergencies. \citet{clark2019framework} develops a framework to allow governments to use crowdsourcing bases approaches to solve problems when interacting with their citizens. \citet{brito2015towards} propose a conceptual framework to guide the design of gamification-based approaches within crowdsourcing platforms. \citet{li2018crowdbc} conceptualize a blockchain-based decentralized framework in which a requester can propose a task without relying on any third trusted institution. \citet{ye2018crowdsourcing} introduce a crowdsourcing framework to support the annotation of medical data sets. \citet{hamrouni2020spatial} propose a framework for spatial mobile crowdsourcing, where workers are required to be physically present at a particular location. The requesters solicit workers to provide photos of ongoing events for event reporting purposes.  CloudResearch \cite{Litman2017} is a research platform that integrates with \mturk that aims to improve the quality of the crowdsourcing data collection process. CrowdForge \cite{kittur2011crowdforge} is a general-purpose framework for accomplishing complex human computation tasks. It is based on the idea that complex work can be broken up into small and independent pieces while the system manages its coordination dependencies. iCrowd \cite{fan2015icrowd} is an adaptive crowdsourcing framework that estimates in real-time the accuracy of workers by evaluating their performances on completed tasks. It can be used before the task's launch to choose the best workers available. CrowdTruth 2.0 \cite{CrowdTruth2} is a method to aggregate crowdsourcing responses after the task using disagreement-aware metrics. It can be used to leverage workers' data processing after the task. CrowdEIM \cite{SHEN2021102024} is a tool based on mobile social media platforms which allow crowdsourcing information during emergencies.

Researchers proposed in the past approaches to model and predict user behaviour when interacting with web applications. Historically, these approaches relied heavily on Markov models, which have been widely popular for this type of task. \citet{ching2004higher} study the usage of such models to analyze categorical data sequences. The research community later focused on different challenges. \citet{borges2007evaluating} describe user web navigation sessions up to a given length. \citet{shukla2011analysis} model the behaviour of users that chooses to switch the browser used to surf the web. \citet{1250921} aims to generate probabilistic browsing behaviour for users on the web. \citet{10.1145/1644893.1644900} analyze user behaviour on social networks. \citet{10.1145/1592748.1592750} integrate user behaviour into contextual advertising. \citet{6163417}, \citet{10.1145/990301.990304}, and \citet{1046594} aim predicting web page accesses. \citet{10.1145/775047.775068} take advantage of user behaviour to personalize websites. \citet{10.2307/1391012} and \citet{REN201752} use behavioural data to build systems that can detect and prevent access to malicious users. More recent approaches involve the usage of behavioural data to represent user interaction using embeddings. \citet{10.1145/3269206.3271730}, for instance, propose a recommendation model that learns user and item attributes represented using embeddings in the context of a recommender system. Learning vector representations for texts have been studied by the research community in depth. For example, \citet{10.5555/3044805.3045025} and \citet{10.5555/2999792.2999959} analyze the usage of sentences as a better way to learn word semantics. Embedding can be helpful also when considering information retrieval tasks such as query rewriting. \citet{10.1145/2783258.2788627} propose a query rewriting method based on a query embedding algorithm. Other approaches include content advertising, as the one by \citet{10.1145/2766462.2767709}. They propose a neural language-based algorithm specifically tailored for delivering effective product recommendations. The embedding representations learnt can be used to predict and understand user behaviour across different scenarios, as done by \citet{10.1145/3269206.3272032} and \citet{10.1145/3340531.3411985}. Crowdsourcing-based approaches can provide a massive amount of behavioural data due to the availability of a large human workforce. \citet{10.1371/journal.pone.0226394} find out that there are more than 250.000 workers around the whole world whose potential is largely untapped. \citet{neil2015average} explains that a task requester can reach more than 7.000 workers each quarter year. \citet{difallah2018logging} analyze the population dynamics and demographics of \mturk workers based on the results of a survey that they conducted over 28 months. They also discover that during the day the peak of active US workers is 90\% around 11 PM UTC. \citet{han2019all} perform a data-driven analysis of logs collected during a large-scale relevance judgment experiment to study the phenomenon of crowdsourcing task abandonment. 

Compared to previous work, this thesis describes in Appendix~\ref{cap:paper_wsdm2022} a software system that supports multiple crowdsourcing platforms and enables task requesters to recruit general-purpose crowd workers. Its configuration mechanism allows complex tasks to be easily decomposed into intermediate steps without the need for additional coordination infrastructure. Furthermore, the system supports the collection of detailed behavioral data as users complete tasks across platforms.

The next chapter describes the data sources used in the experiments presented throughout the remainder of the thesis.

\chapter{Dataset}

\label{cap:dataset}

The experiments described in this thesis either use different subsets of the same dataset across experiments or rely on multiple datasets within a single experiment. It is therefore useful to describe all datasets before detailing each experimental setup. The \politifact dataset is introduced in Section~\ref{cap:dataset-sec:politifact}, and the \abc dataset in Section~\ref{cap:dataset-sec:abc}. Section~\ref{cap:dataset-sec:fever} presents the \fever dataset, while its extended version, \efever, is described in Section~\ref{cap:dataset-sec:e-fever}. Finally, Section~\ref{cap:dataset-sec:e-snli} introduces the \esnli dataset.

\section{PolitiFact}

\label{cap:dataset-sec:politifact}

The \politifact organization, established in 2007 within the Tampa Bay Times\footnote{\url{https://www.tampabay.com/}} and now operated by the Poynter Institute,\footnote{\url{https://www.poynter.org/politifact/}} hosts an online archive of more than \num{25,000} fact-checks covering claims by U.S. politicians, political organizations, public figures, and social-media users (Section \ref{cap:intro-sec:fact-checking}). The site is updated on a rolling basis, and thematic collections are created for major events such as the COVID-19 pandemic\footnote{\url{https://www.politifact.com/coronavirus/}} and the 2020 U.S. presidential election.\footnote{\url{https://www.politifact.com/2020/}} Table \ref{cap:dataset-sec:politifact-tab:statements} lists two example statements fact-checked in 2022.

Each statement receives one of six Truth-O-Meter\footnote{\url{https://www.politifact.com/truth-o-meter/}} ratings: \politifacttrue, \politifactmostlytrue, \politifacthalftrue, \politifactmostlyfalse, \politifactfalse, and \politifactpantsfire. These levels are collectively referred to in Chapter \ref{cap:paper_pauc2021} as \expertsix. The level now called \politifactmostlyfalse was labelled \texttt{Barely-True} until 2011, when it was renamed to reduce potential misinterpretation.\footnote{\url{https://www.politifact.com/article/2011/jul/27/-barely-true-mostly-false/}} Samples used in this thesis include statements published before that change; consequently, the legacy label may still appear, although its semantics remain identical. Researchers and practitioners should bear this in mind when interpreting the results.

\begin{table}[tbp]
\centering
\caption{Statements fact-checked during 2022 by \politifact.}
\label{cap:dataset-sec:politifact-tab:statements}
\begin{tabular}{p{4.1cm}p{2.2cm}p{1.5cm}p{1.8cm}p{2.3cm}}
\toprule
\textbf{Statement} & \textbf{Speaker} & \textbf{Party} & \textbf{Date} & \textbf{Ground Truth}\\
 \midrule
 The United States spends \lq\lq almost three times per capita what they spend in the U.K.\rq\rq{} on health care and \lq\lq 50 percent more than they pay in France.\rq\rq{} & Bernie Sanders & \democratic & 2022-12-19 & \politifacthalftrue \\
 \midrule
  \lq\lq There hasn't been a single of these mass shootings that have been purchased at a gun show or on the internet.\rq\rq{} & Marco Rubio & \republican & 2022-05-25 & \politifactfalse \\
\bottomrule
\end{tabular}
\end{table}

\section{RMIT ABC Fact Check}

\label{cap:dataset-sec:abc}

The \abc dataset\footnote{\url{https://apo.org.au/collection/302996/rmit-abc-fact-check}} consists of 561 verified statements covering the period 2013–2022. It is produced through a partnership between RMIT University\footnote{\url{https://www.rmit.edu.au/}} and the Australian Broadcasting Corporation,\footnote{\url{https://www.abc.net.au/}} which aims at \lq\lq combining academic excellence and the best of Australian journalism to inform the public through an independent non-partisan voice\rq\rq{}. Professional fact-checkers solicit expert opinions and gather evidence before a team assigns a verdict to each statement (Section \ref{cap:intro-sec:fact-checking}). A fine-grained truthfulness scale is employed; individual verdicts include, for example, \abccorrect, \abcchecksout, \abcmisleading, \abcnotthefullstory, \abcoverstated, and \abcwrong. These verdicts are subsequently mapped onto a three-level scale: \abcpositive, \abcinbetween, and \abcnegative. This scale serves as ground truth in the experiments reported in this thesis. Table~\ref{cap:dataset-sec:abc-tab:statements} presents two sample statements fact-checked in 2019 and published on the organization’s website.

As with the \politifact dataset, the labels \abcnegative and \abcpositive occasionally appear as \texttt{Negative}\index{Negative} and \texttt{Positive}\index{Positive}. No official rationale for this alternative notation is provided, but the underlying semantics remain unchanged. Researchers and practitioners should therefore treat the two forms as equivalent when interpreting the results presented in this thesis.

\begin{table}[tbp]
\centering
\caption{Statements fact-checked during 2019 by \abc.}
\label{cap:dataset-sec:abc-tab:statements}
\begin{tabular}{p{4cm}p{2.2cm}p{1.5cm}p{1.8cm}p{2.3cm}}
\toprule
\textbf{Statement} & \textbf{Speaker} & \textbf{Party} & \textbf{Date} & \textbf{Ground Truth}\\
 \midrule
 \lq\lq Labor has more than double the number of women the Liberals have in the Parliament and about twice the number of women on our front bench — that speaks for itself.\rq\rq{} & Tanya Plibersek & \labor & 2019-02-07 & \abcpositive \\
 \midrule
  \lq\lq Labor's proposal is to dismantle offshore detention and will essentially give the ability for two doctors — as has been pointed out, doctors including Dr Brown, Bob Brown, and Dr Richard Di Natale — potentially can provide the advice\rq\rq{} & Peter Dutton &  \liberal & 2019-02-12 & \abcinbetween \\
\bottomrule
\end{tabular}
\end{table}

\section{FEVER}

\label{cap:dataset-sec:fever}

The \fever dataset\footnote{\url{https://fever.ai/dataset/fever.html}} \cite{thorne2018fact} consists of \num{185,445} statements, associated evidence, and truthfulness judgments. The examples for the training set includes \num{165,447} examples, while the development set includes \num{19,998}. The statements have been generated by human annotators in 2017. The annotators extract sentences from Wikipedia thus mutating them in a variety of ways, some of which are meaning-altering. They are subsequently verified without knowledge of the sentence they were derived from and labelled with either \supports, \refutes, or \notenoughinfo based on whether the evidence entails the statement. The annotators also record the sentence(s) forming the necessary evidence for their judgment, for the first two classes. The data is distributed using the \texttt{JSONL}\footnote{\url{https://jsonlines.org/}} \index{JSONL} format. Such a format enforces the usage of the \texttt{UTF-8} encoding to obtain a valid file. 

Furthermore, each line separator must be the \spverb|\n| character and, most importantly, each line must contain a valid \index{JSON}\texttt{JSON} value, such as an object or an array. Each line of the dataset thus contains a single example of a statement generated using Wikipedia. More specifically, the training and development data include four different fields. The term \lq\lq claim\rq\rq{} is used in the dataset instead of \lq\lq statement\rq\rq{}. In this thesis, the latter term is adopted to ensure consistent notation. The four fields are:
\begin{itemize}
	\item \spverb|id|: the ID of the statement;
	\item \spverb|claim|: the text of the statement;
	\item \spverb|label|: the label associated to the statement (\supports, \refutes, \notenoughinfo);
	\item \spverb|evidence|: a list of evidence sets, where each element of the list is in the form:
	\begin{itemize}
	\item \spverb|Annotation ID|, \spverb|Evidence ID|, \spverb|Wikipedia URL|, \spverb|Sentence ID|, if the label is either \supports or \refutes;
	\item \spverb|Annotation ID|, \spverb|Evidence ID|, \spverb|null|, \spverb|null|, if the label is \notenoughinfo.
	\end{itemize}
\end{itemize}
	
Table~\ref{cap:dataset-sec:fever-tab:statements} shows a sample of three statements, one for each class. Each statement has an evidence set made of a single element. For each piece of evidence, the first attribute identifies the human annotator activity, while the second is the internal identifier of the overall set. The third attribute indicates the Wikipedia page, while the fourth and last attribute identifies a sentence within the current piece of evidence. \citeauthor{thorne2018fact} provide also a dump containing all the Wikipedia pages processed.\footnote{\url{https://fever.ai/download/fever/wiki-pages.zip}}

\begin{table}[tbp]
\centering
\caption{Statements sampled from the \fever dataset.}
\label{cap:dataset-sec:fever-tab:statements}
\begin{tabular}{p{0.8cm}p{4.2cm}p{2.3cm}p{5cm}}
\toprule
\textbf{ID} & \textbf{Statement} & \textbf{Label} & \textbf{Evidence}\\
\midrule
77712 & Newfoundland and Labrador is the most linguistically homogeneous of Canada. & \supports & (94661, 107645, "Newfoundland\allowbreak_and_Labrador", 4) \\
 \midrule
73170 & Puerto Rico is not an unincorporated territory of the United States. & \refutes & (89957, 102650, "Puerto_Rico", 0) \\
 \midrule
210010 & Afghanistan is the source of the Kushan dynasty. & \notenoughinfo & (248748, null, null, null) \\
\bottomrule
\end{tabular}
\end{table}

\section{e-FEVER}

\label{cap:dataset-sec:e-fever}

The \efever dataset \cite{stammbach2020fever} augments the original \fever dataset (Section~\ref{cap:dataset-sec:fever}) with explanations generated by their framework. The underlying motivation is that in 16.82\% of cases in the \fever dataset, a statement requires the combination of more than one sentence to be able to support or refute it. Furthermore, they find that sometimes the evidence is not only conditioned by the statement but also by the evidence already retrieved. In light of this, \citeauthor{stammbach2020fever} propose a two-staged selection process based on the \lq\lq two-hop\rq\rq{} evidence enhancement process \cite{10.1609/aaai.v33i01.33016859}. The documents are retrieved by re-using the ukpathene \cite{hanselowski2018ukp} system. The component that performs the final statement verification step employs two strategies, one used by \citet{thorne2018fact}. The statements are labelled with either the \supports or \refutes label. The \notenoughinfo label, present in the original fever dataset, is mapped into one of the other labels. The document retrieval system predicts relevant pages and uses the two-staged process to select relevant evidence for these uncertain statements.

The resulting dataset consists of \num{67,687} examples in total. The examples for the training are \num{50,000}, while those for the development are \num{17,687}. The resulting dataset thus provides a resource with statements, retrieved evidence, truthfulness labels, and explanations. Table~\ref{cap:dataset-sec:efever-tab:statements} shows a sample of two statements, one for each class. Each statement is provided together with the \lq\lq gold\rq\rq{} evidence found, the label assigned and a summary written by a human assessor. The first example shows that the evidence retrieved might not suffice to fully explain the statement, according to the human annotator's opinion. The dataset is available upon request to \citet{stammbach2020fever}. It contains some \nulltext explanations due to the evidence retrieval policy.

\begin{table}[htbp]
\centering
\caption{Statements sampled from the \efever dataset.}
\label{cap:dataset-sec:efever-tab:statements}
\begin{tabular}{p{3.7cm}p{1.8cm}p{4.4cm}p{2.5cm}}
\toprule
\textbf{Statement} & \textbf{Label} & \textbf{Gold Evidence} & \textbf{Summary}\\
\midrule
The Bahamas is a state that's recognized by other states that includes a series of islands that form an archipelago. & \supports & The Bahamas, known officially as the Commonwealth of The Bahamas, is an archipelagic state within the Lucayan Archipelago. An archipelagic state is any internationally recognized state or country that comprises a series of islands that form an archipelago. & The relevant information about the claim is lacking in the context.\\
\midrule
Scandinavia does not contain Greenland. & \refutes & The remote Norwegian islands of Svalbard and Jan Mayen are usually not seen as a part of Scandinavia, nor is Greenland, an overseas territory of Denmark. & Greenland is not a part of Scandinavia. \\
\bottomrule
\end{tabular}
\end{table}

\section{e-SNLI}

\label{cap:dataset-sec:e-snli}

The \esnli dataset\footnote{\url{https://github.com/OanaMariaCamburu/e-SNLI}}~\cite{camburu2018snli} extends the \snli dataset \cite{bowman2015large} by generating human explanations for \num{543,950} out of the \num{570,152} examples of the dataset. The \snli task is to take two sentences and predict whether one entails, contradicts, or is neutral with respect to the other. The examples for the training set are \num{550,152}, while those for the development set are \num{20,000}.

\citet{camburu2018snli} collect the data for the \esnli dataset by publishing a crowdsourcing task on the \mturk platform. The main goal for the workers is to answer a question assessing why a pair of sentences stands in a relation of entailment, neutrality, or contradiction. The workers are instructed to focus on non-obvious elements that induce the given relation, rather than on parts of the premises that are repeated identically in the hypotheses. \citeauthor{camburu2018snli} recruit \num{6,325} workers, who provide an average of 86 explanations each. One explanation is collected for each sentence pair in the training set, while three explanations are collected for each pair in the validation and test sets. Each sentence pair is annotated with the following attributes:
\begin{itemize}[label=--]
	\item \spverb|pairID|: the identifier of the sentence pair;
	\item \spverb|gold_label|: the ground truth relation of the sentence pair (\esnlicontradiction, \esnlineutral, \esnlientail);
	\item \spverb|Sentence1|: the first sentence of the pair;
	\item \spverb|Sentence2|: the second sentence of the pair;
	\item \spverb|Explanation1|: the explanation generated by the worker recruited.
\end{itemize}
There are 5 additional attributes not included, for a total of 10 attributes. Table~\ref{cap:dataset-sec:esnli-tab:statements} shows a sample of three sentence pairs, one for each type of relation.

\begin{table}[tbp]
\centering
\caption{Statements sampled from the \esnli dataset.}
\label{cap:dataset-sec:esnli-tab:statements}
\begin{tabular}{p{1.8cm}p{3cm}p{2.8cm}p{2.1cm}p{2.5cm}}
\toprule
\textbf{Pair ID} & \textbf{Sentence 1} & \textbf{Sentence 2} & \textbf{Gold Label} & \textbf{Explanation}\\
\midrule
3636329461\allowbreak.jpg\#0r1e & The school is having a special event in order to show the american culture on how other cultures are dealt with in parties. & A school is hosting an event. & \esnlientail & An event is a special  occasion so all event is a special event. \\
\midrule
3636329461\allowbreak.jpg\#0r1n & The school is having a special event in order to show the american culture on how other cultures are dealt with in parties. & A high school is hosting an event. & \esnlineutral & The school was never described as a high school \\
\midrule
3636329461\allowbreak.jpg\#0r1n & The school is having a special event in order to show the american culture on how other cultures are dealt with in parties. & A school hosts a basketball game. & \esnlicontradiction & Basketball is american culture. \\
\bottomrule
\end{tabular}
\end{table}

The next chapter presents the first experiment conducted in this thesis. It investigates whether crowd workers can detect and objectively categorize online (mis)information in a sample of political statements drawn from the previously described dataset.

\chapter{The Effect Of Judgment Scales And Workers' Background}

\label{cap:paper_sigir2020}


This chapter is based on the article published at the 43rd International ACM SIGIR Conference on Research and Development in Information Retrieval \cite{roitero2020crowd}. Section~\ref{cap:related_work-sec:crowdsourcing-truthfulness} and Section~\ref{cap:related_work-sec:worker-bias} describe the relevant related work. Section~\ref{cap:paper_sigir2020-sec:research-questions} details the research questions, while Section~\ref{cap:paper_sigir2020-sec:exp-setup} presents the experimental setup. Section~\ref{cap:paper_sigir2020-sec:desc-stat} provides an initial descriptive analysis, while Section~\ref{cap:paper_sigir2020-sec:results} describes the results obtained. Finally, Section~\ref{cap:paper_sigir2020-sec:discussion} summarizes the main findings and concludes the chapter.

\section{Research Questions}

\label{cap:paper_sigir2020-sec:research-questions}

This chapter investigates how non-expert human judges perceive online (mis)information. A large-scale crowdsourcing experiment is conducted for this purpose. Crowd workers are asked to fact-check statements made by politicians and search for evidence of their validity using a custom web search engine. The \politifact and \abc datasets (Section~\ref{cap:intro-sec:fact-checking}) are used to sample statements related to U.S. and Australian politics. U.S.-based crowd workers are recruited to perform the fact-checking task. The experiment involves collecting data on the workers' political background and cognitive abilities. 

For each dataset, expert judgements are compared against the non-expert ones provided by the crowd. This also allows observation of how each crowd worker's bias is reflected in the data they generate. More specifically, U.S.-based workers might be familiar with U.S. politics but are less likely to have knowledge of Australian politics, including political figures and topics of discussion. The following research questions are investigated:
\begin{enumerate}[leftmargin=2.4em, label=RQ\arabic*]
    \item \label{cap:paper_sigir2020-sec:research-questions_1} What are the relationships and levels of agreement between crowd and expert judgments? And between judgments collected using different scales?
    \item \label{cap:paper_sigir2020-sec:research-questions_2} Are the judgment scales used suitable for collecting truthfulness judgments on political statements via crowdsourcing?
    \item \label{cap:paper_sigir2020-sec:research-questions_3} What sources of information do crowd workers use to identify online misinformation? 
    \item \label{cap:paper_sigir2020-sec:research-questions_4} What are the effects and roles of assessors' backgrounds in objectively identifying online misinformation?
\end{enumerate}

\section{Experimental Setting}

\label{cap:paper_sigir2020-sec:exp-setup}

The experimental setup involves statements sampled from \politifact (Section~\ref{cap:dataset-sec:politifact}) and \abc (Section~\ref{cap:dataset-sec:abc}). Specifically, a subset of 20 statements for each truth level from the \politifact dataset, covering the time span from 2007 to 2015, is used. The sample includes statements made by politicians from the two main U.S. parties (Democratic and Republican). The sample from \abc includes 60 randomly selected statements (20 for each truth level) made by politicians from the two main Australian parties (i.e., Liberal and Labor), covering the same time span. 

For both the \politifact and \abc datasets, the sample includes a balanced number of statements per class and per political party. In total, 180 statements are sampled. Table~\ref{cap:paper_sigir2020-sec:exp-setup-tab:statements} shows an example of \politifact and \abc statements. Appendix~\ref{cap:paper_sigir2020-appendix:quest-stat} reports the demographic questionnaire and the \index{Cognitive!Reflection Test}CRT tests used.

\begin{table}[htbp]
\centering
\caption{Example of statements sampled from the \politifact and \abc datasets. 
}
\label{cap:paper_sigir2020-sec:exp-setup-tab:statements}
\begin{tabular}{p{2.5cm}p{2cm}p{4.2cm}p{2cm}p{0.9cm}}
\toprule
\textbf{Dataset} & \textbf{Label} & \textbf{Statement} & \textbf{Speaker} & \textbf{Year}\\
 \midrule
\politifact & \politifactfour & Florida ranks first in the nation for access to free prekindergarten. & Rick Scott  & 2014 \\ 
\midrule
\abc & \abcone & Scrapping the carbon tax means every household will be \$550 a year better off. & Tony Abbott & 2014 \\ 
\bottomrule
\end{tabular}
\end{table}

\subsection{Crowdsourcing Task}

\label{cap:paper_sigir2020-sec:exp-setup-subsec:crowdsourcing-task}

The \mturk platform has been used to collect truthfulness judgments.\footnote{The experimental setup was reviewed and approved by the Human Research Ethics Committee at The University of Queensland.} Each worker accepting a Human Intelligence Task (\index{HIT}HIT) receives a unique input token, which identifies uniquely both the assigned \index{HIT}HIT and the worker. Then, they is redirected to an external application (Appendix~\ref{cap:paper_wsdm2022}) where to complete the task. 

The task is designed as follows: in the first part, the workers are asked to answer an initial survey (Appendix~\ref{cap:paper_sigir2020-appendix:quest-crt-sec:initial}) by providing some details about their background, such as age, family income, political views, the party in which they identify themselves, their opinion on building a wall along the southern border of United States, and on the need for environmental regulations to prevent climate change. Then, workers are asked to answer three modified Cognitive Reflection Test (\index{Cognitive!Reflection Test}CRT) questions to assess their cognitive abilities. In more detail, \index{Cognitive!Reflection Test}CRT questions are used to measure whether a person tends to overturn the incorrect \lq\lq intuitive\rq\rq{} response, and further reflect based on their own cognition to find the correct answer. \citet{Frederick2005} proposed the original version of the \index{Cognitive!Reflection Test}CRT in 2005. The modified questions used are reported in Appendix~\ref{cap:paper_sigir2020-appendix:quest-crt-sec:crt}. 

Workers are asked to provide truthfulness judgments after the initial survey. They judge 11 statements: 6 from \politifact, 3 from \abc, and 2 which serve as gold questions, one obviously true and the other obviously false. All the \politifact statements used are sampled from the most frequent five \emph{contexts} (i.e., the circumstance or media in which the statement was said/written) available in the dataset. To avoid bias, a balanced amount of data from each context is selected. Workers are presented with the following information about each statement to judge its truthfulness:
\begin{itemize}[label=--]
    \item \emph{Statement}: the text of the statement itself.
    \item \emph{Speaker}: the name and surname of whom said the statement.
    \item \emph{Year}: the year in which the statement was made.
\end{itemize}

Each worker is asked to provide both the truthfulness level of the statement and a URL that serves both as justification for their judgment as well as a source of evidence for fact-checking. In order to avoid workers finding and using the original expert labels (which are available on the Web) as the primary source of evidence, workers must use a provided custom web search engine to look for supporting evidence. The custom search engine uses the Bing Web Search API (Section~\ref{cap:paper_wsdm2022-sec:system-design-subsec:search-engine-subsec:bing}) to filter out from the retrieved results from any page from the websites that contain the collection of expert judgments used in the experiment. Workers are allowed to submit the \index{HIT}HIT after judging the whole set of 11 statements. In order to increase the quality of collected data, the following quality check is embedded in the crowdsourcing task:
\begin{itemize}[label=--]
    \item \emph{Gold Questions}: the worker must assign to the obviously false statement a truthfulness value lower than the one assigned to the obviously true statement.
    \item \emph{Time Spent}: the worker must spend at least two seconds on each statement and cognitive question.
\end{itemize}

The \index{HIT}HIT reward is set to \$1.5  after measuring the time and effort taken to successfully complete it. This was computed based on the expected time to complete it and targeting to pay at least the US federal minimum wage of \$7.25  per hour. Several small pilots of the task are performed. The task allows only US-based workers to participate, given the aim of the experiment. Each worker was allowed to complete only one of the \index{HIT}HITs for only one experimental setting (i.e., one judgment scale) to avoid the learning effect. Overall, not including pilot runs whose data was then discarded, the task allowed collecting judgments for 120 (\politifact) $+$ 60 (\abc) = 180 statements,  each one of them judged by 10 distinct workers. Such a setup is repeated over 3 different judgment scales. In total, 1800 (for each scale) * 3 = 5400 judgments are collected. Workers provide a total of 6600 assessments if also gold questions are considered.

\subsection{Judgment Scales And Collections}

\label{cap:paper_sigir2020-sec:exp-setup-subsec:scales}

The experimental design involves three truthfulness scales and five generated collections: two ground truths labeled by experts and three created using crowdsourcing. The expert-labeled collections are:

\begin{itemize}[label=--]
    \item \politifact: uses a six-level scale with labels \politifactzero, \politifactone, \politifacttwo, \politifactthree, \politifactfour, and \politifactfive (Chapter~\ref{cap:dataset-sec:politifact}).
    \item \abc: uses a three-level scale with labels \abczero, \abcone, and \abctwo (Chapter~\ref{cap:dataset-sec:abc}).
\end{itemize}

The remaining three collections are derived from the crowdsourcing task:

\begin{itemize}[label=--]
    \item \three: uses a three-level scale with the same labels as the \abc scale.
    \item \six: uses a six-level scale with the same labels as the \politifact scale, but replacing \politifactzero with \politifactlie.\footnote{The term \politifactlie is used in place of the colloquial \politifactzero for greater clarity.}
    \item \onehundred: uses a one-hundred-and-one-level scale with values in the $[0, 100]$ range.\footnote{Although the scale has 101 levels, it is referred to as \onehundred for simplicity.}
\end{itemize}

The nature and usage of these scales deserve some discussion. The scales used are made of different levels, i.e., categories, but they are not nominal scales. They would be nominal if such categories were independent, which is not the case because they are ordered. This can be seen immediately by considering, for example, that misclassifying a \politifactfive statement as \politifactfour is a smaller error than misclassifying it as \politifactthree. Indeed all of them are ordinal scales. However, they are not mere rankings, as the output of an information retrieval system. Statements are assigned to categories, besides being ranked. Let us suppose having two statements with ground truth labels \politifactfive and \politifactfour respectively. Misclassifying them as \politifactthree and \politifacttwo is an error. It is a smaller error than misclassifying them as \politifactone and \politifactzero. However, the original ranking has been preserved in both two cases. These scales are sometimes named \emph{ordinal categorical scales}~\cite{agresti2010analysis}.

For ordinal categorical scales, it cannot be assumed that the categories are equidistant. For example, misclassifying a \politifactzero statement as \politifactone cannot be assumed to be a smaller error than misclassifying a \politifacttwo statement as \politifactfive, generally. In light of this, taking the arithmetic mean to aggregate individual worker judgments for the same statement into a single label is not correct, since this would assume equidistant categories. On the contrary, the aggregation function for nominal scales (called \emph{majority vote} \cite{10.2307/2981392} by the crowdsourcing community) would discard important information, even if correct. For example, the aggregation of four \politifactzero with six \politifactone judgments should be rather different from --and lower than-- six \politifactone and four \politifactfive, though the aggregation function is the same. The orthodox and correct aggregation function for this kind of scale is the median. However, the situation is not so clear-cut. In the last example, the median would give the exact same result as the mode, thus discarding useful information. A reasonably defined ordinal categorical scale would feature labels which are approximately equidistant. This is particularly true for \onehundred, since it involves numerical labels in the $[0, 100]$ interval for the categories, and the crowd workers had to use a slider to select the values. This makes \onehundred (at least) very similar to an interval scale, for which the usage of the mean is correct. Indeed, it has been already used for \onehundred  \cite{Roitero:2018:FRS:3209978.3210052}. In the field of Information Retrieval people are used to interpreting ordinal scales as interval ones (e.g., when assigning arbitrary gains in the NDCG effectiveness metrics) and/or (ab)using the arithmetic mean (e.g., when taking the mean of ranks in the Mean Reciprocal Rank metric) \cite{Fuhr:2018}. In many practical cases, using the arithmetic mean turns out to be not only adequate but even more useful than the correct aggregation function~\cite{graham2015accurate, aletras2017evaluating, mathur2017sequence}. Even worse, there are no metrics for tasks defined on ordinal categorical scales, like predicting the number of \lq\lq stars\rq\rq{} in a recommendation scenario. Sometimes Accuracy is used, like in NTCIR-7 \cite{Kando-08}, but it is a metric for nominal scales. In some other cases, like in RepLab~2013~\cite{replab2013}, Reliability and Sensitivity~\cite{amigo2013general} are used, which consider only ranking information and no category membership. Even metrics for interval scales, like Mean Average Error (\meanerr), have been used \cite{Ghosh-15}.
 
For these reasons, in the following, the (aggregated) truthfulness labels provided by workers are sometimes treated as if they were expressed on an interval scale. This choice allows the various scales to be handled in a homogeneous manner, enabling the use of the same aggregation method applied to \onehundred also for \three, \six, \politifact, and \abc. Accordingly, the labels of \abc and \three are denoted as \texttt{0}, \texttt{1}, and \texttt{2}, as if they belonged to the $[0, 2]$ range. Similarly, the labels of \politifact and \six are denoted as \texttt{0}, \texttt{1}, $\ldots$, \texttt{5}, corresponding to the $[0, 5]$ range. Finally, the labels of \onehundred are denoted as \texttt{0}, \texttt{1}, $\ldots$, \texttt{100}.

\section{Descriptive Analysis}

\label{cap:paper_sigir2020-sec:desc-stat}

Section~\ref{cap:paper_sigir2020-sec:desc-stat-subsec:demographics} first presents demographic information about the workers. Section~\ref{cap:paper_sigir2020-sec:desc-stat-subsec:task-abandonment} then examines the task abandonment rate. Finally, Section~\ref{cap:paper_sigir2020-sec:desc-stat-subsec:score-distribution} reports the distribution of the crowdsourced truthfulness judgments.

\subsection{Worker Demographics}

\label{cap:paper_sigir2020-sec:desc-stat-subsec:demographics}

The crowdsourcing task was published on \mturk. In total, approximately \num{600} crowd workers residing in the United States\footnote{\mturk workers based in the US must provide evidence of their eligibility to work.} participate across all three experiments. The majority of participants (46.33\%) are between 26 and 35 years old. The workers are generally well-educated, with more than 60.84\% holding at least a four-year college degree. Additionally, around 67.66\% report an annual income below $75,000$.

Nearly half of the participants (47.33\%) identify more closely with the Democratic Party, while only 22.5\% select the Republican Party as their voting preference. In terms of political ideology, the largest groups are \liberal (29.5\%) and \moderate (28.83\%), while the \veryconservative category accounts for only 5.67\%. Regarding specific policy views, 52.33\% of U.S.-based workers oppose the construction of a wall on the southern border, while 36.5\% support it. On environmental issues, 80\% support stronger government regulations to address climate change, whereas 11.33\% are opposed.

\subsection{Task Abandonment}

\label{cap:paper_sigir2020-sec:desc-stat-subsec:task-abandonment}

Table~\ref{cap:paper_sigir2020-sec:desc-stat-subsec:task-abandonment-tab:worker-behavior} reports the proportion of workers who complete the task, abandon it, or fail the quality checks, based on behavioral actions logged as workers proceed through the \index{HIT}HITs. The observed abandonment rates are consistent with prior studies~\cite{8873609}. Particularly, the \onehundred condition shows a higher failure rate and a lower completion rate, which may indicate a reduced level of comfort among workers when using the most fine-grained scale.

\begin{table}[htbp]
\caption{Completion, abandonment, and failure rates of workers in the crowdsourcing tasks for the \three, \six, and \onehundred collections.}
\label{cap:paper_sigir2020-sec:desc-stat-subsec:task-abandonment-tab:worker-behavior}
\centering
\begin{tabular}{lccc}
\toprule
\textbf{Collection} & \textbf{Completion (\%)} & \textbf{Abandonment (\%)} & \textbf{Failure (\%)} \\
\midrule
\three & 35\% & 53\% & 12\% \\
\six & 33\% & 52\% & 14\% \\
\onehundred & 25\% & 53\% & 22\% \\
\bottomrule
\end{tabular}
\end{table}

\subsection{Crowdsourced Judgments Distributions}

\label{cap:paper_sigir2020-sec:desc-stat-subsec:score-distribution}

\begin{figure}[tbp]
  \centering
  \begin{subfigure}{0.49\linewidth}
    \centering
    \includegraphics[width=\linewidth]{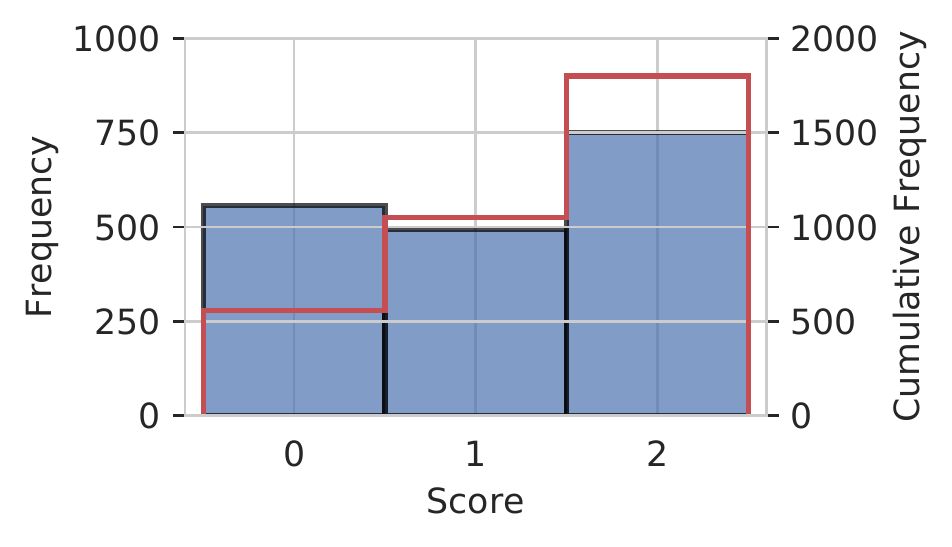}
    \caption{Distribution of individual scores.}
    \label{cap:paper_sigir2020-sec:desc-stat-subsec:score-distribution-fig:scores_distributions_s3_ind}
  \end{subfigure}
  \hfill
  \begin{subfigure}{0.49\linewidth}
    \centering
    \includegraphics[width=.8\linewidth]{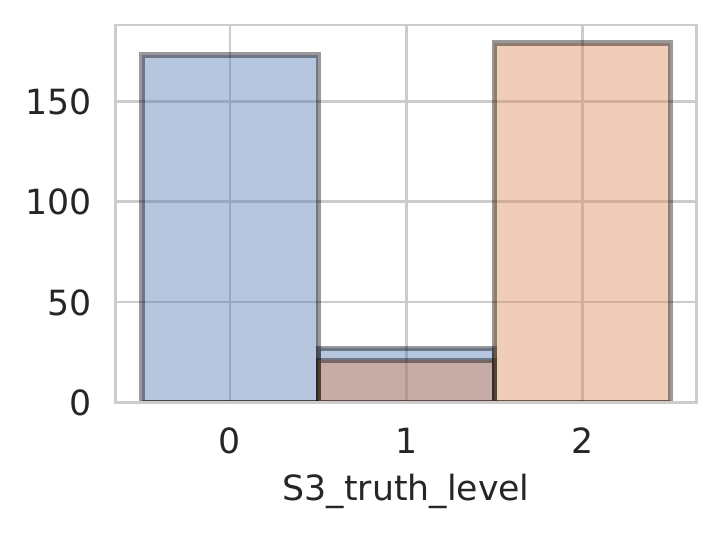}
    \caption{Distribution of expert (gold) judgments.}
    \label{cap:paper_sigir2020-sec:desc-stat-subsec:score-distribution-fig:scores_distributions_s3_gold}
  \end{subfigure}

  \vspace{1em} 

  \begin{subfigure}{0.49\linewidth}
    \centering
    \includegraphics[width=\linewidth]{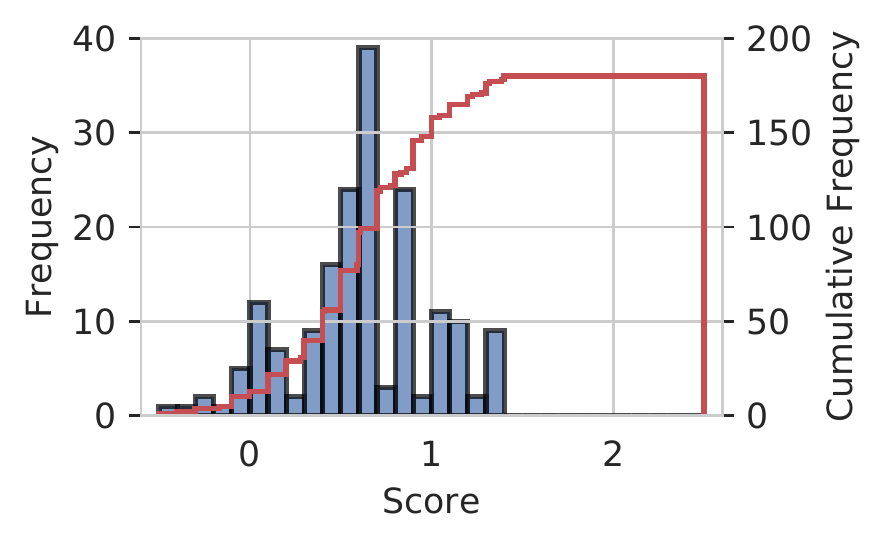}
    \caption{Distribution of aggregated crowd judgments.}
    \label{cap:paper_sigir2020-sec:desc-stat-subsec:score-distribution-fig:scores_distributions_s3_agg}
  \end{subfigure}

  \caption{Distributions of judgment scores for \three: individual responses (top left), expert (gold) labels (top right), and aggregated crowd judgments (bottom).}
  \label{cap:paper_sigir2020-sec:desc-stat-subsec:score-distribution-fig:scores_distributions_s3}
\end{figure}

\begin{figure}[tbp]
  \centering
  \begin{subfigure}{0.49\linewidth}
    \centering
    \includegraphics[width=\linewidth]{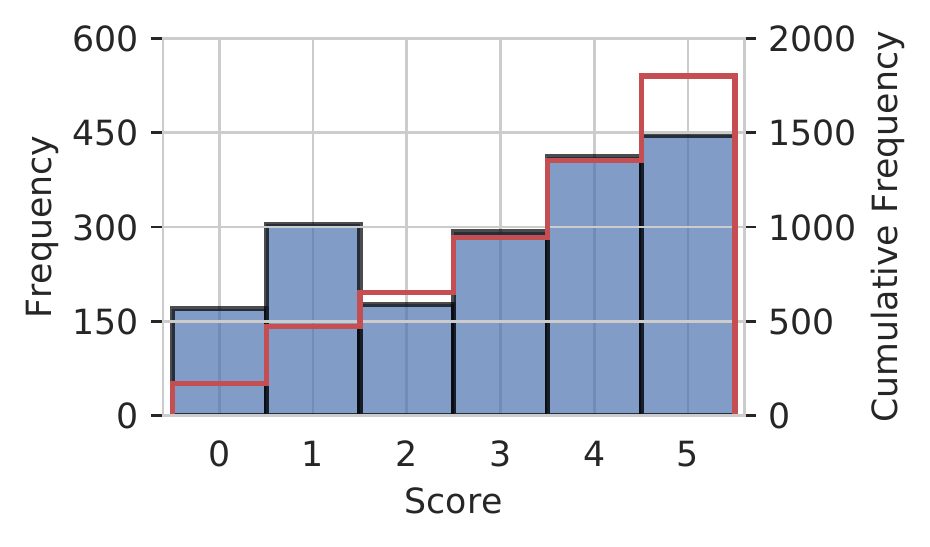}
    \caption{Distribution of individual scores.}
    \label{cap:paper_sigir2020-sec:desc-stat-subsec:score-distribution-fig:scores_distributions_s6_ind}
  \end{subfigure}
  \hfill
  \begin{subfigure}{0.49\linewidth}
    \centering
    \includegraphics[width=.8\linewidth]{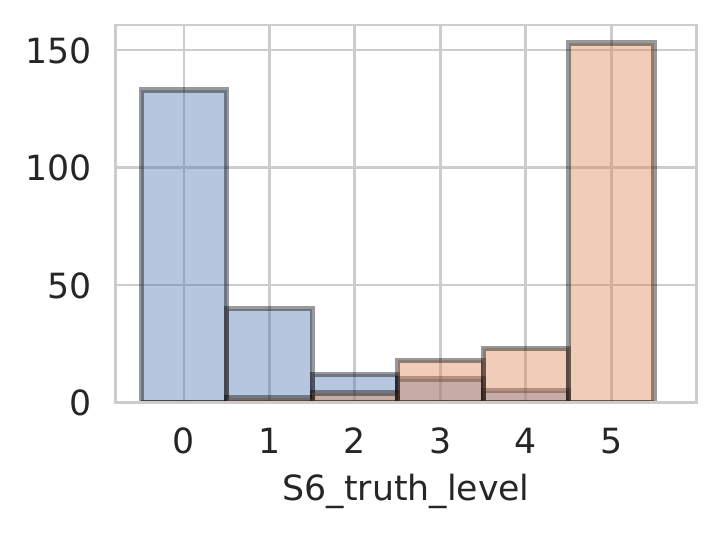}
    \caption{Distribution of expert (gold) judgments.}
    \label{cap:paper_sigir2020-sec:desc-stat-subsec:score-distribution-fig:scores_distributions_s6_gold}
  \end{subfigure}

  \vspace{1em} 

  \begin{subfigure}{0.49\linewidth}
    \centering
    \includegraphics[width=\linewidth]{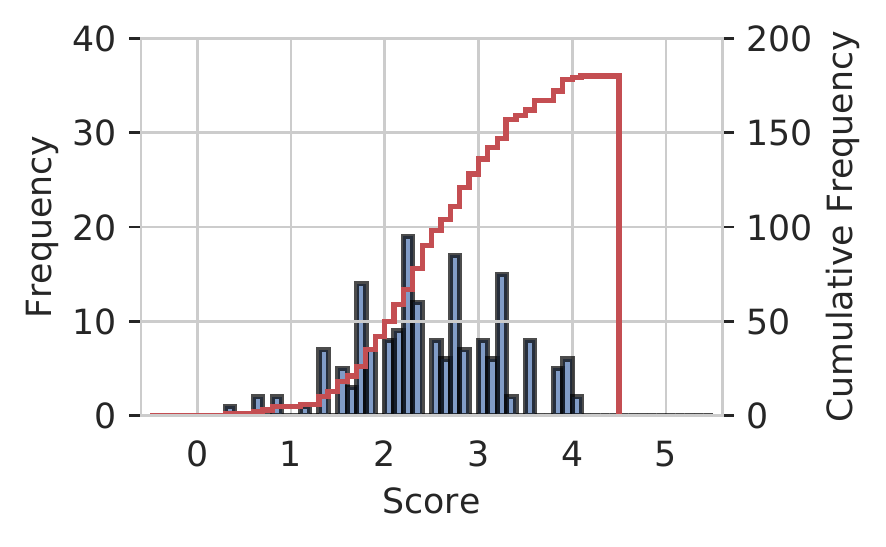}
    \caption{Distribution of aggregated crowd judgments.}
    \label{cap:paper_sigir2020-sec:desc-stat-subsec:score-distribution-fig:scores_distributions_s6_agg}
  \end{subfigure}

  \caption{Distributions of judgment scores for \six: individual responses (top left), expert (gold) labels (top right), and aggregated crowd judgments (bottom).}
  \label{cap:paper_sigir2020-sec:desc-stat-subsec:score-distribution-fig:scores_distributions_s6}
\end{figure}

\begin{figure}[tbp]
  \centering
  \begin{subfigure}{0.49\linewidth}
    \centering
    \includegraphics[width=\linewidth]{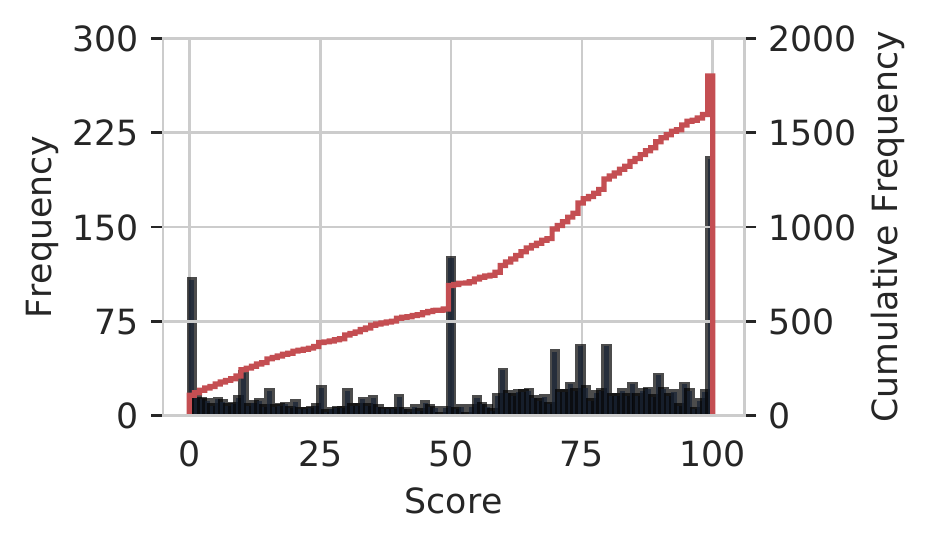}
    \caption{Distribution of individual scores.}
    \label{cap:paper_sigir2020-sec:desc-stat-subsec:score-distribution-fig:scores_distributions_s100_ind}
  \end{subfigure}
  \hfill
  \begin{subfigure}{0.49\linewidth}
    \centering
    \includegraphics[width=.8\linewidth]{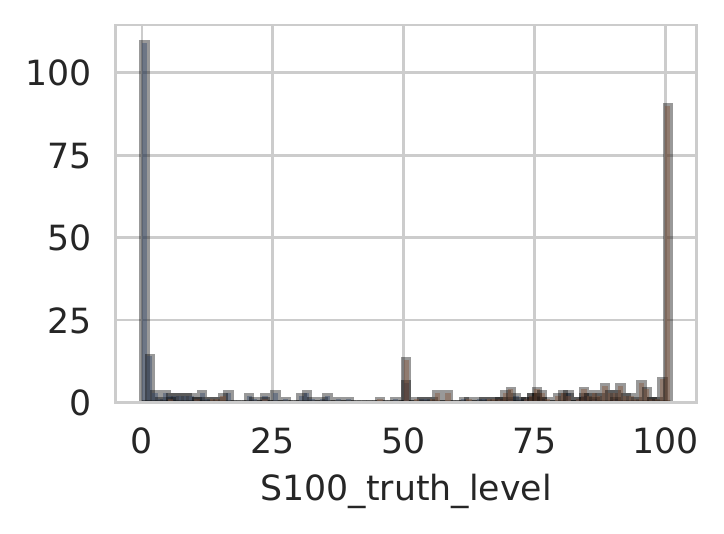}
    \caption{Distribution of expert (gold) judgments.}
    \label{cap:paper_sigir2020-sec:desc-stat-subsec:score-distribution-fig:scores_distributions_s100_gold}
  \end{subfigure}

  \vspace{1em} 

  \begin{subfigure}{0.49\linewidth}
    \centering
    \includegraphics[width=\linewidth]{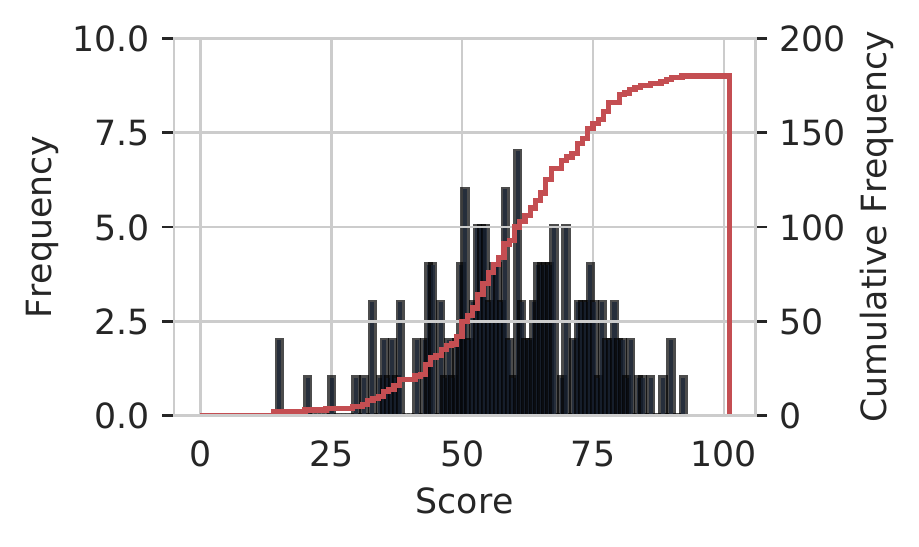}
    \caption{Distribution of aggregated crowd judgments.}
    \label{cap:paper_sigir2020-sec:desc-stat-subsec:score-distribution-fig:scores_distributions_s100_agg}
  \end{subfigure}

  \caption{Distributions of judgment scores for \onehundred: individual responses (top left), expert (gold) labels (top right), and aggregated crowd judgments (bottom).}
  \label{cap:paper_sigir2020-sec:desc-stat-subsec:score-distribution-fig:scores_distributions_s100}
\end{figure}

Figure~\ref{cap:paper_sigir2020-sec:desc-stat-subsec:score-distribution-fig:scores_distributions_s3}, Figure~\ref{cap:paper_sigir2020-sec:desc-stat-subsec:score-distribution-fig:scores_distributions_s6}, and Figure~\ref{cap:paper_sigir2020-sec:desc-stat-subsec:score-distribution-fig:scores_distributions_s100} show the distribution (and the cumulative distribution in red) for the individual judgments provided by workers over the three crowd collections (i.e., \three, \six, and \onehundred) for all the statements considered in the experiment. The behavior is consistent when considering separately \politifact and \abc statements.

Figure~\ref{cap:paper_sigir2020-sec:desc-stat-subsec:score-distribution-fig:scores_distributions_s3_ind}, Figure~\ref{cap:paper_sigir2020-sec:desc-stat-subsec:score-distribution-fig:scores_distributions_s6_ind}, and Figure~\ref{cap:paper_sigir2020-sec:desc-stat-subsec:score-distribution-fig:scores_distributions_s100_ind} show the raw judgments distributions for the three collections. The distribution is skewed towards the right part of the scale representing higher truthfulness values, for the whole set of collections. This can be also seen by looking at the cumulative distribution, which is steeper on the right-hand side of the charts. It can also be seen that all three distributions are multimodal. The \onehundred collection shows a mild round number tendency that is, the tendency of workers to provide truthfulness judgments which are multiple of 10 (35\% of \onehundred scores are multiple of 10; 23\% are 0, 50, or 100); such behavior was already noted by  \citet{Maddalena:2017:CRM:3026478.3002172}, and \citet{Roitero:2018:FRS:3209978.3210052}. The behavior is consistent when considering separately \politifact and \abc statements.

The gold judgments are those provided for the special \high and \low statements used to perform quality checks during the task. The second row of Figure~\ref{cap:paper_sigir2020-sec:desc-stat-subsec:score-distribution-fig:scores_distributions_s3_gold}, Figure~\ref{cap:paper_sigir2020-sec:desc-stat-subsec:score-distribution-fig:scores_distributions_s6_gold}, and Figure~\ref{cap:paper_sigir2020-sec:desc-stat-subsec:score-distribution-fig:scores_distributions_s100_gold} show the distribution of the scores for the three crowd collections considered. The large majority of workers (44\% for \low and 45\% for \high in \three, 34\% for \low and 39\% for \high in \six, and 27\% for \low and 24\% for \high in \onehundred) provide as truthfulness judgment for the gold statements the extreme value of the scales (respectively the lower bound of the scale for \low and the upper bound of the scale for \high). This can be interpreted as a signal that the data gathered is of good quality. However, some workers provide judgments inconsistent with the gold labels.

Figure~\ref{cap:paper_sigir2020-sec:desc-stat-subsec:score-distribution-fig:scores_distributions_s3_agg}, Figure~\ref{cap:paper_sigir2020-sec:desc-stat-subsec:score-distribution-fig:scores_distributions_s6_agg}, and Figure~\ref{cap:paper_sigir2020-sec:desc-stat-subsec:score-distribution-fig:scores_distributions_s100_agg} shows the distributions of \three, \six, and \onehundred judgments aggregated by taking the average of the 10 scores obtained independently for each statement. The distribution of the judgments aggregated for \three, \six, and \onehundred are similar, they are no longer multimodal, and they are roughly bell-shaped. It is worth noting that the judgments for \six and \onehundred (Figure~\ref{cap:paper_sigir2020-sec:desc-stat-subsec:score-distribution-fig:scores_distributions_s6_agg} and Figure~\ref{cap:paper_sigir2020-sec:desc-stat-subsec:score-distribution-fig:scores_distributions_s100_agg}) are skewed to higher/positive scores. On the other hand, the judgments aggregated for \three (Figure~\ref{cap:paper_sigir2020-sec:desc-stat-subsec:score-distribution-fig:scores_distributions_s3_agg}) are skewed towards lower/negative (i.e., \abczero and \abcone scores). This shows how different judgment scales are used differently by the crowd. For \onehundred the round number tendency effect also disappears when judgments from different workers are aggregated together, as expected \cite{Maddalena:2017:CRM:3026478.3002172, RSDM:2018, Roitero:2018:FRS:3209978.3210052}.

\section{Results}

\label{cap:paper_sigir2020-sec:results}

Section~\ref{cap:paper_sigir2020-sec:results-subsec:agreement} analyzes both the external agreement between crowd judgments and expert labels and the internal agreement among workers. Section~\ref{cap:paper_sigir2020-sec:results-subsec:scale-adequacy-subsec:merge} examines how the collected judgments behave when aggregated into more coarse-grained scales. Section~\ref{cap:paper_sigir2020-sec:results-subsec:justification-distribution} investigates the sources of information used by workers when justifying their answers. Finally, Section~\ref{cap:paper_sigir2020-sec:results-subsec:worker-bias} explores the relationship between workers’ backgrounds and their performance.

\subsection{\ref{cap:paper_sigir2020-sec:research-questions_1}: Crowd Workers Accuracy}

\label{cap:paper_sigir2020-sec:results-subsec:agreement}

The quality of the judgments provided by the crowd workers can be assessed by analyzing two forms of agreement. The first is external agreement, i.e., the agreement between crowd-collected judgments and expert ground truth (Section~\ref{cap:paper_sigir2020-sec:results-subsec:agreement-external}). A second standard approach is to measure internal agreement, i.e., the consistency among the workers themselves (Section~\ref{cap:paper_sigir2020-sec:results-subsec:agreement-internal}).

\subsubsection{External Agreement}

\label{cap:paper_sigir2020-sec:results-subsec:agreement-external}

\begin{figure}[tbp]
  \centering

  \begin{subfigure}{0.49\linewidth}
    \centering
    \includegraphics[width=\linewidth]{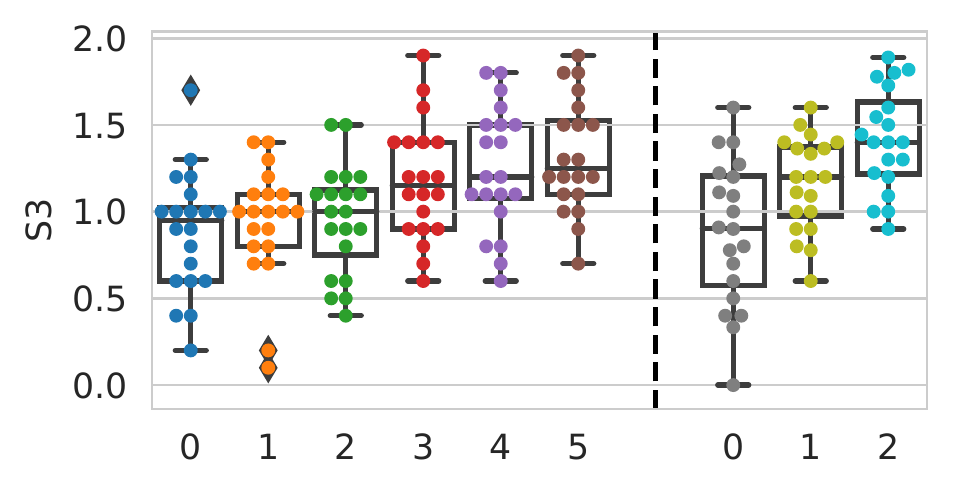}
    \caption{\three.}
    \label{cap:paper_sigir2020-sec:results-subsec:agreement-external-fig:scale-comparison-ground_s3}
  \end{subfigure}
  \hfill
  \begin{subfigure}{0.49\linewidth}
    \centering
    \includegraphics[width=\linewidth]{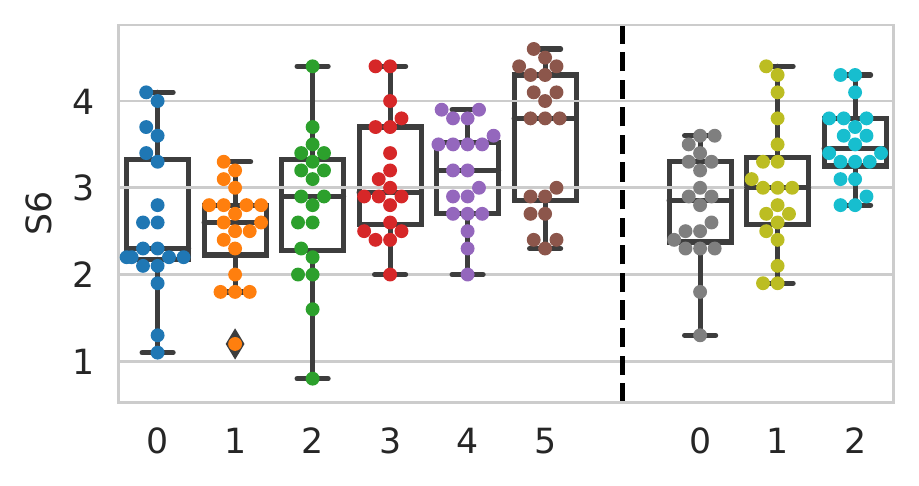}
    \caption{\six.}
    \label{cap:paper_sigir2020-sec:results-subsec:agreement-external-fig:scale-comparison-ground_s6}
  \end{subfigure}

  \vspace{1em}

  \begin{subfigure}{0.49\linewidth}
    \centering
    \includegraphics[width=\linewidth]{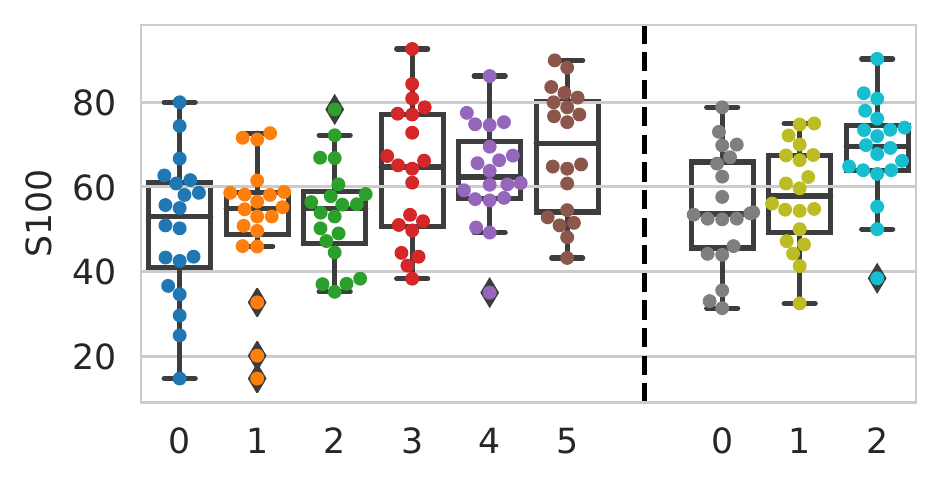}
    \caption{\onehundred.}
    \label{cap:paper_sigir2020-sec:results-subsec:agreement-external-fig:scale-comparison-ground_s100}
  \end{subfigure}

  \caption{External agreement with \politifact and \abc statements, separated by the vertical dashed line.}
  \label{cap:paper_sigir2020-sec:results-subsec:agreement-external-fig:scale-comparison-ground}
\end{figure}

Figure~\ref{cap:paper_sigir2020-sec:results-subsec:agreement-external-fig:scale-comparison-ground} shows the agreement between ground truth (i.e., the expert labels provided for \politifact and \abc) and the crowd judgments aggregated for the \three, \six, and \onehundred crowd collections. The behavior over all three scales is similar, both on \politifact and \abc statements.

Focusing on \politifact statements (shown on the left-hand side of each chart) reveals that the \texttt{0} and \texttt{1} boxplots are very similar. This may indicate that workers have difficulty distinguishing between the \politifactzero and \politifactone labels. The same behavior, even if less evident, is present between the \politifactone and \politifacttwo labels; this behavior is consistent across all the scales. On the contrary, focusing on the remaining \politifact labels and the \abc ones shows that the median lines of each boxplot are increasing while going towards labels representing higher truthfulness values (i.e., going towards the right-hand side of each chart), indicating that workers have a higher agreement with the ground truth for those labels. Again, this behavior is consistent and similar for all the \three, \six, and \onehundred scales. 

The statistical significance of the differences between the mean-aggregated judgments across categories in the \six, \three, and \onehundred collections is assessed using the Mann–Whitney rank test and the t-test. For \abc, adjacent categories are significantly different in 5 out of 12 cases, while all differences between non-adjacent categories are significant at the \index{$p$}$p<.01$ level.
For \politifact, differences between adjacent categories and categories at a distance of 2 (e.g., \texttt{0} vs. \texttt{2}) are generally not significant, with only one exception in the latter group. Differences between categories at a distance of 3 are significant in 4 out of 18 cases, while those at a distance of 4 are significant in 5 out of 12 cases. Finally, comparisons between the most extreme categories (distance 5, i.e., \texttt{0} vs. \texttt{5}) are significant in 4 out of 6 cases.
Although some signal emerges, these results indicate that the answer to \ref{cap:paper_sigir2020-sec:research-questions_1} cannot be positive based on this analysis. Further discussion is provided in Section~\ref{cap:paper_sigir2020-sec:results-subsec:scale-adequacy-subsec:merge}.

\begin{figure}[tbp]
  \centering

  \begin{subfigure}{0.49\linewidth}
    \centering
    \includegraphics[width=\linewidth]{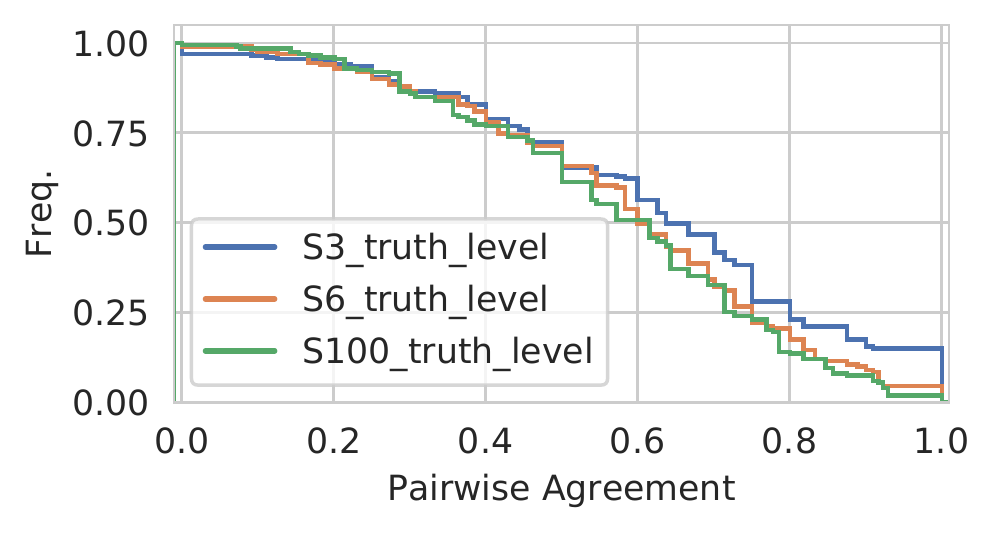}
    \caption{\politifact.}
    \label{cap:paper_sigir2020-sec:results-subsec:agreement-external-fig:pairwise-unit-agreement_pol}
  \end{subfigure}
  \hfill
  \begin{subfigure}{0.49\linewidth}
    \centering
    \includegraphics[width=\linewidth]{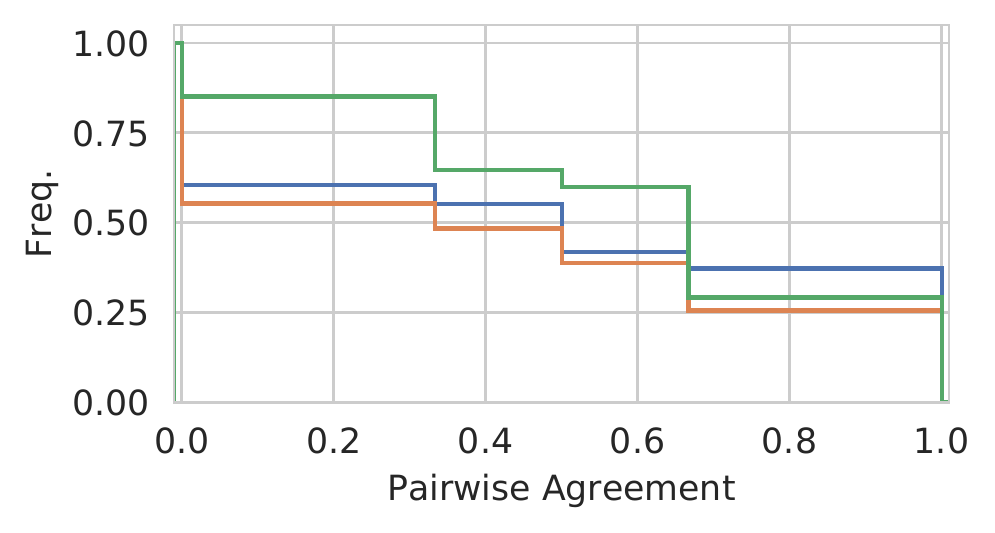}
    \caption{\abc.}
    \label{cap:paper_sigir2020-sec:results-subsec:agreement-external-fig:pairwise-unit-agreement_abc}
  \end{subfigure}

  \caption{Pairwise agreement and relative frequency, computed for each \index{HIT}HIT in the task.}
  \label{cap:paper_sigir2020-sec:results-subsec:agreement-external-fig:pairwise-unit-agreement}
\end{figure}

Figure~\ref{cap:paper_sigir2020-sec:results-subsec:agreement-external-fig:pairwise-unit-agreement} inspects the agreement between the workers and the ground truth by looking at each \index{HIT}HIT. The pairwise agreement \cite{maddalena2017considering} between the truthfulness judgments expressed by workers and the ground truth labels is computed for all the \three, \six, and \onehundred collections,  with a breakdown over \politifact (Figure~\ref{cap:paper_sigir2020-sec:results-subsec:agreement-external-fig:pairwise-unit-agreement_pol}) and \abc (Figure~\ref{cap:paper_sigir2020-sec:results-subsec:agreement-external-fig:pairwise-unit-agreement_abc}) statements. A slightly modified version of the pairwise agreement measure \cite{maddalena2017considering} is used. In the attempt to make the pairwise agreement measure fully comparable across the different scales, all the ties are removed. Intuitively, the pairwise agreement described by \citet{maddalena2017considering} measures the fraction of pairs in the agreement between a \lq\lq ground truth\rq\rq{} scale and a \lq\lq crowd\rq\rq{} scale. Specifically, a pair of crowd judgments (crowd-judgment$_1$, crowd-judgment$_2$) is considered to be in agreement if crowd-judgment$_1$ $\leq$ crowd-judgment$_2$ and the ground truth for crowd-judgment$_1$ is $<$ the ground truth for crowd-judgment$_2$. In this thesis' measurement\footnote{The code used to compute the pairwise agreement as defined is available at \url{https://github.com/KevinRoitero/PairwiseAgreement}.} all the ties (i.e., crowd-judgment$_1$ $=$ crowd-judgment$_2$) are removed and $<$ is used in place of $\leq$. 

In more detail, Figure~\ref{cap:paper_sigir2020-sec:results-subsec:agreement-external-fig:pairwise-unit-agreement} shows the \index{CCDF}CCDF (Complementary Cumulative Distribution Function) of the relative frequencies of agreement across \index{HIT}HITs. The \three, \six, and \onehundred scales show very similar levels of external agreement. This behavior is consistent across both the \politifact and \abc datasets. Overall, the results confirm that all considered scales lead to comparable levels of agreement with the expert ground truth, with the sole exception of \onehundred in the \abc dataset. This deviation is likely due to how ties are treated in the agreement measure, which results in the removal of a different number of units across the three scales.

\subsubsection{Internal Agreement}

\label{cap:paper_sigir2020-sec:results-subsec:agreement-internal}

The Krippendorff's \index{$\upalpha$}$\upalpha$ coefficient \cite{krippendorff2011computing} is computed as a metric to measure the level of internal agreement in a dataset. All \index{$\upalpha$}$\upalpha$ values within each of the three scales \three, \six, \onehundred and on both \politifact and \abc collections are in the $0.066$--$0.131$ range. These results show that there is a rather low agreement among the workers \cite{checco2017let, krippendorff2011computing}. Furthermore, all the possible transformations of judgments from one scale to another are performed to investigate whether the low agreement found depends on the specific scale used to judge the statements, following the methodology described by \citet{scale}.

\begin{figure}[tbp]
  \centering
  \begin{subfigure}{0.49\linewidth}
    \centering
    \includegraphics[width=.9\linewidth]{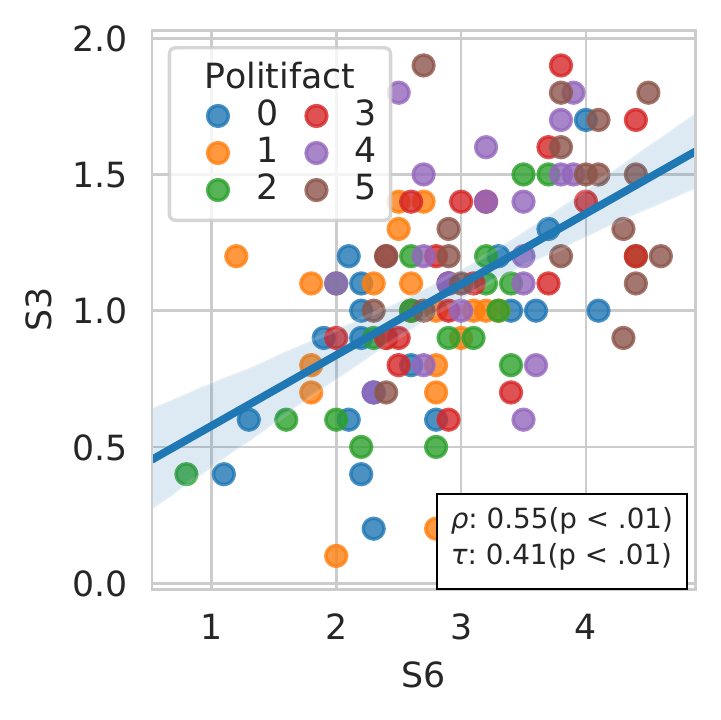}
    \caption{\six vs. \three.}
    \label{cap:paper_sigir2020-sec:results-subsec:agreement-internal-fig:scale-comparisons_pol_six_three}
  \end{subfigure}
  \hfill
  \begin{subfigure}{0.49\linewidth}
    \centering
    \includegraphics[width=.9\linewidth]{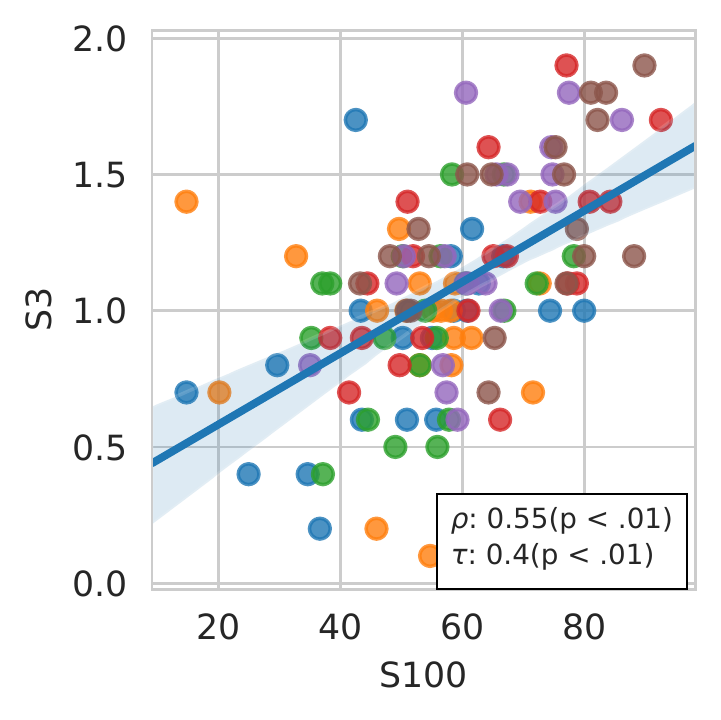}
    \caption{\onehundred vs. \three.}
    \label{cap:paper_sigir2020-sec:results-subsec:agreement-internal-fig:scale-comparisons_pol_onehundred_three}
  \end{subfigure}

  \vspace{1em}

  \begin{subfigure}{0.49\linewidth}
    \centering
    \includegraphics[width=.9\linewidth]{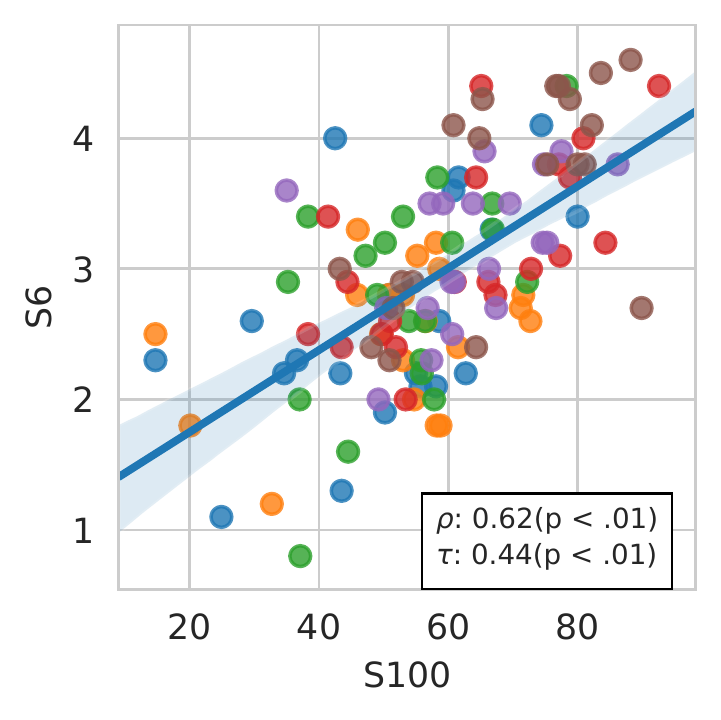}
    \caption{\onehundred vs. \six.}
    \label{cap:paper_sigir2020-sec:results-subsec:agreement-internal-fig:scale-comparisons_pol_onehundred_six}
  \end{subfigure}

  \caption{Agreement between judgment scales for \politifact statements.}
  \label{cap:paper_sigir2020-sec:results-subsec:agreement-internal-fig:scale-comparisons_pol}
\end{figure}

\begin{figure}[tbp]
  \centering
  \begin{subfigure}{0.49\linewidth}
    \centering
    \includegraphics[width=.9\linewidth]{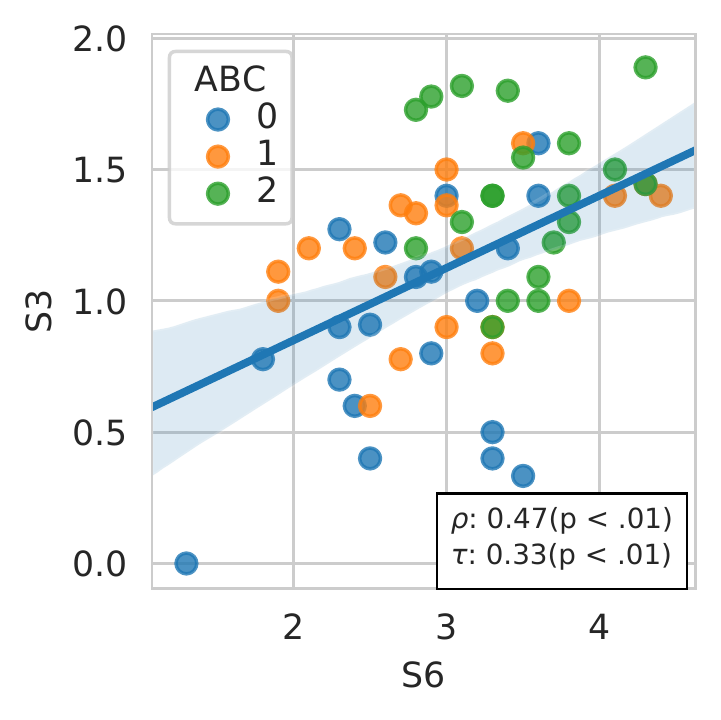}
    \caption{\six vs. \three.}
    \label{cap:paper_sigir2020-sec:results-subsec:agreement-internal-fig:scale-comparisons_abc_six_three}
  \end{subfigure}
  \hfill
  \begin{subfigure}{0.49\linewidth}
    \centering
    \includegraphics[width=.9\linewidth]{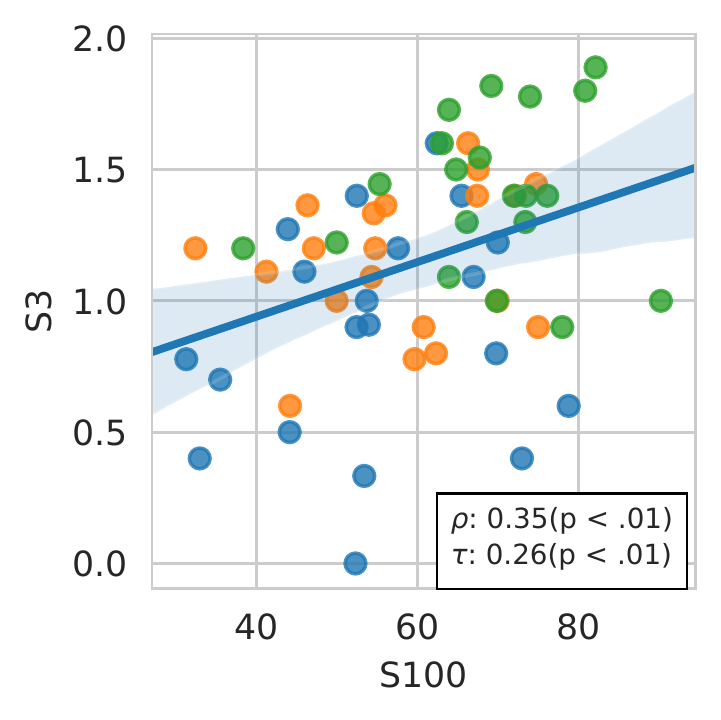}
    \caption{\onehundred vs. \three.}
    \label{cap:paper_sigir2020-sec:results-subsec:agreement-internal-fig:scale-comparisons_abc_onehundred_three}
  \end{subfigure}

  \vspace{1em}

  \begin{subfigure}{0.49\linewidth}
    \centering
    \includegraphics[width=.9\linewidth]{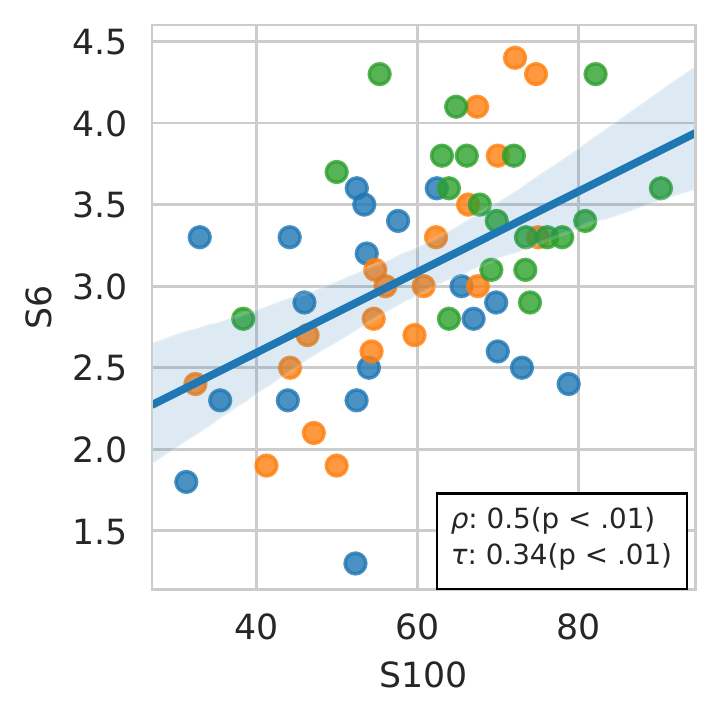}
    \caption{\onehundred vs. \six.}
    \label{cap:paper_sigir2020-sec:results-subsec:agreement-internal-fig:scale-comparisons_abc_onehundred_six}
  \end{subfigure}

  \caption{Agreement between judgment scales for \abc statements.}
  \label{cap:paper_sigir2020-sec:results-subsec:agreement-internal-fig:scale-comparisons_abc}
\end{figure}

Figure~\ref{cap:paper_sigir2020-sec:results-subsec:agreement-internal-fig:scale-comparisons_pol} shows the scatterplots, as well as the correlations, between the different scales on the \politifact statements, while Figure~\ref{cap:paper_sigir2020-sec:results-subsec:agreement-internal-fig:scale-comparisons_abc} reports the same for the \abc statements. The correlation values are around \index{$\uprho$}$\uprho = 0.55$--$0.6$ for \politifact and \index{$\uprho$}$\uprho = 0.35$--$0.5$ for \abc, across all scales. The rank correlation coefficient \index{$\uptau$}$\uptau$ is approximately $\uptau = 0.4$ for \politifact and \index{$\uptau$}$\uptau = 0.3$ for \abc. These values indicate a low correlation between the scales, suggesting that the same statements tend to be judged differently depending on the scale, both in terms of absolute values (i.e., \index{$\uprho$}$\uprho$) and relative ordering (i.e., \index{$\uptau$}$\uptau$).

\begin{figure}[tbp]
  \centering
  \begin{subfigure}{0.49\linewidth}
    \centering
    \includegraphics[width=.9\linewidth]{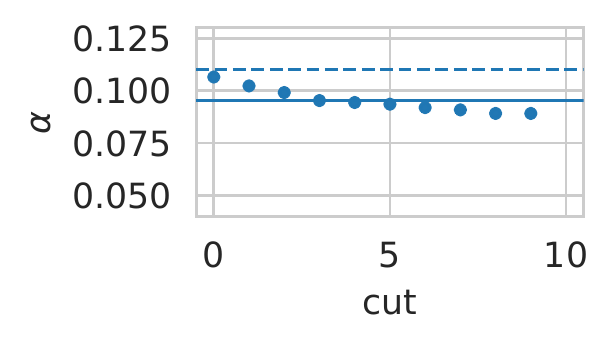}
    \caption{\six cut into \three.}
    \label{cap:paper_sigir2020-sec:results-subsec:agreement-internal-fig:alpha-cuts_6to3}
  \end{subfigure}
  \hfill
  \begin{subfigure}{0.49\linewidth}
    \centering
    \includegraphics[width=.9\linewidth]{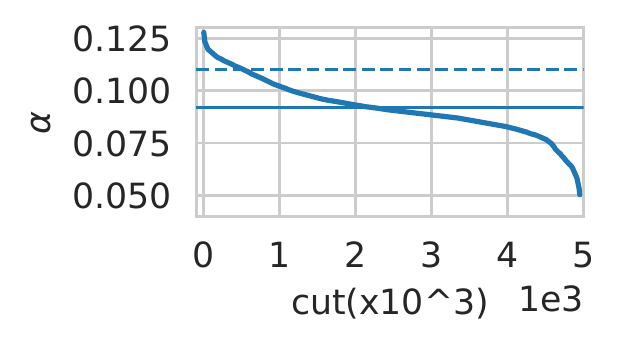}
    \caption{\onehundred cut into \three.}
    \label{cap:paper_sigir2020-sec:results-subsec:agreement-internal-fig:alpha-cuts_100to3}
  \end{subfigure}

  \vspace{1em}

  \begin{subfigure}{0.49\linewidth}
    \centering
    \includegraphics[width=.9\linewidth]{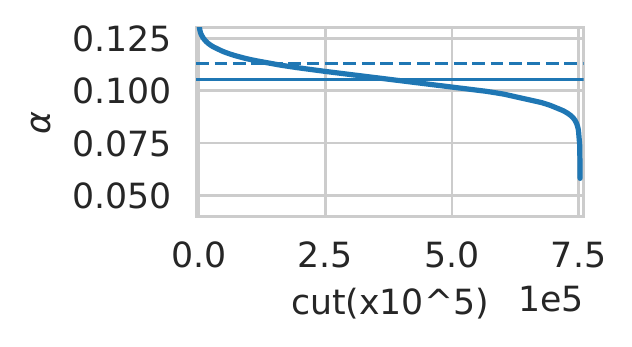}
    \caption{\onehundred cut into \six (1\% stratified sampling).}
    \label{cap:paper_sigir2020-sec:results-subsec:agreement-internal-fig:alpha-cuts_100to6}
  \end{subfigure}

  \caption{$\upalpha$ cuts sorted by decreasing values.}
  \label{cap:paper_sigir2020-sec:results-subsec:agreement-internal-fig:alpha-cuts}
\end{figure}

Figure~\ref{cap:paper_sigir2020-sec:results-subsec:agreement-internal-fig:alpha-cuts} shows the distribution of the \index{$\upalpha$}$\upalpha$ values when transforming one scale into another. The total number of possible cuts from \onehundred to \six is 75,287,520. Thus, a sub-sample of all the possible cuts is selected. Both stratified and random sampling has been used, getting indistinguishable results. The dotted horizontal line in the plot represents \index{$\upalpha$}$\upalpha$ on the original dataset, the dashed line is the mean value of the (sampled) distribution. The values on the y-axis are very concentrated and all \index{$\upalpha$}$\upalpha$ values are close to zero ($[0,0.15]$ range). It can be concluded that across all collections there is a low level of internal agreement among workers, both within the same scale and across different scales. 

\subsection{\ref{cap:paper_sigir2020-sec:research-questions_2}: Judgment Scales Adequacy}

\label{cap:paper_sigir2020-sec:results-subsec:scale-adequacy}

Section~\ref{cap:paper_sigir2020-sec:results--subsec:scale-adequacy-subsec:alternative-aggregation} examines the effect of using aggregation functions other than the arithmetic mean, and compares worker behavior across different scales. Section~\ref{cap:paper_sigir2020-sec:results-subsec:scale-adequacy-subsec:merge} analyzes the impact of merging adjacent ground truth categories.

\subsubsection{Alternative Aggregation Functions}

\label{cap:paper_sigir2020-sec:results--subsec:scale-adequacy-subsec:alternative-aggregation}

\begin{figure}[tbp]
  \centering
  \begin{subfigure}{0.49\linewidth}
    \centering
    \includegraphics[width=\linewidth]{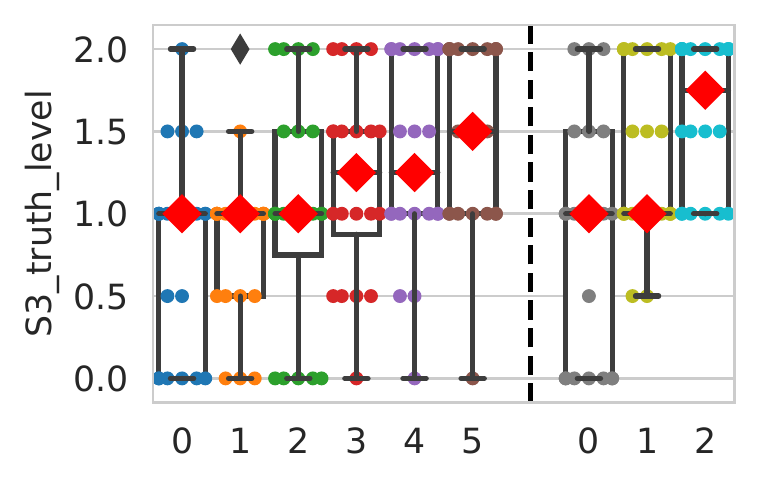}
    \caption{\three}
    \label{cap:paper_sigir2020-sec:results-subsec:scale-adequacy-subsec:alternative-aggregation-fig:alternative-aggregation-median_3}
  \end{subfigure}
  \hfill
  \begin{subfigure}{0.49\linewidth}
    \centering
    \includegraphics[width=\linewidth]{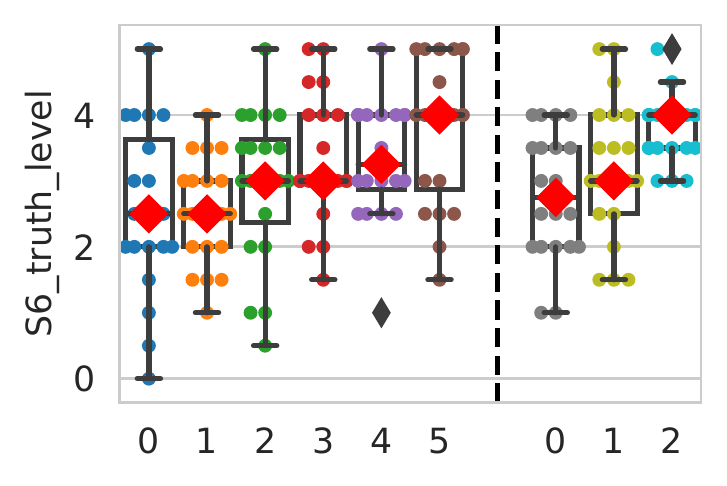}
    \caption{\six}
    \label{cap:paper_sigir2020-sec:results-subsec:scale-adequacy-subsec:alternative-aggregation-fig:alternative-aggregation-median_6}
  \end{subfigure}

  \vspace{1em}

  \begin{subfigure}{0.49\linewidth}
    \centering
    \includegraphics[width=\linewidth]{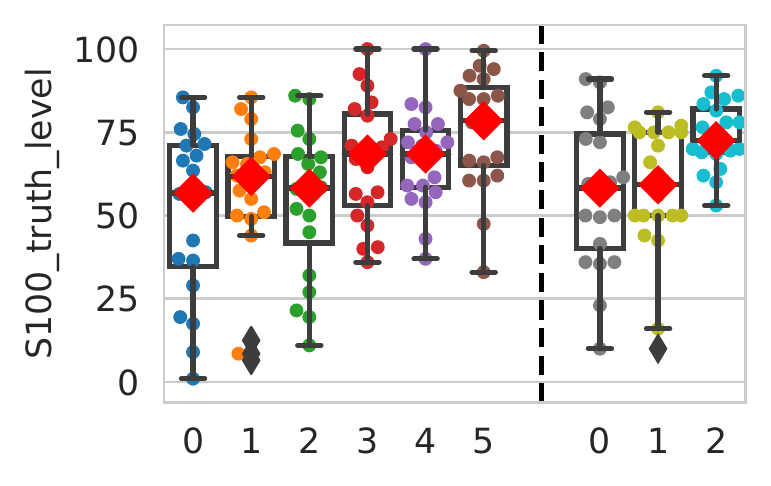}
    \caption{\onehundred}
    \label{cap:paper_sigir2020-sec:results-subsec:scale-adequacy-subsec:alternative-aggregation-fig:alternative-aggregation-median_100}
  \end{subfigure}

  \caption{Agreement with the ground truth using the median as the aggregation function (indicated by the red diamond). See also Figure~\ref{cap:paper_sigir2020-sec:results-subsec:agreement-external-fig:scale-comparison-ground} for comparison.}
  \label{cap:paper_sigir2020-sec:results-subsec:scale-adequacy-subsec:alternative-aggregation-fig:alternative-aggregation-median}
\end{figure}

Figure~\ref{cap:paper_sigir2020-sec:results-subsec:scale-adequacy-subsec:alternative-aggregation-fig:alternative-aggregation-median} shows the results of using the median. In this case, the final truthfulness judgment for each statement has been computed by considering the median of the judgments expressed by the workers. It is clear that the median produces the worst results, by comparing the charts to those in Figure~\ref{cap:paper_sigir2020-sec:results-subsec:agreement-external-fig:scale-comparison-ground} attempts grouping of adjacent categories to improve results quality. 

\begin{figure}[tbp]
  \centering
  \begin{subfigure}{0.49\linewidth}
    \centering
    \includegraphics[width=0.6\linewidth]{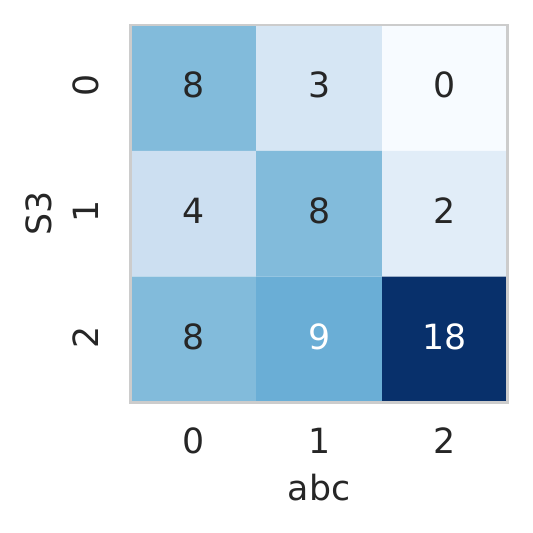}
  \end{subfigure}
  \hfill
  \begin{subfigure}{0.49\linewidth}
    \centering
    \includegraphics[width=\linewidth]{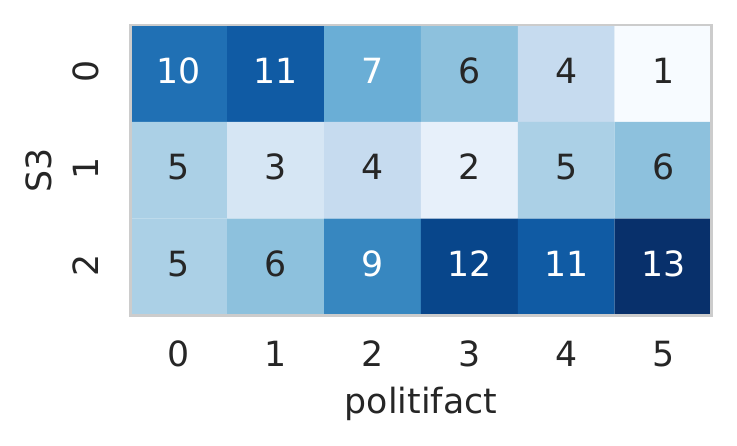}
  \end{subfigure}

  \vspace{1em}

  \begin{subfigure}{0.49\linewidth}
    \centering
    \includegraphics[width=0.6\linewidth]{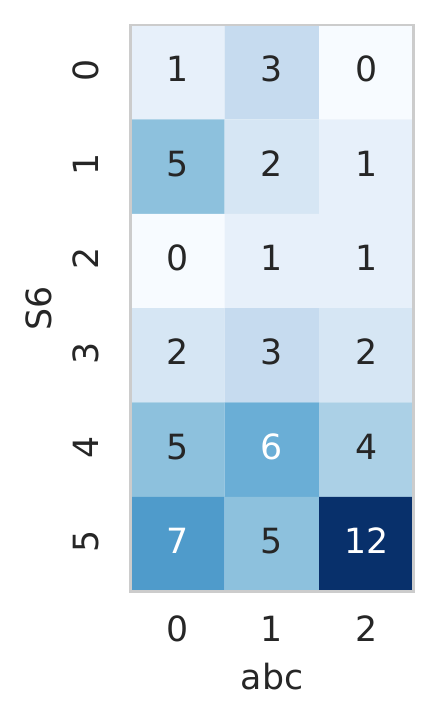}
  \end{subfigure}
  \hfill
  \begin{subfigure}{0.49\linewidth}
    \centering
    \includegraphics[width=\linewidth]{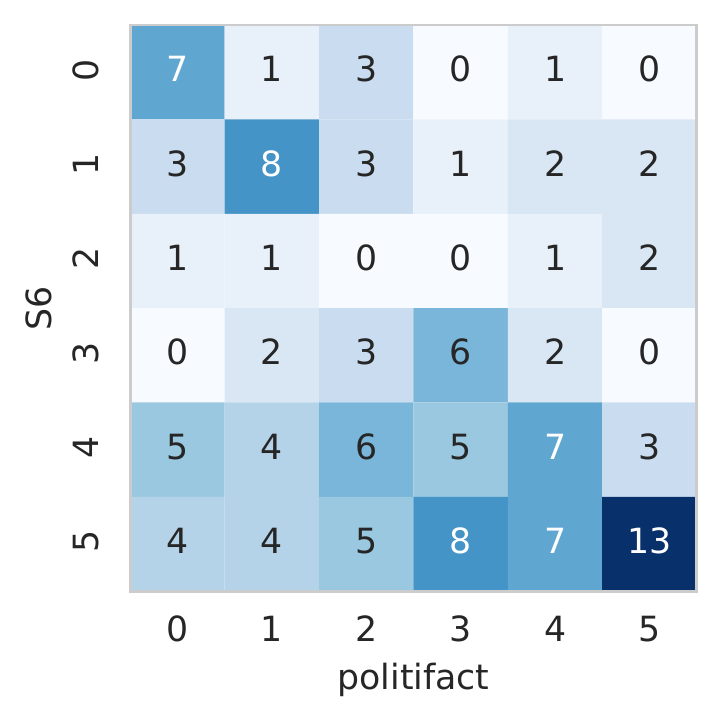}
  \end{subfigure}

  \caption{Agreement between \three (first row) and \six (second row), across \abc (left column) and \politifact (right column). Aggregation function: majority vote.}
  \label{cap:paper_sigir2020-sec:results--subsec:scale-adequacy-subsec:alternative-aggregation-fig:alternative-aggregration}
\end{figure}

The heatmaps in Figure~\ref{cap:paper_sigir2020-sec:results--subsec:scale-adequacy-subsec:alternative-aggregation-fig:alternative-aggregration} show the results of using the majority vote (i.e., the mode) as the alternative aggregation function. The mode is more difficult to compare with the mean, but it is again clear that the overall quality is rather low. Although the squares around the diagonal tend to be darker and contain higher values, there are many exceptions. These are mainly in the lower-left corners, indicating false positives. In other words, statements whose truthfulness value is over-evaluated by the crowd. This tendency to false positives is absent when using the mean (see Figure~\ref{cap:paper_sigir2020-sec:results-subsec:agreement-external-fig:scale-comparison-ground}). Overall, these results confirm that the mean is the most effective aggregation function among those considered.

\subsubsection{Merging Assessment Levels}

\label{cap:paper_sigir2020-sec:results-subsec:scale-adequacy-subsec:merge}

The answer to \ref{cap:paper_sigir2020-sec:research-questions_1} cannot be considered entirely positive, in light of the results presented in Section~\ref{cap:paper_sigir2020-sec:results-subsec:agreement} and Section~\ref{cap:paper_sigir2020-sec:results--subsec:scale-adequacy-subsec:alternative-aggregation}. This conclusion is supported by Figure~\ref{cap:paper_sigir2020-sec:results-subsec:agreement-external-fig:scale-comparison-ground}, Figure~\ref{cap:paper_sigir2020-sec:results-subsec:scale-adequacy-subsec:alternative-aggregation-fig:alternative-aggregation-median}, and Figure~\ref{cap:paper_sigir2020-sec:results--subsec:scale-adequacy-subsec:alternative-aggregation-fig:alternative-aggregration}, as well as by the generally low agreement and correlation values. Although there is a clear signal that the aggregated values resemble the ground truth, several exceptions and misjudged statements remain.
However, some further considerations can be made. First, results tend to be better for \abc than for \politifact. Second, the choice of which scale to adopt remains unclear: the two expert collections used as ground truth are based on different scales, while the experimental setup involves the crowd with \three, \six, and \onehundred, allowing for cross-scale comparisons. Moreover, in many real-world applications, a binary true/false distinction may be sufficient and practically useful. Third, the observed confusion between \politifactzero and \politifactone suggests that these two categories could be effectively merged—an approach already adopted, for example, by \citet{tchechmedjiev2019claimskg}.

\begin{figure}[tbp]
  \centering
  \begin{subfigure}{0.75\linewidth}
    \centering
    \begin{tabular}{@{}c@{}c@{}c@{}}
      \includegraphics[width=.33\linewidth]{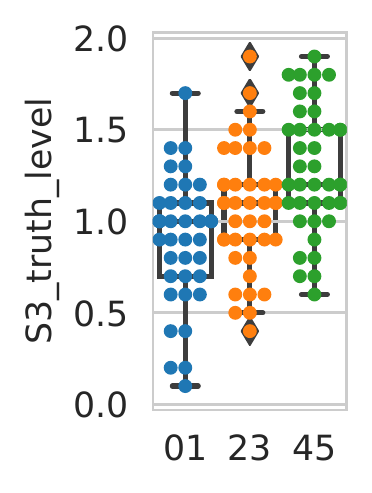} &
      \includegraphics[width=.33\linewidth]{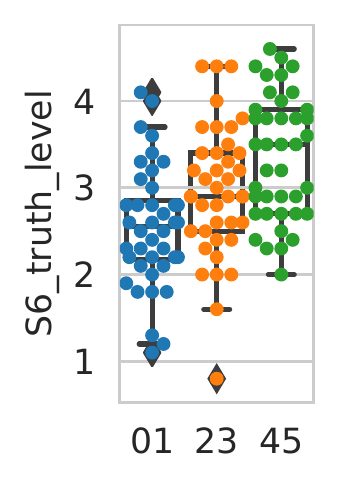} &
      \includegraphics[width=.33\linewidth]{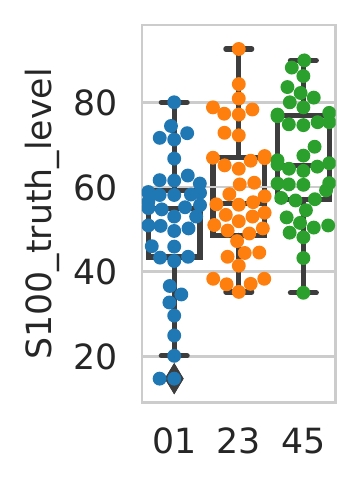}
    \end{tabular}
    \caption{Mean for \six, \three, and \onehundred.}
    \label{cap:paper_sigir2020-sec:results-subsec:scale-adequacy-subsec:merge-fig:alternative-aggregation-binned-3_mean-3}
  \end{subfigure}

  \begin{subfigure}{0.75\linewidth}
    \centering
    \begin{tabular}{@{}c@{}c@{}c@{}}
      \includegraphics[width=.33\linewidth]{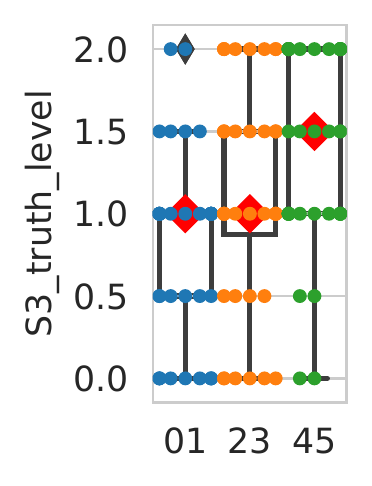} &
      \includegraphics[width=.33\linewidth]{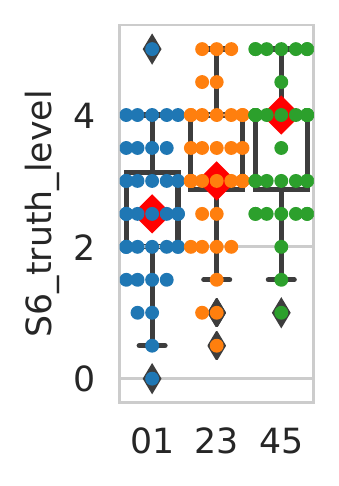} &
      \includegraphics[width=.33\linewidth]{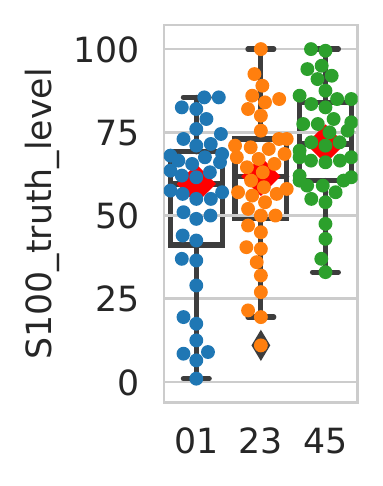}
    \end{tabular}
    \caption{Median for \six, \three, and \onehundred.}
    \label{cap:paper_sigir2020-sec:results-subsec:scale-adequacy-subsec:merge-fig:alternative-aggregation-binned-3_median-3}
  \end{subfigure}

  \begin{subfigure}{0.75\linewidth}
    \centering
    \begin{tabular}{@{}c@{}c@{}c@{}}
      \includegraphics[width=.33\linewidth]{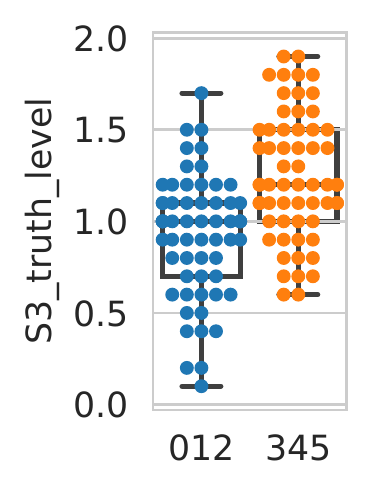} &
      \includegraphics[width=.33\linewidth]{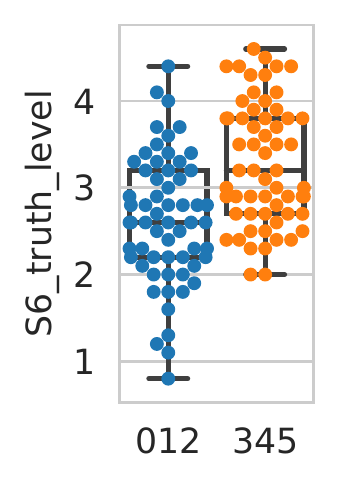} &
      \includegraphics[width=.33\linewidth]{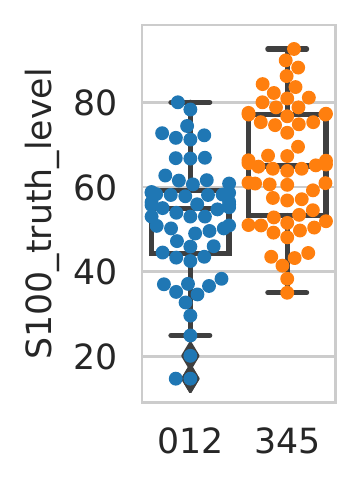}
    \end{tabular}
    \caption{Mean for \six, \three, and \onehundred.}
    \label{cap:paper_sigir2020-sec:results-subsec:scale-adequacy-subsec:merge-fig:alternative-aggregation-binned-2_mean-2}
  \end{subfigure}

  \begin{subfigure}{0.75\linewidth}
    \centering
    \begin{tabular}{@{}c@{}c@{}c@{}}
      \includegraphics[width=.33\linewidth]{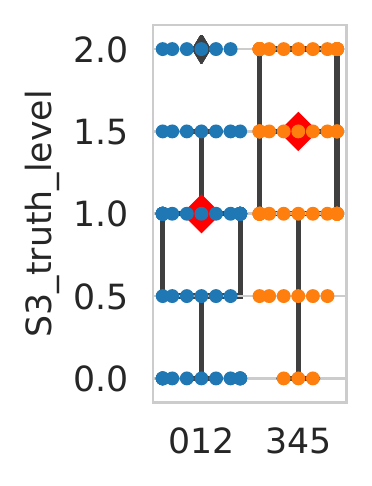} &
      \includegraphics[width=.33\linewidth]{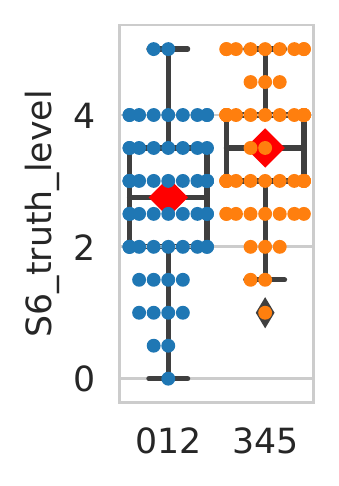} &
      \includegraphics[width=.33\linewidth]{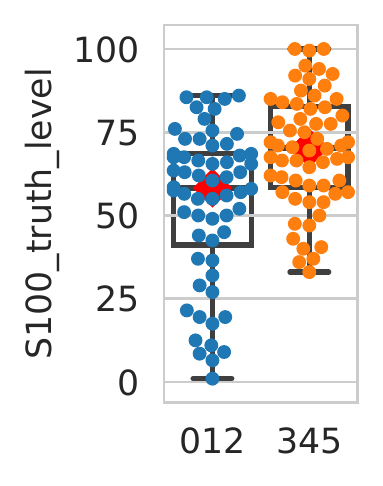}
    \end{tabular}
    \caption{Median for \six, \three, and \onehundred.}
    \label{cap:paper_sigir2020-sec:results-subsec:scale-adequacy-subsec:merge-fig:alternative-aggregation-binned-2_median-3}
  \end{subfigure}
  
  \caption{Agreement with ground truth for merged categories for \politifact. From top to bottom: three and two resulting categories. The median is highlighted by the red diamond.}
  \label{cap:paper_sigir2020-sec:results-subsec:scale-adequacy-subsec:merge-fig:alternative-aggregation-binned-3}
\end{figure}

All these remarks suggest attempting some grouping of adjacent categories, to check if by looking at the data on a more coarse-grained ground truth the results improve. Therefore, the six \politifact categories are grouped into either three (i.e., \politifactthreebinszero, \politifactthreebinsone, and \politifactthreebinstwo) or two (i.e., \politifacttwoebinszero and \politifacttwobinsone) resulting ones, adopting the approach discussed by \citet{scale}. Figure~\ref{cap:paper_sigir2020-sec:results-subsec:scale-adequacy-subsec:merge-fig:alternative-aggregation-binned-3} shows the results. The agreement with the ground truth can now be seen more clearly. The boxplots also seem quite well separated, especially when using the mean (the first three charts on the left). This is confirmed by analyses of statistical significance. 

All the differences in the boxplots on the bottom row are statistically significant at the \index{$p$}$p<.01$ level for both the t-test and Mann–Whitney, both with \index{Bonferroni Correction}Bonferroni correction. The same holds for all the differences between \politifactthreebinszero and \politifactthreebinstwo (the not adjacent categories) in the first row. For the other cases (i.e., concerning the adjacent categories), further statistical significance is found at the \index{$p$}$p<.05$ level in 8 out of 24 possible cases. These results are considerably stronger than those previously reported. They indicate that the crowd is able to distinguish true from false statements with reasonably high accuracy. However, accuracy decreases when evaluating statements that exhibit an intermediate degree of truthfulness or falsehood.

\subsection{\ref{cap:paper_sigir2020-sec:research-questions_3}: Sources Of Information}

\label{cap:paper_sigir2020-sec:results-subsec:justification-distribution}

Table~\ref{cap:paper_sigir2020-sec:results-subsec:justification-distribution-tab:justification-distribution-sheet} reports the distribution of websites used by workers to justify the truthfulness judgments they provide for each statement. The most frequently used sources across all scales are \lq\lq Wikipedia\rq\rq{} and \lq\lq YouTube\rq\rq{}, followed by widely recognized news outlets such as \lq\lq The Guardian\rq\rq{} and \lq\lq The Washington Post\rq\rq{}. Notably, one fact-checking website (i.e., \textit{FactCheck}) also ranks among the most frequently cited sources. URLs from \url{abc.com.au} and \url{politifact.com} were intentionally excluded from the list of selectable sources during the crowdsourcing task. These results indicate that workers, supported by the search engine, tend to select credible information sources to justify their judgments.

Table~\ref{cap:paper_sigir2020-sec:results-subsec:justification-distribution-tab:justification-distribution-rank} presents the distribution of the ranks in the search engine results for the URLs selected by workers (excluding gold questions), for \three, \six, and \onehundred. As expected~\cite{craswell2008experimental, joachims2017accurately, kelly2015how}, workers tend to prefer higher-ranked results, particularly those in the top three positions. However, the distribution also shows that workers consider the top ten results, rather than selecting only the first link, which suggests a degree of effort in identifying a reliable source or justification. Notably, across all scales, no worker selects URLs ranked beyond the tenth position; all activity is confined to the first page of search results.

\begin{table}[tbp]
\centering
\caption{Websites selected by workers to justify their judgments, excluding gold questions, for \three, \six, and \onehundred. Only websites with a percentage $\geq 1\%$ are reported.}
\begin{tabular}{lccc}
\toprule
\textbf{Source} & \three & \six & \onehundred \\
\midrule
Wikipedia          & 17\% & 19\% & 23\% \\  
YouTube            & 13\% & 13\% & 12\% \\  
The Guardian       & 11\% & 12\% & 13\% \\
FactCheck          & 8\%  & 8\%  & 9\% \\
Smh                & 6\%  & 3\%  & 6\% \\
Cleveland          & 6\%  & 6\%  & 5\% \\
Washington Post    & 6\%  & 6\%  & 6\% \\
News               & 5\%  & 4\%  & 5\% \\
Blogspot           & 5\%  & 6\%  & 5\% \\
On The Issues      & 4\%  & 0\%  & 0\% \\
Quizlet            & 4\%  & 3\%  & 4\% \\
New York Times     & 3\%  & 3\%  & 0\% \\
CBS News           & 3\%  & 4\%  & 5\% \\
Forbes             & 3\%  & 3\%  & 0\% \\  
House              & 3\%  & 3\%  & 3\% \\  
Madison            & 3\%  & 0\%  & 0\% \\  
The Australian     & 0\%  & 4\%  & 0\% \\  
Milwaukee Journal  & 0\%  & 3\%  & 0\% \\    
Yahoo              & 0\%  & 0\%  & 3\% \\  
\bottomrule
\end{tabular}
\label{cap:paper_sigir2020-sec:results-subsec:justification-distribution-tab:justification-distribution-sheet}
\end{table}

\begin{table}[tbp]
\centering
\caption{Distribution of search result ranks for URLs selected by workers in \three, \six, and \onehundred.}
\begin{tabular}{ccccc}
\toprule
\textbf{Rank} & \three & \six & \onehundred & Avg. \\
\midrule
1  & 17\% & 13\% & 15\% & 15\% \\
2  & 12\% & 13\% & 15\% & 13\% \\
3  & 13\% & 16\% & 15\% & 15\% \\
4  & 14\% & 12\% & 12\% & 11\% \\
5  & 12\% & 11\% & 8\% & 10\% \\
6  & 9\% & 9\% & 12\% & 10\% \\
7  & 7\% & 8\% & 7\% & 7\% \\
8  & 6\% & 7\% & 6\% & 6\% \\
9  & 4\% & 5\% & 5\% & 5\% \\
10 & 3\% & 4\% & 2\% & 3\% \\
11 & 1\% & 1\% & 1\% & 1\% \\
\bottomrule
\end{tabular}
\label{cap:paper_sigir2020-sec:results-subsec:justification-distribution-tab:justification-distribution-rank}
\end{table}

\subsection{\ref{cap:paper_sigir2020-sec:research-questions_4}: Effect Of Worker Background and Bias} 

\label{cap:paper_sigir2020-sec:results-subsec:worker-bias}

The role of assessors' background in objectively identifying online misinformation can be understood by assessing the relationships that exist between workers' cognitive performances (Section~\ref{cap:paper_sigir2020-sec:results-subsec:worker-bias-subsec:crt}) and their background (Section~\ref{cap:paper_sigir2020-sec:results-subsec:worker-bias-subsec:pol-background}).

\subsubsection{Cognitive Reflection Tests} 

\label{cap:paper_sigir2020-sec:results-subsec:worker-bias-subsec:crt}

Performance on the Cognitive Reflection Test (\index{Cognitive!Reflection Test}CRT) is measured as the percentage of correct answers provided by each worker, serving as an estimate of cognitive skills. A higher \index{Cognitive!Reflection Test}CRT score indicates greater analytical thinking ability~\cite{Frederick2005}. Performance is compared across the three scales using the standardized \index{$z$-score}$z$-score, computed for each worker and each truthfulness level. The \index{$z$-score}$z$-score for each statement reflects how a worker's performance compares to others. A lower \index{$z$-score}$z$-score for false statements indicates a stronger ability to identify misinformation, while a higher \index{$z$-score}$z$-score for true statements suggests a better capacity to recognize accurate information. \index{Discernment}\lq\lq Discernment\rq\rq{} is then defined as the difference between the \index{$z$-score}$z$-score for true and false statements, representing the crowd's ability to distinguish truth from falsehood~\cite{pennycook2019lazy}. This analysis focuses on statements labeled as either true or false in the ground truth, excluding \abcinbetween statements, which do not contribute additional evidence regarding the crowd's discernment.

Table~\ref{cap:paper_sigir2020-sec:results-subsec:worker-bias-subsec:crt-tab:correlation} shows the results. First, there is a statistically significant (Spearman's rank-order test), moderate positive correlation between Discernment and \index{Cognitive!Reflection Test}CRT score on statements from \politifact and \abc (\index{$r_s$}$r_s(598)= 0.128$, \index{$p$}$p = 0.002$ and \index{$r_s$}$r_s(598)= 0.11$, \index{$p$}$p = 0.007$ respectively). This shows that workers who ponder more perform better in identifying \politifactlie statements of US (local) politicians (\index{$r_s$}$r_s(598)= -0.098$, \index{$p$}$p = 0.017$), and identifying true statements of AU (not local) politicians (\index{$r_s$}$r_s(598)= 0.11$, \index{$p$}$p = 0.007$). In general, people with strong analytical abilities (as determined by the \index{Cognitive!Reflection Test}CRT test) can better recognize true statements from false (\index{$r_s$}$r_s(598)= 0.154$, \index{$p$}$p<0.0005$). Besides, the ability to distinguish truthfulness from falsehood increases with age (\index{$r_s$}$r_s(598)= 0.125$, $ p=0.002$). Indeed, older workers perform better in recognizing true statements by US politicians (\index{$r_s$}$r_s(598)= 0.127$, \index{$p$}$p=0.02$). The level of education and income do not have a statistically significant correlation with their judgments.

\begin{table}[tbp]
\centering
\caption{Correlation between Cognitive Reflection Test (\index{Cognitive!Reflection Test}CRT) performance and \index{$z$-score}$z$-scores across different truthfulness levels, along with the correlation between worker age and \index{$z$-score}$z$-scores.}
\label{cap:paper_sigir2020-sec:results-subsec:worker-bias-subsec:crt-tab:correlation}
\begin{threeparttable}
\begin{tabular}{llSS}
\toprule
{\textbf{Dataset}} & {\begin{tabular}[c]{@{}c@{}}\textbf{Correlation} \\ \textbf{With}\end{tabular}} & {\textbf{Age}} & {\begin{tabular}[c]{@{}c@{}}\textbf{CRT} \\ \textbf{Performance}\end{tabular}} \\ 
\midrule
\multirow{4}{*}{\politifact}& \politifactlie & -0.038 & -0.098$^{\ast}$ \\
& \politifactfalse & -0.022 & -.072 \\
& \politifacttrue & .127$^{\ast} $ & 0.062 \\
& Discernment & .113$^{\ast\ast} $ & .128$^{\ast\ast}$ \\
\midrule
\multirow{3}{*}{\abc} & \abcnegative & -0.075 & -0.021 \\
& \abcpositive & 0.048 & .110$^{\ast\ast}$ \\
& Discernment & 0.088$^\ast $ & .110$^{\ast\ast}$ \\
\midrule
Total & Discernment & .125$^{\ast\ast}$  & .154$^{\ast\ast}$ \\ 
\bottomrule
\end{tabular}
\begin{tablenotes}
{\small \item $^{\ast\ast}$: $p < 0.01 $, $^\ast$: $p <0.05$.}
\end{tablenotes}
\end{threeparttable}
\end{table}

\subsubsection{Political Background} 

\label{cap:paper_sigir2020-sec:results-subsec:worker-bias-subsec:pol-background}

Concerning the effect of the political background, the Shapiro-Wilk test  \cite{shapiro1972approximate} confirms that discernment scores are normally distributed (\index{$p$}$p > 0.05$) for groups with diverse political views. In light of this, a one-way \index{ANOVA}ANOVA analysis can be performed to determine whether the ability to distinguish true from false statements is different across groups. Levene's test~\cite{schultz1985levene} shows that the homogeneity of variances is violated (\index{$p$}$p = 0.034$). Thus, a Welch-Satterthwaite correction \cite{satterthwaite1946approximate} is used to calculate the degrees of freedom and the Games-Howell posthoc test \cite{ruxton2008time} to show multiple comparisons. Discernment score is statistically significantly different between different political views (Welch's \index{$F$}$F(4, 176.735) = 3.451$, \index{$p$}$p = 0.01$). A Games-Howell posthoc test confirms that the increase of discernment score (\num{0.453}, 95\% CI (\num{0.028} to \num{0.879}) from \conservative (\num{-0.208} ± \num{1.293}) to \liberal (\num{0.245} ± \num{1.497}) is statistically significant (\index{$p$}$p = 0.03$). In light of these results, the crowd workers who have liberal views can better differentiate between false and true statements. Furthermore, there is no statistically significant difference in discernment scores based on the political party with which crowd workers explicitly identified themselves ($x^2(3)= 3.548$, $ p= 0.315$). Since a Shapiro-Wilk test \cite{shapiro1972approximate} shows a non-normal distribution \index{$p$}$p < 0.05$, a Kruskal-Wallis H test \cite{kruskal1952use} is used. This shows there is no difference in judgment quality based on workers' explicit political stances. 

However, an analysis of workers' implicit political views rather than their explicit party identification shows a different result. The partisan gap on the immigration issue is apparent in the US. According to the survey conducted by Pew Research Center in January 2019, \cite{BorderWall}, about 82\% of Republicans and Republican-leaning independents support developing the barrier along the southern border of the United States, while 93\% of Democrats and Democratic leaners oppose it. Therefore, asking workers' opinions on this matter can be a way to know their implicit political orientation. A Kruskal-Wallis H test \cite{kruskal1952use} is used --Shapiro-Wilk test's \index{$p$}$p < 0.05$-- in workers' discernment between immigration policy groups defined based on their answer to the wall question. Statistically significant differences are observed for \politifact statements ($x_2 (2) = 10.965, p = 0.004$) and for the whole set of statements ($x_2 (2) = 11.966, p = 0.003$). A posthoc analysis using Dunn's procedure with a \index{Bonferroni Correction}Bonferroni correction  \cite{dunn1964multiple} reveals statistically significant differences in discernment scores on \politifact statements between agreeing (\num{-0.335}) and disagreeing (\num{0.245}) (\index{$p$}$p = 0.007$) on building a wall. Similar results are obtained when discernment scores are compared on the whole set of statements. There are statistically significant differences in discernment scores between agreeing (\num{-0.377}) and disagreeing (\num{0.173}) (\index{$p$}$p = 0.002$) on building a wall along the southern border of the USA. These results show how, in this experimental setup, crowd workers who do not want a wall on the southern US border perform better in distinguishing between true and false statements. Lastly, there are no significant differences concerning workers' stances on climate issues.

\section{Summary}

\label{cap:paper_sigir2020-sec:discussion}

This chapter presents an extensive crowdsourcing experiment designed to study how crowd workers identify misinformation online. The dataset includes statements made by politicians from the United States and Australia. US-based crowd workers are asked to perform fact-checking tasks using a customized and controllable Internet search engine to find evidence supporting or refuting the statements. 

The setup enables the collection and analysis of data related to workers' political backgrounds and cognitive abilities. In addition, it allows for control over several experimental factors, including political alignment of the statements, their geographical relevance, the granularity of the judgment scale, and the truthfulness level. The answers to the research questions can be summarized as follows.

\myparagraph{\ref{cap:paper_sigir2020-sec:research-questions_1}} The behavior across all three scales is similar for both \politifact and \abc statements in terms of agreement with the ground truth. There is a low level of internal agreement among workers on the \three, \six, and \onehundred scales, across all judgments collected via crowdsourcing.

\myparagraph{\ref{cap:paper_sigir2020-sec:research-questions_2}} The grouping of adjacent categories reveals that crowdsourced truthfulness judgments are useful for accurately distinguishing true from false statements.

\myparagraph{\ref{cap:paper_sigir2020-sec:research-questions_3}} Workers put effort into finding reliable sources to justify their judgments and tend to choose a source from the first search engine results page, though not necessarily the top-ranked result.

\myparagraph{\ref{cap:paper_sigir2020-sec:research-questions_4}} Assessors' backgrounds affect their ability to objectively identify online misinformation.

\myparagraph{}

The next chapter investigates whether crowd workers are able to detect and objectively categorize recent (mis)information. To this end, statements related to the \covid pandemic are used, and a longitudinal study is conducted.

\chapter{A Longitudinal Study On Misinformation About \covid}

\label{cap:paper_pauc2021}

This chapter is based on the article published in the \lq\lq Personal and Ubiquitous Computing\rq\rq{} journal~\cite{roitero2021crowd}. It is an extension of the work presented at the 29th ACM International Conference on Information and Knowledge Management~\cite{roitero2020covid}. Section~\ref{cap:related_work-sec:crowdsourcing-truthfulness}, Section~\ref{cap:related_work-sec:info-recency}, and Section~\ref{cap:related_work-sec:worker-bias} describe the relevant related work. Section~\ref{cap:paper_pauc2021-sec:research-questions} details the research questions, which are addressed using the experimental setting described in Section~\ref{cap:paper_pauc2021-sec:exp-setup}. Section~\ref{cap:paper_pauc2021-sec:exp-setup-subsec:descriptive-statistics} provides an initial descriptive analysis, while Section~\ref{cap:paper_pauc2021-sec:results} presents the results obtained and reports on the longitudinal study conducted. Finally, Section~\ref{cap:paper_pauc2021-sec:discussion} summarizes the main findings and concludes the chapter.

\section{Research Questions}

\label{cap:paper_pauc2021-sec:research-questions}

This chapter studies how non-expert human judges perceive online (mis)information that is recent in time, by focusing on \covid-related statements. The pandemic was a hot topic in 2020, yet no studies had employed crowdsourcing to assess the truthfulness of related statements, despite the vast number of research efforts devoted to the topic worldwide. Moreover, the health domain is particularly sensitive, making it relevant to assess whether crowdsourcing-based approaches are suitable in such a context.

In previous work~\cite{la2020crowdsourcing, RSDM:2018, roitero2020crowd}, the statements judged by the crowd were not recent, meaning that evidence about their truthfulness was often already available online. While the experimental design limited easy access to that evidence, it is still possible that workers were familiar with the statements—e.g., due to prior media coverage. By focusing on \covid-related statements, we naturally target more recent content. In some cases, evidence may still be available, but this occurs less frequently.
An almost ideal tool to counter misinformation would be a crowd capable of judging truthfulness in real time, immediately after a statement is made public. Targeting recent statements is a step in that direction, although significant work remains. The experimental design differs in some aspects from prior studies and enables the investigation of new research questions. Lastly, a longitudinal study is conducted. It involves collecting data at multiple time points by launching the task at different timestamps and engaging both novice workers—i.e., those who have never completed the task before—and experienced workers—i.e., those who participated in previous batches and were invited to take part again. This setup allows us to study various behavioral aspects of workers engaged in truthfulness assessment.

The experiments focus on statements about \covid, as they are both recent and of interest to the research community, and arguably concern a more relevant and sensitive topic for the crowd workers. They investigate whether the health domain influences the ability of crowd workers to identify and correctly classify (mis)information, and whether the recency of \covid-related statements has an impact as well. The experiments adopt a single truthfulness scale, based on prior evidence that the specific scale used does not significantly affect the results.
Another important difference compared to the setting described in Section~\ref{cap:paper_sigir2020} and previous work~\cite{la2020crowdsourcing, RSDM:2018} is that workers are asked to provide a textual justification for their decisions. These justifications are analyzed to better understand the verification strategies employed by the workers and to assess whether they can be leveraged to extract useful information.
The longitudinal study consists of three crowdsourcing experiments conducted over a period of four months. It enables the collection of additional data, including new responses from both novice and returning workers. The behavior of the workers is also analyzed, as in Section~\ref{cap:paper_sigir2020-sec:results}. The following research questions are investigated:

\begin{enumerate}[start=5, leftmargin=2.4em, label=RQ\arabic*]
\item \label{cap:paper_pauc2021-sec:research-questions_1} Are crowd workers able to detect and objectively categorize online (mis)\-in\-for\-ma\-tion related to the medical domain, and specifically to recent \covid content? What are the relationships and levels of agreement between crowd and expert labels?
\item \label{cap:paper_pauc2021-sec:research-questions_2} Can crowdsourced and/or expert judgments be transformed or aggregated to improve workers' ability to detect and objectively categorize online (mis)information?
\item \label{cap:paper_pauc2021-sec:research-questions_3} What is the effect of workers' political bias and cognitive abilities on their performance?
\item \label{cap:paper_pauc2021-sec:research-questions_4} What signals do workers generate while performing the task that can be recorded? To what extent are these signals related to workers' accuracy? Can these signals be exploited to improve accuracy, for instance, by enabling more effective judgment aggregation?
\item \label{cap:paper_pauc2021-sec:research-questions_5} Which sources of information do crowd workers consider when identifying online misinformation? Are some sources more useful? Do some sources lead to more accurate and reliable judgments?
\item \label{cap:paper_pauc2021-sec:research-questions_6} What is the effect of re-running the experiment and re-collecting all data at different time intervals? Do the findings from the previous research questions still hold?
\item \label{cap:paper_pauc2021-sec:research-questions_7} How does including judgments from workers who completed the task multiple times affect the findings of those in \ref{cap:paper_pauc2021-sec:research-questions_6}? Do these workers differ from those who participated only once?
\item \label{cap:paper_pauc2021-sec:research-questions_8} For which statements does crowdsourcing fail to produce accurate truthfulness judgments? What are the features of the statements that are misjudged by crowd workers?
\end{enumerate}

\section{Experimental Setting}

\label{cap:paper_pauc2021-sec:exp-setup}

The experimental setup involves statements sampled from \politifact (Section~\ref{cap:dataset-sec:politifact}). In more detail, 10 statements for each of the six \politifact categories are selected. Such statements belong to the \covid section and with dates ranging from February 2020 to early April 2020. The sample includes statements by politicians belonging to the two main US parties (Democratic and Republican). A balanced number of statements per class and per political party is included in the sample. Appendix~\ref{cap:paper_pauc2021:-appendix:statements} contains the full list of the statements used.

\subsection{Crowdsourcing Task}

\label{cap:paper_pauc2021-sec:exp-setup-subsec:crowdsourcing-task}

The task design used to collect truthfulness judgments about \covid, and subsequently to perform the longitudinal study, is similar to the one described in Section~\ref{cap:paper_sigir2020-sec:exp-setup-subsec:crowdsourcing-task}. The crowdsourcing platform \mturk is used to collect the judgments. Each worker is assigned a unique pair of values (input token, output token). Then, they are redirected to an external website (Appendix~\ref{cap:paper_wsdm2022}) to complete the task.

The task itself is as follows. First, a demographic questionnaire (Appendix~\ref{cap:paper_sigir2020-appendix:quest-crt-sec:initial}) is shown to the worker to collect background information such as age and political views. Next, the worker is asked to answer three Cognitive Reflection Test (CRT) questions (Appendix~\ref{cap:paper_sigir2020-appendix:quest-crt-sec:crt}). These questionnaires are the same as those described in Section~\ref{cap:paper_sigir2020-sec:exp-setup-subsec:crowdsourcing-task}.
The worker is then asked to judge the truthfulness of 8 statements: 6 drawn from the dataset described in Section~\ref{cap:dataset-sec:politifact} (one for each of the six \politifact categories), and 2 special statements referred to as \lq\lq gold questions\rq\rq{} (one clearly true and one clearly false), which are manually written and used as quality checks. A randomization process is applied when building the \index{HIT}HITs to minimize all possible sources of bias, both within each \index{HIT}HIT and across the overall task.
For each statement, the worker is shown the \emph{Statement}, the \emph{Speaker/Source}, and the \emph{Year} in which the statement was made. Workers are then asked to provide the following information:

\begin{itemize}[label=--]
\item The truthfulness judgment for the statement, using the six-level scale adopted by \politifact, hereafter referred to as \crowdsix. The scale is presented to the worker using radio buttons, with each option labeled according to the category descriptions from the official \politifact website.
\item The URL used as a source of information for the fact-checking.
\item A textual justification for their judgment. The justification must not include the URL and should contain at least 15 words.
\end{itemize}
To prevent workers from using \politifact as their primary source of evidence, its domain is filtered out from the returned search results. The following quality checks are implemented in the task:
\begin{itemize}[label=--]
\item The judgments provided for the gold questions must be coherent; specifically, the judgment for the clearly false statement must be lower than that for the clearly true one.
\item The cumulative time spent on each judgment must be at least 10 seconds.
\end{itemize}

The \index{Cognitive!Reflection Test}CRT and the questionnaire answers are not used for quality checks, although the workers were not informed of this. If a worker successfully completes the assigned \index{HIT}HIT, they are shown the output token. This token is used to submit the \index{HIT}HIT and receive payment, which is set at \$1.50 for evaluating 8 statements. The average time required to complete the task was investigated prior to publishing it, to ensure alignment with the U.S. federal minimum hourly wage.

Overall, the task involves 60 statements in total, and each statement is evaluated by 10 distinct workers. Thus, 100 MTurk \index{HIT}HITs are deployed for the main experiment, leading to the collection of 800 judgments in total (600 statement judgments plus 200 gold question answers). In total, over \num{4300} judgments from \num{542} workers across 7 batches of the crowdsourcing task are collected, considering both the main experiment and the longitudinal study.
The choice to have each statement evaluated by 10 distinct workers deserves further discussion. This number is consistent with previous studies that used crowdsourcing to assess truthfulness~\cite{la2020crowdsourcing, RSDM:2018, roitero2020crowd, roitero2020covid} as well as related tasks such as relevance assessment~\cite{Maddalena:2017:CRM:3026478.3002172, Roitero:2018:FRS:3209978.3210052}. It represents a reasonable trade-off between evaluating fewer statements with higher redundancy and evaluating more statements with lower redundancy.
 
\subsection{Longitudinal Study}

\label{cap:paper_pauc2021-sec:exp-setup-subsec:longitudinal-study}

The longitudinal study is based on the same dataset and experimental setting as the main experiment~\cite{roitero2020covid}. The data for the main experiment (hereafter denoted as \batchone) were collected in May 2020. The \index{HIT}HITs from \batchone were republished in June 2020 for a new set of workers (i.e., workers from \batchone were prevented from participating again), resulting in the dataset denoted as \batchtwo.
Additional judgments were collected in July 2020. The \index{HIT}HITs from \batchone were again republished, this time ensuring that workers from both \batchone and \batchtwo could not participate. The resulting dataset is denoted as \batchthree. Finally, a fourth round of data collection took place in August 2020. As before, workers from the previous three batches were prevented from participating. The resulting dataset is denoted as \batchfour.

Then, another set of experiments is conducted. For each batch, the workers who participated in the previous one are contacted by sending them a \$0.01 bonus along with an invitation to perform the task again. Table~\ref{cap:paper_pauc2021-sec:exp-setup-subsec:longitudinal-study-tab:longitudinal_nomenclature} describes the resulting datasets, where \texttt{BatchX$_{\mbox{\scriptsize{fromY}}}$} denotes the subset of workers who completed \texttt{BatchX} and had previously participated in \texttt{BatchY}. Note that an experienced (i.e., returning) worker completing the task for a second time is generally assigned a new \index{HIT}HIT, different from the one originally completed. At the time of data collection, this assignment process could not be controlled, as \index{HIT}HITs were distributed by the \mturk platform.
Finally, the union of the data from \batchone, \batchtwo, \batchthree, and \batchfour is also considered. The resulting dataset is denoted as \batchall.

\begin{table}[tbp]
\centering
\caption{Experimental setting for the longitudinal study. All dates refer to the year 2020. The table reports absolute values.}
\label{cap:paper_pauc2021-sec:exp-setup-subsec:longitudinal-study-tab:longitudinal_nomenclature}
\begin{tabular}{llrrrrr}
\toprule
 & & \multicolumn{4}{c}{\textbf{Number of Workers}} & \\
\cmidrule(lr){3-6}
\textbf{Date} & \textbf{Acronym} & \textbf{\batchone} & \textbf{\batchtwo} & \textbf{\batchthree} & \textbf{\batchfour} & \textbf{Total} \\
\midrule
May    & \batchone                     & 100 & --  & --  & --  & 100 \\
\midrule
June   & \batchtwo                     & --  & 100 & --  & --  & 100 \\
       & \batchtwofromone              & 29  & --  & --  & --  & 29  \\
\midrule
July   & \batchthree                   & --  & --  & 100 & --  & 100 \\
       & \batchthreefromone           & 22  & --  & --  & --  & 22  \\
       & \batchthreefromtwo           & --  & 20  & --  & --  & 20  \\
       & \batchthreefromoneortwo      & 22  & 20  & --  & --  & 42  \\
\midrule
August & \batchfour                   & --  & --  & --  & 100 & 100 \\
       & \batchfourfromone            & 27  & --  & --  & --  & 27  \\
       & \batchfourfromtwo            & --  & 11  & --  & --  & 11  \\
       & \batchfourfromthree          & --  & --  & 33  & --  & 33  \\
       & \batchfourfromoneortwoorthree& 27  & 11  & 33  & --  & 71  \\
\midrule
       & \batchall                    & 100 & 100 & 100 & 100 & 400 \\
\bottomrule
\end{tabular}
\end{table}

\section{Descriptive Analysis}

\label{cap:paper_pauc2021-sec:exp-setup-subsec:descriptive-statistics}

Section~\ref{cap:paper_pauc2021-sec:exp-setup-subsec:demographics} presents the demographic characteristics of the workers, while Section~\ref{cap:paper_pauc2021-sec:exp-setup-subsec:task-abandonment} analyzes the task abandonment rate.

\subsection{Worker Demographics}

\label{cap:paper_pauc2021-sec:exp-setup-subsec:demographics}

Overall, \num{334} workers residing in the United States participated in the main experiment. The following demographic statistics are derived from the questionnaire responses of the workers who successfully completed the task.

The majority of workers fall within the 26--35 age range (39\%), followed by the 19--25 (27\%) and 36--50 (22\%) age ranges. Most workers are well educated: 48\% hold a four-year college or bachelor's degree, 26\% have a college degree, and 18\% possess a postgraduate or professional degree. Only about 4\% of workers have a high school diploma or less.

Regarding political views, 33\% of workers identify as \liberal, 26\% as \moderate, 17\% as \veryliberal, 15\% as \conservative, and 9\% as \veryconservative. In terms of party affiliation, 52\% identify as \democratic, 24\% as \republican, and 23\% as \independent. Additionally, 50\% of workers disagree with building a wall on the southern U.S. border, while 37\% agree. Overall, the sample is relatively well balanced.

The analysis of the \index{Cognitive!Reflection Test}CRT scores shows that 31\% of workers answered zero questions correctly, 34\% answered one correctly, 18\% answered two correctly, and only 17\% answered all three correctly. The results of the \index{Cognitive!Reflection Test}CRT and worker quality are analyzed in relation to research question~\ref{cap:paper_pauc2021-sec:research-questions_3}.

\subsection{Task Abandonment}

\label{cap:paper_pauc2021-sec:exp-setup-subsec:task-abandonment}

The abandonment rate is measured according to the definition provided by \citet{8873609}. Among the \num{334} workers who accepted the task, 100 (approximately 30\%) successfully completed it, 188 (approximately 56\%) abandoned it (i.e., voluntarily terminated the task before completion), and 46 (approximately 14\%) failed (i.e., were unable to complete the task due to repeatedly failing quality checks).

Furthermore, 115 out of the 188 workers who abandoned the task (approximately 61\%) did so before judging the first statement, i.e., before effectively starting the task. This early abandonment rate is consistent with that observed in similar tasks, as reported in Section~\ref{cap:paper_sigir2020-sec:desc-stat} and Section~\ref{cap:paper_ipm2021-sec:exp-setup-subsec:descriptive-statistics}.

\section{Results}

\label{cap:paper_pauc2021-sec:results}

Crowd accuracy is addressed in Section~\ref{cap:paper_pauc2021-sec:results-subsec:crowd-accuracy}.
Section~\ref{cap:paper_pauc2021-sec:results-subsec:transforming-scales} discusses the transformation of judgment scales.
Section~\ref{cap:paper_pauc2021-sec:results-subsec:worker-background-bias} examines the impact of workers' backgrounds and biases, while worker behavior is analyzed in Section~\ref{cap:paper_pauc2021-sec:results-subsec:worker-behavior}.
Section~\ref{cap:paper_pauc2021-sec:results-subsec:info-source} explores workers' use of information sources.
The following sections focus on the longitudinal study.
Section~\ref{cap:paper_pauc2021-sec:results-subsec:novice-workers} investigates the impact of repeating the experiment with novice workers,
Section~\ref{cap:paper_pauc2021-sec:results-subsec:returning-workers} analyzes the effect of returning workers,
and Section~\ref{cap:paper_pauc2021-sec:results-subsec:misjudged} studies whether the same statements tend to be misjudged across different batches.

Before presenting the detailed discussion of the results, three considerations drawn from Section~\ref{cap:paper_pauc2021-sec:results-subsec:crowd-accuracy} are summarised to clarify the rationale for the order of the remaining subsections.
The answer to \ref{cap:paper_pauc2021-sec:research-questions_1} is largely positive; nevertheless, important caveats remain. First, workers frequently confuse the \politifactpantsfire and \politifactfalse categories. Second, although \politifact employs a six-level scale, many truthfulness studies adopt two- or three-level scales (see Chapter~\ref{cap:paper_sigir2020}). Third, merging adjacent categories improves accuracy, as demonstrated by
\citet{tchechmedjiev2019claimskg}. These observations motivate the scale-transformation analyses undertaken for \ref{cap:paper_pauc2021-sec:research-questions_2} and provide context for the subsequent exploration of information sources in \ref{cap:paper_pauc2021-sec:research-questions_5}.

\subsection{\ref{cap:paper_pauc2021-sec:research-questions_1}: Crowd Workers Accuracy}

\label{cap:paper_pauc2021-sec:results-subsec:crowd-accuracy}

Assessing whether workers can identify misinformation about \covid requires a\-na\-ly\-sing the quality of the judgments provided. Accordingly, Section~\ref{cap:paper_pauc2021-sec:results-subsec:crowd-accuracy-subsec:ext-agreement} examines the external agreement, while Section~\ref{cap:paper_pauc2021-sec:results-subsec:crowd-accuracy-subsec:int-agreement} addresses the internal agreement.

\subsubsection{External Agreement}

\label{cap:paper_pauc2021-sec:results-subsec:crowd-accuracy-subsec:ext-agreement}

\begin{figure}[tbp]
  \centering
  \begin{subfigure}{.49\linewidth}
    \centering
    \includegraphics[width=.95\linewidth]{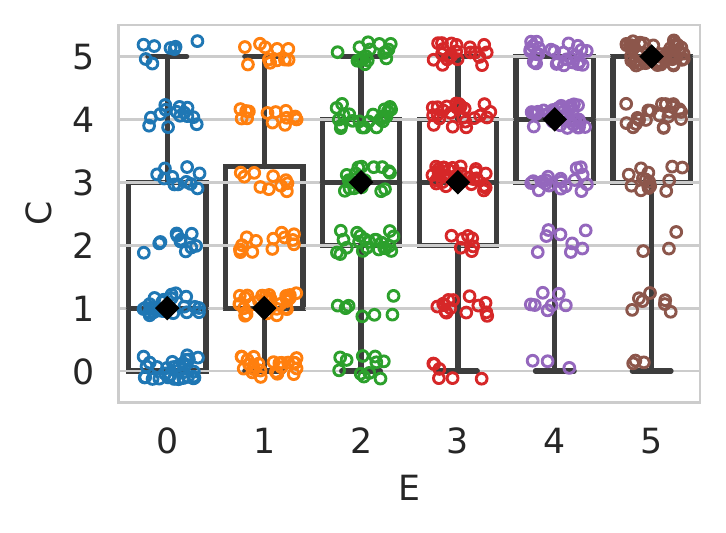}
    \caption{Individual judgments.}
    \label{cap:paper_pauc2021-sec:results-subsec:crowd-accuracy-subsec:ext-agreement-fig:agreement-ground-truth_6}
  \end{subfigure}
  \hfill
  \begin{subfigure}{.49\linewidth}
    \centering
    \includegraphics[width=.95\linewidth]{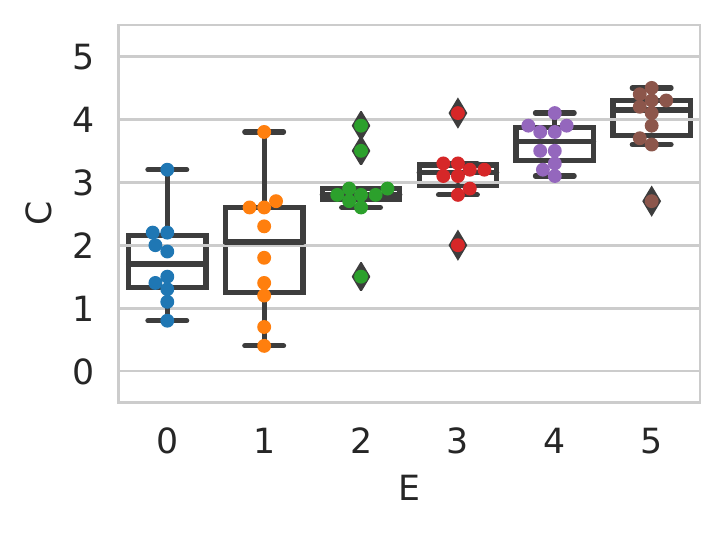}
    \caption{Judgments aggregated using the mean.}
    \label{cap:paper_pauc2021-sec:results-subsec:crowd-accuracy-subsec:ext-agreement-fig:agreement-ground-truth_6-mean}
  \end{subfigure}
  
  \vspace{1em}
  
  \begin{subfigure}{.49\linewidth}
    \centering
    \includegraphics[width=.95\linewidth]{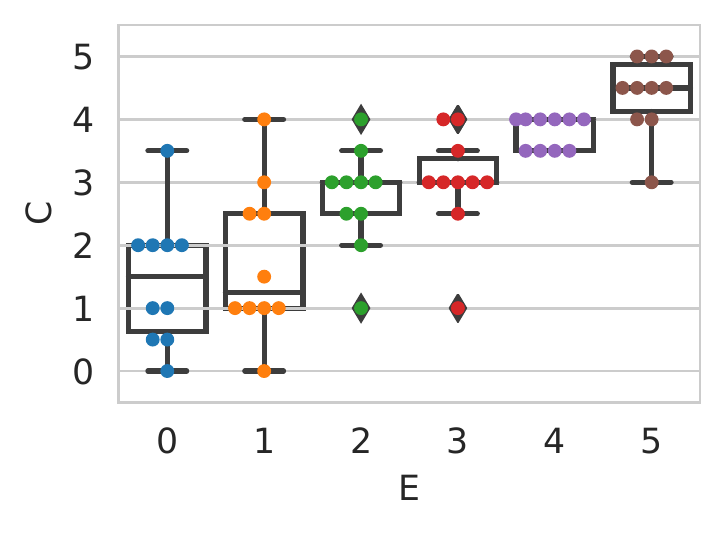}
    \caption{Judgments aggregated using the median.}
    \label{cap:paper_pauc2021-sec:results-subsec:crowd-accuracy-subsec:ext-agreement-fig:agreement-ground-truth_6-median}
  \end{subfigure}
  \hfill
  \begin{subfigure}{.49\linewidth}
    \centering
    \includegraphics[width=.95\linewidth]{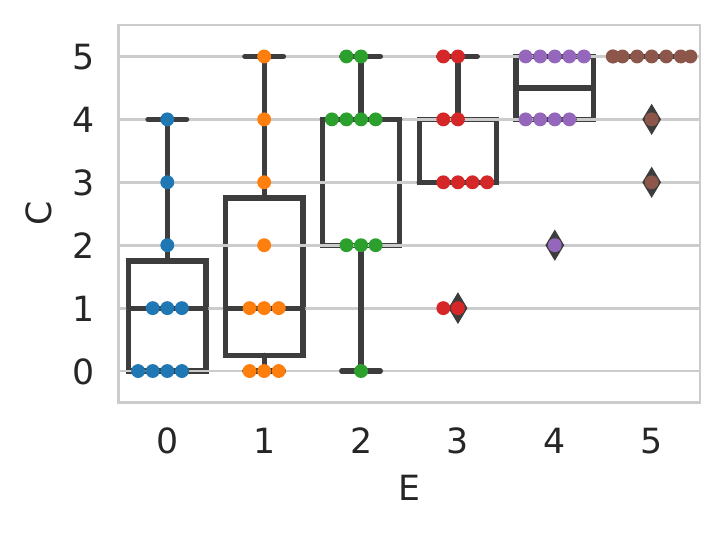}
    \caption{Judgments aggregated using the majority vote.}
    \label{cap:paper_pauc2021-sec:results-subsec:crowd-accuracy-subsec:ext-agreement-fig:agreement-ground-truth_6-maj}
  \end{subfigure}
  
  \caption{Agreement between the \politifact experts (x-axis) and the crowd judgments (y-axis) across different aggregation strategies.}
  \label{cap:paper_pauc2021-sec:results-subsec:crowd-accuracy-subsec:ext-agreement-fig:agreement-ground-truth}
\end{figure}

Figure~\ref{cap:paper_pauc2021-sec:results-subsec:crowd-accuracy-subsec:ext-agreement-fig:agreement-ground-truth} shows the agreement between the \politifact experts (x-axis) and the crowd judgments (y-axis). Figure~\ref{cap:paper_pauc2021-sec:results-subsec:crowd-accuracy-subsec:ext-agreement-fig:agreement-ground-truth_6} presents the raw data, where each point corresponds to a single worker’s judgment on a statement, without any aggregation. In the remaining charts, judgments from multiple workers assigned to the same statement are aggregated using the mean (Figure~\ref{cap:paper_pauc2021-sec:results-subsec:crowd-accuracy-subsec:ext-agreement-fig:agreement-ground-truth_6-mean}), the median (Figure~\ref{cap:paper_pauc2021-sec:results-subsec:crowd-accuracy-subsec:ext-agreement-fig:agreement-ground-truth_6-median}), and the majority vote (Figure~\ref{cap:paper_pauc2021-sec:results-subsec:crowd-accuracy-subsec:ext-agreement-fig:agreement-ground-truth_6-maj}).

Focusing on Figure~\ref{cap:paper_pauc2021-sec:results-subsec:crowd-accuracy-subsec:ext-agreement-fig:agreement-ground-truth_6} (i.e., the chart without any aggregation), individual judgments generally align with the expert labels. This is shown by the median values of the boxplots, which increase as the ground truth truthfulness level increases. In the aggregated results, across all aggregation functions, the \politifactpantsfire and \politifactfalse categories are perceived similarly by workers. This behavior, already reported by \citet{roitero2020crowd, la2020crowdsourcing}, indicates a persistent difficulty in distinguishing between these two categories, despite the interface displaying textual descriptions for each label on every task page. Examining each chart as a whole shows that the median values of the boxplots increase from \politifactpantsfire to \politifacttrue (i.e., from left to right on the x-axis), confirming a general agreement between crowd judgments and the \politifact ground truth. These findings suggest that workers are overall capable of recognising and correctly classifying misinformation related to the \covid pandemic, an important and non-trivial result, given that the crowd is often cited as a major source of misinformation and disinformation on social media platforms~\cite{chen2015students}.

Among the aggregation strategies, the mean (Figure~\ref{cap:paper_pauc2021-sec:results-subsec:crowd-accuracy-subsec:ext-agreement-fig:agreement-ground-truth_6-mean}) results in the highest agreement with expert labels, followed by the median (Figure~\ref{cap:paper_pauc2021-sec:results-subsec:crowd-accuracy-subsec:ext-agreement-fig:agreement-ground-truth_6-median}) and the majority vote (Figure~\ref{cap:paper_pauc2021-sec:results-subsec:crowd-accuracy-subsec:ext-agreement-fig:agreement-ground-truth_6-maj}). This outcome is consistent with previous findings~\cite{la2020crowdsourcing, Roitero:2018:FRS:3209978.3210052, roitero2020crowd}, where the mean typically emerges as the preferred aggregation method.

The statistical significance of the differences between aggregated judgments across the six \politifact categories is assessed to validate the external agreement. Both the Mann--Whitney rank test and the t-test are employed, with the \index{Bonferroni Correction}Bonferroni correction applied to account for multiple comparisons.
No statistically significant differences are found between adjacent categories (e.g., \politifactpantsfire and \politifactfalse) for any aggregation function or test. Similarly, comparisons between categories at distance 2 (e.g., \politifactpantsfire and \politifactmostlyfalse) are not significant, except when using the median aggregation, where significance at the \index{$p$}$p < .05$ level is observed in 2 out of 4 cases for both tests.
For categories at distance 3, significant differences are observed in the following cases: using the mean, all comparisons are significant ($3/3$ for both Mann--Whitney and t-test); using the median, significance is found in $2/3$ and $3/3$ comparisons, respectively; using the majority vote, only $0/3$ and $1/3$ comparisons are significant.
For distances 4 and 5, statistically significant differences are consistently observed at the \index{$p$}$p < .01$ level for all aggregation functions and both tests. The only exception is the majority vote with the Mann--Whitney test, where the result is not significant ($p > .05$).
In the remainder of the analysis, the mean is adopted as the aggregation function, as it is the most commonly used approach for this type of data.

\subsubsection{Internal Agreement}

\label{cap:paper_pauc2021-sec:results-subsec:crowd-accuracy-subsec:int-agreement}

Internal agreement is measured using Krippendorff’s \index{$\upalpha$}$\upalpha$~\cite{krippendorff2011computing} and \index{$\upphi$}$\upphi$, as defined by \citet{checco2017let}, two widely used metrics for assessing worker agreement in crowdsourcing tasks~\cite{maddalena2017considering, RSDM:2018, Roitero:2018:FRS:3209978.3210052, roitero2020crowd}.

The overall agreement consistently falls within the $[0.15, 0.3]$ range. Values obtained from the two metrics are largely consistent across the \politifact categories, with the exception of \index{$\upphi$}$\upphi$, which shows higher agreement for the \politifactmostlytrue and \politifacttrue categories. This observation is supported by the fact that \index{$\upalpha$}$\upalpha$ always lies within the confidence interval of \index{$\upphi$}$\upphi$, suggesting that small fluctuations may not indicate meaningful changes in agreement, particularly when using \index{$\upalpha$}$\upalpha$~\cite{checco2017let}.
Nonetheless, \index{$\upphi$}$\upphi$ reinforces the pattern observed in Figure~\ref{cap:paper_pauc2021-sec:results-subsec:crowd-accuracy-subsec:ext-agreement-fig:agreement-ground-truth}: workers are more effective at identifying and categorizing statements with higher truthfulness levels. This conclusion is also consistent with \citet{checco2017let}, who argue that \index{$\upphi$}$\upphi$ is better suited for distinguishing levels of agreement in crowdsourcing scenarios, whereas \index{$\upalpha$}$\upalpha$ is more appropriate for measuring data reliability in non-crowdsourced contexts.

\subsection{\ref{cap:paper_pauc2021-sec:research-questions_2}: Transforming Judgment Scales}

\label{cap:paper_pauc2021-sec:results-subsec:transforming-scales}

Section~\ref{cap:paper_pauc2021-sec:results-subsec:transforming-scales-subsec:gt-merge} examines how merging the ground-truth levels affects accuracy, while Section~\ref{cap:paper_pauc2021-sec:results-subsec:transforming-scales-subsec:cw-merge} focuses on transformations applied to the crowdsourced judgments. Finally, Section~\ref{cap:paper_pauc2021-sec:results-subsec:transforming-scales-subsec:gt-cw-merge} considers a combined approach involving both ground-truth and crowd levels.

\subsubsection{Merging Ground Truth Levels}

\label{cap:paper_pauc2021-sec:results-subsec:transforming-scales-subsec:gt-merge}

\begin{figure}[tbp]
  \centering
  \begin{subfigure}{.32\linewidth}
    \centering
    \includegraphics[width=\linewidth]{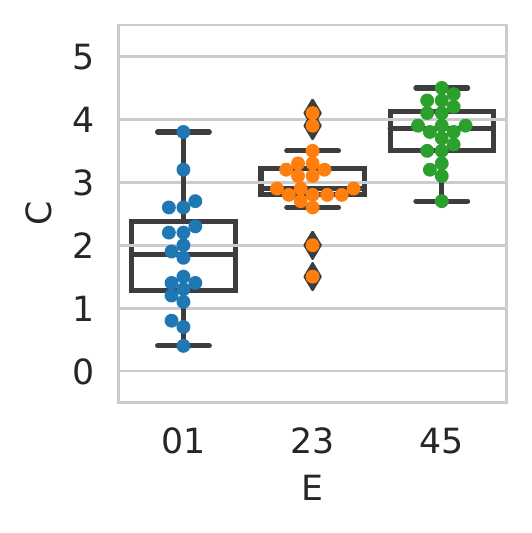}
    \caption{\crowdsix agg. using the mean.}
    \label{cap:paper_pauc2021-sec:results-subsec:transforming-scales-subsec:gt-merge-fig:binning_3-mean}
  \end{subfigure}
  \hfill
  \begin{subfigure}{.32\linewidth}
    \centering
    \includegraphics[width=\linewidth]{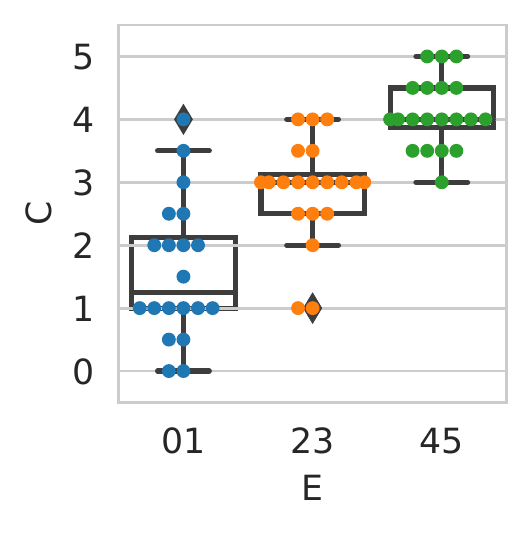}
    \caption{\crowdsix agg. using the median.}
    \label{cap:paper_pauc2021-sec:results-subsec:transforming-scales-subsec:gt-merge-fig:binning_3-median}
  \end{subfigure}
  \hfill
  \begin{subfigure}{.32\linewidth}
    \centering
    \includegraphics[width=\linewidth]{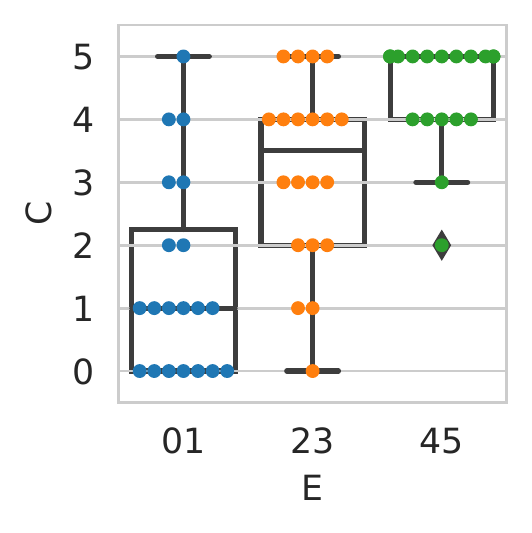}
    \caption{\crowdsix agg. using majority vote.}
    \label{cap:paper_pauc2021-sec:results-subsec:transforming-scales-subsec:gt-merge-fig:binning_3-maj}
  \end{subfigure}

  \begin{subfigure}{.32\linewidth}
    \centering
    \includegraphics[width=\linewidth]{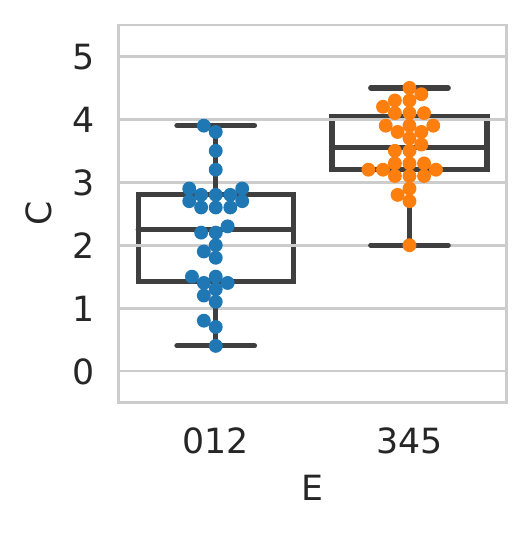}
    \caption{\crowdsix agg. using the mean.}
    \label{cap:paper_pauc2021-sec:results-subsec:transforming-scales-subsec:gt-merge-fig:binning_2-mean}
  \end{subfigure}
  \hfill
  \begin{subfigure}{.32\linewidth}
    \centering
    \includegraphics[width=\linewidth]{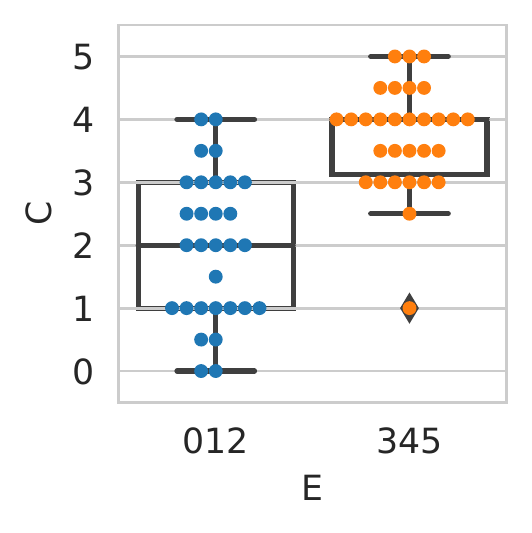}
    \caption{\crowdsix agg. using the median.}
    \label{cap:paper_pauc2021-sec:results-subsec:transforming-scales-subsec:gt-merge-fig:binning_2-median}
  \end{subfigure}
  \hfill
  \begin{subfigure}{.32\linewidth}
    \centering
    \includegraphics[width=\linewidth]{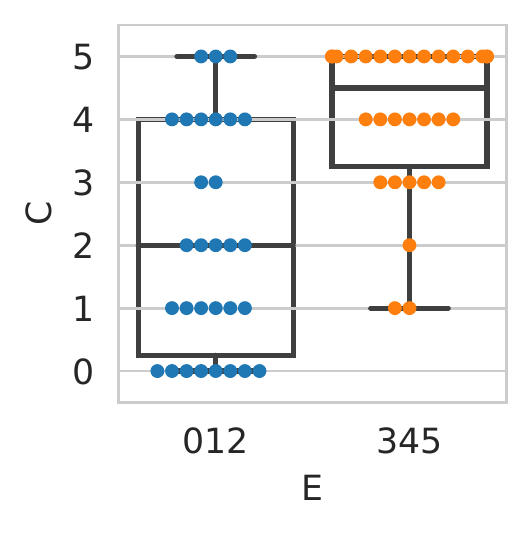}
    \caption{\crowdsix agg. using majority vote.}
    \label{cap:paper_pauc2021-sec:results-subsec:transforming-scales-subsec:gt-merge-fig:binning_2-maj}
  \end{subfigure}

  \caption{Agreement between the \politifact experts (x-axis) and the crowd judgments (y-axis). First row: transformation from \expertsix to \expertthree. Second row: transformation from \expertsix to \experttwo. Compare with Figure~\ref{cap:paper_pauc2021-sec:results-subsec:crowd-accuracy-subsec:ext-agreement-fig:agreement-ground-truth}.}
  \label{cap:paper_pauc2021-sec:results-subsec:transforming-scales-subsec:gt-merge-fig:binning}
\end{figure}

The six \politifact categories (i.e., \expertsix) are grouped into three (\expertthree) or two (\experttwo) broader categories, referred to as \politifactthreebinszero, \politifactthreebinsone, and \politifactthreebinstwo for the three-level scale, and \politifacttwoebinszero and \politifacttwobinsone for the two-level scale. Figure~\ref{cap:paper_pauc2021-sec:results-subsec:transforming-scales-subsec:gt-merge-fig:binning} illustrates the results of this transformation. The agreement between crowd and expert judgments is more clearly visible. As in Figure~\ref{cap:paper_pauc2021-sec:results-subsec:crowd-accuracy-subsec:ext-agreement-fig:agreement-ground-truth}, the median values of the boxplots increase with the ground-truth truthfulness levels (i.e., from left to right within each plot). This trend holds across all aggregation functions and for both transformations of the \expertsix scale into three and two levels.

As in the previous analysis, statistical significance between categories is assessed using the Mann–Whitney and t-tests, with the \index{Bonferroni Correction}Bonferroni correction applied to account for multiple comparisons. For the three-group transformation, categories at distances one and two are consistently significant at the \index{$p$}$p < .01$ level across all aggregation functions and both tests. The same pattern is observed in the two-group transformation, where categories at distance one are always significant at the \index{$p$}$p < .01$ level. 

In summary, merging the ground-truth levels produces a much stronger signal: the crowd is able to effectively detect and classify misinformation statements related to the \covid pandemic.

\subsubsection{Merging Crowd Levels}

\label{cap:paper_pauc2021-sec:results-subsec:transforming-scales-subsec:cw-merge}

The crowd judgments (i.e., \crowdsix) can be merged in the same way as the ground-truth labels. Specifically, judgments are grouped into either three categories (\crowdthree) or two (\crowdtwo). The transformation process follows the method proposed by \citet{scale}, which offers several advantages. It enables simulating the effect of using a more coarse-grained scale (rather than \crowdsix), allowing the generation of new data without rerunning the entire experiment on \mturk.

The following procedure is adopted. All possible cuts\footnote{\crowdsix can be transformed into \crowdthree in 10 different ways, and into \crowdtwo in 5 different ways.} from \crowdsix to \crowdthree and from \crowdsix to \crowdtwo are applied. For each transformation, the internal agreement is measured on both the original and transformed scales using \index{$\upalpha$}$\upalpha$ and \index{$\upphi$}$\upphi$. These values are then compared to identify, among all possible cuts, the one that yields the highest internal agreement. For the \crowdsix to \crowdthree transformation, a single cut achieves higher agreement than the original \crowdsix scale with both \index{$\upalpha$}$\upalpha$ and \index{$\upphi$}$\upphi$. In contrast, for the \crowdsix to \crowdtwo transformation, only one cut yields similar agreement to the original scale with \index{$\upalpha$}$\upalpha$, and no such cut is found when using \index{$\upphi$}$\upphi$. Once the optimal cuts are identified for both transformations and both agreement metrics, external agreement between the crowd and expert judgments is measured using the selected cut.

\begin{figure}[tbp]
  \centering
  \begin{subfigure}{.49\linewidth}
    \centering
    \includegraphics[width=\linewidth]{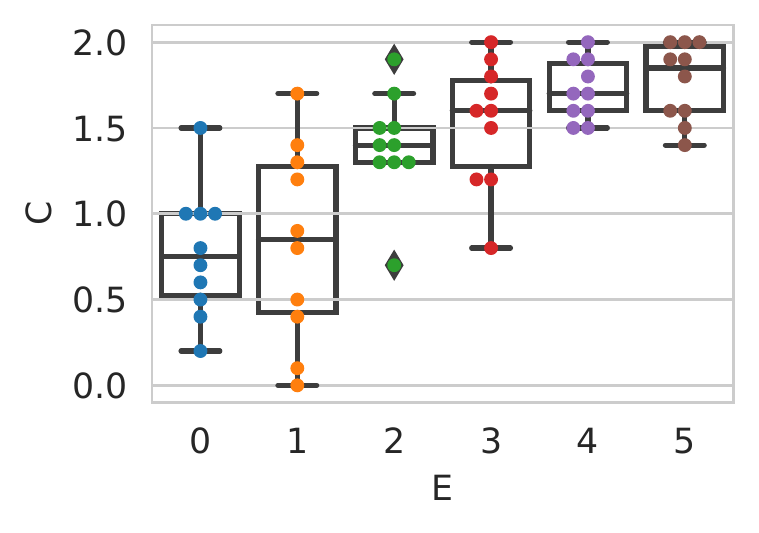}
    \caption{\crowdsix to \crowdthree. Best cut selected according to $\upalpha$.}
    \label{cap:paper_pauc2021-sec:results-subsec:transforming-scales-subsec:cw-merge-fig:agreement-gt_6to3-alpha}
  \end{subfigure}
  \begin{subfigure}{.49\linewidth}
    \centering
    \includegraphics[width=\linewidth]{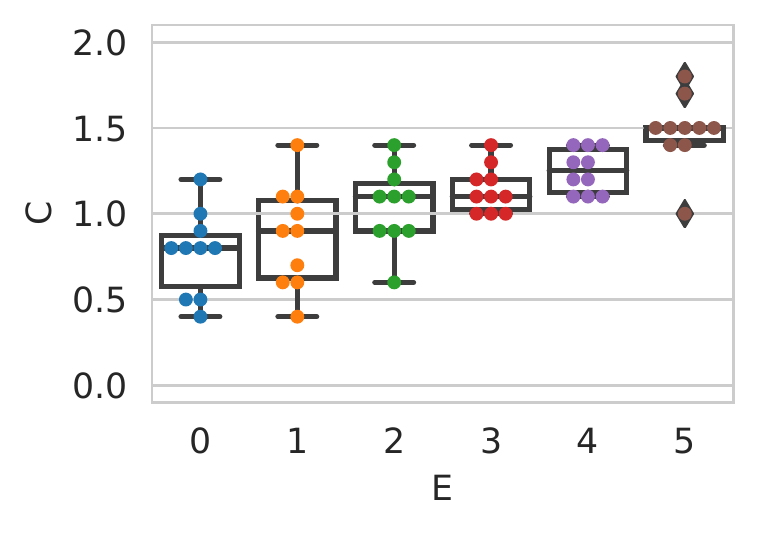}
    \caption{\crowdsix to \crowdthree. Best cut selected according to $\upphi$.}
    \label{cap:paper_pauc2021-sec:results-subsec:transforming-scales-subsec:cw-merge-fig:agreement-gt_6to3-phi}
  \end{subfigure}
  \begin{subfigure}{.49\linewidth}
    \centering
    \includegraphics[width=\linewidth]{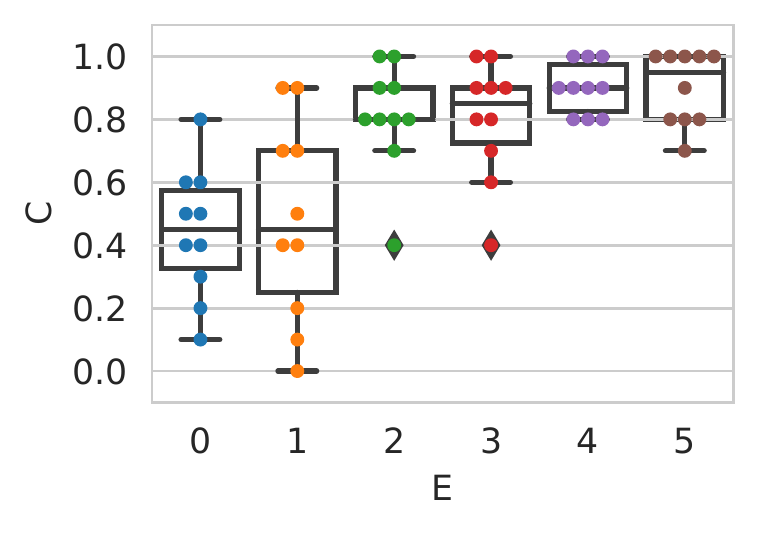}
    \caption{\crowdsix to \crowdtwo. Best cut selected according to $\upalpha$.}
    \label{cap:paper_pauc2021-sec:results-subsec:transforming-scales-subsec:cw-merge-fig:agreement-gt_6to2-alpha}
  \end{subfigure}
  \begin{subfigure}{.49\linewidth}
    \centering
    \includegraphics[width=\linewidth]{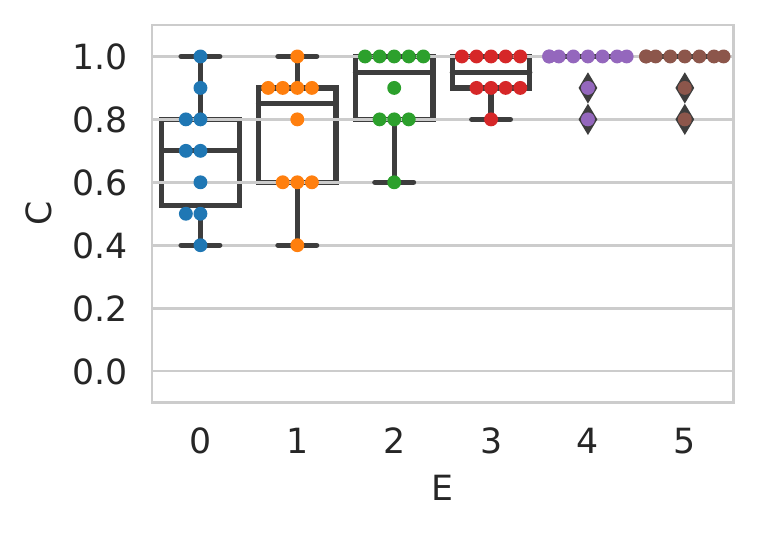}
    \caption{\crowdsix to \crowdtwo. Best cut selected according to $\upphi$.}
    \label{cap:paper_pauc2021-sec:results-subsec:transforming-scales-subsec:cw-merge-fig:agreement-gt_6to2-phi}
  \end{subfigure}
  \caption{Crowd judgments transformed into coarser categories and then aggregated using the mean. Agreement with \expertsix ground-truth labels. Compare with Figure~\ref{cap:paper_pauc2021-sec:results-subsec:crowd-accuracy-subsec:ext-agreement-fig:agreement-ground-truth}.}
  \label{cap:paper_pauc2021-sec:results-subsec:transforming-scales-subsec:cw-merge-fig:agreement-gt}
\end{figure}

Figure~\ref{cap:paper_pauc2021-sec:results-subsec:transforming-scales-subsec:cw-merge-fig:agreement-gt} presents the results obtained using the mean aggregation function. As in previous analyses, the median values of the boxplots consistently increase across truthfulness levels for all transformations. However, visual inspection of the plots suggests that the overall external agreement is lower compared to Figure~\ref{cap:paper_pauc2021-sec:results-subsec:crowd-accuracy-subsec:ext-agreement-fig:agreement-ground-truth}. In addition, even after applying these transformations, the \politifactpantsfire and \politifactfalse categories remain indistinguishable.
In summary, it is feasible to transform the judgments collected on the \crowdsix scale into coarser \crowdthree and \crowdtwo scales. The resulting judgments achieve internal agreement comparable to the original, though with slightly reduced external agreement relative to expert labels.

\subsubsection{Merging Both Ground Truth and Crowd Levels}

\label{cap:paper_pauc2021-sec:results-subsec:transforming-scales-subsec:gt-cw-merge}

\begin{figure}[tbp]
  \centering
  \begin{subfigure}{.49\linewidth}
\centering
  \includegraphics[width=\linewidth]{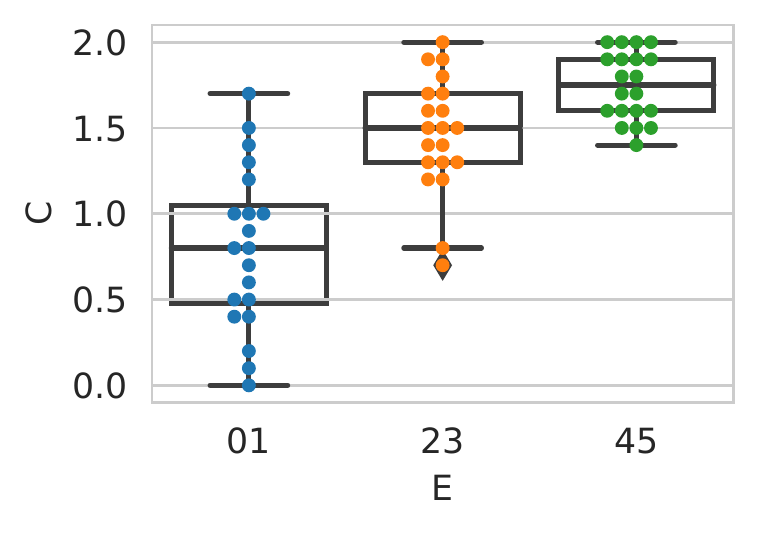}
  \caption{\expertsix to \expertthree. Best cut selected according to $\upalpha$.}
  \label{cap:paper_pauc2021-sec:results-subsec:transforming-scales-subsec:gt-cw-merge-fig:transf-binning_3-alpha}
\end{subfigure}
\centering
  \begin{subfigure}{.49\linewidth}
\centering
  \includegraphics[width=\linewidth]{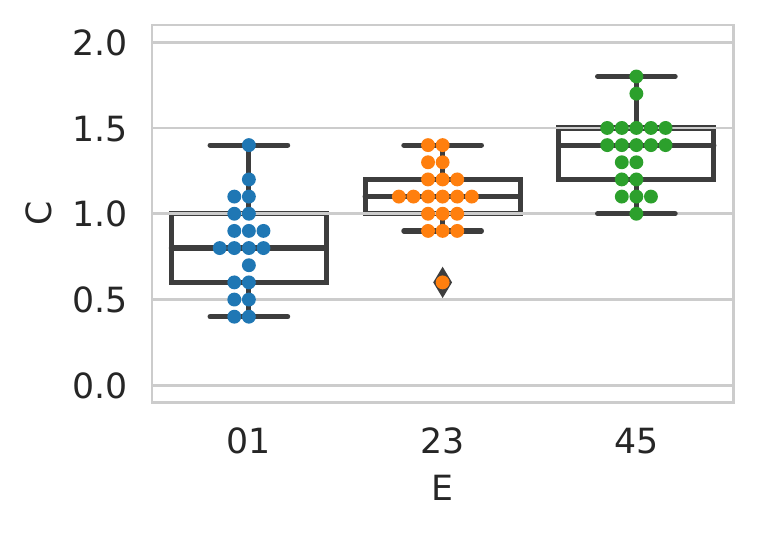}
  \caption{\expertsix to \expertthree. Best cut selected according to $\upphi$.}
  \label{cap:paper_pauc2021-sec:results-subsec:transforming-scales-subsec:gt-cw-merge-fig:transf-binning_3-phi}
\end{subfigure}
  \centering
  \begin{subfigure}{.49\linewidth}
\centering
  \includegraphics[width=\linewidth]{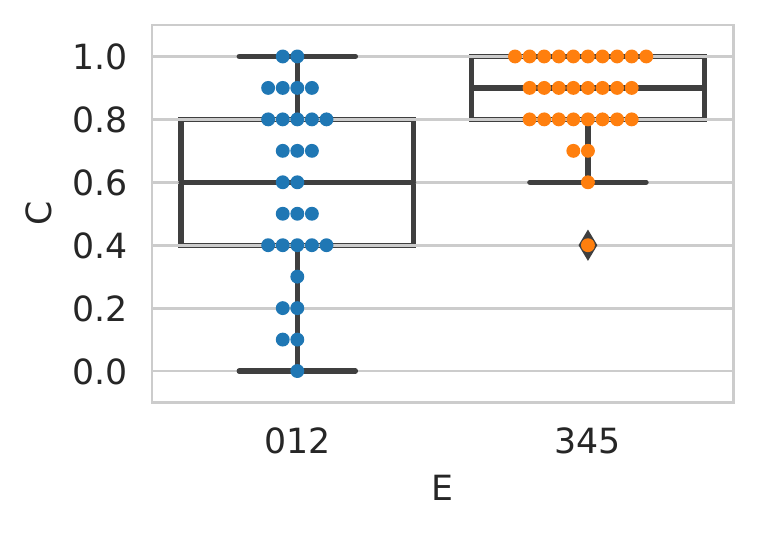}
  \caption{\expertsix to \experttwo. Best cut selected according to $\upalpha$.}
  \label{cap:paper_pauc2021-sec:results-subsec:transforming-scales-subsec:gt-cw-merge-fig:transf-binning_2-alpha}
\end{subfigure}
\centering
  \begin{subfigure}{.49\linewidth}
\centering
  \includegraphics[width=\linewidth]{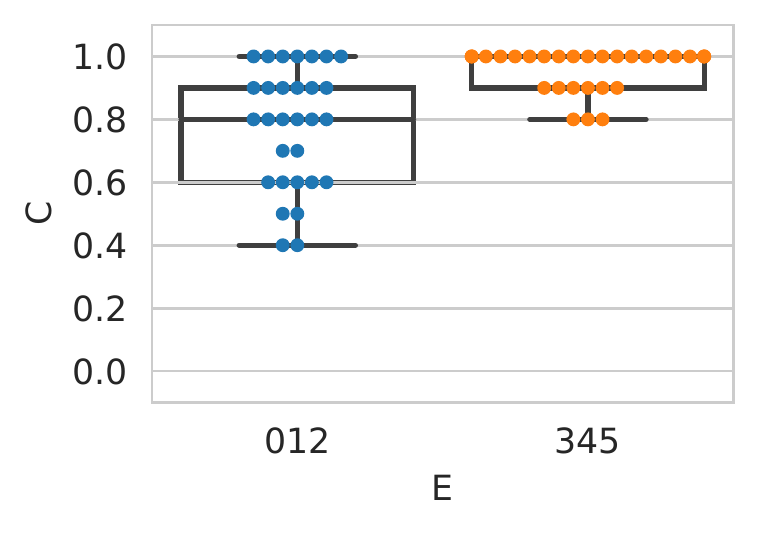}
  \caption{\expertsix to \experttwo. Best cut selected according to $\upphi$.}
  \label{cap:paper_pauc2021-sec:results-subsec:transforming-scales-subsec:gt-cw-merge-fig:transf-binning_2-phi}
\end{subfigure}
\caption{Expert judgments merged into groups and then aggregated with the mean function. Compare with Figure~\ref{cap:paper_pauc2021-sec:results-subsec:transforming-scales-subsec:gt-merge-fig:binning} and Figure~\ref{cap:paper_pauc2021-sec:results-subsec:transforming-scales-subsec:cw-merge-fig:agreement-gt}.}
\label{cap:paper_pauc2021-sec:results-subsec:transforming-scales-subsec:gt-cw-merge-fig:transf-binning}
\end{figure}

It is now natural to combine the two approaches. Figure~\ref{cap:paper_pauc2021-sec:results-subsec:transforming-scales-subsec:gt-cw-merge-fig:transf-binning} shows the comparison between \crowdsix transformed into \crowdthree and \crowdtwo, and \expertsix transformed into \expertthree and \experttwo. As in previous cases, the median values of the boxplots increase consistently with the ground truth levels, particularly for the \expertthree case (Figures~\ref{cap:paper_pauc2021-sec:results-subsec:transforming-scales-subsec:gt-cw-merge-fig:transf-binning_3-alpha} and~\ref{cap:paper_pauc2021-sec:results-subsec:transforming-scales-subsec:gt-cw-merge-fig:transf-binning_3-phi}). Furthermore, external agreement with the ground truth is still present, although in the \experttwo case (Figures~\ref{cap:paper_pauc2021-sec:results-subsec:transforming-scales-subsec:gt-cw-merge-fig:transf-binning_2-alpha} and~\ref{cap:paper_pauc2021-sec:results-subsec:transforming-scales-subsec:gt-cw-merge-fig:transf-binning_2-phi}) the classes appear less separable.

In summary, the results confirm the feasibility of successfully combining the two transformation strategies and applying them to both the crowd and expert judgments, leading to coherent three-level and two-level scales.

\subsection{\ref{cap:paper_pauc2021-sec:research-questions_3}: Worker Background And Bias}

\label{cap:paper_pauc2021-sec:results-subsec:worker-background-bias}

Previous work has shown that political and personal biases as well as cognitive abilities have an impact on the workers' quality \cite{la2020crowdsourcing, roitero2020crowd}. Recent studies have indicated that similar effects may also apply to fake news scenarios \cite{mypartisanship}. For this reason, it is reasonable to investigate whether workers' political views and cognitive abilities influence judgment quality in the context of misinformation related to \covid.

When examining questionnaire responses, only political orientation shows a clear relationship with worker quality. Table~\ref{tab:political-bias-quality} reports both Accuracy (i.e., the fraction of exactly classified statements) and the Closeness Evaluation Measure (\cem), an effectiveness metric designed for ordinal classification \cite{ACL:2020} (see Section~\ref{cap:paper_sigir2020-sec:exp-setup-subsec:scales}). While Accuracy is a coarse indicator, \cem accounts for the distance between predicted and true labels, mitigating the impact of minor misclassification errors.

\begin{table}[tbp]
\centering
\caption{Accuracy and Closeness Evaluation Measure (\cem) by political view. Asterisks denote statistically significant difference in \cem with respect to \veryconservative workers (Bonferroni-corrected t-test, $p < 0.05$).}
\label{tab:political-bias-quality}
\begin{tabular}{l
                S[table-format=1.2]
                S[table-format=1.2,table-space-text-post={*}]}
\toprule
\textbf{Political View} & {\textbf{Accuracy}} & {\textbf{\cem}} \\
\midrule
\veryconservative & 0.13 & 0.46 \\
\conservative     & 0.21 & 0.51{*} \\
\moderate         & 0.20 & 0.50 \\
\liberal          & 0.16 & 0.50 \\
\veryliberal      & 0.21 & 0.51{*} \\
\bottomrule
\end{tabular}
\end{table}

The results indicate that \veryconservative workers provide lower-quality judgments. The Bonferroni-corrected two-tailed t-test confirms that their performance in terms of \cem is significantly worse than that of \conservative and \veryliberal workers. While the effect is relatively small, it highlights the influence of political orientation, especially at the extremes of the scale.

An initial analysis of other questionnaire responses (not shown) does not reveal strong correlations with worker quality. The effect of cognitive ability, as measured by the \index{Cognitive!Reflection Test}CRT, is also investigated. Although small variations in Accuracy and \cem are observed, none are statistically significant. The number of correct CRT answers does not appear to correlate with judgment quality.

\subsection{\ref{cap:paper_pauc2021-sec:research-questions_4}: Worker Behavior}

\label{cap:paper_pauc2021-sec:results-subsec:worker-behavior}

Workers provide several behavioral signals while completing the misinformation assessment task. These signals can be extracted, for example, from the time spent on each statement and from their query patterns, as discussed in Section~\ref{cap:paper_pauc2021-sec:results-subsec:worker-behavior-subsec:time-query}. It is therefore relevant to investigate whether such behavioral signals can be leveraged to improve the quality of the work performed, as examined in Section~\ref{cap:paper_pauc2021-sec:results-subsec:worker-behavior-subsec:exploiting}.

\subsubsection{Time and Queries}
\label{cap:paper_pauc2021-sec:results-subsec:worker-behavior-subsec:time-query}

Table~\ref{cap:paper_pauc2021-sec:results-subsec:worker-behavior-subsec:time-query-tab:query_stats} (first two rows) reports the average time spent by workers on the statements and their corresponding \cem scores. The time spent on the first statement is considerably higher than that on subsequent ones, and overall, the time spent by workers decreases almost monotonically with the progression of statement positions. This pattern, combined with the observation that the quality of judgments (measured by \cem) does not decline for later statements, suggests the presence of a learning effect: workers appear to become more efficient at assessing truthfulness as they proceed through the task.

Table~\ref{cap:paper_pauc2021-sec:results-subsec:worker-behavior-subsec:time-query-tab:query_stats} (third and fourth rows) reports query statistics for the 100 workers who completed the task. The total and average number of queries are 2095 and 262, respectively, while the total and average number of statements used as queries are 245 and 30.6, respectively. A clear trend emerges: as the statement position increases, the number of queries issued decreases. On average, workers issued 3.52\% of all queries for the first statement, dropping to 2.30\% for the last. This behavior may reflect a tendency to issue fewer queries over time, possibly due to fatigue, boredom, or learning effects.

Despite this decrease, the average number of queries per statement remains above one for all positions, suggesting that workers frequently reformulate their initial queries. This reinforces the interpretation that they actively engage with the task. The third row of the table shows how often workers used the entire statement as a query. This proportion remains low, approximately 13\% across all statement positions, further indicating that workers typically invest effort into crafting more targeted queries. This supports the overall high quality of the collected judgments.

\begin{table}[tbp]
\centering
\caption{Statement position in the task versus: time elapsed (first row), \cem (second row), number of queries issued (third row), and number of times the statement was used as a query (fourth row). Percentages are relative to the total.}
\label{cap:paper_pauc2021-sec:results-subsec:worker-behavior-subsec:time-query-tab:query_stats}
\begin{tabular}{m{3.2cm}m{0.8cm}m{0.8cm}m{0.8cm}m{0.8cm}m{0.8cm}m{0.8cm}m{0.8cm}m{0.8cm}}
\toprule
\textbf{Statement Position} & \textbf{1} & \textbf{2} & \textbf{3}  & \textbf{4} & \textbf{5} & \textbf{6} & \textbf{7} & \textbf{8}\\
\midrule
\textbf{Time (sec)}& 299 & 282 & 218 & 216 & 223 & 181 & 190 & 180  \\
\midrule
\textbf{\cem}& 0.63 & 0.618 & 0.657 & 0.611 & 0.614 & 0.569 & 0.639 & 0.655 \\
\midrule
\textbf{Queries Number}& 352 16.8\% & 280 13.4\% & 259 12.4\% & 255 12.1\% & 242 11.6\% & 238 11.3\% &230 11.0\% & 230 11.4\% \\
\midrule
\textbf{Statement as Query}& 22 9\% & 32 13\% & 31 12.6\% & 33 13.5\%  & 34 13.9\% & 30 12.2\% & 29 11.9\% & 34 13.9\% \\
\bottomrule
\end{tabular}
\end{table}

\subsubsection{Exploiting Worker Signals to Improve Quality}

\label{cap:paper_pauc2021-sec:results-subsec:worker-behavior-subsec:exploiting}

Workers emit several behavioral and background signals that may correlate with the quality of their judgments during the task. These signals can potentially be exploited to improve aggregation, for instance by giving more weight to workers whose characteristics are associated with higher quality. As discussed in Section~\ref{cap:paper_pauc2021-sec:results-subsec:worker-background-bias}, political orientation and performance on cognitive reflection tasks are among the factors that show such correlations.

To explore this idea, an additional experiment is conducted. The \crowdsix individual scores are aggregated using a weighted mean, where the weights are based on either political views or the number of correct answers to the \index{Cognitive!Reflection Test}CRT. All weights are normalized within the $[0.5,1]$ range. The results show a pattern similar to the one shown in Figure~\ref{cap:paper_pauc2021-sec:results-subsec:transforming-scales-subsec:gt-cw-merge-fig:transf-binning_3-phi}.

Overall, leveraging quality-related signals (such as questionnaire responses or \index{Cognitive!Reflection Test}CRT scores) for weighted aggregation does not result in a significant improvement in external agreement. However, no adverse effect is observed either, suggesting that this strategy is neutral in terms of effectiveness.

\subsection{\ref{cap:paper_pauc2021-sec:research-questions_5}: Sources Of Information}

\label{cap:paper_pauc2021-sec:results-subsec:info-source}

The sources of information consulted by the workers during the task are examined by analyzing the distribution of URLs (Section~\ref{cap:paper_pauc2021-sec:results-subsec:info-source-subsec:url-analysis}) and the textual justifications provided to support the evidence found (Section~\ref{cap:paper_pauc2021-sec:results-subsec:info-source-subsec:justifications}).

\subsubsection{URLs Analysis}

\label{cap:paper_pauc2021-sec:results-subsec:info-source-subsec:url-analysis}

\begin{figure}[tbp]
\centering
\includegraphics[width=0.45\linewidth]{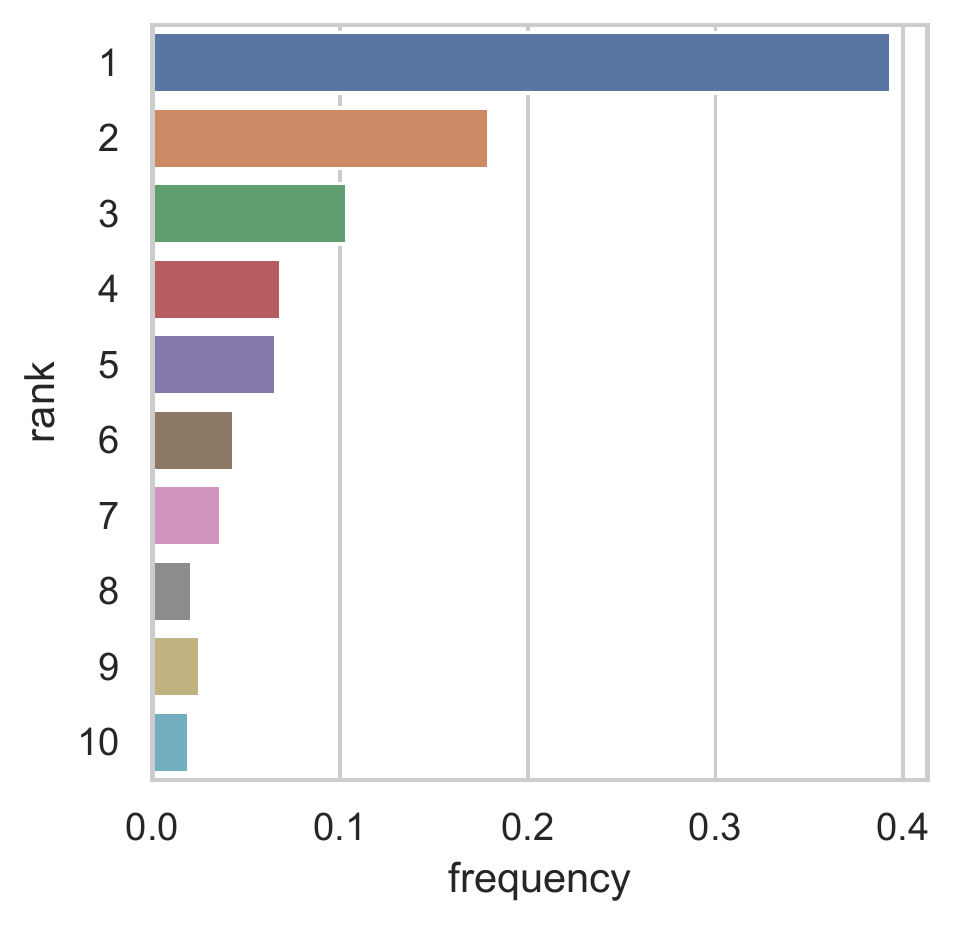}
\caption{Distribution of the ranks of the URLs selected by workers.}
\label{cap:paper_pauc2021-sec:results-subsec:info-source-subsec:url-analysis-fig:url-ranks}
\end{figure}

Figure~\ref{cap:paper_pauc2021-sec:results-subsec:info-source-subsec:url-analysis-fig:url-ranks} shows the distribution of the ranks of the URLs selected by workers when providing each judgment. URLs selected less than 1\% of the time are filtered out. About 40\% of the workers choose the top-ranked result returned by the custom search engine, with lower-ranked URLs being selected less frequently. The distribution follows an almost monotonic decrease, with the only exception being rank 8. No significant differences are observed when stratifying the distribution by \politifact truthfulness levels.

Table~\ref{cap:paper_pauc2021-sec:results-subsec:info-source-subsec:url-analysis-tab:url-ranks} reports the top 10 websites from which the workers selected URLs to justify their judgments. Websites with percentage $\leq 3.9\%$ are filtered out. Several fact-checking websites appear prominently, including \textit{snopes.com} (11.79\%) and \textit{factcheck.org} (6.79\%). Medical sources such as \textit{cdc.gov} (4.29\%) are also represented. This variety of domains shows that workers draw on different types of sources, suggesting that they make a deliberate effort to gather evidence before providing a truthfulness judgment.

\begin{table}[tbp]
\centering
\caption{Top websites from which workers selected URLs to justify their judgments. Only sources used in more than 3.9\% of cases are shown.}
\label{cap:paper_pauc2021-sec:results-subsec:info-source-subsec:url-analysis-tab:url-ranks}
\begin{tabular}{l r}
\toprule
\textbf{Domain} & \textbf{Usage (\%)} \\
\midrule
snopes.com                & 11.79 \\
msn.com                   & 8.93 \\
factcheck.org             & 6.79 \\
wral.com                  & 6.79 \\
usatoday.com              & 5.36 \\
statesman.com             & 4.64 \\
reuters.com               & 4.64 \\
cdc.gov                   & 4.29 \\
mediabiasfactcheck.com    & 4.29 \\
businessinsider.com       & 3.93 \\
\bottomrule
\end{tabular}
\end{table}

\subsubsection{Justifications}

\label{cap:paper_pauc2021-sec:results-subsec:info-source-subsec:justifications}

The textual justifications provided by the workers, their relation to the web pages at the selected URLs, and their connection to worker quality are also analyzed. In 54\% of the cases, the justification contains text copied from the web page selected as evidence, while in 46\% it does not. Additionally, 48\% of the justifications include some "free text" (i.e., text generated and written by the worker), while 52\% do not. Considering all possible combinations:
\begin{itemize}[label=--]
  \item 6\% of the justifications use both free text and text copied from web pages.
  \item 42\% use free text but no text from web pages.
  \item 48\% use only copied text from web pages.
  \item 4\% use neither free text nor copied text, instead inserting content from unrelated pages, the task interface, or the instructions.
\end{itemize}

Each worker shows a clear and consistent behavior regarding how justifications are formulated:
\begin{itemize}[label=--]
  \item 48\% of workers use only text copied from selected web pages.
  \item 46\% use only free text.
  \item 4\% combine both types.
  \item 2\% consistently provide unrelated or interface-derived text.
\end{itemize}

\begin{figure}[tbp]
  \centering
  \begin{subfigure}[t]{.49\linewidth}
    \centering
    \includegraphics[width=\linewidth]{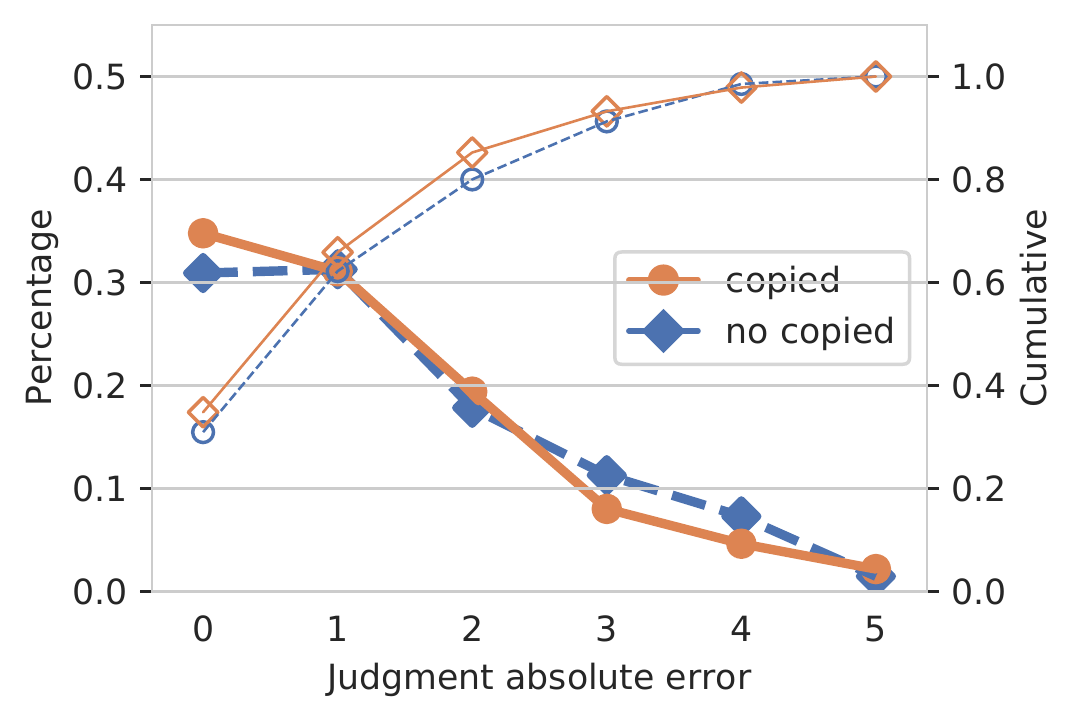}
    \caption{Absolute error.}
    \label{cap:paper_pauc2021-sec:results-subsec:info-source-subsec:justifications-fig:error_cumulative}
  \end{subfigure}
  \hfill
  \begin{subfigure}[t]{.49\linewidth}
    \centering
    \includegraphics[width=\linewidth]{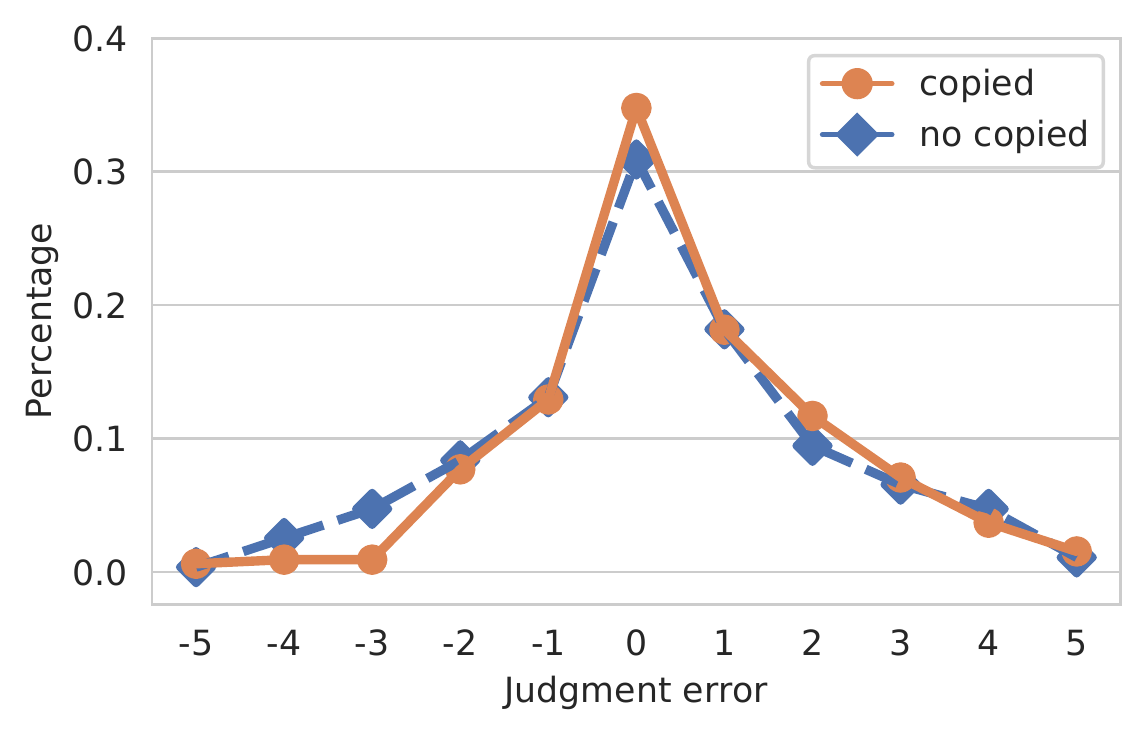}
    \caption{Prediction error.}
    \label{cap:paper_pauc2021-sec:results-subsec:info-source-subsec:justifications-fig:error_error}
  \end{subfigure}
  \caption{Effect of justification origin on worker accuracy. The left plot shows the absolute value of the prediction error, with cumulative distributions indicated by thinner lines and empty markers. The right plot shows the prediction error. Both plots compare justifications that include copied text from the selected URL against those that do not.}
  \label{cap:paper_pauc2021-sec:results-subsec:info-source-subsec:justifications-fig:error}
\end{figure}

This behavior can be correlated with worker quality. Figure~\ref{cap:paper_pauc2021-sec:results-subsec:info-source-subsec:justifications-fig:error} shows the relationship between justification type and worker accuracy. Figure~\ref{cap:paper_pauc2021-sec:results-subsec:info-source-subsec:justifications-fig:error_cumulative} presents the absolute value of the prediction error. Figure~\ref{cap:paper_pauc2021-sec:results-subsec:info-source-subsec:justifications-fig:error_error} shows the prediction error itself. The figures distinguish between text copied or not from the selected web pages. A similar analysis was conducted based on whether free text was used, but the results were nearly identical.

As can be seen, statements on which workers make fewer errors (i.e., where the x-axis is equal to 0) tend to include text copied from the selected web page. Conversely, statements with larger errors (values close to 5 in Figure~\ref{cap:paper_pauc2021-sec:results-subsec:info-source-subsec:justifications-fig:error_cumulative} and values close to $+/-5$ in Figure~\ref{cap:paper_pauc2021-sec:results-subsec:info-source-subsec:justifications-fig:error_error}) are more often associated with justifications that do not contain copied text.
To support this observation, the \cem scores are computed for the two justification types. The "copied" class has \cem$=0.640$, while the "not copied" class has a lower score of \cem$=0.600$.
A similar trend is observed for free-text justifications: statements with fewer errors tend to include free text, possibly rephrased from the source URL, whereas those with more errors tend to lack it. The corresponding \cem scores are very similar, with the "free text" class at \cem$=0.624$ and the "not free text" class at \cem$=0.621$.

The right side of both plots in Figure~\ref{cap:paper_pauc2021-sec:results-subsec:info-source-subsec:justifications-fig:error} shows that prediction errors are more frequent on the positive side of the x-axis ($[0,5]$), indicating a tendency to overestimate the truthfulness of statements (e.g., labeling a \politifactpantsfire statement as \politifacttrue). Justifications containing copied text also lead to fewer underestimations, as shown by a lower error frequency on the negative x-axis range.

\subsection{\ref{cap:paper_pauc2021-sec:research-questions_6}: Repeating The Experiment With Novice Workers}

\label{cap:paper_pauc2021-sec:results-subsec:novice-workers}

Repeating the experiment in the longitudinal study with newly recruited novice workers involves replicating the full set of analyses conducted for the base crowdsourcing task, described in Section~\ref{cap:paper_pauc2021-sec:exp-setup-subsec:crowdsourcing-task}. 
Section~\ref{cap:paper_pauc2021-sec:results-subsec:novice-workers-subsec:background} examines changes in the composition of the worker population. Section~\ref{cap:paper_pauc2021-sec:results-subsec:novice-workers-subsec:agreement} evaluates the quality of both individual and aggregated judgments. Section~\ref{cap:paper_pauc2021-sec:results-subsec:novice-workers-subsec:external} and Section~\ref{cap:paper_pauc2021-sec:results-subsec:novice-workers-subsec:internal} analyze external and internal agreement, respectively. Section~\ref{cap:paper_pauc2021-sec:results-subsec:novice-workers-subsec:time-query} addresses the time spent and querying behavior of novice workers. Section~\ref{cap:paper_pauc2021-sec:results-subsec:novice-workers-subsec:urls} analyzes the distribution of URLs used. Finally, Section~\ref{cap:paper_pauc2021-sec:results-subsec:novice-workers-subsec:justifications} investigates how different types of justifications affect worker accuracy.

\subsubsection{Worker Background, Behavior, Bias, and Abandonment}

\label{cap:paper_pauc2021-sec:results-subsec:novice-workers-subsec:background}

The variation in the composition of the worker population across different batches is studied using a General Linear Mixture Model (GLMM) \index{GLMM} \cite{mccullagh2018generalized}, together with Analysis of Variance (ANOVA) \index{ANOVA} \cite{morrison2005multivariate}. This methodology enables an assessment of how worker behavior evolves across batches and quantifies the impact of such variations.
In particular, the ANOVA effect size \index{$\upomega^2$} $\upomega^2$ is considered. It is an unbiased index that provides insights into the population-wide relationship between a set of factors and the studied outcomes \cite{ferro2019using, ferro2016general, ferro2018toward, roitero2020leveraging, zampieri2019topic}. A linear model is fitted to measure the effects of age, education, and all other questionnaire responses.

Inspection of the $\upomega^2$ index reveals that the strongest effects are associated with responses to the questions about taxes and the southern border. All other effects are either small or negligible \cite{olejnikAnova}. The effect of the batch itself is also small but not negligible and is of similar magnitude to the effects of other individual factors.
Interaction plots (see e.g., \cite{de2007interpretation}) are computed to study the variation of these factors across batches. The results suggest only small or statistically insignificant \cite{embretson1996item} interactions between batch and the other factors. This analysis indicates that, although some differences exist across batches, the population of workers is largely homogeneous. Consequently, the different datasets (i.e., batches) can be considered comparable.

Table~\ref{cap:paper_pauc2021-sec:results-subsec:novice-workers-subsec:background-tab:longitudinal-abandonment} reports the abandonment statistics for each batch of the longitudinal study, indicating the number of workers who completed the task, abandoned it, or failed due to not satisfying the quality checks. Overall, the abandonment ratio remains relatively consistent across batches, with the exception of \batchthree, which shows a higher failure rate. This variation is modest and likely attributable to a slightly lower average quality among the workers who participated in \batchthree. On average, 31\% of workers completed the task, 50\% abandoned it, and 19\% failed the quality checks. These values are consistent with those reported in Section~\ref{cap:paper_sigir2020-sec:desc-stat}.

\begin{table}[tbp]
    \centering
    \caption{Abandonment data for each batch of the longitudinal study.}
    \label{cap:paper_pauc2021-sec:results-subsec:novice-workers-subsec:background-tab:longitudinal-abandonment}
    \begin{tabular}{lrrrrr}
    \toprule
    & \multicolumn{4}{c}{\textbf{Number of Workers}} \\
    \cmidrule(lr){2-5}
    \textbf{Batch} & \textbf{Completed} & \textbf{Abandoned} & \textbf{Failed} & \textbf{Total} \\
    \midrule
    \batchone   & 100 (30\%) & 188 (56\%) & 46 (14\%)  & 334 \\
    \batchtwo   & 100 (37\%) & 129 (48\%) & 40 (15\%)  & 269 \\
    \batchthree & 100 (23\%) & 220 (51\%) & 116 (26\%) & 436 \\
    \batchfour  & 100 (36\%) & 124 (45\%) & 54 (19\%)  & 278 \\
    \midrule
    \textbf{Average} & 100 (31\%) & 165 (50\%) & 64 (19\%) & 1317 \\
    \bottomrule
    \end{tabular}
\end{table}

\subsubsection{Agreement Across Batches}

\label{cap:paper_pauc2021-sec:results-subsec:novice-workers-subsec:agreement}

The correlation between individual judgments across batches is generally low:
\begin{itemize}[label=--]
  \item Between \batchone and \batchtwo: \index{$\uprho$}$\uprho = 0.33$, \index{$\uptau$}$\uptau = 0.25$.
  \item Between \batchone and \batchthree: \index{$\uprho$}$\uprho = 0.20$, \index{$\uptau$}$\uptau = 0.14$; between \batchone and \batchfour: \index{$\uprho$}$\uprho = 0.10$, \index{$\uptau$}$\uptau = 0.074$.
  \item Between \batchtwo and \batchthree: \index{$\uprho$}$\uprho = 0.21$, \index{$\uptau$}$\uptau = 0.15$; between \batchtwo and \batchfour: \index{$\uprho$}$\uprho = 0.10$, \index{$\uptau$}$\uptau = 0.085$.
  \item Between \batchthree and \batchfour: \index{$\uprho$}$\uprho = 0.08$, \index{$\uptau$}$\uptau = 0.06$.
\end{itemize}

Among all pairs, \batchfour consistently shows the lowest correlation values relative to the other batches, followed by \batchthree. The highest correlation is observed between \batchone and \batchtwo. This suggests that the temporal proximity of batch execution may influence the consistency of judgments across batches: batches launched closer in time tend to be more aligned in their outputs.

\begin{figure}[tpb]
  \centering
  \includegraphics[width=.9\linewidth]{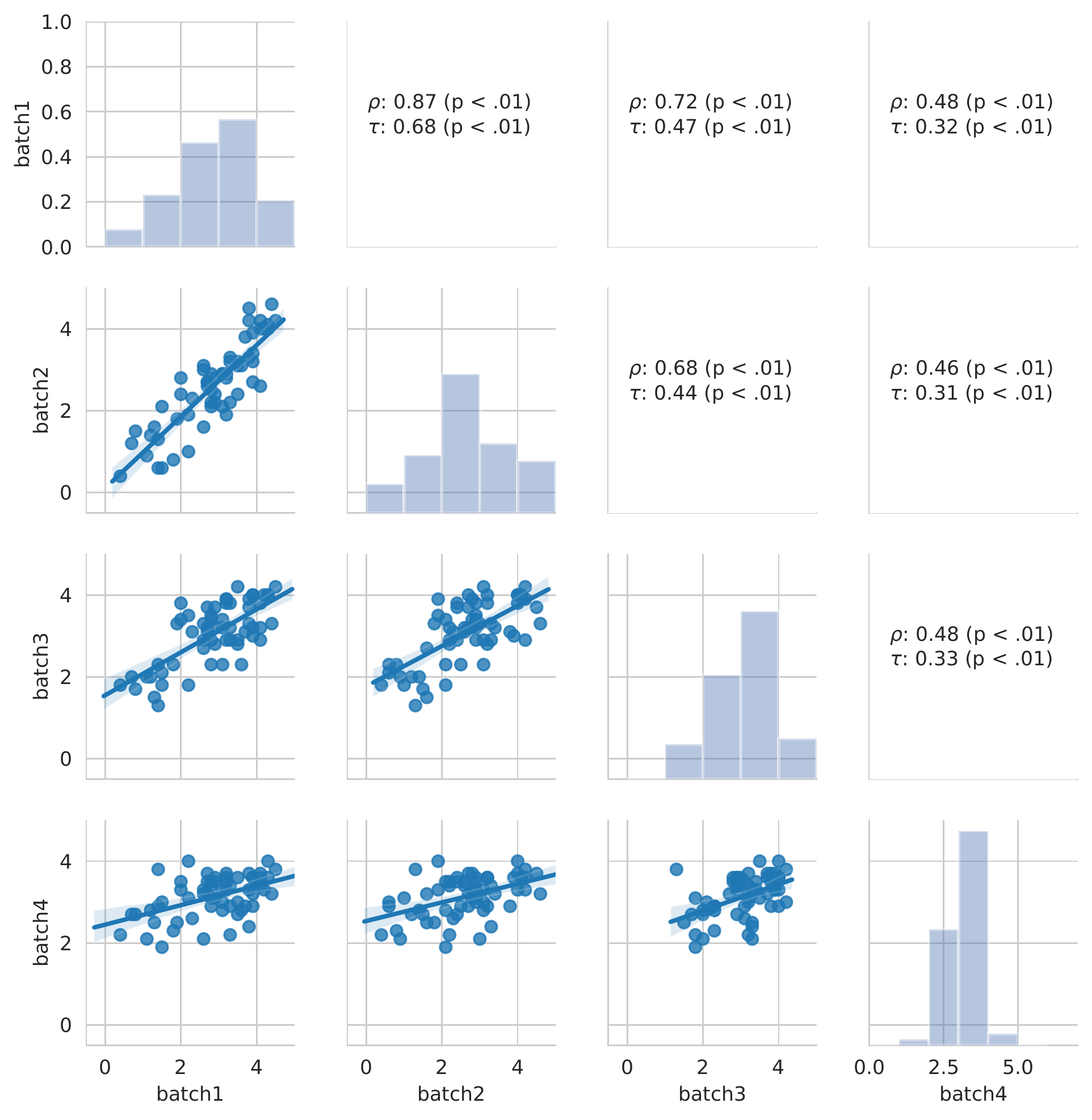}
  \caption{Correlation between aggregated judgments (mean) across \batchone, \batchtwo, \batchthree, and \batchfour. Diagonal plots show the distribution of scores per batch; lower triangle shows scatterplots; upper triangle reports $\uprho$ and $\uptau$ values.}
  \label{cap:paper_pauc2021-sec:results-subsec:novice-workers-subsec:agreement-fig:correlation-batches13}
\end{figure}

The aggregated judgments are also analyzed to determine whether the trends observed in individual judgments persist after aggregation. Figure~\ref{cap:paper_pauc2021-sec:results-subsec:novice-workers-subsec:agreement-fig:correlation-batches13} shows the agreement between the aggregated judgments for \batchone, \batchtwo, \batchthree, and \batchfour. The diagonal of the figure reports the distribution of the aggregated scores within each batch. The lower triangle presents scatterplots comparing aggregated judgments between batches, while the upper triangle reports the corresponding \index{$\uprho$}$\uprho$ and \index{$\uptau$}$\uptau$ correlation values.

The results show that the correlation values for aggregated judgments are consistently higher than those observed for individual judgments across all batches. In particular, the agreement between \batchone and \batchtwo (\index{$\uprho$}$\uprho = 0.87$, \index{$\uptau$}$\uptau = 0.68$) is the highest. The correlations between \batchone and \batchthree, and between \batchtwo and \batchthree, are similar to each other. Once again, \batchfour shows lower correlation values with all other batches.
Overall, these results suggest that:
\begin{enumerate}
  \item Individual judgments differ across batches, but aggregation increases consistency.
  \item Correlation tends to degrade over time, with earlier batches being more consistent with one another than later ones.
  \item Batches launched closer in time produce more similar results.
\end{enumerate}

\subsubsection{Crowd Workers Accuracy: External Agreement}

\label{cap:paper_pauc2021-sec:results-subsec:novice-workers-subsec:external}

\begin{figure}[tpb]
  \centering
  \begin{subfigure}{.49\linewidth}
    \centering
    \includegraphics[width=\linewidth]{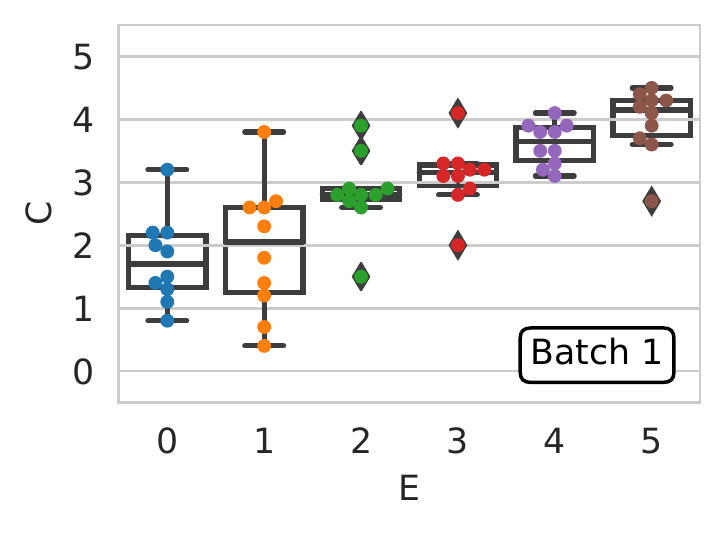}
    \caption{\batchone.}
    \label{cap:paper_pauc2021-sec:results-subsec:novice-workers-subsec:external-fig:long-gt_b1}
  \end{subfigure}
  \begin{subfigure}{.49\linewidth}
    \centering
    \includegraphics[width=\linewidth]{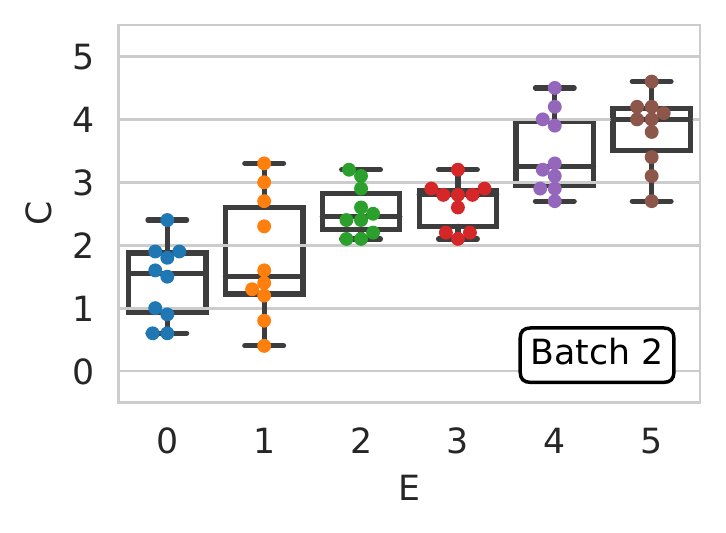}
    \caption{\batchtwo.}
    \label{cap:paper_pauc2021-sec:results-subsec:novice-workers-subsec:external-fig:long-gt_b2}
  \end{subfigure}
  \begin{subfigure}{.49\linewidth}
    \centering
    \includegraphics[width=\linewidth]{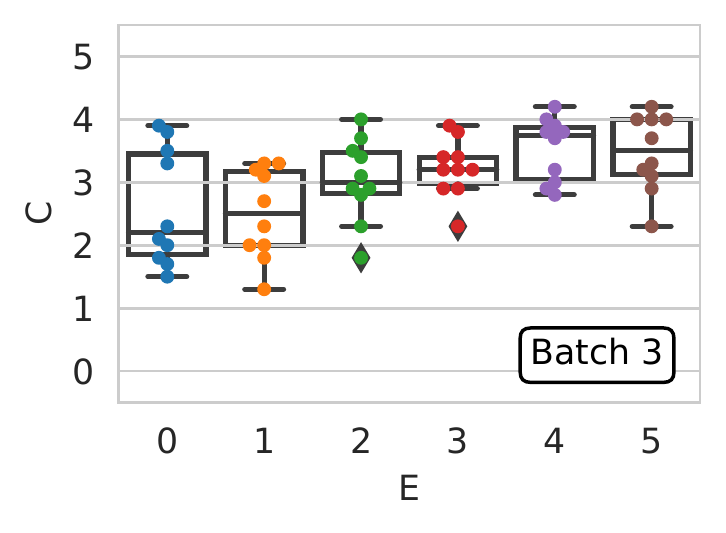}
    \caption{\batchthree.}
    \label{cap:paper_pauc2021-sec:results-subsec:novice-workers-subsec:external-fig:long-gt_b3}
  \end{subfigure}
  \begin{subfigure}{.49\linewidth}
    \centering
    \includegraphics[width=\linewidth]{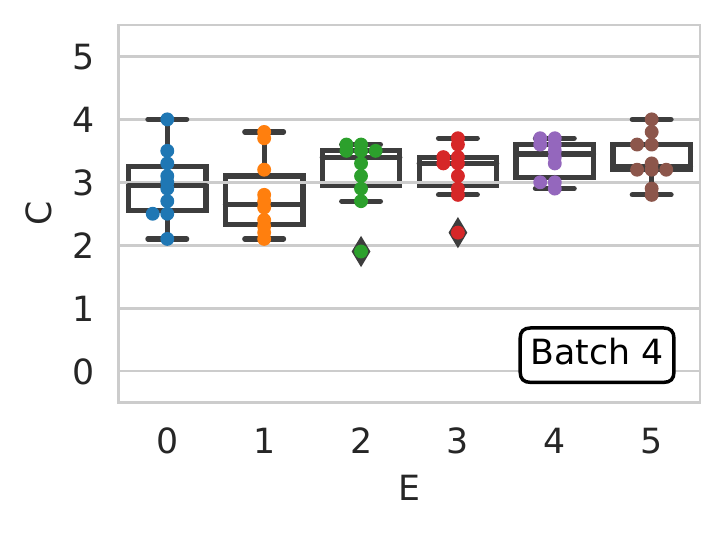}
    \caption{\batchfour.}
    \label{cap:paper_pauc2021-sec:results-subsec:novice-workers-subsec:external-fig:long-gt_b4}
  \end{subfigure}
  \begin{subfigure}{.49\linewidth}
    \centering
    \includegraphics[width=\linewidth]{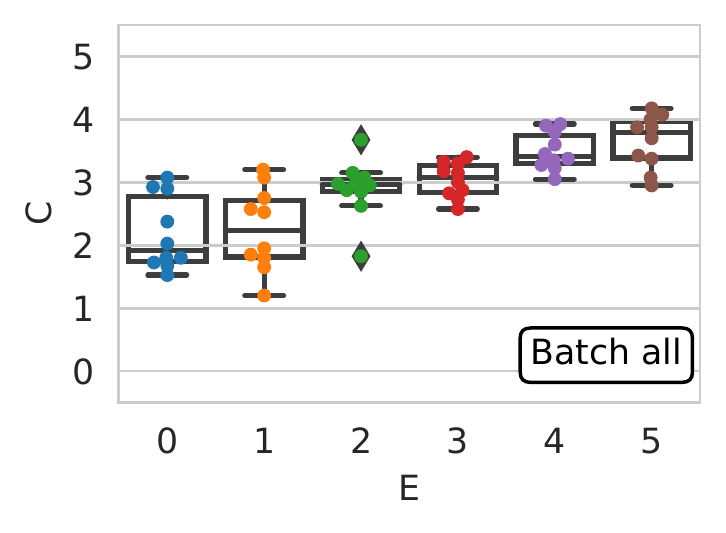}
    \caption{\batchall.}
    \label{cap:paper_pauc2021-sec:results-subsec:novice-workers-subsec:external-fig:long-gt_b_all}
  \end{subfigure}
  \caption{Agreement between the \politifact expert labels (x-axis) and crowd judgments (y-axis) across batches. Each subfigure shows the distribution of crowd-aggregated judgments (mean) for one batch. Figure~\ref{cap:paper_pauc2021-sec:results-subsec:novice-workers-subsec:external-fig:long-gt_b1} replicates Figure~\ref{cap:paper_pauc2021-sec:results-subsec:crowd-accuracy-subsec:ext-agreement-fig:agreement-ground-truth_6-mean} for comparison.}
  \label{cap:paper_pauc2021-sec:results-subsec:novice-workers-subsec:external-fig:long-gt}
\end{figure}

Figure~\ref{cap:paper_pauc2021-sec:results-subsec:novice-workers-subsec:external-fig:long-gt} shows the agreement between the \politifact experts (x-axis) and the crowd judgments (y-axis) for \batchone, \batchtwo, \batchthree, \batchfour, and \batchall. The judgments are aggregated using the mean. Overall, individual judgments align with expert labels, as shown by the increasing median values in the boxplots along the ground truth truthfulness scale.

\batchone and \batchtwo display higher agreement levels with expert labels compared to \batchthree and \batchfour. As also observed in Figure~\ref{cap:paper_pauc2021-sec:results-subsec:crowd-accuracy-subsec:ext-agreement-fig:agreement-ground-truth_6-mean}, the \politifactpantsfire and \politifactfalse categories are perceived similarly by workers across all aggregation functions, indicating difficulties in distinguishing between the two.
The median values in each boxplot increase consistently from \politifactpantsfire to \politifacttrue for all batches, except for \batchthree and more notably \batchfour. This pattern confirms a general alignment between crowd and expert judgments, even when the experiment is repeated at different time spans.

However, a counterintuitive pattern emerges. Since data for each batch was collected at different time points, one might expect that the more time passes, the better workers would recognize the actual label of each statement (e.g., from online exposure or news coverage). Figure~\ref{cap:paper_pauc2021-sec:results-subsec:novice-workers-subsec:external-fig:long-gt} instead suggests the opposite: agreement with expert labels tends to decrease over time. This phenomenon may be influenced by multiple factors, which are discussed in the following sections.

Finally, Figure~\ref{cap:paper_pauc2021-sec:results-subsec:novice-workers-subsec:external-fig:long-gt_b_all} shows that \batchall behaves similarly to \batchone and \batchtwo. The median values of the boxplots increase from left to right on the x-axis, except between \politifactpantsfire and \politifactfalse. This confirms that even when combining all batches, the crowd remains generally aligned with the \politifact ground truth.

The presence of differences in how the statements are evaluated across different batches is evident from the previous analyses. To investigate whether the same statements are consistently ranked across batches, the \index{$\uprho$}$\uprho$, \index{$\uptau$}$\uptau$, and rank-biased overlap (\index{RBO}RBO) \cite{webber2010similarity} correlation coefficients are computed. These are calculated between the scores aggregated using the mean across batches, within each \politifact category. The RBO parameter is configured so that the top-5 results receive approximately 85\% of the evaluation weight.

Table~\ref{cap:paper_pauc2021-sec:results-subsec:novice-workers-subsec:agreement-tab:longitudinal_agreement_ground_truth_corrs} reports the \index{$\uprho$}$\uprho$ and \index{$\uptau$}$\uptau$ correlation scores, while Table~\ref{cap:paper_pauc2021-sec:results-subsec:novice-workers-subsec:agreement-tab:longitudinal_agreement_ground_truth_corrs-2} presents the bottom-heavy and top-heavy RBO correlation scores. Since statements are ranked by their aggregated scores in decreasing order, the top-heavy RBO emphasizes agreement on misjudged statements from the \politifactpantsfire and \politifactfalse categories, while the bottom-heavy RBO emphasizes agreement on misjudged statements from the \politifacttrue category.

As shown in both tables, agreement across batches is generally low. This holds for both absolute values (as indicated by \index{$\uprho$}$\uprho$) and relative rankings (as measured by \index{$\uptau$}$\uptau$ and \index{RBO}RBO). The RBO values, in particular, reveal that the specific statements misjudged by workers tend to differ across batches. Notable exceptions are found in the \politifactfalse category for \batchone and \batchtwo (top-heavy RBO = $0.85$), and in the \politifacttrue category for the same pair of batches (bottom-heavy RBO = $0.92$). A similar pattern is observed for correctly judged statements: for instance, a bottom-heavy RBO value of $0.81$ is found for \politifactfalse, and a top-heavy RBO value of $0.50$ for \politifacttrue. These findings provide further evidence of the similarities between \batchone and \batchtwo.

\begin{table}[tbp]
    \caption{$\uprho$ (lower triangle) and $\uptau$ (upper triangle) correlation values among batches for the aggregated scores of Figure~\ref{cap:paper_pauc2021-sec:results-subsec:novice-workers-subsec:external-fig:long-gt}.}
    \label{cap:paper_pauc2021-sec:results-subsec:novice-workers-subsec:agreement-tab:longitudinal_agreement_ground_truth_corrs}

    \begin{minipage}{.49\linewidth}
    \centering
    \begin{tabular}{ccccc}
    \multicolumn{5}{c}{\politifactpantsfire (\texttt{0})} \\
    \toprule
    {} & B1 & B2 & B3 & B4 \\
    \midrule
    B1 & --  & 0.37 & 0.58 & 0.54 \\
    B2 & 0.44 & --  & 0.30 & 0.25 \\
    B3 & 0.74 & 0.69 & --  & 0.42 \\
    B4 & 0.58 & 0.24 & 0.46 & --  \\
    \bottomrule
    \end{tabular}
    \end{minipage}%
    \begin{minipage}{.49\linewidth}
    \centering
    \begin{tabular}{ccccc}
    \multicolumn{5}{c}{\politifactfalse (\texttt{1})} \\
    \toprule
    {} & B1 & B2 & B3 & B4 \\
    \midrule
    B1 & --  & 0.72 & 0.74 & 0.04 \\
    B2 & 0.87 & --  & 0.75 & 0.02 \\
    B3 & 0.84 & 0.85 & --  & -0.20 \\
    B4 & -0.01 & -0.07 & -0.29 & --  \\
    \bottomrule
    \end{tabular}
    \end{minipage}
    
    \vspace{0.7em}

    \begin{minipage}{.49\linewidth}
    \centering
    \begin{tabular}{ccccc}
    \multicolumn{5}{c}{\politifactmostlyfalse (\texttt{2})} \\
    \toprule
    {} & B1 & B2 & B3 & B4 \\
    \midrule
    B1 & --  & 0.07 & 0.47 & 0.51 \\
    B2 & 0.46 & --  & 0.37 & 0.09 \\
    B3 & 0.72 & 0.49 & --  & 0.58 \\
    B4 & 0.82 & 0.36 & 0.83 & --  \\
    \bottomrule
    \end{tabular}
    \end{minipage}%
    \begin{minipage}{.49\linewidth}
    \centering
    \begin{tabular}{ccccc}
    \multicolumn{5}{c}{\politifacthalftrue (\texttt{3})} \\
    \toprule
    {} & B1 & B2 & B3 & B4 \\
    \midrule
    B1 & --  & 0.12 & 0.12 & 0.00 \\
    B2 & -0.03 & --  & 0.52 & 0.22 \\
    B3 & 0.01 & 0.70 & --  & 0.10 \\
    B4 & 0.09 & 0.28 & 0.20 & --  \\
    \bottomrule
    \end{tabular}
    \end{minipage}
    
    \vspace{0.7em}

    \begin{minipage}{.49\linewidth}
    \centering
    \begin{tabular}{ccccc}
    \multicolumn{5}{c}{\politifactmostlytrue (\texttt{4})} \\
    \toprule
    {} & B1 & B2 & B3 & B4 \\
    \midrule
    B1 & --  & 0.35 & 0.16 & 0.24 \\
    B2 & 0.60 & --  & -0.07 & 0.69 \\
    B3 & 0.31 & 0.03 & --  & -0.28 \\
    B4 & 0.24 & 0.62 & -0.22 & --  \\
    \bottomrule
    \end{tabular}
    \end{minipage}%
    \begin{minipage}{.49\linewidth}
    \centering
    \begin{tabular}{ccccc}
    \multicolumn{5}{c}{\politifacttrue (\texttt{5})} \\
    \toprule
    {} & B1 & B2 & B3 & B4 \\
    \midrule
    B1 & --  & 0.74 & 0.51 & 0.48 \\
    B2 & 0.90 & --  & 0.26 & 0.28 \\
    B3 & 0.33 & 0.31 & --  & 0.67 \\
    B4 & 0.51 & 0.45 & 0.69 & --  \\
    \bottomrule
    \end{tabular}
    \end{minipage}
\end{table}

\begin{table}[tbp]
\centering
\caption{RBO bottom-heavy (lower triangle) and RBO top-heavy (upper triangle) correlation values among batches for the aggregated scores of Figure~\ref{cap:paper_pauc2021-sec:results-subsec:novice-workers-subsec:external-fig:long-gt}. Documents sorted by increasing aggregated score.}
\label{cap:paper_pauc2021-sec:results-subsec:novice-workers-subsec:agreement-tab:longitudinal_agreement_ground_truth_corrs-2}

\begin{minipage}{.49\linewidth}
\centering
\begin{tabular}{ccccc}
\multicolumn{5}{c}{\politifactpantsfire (\texttt{0})} \\
\toprule
{} & B1 & B2 & B3 & B4 \\
\midrule
B1 & --   & 0.47 & 0.79 & 0.51 \\
B2 & 0.31 & --   & 0.54 & 0.60 \\
B3 & 0.49 & 0.27 & --   & 0.51 \\
B4 & 0.50 & 0.28 & 0.32 & --   \\
\bottomrule
\end{tabular}
\end{minipage}%
\begin{minipage}{.49\linewidth}
\centering
\begin{tabular}{ccccc}
\multicolumn{5}{c}{\politifactfalse (\texttt{1})} \\
\toprule
{} & B1 & B2 & B3 & B4 \\
\midrule
B1 & --   & 0.85 & 0.86 & 0.36 \\
B2 & 0.81 & --   & 0.98 & 0.24 \\
B3 & 0.53 & 0.47 & --   & 0.23 \\
B4 & 0.34 & 0.41 & 0.33 & --   \\
\bottomrule
\end{tabular}
\end{minipage}

\vspace{0.7em}

\begin{minipage}{.49\linewidth}
\centering
\begin{tabular}{ccccc}
\multicolumn{5}{c}{\politifactmostlyfalse (\texttt{2})} \\
\toprule
{} & B1 & B2 & B3 & B4 \\
\midrule
B1 & --   & 0.62 & 0.70 & 0.43 \\
B2 & 0.62 & --   & 0.74 & 0.34 \\
B3 & 0.71 & 0.74 & --   & 0.59 \\
B4 & 0.71 & 0.64 & 0.76 & --   \\
\bottomrule
\end{tabular}
\end{minipage}%
\begin{minipage}{.49\linewidth}
\centering
\begin{tabular}{ccccc}
\multicolumn{5}{c}{\politifacthalftrue (\texttt{3})} \\
\toprule
{} & B1 & B2 & B3 & B4 \\
\midrule
B1 & --   & 0.26 & 0.26 & 0.22 \\
B2 & 0.26 & --   & 0.47 & 0.25 \\
B3 & 0.29 & 0.75 & --   & 0.64 \\
B4 & 0.22 & 0.51 & 0.36 & --   \\
\bottomrule
\end{tabular}
\end{minipage}

\vspace{0.7em}

\begin{minipage}{.49\linewidth}
\centering
\begin{tabular}{ccccc}
\multicolumn{5}{c}{\politifactmostlytrue (\texttt{4})} \\
\toprule
{} & B1 & B2 & B3 & B4 \\
\midrule
B1 & --   & 0.33 & 0.28 & 0.43 \\
B2 & 0.48 & --   & 0.22 & 0.78 \\
B3 & 0.28 & 0.18 & --   & 0.15 \\
B4 & 0.39 & 0.88 & 0.17 & --   \\
\bottomrule
\end{tabular}
\end{minipage}%
\begin{minipage}{.49\linewidth}
\centering
\begin{tabular}{ccccc}
\multicolumn{5}{c}{\politifacttrue (\texttt{5})} \\
\toprule
{} & B1 & B2 & B3 & B4 \\
\midrule
B1 & --   & 0.50 & 0.79 & 0.49 \\
B2 & 0.92 & --   & 0.29 & 0.38 \\
B3 & 0.49 & 0.41 & --   & 0.49 \\
B4 & 0.49 & 0.44 & 0.79 & --   \\
\bottomrule
\end{tabular}
\end{minipage}
\end{table}

\subsubsection{Crowd Workers Accuracy: Internal Agreement}

\label{cap:paper_pauc2021-sec:results-subsec:novice-workers-subsec:internal}

Table~\ref{cap:paper_pauc2021-sec:results-subsec:novice-workers-subsec:internal-tab:longitudinal_alphaphi} reports the internal agreement for each batch, measured using \index{$\upalpha$}$\upalpha$ \cite{krippendorff2011computing} and \index{$\upphi$}$\upphi$ \cite{checco2017let}. The lower triangle of the table shows the pairwise correlations based on \index{$\uprho$}$\uprho$, while the upper triangle reports those based on \index{$\uptau$}$\uptau$. Both correlation measures are computed across all \politifact categories, using \index{$\upalpha$}$\upalpha$ and the mean value of \index{$\upphi$}$\upphi$, without considering the 3\% lower and 97\% upper confidence bounds for \index{$\upphi$}$\upphi$.

The highest correlation for \index{$\upalpha$}$\upalpha$ is observed between \batchone and \batchthree, while the highest correlation for \index{$\upphi$}$\upphi$ is found between \batchone and \batchtwo. Notably, \index{$\upphi$}$\upphi$ yields generally lower correlation values compared to \index{$\upalpha$}$\upalpha$, particularly for \batchfour, which shows almost no correlation with the other batches. This suggests that \batchone and \batchtwo are the most similar in terms of internal agreement, whereas \batchthree and especially \batchfour include judgments that differ more in agreement levels.

\begin{table}[tbp]
  \centering
  \caption{Correlation between $\upalpha$ and $\upphi$ values across batches. $\uprho$ values are shown in the lower triangle, and $\uptau$ values in the upper triangle.}
  \label{cap:paper_pauc2021-sec:results-subsec:novice-workers-subsec:internal-tab:longitudinal_alphaphi}
  \begin{minipage}{.48\linewidth}
    \centering
    \begin{tabular}{@{}ccccc@{}}
      \multicolumn{5}{c}{$\upalpha$} \\
      \toprule
      {} & B1 & B2 & B3 & B4 \\
      \midrule
      B1 & --  & 0.49 & 0.61 & 0.52 \\
      B2 & 0.72 & --  & 0.42 & 0.39 \\
      B3 & 0.79 & 0.67 & --  & 0.57 \\
      B4 & 0.67 & 0.55 & 0.78 & --  \\
      \bottomrule
    \end{tabular}
  \end{minipage}%
  \hfill
  \begin{minipage}{.48\linewidth}
    \centering
    \begin{tabular}{@{}ccccc@{}}
      \multicolumn{5}{c}{$\upphi$} \\
      \toprule
      {} & B1 & B2 & B3 & B4 \\
      \midrule
      B1 & --   & 0.25 & 0.13 & -0.03 \\
      B2 & 0.38 & --   & 0.15 &  0.04 \\
      B3 & 0.19 & 0.23 & --   &  0.06 \\
      B4 & -0.06 & 0.05 & 0.09 & --   \\
      \bottomrule
    \end{tabular}
  \end{minipage}
\end{table}

\subsubsection{Worker Behavior: Time And Queries}

\label{cap:paper_pauc2021-sec:results-subsec:novice-workers-subsec:time-query}

Analyzing the amount of time spent by the workers for each position of the statement in the task confirms an observation already highlighted in Section~\ref{cap:paper_pauc2021-sec:results-subsec:worker-behavior}. The amount of time spent on average by the workers on the first statements is considerably higher than the time spent on the last statements, for all batches. This confirms the presence of a learning effect: workers learn how to assess truthfulness more efficiently as they proceed with the task.

Furthermore, the average time spent on all documents decreases substantially as the number of batches increases. The average time spent on each document for the four batches is, respectively, 222, 168, 182, and 140 seconds. A statistical test is performed between each pair of batches and is significant in all comparisons, with the only exception of \batchtwo versus \batchthree. This decreasing trend might indeed contribute to the degradation in quality observed across later batches. If workers spend less time on each statement, it can be assumed that they either dedicate less attention to the judgment or invest less effort in searching for appropriate and relevant sources of evidence.

To further investigate this quality degradation, the querying behavior of workers is analyzed across batches. The number of queries issued by each worker shows a decreasing trend with respect to statement position. This effect is still present in all batches, although less pronounced (but not in a significant way) in \batchtwo and \batchthree. Therefore, it can still be stated that workers tend to issue fewer queries as they progress through the task, likely due to fatigue, boredom, or learning effects.

Furthermore, it is again the case that, on average and across all statement positions, each worker issues more than one query. In other words, workers often reformulate their initial query. This provides further evidence that they invest effort in performing the task and suggests an overall high quality of the collected judgments. Only a small fraction of queries, less than 2\% for all batches, corresponds exactly to the statement text. This indicates that the vast majority of workers put meaningful effort into writing queries, which can be taken as an indication of their willingness to produce high-quality work.

\subsubsection{Sources of Information: URL Analysis}

\label{cap:paper_pauc2021-sec:results-subsec:novice-workers-subsec:urls}

\begin{figure}[tbp]
  \centering
  \includegraphics[width=.6\linewidth]{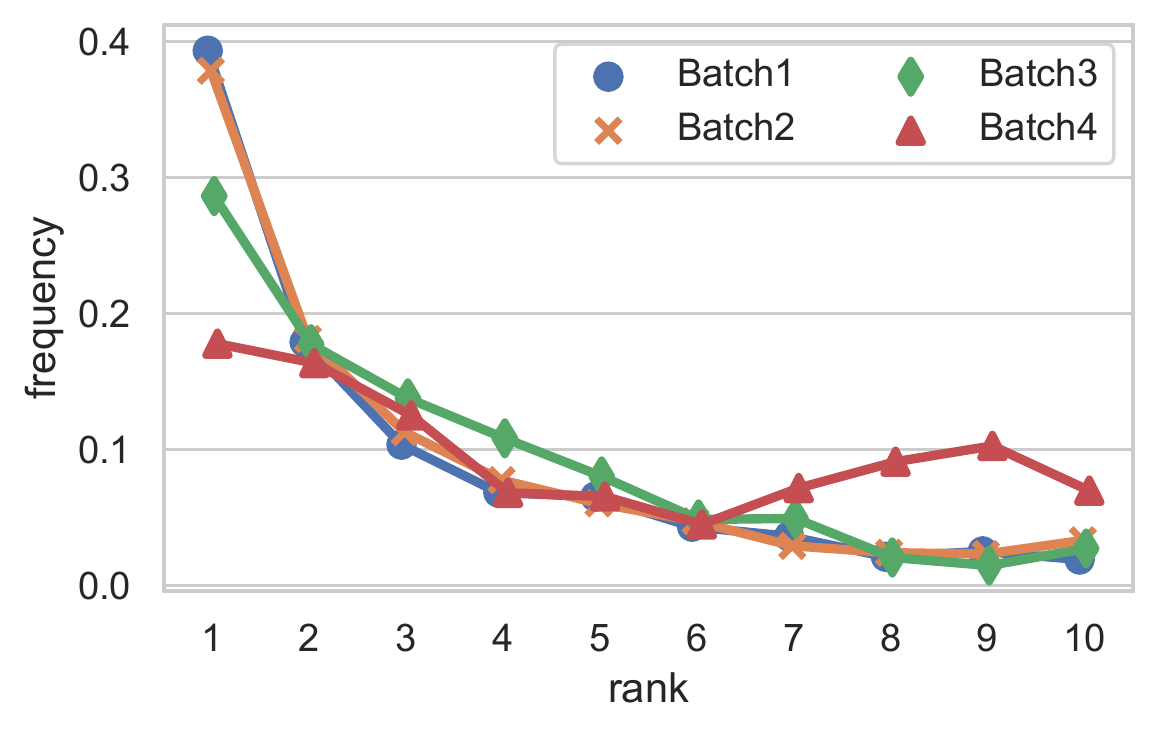}
  \caption{Distribution of the ranks of the URLs selected by workers for all the batches.}
  \label{cap:paper_pauc2021-sec:results-subsec:novice-workers-subsec:urls-fig:longitudinal_ranks}
\end{figure}

Figure~\ref{cap:paper_pauc2021-sec:results-subsec:novice-workers-subsec:urls-fig:longitudinal_ranks} shows the rank distributions of the URLs selected as evidence by the workers when performing each judgment. As for Figure~\ref{cap:paper_pauc2021-sec:results-subsec:info-source-subsec:url-analysis-fig:url-ranks}, URLs selected less than 1\% of the times are filtered out from the results. The trend is similar for \batchone and \batchtwo, while \batchthree and \batchfour display different behavior. For \batchone and \batchtwo, about 40\% of workers select the first result retrieved by the search engine, and select the results down the rank less frequently: about 30\% of workers from \batchtwo and less than 20\% of workers from \batchthree select the first result retrieved by the search engine. The behavior of workers from \batchthree and \batchfour is more oriented towards a model where the user clicks randomly on the retrieved list of results; moreover, the spike which occurs in correspondence of ranks 8, 9, and 10 for \batchfour can be caused by the fact that workers from such batch scroll directly down the user interface with the aim of finishing the task as fast as possible, without putting any effort into providing meaningful sources of evidence.

To provide further insights into the observed change in worker behavior associated with the use of the custom search engine, the sources of information provided by workers as justifications for their judgments are analyzed. Examining the top 10 websites from which workers select URLs shows that, similarly to Table~\ref{cap:paper_pauc2021-sec:results-subsec:info-source-subsec:url-analysis-tab:url-ranks}, many fact-checking websites appear among the top-ranked domains. In particular, \texttt{snopes} is consistently the top-ranked website, and \texttt{factcheck} is always included in the list, except for \batchfour, where all fact-checking websites appear in lower-ranking positions. Additionally, medical websites such as \texttt{cdc} appear only in two out of the four batches (i.e., \batchone and \batchtwo), while the Raleigh-based news website \texttt{wral} is among the top sources in all batches except \batchthree. This is likely due to variations in the geographic location of workers across batches, which may influence their choice of information sources. Overall, this analysis confirms that workers tend to use a variety of sources when selecting URLs. This suggests that they generally put effort into finding evidence to support reliable truthfulness judgments.

As a further analysis, the amount of change in the URLs retrieved by the custom search engine is investigated, focusing in particular on inter- and intra-batch similarity. To this end, the subset of judgments in which the statement is used as a query is selected. The remaining judgments are excluded, as differences in the retrieved URLs would result from the use of different queries. The \meanerr of the two populations of workers (i.e., those who used the statements as queries and those who did not) is computed to ensure that a representative and unbiased subset of workers is selected. In both cases, the \meanerr is nearly the same: $1.41$ for the former and $1.46$ for the latter.

Next, for each statement, all possible pairs of workers who used the statement as a query are considered. For each pair, the overlap between their top 10 retrieved URLs is measured. Three different metrics are used: the rank-based fraction of common documents between the two lists, the number of elements in common, and \index{RBO}RBO. Each metric yields a value in the $[0,1]$ range, indicating the degree of overlap between the two workers' results.

Since the query issued is the same for both workers, any variation in the ranked lists must result from internal policies of the search engine (e.g., considering the worker’s IP address or applying load-balancing strategies). Similarity is measured using either the full URL or the domain only, with a focus on the latter. This allows the detection of cases where an article has moved from one section of a website to another. The findings remain consistent even when full URLs are considered.

The average similarity score for each statement is then computed across all worker pairs to normalize for differences in the number of workers using the same query. This normalization is optional, as the overall findings do not change. Finally, the average similarity score is calculated for each of the three metrics. The similarity of lists within the same batch is higher than that of lists from different batches. In the former case, the similarity scores are $0.45$, $0.64$, and $0.72$, whereas in the latter they are $0.14$, $0.42$, and $0.49$.

\subsubsection{Sources of Information: Justifications}

\label{cap:paper_pauc2021-sec:results-subsec:novice-workers-subsec:justifications}

The textual justifications provided by workers, their relationship with the web pages at the selected URLs, and their association with worker quality are analyzed to assess the impact of different types of justifications on accuracy. This analysis follows the same approach as the main study (Section~\ref{cap:paper_pauc2021-sec:results-subsec:info-source-subsec:justifications}).

\begin{figure}[tpb]
  \centering
  \includegraphics[width=.8\linewidth]{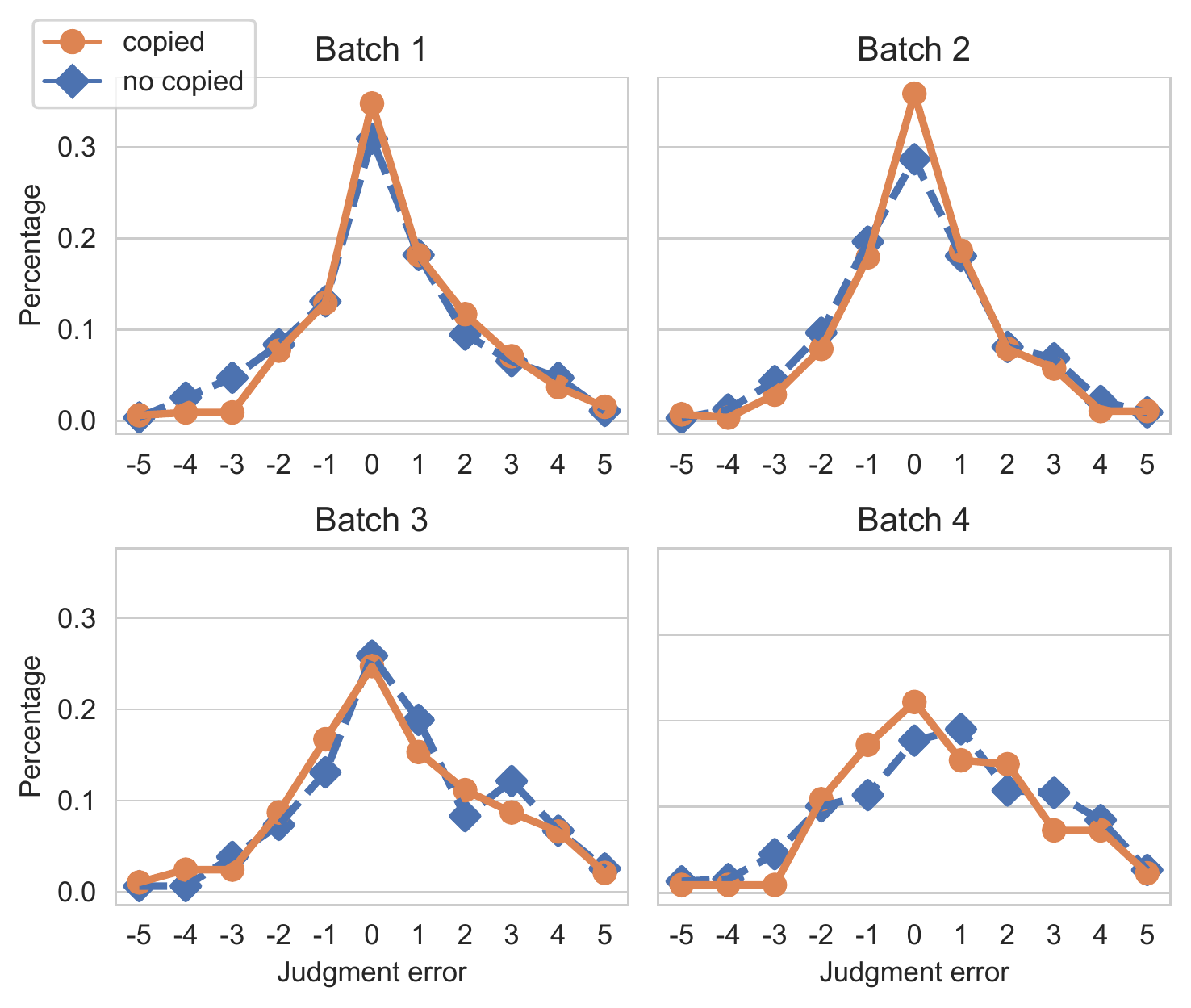}
  \caption{Effect of justification origin on labeling error. Text copied or not copied from the selected URL.}
  \label{cap:paper_pauc2021-sec:results-subsec:novice-workers-subsec:justifications-fig:lg-justification-error}
\end{figure}

Figure~\ref{cap:paper_pauc2021-sec:results-subsec:novice-workers-subsec:justifications-fig:lg-justification-error} shows the relationship between different types of justifications and worker accuracy, following the same approach as in Figure~\ref{cap:paper_pauc2021-sec:results-subsec:info-source-subsec:justifications-fig:error}. The charts display the prediction error for each batch, calculated at each point of disagreement between expert and crowd judgments. They also indicate whether the text entered by the worker was copied from the selected web page or not.

While \batchone and \batchtwo are very similar, \batchthree and \batchfour exhibit notable differences. Statements on which workers make fewer errors (i.e., where $\mbox{x-axis} = 0$) tend to include text copied from the selected web page. Conversely, statements associated with higher error levels (i.e., values close to +/-5) tend to feature justifications not copied from the web page. Overall, workers in \batchthree and \batchfour tend to make more errors than those in \batchone and \batchtwo. As in Figure~\ref{cap:paper_pauc2021-sec:results-subsec:info-source-subsec:justifications-fig:error}, the differences between the two groups of workers are small, but the pattern suggests that higher-quality workers are more likely to read the selected web page and incorporate its content into their justification. Additionally, the distribution of prediction error is not symmetrical. For \batchone, \batchtwo, and \batchthree, the frequency of errors is higher on the positive side of the x-axis (i.e., $[0,5]$), indicating that workers tend to overestimate the truthfulness of the statements. In contrast, \batchfour shows a different pattern. The right-hand side of the chart is noticeably higher for \batchthree compared to \batchone and \batchtwo, reinforcing the observation that workers in \batchthree are of lower quality.

\subsection{\ref{cap:paper_pauc2021-sec:research-questions_7}: Analysis Of Returning Workers}

\label{cap:paper_pauc2021-sec:results-subsec:returning-workers}

The impact of returning workers on the dataset is analyzed. Specifically, the goal is to determine whether workers who performed the task more than once are of higher quality than those who completed it only once. To this end, each possible pair of datasets is considered: one containing returning workers and the other containing only workers who participated once. For each pair, only the subset of \index{HIT}HITs completed by returning workers is analyzed. For this subset, the \meanerr and \cem scores of the two groups of workers are compared.

\begin{figure}[tbp]
  \centering
  \begin{subfigure}{0.49\linewidth}
    \centering
    \includegraphics[width=\linewidth]{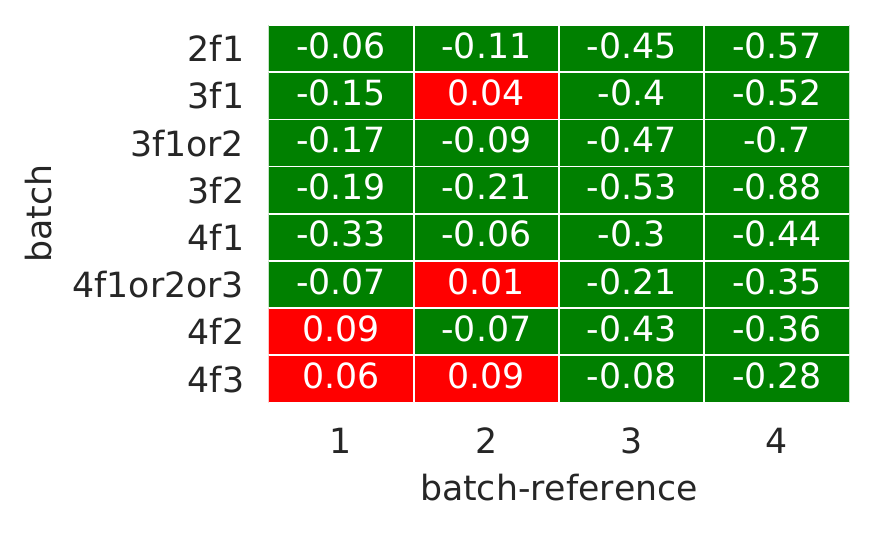}
    \caption{\meanerr}
    \label{cap:paper_pauc2021-sec:results-subsec:returning-workers-fig:mae-cem_mae}
  \end{subfigure}
  \hfill
  \begin{subfigure}{0.49\linewidth}
    \centering
    \includegraphics[width=\linewidth]{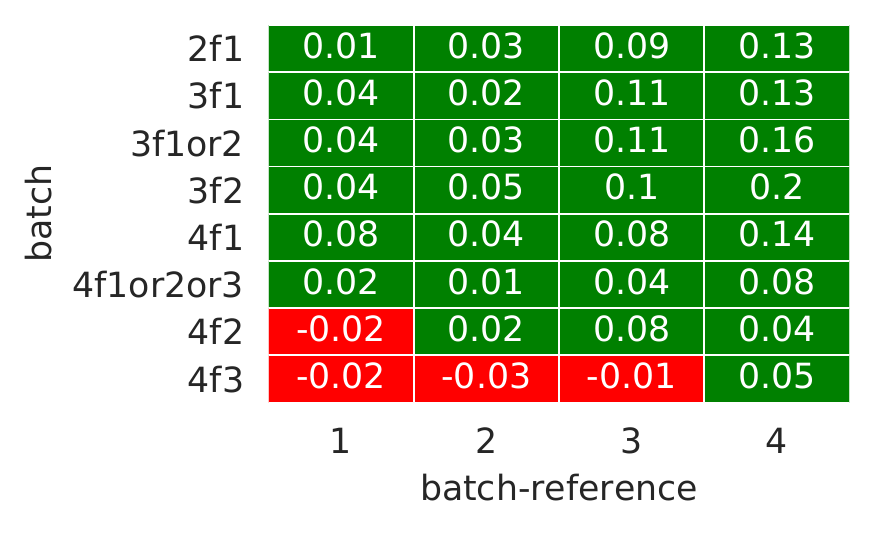}
    \caption{\cem}
    \label{cap:paper_pauc2021-sec:results-subsec:returning-workers-fig:mae-cem_cem}
  \end{subfigure}
  \caption{Comparison of \meanerr and \cem for individual judgments from returning workers. Green indicates that the group on the y-axis has higher quality than the one on the x-axis; red indicates the opposite.}
  \label{cap:paper_pauc2021-sec:results-subsec:returning-workers-fig:mae-cem}
\end{figure}

Figure~\ref{cap:paper_pauc2021-sec:results-subsec:returning-workers-fig:mae-cem} shows the four batches on the x-axis, while the y-axis lists the batches containing returning workers (\texttt{2f1} denotes \batchtwofromone, and so on). Each value represents the difference in \meanerr (Figure~\ref{cap:paper_pauc2021-sec:results-subsec:returning-workers-fig:mae-cem_mae}) and \cem (Figure~\ref{cap:paper_pauc2021-sec:results-subsec:returning-workers-fig:mae-cem_cem}). A cell is colored green if the set of workers on the y-axis has higher quality than the one on the x-axis, and red otherwise.

The trend is consistent across both metrics. With the exception of a few cases involving \batchfour (and with only small differences), each group of returning workers shows similar or higher quality than the others. This effect is particularly evident when the reference batch is \batchthree or \batchfour and the returning workers are from \batchone or \batchtwo, highlighting the high quality of data collected in the first two batches. This outcome is somewhat expected, as people tend to improve at a task through repeated exposure—that is, they learn from experience. However, this behavior should not be taken for granted in a crowdsourcing setting.
An alternative possibility is that returning workers might have focused solely on passing the quality checks to receive the reward, without genuinely trying to perform the task well. The findings indicate that this is not the case, and suggest that the quality checks are effectively designed.

The average time spent on each statement position across all batches is also analyzed. For \batchtwofromone, the average time is 190 seconds (compared to 169 seconds for \batchtwo). For \batchthreefromoneortwo, it is 199 seconds (versus 182 seconds for \batchthree), and for \batchfourfromoneortwoorthree, it is 213 seconds (compared to 140 seconds for \batchfour). Overall, returning workers spend more time on each document than novice workers in the corresponding batch.
A statistical test is performed for each pair of new and returning worker groups. In 12 out of 24 comparisons, the difference is statistically significant (\index{$p$}$p < 0.05$).

\subsection{\ref{cap:paper_pauc2021-sec:research-questions_8}: Qualitative Analysis of Misjudged Statements}
\label{cap:paper_pauc2021-sec:results-subsec:misjudged}

To understand where crowd judgments diverge from expert assessments, the most frequently misjudged statements across all batches are analyzed. Statements are ranked by their \meanerr within each \politifact category to assess error consistency. Section~\ref{cap:paper_pauc2021-sec:results-subsec:misjudged-manual} identifies recurring error patterns through manual inspection, while Section~\ref{cap:paper_pauc2021-sec:results-subsec:misjudged-attributes} examines how statement attributes relate to accuracy.

\subsubsection{Manual Inspection of Misjudged Statements}
\label{cap:paper_pauc2021-sec:results-subsec:misjudged-manual}

\begin{figure}[tbp]
  \centering
  \begin{subfigure}{.49\linewidth}
    \includegraphics[width=\linewidth]{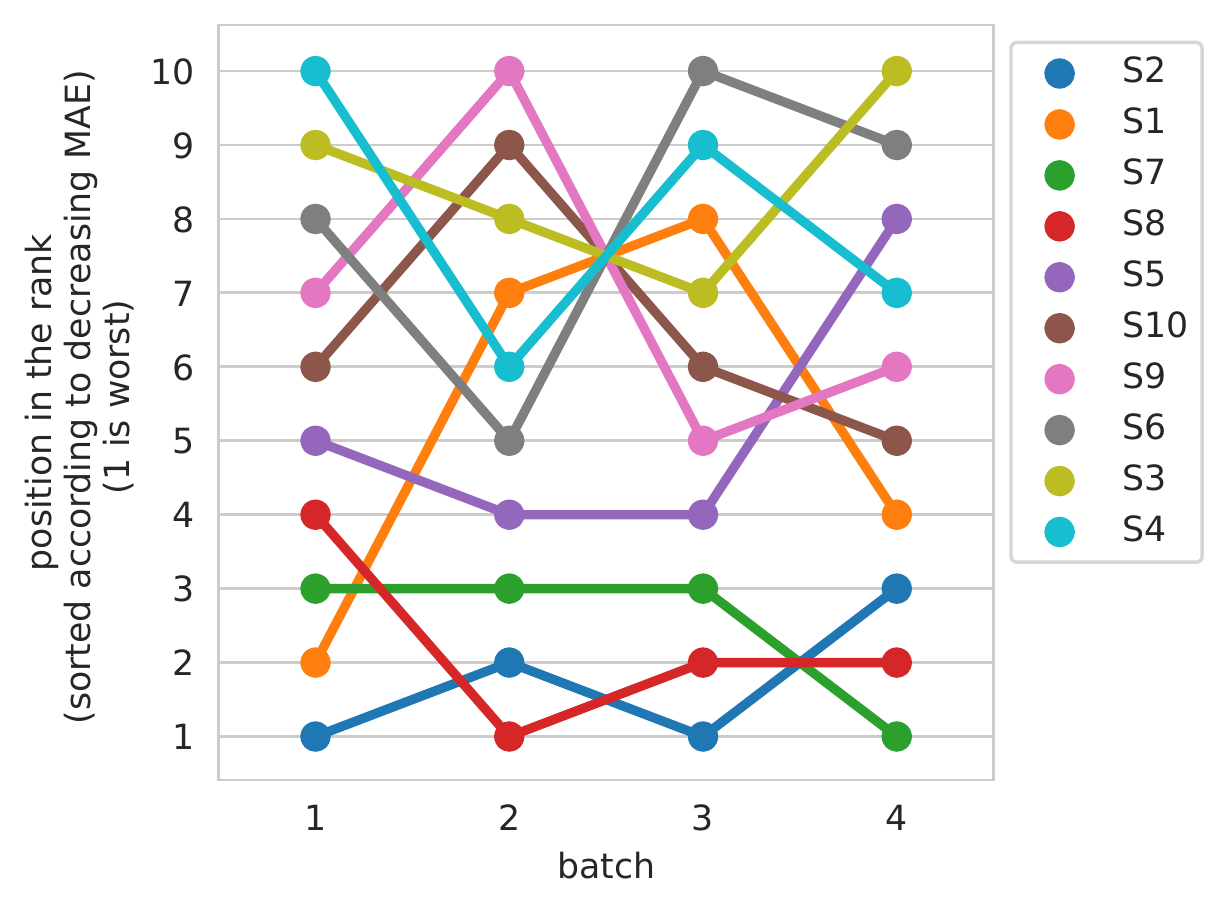}
    \caption{\politifactzero.}
    \label{cap:paper_pauc2021-sec:results-subsec:misjudged-fig:failure-document-parallel-breakdown_0}
  \end{subfigure}
  \hfill
  \begin{subfigure}{.49\linewidth}
    \includegraphics[width=\linewidth]{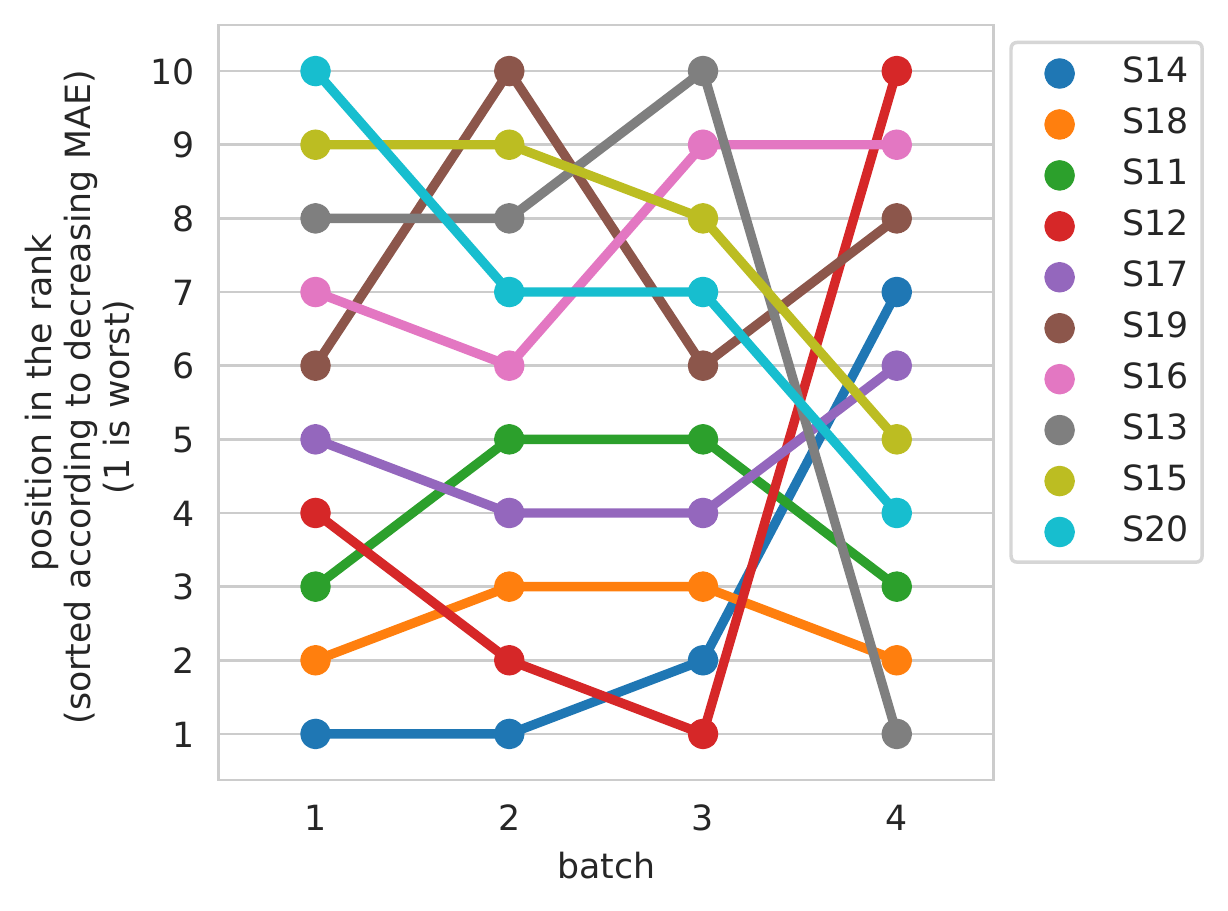}
    \caption{\politifactone.}
    \label{cap:paper_pauc2021-sec:results-subsec:misjudged-fig:failure-document-parallel-breakdown_1}
  \end{subfigure}
  \begin{subfigure}{.49\linewidth}
    \includegraphics[width=\linewidth]{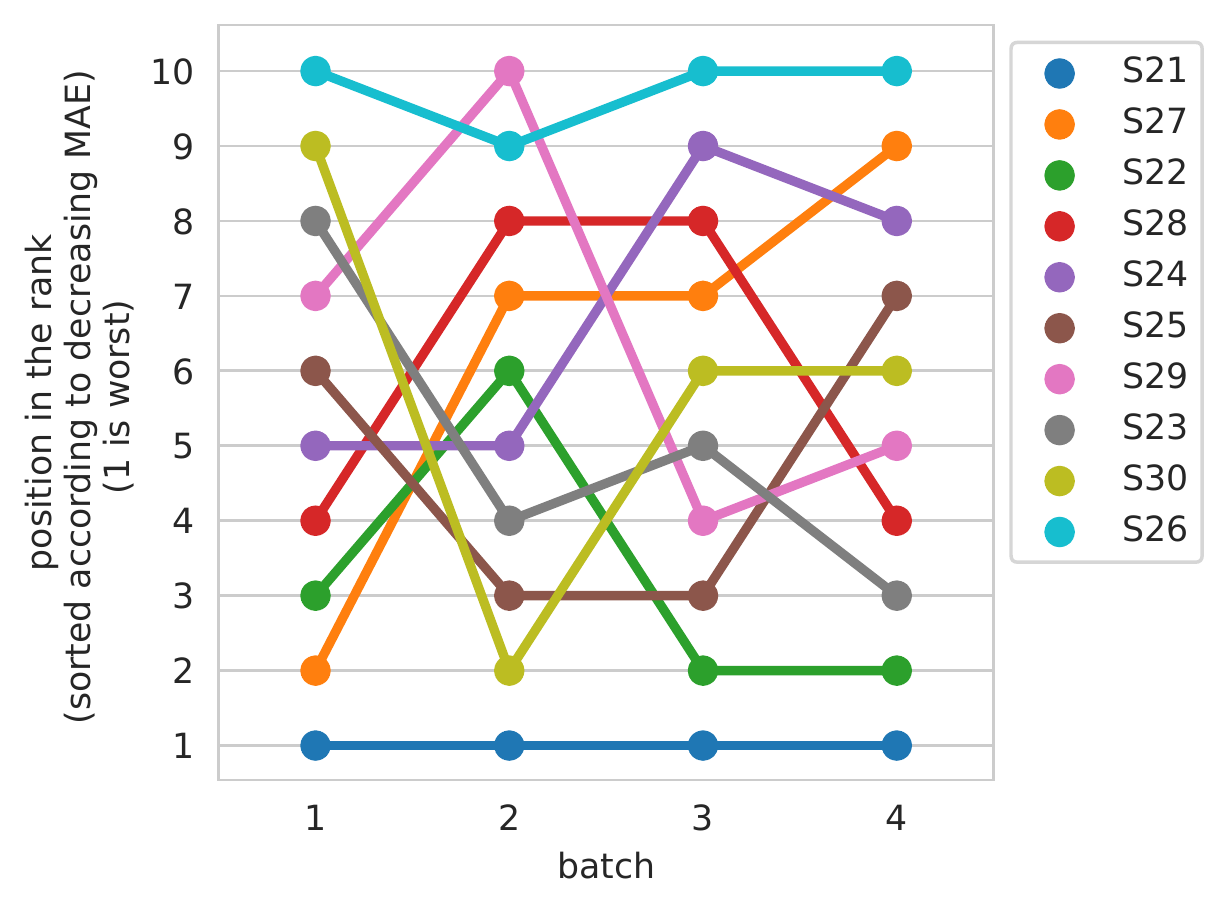}
    \caption{\politifacttwo.}
    \label{cap:paper_pauc2021-sec:results-subsec:misjudged-fig:failure-document-parallel-breakdown_2}
  \end{subfigure}
  \hfill
  \begin{subfigure}{.49\linewidth}
    \includegraphics[width=\linewidth]{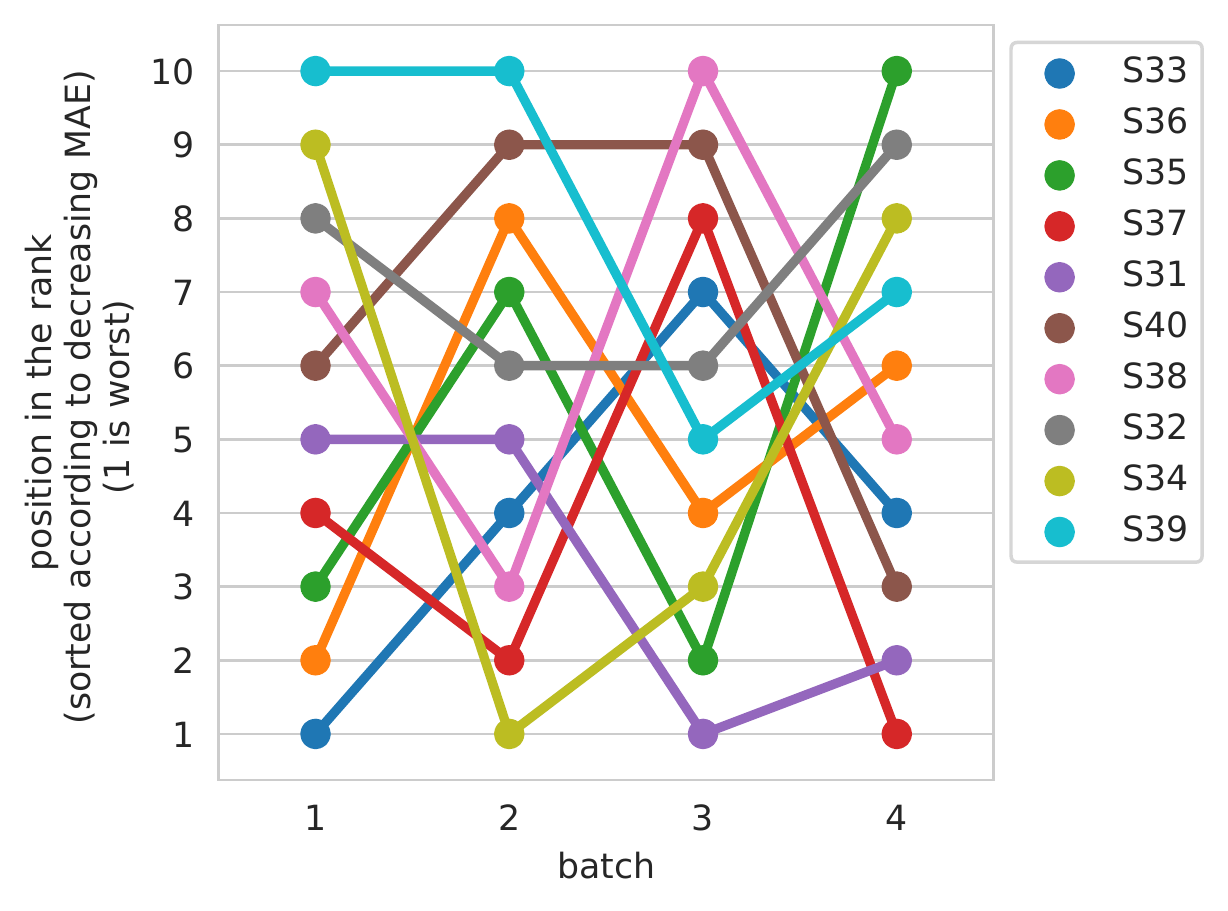}
    \caption{\politifactthree.}
    \label{cap:paper_pauc2021-sec:results-subsec:misjudged-fig:failure-document-parallel-breakdown_3}
  \end{subfigure}
  \begin{subfigure}{.49\linewidth}
    \includegraphics[width=\linewidth]{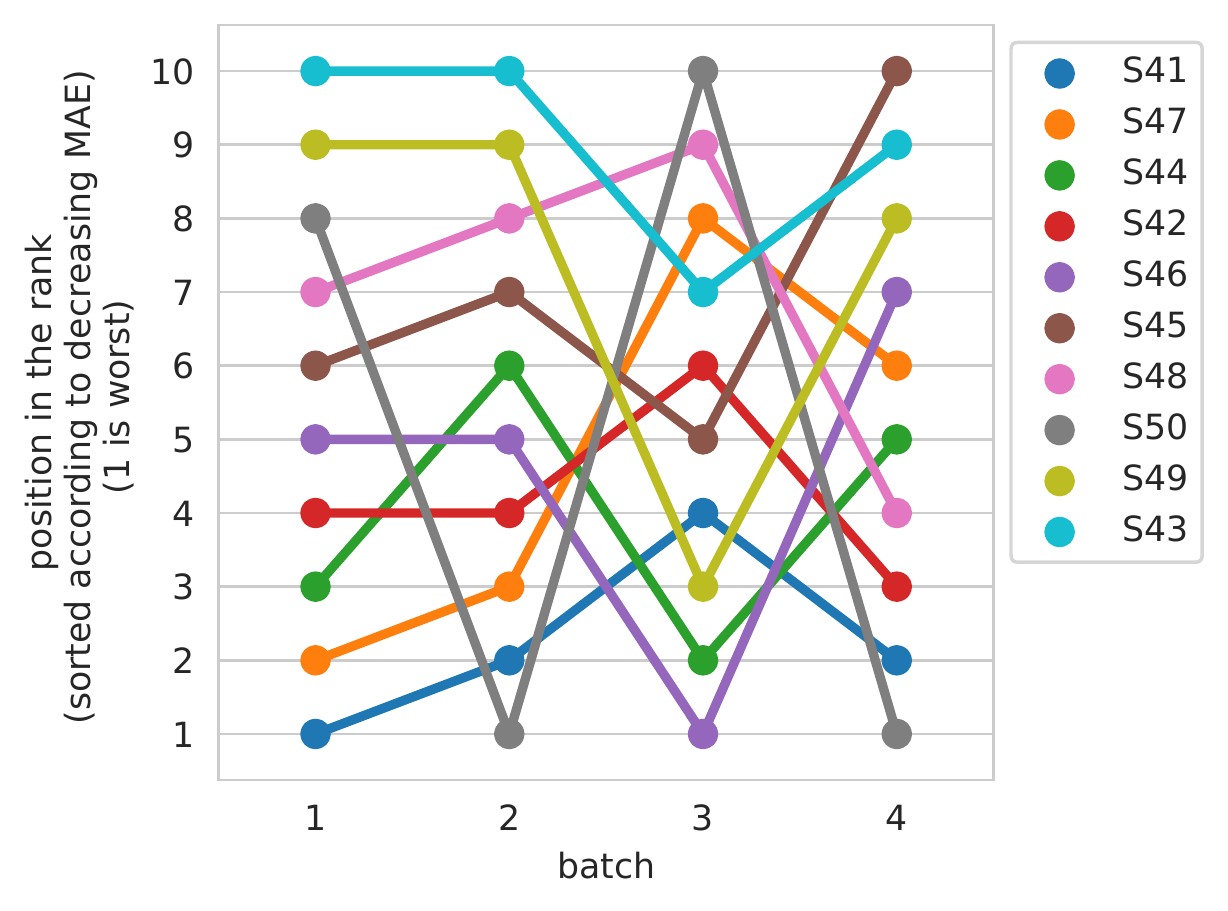}
    \caption{\politifactfour.}
    \label{cap:paper_pauc2021-sec:results-subsec:misjudged-fig:failure-document-parallel-breakdown_4}
  \end{subfigure}
  \hfill
  \begin{subfigure}{.49\linewidth}
    \includegraphics[width=\linewidth]{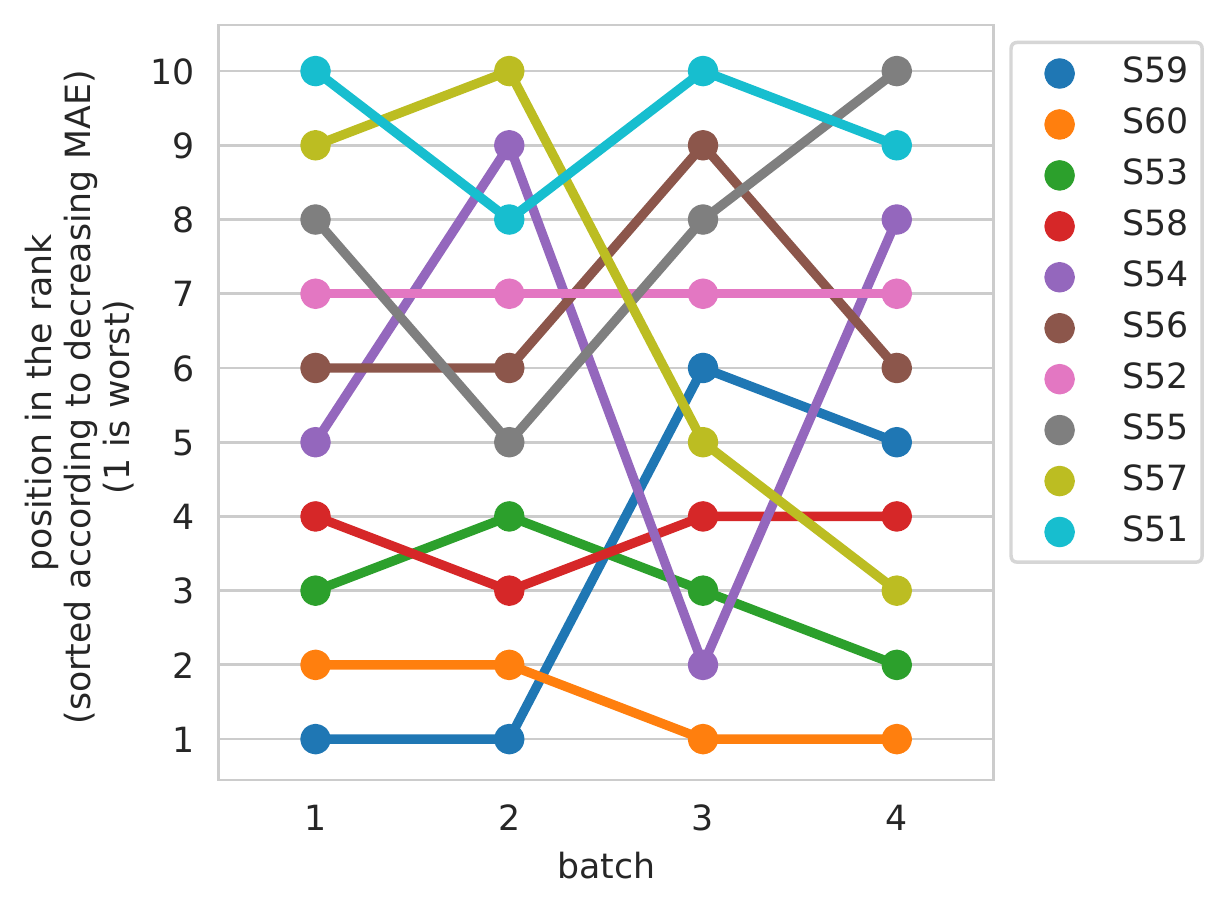}
    \caption{\politifactfive.}
    \label{cap:paper_pauc2021-sec:results-subsec:misjudged-fig:failure-document-parallel-breakdown_5}
  \end{subfigure}
  \caption{Relative ordering of statements across batches according to MAE for each \politifact category. Rank 1 represents the highest MAE.}
  \label{cap:paper_pauc2021-sec:results-subsec:misjudged-fig:failure-document-parallel-breakdown}
\end{figure}

Figure~\ref{cap:paper_pauc2021-sec:results-subsec:misjudged-fig:failure-document-parallel-breakdown} shows, for each \politifact category, the relative ranking of statements in decreasing order of \meanerr. The statement ranked first has the highest \meanerr. Some statements appear consistently among the most misjudged within each category.

\begin{itemize}[label=--]
\item \politifactpantsfire: \statement{2}, \statement{8}, \statement{7}, \statement{5}, \statement{21}.
\item \politifactfalse: \statement{18}, \statement{14}, \statement{11}, \statement{12}, \statement{17}.
\item \politifactmostlyfalse: \statement{21}, \statement{22}, \statement{25}.
\item \politifacthalftrue: \statement{31}, \statement{37}, \statement{33}.
\item \politifactmostlytrue: \statement{41}, \statement{44}, \statement{42}, \statement{46}.
\item \politifacttrue: \statement{60}, \statement{53}, \statement{59}, \statement{58}.
\end{itemize}

The 24 statements selected (Appendix~\ref{cap:paper_pauc2021:-appendix:statements}) are manually inspected to investigate the causes of error. The corresponding justifications provided by workers are also reviewed. Most of the errors in \batchthree and \batchfour are caused by workers who respond randomly, introducing noise into the data. Responses are classified as noise when the following two criteria are both satisfied:
\begin{enumerate}
\item The chosen URL is unrelated to the statement (for example, a Wikipedia page defining the word truthfulness or a website for creating flashcards).
\item The justification text does not explain the selected truthfulness level. This includes both personal explanations and copied content that does not come from a relevant or matching source.
\end{enumerate}

Noisy responses become more frequent in later batches and account for almost all errors in \batchfour. The number of judgments with noisy answers in the four batches is respectively $27$, $42$, $102$, and $166$. In contrast, the number of non-noisy answers is $159$, $166$, $97$, and $54$.
Non-noise errors in \batchone, \batchtwo, and \batchthree appear to depend on the specific statement. The main reasons for failure in identifying the correct label are identified through manual inspection of the justifications provided by the workers.

In four cases (\statement{14}, \statement{25}, \statement{41}, \statement{53}), the statements are objectively difficult to evaluate. This may be due to the need for close attention to medical terminology (\statement{14}), the presence of highly debated issues (\statement{25}), or the requirement of legal knowledge (\statement{53}).

In another four cases (\statement{42}, \statement{46}, \statement{59}, \statement{60}), workers were unable to find relevant information and therefore guessed. The difficulty in retrieving information is justified: some statements are too vague to yield useful results (\statement{59}), others involve topics with little official data available (\statement{46}), or refer to issues that have already been resolved and replaced by newer coverage, making them harder to find online (\statement{60}, \statement{59}, \statement{42}). For example, truck drivers initially faced difficulties obtaining food at fast food restaurants, but once this issue was resolved, news outlets shifted their focus to the new problem of a lack of truck drivers for restocking supermarkets and fast food chains.

In four other cases (\statement{33}, \statement{37}, \statement{59}, \statement{60}), workers retrieved information that covered only part of the statement. This sometimes occurred by accident—for example, retrieving information about Mardi Gras 2021 instead of Mardi Gras 2020 (\statement{60})—or by relying on generic sources that support only part of the claim (\statement{33}, \statement{37}).

Finally, in four cases (\statement{2}, \statement{8}, \statement{7}, \statement{1}), \politifactpantsfire statements were judged as true, likely because they were actually stated by the person in question. In these cases, workers sometimes used fact-checking websites as the selected URL and even wrote in their justification that the statement was false, but still selected \politifacttrue as the label.

In thirteen cases (\statement{7}, \statement{8}, \statement{2}, \statement{18}, \statement{22}, \statement{21}, \statement{33}, \statement{37}, \statement{31}, \statement{42}, \statement{44}, \statement{58}, \statement{60}), the statements are rated as more true or more false than they actually are, often because workers focus on how plausible the statements sound. In most of these cases, workers find a fact-checking website that reports the correct judgment, but they choose to modify their answer based on personal opinion. For example, true statements related to politics are sometimes questioned (\statement{60}, about nobody suggesting to cancel Mardi Gras), while false statements are treated as exaggerations intended to emphasize the seriousness of a situation (\statement{18}, about church services not resuming until everyone is vaccinated).

In five cases (\statement{1}, \statement{5}, \statement{17}, \statement{12}, \statement{11}), the statements are difficult to verify due to a lack of reliable sources or supporting test data. These statements also contain concerning information, particularly about how coronavirus can be transmitted and how long it can survive. Most workers retrieve fact-checking articles that classify the statements as \politifactfalse or \politifactpantsfire, but they opt for an intermediate rating. The justifications in these cases often include personal reflections or excerpts from the selected URL that express uncertainty, such as mentioning inconclusive test results or limited scientific knowledge about the virus. In some cases, workers suggest it is better to act with caution by assuming the worst-case scenario, for instance by avoiding products from China or leaving packages in the sunlight in an attempt to neutralize the virus.

\subsubsection{Effect of Statement Attributes and Time}
\label{cap:paper_pauc2021-sec:results-subsec:misjudged-attributes}

Following the manual inspection, the worst individual judgments (i.e., the ones classified as noise) are removed. This filtering has only a minimal effect on the aggregated judgments. The resulting boxplots remain nearly identical to those shown in Figure~\ref{cap:paper_pauc2021-sec:results-subsec:crowd-accuracy-subsec:ext-agreement-fig:agreement-ground-truth}, where no judgments were excluded.

This section investigates how the correctness of the judgments relates to the attributes of the statements—namely position, speaker, and context—as well as to the passage of time. For each batch, the absolute distance from the correct truthfulness value is computed for each judgment and then aggregated by statement, yielding the mean absolute error (\meanerr) and the standard deviation (\index{$\upsigma$}$\upsigma$). The statements are then sorted in descending order of \meanerr and \index{$\upsigma$}$\upsigma$, and the top 10 statements are selected for further analysis.

When examining the position of the statement within the task, no specific pattern emerges. The incorrect statements are distributed across all positions in all batches, suggesting that position has no significant effect.
When considering the speaker and the context, the most frequently misjudged statements are often attributed to \lq\lq Facebook User\rq\rq{}, which is also the most common speaker label in the dataset. This pattern may reflect the general ambiguity and lack of trustworthiness associated with anonymous or crowd-sourced statements on social media.
To assess the effect of time, the \meanerr of each statement is plotted against the number of days between the date the statement was made and the date it was evaluated by the workers. This is done separately for novice workers (\batchone to \batchfour) and returning workers (\batchtwofromone, \batchthreefromoneortwo, \batchfourfromoneortwoorthree).

\begin{figure}[tbp]
  \centering
  \begin{subfigure}{.99\linewidth}
    \centering
    \includegraphics[width=\linewidth]{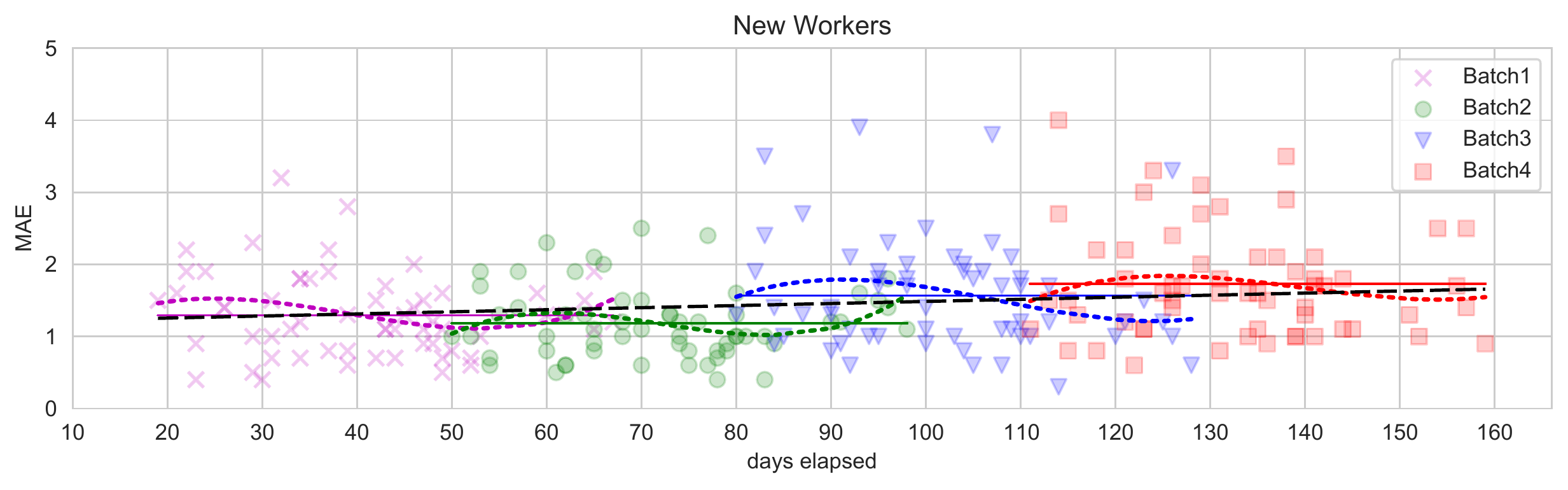}
    \caption{Novice workers.}    
    \label{cap:paper_pauc2021-sec:results-subsec:misjudged-fig:deltaT-mae_new}
  \end{subfigure}
  \begin{subfigure}{.99\linewidth}
    \centering
    \includegraphics[width=\linewidth]{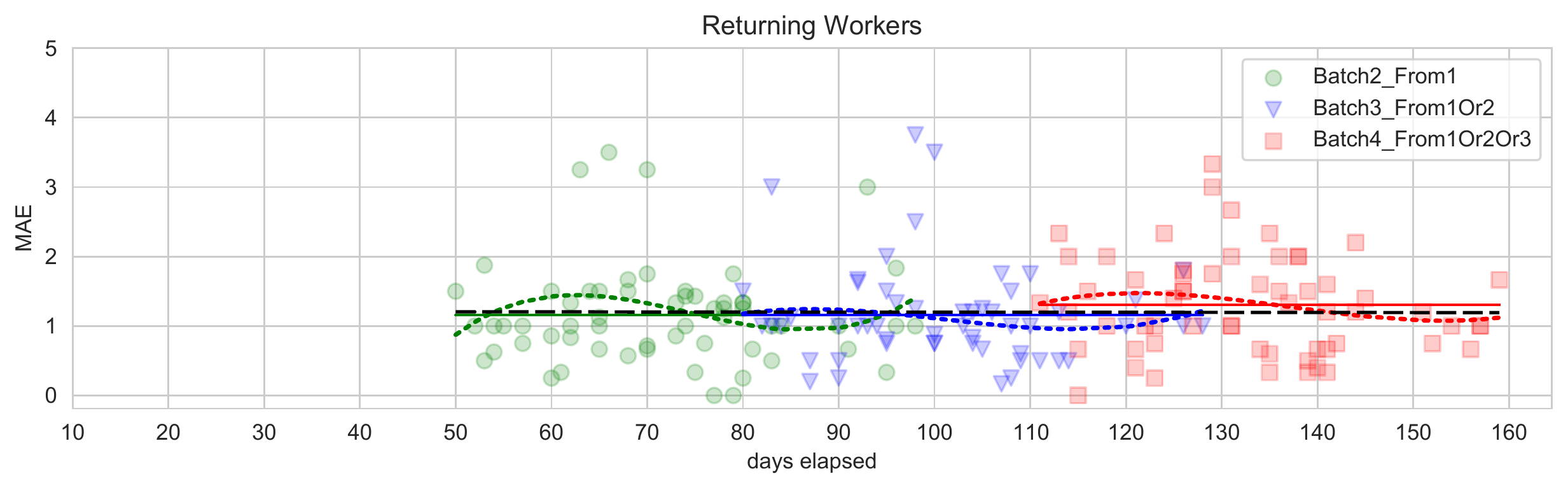}
    \caption{Returning workers.}   
    \label{cap:paper_pauc2021-sec:results-subsec:misjudged-fig:deltaT-mae_all}
  \end{subfigure}
  \caption{\meanerr (aggregated by statement) plotted against the number of days elapsed between the date the statement was made and the date it was evaluated. Each point represents the \meanerr of a single statement within a batch. Dotted lines indicate the trend of \meanerr over time for each batch, straight lines show the mean \meanerr for each batch, and the black dashed line represents the global trend of \meanerr across all batches.}
  \label{cap:paper_pauc2021-sec:results-subsec:misjudged-fig:deltaT-mae}
\end{figure}

Figure~\ref{cap:paper_pauc2021-sec:results-subsec:misjudged-fig:deltaT-mae} shows that the \meanerr trends for novice workers (panel a) are consistent across batches: statements made in April (on the left side of each plot) tend to have higher error than those from earlier in the pandemic. Figure~\ref{cap:paper_pauc2021-sec:results-subsec:misjudged-fig:deltaT-mae_new} confirms that overall error increases across batches, as shown by the black dashed trend line. This effect appears not to be driven by the time elapsed since the statement was made, but rather by decreasing worker quality.

This interpretation is supported by Figure~\ref{cap:paper_pauc2021-sec:results-subsec:misjudged-fig:deltaT-mae_all}, which shows that \meanerr for returning workers remains stable over time. For these workers, the internal trend within each batch also matches what was observed earlier: statements from April and late March are consistently harder to evaluate. Overall, the elapsed time between the origin and evaluation of a statement does not seem to significantly affect judgment accuracy.

\section{Summary}

\label{cap:paper_pauc2021-sec:discussion}

This chapter presents a comprehensive investigation into the ability and behavior of crowd workers when asked to identify and assess the truthfulness of recent health-related statements concerning the \covid pandemic. Workers are tasked with evaluating the truthfulness of eight statements using a customized search engine, which enables control over their behavior. Workers’ backgrounds, biases, and cognitive abilities are also analyzed, and this information is correlated with their performance quality.

The experiment is conducted in four separate batches, each spaced one month apart, and includes both novice (new) and experienced (returning) workers. The longitudinal design serves two distinct purposes:
\begin{enumerate}
\item To examine how novice workers assess the truthfulness of \covid-related news over time.
\item To analyze how returning workers reassess the same set of statements after some time.
\end{enumerate}

The underlying hypothesis is that, over time, workers become more aware of the truthfulness of \covid-related information. This hypothesis does not hold for novice workers, and the result aligns with the findings of \citet{Qarout2019}.
However, when a batch includes only returning workers, an increase in agreement is observed, indicating that workers tend to improve through experience. Moreover, returning workers do not simply focus on passing quality checks, which confirms the overall high quality of the data collected. These findings suggest that running additional batches with returning workers can be expected to provide higher-quality judgments over time. The answers to the research questions can be summarized as follows.

\myparagraph{\ref{cap:paper_pauc2021-sec:research-questions_1}}
There is evidence that workers can detect and objectively categorize recent online (mis)information related to the \covid pandemic. Aggregated worker judgments show high levels of agreement with expert labels, with the only exception being the two truthfulness categories at the lower end of the scale (\politifactpantsfire and \politifactfalse). However, agreement among workers alone does not provide a strong signal.

\myparagraph{\ref{cap:paper_pauc2021-sec:research-questions_2}}
Both crowdsourced and expert judgments can be transformed and aggregated to improve label quality.

\myparagraph{\ref{cap:paper_pauc2021-sec:research-questions_3}}
The effectiveness of workers is slightly correlated with their questionnaire responses, although the correlation is never statistically significant.

\myparagraph{\ref{cap:paper_pauc2021-sec:research-questions_4}}
The relationship between workers' background or bias and their quality has been used to enhance the aggregation of individual judgments. However, this does not lead to a noticeable increase in external agreement. It is possible that these signals could inform new aggregation strategies, which should be explored using more complex methods in future work.

\myparagraph{\ref{cap:paper_pauc2021-sec:research-questions_5}}
Workers rely on multiple sources of information, including both fact-checking and health-related websites. There are notable relationships between the justifications provided and the quality of the corresponding judgments.

\myparagraph{\ref{cap:paper_pauc2021-sec:research-questions_6}}
Re-collecting data at different time intervals has a significant impact on judgment quality. Among novice workers, early batches are more consistent with each other than later ones. Additionally, batches that are temporally closer tend to be more similar in terms of worker quality. Novice workers appear to put considerable effort into the task, using various sources and frequently reformulating queries.

\myparagraph{\ref{cap:paper_pauc2021-sec:research-questions_7}}
Experienced (returning) workers spend more time on each statement compared to novice workers in the corresponding batch. They also show similar or higher quality. Moreover, as the number of batches increases, the average time spent per statement decreases substantially.

\myparagraph{\ref{cap:paper_pauc2021-sec:research-questions_8}}
An extensive analysis of the features and peculiarities of statements misjudged across datasets is provided. The time elapsed since a statement was made does not appear to affect the quality of workers' judgments.

\myparagraph{}
The next chapter explores the barriers to conducting longitudinal studies on crowdsourcing platforms, using the study presented above as a motivating example. To this end, a large-scale survey is conducted across three major commercial crowdsourcing platforms.

\chapter[The Barriers To Longitudinal Studies On Crowdsourcing Platforms]{Understanding The Barriers To Running Longitudinal Studies On Crowdsourcing Platforms}

\label{cap:paper_tsc2024}

This chapter is based on the article published in the \lq\lq Transactions on Social Computing\rq\rq{} journal~\cite{soprano2023loyalty}. Section~\ref{cap:related_work-sec:longitudinal-studies} reviews the relevant related work. Section~\ref{cap:paper_tsc2024-sec:research_questions} outlines the research questions. Section~\ref{cap:paper_tsc2024-sec:exp-setup} describes the experimental setup. Section~\ref{cap:paper_tsc2024-sec:results} presents the results. Finally, Section~\ref{cap:paper_tsc2024-sec:discussion} summarizes the main findings and concludes the chapter.

\section{Research Questions}

\label{cap:paper_tsc2024-sec:research_questions}

This chapter addresses the research gap concerning worker perception in longitudinal studies. While previous research has primarily focused on short-term micro-task crowdsourcing, it has provided limited exploration of longitudinal studies, an area that remains under-explored. Although considerations and suggestions for designing and conducting such studies have been proposed, there has been limited characterization from the worker perspective.

The novelty of this research lies in its experimental data, with surveys conducted across three diverse crowdsourcing platforms to capture a broad spectrum of experiences and perspectives on longitudinal studies. Additionally, the study investigates the specific dynamics of longitudinal studies on crowdsourcing platforms using a mixed-methods approach. While previous works have examined various aspects of crowdsourcing, this study seeks to provide new insights into the unique challenges and opportunities associated with conducting longitudinal studies. Understanding key aspects of longitudinal study design is crucial for identifying barriers experienced by workers and for providing recommendations for practitioners and researchers conducting such studies on micro-task crowdsourcing platforms. This also enables the proposal of best practices for platforms supporting longitudinal research. The research focuses on those involved in designing and enabling longitudinal studies, with considerations based on both the worker perspective and prior experience as task requesters.
The following research questions are investigated:

\begin{enumerate}[start=13, leftmargin=2.92em, label=RQ\arabic*]
\item \label{cap:paper_tsc2024-sec:research-questions_1} What is the current perception of workers regarding longitudinal studies on commercial micro-task crowdsourcing platforms? What were their previous experiences like? What are their opinions on participation and commitment to future longitudinal studies? What are their preferred characteristics for such studies?
\item \label{cap:paper_tsc2024-sec:research-questions_2} What recommendations should researchers and practitioners follow when designing and conducting longitudinal studies on commercial micro-task crowdsourcing platforms?
\item \label{cap:paper_tsc2024-sec:research-questions_3} What best practices should commercial micro-task crowdsourcing platforms adopt to effectively support longitudinal studies and improve their general suitability for such research?
\end{enumerate}

\section{Experimental Setting}

\label{cap:paper_tsc2024-sec:exp-setup}

A survey is designed to characterize longitudinal studies from the perspective of crowd workers (Section~\ref{cap:paper_tsc2024-sec:exp-setup-subsec:survey-design}). Responses are collected through a crowdsourcing task (Section~\ref{cap:paper_tsc2024-sec:exp-setup-subsec:task}) conducted on three popular commercial crowdsourcing platforms: \mturk (Section~\ref{cap:paper_wsdm2022-sec:crowdsourcing_platforms-subsec:mturk}), \prolific (Section~\ref{cap:paper_wsdm2022-sec:crowdsourcing_platforms-subsec:prolific}), and \toloka (Section~\ref{cap:paper_wsdm2022-sec:crowdsourcing_platforms-subsec:toloka}). The responses are analyzed using both quantitative and qualitative methods (Section~\ref{cap:paper_tsc2024-sec:exp-setup-subsec:responses-analysis}), and statistical significance tests are performed to support the findings (Section~\ref{cap:paper_tsc2024-sec:exp-setup-subsec:stat-test}).

The complete survey, including workers' responses and the dataset used for both quantitative and qualitative analyses, has been released. The qualitative component presents a thematic analysis and includes a detailed description of the coding scheme, codes, and resulting themes. The full text of the survey is provided in Appendix~\ref{cap:paper_tsc2024-appendix:survey-questions}.

\subsection{Survey Design}

\label{cap:paper_tsc2024-sec:exp-setup-subsec:survey-design}

The survey consists of two parts: \pone\ and \ptwo. The first part of the survey (\pone), reported in Appendix~\ref{cap:paper_tsc2024-appendix:survey-questions-pone}, aims to explore the current perception of longitudinal studies in crowdsourcing. It focuses on workers' prior experience, the perceived suitability of platforms for hosting longitudinal studies, possible reasons limiting the popularity of longitudinal studies and their availability on crowdsourcing platforms. 

The second part (\ptwo), reported in Appendix~\ref{cap:paper_tsc2024-appendix:survey-questions-ptwo}, on the other hand, investigates workers' thoughts, opinions, and ideas about the design of, and their underlying motivations to participate in future longitudinal studies.

More specifically, the survey comprises 16 multiple-choice questions, 4 text-based questions (i.e., questions requiring a mandatory textual answer), and 6 numerical questions. Additionally, there are 11 questions that allowed workers to provide custom free-text responses to elaborate on their answers. Among the multiple-choice questions, 9 of them are implemented using radio buttons, as only a single answer was possible. In contrast, checkboxes are employed for the remaining 7 questions, as they allow for multiple answers, thus permitting a broader range of responses. The naming convention reported in Appendix~\ref{cap:paper_tsc2024-appendix:survey-questions} is used throughout the rest of this thesis.

\subsection{Crowdsourcing Task}

\label{cap:paper_tsc2024-sec:exp-setup-subsec:task}

The crowdsourcing task aimed to recruit \numworkers workers from three platforms: \mturk, \prolific, and \toloka, with 100 participants from each. Participation criteria required completing at least $4000$ tasks on \mturk and $2000$ tasks on \prolific. On \toloka, participants were directly asked about their prior experiences with longitudinal studies. Recruitment continued on each platform until 100 participants with at least one previous longitudinal study experience were obtained.

Fifty workers were initially recruited from each of the three platforms. However, analysis revealed that only a subset had prior experience with longitudinal studies. As a result, the recruitment process was repeated up to four times per platform until a total of 300 workers with at least one such experience was obtained. In total, 729 workers successfully completed the task: 153 from \mturk, 160 from \prolific, and 412 from \toloka. This means that, for example, the required 100 experienced workers on \mturk were selected from among 153 participants. The task was deployed across multiple time periods on each platform:
\begin{itemize}[label=--]
\item \mturk: April 14–15, 2022; August 29–September 1, 2022; September 12, 2022; and March 10–13, 2023
\item \prolific: April 14, 2022; September 15, 2022; March 16–17, 2023; and April 11, 2023
\item \toloka: September 12–15, 2022; March 10, 2023; March 13, 2023; and March 15–17, 2023
\end{itemize}
The first iteration was launched on \mturk on April 14, 2022, and the final one on \prolific on April 11, 2023. The task workflow and layout remained unchanged throughout and were consistently available during the specified periods.

The task proceeded as follows. Workers were first provided with general instructions and background on the study, including the definition of longitudinal studies (Section~\ref{cap:intro-sec:crowdsourcing-activity}). They then completed the first part of the survey (\pone), followed by the second part (\ptwo). In \pone, workers were asked to report their experiences with up to three longitudinal studies they had participated in. This cap was introduced to ensure a reasonable completion time for the crowdsourcing task.

Each experience was reported and described by responding to a subset of 11-13 questions, with the total number of questions shown depending on the answer provided for question 1.1 (Appendix~\ref{cap:paper_tsc2024-appendix:survey-questions-pone}). Conditional logic was used to determine whether certain sub-questions needed to be asked. Specifically, if a worker reported between $0\leq X\leq3$ experiences (denoted as $X$), the number of questions ranged from $1+(11*X)+2$ to $1+(13*X)+2$, as the block of questions 1.1.X was repeated $X$ times, once for each experience. Additionally, only one question from either 1.1.X.9.1 or 1.1.X.9.2 was shown, depending on the answer provided for question 1.1.X.9. Conversely, the \ptwo part comprised 11 questions, asked only once. Thus, the total number of questions in the entire survey ranged from $1+(11*X)+13$ to $1+(13*X)+13$.

After completing \pone and \ptwo, workers could submit their responses and receive payment. They also had the opportunity to provide final comments. To ensure response quality, a criterion required workers to spend a minimum of 3 seconds on each question. Workers received \$2 USD for their participation, based on an hourly rate derived from the US minimum wage and task completion time. The median reward ranged from \$10-13 per hour, with an average completion time of 700 seconds, a standard deviation of 593, and a median of 548 seconds.

\subsection{Analysis Of Workers' Responses}

\label{cap:paper_tsc2024-sec:exp-setup-subsec:responses-analysis}

Each survey question is addressed from a quantitative or qualitative viewpoint, depending on the question type.

Initially, general remarks concerning the results obtained are provided (Section~\ref{cap:paper_tsc2024-sec:exp-setup-subsec:responses-analysis-subsec:remarks}). Then, the focus shifts specifically to the quantitative analysis (Section~\ref{cap:paper_tsc2024-sec:exp-setup-subsec:responses-analysis-subsec:quantitative-analysis}) and the qualitative approach followed (Section~\ref{cap:paper_tsc2024-sec:exp-setup-subsec:responses-analysis-subsec:qualitative-analysis}).

\subsubsection{General Remarks}

\label{cap:paper_tsc2024-sec:exp-setup-subsec:responses-analysis-subsec:remarks}

Accurate interpretation of the results requires noting that some survey questions allowed multiple responses based on workers' past experiences with longitudinal studies, while others required only a single response, as detailed in Section~\ref{cap:paper_tsc2024-sec:exp-setup-subsec:task}.

Most questions in the \pone part required responses for each past experience. In contrast, questions in the \ptwo part and one question from the \pone part required only a single response. With \numworkers workers participating, the maximum number of responses for the former category was 900 (assuming three experiences per worker), while the maximum for the latter was \numworkers. Section~\ref{cap:paper_tsc2024-sec:results-subsec:rq1-analysis-p1-first} indicates that the total number of reported experiences was~\numexperiences.

For result analysis, breakdowns are often provided based on the platform used to recruit workers who responded to the survey. For instance, a worker recruited from \mturk but reporting on a longitudinal study conducted on \prolific is categorized under the \mturk breakdown..

\subsubsection{Quantitative Analysis}

\label{cap:paper_tsc2024-sec:exp-setup-subsec:responses-analysis-subsec:quantitative-analysis}

Bar charts are used for closed-ended multiple-choice questions, while univariate distribution charts are applied for numerical questions in the quantitative analysis. Results are categorized by crowdsourcing platform to visually highlight differences. A color scheme, with blue for \mturk, orange for \prolific, and green for \toloka, is introduced in the legend of the first figure and applied consistently in subsequent figures to minimize repetition and avoid information overload.

In bar charts, the x-axis displays the available answers, while the second row shows their relative frequencies across platforms. The y-axis represents answer frequencies, with absolute frequencies shown above each bar. Total absolute frequencies correspond to \numexperiences or \numworkers, depending on the question requirements, and are denoted as $E$ or $W$, respectively. These values appear in the lower left corner of the chart. For questions allowing non-mutually-exclusive answers, the top chart includes the label $A$. Additionally, the total number of answers and experiences or workers is provided in the lower left corner of the chart.

In univariate distribution charts, the y-axis represents the probability density function for three continuous random variables corresponding to the answers provided by workers across each considered crowdsourcing platform. The x-axis spans from the minimum value to a cutoff to exclude outliers. Dashed lines indicate the mean values for each platform, using the predefined color scheme. The total data used is displayed in the lower left corner, marked with a corresponding letter. In cases where outliers are excluded, an additional label beneath the data count highlights this adjustment.

\subsubsection{Qualitative Analysis}
\label{cap:paper_tsc2024-sec:exp-setup-subsec:responses-analysis-subsec:qualitative-analysis}

A conventional qualitative content analysis approach \cite{doi:10.1177/1049732305276687} was applied to analyze open-ended responses. This inductive method is used to describe phenomena with limited existing research or theory, in contrast to deductive qualitative analysis, which depends on predetermined themes from literature.

Two expert researchers reviewed all responses to the open-ended mandatory questions and those allowing free-text input. For each response, a custom "code" was created by identifying and highlighting key phrases that captured significant insights, using a predefined keyword. For example, if a worker mentioned participating in the longitudinal study due to its engaging nature and opportunities for self-discovery, the initial code might be the keyword \taskinterest. As the analysis advanced, multiple core concepts emerged, forming the basis of the initial overall coding scheme.

The qualitative analysis phase involved merging initially identified codes based on their interdependencies through multiple iterations and discussions. For example, codes such as \taskinterest, \taskpayment, and \taskeasiness were merged into the overarching theme of \taskfeatures. This process led to the emergence of seven themes, detailed in Table~\ref{cap:paper_tsc2024-sec:qualitative-analysis-subsec:remarks-tab:theme-dist}, along with sample answers and initial codes. Due to expert involvement and iterative refinement, internal agreement is not reported here; interested readers can refer to \citet{mcdonald2019reliability}.

Table~\ref{cap:paper_tsc2024-sec:results-subsec:rq1-analysis-subsec:remarks-tab:free-texts} details the distribution of additional free-text responses provided by workers when answering each mandatory non-text-based question. For \pone part questions, the table reports both the total number of experiences with text and the number of workers providing them. For \ptwo part questions, only the latter is provided, as workers are asked only once per question. These texts complement the thematic analysis, offering additional insights to the quantitative analysis of the provided answers.

Finally, it is important to note that if a question, such as question 1.1.X.9.2, is not included in the result analysis, it is because it specifically required text-based answers, and the workers did not provide any useful responses for analysis.

\begin{table}[htpb]
    \centering
    \caption{Themes emerged while reading each text-based answer provided by the workers.}
\label{cap:paper_tsc2024-sec:qualitative-analysis-subsec:remarks-tab:theme-dist}
\begin{tabular}{p{2.5cm}p{4.8cm}p{4.65cm}p{2.5cm}}
\toprule
  \textbf{Theme} & \textbf{Description} & \textbf{Sample Answer}  & \textbf{Initial Code}  \\
  \midrule
  \taskfeatures & Aspects related to the task to be performed during a given session of the longitudinal study, such as its design, easiness, etc. & \lq\lq It was easy to complete.\rq\rq & \taskeasiness \\
 \midrule
  \workerfeatures & Aspects related to workers' own beliefs and motivations, their satisfaction after participating in the longitudinal study, etc. & \lq\lq It gave me the chance to be a part of change and real scientific study and know that my part contributed.\rq\rq  & \workermotivation  \\
  \midrule
  \requesterfeatures & Aspects related to the requester who is publishing the longitudinal study, such as reliability, communicativeness, etc. & \lq\lq Be reliable - offer a reasonable window during which the study can be completed and respond promptly to any messages from participants.\rq\rq  & \requesterreliability  \\
  \midrule
  \lsfeatures & Aspects related to the longitudinal study as a whole, such as session scheduling, reward mechanism, etc. & \lq\lq Performance rewards are a good way to maintain interest, as it feels like your time and effort are being rewarded.\rq\rq  & \lsprogress \\
  \midrule
  \platformfeatures & Aspects related to the crowdsourcing platform on which the longitudinal study is conducted such as its features, interface, general design, etc. & \lq\lq Yes. I think there is a large enough pool to pull from and if set up properly and rewarded, people will respond.\rq\rq  & \platformadequacy \\
  \midrule
  \nosuggestion & Answers provided by workers that acknowledge by explaining explicitly that they do not have any additional suggestions. & \lq\lq Nothing comes to mind.\rq\rq  & \nosuggestion  \\
  \midrule
  \answeruseless & Answers that do not convey anything related to the question proposed or that are made of random words and digits. & \lq\lq Unique crowdsourcing business model.\rq\rq  & \answeruseless  \\
\bottomrule
\end{tabular}
\end{table}

\begin{table}[htpb]
    \centering
    \caption{Distribution of the additional free texts provided by the workers while answering non-text-based questions.}
\label{cap:paper_tsc2024-sec:results-subsec:rq1-analysis-subsec:remarks-tab:free-texts}
\begin{tabular}{ccp{3.8cm}p{1.8cm}p{1.8cm}}
\toprule
  \textbf{Part} & \textbf{Section} & \textbf{Question} & \textbf{Experiences} & \textbf{Workers}  \\
  \midrule
  \pone & \ref{cap:paper_tsc2024-sec:results-subsec:rq1-analysis-p1-question:interval-between-sessions} & \emph{Interval Between Sessions} & 22 (4.02\%) & 18 (6.00\%) \\
  \midrule
  \pone & \ref{cap:paper_tsc2024-sec:results-subsec:rq1-analysis-p1-question:session-duration}  & \emph{Session Duration} & 22 (4.02\%) & 16 (5.33\%) \\
  \midrule
  \pone & \ref{cap:paper_tsc2024-sec:results-subsec:rq1-analysis-p1-question:crowdsourcing-platform}  & \emph{Crowdsourcing Platform} & 33 (6.10\%) & 26 (8.67\%) \\
  \midrule
  \pone & \ref{cap:paper_tsc2024-sec:results-subsec:rq1-analysis-p1-question:payment-model}  & \emph{Payment Model} & 35 (6.40\%) & 30 (10.00\%) \\
  \midrule
  \pone & \ref{cap:paper_tsc2024-sec:results-subsec:rq1-analysis-p1-question:participation-incentives-previous}  & \emph{Participation Incentives (In Prev. Experiences)} & 27 (4.94\%) & 22 (7.33\%) \\
  \midrule
  \pone & \ref{cap:paper_tsc2024-sec:results-subsec:rq1-analysis-p1-question:platform-availability}  & \emph{Reasons That Limit Availability On Platforms} & 48 (8.78\%) & 48 (7.67\%) \\
  \midrule
  \ptwo & \ref{cap:paper_tsc2024-sec:results-subsec:rq1-analysis-p2-question:participation-decline} & \emph{Reasons For Declining Participation} & -- & 50 (16.67\%) \\
  \midrule
  \ptwo & \ref{cap:paper_tsc2024-sec:results-subsec:rq1-analysis-p2-question:participation-incentives-new} & \emph{Participation Incentives (In New Experiences)} & -- & 17 (5.67\%) \\
  \midrule
  \ptwo & \ref{cap:paper_tsc2024-sec:results-subsec:rq1-analysis-p2-question:tasks-type} & \emph{Tasks Type} & -- & 14 (4.67\%) \\
  \midrule
  \ptwo & \ref{cap:paper_tsc2024-sec:results-subsec:rq1-analysis-p2-question:involvement-downsides} & \emph{Involvement Downsides} & -- & 23 (7.67\%) \\
\bottomrule
\end{tabular}
\end{table}

\subsection{Statistical Testing}
\label{cap:paper_tsc2024-sec:exp-setup-subsec:stat-test}

Statistical significance tests were conducted on survey responses for closed-ended questions to examine relationships across variables of interest.

The approach followed for each type of question is described as follows: questions requiring numerical answers are addressed first (Section~\ref{cap:paper_tsc2024-sec:exp-setup-subsec:stat-test-subsec:numerical}), followed by those requiring a mutually exclusive answer (Section~\ref{cap:paper_tsc2024-sec:exp-setup-subsec:stat-test-subsec:mutually-exclusive}), and finally those requiring a non-mutually exclusive answer (Section~\ref{cap:paper_tsc2024-sec:exp-setup-subsec:stat-test-subsec:non-mutually-exclusive}).

\subsubsection{Numerical Answers}
\label{cap:paper_tsc2024-sec:exp-setup-subsec:stat-test-subsec:numerical}

In six cases where the answers provided by workers were numeric, such as for question 1.1.X.1 of the \pone part (Section~\ref{cap:paper_tsc2024-sec:results-subsec:rq1-analysis-p1-question:time-elapsed}), ANOVA~\cite{olejnikAnova} was used to determine whether a statistically significant difference ($p < 0.05$) existed between the means of the groups.

A one-way ANOVA was applied to compare the means of the three groups of workers (\mturk, \prolific, and \toloka). In cases where a statistically significant difference was identified at the $p < 0.05$ level, posthoc tests were conducted using Tukey's HSD method \cite{abdi2010tukey} to identify which groups differed significantly. Tukey's HSD is a multiple comparison test that adjusts the significance level based on the number of pairwise comparisons to control for the Type I error rate.

\subsubsection{Mutually-Exclusive Answers}
\label{cap:paper_tsc2024-sec:exp-setup-subsec:stat-test-subsec:mutually-exclusive}

For the nine closed-ended questions requiring a mutually exclusive answer from a predefined set, such as question 1.1.X.5 of the \pone part (Section~\ref{cap:paper_tsc2024-sec:results-subsec:rq1-analysis-p1-question:crowdsourcing-platform}), chi-squared tests were used to identify statistically significant differences between the groups.

The observed contingency table of frequencies was calculated and compared to the expected contingency table under the null hypothesis of no difference between the groups. To correct for multiple comparisons, the false discovery rate (FDR) correction~\cite{Rouam2013} was applied to control the expected proportion of false discoveries among the rejected null hypotheses. Any comparisons involving zero expected frequencies during the chi-squared test were excluded from the analysis.

\subsubsection{Non-Mutually-Exclusive Answers}
\label{cap:paper_tsc2024-sec:exp-setup-subsec:stat-test-subsec:non-mutually-exclusive}

For the seven questions allowing the selection of multiple non-mutually-exclusive answers, such as question 7 of the \ptwo part (Section~\ref{cap:paper_tsc2024-sec:results-subsec:rq1-analysis-p2-question:participation-incentives-new}), chi-squared tests were used to identify statistically significant differences between the groups.

Unlike the previous case, where responses were mutually exclusive, this scenario required addressing the possibility of overlapping categories due to multiple selections by respondents. To account for this, the observed contingency table of frequencies was calculated using a modified approach accommodating overlapping categories. The chi-squared test and FDR correction were then applied, as in the previous case, to determine if significant differences existed between the groups.

\section{Results}
\label{cap:paper_tsc2024-sec:results}

Section~\ref{cap:paper_tsc2024-sec:results-subsec:rq1-analysis} analyzes the responses to the survey questions related to each part of the study (\ref{cap:paper_tsc2024-sec:research-questions_1}). Section~\ref{cap:paper_tsc2024-sec:results-subsec:rq2-recommendations} provides recommendations for practitioners and researchers interested in conducting longitudinal studies, based on the findings (\ref{cap:paper_tsc2024-sec:research-questions_2}). Finally, Section~\ref{cap:paper_tsc2024-sec:results-subsec:rq3-best-practices} outlines best practices for crowdsourcing platforms to support similar experiments (\ref{cap:paper_tsc2024-sec:research-questions_3}).

\subsection{\ref{cap:paper_tsc2024-sec:research-questions_1}: Analysis Of Workers' Responses}
\label{cap:paper_tsc2024-sec:results-subsec:rq1-analysis}

The analysis begins with the answers provided by workers for the \pone part of the survey, detailed from Section~\ref{cap:paper_tsc2024-sec:results-subsec:rq1-analysis-p1-first} to Section~\ref{cap:paper_tsc2024-sec:results-subsec:rq1-analysis-p1-last}, followed by the answers for the \ptwo part, covered from Section~\ref{cap:paper_tsc2024-sec:results-subsec:rq1-analysis-p2-first} to Section~\ref{cap:paper_tsc2024-sec:results-subsec:rq1-analysis-p2-last}. A summary of all findings is presented in Section~\ref{cap:paper_tsc2024-sec:results-subsec:rq1-analysis-subsec:summary}.

\subsubsection{Previous Experiences}
\label{cap:paper_tsc2024-sec:results-subsec:rq1-analysis-p1-first}
\label{cap:paper_tsc2024-sec:results-subsec:rq1-analysis-p1-question:previous-experiences}

The investigation begins by analyzing the previous experiences with longitudinal studies reported by each worker, as detailed in Table~\ref{cap:paper_tsc2024-sec:results-subsec:quantitative-analysis-subsec:prev-exp-tab:exp-data}. The charts in the following figures (Figure~\ref{cap:paper_tsc2024-sec:results-subsec:rq1-analysis-fig:how_many}--Figure~\ref{cap:paper_tsc2024-sec:results-subsec:rq1-analysis-fig:downsides}) are interpreted according to the guidelines provided in Section~\ref{cap:paper_tsc2024-sec:exp-setup-subsec:responses-analysis-subsec:quantitative-analysis}.

A total of \numworkers workers were recruited, with each platform contributing 100 workers. Collectively, they reported \numexperiences previous experiences with longitudinal studies, averaging 1.82 experiences per worker. \prolific workers reported the highest number of experiences (193), followed by \mturk (187) and \toloka (167). The highest proportion of workers with previous experience was observed on \prolific (35.28\%), followed by \mturk (34.19\%), while \toloka had the lowest proportion (30.53\%). Additionally, 97 workers (32.3\%) reported experiences from a platform other than their recruitment platform (see also Figure~\ref{cap:paper_tsc2024-sec:results-subsec:rq1-analysis-fig:platform}).

\begin{table}[t]
    \centering
    \caption{Previous experiences with longitudinal studies reported by the workers recruited.}
\label{cap:paper_tsc2024-sec:results-subsec:quantitative-analysis-subsec:prev-exp-tab:exp-data}
\begin{tabular}{p{4cm}ccc}
\toprule
\textbf{Platform} & \textbf{Experiences} & \textbf{Percentage} & \textbf{Mean} \\
\midrule
\mturk    & 187 & 34.19\% & 1.85 \\
\midrule
\prolific & 193 & 35.28\% & 1.89\\
\midrule
\toloka   & 167 & 30.53\% & 1.67 \\
\midrule
 Total    & \numexperiences & 100\% & 1.82 \\
 \bottomrule
\end{tabular}
\end{table}

Figure~\ref{cap:paper_tsc2024-sec:results-subsec:rq1-analysis-fig:how_many} illustrates workers' previous experiences with longitudinal studies, as summarized in Table~\ref{cap:paper_tsc2024-sec:results-subsec:quantitative-analysis-subsec:prev-exp-tab:exp-data}. Overall, 45\% of workers reported one experience, while 27.67\% and 27.33\% reported two and three experiences, respectively. The distribution varied across platforms. On \mturk, 42\% of workers reported one experience, 29\% reported two, and 29\% reported three. On \prolific, 43\% reported one experience, 21\% reported two, and 36\% reported three. On \toloka, 50\% reported one experience, 33\% reported two, and 17\% reported three. No statistically significant differences were detected across platforms.

The analysis indicates that workers on \prolific tend to report multiple previous experiences more frequently compared to those on other platforms, aligning with the recruitment criteria outlined in Section~\ref{cap:paper_tsc2024-sec:exp-setup-subsec:task}. Workers on \mturk and \toloka demonstrate familiarity with longitudinal studies, highlighting the necessity of a higher HIT completion threshold for effective recruitment.

\begin{figure}[tpb]
    \centering
    \includegraphics[width=\linewidth]{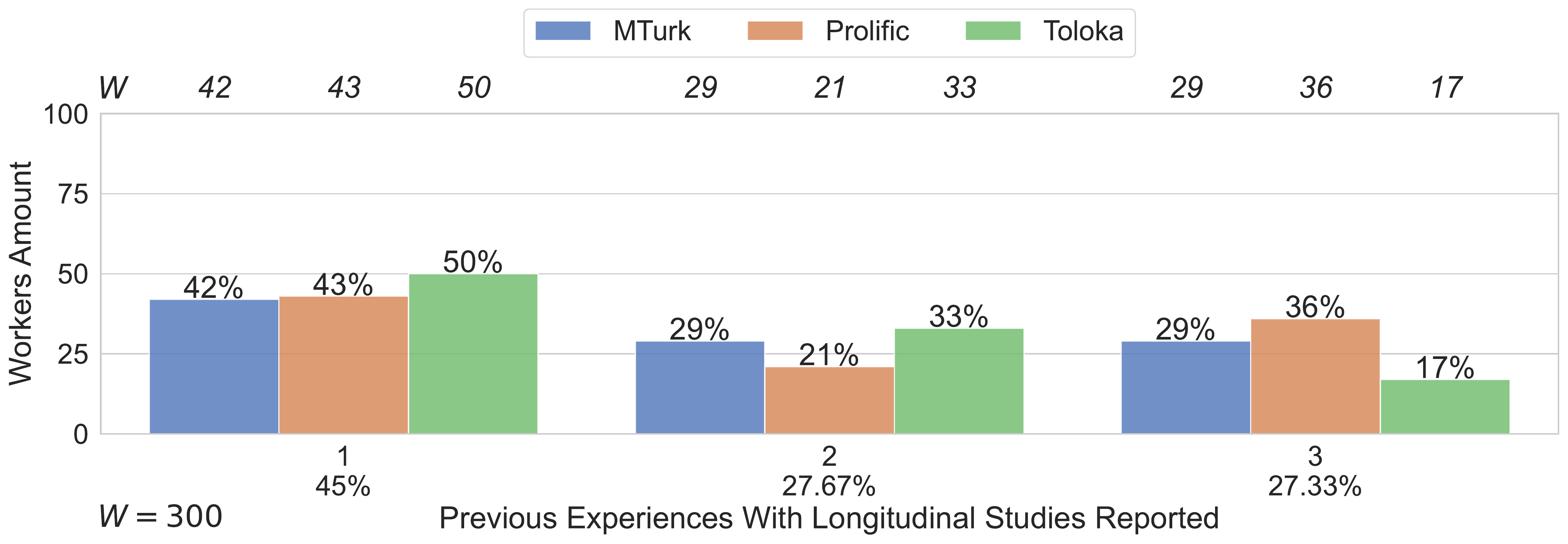}
    \caption{Number of workers who report 1, 2, or 3 previous experiences with longitudinal studies.}
    \label{cap:paper_tsc2024-sec:results-subsec:rq1-analysis-fig:how_many}
\end{figure} 

\subsubsection{Time Elapsed} 
\label{cap:paper_tsc2024-sec:results-subsec:rq1-analysis-p1-question:time-elapsed}

\begin{figure}[tpb]
	\centering
	\includegraphics[width=\linewidth]{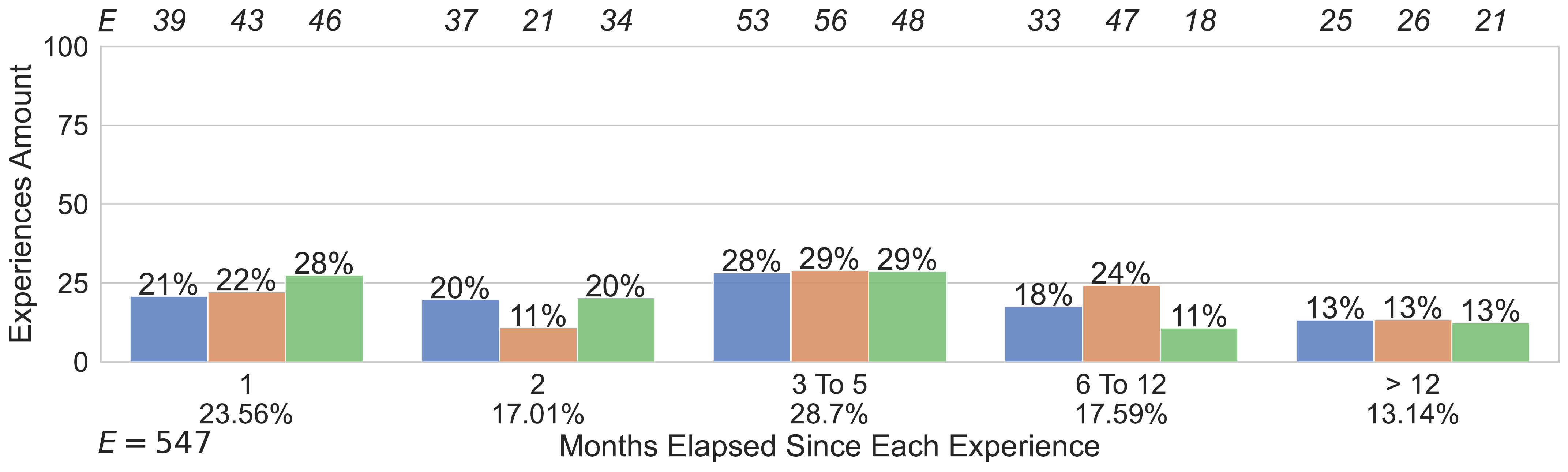}
		\caption{Time elapsed in months since each previous experience with longitudinal studies reported.}
    \label{cap:paper_tsc2024-sec:results-subsec:rq1-analysis-fig:when}
\end{figure}

Figure~\ref{cap:paper_tsc2024-sec:results-subsec:rq1-analysis-fig:when} illustrates the time elapsed in months since each previous experience reported, with a focus on participation in longitudinal studies occurring within the past 12 months.

The majority of reported experiences (87\%) took place within the 12 months preceding survey participation, while the remaining 13\% occurred earlier. The distribution of participation within the previous year is fairly consistent, with approximately 13\% of participation from each crowdsourcing platform occurring more than 12 months prior. This suggests that on \mturk and \prolific, workers were able to engage in longitudinal studies throughout the entire year before the survey, while on \toloka, the reported experiences were more recent (with a statistically significant difference between \mturk and \toloka, adjusted p-value~<~0.05).

\subsubsection{Number Of Sessions} 
\label{cap:paper_tsc2024-sec:results-subsec:rq1-analysis-p1-question:number-of-session}

\begin{figure}[tpb]
 \centering
 \includegraphics[width=.9\linewidth]{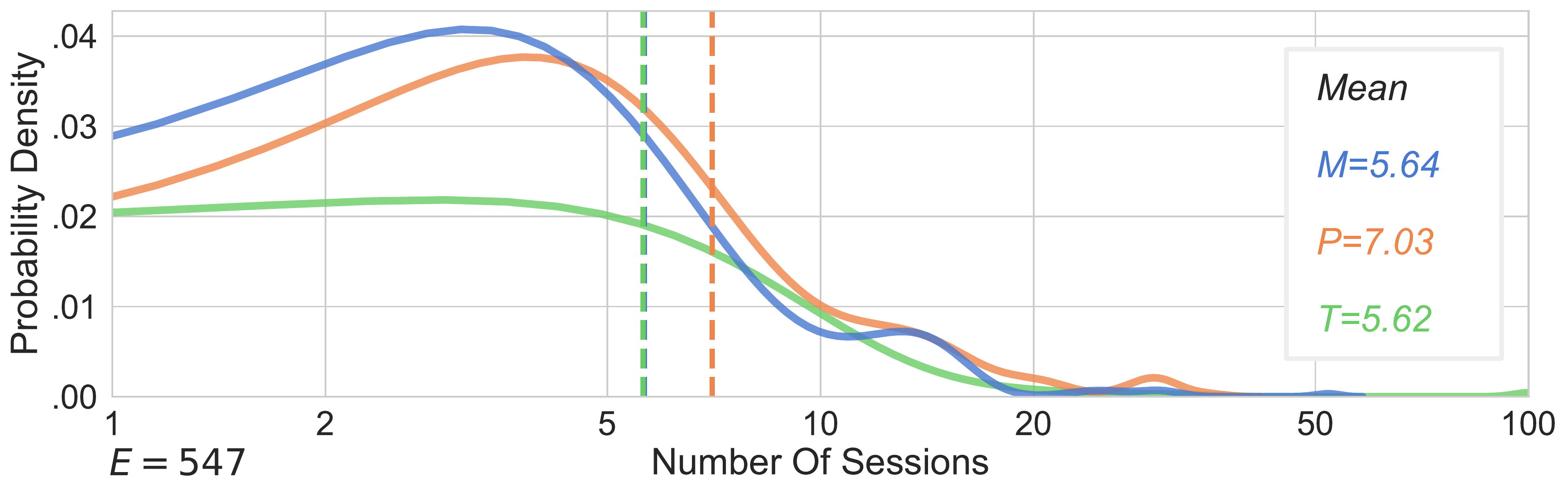}
 \caption{Number of sessions of the longitudinal study to which each reported experience refers.}  
\label{cap:paper_tsc2024-sec:results-subsec:rq1-analysis-fig:sessions}
\end{figure}

Figure~\ref{cap:paper_tsc2024-sec:results-subsec:rq1-analysis-fig:sessions} shows, for each previous experience with longitudinal studies reported, the number of sessions that comprised the overall study referred to.

The longitudinal studies in which workers participated on \mturk and \toloka have an average of approximately 6 sessions, while those on \prolific average 7 sessions. In general, it appears that task requesters tend to publish slightly longer longitudinal studies on \prolific, although no statistically significant differences were observed across platforms.

\subsubsection{Interval Between Sessions} 
\label{cap:paper_tsc2024-sec:results-subsec:rq1-analysis-p1-question:interval-between-sessions}

\begin{figure}[tpb]
	\centering
	\includegraphics[width=\linewidth]{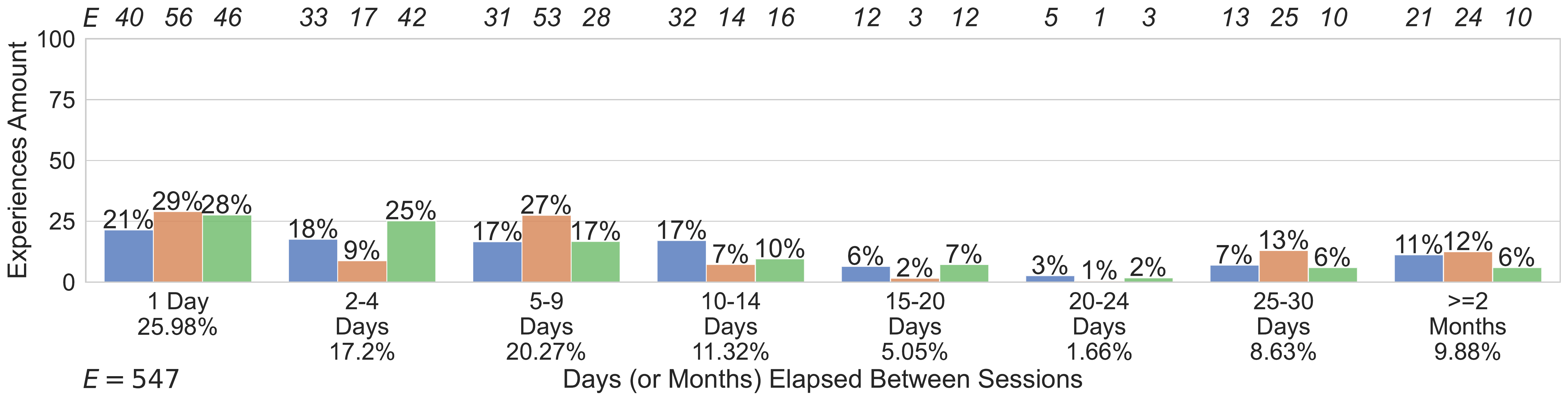}
		\caption{Time elapsed in days or months between the sessions of the longitudinal study to which each reported experience refers.}
	\label{cap:paper_tsc2024-sec:results-subsec:rq1-analysis-fig:time_interval}
\end{figure}

Figure~\ref{cap:paper_tsc2024-sec:results-subsec:rq1-analysis-fig:time_interval} illustrates the time elapsed, in terms of days, between sessions of the longitudinal study referred to by the reported experiences, focusing on intervals from 1 day to more than 30 days.

The time intervals ranging from 1 day to 9 days account for the majority of the longitudinal studies mentioned in the reported experiences (63.45\%). Expanding the range up to 30 days captures the vast majority of previous experiences (90\%). In summary, most requesters schedule the next session of a study within a period of 1 day to 30 days, with 10 days being the most common interval (\mturk vs. \toloka statistically significant, adjusted p-value~<~0.01).

\subsubsection{Session Duration} 
\label{cap:paper_tsc2024-sec:results-subsec:rq1-analysis-p1-question:session-duration}

\begin{figure}[tpb]
	\centering
	\includegraphics[width=\linewidth]{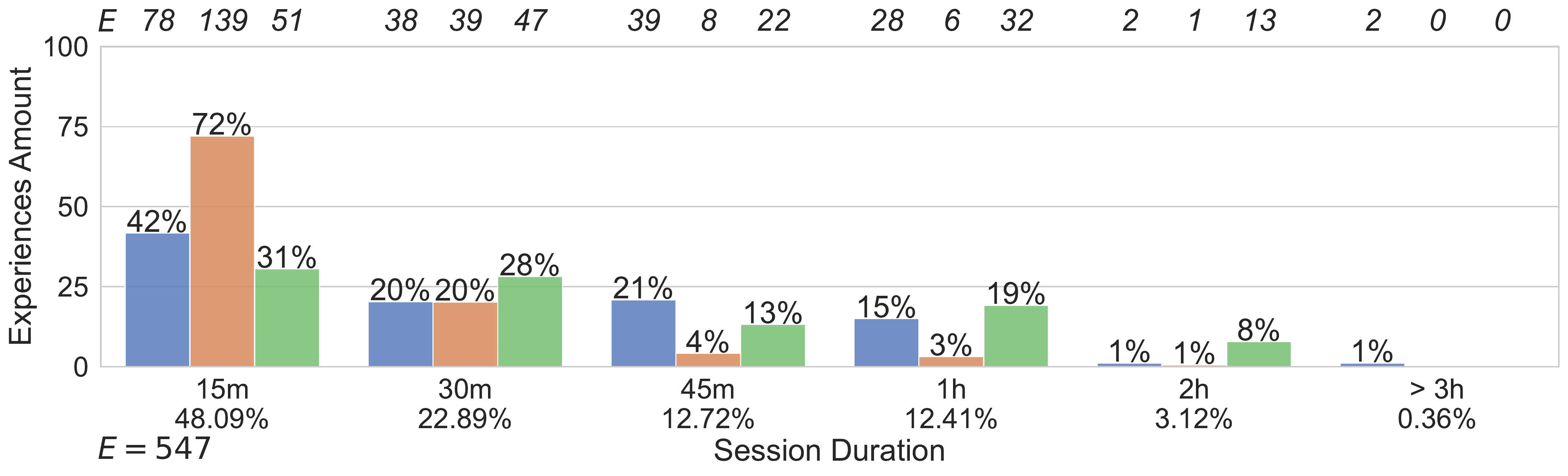}
		\caption{Duration in minutes or hours of the sessions of the longitudinal study to which each reported experience refers.}  
	\label{cap:paper_tsc2024-sec:results-subsec:rq1-analysis-fig:time_duration}
\end{figure}

Figure~\ref{cap:paper_tsc2024-sec:results-subsec:rq1-analysis-fig:time_duration} presents the duration of sessions in the longitudinal study referenced by the reported experiences, measured in minutes or hours.

Nearly half of the longitudinal studies had sessions lasting 15 minutes (48.09\%), followed by 22.89\% of sessions lasting 30 minutes, 12.72\% lasting 45 minutes, and 12.41\% lasting 60 minutes. As a result, the majority of sessions (96.11\%) are completed within one hour. A small, but notable, proportion of sessions on \toloka lasted two hours (13 sessions), with 2 sessions on \mturk and a single session on \prolific. Additionally, two workers reported \mturk sessions lasting three hours or more.

In general, the majority of task requesters on \prolific tend to schedule shorter sessions, primarily 15 minutes (72\%) or 30 minutes (20\%), compared to other platforms. The distribution of answers is more varied on \mturk and \toloka, though \toloka requesters tend to publish studies with longer sessions (\mturk vs. \prolific, \mturk vs. \toloka, and \toloka vs. \prolific statistically significant; adjusted p-value~<~0.01).

\subsubsection{Crowdsourcing Platform} 
\label{cap:paper_tsc2024-sec:results-subsec:rq1-analysis-p1-question:crowdsourcing-platform}

\begin{figure}[tpb]
\centering
\includegraphics[width=\linewidth]{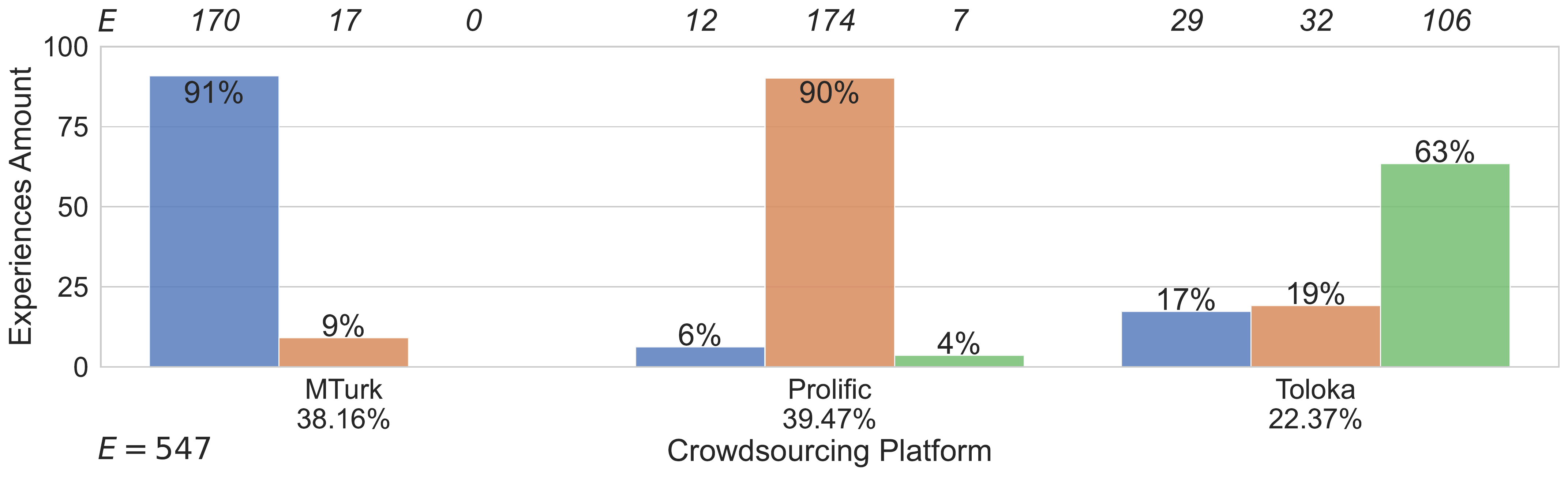}
	\caption{Crowdsourcing platforms where the longitudinal study to which each reported experience refers took place.}  
\label{cap:paper_tsc2024-sec:results-subsec:rq1-analysis-fig:platform}
\end{figure}

Figure~\ref{cap:paper_tsc2024-sec:results-subsec:rq1-analysis-fig:platform} shows the platforms where workers conducted their previous experiences with longitudinal studies, as workers recruited on one platform might have also worked elsewhere. A roughly equal number of experiences occurred on \mturk (38.16\%) and \prolific (39.47\%), while fewer experiences (22.37\%) were reported on \toloka.

Breaking down the responses by platform, most experiences reported by \mturk and \prolific workers took place on their respective platforms (around 90\%). However, cross-platform participation occurred: 9\% of \mturk workers reported experiences on \prolific, while 6\% of \prolific workers reported experiences on \mturk and 4\% on \toloka. Although \toloka workers primarily participated on \toloka (63\%), a notable portion also participated on \mturk (17\%) and \prolific (19\%).

In summary, the distribution of responses shows that \toloka workers tend to participate on other platforms more frequently than those recruited from \mturk and \prolific, especially in the context of longitudinal studies. However, this trend also appears on the other platforms (\mturk vs. \prolific, \mturk vs. \toloka, and \prolific vs. \toloka are statistically significant with an adjusted p-value~<~0.01).

\subsubsection{Payment Model} 
\label{cap:paper_tsc2024-sec:results-subsec:rq1-analysis-p1-question:payment-model}

\begin{figure}[tpb]
	\centering
	\includegraphics[width=\linewidth]{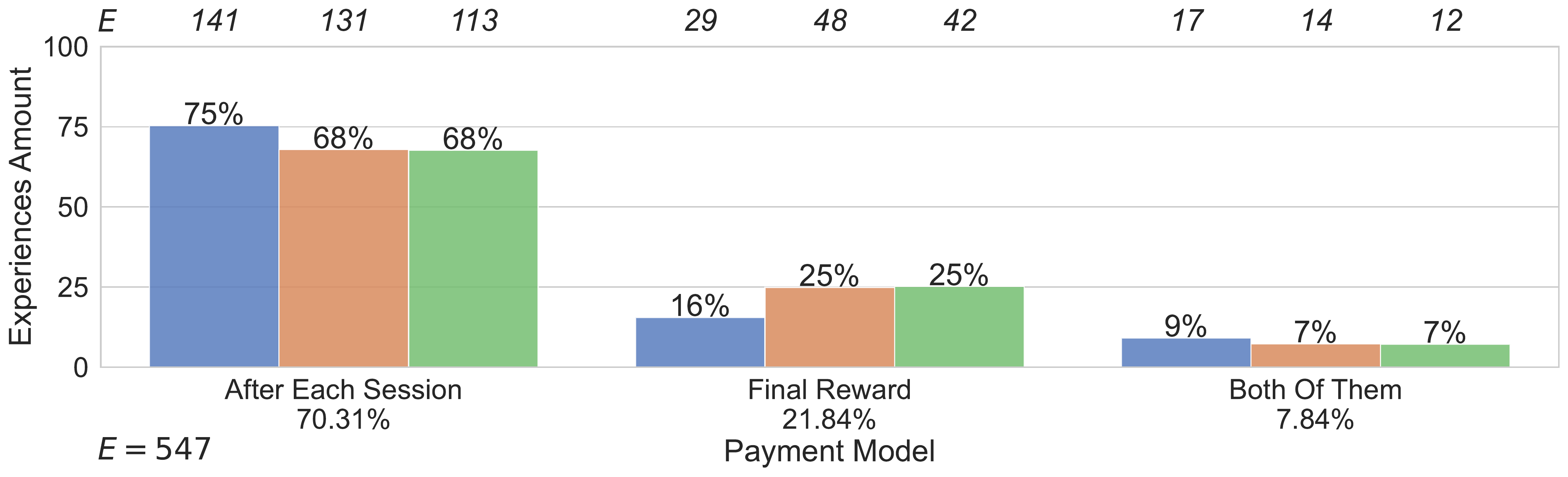}
		\caption{Payment model of the longitudinal study to which each reported experience refers (i.e., when the reward was provided).}  
	\label{cap:paper_tsc2024-sec:results-subsec:rq1-analysis-fig:payment_model}
\end{figure}

Figure~\ref{cap:paper_tsc2024-sec:results-subsec:rq1-analysis-fig:payment_model} investigates the payment models used in the longitudinal studies that workers reported participating in.

Most workers (70.31\%) reported participating in longitudinal studies where they received payment after each session, while 21.84\% reported studies where a final reward was the only form of payment. Only 7.84\% of the workers described studies that combined both payment models.

The distribution of answers shows that most workers reported studies where they received payment after each session, particularly on \mturk (75\%). Workers on \prolific and \toloka reported that 25\% of their experiences involved final rewards. Additionally, 9\% of the \mturk workers and 7\% of workers on other platforms reported studies that used a combination of both payment models. Statistically significant differences appeared between platforms (\mturk vs. \prolific, \mturk vs. \toloka, and \prolific vs. \toloka; adjusted p-value~<~0.01).

\subsubsection{Participation In Same Study} 
\label{cap:paper_tsc2024-sec:results-subsec:rq1-analysis-p1-question:participation-same-study}

\begin{figure}[tpb]
	\centering
	\includegraphics[width=\linewidth]{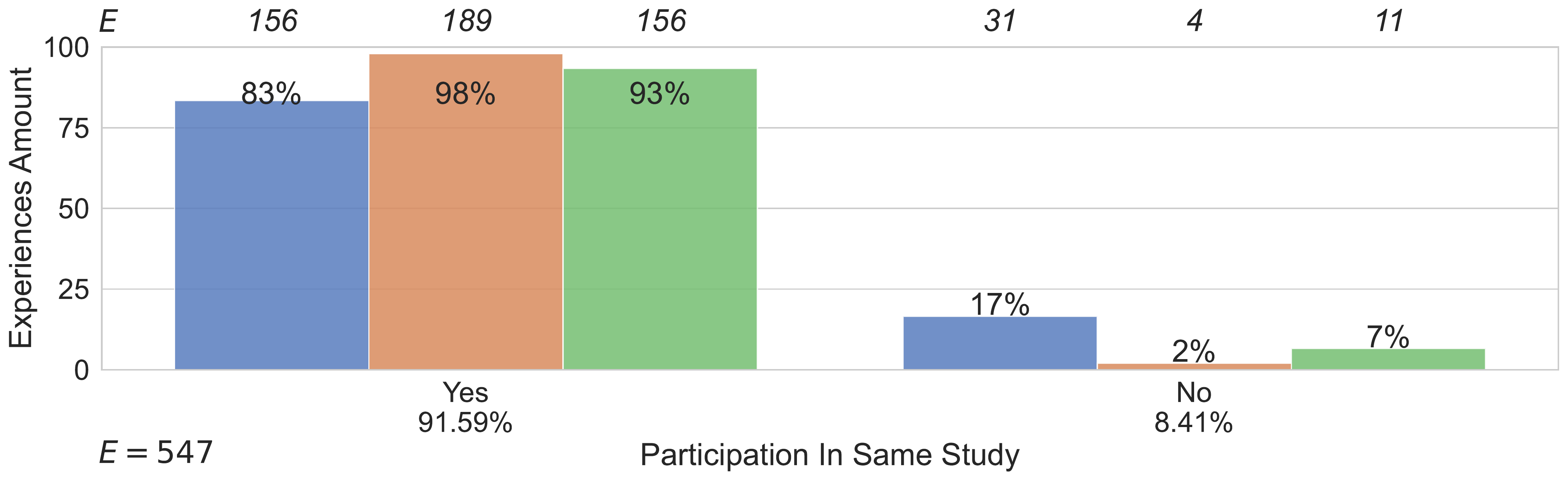}
		\caption{Workers willingness to participate again in the longitudinal study to which each reported experience refers.}  
	\label{cap:paper_tsc2024-sec:results-subsec:rq1-analysis-fig:same_study}
\end{figure}

Figure~\ref{cap:paper_tsc2024-sec:results-subsec:rq1-analysis-fig:same_study} explores workers' satisfaction after participating in the longitudinal study referred to by each reported experience.

The vast majority of workers (91.59\%) express interest in participating in the same longitudinal study again. When analyzing the data across platforms, \prolific and \toloka workers consistently show high levels of interest, with 98\% and 93\% positive responses, respectively. However, the percentage drops to 83\% for \mturk workers (\mturk vs. \prolific and \prolific vs. \toloka are statistically significant; adjusted p-value~<~0.01).

\subsubsection{Loyalty And Commitment}
\label{cap:paper_tsc2024-subsec:qualitative-analysis-subsec:worker-commitment}
\label{cap:paper_tsc2024-sec:results-subsec:rq1-analysis-p1-question:loyalty-and-commitment}

The mandatory open question 1.1.X.7.2 (\pone part) asks workers to specify what motivated them to return for a second session after completing the first one in the longitudinal study referred to by their reported experience. Additionally, workers explain why they would refuse to participate in the same study altogether.

Workers provided 485 answers among the \numexperiences previous experiences with longitudinal studies reported (88.66\%). The distribution of answers across different themes is as follows: 272 out of 485 (56.08\%) focus on aspects related to the task performed (\taskfeatures), while 101 (20.82\%) center on workers' beliefs and motivations (\workerfeatures). Furthermore, 10 (2.06\%) address the longitudinal study as a whole (\lsfeatures), 9 (1.86\%) pertain to the requester (\requesterfeatures), and 2 (0.41\%) relate to the platform (\platformfeatures). Lastly, 91 (18.76\%) answers are considered unusable (\answeruseless). Table~\ref{cap:paper_tsc2024-subsec:qualitative-analysis-subsec:worker-commitment-tab:worker-commitment} provides a sample of these answers.

The majority of responses (272 out of 485, 56.08\%) highlight how task attributes influence workers' decisions. Many workers find tasks interesting (100 out of 272, 36.76\%), easy (54 out of 272, 19.85\%), or well-paid (112 out of 272, 41.58\%), which motivates them to return. Additionally, 15 workers (5\%) mention the reliability of securing rewards in subsequent sessions as a key factor. Some workers (58 out of 272, 21.32\%) appreciate the opportunity to express their views and receive payment in return. Conversely, workers often cite low or unfair rewards, unavailability during follow-up sessions, or device-specific requirements as reasons for abandoning or refusing to participate in longitudinal studies after a session. About 20.82\% of responses (101 out of 485) reflect workers' belief that their preferences and attributes play a significant role in their decision to return for subsequent sessions.

A few workers (4 out of 101, 3.96\%) identified the sunk costs of completing the first session as a motivating factor to return~\cite{arkes1985psychology}. Additionally, 45 workers (44.55\%) expressed satisfaction with completing the initial session, highlighting the commitment it required (12 out of 101, 11.88\%) and the overall involvement. Some also appreciated the opportunity to gain insights, learn, and develop skills throughout the studies (15 out of 101, 15\%).

A small number of workers (9 out of 485, 1.86\%) focused on the task requester's characteristics, noting that communication and reminders for subsequent study sessions significantly impact their loyalty and commitment to longitudinal studies. Furthermore, 10 workers (2\%) discussed the longitudinal study itself, emphasizing the appeal of guaranteed work and the lack of competition for tasks.

\begin{longtable}{p{13cm}}
\caption{Sample of answers provided by workers concerning loyalty to longitudinal studies.}
\label{cap:paper_tsc2024-subsec:qualitative-analysis-subsec:worker-commitment-tab:worker-commitment} \\
\toprule
\textbf{Worker Responses} \\
\midrule
\endfirsthead
\toprule
\textbf{Worker Responses} \\
\midrule
\endhead

\footnotesize\itshape Continues in the next page \\
\endfoot
\endlastfoot

\emph{It was a well-designed study and the requester was very specific about when the follow-up tasks would be posted, and they sent reminders as well.} \\
\midrule
\emph{I felt the study was interesting and the reward was excellent so happy to do it again} \\
\midrule
\emph{It was very well organized and efficient. I didn't have to wait much between sessions.} \\
\midrule
\emph{Because I find interesting seeing how differently sometimes my answers can be just after a few days due to changes in the circumstances.} \\
\midrule
\emph{I dont likes that participating in same studies again because of im afraid of getting rejected} \\
\midrule
\emph{As long as the daily tasks are short and do not require an app download of any sort, I'll do them. I don't like downloading software or committing much time. I also don't like time windows. I like doing studies when I have free time, not during required blocks of time.} \\
\midrule
\emph{The individual studies were well-compensated and there was a generous bonus for completing all sessions of the study. Other than that, the study itself was quite unique and enjoyable to complete.}\\
\midrule
\emph{It's interesting to participate in longitudinal studies because it's pleasant to help with a research that monitors our learning/evolution over time in a given subject. This particular study was a monitored study that checked my performance on a repetitive memory task over the weeks. Also, the reward was excellent.} \\
\midrule
\emph{There would be random alerts on my phone (the study work took place within an app but was paid via Prolific) and I really struggled over the course of the fortnight duration - I was effectively a slave to my phone.}\\
\midrule
\emph{I don't find them any different to normal single part studies other than they can be more repetitive but so long as they meet the minimum payment reward on Prolific then I don't have any issue and I don't even care about bonuses for completing all parts because I complete all studies that I am invited to anyway and with Prolific I get instant alerts but you also get e-mail invitations when you aren't available so you can always complete them later on, it is really impossible to miss them and because each\ part is paid separately and approved individually it is more trustworthy for both participant and researcher.} \\
\bottomrule

\end{longtable}

\subsubsection{Participation Incentives (In Previous Experiences)} 
\label{cap:paper_tsc2024-sec:results-subsec:rq1-analysis-p1-question:participation-incentives-previous}

\begin{figure}[tpb]
\centering
\includegraphics[width=\linewidth]{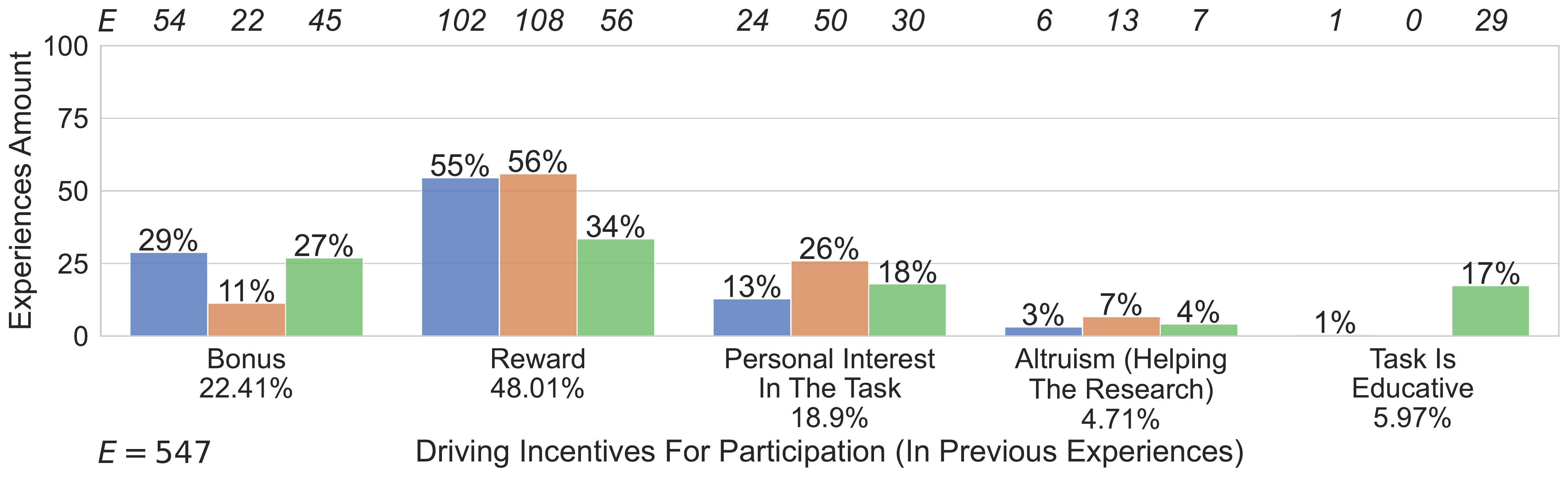}
    \caption{Incentives that drive workers to participate in the longitudinal study to which each reported experience refers.}
\label{cap:paper_tsc2024-sec:results-subsec:rq1-analysis-fig:incentives}
\end{figure}

Figure~\ref{cap:paper_tsc2024-sec:results-subsec:rq1-analysis-fig:incentives} examines the motivations that drive workers' participation in longitudinal studies based on the reported experiences.

Monetary incentives, such as rewards and bonuses, play the most significant role, driving participation in 70.42\% of reported experiences. Approximately 19\% of workers participate due to personal interest in the task proposed by the requester. Additionally, 6\% of workers cite the task's educational value as a motivating factor, while altruism, expressed as helping advance research, motivates 4.71\% of respondents.

Analyzing responses by platform reveals notable differences. On \toloka, 17\% of workers find tasks educational, which is nearly absent on \mturk (1\%) and \prolific. \prolific stands out for fostering participation driven by personal interest (26\%) or a willingness to support research (7\%). These trends likely reflect \prolific's focus on academic research projects and its frequent use by researchers~\cite{PALAN201822}.

While monetary incentives remain the primary motivator, the influence of personal interest, educational value, and altruism should not be underestimated when designing longitudinal studies. The differences across platforms (\mturk vs. \prolific, \mturk vs. \toloka, and \prolific vs. \toloka) are statistically significant, with an adjusted p-value~<~0.01.

\subsubsection{Study Completion} 
\label{cap:paper_tsc2024-sec:results-subsec:rq1-analysis-p1-question:study-completion}

\begin{figure}[tpb]
	\centering
	\includegraphics[width=\linewidth]{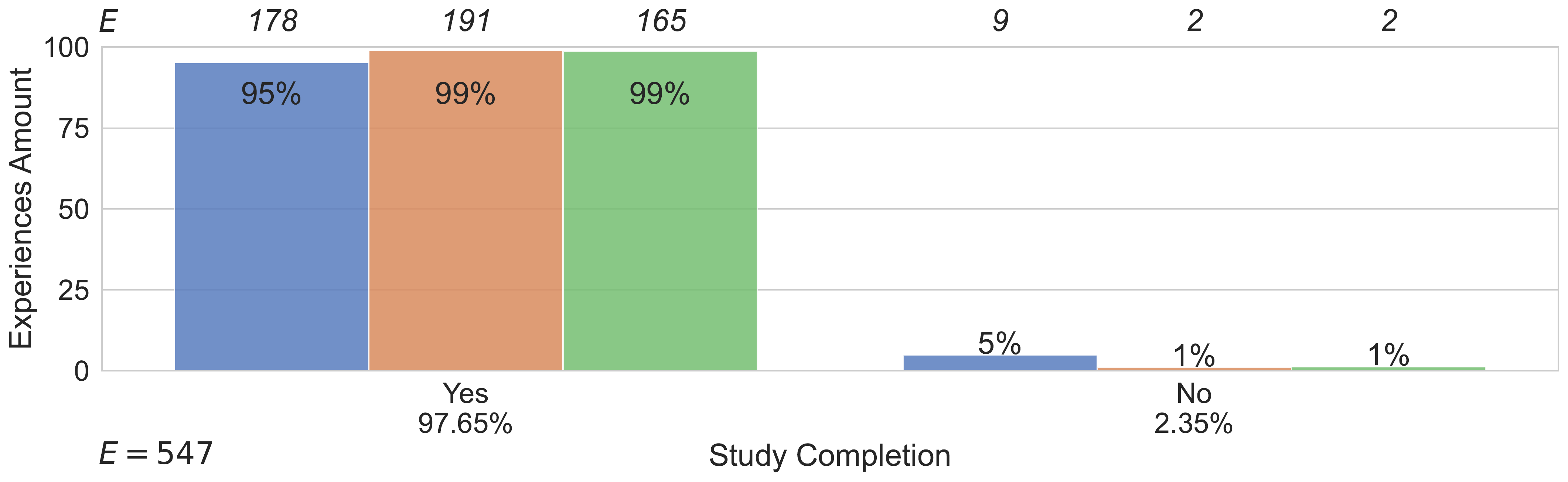}
		\caption{Completion claimed by workers of the longitudinal study to which each reported experience refers.}
	\label{cap:paper_tsc2024-sec:results-subsec:rq1-analysis-fig:task_completion}
\end{figure}

Figure~\ref{cap:paper_tsc2024-sec:results-subsec:rq1-analysis-fig:task_completion} examines whether workers completed the longitudinal studies referred to in the reported experiences. Workers report completing nearly all the previous experiences (97.65\%), with only 13 out of 547 experiences (2.35\%) left incomplete.

Breaking the data down by platform, workers report completing almost every experience on \prolific and \toloka (99\%), while the completion rate on \mturk is slightly lower at 95\%. The comparisons across platforms do not show statistically significant differences. The study recruited workers with specific task completion rates (i.e., experienced workers), as detailed in Section~\ref{cap:paper_tsc2024-sec:exp-setup-subsec:task}. This criterion suggests that workers have little motivation to provide inaccurate information about their past completions.

\subsubsection{Completion Incentives (In Previous Experiences)} 
\label{cap:paper_tsc2024-sec:results-subsec:rq1-analysis-p1-question:completion-incentives-previous}

\begin{figure}[tpb]
	\centering
	\includegraphics[width=\linewidth]{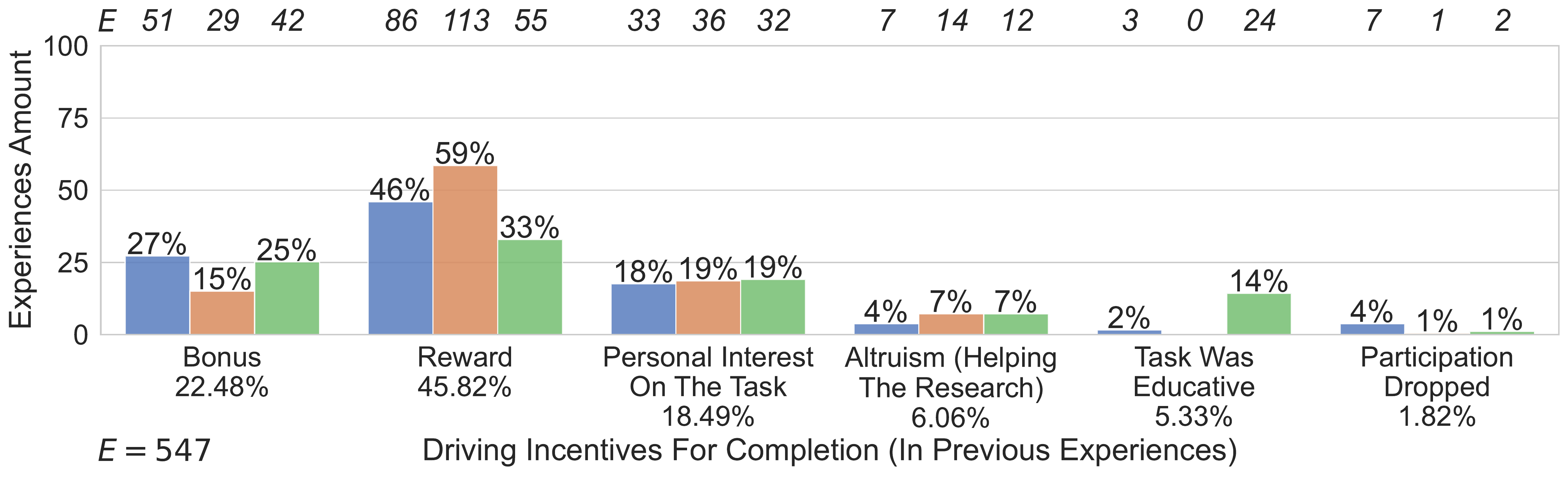}
		\caption{Incentives that drive workers to completing the longitudinal study to which each reported experience refers.}
	\label{cap:paper_tsc2024-sec:results-subsec:rq1-analysis-fig:yes}
\end{figure}

Figure~\ref{cap:paper_tsc2024-sec:results-subsec:rq1-analysis-fig:yes} examines the motivations that drive workers to complete the longitudinal studies referenced in the reported experiences. This analysis compares directly with the responses to question 1.1.X.8, discussed in Section~\ref{cap:paper_tsc2024-sec:results-subsec:rq1-analysis-p1-question:participation-incentives-previous}, which focuses on the incentives for initial participation. Although the set of possible answers remains the same, this question highlights completed experiences and identifies changes in workers' perceptions of the overall experience. The 11 experiences where workers dropped participation (those shown in the right half of Figure~\ref{cap:paper_tsc2024-sec:results-subsec:rq1-analysis-fig:task_completion}) appear under the label \lq\lq Participation Dropped\rq\rq{} to enable a direct comparison of the bar charts.

Monetary aspects, such as rewards and bonuses, remain the most important factors for the majority of the reported experiences (68.3\%), showing a slight decrease of 2.12\%. Workers' personal interest in the task proposed by the requester (18.49\%) remains almost unchanged, as does their view of the task being educative. Most of the shift away from monetary aspects is attributed to workers' increased willingness to help with the overall research, rising from 4.71\% to 6.06\%.

When considering each platform, the overall distribution of answers remains consistent in terms of relative comparisons. The most noticeable change occurs on \prolific, where workers' personal interest in the proposed task drops from 26\% to 19\%, aligning with the levels observed on other platforms. A similar trend is seen on \mturk, where interest in the final reward shifts from 49\% to 55\% (\mturk vs. \prolific, \mturk vs. \toloka, and \prolific vs. \toloka statistically significant; adjusted p-value~<~0.01).

\subsubsection{Crowdsourcing Platforms Suitability}
\label{cap:paper_tsc2024-subsec:qualitative-analysis-subsec:platform-suitability}
\label{cap:paper_tsc2024-sec:results-subsec:rq1-analysis-p1-question:platform-suitability}

The mandatory open-ended question 2 (\pone part) asks workers about the adequacy and suitability of the crowdsourcing platform they used in supporting longitudinal studies.

The majority of workers (273 out of 300, 91\%) provided answers that allow for further consideration. The distribution of responses across different themes is as follows: 244 out of 273 (89.34\%) addressed aspects related to the crowdsourcing platform (\platformfeatures), 11 (4.03\%) focused on workers' own beliefs and motivations (\workerfeatures), and 18 (6.59\%) answers were considered unusable (\answeruseless). Table~\ref{cap:paper_tsc2024-subsec:qualitative-analysis-subsec:platform-suitability-tab:platform-suitability} shows a sample of these answers.

The vast majority of answers directly relate to the crowdsourcing platform of origin (244 out of 273, 89.34\%). Breaking down the respondents across each platform reveals 98 workers from \mturk, 100 from \prolific, and 76 from \toloka.

The majority of \mturk workers (70 out of 98, 71.43\%) believe the platform is generally adequate, with few providing additional details. Three of them (3.06\%) specifically mention the ease of sending reminders for upcoming longitudinal study sessions. Only seven (7.14\%) find the platform inadequate in supporting longitudinal studies. One worker suggests that the platform needs design improvements to facilitate scheduling tasks for longitudinal studies, while another highlights the challenge for requesters to ensure worker honesty.

Nearly all \prolific workers (97 out of 100, 97\%) consider the platform adequate for supporting longitudinal studies, with many providing detailed responses. Some mention the platform's detailed task reports, which facilitate tracking throughout the study. Others (7 out of 100, 7\%) highlight the diverse backgrounds and skills of available individuals. Sixteen workers (16\%) note factors such as ease of contacting or sending reminders to workers using their identifier. Additionally, two workers (2 out of 100, 2\%) emphasize worker motivation and reliability as important considerations for researchers. Notably, one worker mentions being recruited from the platform via a third-party application that relies on the platform's API.

The majority of \toloka workers (68 out of 76, 89.47\%) consider the platform adequate overall, with few providing specific details. Two workers (2 out of 76, 2.63\%) mention worker availability and the ease of contacting them using their identifier. One worker’s response is notable: they believe the platform cannot adequately support a longitudinal study due to residing in a country with poor network infrastructure.

When considering workers who are uncertain or outright deny the adequacy of the platform, several cross-platform factors emerge. These workers are more likely to drop out of longitudinal studies due to perceived inadequacies. They express difficulties in assessing requester honesty, which leads to skepticism about participating in such studies. Additionally, respondents believe that workers typically do not actively seek out longitudinal studies, suggesting a need for platforms to better distinguish these studies from standard crowdsourcing tasks.

\begin{longtable}{p{11.5cm}p{2cm}}

\caption{Sample of answers provided by workers concerning the adequacy of crowdsourcing platforms in supporting longitudinal studies.}
\label{cap:paper_tsc2024-subsec:qualitative-analysis-subsec:platform-suitability-tab:platform-suitability} \\
\toprule
\textbf{Worker Responses} & \textbf{Platform} \\
\midrule
\endfirsthead
\toprule
\textbf{Worker Responses} & \textbf{Platform} \\
\midrule
\endhead

\footnotesize\itshape Continues in the next page \\
\endfoot
\endlastfoot

\emph{I think that this platform is good for longitudinal studies, especially when a Requester can send email reminders to the Workers about when the follow-up tasks are available to be completed.} & MTurk \\
\midrule
\emph{Yes, I have done tasks like that on this platform before and it went well for me.} & MTurk \\
\midrule
\emph{I don't think so because everything that gets released gets snatched up quickly. Also, the requesters on this platform don't respond much. Before, yes but not most likely not.} & MTurk \\
\midrule
\emph{Yes but it need further improvements for this specific type of tasks such as scheduling improvements etc.} & MTurk \\
\midrule 
\emph{Yes, I think it is perfectly suitable given its nature. I do think coordinating longer studies can be more difficult on mturk compared to other platforms, as there are many other studies constantly on the platform and remembering longitudinal studies can be difficult while also keeping up with regular studies. To remedy this, requestors must often use e-mail reminders and other types of reminders, which I have no issues with at all.} & MTurk \\
\midrule
\emph{Yes, I believe it is. This platform is the host of many other studies all of which provide for a professional and safe environment (on both sides, for the requester and surveyee with full disclosure of all procedures. I've had previous experience with a longitudinal study on this platform and I have zero complaints.} & \prolific \\
\midrule
\emph{Yes. The messaging system on Prolific is very useful in this regard, the platform itself can easily be tailored to longitudinal studies, and both the researcher and the participant can rely on Prolific for any support required around the task.} & \prolific \\
\midrule
\emph{Yes I think Prolific works very well, I have Prolific Assistant so get the alerts if I'm on my PC so usually I start them just like any other study but even if you don't then you would be sent an e-mail invitation to remind you so you are very unlikely to ever miss any part of a study and I have completed all parts of any longitudinal studies that I have been part of. I think so long as all of the details are explained in the first part and the participant agrees to complete all of the following parts then they should have very high success rates and if anyone does drop out or has any reason to you can also communicate this via Prolific messaging.} & \prolific \\
\midrule
\emph{Not really, there should be an option to separate normal from longitudinal studies.} & \prolific \\
\midrule
\emph{Yes, but Prolific does not email you outside of itself. This can be a problem if the study requires out-of-band responses. With Mechanical Turk your requests hit email so I get message reminders when I am not at my desk.} & \prolific \\
\midrule
\emph{Yes, it's a nice platform to work, to earn rewards and to learn some new things so it would be a great platform for longitudinal studies too.} & \toloka 
\\
\midrule
\emph{Yes it fits. I think there is a large number of participants, which makes the study more accurate.} & \toloka \\
\midrule
\emph{I have had good experiences with tasks offered by Toloka. Proper instructions are provided.}  & \toloka \\
\midrule
\emph{Yes, it has participants which login every or almost every day, they are interested in completing tasks they are already acquainted with.}  & \toloka \\
\midrule
\emph{Yes, it is suitable because most people in this platform work more than five hours everyday} & \toloka \\
\bottomrule

\end{longtable}

\subsubsection{Reasons That Limit Availability On Platforms} 
\label{cap:paper_tsc2024-sec:results-subsec:rq1-analysis-p1-last}
\label{cap:paper_tsc2024-sec:results-subsec:rq1-analysis-p1-question:platform-availability}

\begin{figure}[tpb]
	\centering
	\includegraphics[width=.95\linewidth]{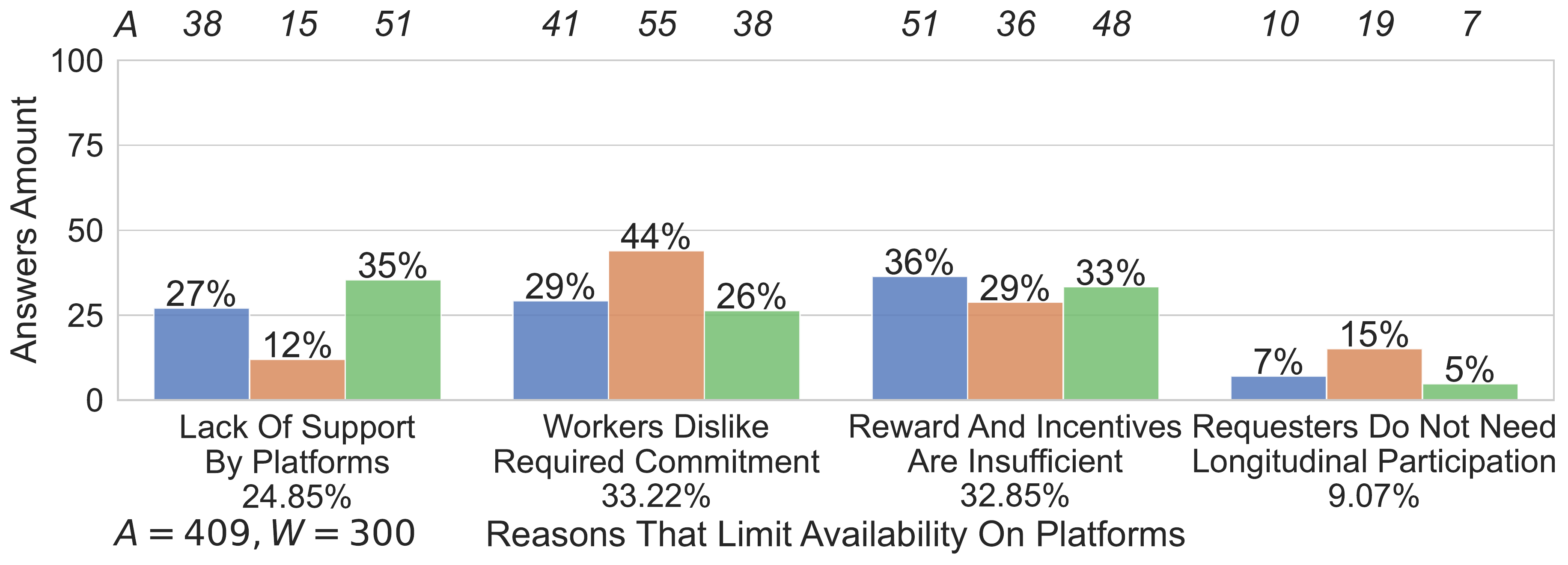}
		\caption{Reasons that limit the availability of longitudinal studies on crowdsourcing platforms, according to workers.}
	\label{cap:paper_tsc2024-sec:results-subsec:rq1-analysis-fig:common}
\end{figure}

Figure~\ref{cap:paper_tsc2024-sec:results-subsec:rq1-analysis-fig:common} investigates the reasons that limit the availability of longitudinal studies on crowdsourcing platforms according to workers' opinions. The most prevalent reasons, chosen roughly the same number of times, are that workers dislike the required commitment (32.85\%) and that the provided rewards and incentives are insufficient. Several answers indicate that current popular crowdsourcing platforms do not optimally support longitudinal studies (24.85\%), and 9.07\% of answers point out that requesters usually do not need longitudinal participation since most tasks involve annotating static data.

The distribution of answers changes when considering each platform. Specifically, 44\% of the answers provided by \prolific workers indicate their dislike of the required commitment, while this factor is less important for \mturk workers (29\%) and \toloka workers (26\%). The lack of adequate technical support is prevalent among the answers provided by \toloka workers (35\%), while for \prolific, only 12\% of answers report this issue. The percentage of answers indicating that rewards and incentives are insufficient is slightly higher for \mturk (36\%) compared to \toloka (33\%), which in turn is slightly higher than \prolific (29\%). Among the answers describing that crowdsourcing tasks often do not need longitudinal participation, those from \prolific are prevalent (15\%).

In summary, workers dislike the required commitment and find monetary aspects and related incentives insufficient. They also believe that crowdsourcing platforms do not adequately support longitudinal studies (\mturk vs. \prolific, \mturk vs. \toloka, and \prolific vs. \toloka statistically significant; adjusted p-value < 0.01).

\subsubsection{Preferred Commitment Duration} 
\label{cap:paper_tsc2024-sec:results-subsec:rq1-analysis-p2-first}
\label{cap:paper_tsc2024-sec:results-subsec:rq1-analysis-p2-question:preferred-commitment-duration}

\begin{figure}[tpb]
	\centering
	\includegraphics[width=.95\linewidth]{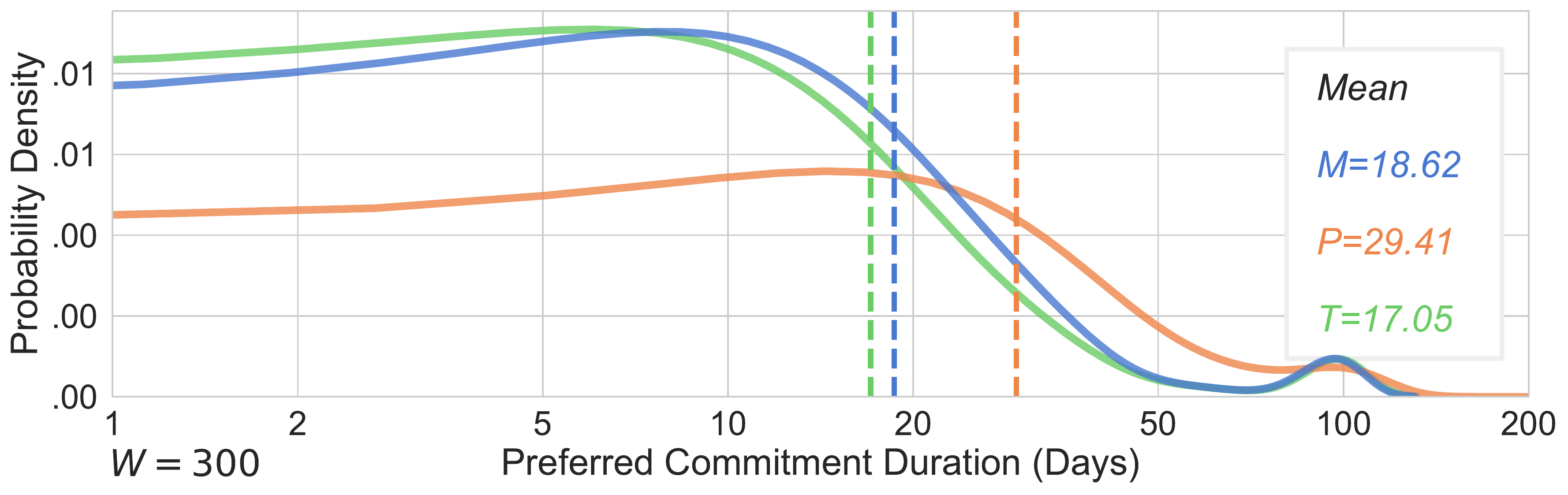}
		\caption{Number of days workers would be happy to commit for a longitudinal study, hypothesizing a single session of 15 minutes per day.}
	\label{cap:paper_tsc2024-sec:results-subsec:rq1-analysis-fig:daily_commitment}
\end{figure}

Figure~\ref{cap:paper_tsc2024-sec:results-subsec:rq1-analysis-fig:daily_commitment} investigates the number of days workers would be willing to commit to a longitudinal study, assuming each session lasts 15 minutes per day.

Considering each platform, \mturk and \toloka workers exhibit similar trends, with mean numbers of days around 19 and 17, respectively. For \prolific, this number increases to an average of nearly a month (29 days).
Overall, \prolific is the platform that finds workers willing to commit to longitudinal studies for longer periods, at least compared with \toloka (\prolific vs \toloka statistically significant with adjusted p-value < 0.05).

\subsubsection{Reasons For Declining Participation} 
\label{cap:paper_tsc2024-sec:results-subsec:rq1-analysis-p2-question:participation-decline}

\begin{figure}[tpb]
	\centering
	\includegraphics[width=.9\linewidth]{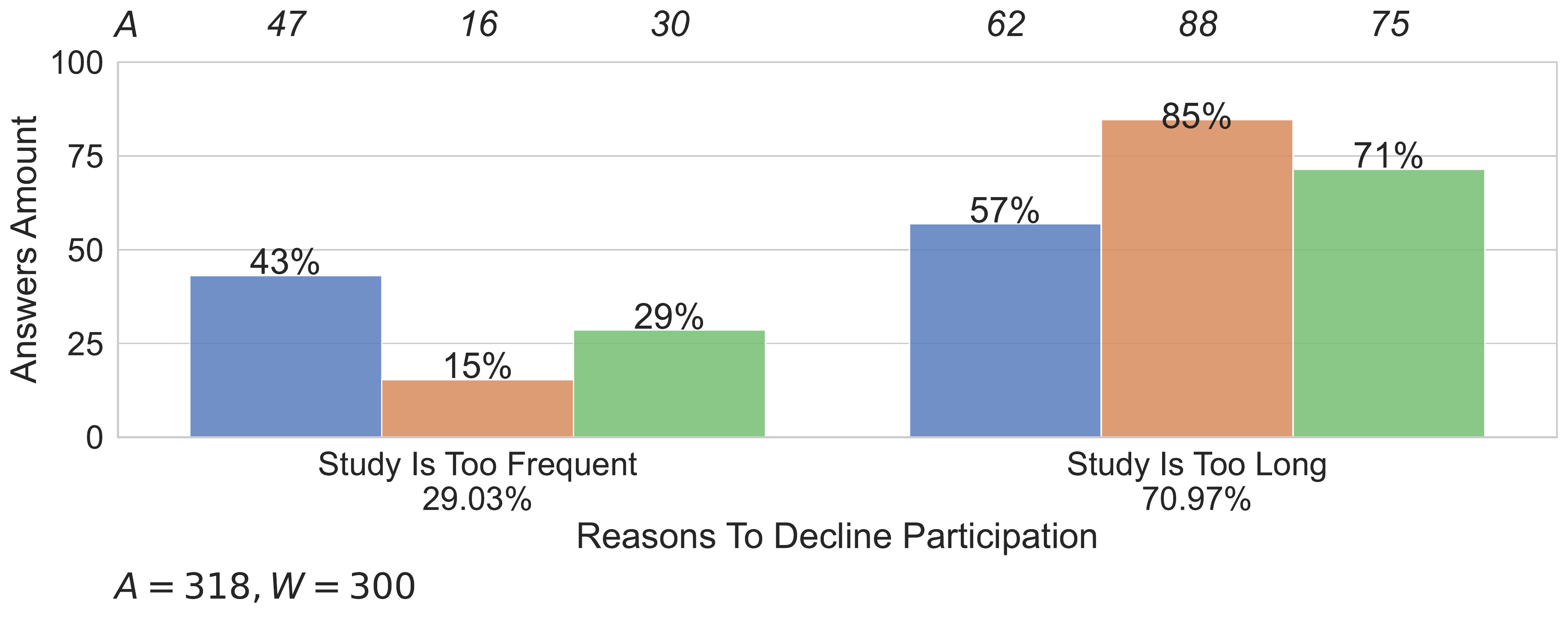}
		\caption{Reasons that drive workers to decline participation in longitudinal studies.}
	\label{cap:paper_tsc2024-sec:results-subsec:rq1-analysis-fig:participation_decline}
\end{figure}

Figure~\ref{cap:paper_tsc2024-sec:results-subsec:rq1-analysis-fig:participation_decline} investigates the reasons that drive workers to decline participation in longitudinal studies.

The majority of answers indicate that the length of the longitudinal study, in terms of the number of sessions and the time elapsed (days or even months since its start), is the most important factor (70.79\%). The remaining answers (29.03\%) point to the frequency of sessions as another factor that can lead to declining participation and should not be overlooked.

Considering each platform, the vast majority of Prolific workers (85\%) view study length as a major concern. This also holds for \toloka, albeit to a lesser extent (71\%). For \mturk, the trend is more nuanced, with the gap between answers that consider study length (57\%) and study frequency (43\%) being smaller (\mturk vs. \prolific, \mturk vs. \toloka, and \prolific vs. \toloka statistically significant; adjusted p-value < 0.01).

\subsubsection{Preferred Participation Frequency} 
\label{cap:paper_tsc2024-sec:results-subsec:rq1-analysis-p2-question:preferred-participation-frequency}

\begin{figure}[tpb]
	\centering
	\includegraphics[width=\linewidth]{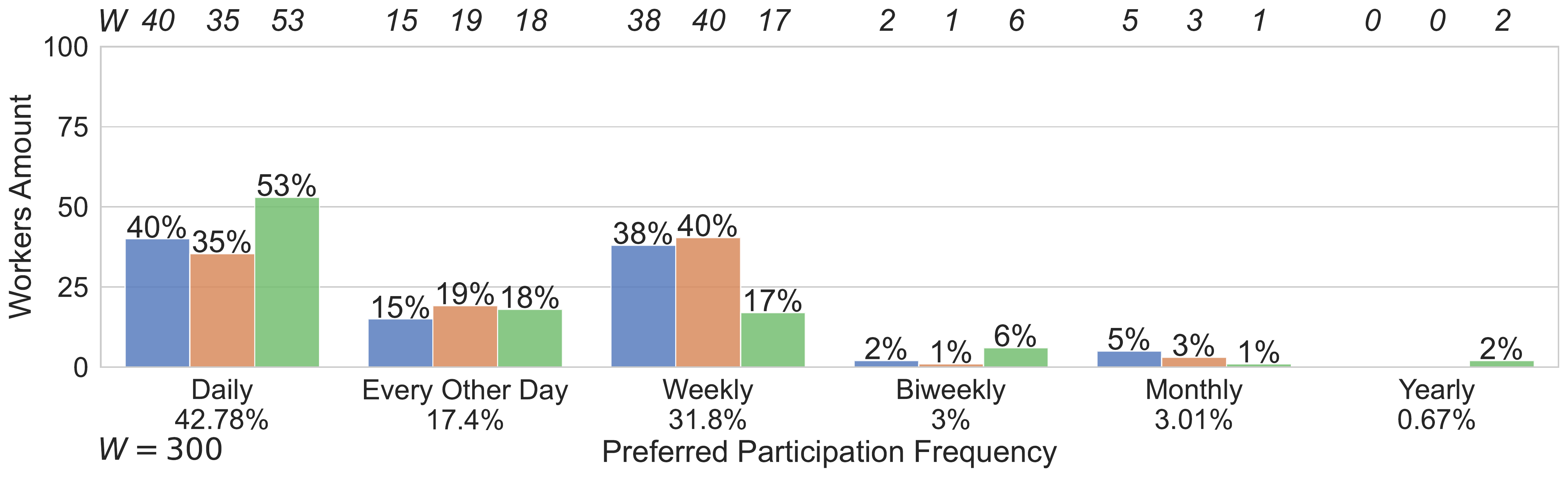}
		\caption{Preferred participation frequency in a longitudinal study according to workers.}
	\label{cap:paper_tsc2024-sec:results-subsec:rq1-analysis-fig:which_commitment}
\end{figure}

Figure~\ref{cap:paper_tsc2024-sec:results-subsec:rq1-analysis-fig:which_commitment} investigates the preferred participation frequency in longitudinal studies according to workers, based on time periods.

The vast majority of workers prefer frequent studies, with a daily to weekly participation commitment. Specifically, daily participation is the most popular option overall (42.78\%). Only a small group of 11 workers (6.68\%) would prefer longer time periods.

There are some nuances in the preferences of workers recruited from each platform. \toloka workers primarily prefer daily participation (53\%). \prolific workers, on the other hand, slightly favor a weekly frequency (40\%), followed by a daily frequency (35\%). For \mturk workers, the trend is reversed, with daily participation as the top preference (40\%), closely followed by weekly participation (38\%). Regarding longer time periods, it is noteworthy that 6 \toloka workers (6\%) prefer a biweekly frequency, while 5 \mturk workers (5\%) and 3 \prolific workers (3\%) prefer a monthly frequency.

These findings align with those in Figure~\ref{cap:paper_tsc2024-sec:results-subsec:rq1-analysis-fig:participation_decline}, where study length is identified as a major concern for workers (\mturk vs. \prolific, \mturk vs. \toloka, and \prolific vs. \toloka statistically significant; adjusted p-value < 0.01).

\subsubsection{Preferred Session Duration} 
\label{cap:paper_tsc2024-sec:results-subsec:rq1-analysis-p2-question:preferred-session-duration}

\begin{figure}[tpb]
	\centering
	\includegraphics[width=.9\linewidth]{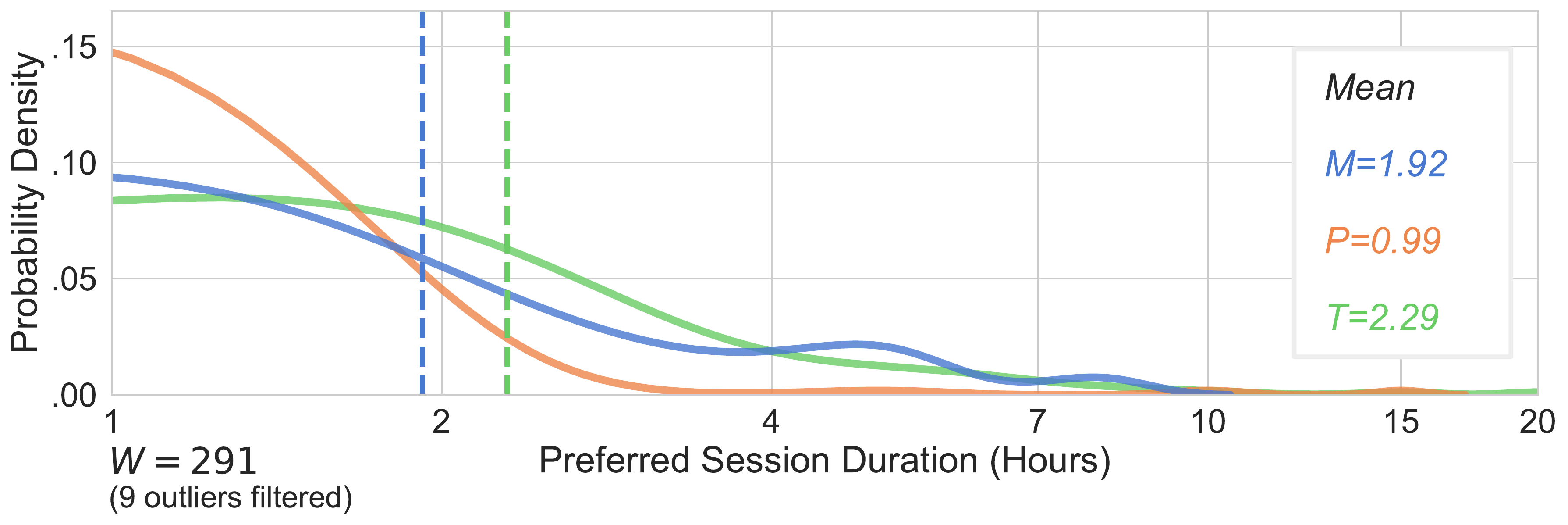}
		\caption{Preferred session duration in hours for longitudinal studies according to workers.}
	\label{cap:paper_tsc2024-sec:results-subsec:rq1-analysis-fig:ideal_session}
\end{figure}

Figure~\ref{cap:paper_tsc2024-sec:results-subsec:rq1-analysis-fig:ideal_session} investigates the preferred session duration in hours for longitudinal studies according to workers.

\prolific workers prefer short sessions of less than 1 hour on average, while \mturk and \toloka workers exhibit a more uniform preference, with an average session duration of about two hours. The figure excludes 9 outliers who provided unreasonable durations (i.e., between 15 and 50 hours), as they were removed from the analysis.

Overall, \mturk and \toloka workers are more willing to work for a longer period within a single session compared to \prolific workers (\mturk vs. \prolific and \prolific vs. \toloka statistically significant; adjusted p-value < 0.05).

\subsubsection{Acceptable Hourly Payment} 
\label{cap:paper_tsc2024-sec:results-subsec:rq1-analysis-p2-question:acceptable-hourly-payment}

\begin{figure}[tpb]
	\centering
	\includegraphics[width=.9\linewidth]{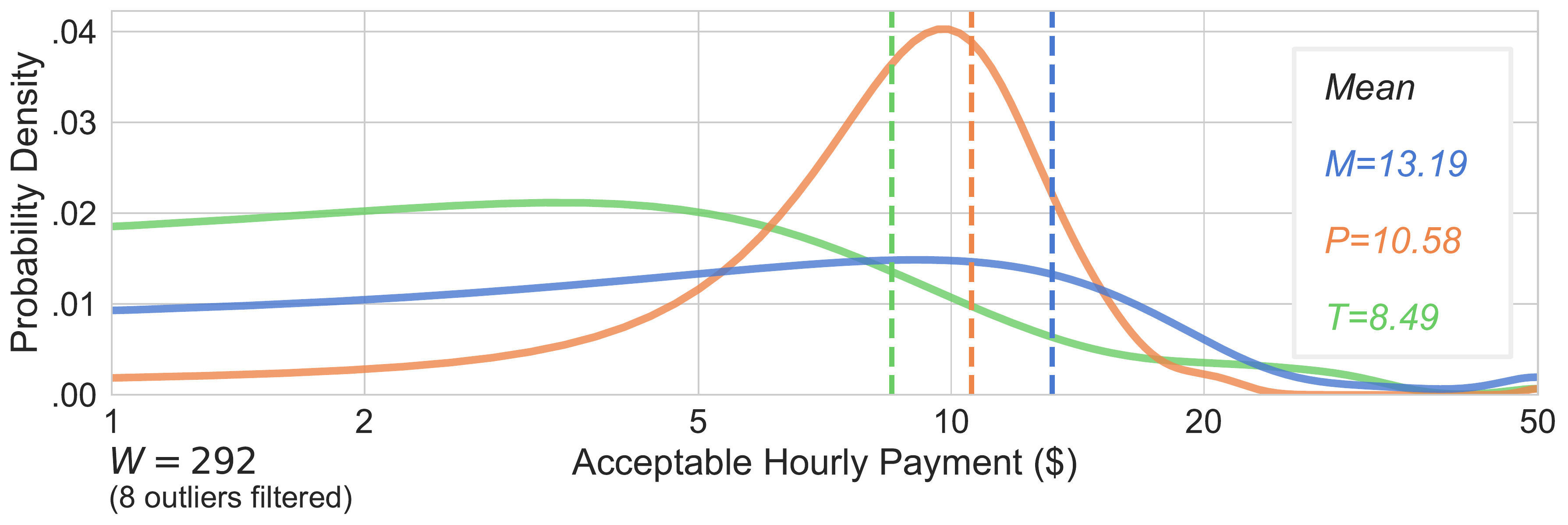}
		\caption{Acceptable hourly payment in USD\$ for participation in longitudinal studies according to the workers.}
	\label{cap:paper_tsc2024-sec:results-subsec:rq1-analysis-fig:ideal_hourly_payment}
\end{figure}

Figure~\ref{cap:paper_tsc2024-sec:results-subsec:rq1-analysis-fig:ideal_hourly_payment} investigates the acceptable hourly payment rate in USD\$ for participating in longitudinal studies on the recruitment platform, as reported by workers.

\mturk workers request the highest hourly payment on average (about \$13), while \prolific workers request slightly lower, at about \$10.50. In contrast, \toloka workers indicate the lowest acceptable amount (about \$8.50). The figure excludes 8 outliers who provided unreasonable amounts (i.e., ranging from \$80 to \$100), which were removed from the analysis.

To interpret the provided answers, one must consider that the payment models of \mturk and \toloka differ from that of \prolific. On the first two platforms, a task requester proposes a fixed amount of money for each work unit, which can vary significantly. In contrast, the \prolific platform requires requesters to estimate the task completion time and propose a minimum payment based on the hourly estimate. This difference may influence workers' perception of the acceptable payment amount (\mturk vs. \prolific, \mturk vs. \toloka statistically significant; adjusted p-value < 0.05).

\subsubsection{Preferred Time To Allocate Daily} 
\label{cap:paper_tsc2024-sec:results-subsec:rq1-analysis-p2-question:preferred-time-to-allocate-daily}

\begin{figure}[tpb]
	\centering
	\includegraphics[width=.9\linewidth]{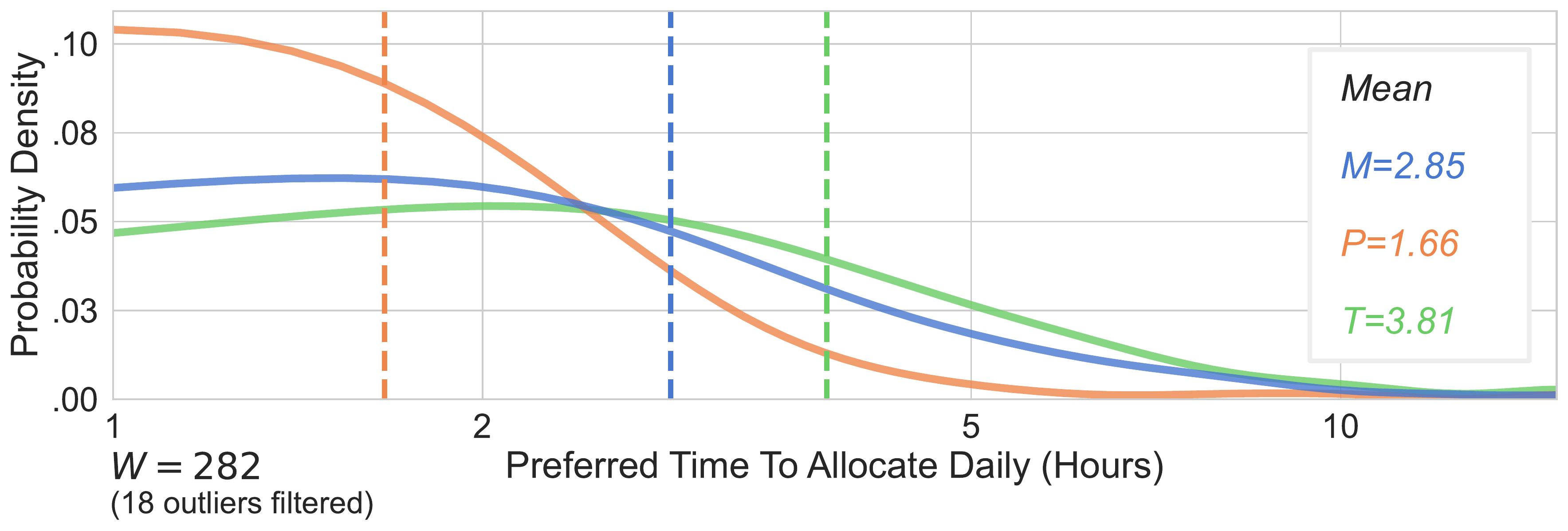}
		\caption{Preferred amount of time in hours to allocate on a daily basis for participating in longitudinal studies according the workers.}
	\label{cap:paper_tsc2024-sec:results-subsec:rq1-analysis-fig:time_per_day}
\end{figure}

Figure~\ref{cap:paper_tsc2024-sec:results-subsec:rq1-analysis-fig:time_per_day} investigates the preferred amount of time in hours that workers are available to allocate for participating in longitudinal studies on a daily basis.

Workers recruited on \toloka are the most willing to work per day, with an average of 3.81 hours. \mturk workers prefer working up to nearly three hours (2.85), while \prolific workers expect to work less, averaging around one and a half hours (1.66). The figure excludes 18 outliers who provided unreasonable daily hours (i.e., between 20 and 25), which were removed.

In general, \toloka workers are the most willing to allocate time per day and expect lower rewards. This is evident not only in their daily participation time, as shown in Figure~\ref{cap:paper_tsc2024-sec:results-subsec:rq1-analysis-fig:time_per_day}, but also when asked about their preferred session duration (Figure~\ref{cap:paper_tsc2024-sec:results-subsec:rq1-analysis-fig:ideal_session}) and their ideal daily payment (Figure~\ref{cap:paper_tsc2024-sec:results-subsec:rq1-analysis-fig:ideal_hourly_payment}). As for \mturk and \prolific workers, they expect to work less on average, particularly \prolific workers (\mturk vs. \prolific, \prolific vs. \toloka statistically significant; adjusted p-value < 0.05).

\subsubsection{Participation Incentives (In New Experiences)} 
\label{cap:paper_tsc2024-sec:results-subsec:rq1-analysis-p2-question:participation-incentives-new}

\begin{figure}[tpb]
	\centering
	\includegraphics[width=\linewidth]{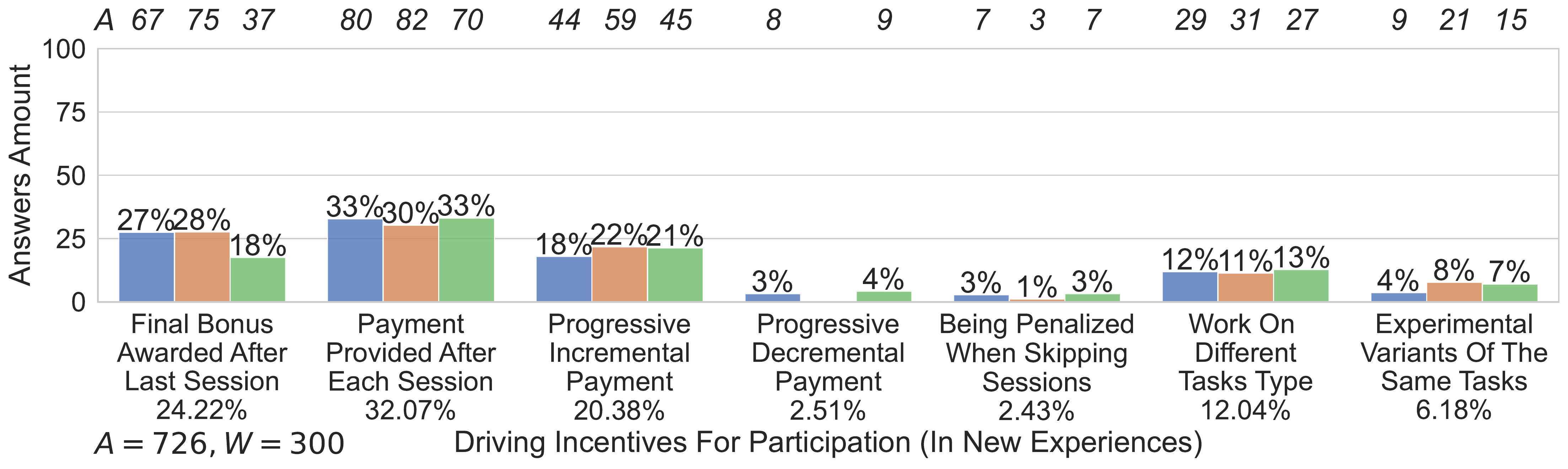}
		\caption{Incentives that drive workers to participate in new longitudinal studies according to those who answered the survey.}
	\label{cap:paper_tsc2024-sec:results-subsec:rq1-analysis-fig:incentives_motivation}
\end{figure}

Figure~\ref{cap:paper_tsc2024-sec:results-subsec:rq1-analysis-fig:incentives_motivation} investigates the underlying motivations that drive participation in new longitudinal studies.

The type of reward/payment mechanism is the most important incentive for the vast majority of workers (81.86\%). Among them, the most preferred option is payment after each session (32.07\%). Other alternatives include a final bonus awarded after the last session (24.22\%) and progressive incremental payments after each session (20.38\%). Progressive decremental payments (2.51\%) or penalization for skipping sessions (2.43\%) have a smaller but still notable influence on participation in new studies.

Beyond the reward/payment mechanism, 12.04\% of workers indicate that working on different task types to increase engagement diversity is important, while 6.18\% suggest experimental variants of the same tasks to reduce repeatability. When considering each crowdsourcing platform, no particular trends emerge (\mturk vs. \prolific, \mturk vs. \toloka, and \prolific vs. \toloka statistically significant; adjusted p-value~<~0.01).

\subsubsection{Tasks Type} 
\label{cap:paper_tsc2024-sec:results-subsec:rq1-analysis-p2-question:tasks-type}

\begin{figure}[tpb]
	\centering
	\includegraphics[width=\linewidth]{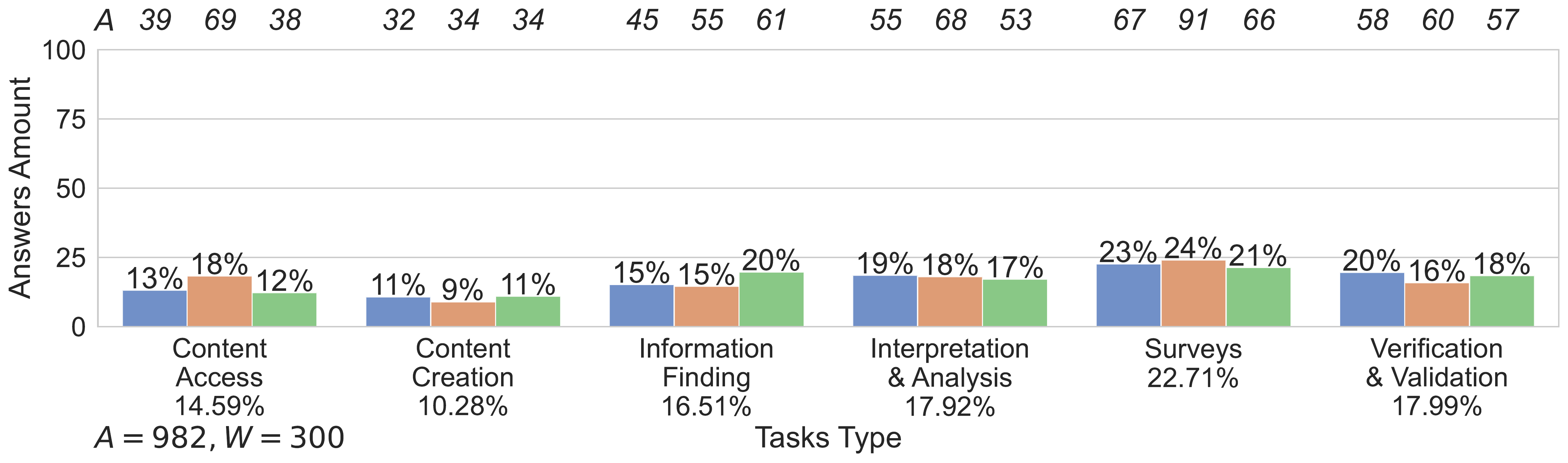}
		\caption{Tasks type that workers would like to perform in a longitudinal studies according to those who answered the survey.}
	\label{cap:paper_tsc2024-sec:results-subsec:rq1-analysis-fig:task_type}
\end{figure}

Figure~\ref{cap:paper_tsc2024-sec:results-subsec:rq1-analysis-fig:task_type} investigates the tasks that workers would like to perform in a longitudinal study. We acknowledge that the predefined set of answers we provided might not have been perceived as exhaustive, and workers were given the opportunity to provide free-text responses to elaborate further.

Surveys, which focus on various aspects usually crowdsourced, such as demographics, make up 22.71\%. Verification and validation tasks require workers to verify aspects or confirm the validity of content (17.99\%). Interpretation and analysis tasks rely on the crowd's ability to interpret and analyze during task completion (17.92\%). Information finding tasks involve workers searching to satisfy specific information needs (16.51\%). Content access tasks simply require workers to access content (14.59\%), while content creation tasks require generating new content for a document or website (10.28\%). Two workers mentioned other task types, including gamified tasks and content editing, which were not initially considered, along with content access and creation.

In summary, workers across all platforms are open to the proposed task types, with a fairly even distribution of responses. Nonetheless, the differences remain statistically significant (\mturk vs. \prolific, \mturk vs. \toloka, and \prolific vs. \toloka statistically significant; adjusted p-value~<~0.01).

\subsubsection{Involvement Benefits} 
\label{cap:paper_tsc2024-sec:results-subsec:rq1-analysis-p2-question:involvement-benefits}

\begin{figure}[tpb]
	\centering
	\includegraphics[width=\linewidth]{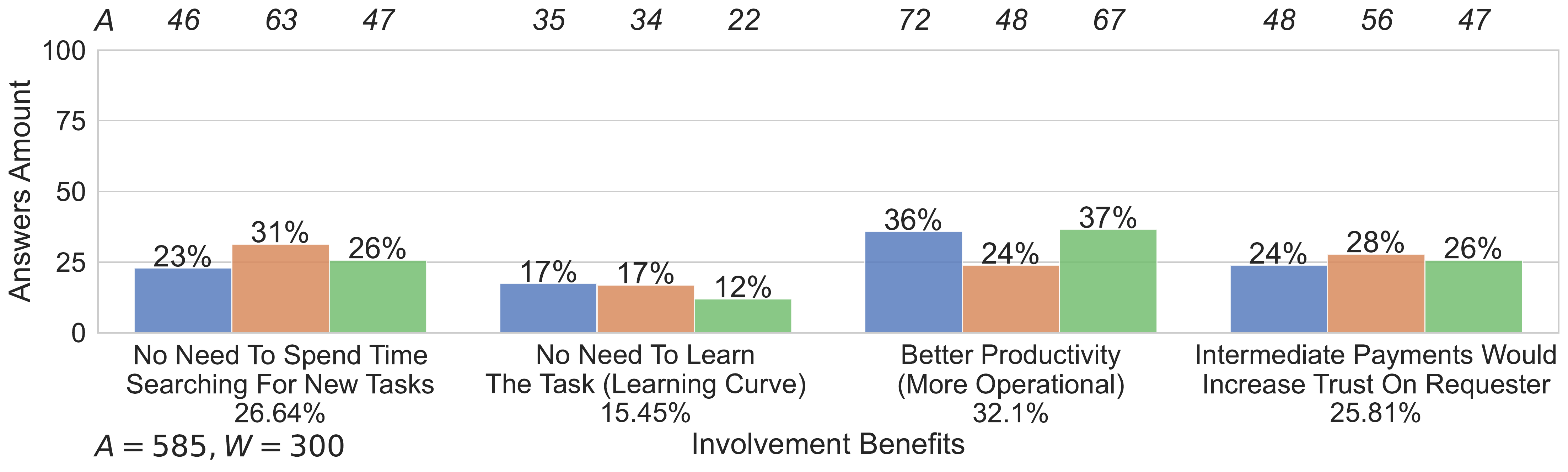}
		\caption{Benefits of being involved in longitudinal studies according to the workers.}
	\label{cap:paper_tsc2024-sec:results-subsec:rq1-analysis-fig:benefits}
\end{figure}

Figure~\ref{cap:paper_tsc2024-sec:results-subsec:rq1-analysis-fig:benefits} investigates the benefits of participating in longitudinal studies, according to workers.

In general, workers believe that the most significant benefit of longitudinal studies is increased productivity due to their more operational nature (32.1\%). They also appreciate the time-saving aspect, as longitudinal studies eliminate the need for regular task searching (26.64\%). Additionally, workers feel that receiving intermediate payments after each session would increase trust in the requester (25.81\%). Some workers value avoiding the need to re-learn tasks when participating in longitudinal studies (15.45\%).

Trends are generally homogeneous across platforms, with no factor considered significantly more important than others. However, increased productivity stands out more for \mturk workers (36\%) and \toloka workers (37\%) compared to \prolific (24\%). This distribution is statistically significant (\mturk vs. \prolific, \mturk vs. \toloka, and \prolific vs. \toloka statistically significant; adjusted p-value~<~0.01).

\subsubsection{Involvement Downsides} 
\label{cap:paper_tsc2024-sec:results-subsec:rq1-analysis-p2-question:involvement-downsides}

\begin{figure}[tpb]
    \centering
    \includegraphics[width=\linewidth]{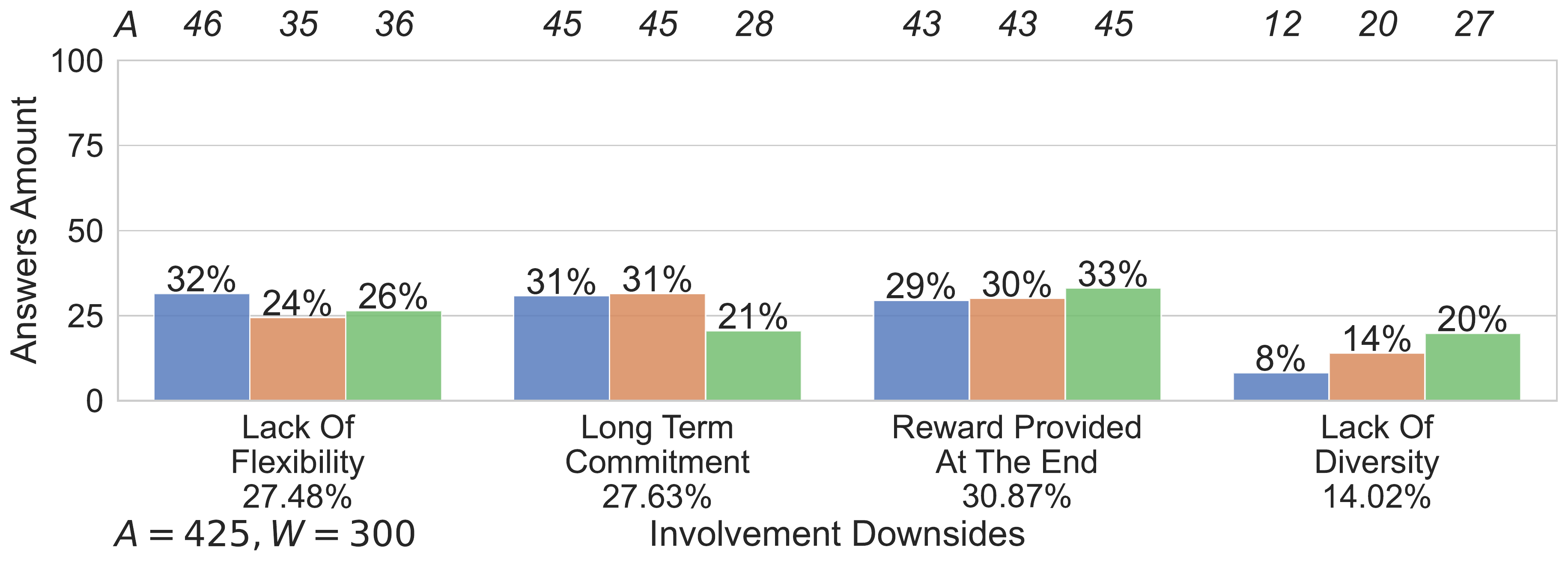}
        \caption{Downsides of being involved in longitudinal studies according to the workers.}
    \label{cap:paper_tsc2024-sec:results-subsec:rq1-analysis-fig:downsides}
\end{figure}

Figure~\ref{cap:paper_tsc2024-sec:results-subsec:rq1-analysis-fig:downsides} examines the downsides of participating in longitudinal studies, as reported by workers.

Workers indicate that the most significant downside is receiving a reward only at the end of the longitudinal study (30.87\%). The lack of flexibility in the study schedule and the long-term commitment required are reported with similar frequency, at 27.48\% and 27.63\%, respectively. The lack of diversity in the tasks performed during each session plays a smaller role (14.02\%).

When considering each platform, the trends are largely consistent for \mturk and \prolific. However, an interesting distinction emerges for \toloka workers, who report the lack of diversity as a more significant downside (20\%), while the long-term commitment is less of an issue for them (21\%) compared to workers on the other platforms. These differences are statistically significant (\mturk vs. \prolific, \mturk vs. \toloka, and \prolific vs. \toloka statistically significant; adjusted p-value~<~0.01).

\subsubsection{Suggestions About Longitudinal Study Design}
\label{cap:paper_tsc2024-sec:results-subsec:rq1-analysis-p2-last}
\label{cap:paper_tsc2024-sec:results-subsec:rq1-analysis-p2-question:design-suggestions}

The last optional question (question 11, part \ptwo) asked workers to provide suggestions for requesters designing a longitudinal study.

Out of \numworkers, 201 workers (67\%) provided some form of answer. The distribution of responses across different themes is as follows: 139 out of 201 answers (69.15\%) addressed aspects related to the task performed (\taskfeatures), 9 (4.48\%) focused on requesters' characteristics (\requesterfeatures), 7 (3.48\%) discussed workers' own beliefs and motivations (\workerfeatures), 5 (2.49\%) referred to the longitudinal study as a whole (\lsfeatures), and 2 (1\%) concerned the platform (\platformfeatures). Additionally, 2 (1\%) answers were considered unusable (\answeruseless). Finally, 37 workers (18.41\%) explicitly stated that they did not have any suggestions (\nosuggestion). A sample of these answers is shown in Table~\ref{cap:paper_tsc2024-sec:results-subsec:rq1-analysis-p2-last-tab:design-suggestions}.

The majority of workers (139 out of 201, 69.15\%) suggest improvements related to the features of the tasks performed within each session of the longitudinal study, including aspects like task design, scheduling, and participant filtering. Among them, six workers (4.32\%) propose allowing a reasonable window for task completion, accounting for other activities in workers' schedules. One worker suggests offering the option to skip a session occasionally if unable to commit. Additionally, a few workers (3 out of 139, 2.16\%) stress the importance of conducting pilot tests, which help requesters identify suitable participants and retain workers throughout the study. One worker suggests offering different systems for participation (e.g., desktop devices, smartphones), while another advises against requiring downloads. This aligns with prior research highlighting the diverse work environments in which workers are embedded~\cite{gadiraju2017modus}. Workers also emphasize the need for clear instructions, a well-designed user interface, an understandable sequence of events, the ability to identify changes over time, and insight into cause-and-effect relationships. Some workers believe that variability in tasks could help maintain interest in the study.

Regarding the overall structure of a longitudinal study (5 out of 201, 2.49\%), workers suggest planning all sessions in advance while maintaining flexibility with the schedule, particularly when involving multiple geographic time zones. They also recommend establishing a sense of progression, such as highlighting differences in previous responses at the end of each session.

A few workers (7 out of 201, 3.48\%) provide personal insights. One worker notes that many participants are self-employed and must pay taxes on earnings from crowdsourcing platforms, so rewards should reflect this. Another worker prefers small payments with a bonus for completing all sessions.

Concerning the task requesters (9 out of 201, 4.48\%), workers suggest that regular feedback from requesters is important. They recommend that requesters be communicative and friendly, leave space for feedback in each study, send reminders when necessary, and provide clear upfront information.

\begin{longtable}{p{14cm}}
\caption{Sample of suggestions provided by workers concerning longitudinal studies.}
\label{cap:paper_tsc2024-sec:results-subsec:rq1-analysis-p2-last-tab:design-suggestions} \\
\toprule
\textbf{Worker Responses} \\
\midrule
\endfirsthead
\toprule
\textbf{Worker Responses} \\
\midrule
\endhead

\footnotesize\itshape Continues in the next page \\
\endfoot
\endlastfoot

\emph{Establish the correct sequence of events, identify changes over time, and provide insight into cause-and-effect relationships.} \\
\midrule
\emph{Plan each session in a way that it makes the surveyee feel like the're making progress. Maybe at the end of each session highlight the differences in their previous answer to accentuate that feeling of progression.} \\
\midrule
\emph{Beside all of the aspects regarding time and money, fast communication between requester and worker and also regular feedbacks regarding workers task quality would be great to increase their (our :) ) commitment.} \\
\midrule
\emph{Maybe offer different platforms on which to take the study (ie android, PC, mac, etc).} \\
\midrule
\emph{Just don't require downloads. Keep tasks short. No time frames.} \\
\midrule
\emph{A lot of us work from home and are self employed so we have to pay tax on these earnings. As long as it pays a decent amount for the time taken (at least £6 per hour), I would be more than happy to take part.} \\
\midrule
\emph{It is useful to allow one or two sessions to be skipped if the responder can't commit to absolutely every session.}\\
\midrule
\emph{Be reasonable with what you expect people to do. People who work full time and have caring responsibilities won't necessarily have the capacity/flexibility to do daily tasks that last an hour or more. If your study makes those demands then you're going to only be getting a certain kind of participant (e.g. unemployed).} \\
\midrule
\emph{Keep them to the point, don't give long, fatigued instructions, try not to ask the same question fifty different ways. Also, if you have a game, games are very attractive for me; I'd be interested in longitudinal studies where we have to play a game and collect something, like points, or something. And gives a good bonus! Good base pay, as well. At least 12 dollars an hour.} \\
\midrule
\emph{Ensure the timings are not onerous when considering participants from multiple geographic zones - they need adequate time to complete. A final bonus payment \ completion incentive helps reduce attrition - and on that note, keep the study shorter (say 2 weeks) to minimise participant drop-off.} \\
\midrule
\emph{I think you have to be as revealing as possible in the first part of the study so the participant knows in advance what they are signing up for, it would help if the participant gets a good idea or sampling of the task in full so there are no surprises if that is possible so it would be good to have them complete the worst part of it if there is one and if it is repetitive and hard to complete over a longer period then to explain that so they can make a judgement. So long as they know what is involved and what is expected of them in advance before they then agree to take part because then so long as they understand the commitment they are making and the schedule and timing they should be able to complete it.} \\
\midrule
\emph{Using good screeners can both help requesters find participants that fit the needs of the study, as well as participants that are less likely to quit part-way through. Also, compensation schemes that reward consistent participation are likely to increase the odds that participants complete all required sessions of the study.} \\
\bottomrule

\end{longtable}

\subsubsection{Summary}
\label{cap:paper_tsc2024-sec:results-subsec:rq1-analysis-subsec:summary}

The workers' answers for the \pone part of the survey are summarized in Table~\ref{cap:paper_tsc2024-sec:results-subsec:rq1-analysis-subsec:summary-tab:summary-pone}, and those for \ptwo are presented in Table~\ref{cap:paper_tsc2024-sec:results-subsec:rq1-analysis-subsec:summary-tab:summary-ptwo}. Both tables provide a detailed summary of the answers, along with the code used to classify each question and a breakdown of responses across the crowdsourcing platforms considered.

Table~\ref{cap:paper_tsc2024-sec:results-subsec:rq1-analysis-subsec:summary-tab:summary-stat-sig} presents the results of statistical tests comparing the groups of answers provided across each platform. The table includes the name and answer type for each question. A checkmark (\checkmark) indicates a statistically significant comparison with the adjusted p-value provided, while its absence means that the comparison was not statistically significant.

Finally, the key findings are summarized with a list of take-home messages. The first set of messages (1-9) covers the perception of longitudinal studies based on workers' previous experiences (\pone part questions), while the second set (10-17) addresses workers' opinions on future longitudinal studies (\ptwo part questions). Each message is referenced to the corresponding section where the analysis is reported.

\begin{enumerate}[label=\arabic*.] 
\item Workers with more experience in longitudinal studies are more easily found on the \prolific platform (Section~\ref{cap:paper_tsc2024-sec:results-subsec:rq1-analysis-p1-question:previous-experiences}), and the studies available on this platform tend to have more sessions compared to other platforms (Section~\ref{cap:paper_tsc2024-sec:results-subsec:rq1-analysis-p1-question:number-of-session}). 
\item Most workers report that their experiences took place within one year prior to participating in the survey (Section~\ref{cap:paper_tsc2024-sec:results-subsec:rq1-analysis-p1-question:time-elapsed}). 
\item The majority of sessions in the reported longitudinal studies lasted up to 2 hours, with about half of them lasting only 15 minutes (Section~\ref{cap:paper_tsc2024-sec:results-subsec:rq1-analysis-p1-question:session-duration}). 
\item The time intervals between sessions in the reported longitudinal studies most often range from 1 to 30 days (Section~\ref{cap:paper_tsc2024-sec:results-subsec:rq1-analysis-p1-question:interval-between-sessions}). 
\item Most longitudinal studies reported provide partial rewards after each session (Section~\ref{cap:paper_tsc2024-sec:results-subsec:rq1-analysis-p1-question:payment-model}). 
\item The main motivation for workers to participate in and complete the reported longitudinal studies is the monetary aspect (Section~\ref{cap:paper_tsc2024-sec:results-subsec:rq1-analysis-p1-question:participation-incentives-previous} and Section~\ref{cap:paper_tsc2024-sec:results-subsec:rq1-analysis-p1-question:completion-incentives-previous}).
\item Nearly all workers claim to have completed the reported longitudinal studies (Section~\ref{cap:paper_tsc2024-sec:results-subsec:rq1-analysis-p1-question:study-completion}). 
\item Most workers express a desire to continue participating in similar longitudinal studies in the future (Section~\ref{cap:paper_tsc2024-sec:results-subsec:rq1-analysis-p1-question:participation-same-study}). 
\item The main factors limiting the availability of longitudinal studies on crowdsourcing platforms are workers' dislike for the required commitment and insufficient rewards (Section~\ref{cap:paper_tsc2024-sec:results-subsec:rq1-analysis-p1-question:platform-availability}). 
\item In a hypothetical longitudinal study where workers engage in a single session lasting 15 minutes daily, workers would be willing to commit to participating for an average of 21 days (Section~\ref{cap:paper_tsc2024-sec:results-subsec:rq1-analysis-p2-question:preferred-commitment-duration}). However, when considering session duration, workers are generally willing to work for up to an average of 103 minutes per session (Section~\ref{cap:paper_tsc2024-sec:results-subsec:rq1-analysis-p2-question:preferred-session-duration}). 
\item Most workers prefer daily participation over weekly participation in longitudinal studies (Section~\ref{cap:paper_tsc2024-sec:results-subsec:rq1-analysis-p2-question:preferred-participation-frequency}). \item On average, workers would prefer to allocate 2.7 hours daily for participating in longitudinal studies (Section~\ref{cap:paper_tsc2024-sec:results-subsec:rq1-analysis-p2-question:preferred-time-to-allocate-daily}). 
\item Workers report that the acceptable hourly payment for participating in longitudinal studies is approximately \$10.75 on average (Section~\ref{cap:paper_tsc2024-sec:results-subsec:rq1-analysis-p2-question:acceptable-hourly-payment}). This amount should be adjusted for inflation. \item Workers report that the primary incentives driving participation in new longitudinal studies are related to the rewards provided (Section~\ref{cap:paper_tsc2024-sec:results-subsec:rq1-analysis-p2-question:participation-incentives-new}). 
\item Most workers believe that the length of a longitudinal study plays a critical role in their decision to refuse participation (Section~\ref{cap:paper_tsc2024-sec:results-subsec:rq1-analysis-p2-question:participation-decline}). 
\item Workers report that the main benefits of being involved in longitudinal studies are increased productivity due to their operational nature and the elimination of the need for regular task searching (Section~\ref{cap:paper_tsc2024-sec:results-subsec:rq1-analysis-p2-question:involvement-benefits}). 
\item Workers report that the main downsides of being involved in longitudinal studies are the long-term commitment required, the lack of flexibility, and the reward provided only at the end of the study (Section~\ref{cap:paper_tsc2024-sec:results-subsec:rq1-analysis-p2-question:involvement-downsides}). 
\end{enumerate}

\begin{table}[htbp]
    \centering
    \caption{Summary of the key findings for the \pone part of the survey presented in the quantitative analysis.}
\label{cap:paper_tsc2024-sec:results-subsec:rq1-analysis-subsec:summary-tab:summary-pone}
\begin{tabular}{C{1.2cm}p{2.9cm}p{3.6cm}p{3.6cm}p{3.6cm}}
\toprule
 \textbf{Section} & \textbf{Question} & \textbf{\mturk} & \textbf{\prolific} & \textbf{\toloka} \\
\midrule
\ref{cap:paper_tsc2024-sec:results-subsec:rq1-analysis-p1-question:previous-experiences} &
\emph{Previous Experiences}  
& 42\% 1 experience, 29\% 2 experiences, 29\% 3 experiences 
& 43\% 1 experience, 21\% 2 experiences, 36\% 3 experiences 
& 50\% 1 experience, 33\% 2 experiences, 17\% 3 experiences \\
\midrule
\ref{cap:paper_tsc2024-sec:results-subsec:rq1-analysis-p1-question:time-elapsed} &
\emph{Time Elapsed} 
& 87\% up to 1 year before, 13\% later 
& 87\% up to 1 year before, 13\% later 
& 87\% up to 1 year before, 13\% later \\
\midrule
\ref{cap:paper_tsc2024-sec:results-subsec:rq1-analysis-p1-question:number-of-session} &
\emph{Sessions} 
& $\sim$6 on average 
& $\sim$7 on average 
& $\sim$6 on average \\
\midrule
\ref{cap:paper_tsc2024-sec:results-subsec:rq1-analysis-p1-question:interval-between-sessions} &
\emph{Interval Between Sessions}  
& 89\% up to 1 month, 11\% later 
& 88\% up to 1 month, 12\% later 
& 97\% up to 1 month, 6\% later \\
\midrule
\ref{cap:paper_tsc2024-sec:results-subsec:rq1-analysis-p1-question:session-duration} &
\emph{Session Duration} 
& 98\% up to 1 hour, 3\% more 
& 99\% up to 1 hour, 1\% more 
& 91\% up to 1 hour, 8\% more \\
\midrule
\ref{cap:paper_tsc2024-sec:results-subsec:rq1-analysis-p1-question:crowdsourcing-platform} &
\emph{Crowdsourcing Platform} 
& 91\% MTurk, 9\% Prolific, 0\% Toloka 
& 6\% MTurk, 90\% Prolific, 4\% Toloka 
& 17\% MTurk, 19\% Prolific, 63\% Toloka \\
\midrule
\ref{cap:paper_tsc2024-sec:results-subsec:rq1-analysis-p1-question:payment-model} &
\emph{Payment Model} 
& 75\% after each session, 16\% final reward, 9\% both 
& 68\% after each session, 25\% final reward, 7\% both 
& 68\% after each session, 25\% final reward, 7\% both \\
\midrule
\ref{cap:paper_tsc2024-sec:results-subsec:rq1-analysis-p1-question:participation-same-study} &
\emph{Participation in Same Study} 
& 83\% yes, 17\% no 
& 98\% yes, 2\% no 
& 93\% yes, 7\% no \\
\midrule
\ref{cap:paper_tsc2024-sec:results-subsec:rq1-analysis-p1-question:participation-incentives-previous} &
\emph{Participation Incentives (in Previous Experiences)} 
& 29\% bonus, 55\% reward, 13\% personal interest, 3\% altruism, 1\% educative task 
& 11\% bonus, 56\% reward, 26\% personal interest, 7\% altruism, 0\% educative task  
& 27\% bonus, 34\% reward, 18\% personal interest, 4\% altruism, 17\% educative task \\
\midrule
\ref{cap:paper_tsc2024-sec:results-subsec:rq1-analysis-p1-question:study-completion} &
\emph{Study Completion} 
& 95\% yes, 5\% no 
& 99\% yes, 1\% no 
& 99\% yes, 1\% no \\
\midrule
\ref{cap:paper_tsc2024-sec:results-subsec:rq1-analysis-p1-question:completion-incentives-previous} &
\emph{Completion Incentives (In Previous Experiences)} 
& 27\% bonus, 46\% reward, 18\% personal interest, 4\% altruism, 2\% educative task, 4\% participation dropped
& 15\% bonus, 59\% reward, 19\% personal interest, 7\% altruism, 0\% educative task, 1\% participation dropped
& 25\% bonus, 33\% reward, 19\% personal interest, 7\% altruism, 14\% educative task, 1\% participation dropped  \\
\midrule
\ref{cap:paper_tsc2024-sec:results-subsec:rq1-analysis-p1-question:platform-availability} &
\emph{Reasons that Limit Availability on Platforms} 
& 27\% lack of support, 29\% dislike commitment, 36\% reward and incentives insufficient, 7\% no need longitudinal participation
& 12\% lack of support, 44\% dislike commitment, 29\% reward and incentives insufficient, 15\% no need longitudinal participation
& 35\% lack of support, 26\% dislike commitment, 33\% reward and incentives insufficient, 5\% no need longitudinal participation  \\
\bottomrule                  
\end{tabular}
\end{table}

\begin{table}[htbp]
    \centering
    \caption{Summary of the key findings for the \ptwo part of the survey presented in the quantitative analysis.}
\label{cap:paper_tsc2024-sec:results-subsec:rq1-analysis-subsec:summary-tab:summary-ptwo}
\begin{tabular}{C{1.2cm}p{2.9cm}p{3.6cm}p{3.6cm}p{3.6cm}}
\toprule
 \textbf{Section} & \textbf{Question} & \textbf{\mturk} & \textbf{\prolific} & \textbf{\toloka} \\
\midrule
\ref{cap:paper_tsc2024-sec:results-subsec:rq1-analysis-p2-question:preferred-commitment-duration} &
\emph{Preferred Commitment Duration} 
& $\sim$19 days on average 
& $\sim$29 days on average 
& $\sim$17 days on average \\
\midrule
\ref{cap:paper_tsc2024-sec:results-subsec:rq1-analysis-p2-question:participation-decline} &
\emph{Reasons for Declining Participation} 
& 42\% study is too frequent, 58\% study is too long 
& 15\% study is too frequent, 85\% study is too long 
& 29\% study is too frequent, 71\% study is too long \\
\midrule
\ref{cap:paper_tsc2024-sec:results-subsec:rq1-analysis-p2-question:preferred-participation-frequency} &
\emph{Preferred Participation Frequency}
& 92\% up to 1 week, 3\% biweekly, 5\% monthly, 0\% yearly
& 95\% up to 1 week, 1\% biweekly, 2\% monthly, 0\% yearly
& 88\% up to 1 week, 5\% biweekly, 1\% monthly, 2\% yearly \\
\midrule
\ref{cap:paper_tsc2024-sec:results-subsec:rq1-analysis-p2-question:preferred-session-duration} &
\emph{Preferred Session Duration} 
& $\sim$115 minutes on average 
& $\sim$60 minutes on average 
& $\sim$137 minutes on average \\
\midrule
\ref{cap:paper_tsc2024-sec:results-subsec:rq1-analysis-p2-question:acceptable-hourly-payment} &
\emph{Acceptable Hourly Payment} 
& 13.19 USD\$ on average 
& 10.58 USD\$ on average 
& 8.49 USD\$ on average \\
\midrule
\ref{cap:paper_tsc2024-sec:results-subsec:rq1-analysis-p2-question:preferred-time-to-allocate-daily} &
\emph{Preferred Time to Allocate Daily} 
& $\sim$171 minutes on average 
& $\sim$100 minutes on average 
& $\sim$228 minutes on average \\
\midrule
\ref{cap:paper_tsc2024-sec:results-subsec:rq1-analysis-p2-question:participation-incentives-new} &
\emph{Participation Incentives (in New Experiences)} 
& 27\% final bonus, 33\% pay after each session, 18\% prog. incr. payment, 3\% progr. decr. payment, 3\% penalization for skipping, 12\% different task types, 4\% experimental variants
& 28\% final bonus, 30\% pay after each session, 22\% prog. incr. payment, 0\% prog. decr. payment, 1\% penalization for skipping, 11\% different task types, 8\% experimental variants
& 18\% final bonus, 33\% pay after each session, 21\% progr. incr. payment, 4\% progr. decr. payment, 3\% penalization for skipping, 13\% different task types, 7\% experimental variants \\
\midrule
\ref{cap:paper_tsc2024-sec:results-subsec:rq1-analysis-p2-question:tasks-type} &
\emph{Tasks Type} 
& 13\% content access, 11\% content creation, 15\% information finding, 18\% interpretation and analysis, 23\% surveys, 19\% verification and validation 
& 18\% content access, 9\% content creation, 15\% information finding, 18\% interpretation and analysis, 24\% surveys, 16\% verification and validation 
& 12\% content access, 9\% content creation, 20\% information finding, 18\% interpretation and analysis, 24\% surveys, 16\% verification and validation \\
\midrule
\ref{cap:paper_tsc2024-sec:results-subsec:rq1-analysis-p2-question:involvement-benefits} &
\emph{Involvement Benefits} 
& 23\% no need to search, 17\% no need to learn, 36\% better productivity, 24\% increase trust 
& 31\% no need to search, 17\% no need to learn, 34\% better productivity, 28\% increase trust 
& 26\% no need to search, 12\% no need to learn, 37\% better productivity, 26\% increase trust \\
\midrule
\ref{cap:paper_tsc2024-sec:results-subsec:rq1-analysis-p2-question:involvement-downsides} &
\emph{Involvement Downsides} 
& 32\% lack of flexibility, 31\% long term commitment, 29\% reward at the end, 8\% lack of diversity 
& 24\% lack of flexibility, 31\% long term commitment, 30\% reward at the end, 14\% lack of diversity
& 26\% lack of flexibility, 21\% long term commitment, 33\% reward at the end, 20\% lack of diversity \\
\bottomrule                    
\end{tabular}
\end{table}

\begin{table}[htpb]
    \centering
    \caption{Summary of statistical tests comparing answer groups of each platform. Questions without any statistically significant comparisons are not reported. Statistical significance is computed using adjusted p-values according to Section~\ref{cap:paper_tsc2024-sec:exp-setup-subsec:stat-test}.}
\label{cap:paper_tsc2024-sec:results-subsec:rq1-analysis-subsec:summary-tab:summary-stat-sig}
\begin{tabular}{C{0.7cm}C{1.2cm}p{3.5cm}C{1.2cm}C{1.4cm}C{1.4cm}C{1.4cm}C{1.6cm}}
\toprule
  \textbf{Part} & \textbf{Section} & \textbf{Question} & \textbf{Type} & \textbf{MTurk Vs. \prolific}  & \textbf{MTurk Vs. \toloka} & \textbf{\prolific Vs. \toloka} & \textbf{Significance Level} \\
  \midrule
  \pone & \ref{cap:paper_tsc2024-sec:results-subsec:rq1-analysis-p1-question:time-elapsed} & \emph{Time Elapsed} & mcq &  & \checkmark &  & $p\leq0.05$ \\
  \midrule
  \pone  & \ref{cap:paper_tsc2024-sec:results-subsec:rq1-analysis-p1-question:interval-between-sessions} & \emph{Interval Between Sessions} & mcq &  & \checkmark &  & $p\leq0.01$ \\
  \midrule
  \pone  & \ref{cap:paper_tsc2024-sec:results-subsec:rq1-analysis-p1-question:session-duration} & \emph{Session Duration} & mcq & \checkmark & \checkmark & \checkmark & $p\leq0.01$ \\
  \midrule
  \pone  & \ref{cap:paper_tsc2024-sec:results-subsec:rq1-analysis-p1-question:crowdsourcing-platform} & \emph{Crowdsourcing Platform} & mcq & \checkmark & \checkmark & \checkmark & $p\leq0.01$ \\
  \midrule
  \pone  & \ref{cap:paper_tsc2024-sec:results-subsec:rq1-analysis-p1-question:payment-model} & \emph{Payment Model} & list & \checkmark & \checkmark & \checkmark & $p\leq0.01$ \\
  \midrule
   \pone & \ref{cap:paper_tsc2024-sec:results-subsec:rq1-analysis-p1-question:participation-same-study} & \emph{Participation In Same Study} & mcq & \checkmark & \checkmark & \checkmark & $p\leq0.01$ \\
  \midrule
  \pone & \ref{cap:paper_tsc2024-sec:results-subsec:rq1-analysis-p1-question:participation-incentives-previous} & \emph{Participation Incentives (In Previous Experience)} & mcq & \checkmark & & \checkmark & $p\leq0.01$ \\
 \midrule
  \pone & \ref{cap:paper_tsc2024-sec:results-subsec:rq1-analysis-p1-question:completion-incentives-previous} &  \emph{Completion Incentives (In Previous Experience)} & mcq & \checkmark & \checkmark & \checkmark & $p\leq0.01$ \\
  \midrule
  \pone & \ref{cap:paper_tsc2024-sec:results-subsec:rq1-analysis-p1-question:platform-availability} & \emph{Reasons That Limit Availability On Platforms} & mcq & \checkmark & \checkmark & \checkmark & $p\leq0.01$ \\
  \midrule
  \ptwo & \ref{cap:paper_tsc2024-sec:results-subsec:rq1-analysis-p2-question:preferred-commitment-duration} & \emph{Preferred Commitment Duration} & number &  &  & \checkmark & $p\leq0.05$ \\
  \midrule
  \ptwo & \ref{cap:paper_tsc2024-sec:results-subsec:rq1-analysis-p2-question:participation-decline} & \emph{Reasons For Declining Participation} & list & \checkmark & \checkmark & \checkmark & $p\leq0.01$ \\
  \midrule
  \ptwo & \ref{cap:paper_tsc2024-sec:results-subsec:rq1-analysis-p2-question:preferred-participation-frequency} & \emph{Preferred Participation Frequency} & mcq & \checkmark & \checkmark & \checkmark & $p\leq0.01$ \\
  \midrule
  \ptwo & \ref{cap:paper_tsc2024-sec:results-subsec:rq1-analysis-p2-question:preferred-session-duration} & \emph{Preferred Session Duration} & number &  & \checkmark &  & $p\leq0.05$ \\
  \midrule
  \ptwo & \ref{cap:paper_tsc2024-sec:results-subsec:rq1-analysis-p2-question:acceptable-hourly-payment} & \emph{Acceptable Hourly Payment} & number & \checkmark & \checkmark &  & $p\leq0.05$ \\
  \midrule
  \ptwo & \ref{cap:paper_tsc2024-sec:results-subsec:rq1-analysis-p2-question:preferred-time-to-allocate-daily} & \emph{Preferred Time To Allocate Daily} & number & \checkmark &  & \checkmark & $p\leq0.05$ \\
  \midrule
  \ptwo & \ref{cap:paper_tsc2024-sec:results-subsec:rq1-analysis-p2-question:participation-incentives-new} & \emph{Participation Incentives (In New Experiences)} & list & \checkmark & \checkmark & \checkmark & $p\leq0.01$ \\
  \midrule
  \ptwo & \ref{cap:paper_tsc2024-sec:results-subsec:rq1-analysis-p2-question:tasks-type} & \emph{Tasks Type} & list & \checkmark & \checkmark & \checkmark & $p\leq0.01$ \\
  \midrule
  \ptwo & \ref{cap:paper_tsc2024-sec:results-subsec:rq1-analysis-p2-question:involvement-benefits} & \emph{Involvement Benefits} & list & \checkmark & \checkmark & \checkmark & $p\leq0.01$ \\
  \midrule
  \ptwo & \ref{cap:paper_tsc2024-sec:results-subsec:rq1-analysis-p2-question:involvement-downsides} & \emph{Involvement Downsides} & list & \checkmark & \checkmark & \checkmark & $p\leq0.01$ \\
\bottomrule
\end{tabular}
\end{table}

\subsection{\ref{cap:paper_tsc2024-sec:research-questions_2}: Recommendations For Researchers And Practitioners}
\label{cap:paper_tsc2024-sec:results-subsec:rq2-recommendations}

There is no standard approach for designing and conducting longitudinal studies on a crowdsourcing platform. However, the quantitative and qualitative analyses of workers' responses, combined with observations from deploying crowdsourcings tasks, have led to the development of \numrecommendations recommendations that could serve as a framework (Section~\ref{cap:paper_tsc2024-sec:results-subsec:rq2-recommendations-subsec:r1}--\ref{cap:paper_tsc2024-sec:results-subsec:rq2-recommendations-subsec:r8}). 

These recommendations should be considered by task requesters when designing longitudinal studies, as they provide useful guidelines and address workers' concerns and needs identified in the study.

\subsubsection{\texttt{R1:} Be Communicative And Provide Feedback}
\label{cap:paper_tsc2024-sec:results-subsec:rq2-recommendations-subsec:r1}

Communication is a critical factor in encouraging worker retention and decreasing the abandonment rate, as evident from their answers about what drove them to return to longitudinal studies and their suggestions to task requesters, described in Section~\ref{cap:paper_tsc2024-subsec:qualitative-analysis-subsec:worker-commitment}.

According to the workers, task requesters should inform them about upcoming sessions, progress throughout the study, and, when necessary, explicitly invite them to participate in newly published studies, considering that several days may pass, as shown in Figure~\ref{cap:paper_tsc2024-sec:results-subsec:rq1-analysis-fig:time_interval}. Requesters should also provide information about the overall progress of the longitudinal study and feedback on the quality of the work completed up to the current session.

When asked to provide additional suggestions, as reported in Section~\ref{cap:paper_tsc2024-sec:results-subsec:rq1-analysis-p2-question:design-suggestions}, they also pointed out that alerts, emails, or notifications should be sent on a regular schedule, as sending them randomly could negatively impact the worker experience. Furthermore, they noted that platforms like \prolific provide only an internal notification system, without an option to send a standard email to workers.

\subsubsection{\texttt{R2:} Schedule Each Session Mindfully}

Workers have free time to dedicate to participation in crowdsourcing tasks on different days of their working week, as reported in Section~\ref{cap:paper_tsc2024-sec:results-subsec:rq1-analysis-p2-question:design-suggestions}, and, on average, this amounts to roughly two hours (Figure~\ref{cap:paper_tsc2024-sec:results-subsec:rq1-analysis-fig:time_per_day}). Properly scheduling the required work is particularly important for longitudinal studies, especially considering that they may consist of a high number of sessions, as shown in Figure~\ref{cap:paper_tsc2024-sec:results-subsec:rq1-analysis-fig:sessions}.

Determining the overall number of sessions in advance and explicitly stating it is beneficial, as it allows workers to estimate the level of commitment required. This is particularly relevant given that they are generally willing to commit for up to a month, as shown in Figure~\ref{cap:paper_tsc2024-sec:results-subsec:rq1-analysis-fig:daily_commitment}. Communicating when the next session will take place also provides workers with greater flexibility.

Additionally, task requesters should consider the impact of recruiting workers from multiple geographic time zones. A session might start in the morning for some workers, while for others, it may occur during the night. Splitting the work into multiple batches spread across a 24-hour period could be beneficial. Alternatively, requesters could provide a sufficiently long time frame for session completion, with some workers suggesting 24 to 48 hours. Furthermore, sessions should not be too far apart, as workers prefer a daily participation frequency (Figure~\ref{cap:paper_tsc2024-sec:results-subsec:rq1-analysis-fig:which_commitment}). If too much time passes between sessions, workers may lose interest or fail to recall details of the study, increasing the likelihood of dropping out.

Requesters could also consider allowing workers to skip one or more sessions to provide additional flexibility. This is particularly relevant given that the presence of penalties does not appear to further motivate participation, according to workers (Figure~\ref{cap:paper_tsc2024-sec:results-subsec:rq1-analysis-fig:incentives_motivation}).

\subsubsection{\texttt{R3:} Workers Fear Performance Measurement}

Crowdsourcing platforms assess worker performance and quality using various metrics and indicators, such as the time elapsed between accepting a HIT and its successful submission, as well as the overall completion rate. These indicators can be used by task requesters to filter the pool of available workers, as was done in this study (Section~\ref{cap:paper_tsc2024-sec:exp-setup-subsec:task}).

Workers suggest that they may avoid participating in longitudinal studies due to concerns that doing so could increase the likelihood of being rejected after completing a session, which could negatively impact their completion rates and performance as measured by the platform. In other words, workers are apprehensive about performance measurement, particularly in the context of longitudinal studies (Section~\ref{cap:paper_tsc2024-subsec:qualitative-analysis-subsec:worker-commitment}).
One way to address this issue is by clearly disclosing and explaining the study's workflow, with particular attention to the rejection criteria. These criteria should be described in detail, along with the specific behaviors and conditions that may trigger them.

Crowdsourcing platforms measure worker performances and quality using various metrics and indicators, such as the time elapsed between accepting a given HIT and its successful submission and the overall completion rate. These indicators can be used by task requesters to filter the pool of available workers as, indeed, we ourselves have done (Section~\ref{cap:paper_tsc2024-sec:exp-setup-subsec:task}). 

Workers suggest that they might avoid participating in longitudinal studies because they somehow believe that this could increase the odds of being rejected at any time after a given session, once completed, thus impacting the completion rates and performance as measured by the platform. In other words, workers fear performance measurement, especially in the context of longitudinal studies (Section~\ref{cap:paper_tsc2024-subsec:qualitative-analysis-subsec:worker-commitment}). 
A way to address such an issue is by disclosing and clarifying the whole study's workflow, having a particular focus on the rejection criteria. They should be described accurately along with the behaviors and causes that may trigger them. 

\subsubsection{\texttt{R4:} Longitudinal Studies Boost Reliability And Trustworthiness}

Even though longitudinal studies may heighten concerns about performance indicators, task requesters should consider that workers perceive such studies as more reliable than other types of crowdsourcing-based studies. This perception is reflected in their responses regarding loyalty and commitment, as discussed in Section~\ref{cap:paper_tsc2024-subsec:qualitative-analysis-subsec:worker-commitment}.

This reliability stems from the fact that workers find longitudinal studies more structured, as the same tasks are repeated over time, leading to improved productivity. Additionally, they consider longitudinal studies advantageous because they eliminate the need to spend time searching for new tasks, as shown in Figure~\ref{cap:paper_tsc2024-sec:results-subsec:rq1-analysis-fig:benefits}.

Workers also indicate, in responses analyzed in Section~\ref{cap:paper_tsc2024-sec:results-subsec:rq1-analysis-p2-question:design-suggestions}, that a well-executed longitudinal study demonstrates researcher integrity, thereby increasing overall trustworthiness. To support this, task requesters should implement a well-documented task design that remains as consistent as possible across sessions, following a clear and logical sequence of events. As further illustrated in Figure~\ref{cap:paper_tsc2024-sec:results-subsec:rq1-analysis-fig:benefits}, several workers also suggest that planning intermediate payments can help enhance trust in the requester.

\subsubsection{\texttt{R5:} Worker Provenance Affects Their Availability}

Crowdsourcing platforms enable task requesters to recruit workers from around the world, including regions with inadequate network infrastructure. For example, on the \toloka platform, it is relatively common to find workers from CIS countries \cite{kubicek2009} (Commonwealth of Independent States), as noted by a worker.

Task requesters should carefully consider where to recruit workers, as their location can significantly impact availability, loyalty, and commitment. When asked about the suitability of the platform for longitudinal studies in Section~\ref{cap:paper_tsc2024-subsec:qualitative-analysis-subsec:platform-suitability}, a worker specifically mentioned that improvements are needed for such studies, particularly in scheduling sessions. Additionally, infrastructural limitations may further exacerbate platform-intrinsic issues.

\subsubsection{\texttt{R6:} Design Cross-Device Layouts And Avoid Requiring Additional Software}

Workers may use various devices to perform crowdsourcing tasks. For instance, the \prolific platform provides task requesters with a user interface control to explicitly allow the use of specific device classes. Additionally, a worker may begin a task on one device and switch to another at a later time. This is particularly relevant for longitudinal studies, which consist of multiple sessions spread over an arbitrary number of days.

Task requesters should therefore design and implement layouts that are as cross-platform as possible, ensuring compatibility with different devices. However, workers do not necessarily agree with being required to download additional software to complete a crowdsourcing task, as noted in Section~\ref{cap:paper_tsc2024-subsec:qualitative-analysis-subsec:worker-commitment} and Section~\ref{cap:paper_tsc2024-sec:results-subsec:rq1-analysis-p2-question:design-suggestions}. Whenever possible, task requesters should provide a single, preferably web-based, interface that allows workers to complete tasks seamlessly across devices.

\subsubsection{\texttt{R7:} Provide Partial Payments And Consider Bonuses}

The most significant incentives for increasing the availability of longitudinal studies on crowdsourcing platforms and motivating workers to participate and complete them are monetary, such as rewards and bonuses, as shown in Figure~\ref{cap:paper_tsc2024-sec:results-subsec:rq1-analysis-fig:incentives_motivation}.

While both terms refer to monetary compensation in a crowdsourcing setting, they differ in how they are provided. Typically, the reward is the payment issued upon task completion, whereas a bonus may be implicit or granted based on worker performance or other factors. In the context of crowdsourcing-based longitudinal studies, bonuses are often awarded after completing all sessions of a study or specific parts of it, as demonstrated by \citet{strickland2018feasibility}.

Task requesters should consider providing a reward after each individual session, along with one or more bonuses distributed throughout the study, to minimize worker drop-off. A partial reward could be a fixed amount or initially set at a lower value, increasing as the study progresses to encourage consistent participation. This approach may help reduce abandonment rates by further motivating workers while also allowing an incremental payment structure that helps control costs in the early stages of the study.

\subsubsection{\texttt{R8:} Consider Deploying Pilot And Training Versions}
\label{cap:paper_tsc2024-sec:results-subsec:rq2-recommendations-subsec:r8}

Piloting a task helps reduce worker attrition caused by errors and unexpected scenarios within its business logic, and longitudinal studies are no exception. In Section~\ref{cap:paper_tsc2024-sec:results-subsec:rq1-analysis-p2-question:design-suggestions}, workers suggest that using well-designed screeners can help requesters identify participants who fit the study’s requirements and are less likely to quit partway through.

Additionally, longitudinal studies may involve recruiting novice workers during later sessions, as demonstrated by \citet{roitero2021crowd}. Task requesters may consider deploying a lightweight training version of the task to familiarize first-time participants with the process and prepare them to complete the study as expected.

\subsection{\ref{cap:paper_tsc2024-sec:research-questions_3}: Best Practices For Crowdsourcing Platforms}
\label{cap:paper_tsc2024-sec:results-subsec:rq3-best-practices}

Longitudinal studies have been conducted on crowdsourcing platforms to some extent. However, the support provided by commercial platforms for such studies is not as straightforward as it may seem.

An analysis of workers' responses, combined with observations from deploying a crowdsourcing task, reveals that even simple objectives, such as tracking the overall progress of a study for both requesters and workers, are not easily achievable. Based on these findings, a set of \numpractices best practices has been identified, which crowdsourcing platform designers should adopt and prioritize to better support longitudinal studies (Section~\ref{cap:paper_tsc2024-sec:results-subsec:rq3-best-practices-subsec:bp1}--\ref{cap:paper_tsc2024-sec:results-subsec:rq3-best-practices-subsec:bp5}).

\subsubsection{\texttt{BP1:} Allow Requesters Sending Reminders To Workers}
\label{cap:paper_tsc2024-sec:results-subsec:rq3-best-practices-subsec:bp1}

One of the most pressing issues reported by workers is the need to be reminded of an upcoming session when committing to a longitudinal study (Section~\ref{cap:paper_tsc2024-subsec:qualitative-analysis-subsec:platform-suitability} and Section~\ref{cap:paper_tsc2024-sec:results-subsec:rq1-analysis-p2-question:design-suggestions}). For instance, a worker stated that they enjoyed participating because they had been reminded daily. Several workers also believe that longitudinal studies are not optimally supported in general, which could be part of the problem. Hence, the crowdsourcing platform should provide a way for task requesters to remind workers.

A solution could involve allowing automatic reminders to be scheduled. These reminders could be set to occur after each session or after a fixed period and should include a customizable message if needed. They could be delivered as notifications within the platform's user interface or as simple email messages.

\subsubsection{\texttt{BP2:} Report To Workers The Overall Progress}

Workers often express a desire to perceive and understand their progress within a longitudinal study (Section~\ref{cap:paper_tsc2024-sec:results-subsec:rq1-analysis-p2-question:design-suggestions}). This desire is further reinforced by the fact that some feel incentivized by their personal interest in the task, both in participating, as shown in Figure~\ref{cap:paper_tsc2024-sec:results-subsec:rq1-analysis-fig:incentives}, and in completing it (Figure~\ref{cap:paper_tsc2024-sec:results-subsec:rq1-analysis-fig:yes}).

Similar to reminding workers, enabling them to track their progress within a longitudinal study may seem like a straightforward requirement at first glance. However, it is difficult to achieve on the platforms considered, as they generally provide feedback only within a single crowdsourcing task (i.e., a single session of the overall study).

One way to provide feedback and build a sense of progress is to allow requesters to display the total number of sessions in the study in advance. Additionally, workers have reported enjoying longitudinal studies because they can observe changes in their answers over time. This could be another valuable piece of information to summarize and present as a performance indicator.

\subsubsection{\texttt{BP3:} Support More Advanced Worker Recruitment Strategies}
\label{cap:paper_tsc2024-sec:results-subsec:rq3-best-practices-subsec:bp3}

Chapter~\ref{cap:paper_pauc2021} described a longitudinal study in which workers were asked to fact-check statements related to the COVID-19 pandemic made by public figures, such as politicians. A distinctive aspect of the study was the republishing of a fixed set of HITs four times. Each time, workers who had previously participated were contacted and invited to repeat the fact-checking activity. Additionally, novice workers were recruited to compare their performance with that of returning participants.

Given that workers are willing to commit to a longitudinal study for approximately 22 days on average, as shown in Figure~\ref{cap:paper_tsc2024-sec:results-subsec:rq1-analysis-fig:daily_commitment}, it is reasonable to expect that many will drop out over time. This is further supported by the fact that study length is a major factor influencing participation decline (Figure~\ref{cap:paper_tsc2024-sec:results-subsec:rq1-analysis-fig:participation_decline}), even though most workers report having completed previous longitudinal studies (Figure~\ref{cap:paper_tsc2024-sec:results-subsec:rq1-analysis-fig:task_completion}). Indeed, the task abandonment ratio measured in the study described in Chapter~\ref{cap:paper_pauc2021} is 50\% \cite{han2019all}.

Crowdsourcing platforms should prioritize offering a straightforward method to facilitate worker recruitment based not only on demographic criteria but also on previous participation in the study. Additionally, they should provide a way to compensate for the reduced number of returning workers by allowing requesters to specify whether they want to recruit novice workers as well. As of today, \prolific partially addresses this by enabling requesters to save lists of worker groups, allowing them to select the same participants for new studies.

\subsubsection{\texttt{BP4:} Add Adequate User Interface Filters For The Workers}
\label{cap:paper_tsc2024-sec:results-subsec:rq3-best-practices-subsec:bp4}

When designing and publishing a study on a crowdsourcing platform, it is not possible to indicate that it will be conducted in a longitudinal fashion by publishing additional sessions over time. Given that Figure~\ref{cap:paper_tsc2024-sec:results-subsec:rq1-analysis-fig:benefits} shows that several workers believe longitudinal studies help them avoid spending time searching for new tasks and allow them to be more productive, platforms should provide workers with a user interface filter to distinguish longitudinal studies from standard tasks. Consequently, platforms should offer requesters the option to specify whether their studies will be longitudinal.

While adding appropriate user interface filters may seem like an obvious improvement, workers on every platform considered can currently only guess or rely on study descriptions provided by the requester to determine whether they are participating in a longitudinal study. Implementing this change would enhance worker awareness and facilitate participation, allowing task requesters to optimize the time required to recruit the necessary number of workers.

\subsubsection{\texttt{BP5:} Provide Support For Non-Desktop Devices}
\label{cap:paper_tsc2024-sec:results-subsec:rq3-best-practices-subsec:bp5}

Workers use a variety of devices to participate in crowdsourcing tasks of any kind. Among the platforms considered, only \prolific allows task requesters to specify the required device class (i.e., mobile, desktop, or tablet) for a given task, and workers can filter available tasks accordingly. Specifically, a worker reported participating in a longitudinal study that involved maintaining a log on an Android device while collecting certain health data (e.g., heartbeat, etc.). This study took place on \prolific and was mentioned in responses about platform suitability (Section~\ref{cap:paper_tsc2024-subsec:qualitative-analysis-subsec:platform-suitability}).

Crowdsourcing platforms should provide task requesters with a way to design layouts suitable for each device class. This could be achieved by offering a set of predefined, responsive user interface components, as done to some extent by MTurk with its Crowd HTML elements,\footnote{\url{https://docs.aws.amazon.com/AWSMechTurk/latest/AWSMturkAPI/ApiReference_HTMLCustomElementsArticle.html}} or by \toloka with its template builder. However, both approaches require considerable web development skills. \prolific, on the other hand, began moving in this direction in October 2023 by rolling out a survey builder, which currently supports designing simple polls with 1-5 questions.\footnote{\url{https://researcher-help.prolific.com/en/article/71c7b2}}

To further improve support for as many devices as possible, platforms could allow requesters to design different layouts for the same task, with a version optimized for each supported device class. Workers should then be able to filter and select studies compatible with their preferred device class, similar to the option to choose between longitudinal and standard studies. This best practice is general and applies beyond longitudinal study design.

\section{Summary}

\label{cap:paper_tsc2024-sec:discussion}

\begin{figure}[tpb]
    \centering
    \includegraphics[width=\linewidth]{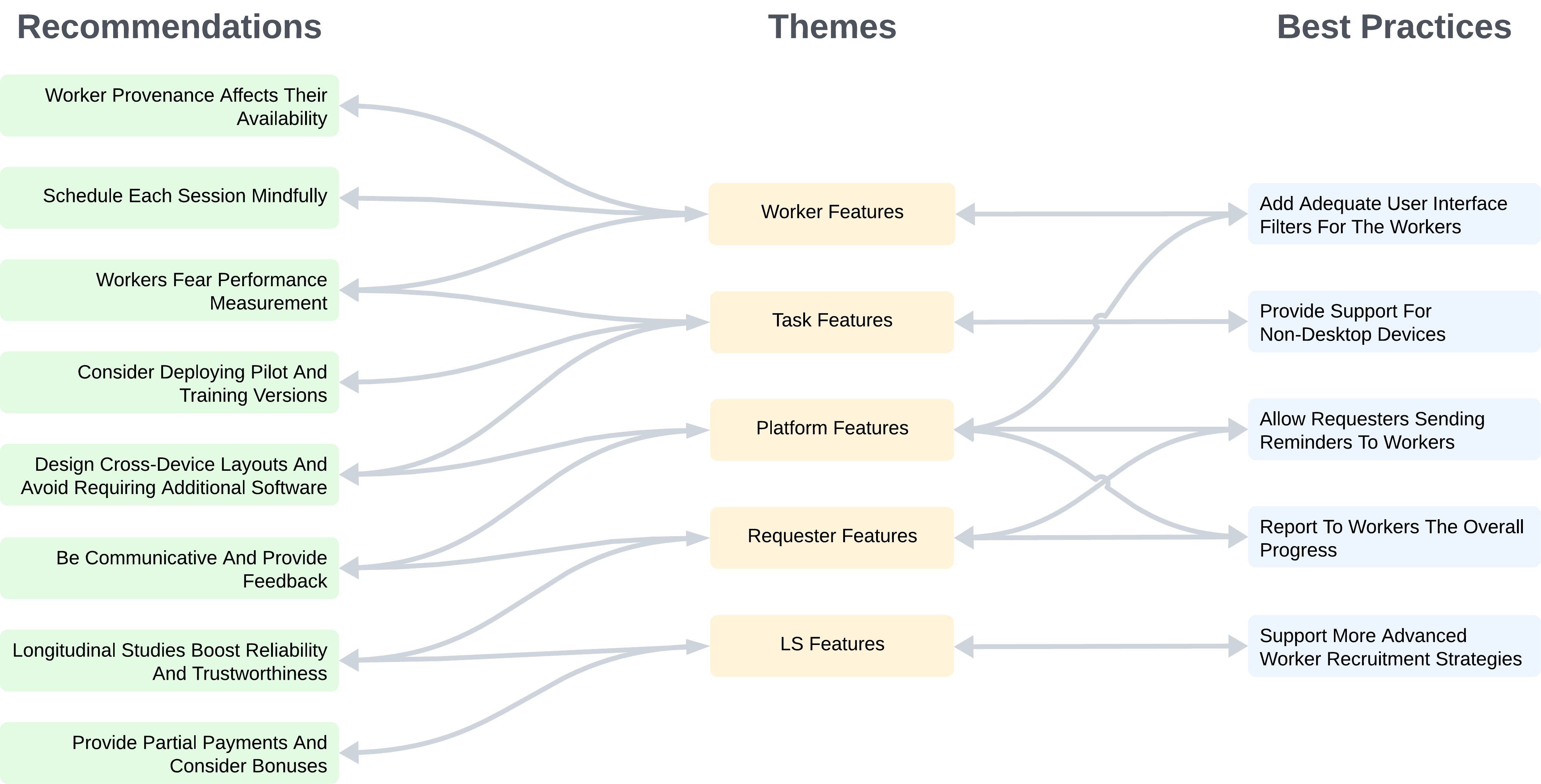}
    \caption{Summary of the barriers emerged from our analyses, along with \numrecommendations recommendations for researchers and practitioners, \numpractices best practices for crowdsourcing platforms, and their interconnections.}
    \label{cap:paper_tsc2024-sec:discussion-subsec:summary-fig:scheme}
\end{figure}

This chapter investigates the barriers encountered when conducting longitudinal tasks on crowdsourcing platforms, emphasizing the worker perspective. A large-scale survey across three major platforms examines the current perception, popularity, motivational factors, strengths, and weaknesses of longitudinal studies within these environments. Both quantitative and qualitative analyses were conducted to gain insights, with an inductive thematic analysis applied for the qualitative data. By integrating these findings with experiences as task requesters, an overview of the results and their interconnections is provided in Figure~\ref{cap:paper_tsc2024-sec:discussion-subsec:summary-fig:scheme}. 

Conducting successful crowdsourcing-based longitudinal studies remains notably difficult, with worker abandonment rates reportedly ranging from 50\% to 80\% \cite{roitero2021crowd, holden2013assessing, buhrmester2016amazon, shapiro2013using, MUN20221234, lanaj2014beginning}. Existing literature provides neither unified platform support nor established best practices. Nonetheless, the recommendations and best practices proposed in this chapter offer guidance for increasing the likelihood of success. Applying these guidelines can help overcome barriers to conducting longitudinal studies and leverage the benefits of crowdsourcing platforms. Moreover, implementing the suggested best practices would enhance the experience for both task requesters and workers.
The answers to the research questions can be summarized as follows.

\myparagraph{\ref{cap:paper_tsc2024-sec:research-questions_1}} Several barriers to longitudinal studies on crowdsourcing platforms are identified, categorized into five main themes, shown in the central part of Figure~\ref{cap:paper_tsc2024-sec:discussion-subsec:summary-fig:scheme}. Workers’ activities are influenced by their needs, expectations, motivations, and fears. Task characteristics, platform design, and requester actions can also impact study outcomes. Additional difficulties may stem from the longitudinal nature of these studies (Table~\ref{cap:paper_tsc2024-sec:qualitative-analysis-subsec:remarks-tab:theme-dist}). For instance, the study found that workers were more likely to engage in longitudinal tasks offering higher payments, improved communication, and clear progress tracking. However, challenges arose from the absence of effective quality control mechanisms and transparent communication channels between requesters and workers. 

\myparagraph{\ref{cap:paper_tsc2024-sec:research-questions_2}} A set of \numrecommendations recommendations is provided to assist researchers and practitioners in effectively designing and conducting longitudinal studies on commercial micro-task crowdsourcing platforms, summarized in the leftmost part of Figure~\ref{cap:paper_tsc2024-sec:discussion-subsec:summary-fig:scheme}.

\myparagraph{\ref{cap:paper_tsc2024-sec:research-questions_3}} A set of \numpractices best practices is proposed for platforms to support successful longitudinal studies conducted via crowdsourcing, outlined in the rightmost part of Figure~\ref{cap:paper_tsc2024-sec:discussion-subsec:summary-fig:scheme}.

\myparagraph{}  
The next chapter introduces the concept of a multidimensional notion of truthfulness, in which crowd workers are asked to assess the truthfulness of information items across seven distinct dimensions identified in the literature. The study is based on the same set of statements used in Chapter~\ref{cap:paper_sigir2020}.

\chapter{The Multidimensionality Of Truthfulness}

\label{cap:paper_ipm2021}

This chapter is based on the article published in the \lq\lq Information Processing \& Management\rq\rq{} journal~\cite{SOPRANO2021102710}. Section~\ref{cap:related_work-sec:crowdsourcing-truthfulness}, Section~\ref{cap:related_work-sec:multidimensionality}, Section~\ref{cap:related_work-sec:worker-bias}, Section~\ref{cap:related_work-sec:argument-mining}, and Section~\ref{cap:related_work-sec:afc-models} describe the relevant related work. Section~\ref{cap:paper_ipm2021-sec:research-questions} addresses the research questions, while Section~\ref{cap:paper_ipm2021-sec:exp-setup} describes the experimental setting. Section~\ref{cap:paper_ipm2021-sec:exp-setup-subsec:descriptive-statistics} provides an initial descriptive analysis, while Section~\ref{cap:paper_ipm2021-sec:results} presents the results obtained. Finally, Section~\ref{cap:paper_ipm2021-sec:discussion} summarizes the main findings and concludes the chapter.

\section{Research Questions}

\label{cap:paper_ipm2021-sec:research-questions}

Several studies have explored the use of crowdsourcing for scalable fact-checking~\cite{la2020crowdsourcing, roitero2020crowd, roitero2020covid}. Comparisons of truthfulness scales with different levels of granularity have shown that coarse-grained options (e.g., three levels) are more suitable for crowdsourced truthfulness judgments~\cite{la2020crowdsourcing}. However, uni-dimensional scales such as those used in Chapter~\ref{cap:paper_sigir2020} and Chapter~\ref{cap:paper_pauc2021} may be too simplistic to capture the full range of truthfulness nuances. This chapter investigates a \emph{multidimensional} approach to crowdsourcing truthfulness judgments, moving beyond a single true--false continuum. A crowdsourcing task was published on \mturk, targeting US-based workers, to evaluate the truthfulness of political statements. Unlike previous studies using a single multi-level scale, this task collects judgments along multiple dimensions: \correctness, \neutrality, \comprehensibility, \precision, \completeness, \speakertrustworthiness, and \informativeness.

A large-scale crowdsourcing experiment is conducted in which crowd workers are asked to evaluate political statements with the aim of identifying online misinformation. The same set of statements described in the experimental setup of Section~\ref{cap:paper_sigir2020-sec:exp-setup} is used. These statements have been fact-checked by experts from \politifact (Section~\ref{cap:dataset-sec:politifact}) and \abc (Section~\ref{cap:dataset-sec:abc}). Differently from the approaches presented in Chapter~\ref{cap:paper_sigir2020} and Chapter~\ref{cap:paper_pauc2021}, this study adopts a multidimensional notion of truthfulness. Independent judgments are collected for each dimension from every worker. In addition, workers are asked to provide an \overalltruthfulness judgment for each statement and to justify their answer by submitting a URL linking to the web page they used to verify the statement's truthfulness. The following research questions are investigated:

\begin{enumerate}[start=16, leftmargin=2.92em, label=RQ\arabic*]
\item \label{cap:paper_ipm2021-sec:research-questions_1} Are crowd workers able to reliably assess multiple dimensions of information truthfulness? How do their judgments correlate with expert judgments?
\item \label{cap:paper_ipm2021-sec:research-questions_2} Are all truthfulness dimensions independent, and thus necessary? Can some dimensions be derived from a combination of others? Is it possible to combine the individual dimensions in a way that improves agreement with expert judgments?
\item \label{cap:paper_ipm2021-sec:research-questions_3} How do workers behave when selecting labels for different truthfulness dimensions? Do their cognitive abilities have any influence?
\item \label{cap:paper_ipm2021-sec:research-questions_4} How meaningful and informative are the individual truthfulness dimensions?
\item \label{cap:paper_ipm2021-sec:research-questions_5} Can multidimensional judgments be used to accurately predict expert verdicts?
\end{enumerate}

\section{Experimental Setting}

\label{cap:paper_ipm2021-sec:exp-setup}

The experimental setting adopts the same 180 political statements ranging from 2007 to 2019 described in Section~\ref{cap:paper_sigir2020-sec:exp-setup}, and sampled from the \politifact and \abc datasets (Chapter~\ref{cap:dataset}). This design enables a direct comparison of results under a multidimensional scale and offers the research community two distinct sets of annotations for the same collection of statements.

Table~\ref{cap:paper_ipm2021-sec:exp-setup-tab:statements} presents a sample of the statements used, mirroring the structure of Table~\ref{cap:paper_sigir2020-sec:exp-setup-tab:statements}.

\begin{table}[tbp]
\centering
\caption{Examples of statements sampled from the \politifact and \abc datasets.}
\label{cap:paper_ipm2021-sec:exp-setup-tab:statements}
\begin{tabular}{p{2.5cm}p{2cm}p{4.2cm}p{2cm}p{0.9cm}}
\toprule
\textbf{Dataset} & \textbf{Label} & \textbf{Statement} & \textbf{Speaker} & \textbf{Year} \\
\midrule
\politifact & \politifacttrue & Washing your hands and covering your mouth when you cough makes a huge difference in reducing transmission of the flu. & Barack Obama & 2009 \\ 
\midrule
\abc & \abcpositive & Under this government, the tax-to-GDP ratio has, in the period we’ve been in office, [been] an average of 22.7 per cent. & Kevin Rudd & 2013 \\
\bottomrule
\end{tabular}
\end{table}

\subsection{The Seven Dimensions Of Truthfulness}

\label{cap:paper_ipm2021-sec:exp-setup-subsec:seven-dim}

The main difference between this experimental setting and the one described in Section~\ref{cap:paper_sigir2020-sec:exp-setup} is that, in addition to assessing the \overalltruthfulness of each statement, workers are asked to independently evaluate seven distinct dimensions of truthfulness. These dimensions, listed below as presented to the workers (each accompanied by an example in the actual task), are:
\begin{enumerate}
\item \correctness: the statement is expressed in an accurate way, as opposed to being incorrect and/or reporting mistaken information.
\item \neutrality: the statement is expressed in a neutral/objective way, as opposed to subjective/biased. 
\item \comprehensibility: the statement is comprehensible/understandable/readable as opposed to difficult to understand. 
\item \precision: the information provided in the statement is precise/specific, as opposed to vague.
\item \completeness: the information reported in the statement is complete as opposed to telling only a part of the story.
\item \speakertrustworthiness: The speaker is generally trustworthy/reliable as opposed to untrustworthy/unreliable/malicious.
\item \informativeness: the statement allows us to derive useful information as opposed to simply stating well known facts and/or tautologies.
\end{enumerate}
A detailed description of each dimension and the examples provided to the workers can be found in Appendix~\ref{cap:paper_ipm2021:-appendix:instructions}. 

The choice of dimensions is informed by prior research. In information systems literature, information quality and user satisfaction are two primary criteria for evaluating system success~\cite{10.1145/505248.506007}. These broad facets can be further divided into various characteristics. Since this work focuses on news truthfulness, particular attention is given to information quality attributes such as accuracy and precision. The \index{ISO 25012 Model}ISO 25012 Model~\cite{ISOmodel}, which draws from several related studies~\cite{10.1145/505248.506007, vanivcek2005software, wang1996beyond}, motivates the selection of dimensions like \correctness, \completeness, \precision, \comprehensibility, and \neutrality as descriptors of information quality. 

Additionally, two further dimensions, \speakertrustworthiness and \informativeness, are incorporated based on literature support. For example, \citet{jowett2018propaganda} emphasizes the impact of the speaker’s trustworthiness on statement judgments, and \citet{BARRY1998219} categorizes source reliability as a key relevance dimension. \citet{ceolin2016capturing} and \citet{INRA:2018} use \informativeness among other factors in crowdsourcing tasks assessing information quality. It is important to note that while the ISO model centers on data quality, this experimental setting evaluates the quality of information conveyed by such data. Therefore, the selected ISO dimensions are adapted and extended with additional, literature-motivated ones to suit the context.

More specifically, the same dimensions used by \citet{INRA:2018} are adopted in this experimental setting. \citeauthor{INRA:2018} conducted a crowdsourcing study to evaluate whether the crowd can reliably assess information quality compared to experts. Their work builds on the dimensions previously defined by \citet{ceolin2016capturing}, who performed user studies on web documents related to the vaccination debate. \citet{INRA:2018} slightly reformulated some dimension descriptions to better suit the crowdsourcing context. Both studies demonstrated that crowd workers and experts achieve a satisfactory level of external agreement using this set of dimensions. In summary, these seven dimensions are chosen because they are theoretically grounded and have proven effective at capturing information accuracy and relevance while allowing reliable crowd-expert comparisons.

\subsection{Crowdsourcing Task}

\label{cap:paper_ipm2021-sec:exp-setup-subsec:crowdsourcing-task}

The \mturk crowdsourcing platform was used to collect data. When a worker accepted a Human Intelligence Task (\index{HIT}HIT), they were shown an input token and a URL linking to an external server hosting the deployed web application (i.e., the actual task). The worker completed the assigned \index{HIT}HIT using this external application (Appendix~\ref{cap:paper_wsdm2022}). Upon successful completion, the worker was shown an output token, which they had to copy back to the platform’s page to receive payment after approval.

The task itself is as follows. First, a (mandatory) questionnaire is shown to the worker, to collect background information such as age and political views. Then, the worker needs to provide answers to three Cognitive Reflection Test (\index{Cognitive!Reflection Test}CRT) questions. These questionnaires are those already described in Section~\ref{cap:paper_sigir2020-sec:exp-setup-subsec:crowdsourcing-task}. The workers are then asked to assess 11 statements selected from \politifact (6 statements) and \abc (3 statements) dataset. Each \index{HIT}HIT contains a statement for each truthfulness label of the \politifact and \abc datasets, plus 2 special statements used for the purpose of quality checks. Each \index{HIT}HIT is built using a randomization process to avoid all possible sources of bias. In more detail, each crowd worker is first asked to provide the \overalltruthfulness of the statement and a \confidence level of the knowledge of the topic. Then, the worker had to provide the URL that they used as a source of information to assess the \overalltruthfulness. Such a URL had to be found using the customized search engine (Section~\ref{cap:paper_wsdm2022-sec:system-design-subsec:search-engine}) which allows to filter out \politifact and ABC websites from search results. Then, each worker is also asked to assess the seven different dimensions of truthfulness described in Section~\ref{cap:paper_ipm2021-sec:exp-setup-subsec:seven-dim}. Each judgment was expressed on the following \index{Likert}Likert scale \cite{likert}: \completelydisagree, \disagree, \neitheraord, \agree, \completelyagree. The quality checks concerning the selected URL, the gold questions, and the time spent on each statement described in Section~\ref{cap:paper_sigir2020-sec:exp-setup-subsec:crowdsourcing-task} are implemented also for this task.
The set of instructions shown to the workers and containing a detailed description of the assessment process is available in Appendix~\ref{cap:paper_ipm2021:-appendix:instructions}.

Each \index{HIT}HIT reward is set at 2 USD\$, covering a set of 11 judgments. This rate is based on the estimated time required to complete the task and the U.S. minimum wage of 7.25 USD\$ per hour. A total of 180 statements are used, as described in Section~\ref{cap:paper_sigir2020-sec:exp-setup}, with each statement evaluated by 10 distinct crowd workers. Consequently, 200 \index{HIT}HITs were published on \mturk, yielding 2200 judgments in total. The crowdsourcing task ran from June 1st to June 4th, 2020.

\section{Descriptive Analysis}

\label{cap:paper_ipm2021-sec:exp-setup-subsec:descriptive-statistics}

First, Section~\ref{cap:paper_ipm2021-sec:exp-setup-subsec:demographics} presents demographic information about the workers, followed by Section~\ref{cap:paper_ipm2021-sec:exp-setup-subsec:task-abandonment} which analyzes the task abandonment rate.

\subsection{Worker Demographics}

\label{cap:paper_ipm2021-sec:exp-setup-subsec:demographics}

A total of \num{200} crowd workers successfully completed the experiment. Using \mturk, recruitment was limited to U.S. residents, who provided personal information such as their home address upon registration. Nearly 49\% of workers (95/200) are aged between 26 and 35. The majority (52\%) hold a college or bachelor’s degree. Regarding pre-tax income, 22\% earn between \$50,000 and less than \$75,000, while 18\% earn between \$40,000 and less than \$50,000.

In terms of political views, 33\% identify as \liberal, 22\% as \moderate, and 16\% as \conservative. Most workers consider themselves \democratic (46\%), with 28\% identifying as \republican and 23\% as \independent. A majority (53\%) disagree with building a wall on the U.S. southern border, while 40\% agree. Finally, 85\% believe the government should increase environmental regulations to combat climate change, with only 9\% disagreeing. Overall, the sample is well balanced aside from a few categories and aligns with demographic distributions observed in previous tasks (see Section~\ref{cap:paper_sigir2020-sec:desc-stat-subsec:task-abandonment} and Section~\ref{cap:paper_pauc2021-sec:exp-setup-subsec:task-abandonment}).

\subsection{Task Abandonment}

\label{cap:paper_ipm2021-sec:exp-setup-subsec:task-abandonment}

To quantify task abandonment, the abandonment rate is measured following the definition provided by \citet{8873609}. Overall, 200 out of 681 workers (approximately 29\%) successfully completed the task, while 355 out of 681 workers (about 52\%) abandoned it—that is, voluntarily terminated the task before completion—and 126 out of 681 workers (around 18\%) failed by terminating due to repeated quality check failures. Additionally, 184 out of 651 workers (about 27\%) abandoned the task before actually starting it, that is, immediately after completing the initial questionnaire.

\begin{figure}[tbp]
  \centering
  \begin{subfigure}{.49\linewidth}
    \centering
    \includegraphics[width=\linewidth]{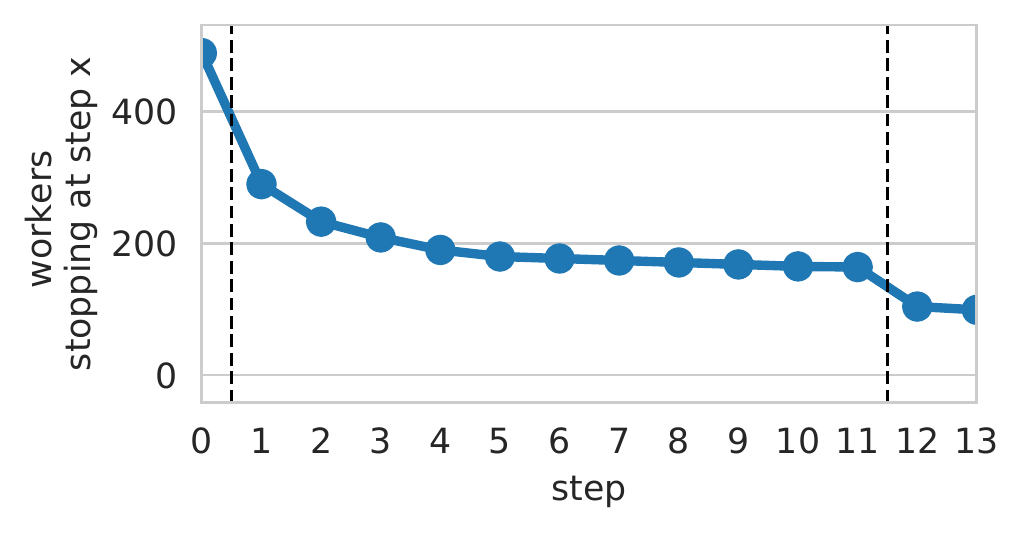}
    \caption{Questionnaire and first attempt.}    
    \label{cap:paper_ipm2021-sec:exp-setup-subsec:task-abandonment-fig:comparison_first}
  \end{subfigure}
  \begin{subfigure}{.49\linewidth}
    \centering
    \includegraphics[width=\linewidth]{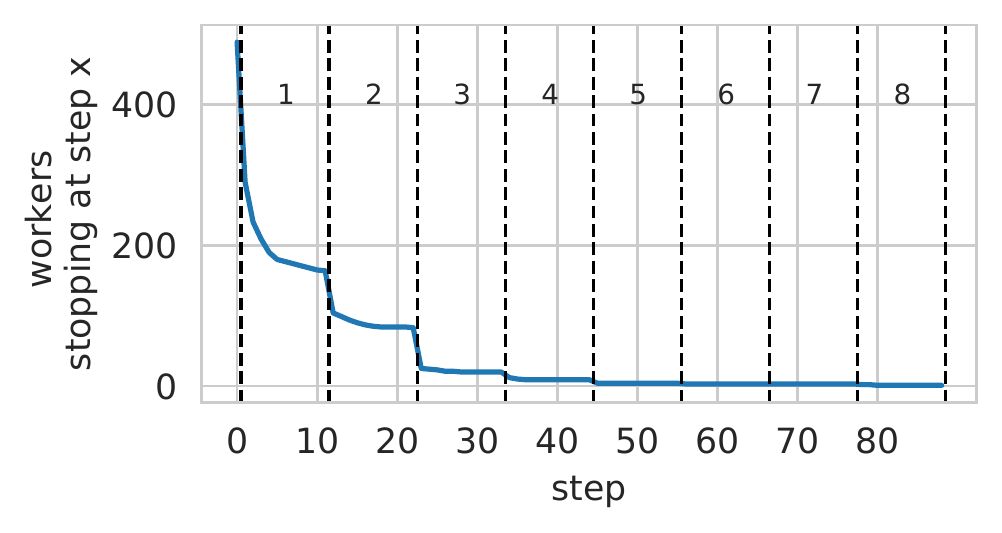}
    \caption{All attempts.}   
    \label{cap:paper_ipm2021-sec:exp-setup-subsec:task-abandonment-fig:comparison_all}
  \end{subfigure}
  \caption{Abandonment rate shown as the number of workers reaching each task step.}
  \label{cap:paper_ipm2021-sec:exp-setup-subsec:task-abandonment-fig:comparison}
\end{figure}

Figure~\ref{cap:paper_ipm2021-sec:exp-setup-subsec:task-abandonment-fig:comparison_first} shows the abandonment rate breakdown across task steps. A worker proceeds to the next step upon completing the judgment of a single statement; thus, a task is considered complete if the worker judges every statement within the current attempt. This definition does not assume task success. Step $0$ corresponds to the questionnaire, and each submission attempt spans 11 steps, as each \index{HIT}HIT consists of 11 statements. The abandonment rate decreases monotonically as the step number increases. Two notable drops occur, highlighted by dashed vertical lines in the figure. The first drop occurs after step $0$, reflecting many workers abandoning immediately after the questionnaire. The second drop happens between steps $11$ and $12$, indicating workers who completed but failed the first attempt, possibly due to boredom or frustration. Some workers made up to eight attempts before abandoning the task. These abandonment patterns align with those reported by \citet{8873609} and match the distributions described in Section~\ref{cap:paper_sigir2020-sec:desc-stat-subsec:task-abandonment} and Section~\ref{cap:paper_pauc2021-sec:exp-setup-subsec:descriptive-statistics}, providing an initial indication of data quality.

\section{Results}

\label{cap:paper_ipm2021-sec:results}

First, Section~\ref{cap:paper_ipm-sec:results-subsec:judgment-reliability} examines the reliability of multidimensional judgments. Next, Section~\ref{cap:paper_ipm-sec:results-subsec:truthfulness-dimensions} explores the relationships and independence among the truthfulness dimensions. Then, Section~\ref{cap:paper_ipm-sec:results-subsec:worker-behavior} investigates the use of worker behavior as a proxy for quality. Section~\ref{cap:paper_ipm-sec:results-subsec:informativeness} assesses the informativeness of the multidimensional judgments (distinct from the truthfulness dimension named \informativeness). Finally, Section~\ref{cap:paper_ipm-sec:results-subsec:machine-learning} analyzes a machine learning approach to evaluate the utility of multidimensional assessments and worker behavior in predicting expert judgments.

\subsection{\ref{cap:paper_ipm2021-sec:research-questions_1}: Reliability Of Multidimensional Judgments}
\label{cap:paper_ipm-sec:results-subsec:judgment-reliability}

First, Section~\ref{cap:paper_ipm-sec:results-subsec:judgment-reliability-subsec:judgment-dist} examines the distributions of individual and aggregated judgments provided by crowd workers. Section~\ref{cap:paper_ipm2021-sec:results-subsec:judgment-reliability-subsec:agreement-external} analyzes external agreement with expert labels, while Section~\ref{cap:paper_ipm-sec:results-subsec:judgment-reliability-subsec:agreement-internal} focuses on internal agreement among workers. Section~\ref{cap:paper_ipm-sec:results-subsec:judgment-reliability-subsec:behavioral} investigates worker behavior when assessing each truthfulness dimension. Finally, Section~\ref{cap:paper_ipm-sec:results-subsec:judgment-reliability-subsec:summary} summarizes the key findings regarding judgment reliability.

\subsubsection{Crowdsourced Judgments Distributions}

\label{cap:paper_ipm-sec:results-subsec:judgment-reliability-subsec:judgment-dist}

\begin{figure}[tbp]
  \centering
  \begin{subfigure}{\linewidth}
\centering
  \includegraphics[width=\linewidth]{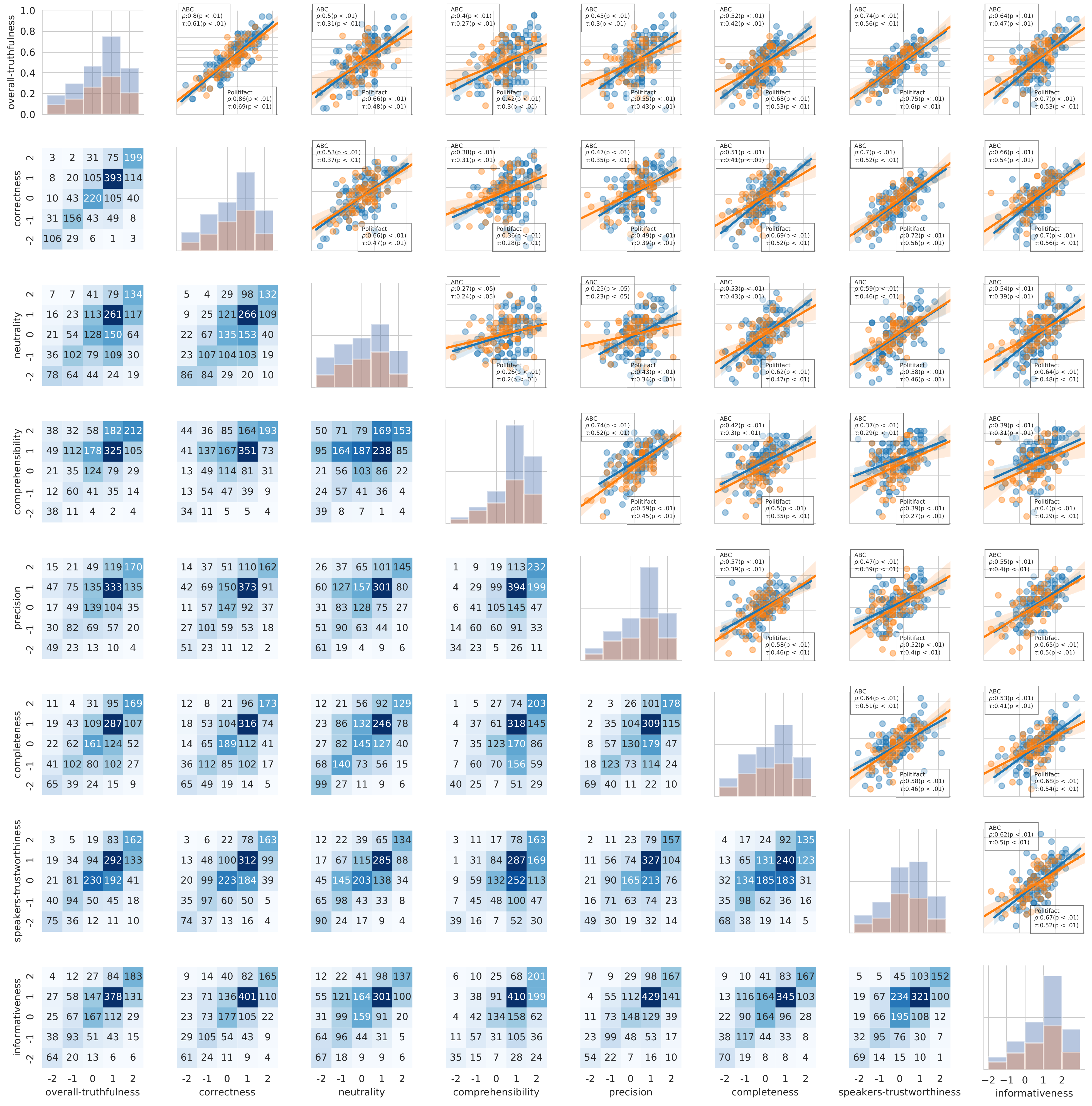}
\end{subfigure}
  \centering
  \begin{subfigure}{\linewidth}
\centering
  \includegraphics[width=\linewidth]{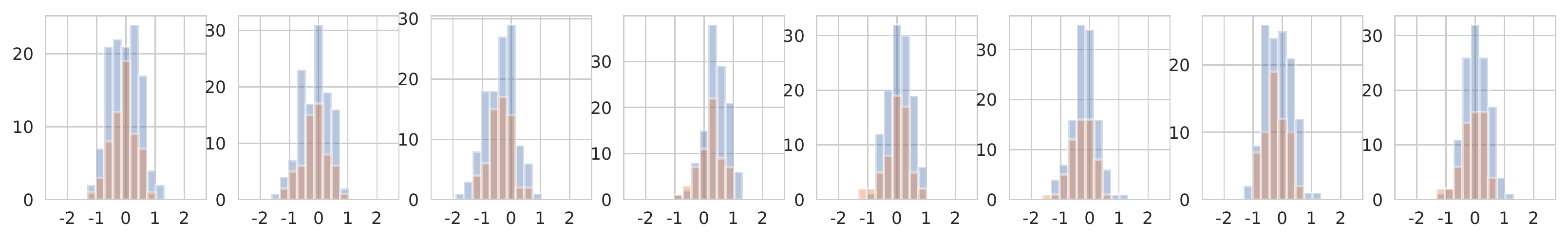}
\end{subfigure}
\caption{Correlation between dimensions: individual judgments are shown in the lower triangle and on the diagonal; aggregated judgments are shown in the upper triangle; the final row shows the distribution of aggregated judgments. Results are broken down by \politifact (in blue) and \abc (in orange). Best viewed on screen with zoom.}
\label{cap:paper_ipm2021-sec:results-subsec:judgment-reliability-subsec:judgment-dist-fig:correlation-between-dimensions}
\end{figure}

Figure~\ref{cap:paper_ipm2021-sec:results-subsec:judgment-reliability-subsec:judgment-dist-fig:correlation-between-dimensions} analyzes the correlation between the different truthfulness dimensions. The heatmaps in the lower triangular matrix display individual judgments collected for each dimension pair, for a total of 28 heatmaps. In each heatmap, cells indicate the number of times workers gave the same rating across the two dimensions being compared.

The histograms on the diagonal show the distributions of individual judgments for \politifact (blue) and \abc (orange) for each dimension. Note that only half as many judgments were collected for \abc as for \politifact. Each distribution is skewed to the right—toward \agree and \completelyagree (as shown in the diagonal and bottom plots), indicating that workers tend to agree with the statements, or at least avoid expressing strong disagreement. Given that the set of statements is balanced in terms of ground truth labels (see Section~\ref{cap:paper_ipm2021-sec:exp-setup-subsec:crowdsourcing-task}), this suggests that workers tend to agree even with false statements. This behavior may be influenced by the response scale used, which differs from that in the original task design (Section~\ref{cap:paper_sigir2020-sec:exp-setup-subsec:crowdsourcing-task}).

Individual judgments are then aggregated using the arithmetic mean, as this method has been shown to yield more reliable results in prior work~\cite{la2020crowdsourcing, RSDM:2018} and in Section~\ref{cap:paper_sigir2020-sec:results-subsec:scale-adequacy}. The upper triangular matrix displays scatterplots representing correlations between aggregated judgments for each pair of dimensions, with one point per statement. Blue and orange denote \politifact and \abc, respectively. Finally, the histograms along the bottom row show the distributions of the aggregated judgments for each dimension. These distributions are approximately bell-shaped and slightly skewed to the right.

Overall, the correlation values shown in Figure~\ref{cap:paper_ipm2021-sec:results-subsec:judgment-reliability-subsec:judgment-dist-fig:correlation-between-dimensions} for both individual judgments (heatmaps in the lower triangular matrix) and aggregated judgments (scatterplots in the upper triangular matrix) are consistently positive, as expected given that all seven dimensions share a positive connotation. In some cases, the correlations are notably strong (e.g., \index{$\uprho$}$\uprho = 0.86$ between aggregated \correctness and \overalltruthfulness for \politifact statements), suggesting a close relationship between those dimensions.

However, other correlations are much weaker (e.g., \index{$\uptau$}$\uptau = 0.24$ and $0.2$ for \neutrality and \comprehensibility), indicating a greater degree of independence between certain dimensions. In summary, correlation strength varies across pairs of dimensions, as reflected in both the upper and lower triangles of the matrix.

\subsubsection{External Agreement}

\label{cap:paper_ipm2021-sec:results-subsec:judgment-reliability-subsec:agreement-external}

\begin{figure}[tbp]
  \centering
  \begin{subfigure}{0.49\linewidth}
    \centering
    \includegraphics[width=.95\linewidth]{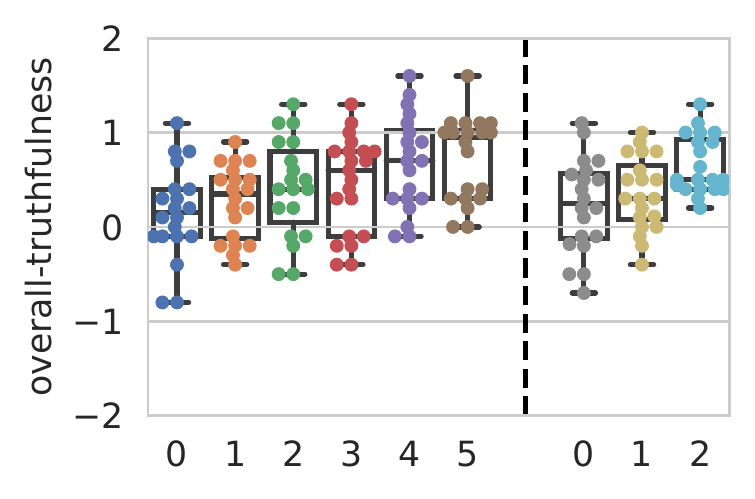}
    \caption{\overalltruthfulness.}
    \label{cap:paper_ipm2021-sec:results-subsec:judgment-reliability-subsec:agreement-external-fig:scale-comparison_overall}
  \end{subfigure}
  \hfill
  \begin{subfigure}{0.49\linewidth}
    \centering
    \includegraphics[width=.95\linewidth]{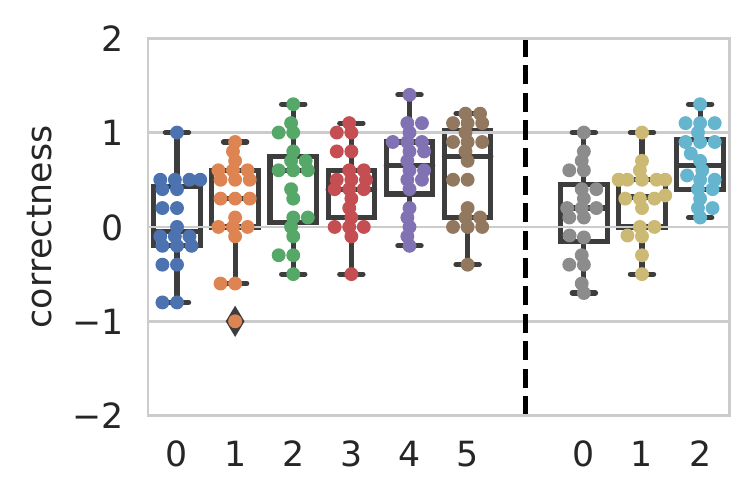}
    \caption{\correctness.}
    \label{cap:paper_ipm2021-sec:results-subsec:judgment-reliability-subsec:agreement-external-fig:scale-comparison_correctness}
  \end{subfigure}

  \vspace{1em} 

  \begin{subfigure}{0.49\linewidth}
    \centering
    \includegraphics[width=.95\linewidth]{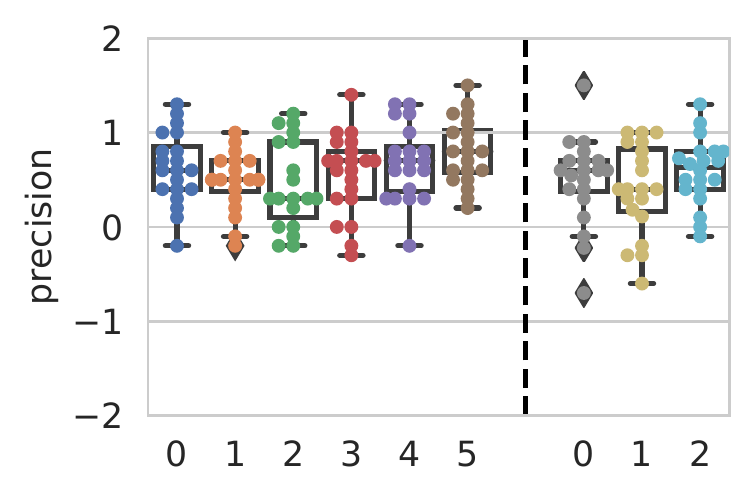}
    \caption{\precision.}
    \label{cap:paper_ipm2021-sec:results-subsec:judgment-reliability-subsec:agreement-external-fig:scale-comparison_precision}
  \end{subfigure}

  \caption{Crowd judgments aggregated by mean for three dimensions: \overalltruthfulness and \correctness (top row), and \precision (bottom row), broken down by \politifact and \abc labels. The plot for \overalltruthfulness allows comparison with the ground truth, while the others show orthogonal dimensions.}
  \label{cap:paper_ipm2021-sec:results-subsec:judgment-reliability-subsec:agreement-external-fig:scale-comparison}
\end{figure}

Figure~\ref{cap:paper_ipm2021-sec:results-subsec:judgment-reliability-subsec:agreement-external-fig:scale-comparison} presents a comparison between aggregated worker scores and expert scores. The analysis focuses on three dimensions: \overalltruthfulness, \correctness, and \precision. Before interpreting the charts, a few clarifications are necessary.

First, expert judgments are available only for the \overalltruthfulness dimension. Consequently, the values reported for \correctness and \precision reflect workers’ perceptions rather than alignment with expert assessments. These scores are broken down by the \politifact and \abc categories. While \overalltruthfulness is expected to correlate with expert judgments, \precision is designed to capture orthogonal and independent aspects of the information. This difference is visible in the distinct trends of the median worker scores shown in the figure.
Second, the judgment scales used by workers and experts for \overalltruthfulness differ. Crowd workers rated truthfulness using a five-level \index{Likert}Likert scale, whereas experts employed either a six-level ordinal scale (for \politifact) or a three-level scale (for \abc). These scales vary not only in the number of levels but also in their psychological interpretation, making direct comparisons non-trivial.
That being said, Figure~\ref{cap:paper_ipm2021-sec:results-subsec:judgment-reliability-subsec:agreement-external-fig:scale-comparison} can now be analyzed. The ground truth labels are on the horizontal axis (\politifact on the left, \abc on the right), and the aggregated crowd judgments are on the vertical axis. Each dot represents a statement, and boxplots display the distribution of aggregated scores for each ground truth level, highlighting medians and quantiles.

In Figure~\ref{cap:paper_ipm2021-sec:results-subsec:judgment-reliability-subsec:agreement-external-fig:scale-comparison_overall}, the median scores for \overalltruthfulness clearly increase with increasing ground-truth truthfulness levels. This trend is not as evident for the other two dimensions, \correctness and \precision, which are not directly tied to the expert ground truth. The increasing trend for \overalltruthfulness indicates a positive alignment between crowd and expert judgments, despite the use of different scales—five-point Likert for workers vs. six- or three-point ordinal scales for experts. This result is comparable to what is shown in Figure~\ref{cap:paper_sigir2020-sec:results-subsec:agreement-external-fig:scale-comparison-ground}, which depicts a similar relationship between crowd and expert scores under three different scale conditions. No major qualitative differences are observed, reinforcing the robustness of the crowd’s performance. Thus, the crowd workers in this task (Section~\ref{cap:paper_ipm2021-sec:exp-setup-subsec:crowdsourcing-task}) demonstrated judgment quality similar to that of workers in the prior task (Section~\ref{cap:paper_sigir2020-sec:exp-setup-subsec:crowdsourcing-task}).

Furthermore, the plots for \correctness and \precision exhibit different patterns, suggesting that these dimensions capture information orthogonal to \overalltruthfulness. Other dimensions (not shown) follow trends similar to either \precision or \overalltruthfulness. Since expert judgments are only available for \overalltruthfulness, it is not meaningful to directly compare the other dimensions to expert labels. These additional dimensions may capture complementary facets of truthfulness (e.g., \precision may be high even for a false but specific statement). Future work could explore combining these dimensions to construct improved truthfulness metrics.

Finally, to assess perceived disagreement between crowd and expert judgments on \overalltruthfulness, we compute how often the aggregated Likert score falls exactly between two categories (i.e., values of the form x.5 for integer x between 0 and 4). This occurs in 20.5\% of statements. Comparing with results from Section~\ref{cap:paper_sigir2020-sec:exp-setup-subsec:scales}, the corresponding percentages are 19.4\%, 18.3\%, and 23.9\% for the three-, six-, and one-hundred-level scales, respectively. This suggests that perceived disagreement is not primarily caused by the choice of scale but rather by other factors.

\begin{figure}[tbp]
  \centering
  \begin{subfigure}{.32\linewidth}
    \centering
    \includegraphics[width=.8\linewidth]{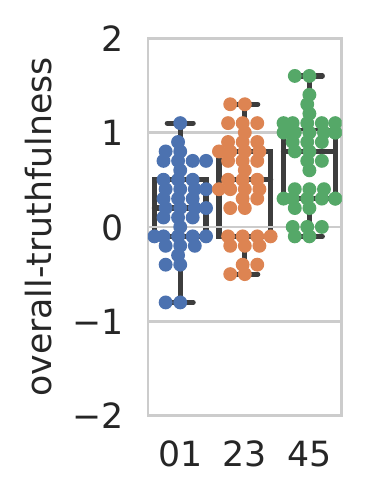}
    \caption{\overalltruthfulness.}
    \label{cap:paper_ipm2021-sec:results-subsec:judgment-reliability-subsec:agreement-external-fig:scale-comparison-bin3_overall}
  \end{subfigure}
  \hfill
  \begin{subfigure}{.32\linewidth}
    \centering
    \includegraphics[width=.8\linewidth]{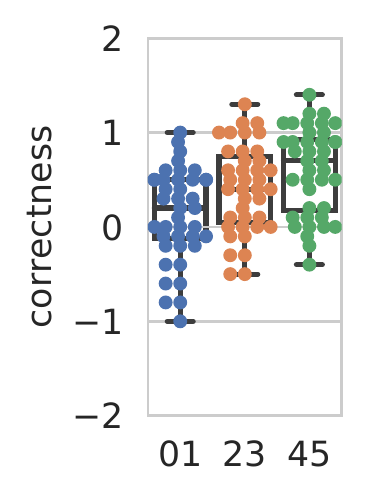}
    \caption{\correctness.}
    \label{cap:paper_ipm2021-sec:results-subsec:judgment-reliability-subsec:agreement-external-fig:scale-comparison-bin3_correctness}
  \end{subfigure}
  \hfill
  \begin{subfigure}{.32\linewidth}
    \centering
    \includegraphics[width=.8\linewidth]{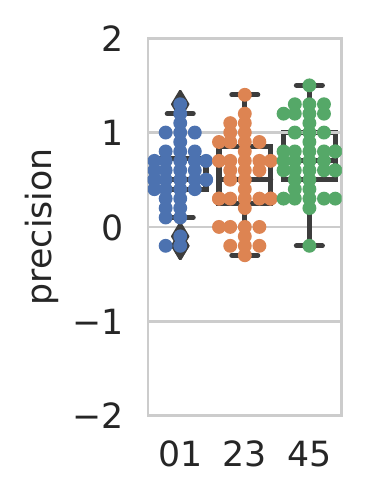}
    \caption{\precision.}
    \label{cap:paper_ipm2021-sec:results-subsec:judgment-reliability-subsec:agreement-external-fig:scale-comparison-bin3_precision}
  \end{subfigure}

  \begin{subfigure}{.32\linewidth}
    \centering
    \includegraphics[width=.8\linewidth]{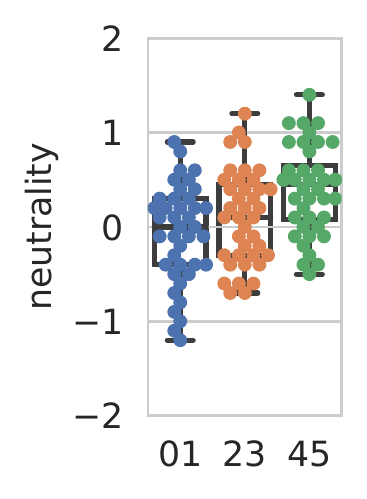}
    \caption{\neutrality.}
    \label{cap:paper_ipm2021-sec:results-subsec:judgment-reliability-subsec:agreement-external-fig:scale-comparison-bin3_neutrality}
  \end{subfigure}
  \begin{subfigure}{.32\linewidth}
    \centering
    \includegraphics[width=.8\linewidth]{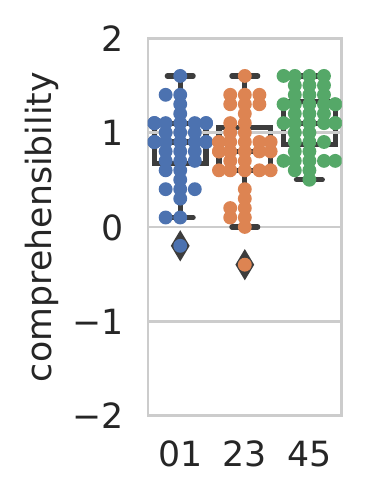}
    \caption{\comprehensibility.}
    \label{cap:paper_ipm2021-sec:results-subsec:judgment-reliability-subsec:agreement-external-fig:scale-comparison-bin3_comprehensibility}
  \end{subfigure}

  \caption{Correlation with the ground truth of \overalltruthfulness and a sample of the other dimensions. \politifact has been grouped into 3 bins. Mean used as aggregation function. Compare to Figure~\ref{cap:paper_ipm2021-sec:results-subsec:judgment-reliability-subsec:agreement-external-fig:scale-comparison}.}
  \label{cap:paper_ipm2021-sec:results-subsec:judgment-reliability-subsec:agreement-external-fig:scale-comparison-bin3}
\end{figure}

To assess whether the agreement between experts and crowd workers improves with a coarser-grained scale, the ground truth labels are grouped, following the same binning strategy described in Section~\ref{cap:paper_sigir2020-sec:results-subsec:scale-adequacy-subsec:merge}. Figure~\ref{cap:paper_ipm2021-sec:results-subsec:judgment-reliability-subsec:agreement-external-fig:scale-comparison-bin3} reports the correlation between \overalltruthfulness and the expert ground truth after merging the \politifact categories into three bins, using the mean as aggregation function. This binning approach reveals more clearly the expected trend of increasing median values and slightly improves some correlation coefficients compared to Figure~\ref{cap:paper_ipm2021-sec:results-subsec:judgment-reliability-subsec:agreement-external-fig:scale-comparison}. The improvement is most evident for \overalltruthfulness, but consistent patterns are also observed across the other dimensions shown (five in total). These results confirm the findings reported in Section~\ref{cap:paper_sigir2020-sec:results-subsec:scale-adequacy-subsec:merge}.

\subsubsection{Internal Agreement}

\label{cap:paper_ipm-sec:results-subsec:judgment-reliability-subsec:agreement-internal} 

Krippendorff's \index{$\upalpha$}$\upalpha$ \cite{krippendorff2011computing} metric is used to measure internal agreement, both across different ground truth levels and at the unit level, following the methodology in Section~\ref{cap:paper_sigir2020-sec:results-subsec:agreement-internal} and Section~\ref{cap:paper_pauc2021-sec:results-subsec:crowd-accuracy-subsec:int-agreement}. 
The use of Krippendorff’s $\upalpha$ is supported by prior work \cite{la2020crowdsourcing} and by theoretical considerations: alternative agreement metrics are not suitable for this setting. For instance, Cohen's $\upkappa$ \cite{cohen2013statistical} \index{$\upkappa$} is applicable only to pairs of assessors, and Fleiss' $\upkappa$ \cite{fleiss1973equivalence}, a generalization of Cohen's metric to multiple assessors, assumes categorical (nominal) ratings. These limitations make them inappropriate for scenarios involving multiple annotators (e.g., 10 workers) and ordinal scales, as in the current study. For this reason, Krippendorff's $\upalpha$ is preferred, as it supports ordinal data and multiple raters. Additional considerations on agreement metrics can be found in \cite{checco2017let, fleiss1973equivalence, kvaalseth1989note}.

The results show an overall low level of agreement. The $\upalpha$ values for the different conditions are summarized below:
\begin{itemize}[label=--]
  \item In the range $[.02, .08]$ when computed across all statements.
  \item In the range $[-0.02, 0.1]$ when computed within each of the three \abc categories.
  \item In the range $[-0.02, 0.1]$ (mean: $0.03$) across \politifact categories, with two exceptions:
  \begin{itemize}[label=--]
    \item \politifactmostlyfalse: range $[-0.02, 0.14]$ (mean: $0.09$).
    \item \politifacthalftrue: range $[-0.02, 0.14]$ (mean: $0.05$).
  \end{itemize}
\end{itemize}
It is well known that Krippendorff’s $\upalpha$ is sensitive to both the amount of data and the scale used for evaluation \cite{checco2017let}. Since both factors are held constant in this experiment, the observed patterns may indicate that workers tend to agree more when judging statements that fall in the middle of the truthfulness spectrum.

\subsubsection{Behavioral Data}

\label{cap:paper_ipm-sec:results-subsec:judgment-reliability-subsec:behavioral} 

\begin{figure}[tbp]
  \centering
  \includegraphics[width=.6\linewidth]{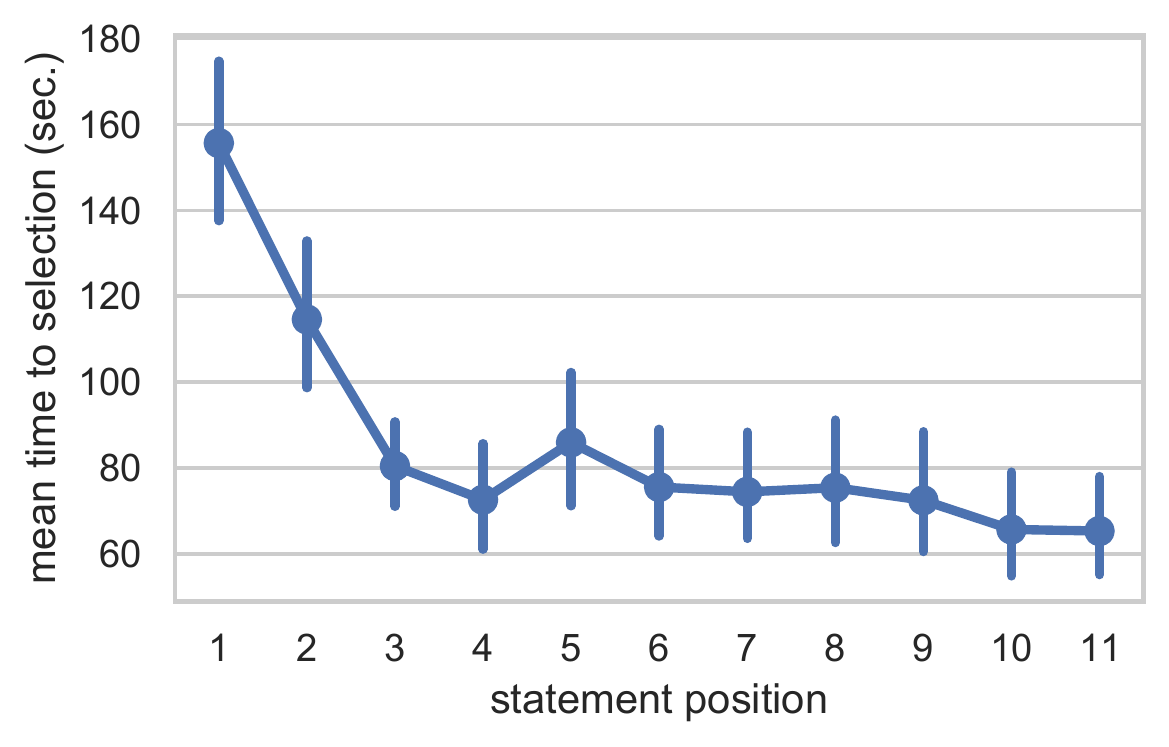}
  \caption{Average time (in seconds) spent by workers to judge the \overalltruthfulness for each statement position.}
  \label{cap:paper_ipm-sec:results-subsec:judgment-reliability-subsec:behavioral-fig:dimensions-values-changes}
\end{figure}

Concerning the analysis of workers' behavior while judging each truthfulness dimension, Figure~\ref{cap:paper_ipm-sec:results-subsec:judgment-reliability-subsec:behavioral-fig:dimensions-values-changes} shows the average time spent by each worker to select a value for \overalltruthfulness at each statement position. The results indicate a clear learning effect: the average time spent decreases as the statement position increases. 

To support this observation, statistical tests are conducted to assess the significance of the differences in time values across statement positions. The differences are statistically significant at the \index{$p$}$p<.01$ level when comparing positions $1$ and $2$ to any other position. For positions $3$ and $4$, statistically significant differences (\index{$p$}$p<.05$) are found only in comparison to the first two and the last two positions. These findings confirm the presence of a learning effect during the initial part of the task, particularly between positions $1$ and $4$. After the fourth statement, the evaluation time stabilizes, suggesting that workers assess the remaining statements with a consistent pace.

Workers spend most of the time assessing \overalltruthfulness because they are also required to provide a URL as justification for their choice. In contrast, workers spend considerably less time on the other dimensions, and no clear time-based trends are visible. This likely indicates that workers reflect on the values for these dimensions while evaluating \overalltruthfulness. Indeed, the time spent on each of the other dimensions is much lower than the average of 85 seconds taken for \overalltruthfulness.

\begin{table}[tbp]
\centering
\caption{Average time (in seconds) spent to assess each dimension, excluding \overalltruthfulness.}
\label{tab:dimension-times}
\begin{tabular}{@{}lc@{}}
\toprule
\textbf{Dimension} & \textbf{Average Time (s)} \\
\midrule
\confidence & 1.7 \\
\correctness & 3.0 \\
\neutrality & 4.1 \\
\comprehensibility & 5.0 \\
\precision & 6.2 \\
\completeness & 7.1 \\
\speakertrustworthiness & 8.3 \\
\informativeness & 9.4 \\
\bottomrule
\end{tabular}
\end{table}

\subsubsection{Summary}

\label{cap:paper_ipm-sec:results-subsec:judgment-reliability-subsec:summary}

The analysis of judgment reliability highlights several important findings. Workers tend to agree with statements more often than they disagree, even when the statements are false. This trend is evident despite the balanced nature of the dataset and suggests a systematic bias toward agreement, as shown by the skewed distributions in Figure~\ref{cap:paper_ipm2021-sec:results-subsec:judgment-reliability-subsec:judgment-dist-fig:correlation-between-dimensions}. 

Moreover, a strong correlation emerges between the crowd’s aggregated judgments and expert truthfulness levels, particularly for the \overalltruthfulness dimension. This pattern holds across different levels of granularity and remains consistent when the expert ground truth is binned, as shown in Figure~\ref{cap:paper_ipm2021-sec:results-subsec:judgment-reliability-subsec:agreement-external-fig:scale-comparison} and Figure~\ref{cap:paper_ipm2021-sec:results-subsec:judgment-reliability-subsec:agreement-external-fig:scale-comparison-bin3}. 
Internal agreement analyses further reveal that workers are more consistent when evaluating statements located in the middle of the truthfulness scale, suggesting that moderate statements are less prone to disagreement. 

Finally, behavioral analysis indicates a learning effect: workers initially spend more time judging \overalltruthfulness, but this time decreases with each additional statement and stabilizes after a few iterations, as shown in Figure~\ref{cap:paper_ipm-sec:results-subsec:judgment-reliability-subsec:behavioral-fig:dimensions-values-changes}. This indicates an adaptation process in which workers become more efficient as they get used to the task.

\subsection{\ref{cap:paper_ipm2021-sec:research-questions_2}: Independence Of The Dimensions}

\label{cap:paper_ipm-sec:results-subsec:truthfulness-dimensions}

The results reported so far indicate that the various dimensions, including \overalltruthfulness, are correlated to some extent. However, it is important to assess whether these dimensions capture distinct aspects of truthfulness or whether some could be derived from others. Figure~\ref{cap:paper_ipm2021-sec:results-subsec:judgment-reliability-subsec:judgment-dist-fig:correlation-between-dimensions} shows both higher and lower correlations among dimensions. The heatmaps in the bottom left (non-aggregated judgments) reveal particularly strong correlations between \correctness and both \overalltruthfulness and \speakertrustworthiness. This pattern is confirmed for aggregated judgments, shown in the upper right, where Pearson’s $\uprho$ and Kendall’s $\uptau$ values are also reported.

Focusing specifically on the correlation between \overalltruthfulness and each of the seven dimensions (i.e., the first row and column of the correlation matrix), it is evident that \neutrality, \comprehensibility, and \precision exhibit relatively low correlations. For aggregated judgments over \politifact statements, the Kendall's $\uptau$ values for these dimensions are 0.48, 0.30, and 0.43 respectively, while for \abc statements the values are 0.31, 0.27, and 0.30. Slightly higher correlations are observed for \completeness, \speakertrustworthiness, and \informativeness, with $\uptau$ values of 0.53, 0.60, and 0.53 for \politifact, and 0.42, 0.56, and 0.40 for \abc, respectively. Among all dimensions, \correctness shows the strongest association with \overalltruthfulness.
In summary, although some correlations exist, each dimension appears to capture a distinct aspect of truthfulness, different from what is measured by \overalltruthfulness. This conclusion holds when analyzing both individual judgments and aggregated assessments across workers.

Reconsidering Figure~\ref{cap:paper_ipm2021-sec:results-subsec:judgment-reliability-subsec:agreement-external-fig:scale-comparison} and Figure~\ref{cap:paper_ipm2021-sec:results-subsec:judgment-reliability-subsec:agreement-external-fig:scale-comparison-bin3} provides further confirmation of the partial independence of the evaluated dimensions. Although similar overall trends can be observed, distinct differences emerge across dimensions. To further explore this aspect, an \index{ANOVA}ANOVA analysis is conducted to assess the extent to which \overalltruthfulness can be explained by the other dimensions. The $\upomega^2$ \index{$\upomega^2$} index \cite{olejnikAnova} is used to quantify the effect size of each dimension after fitting the ANOVA model.
The results indicate that \overalltruthfulness is mostly influenced by the following dimensions, with a difference in magnitude of impact:
\begin{enumerate}
  \item \correctness ($\upomega^2 = 0.228$)
  \item \speakertrustworthiness ($\upomega^2 = 0.019$)
  \item \informativeness ($\upomega^2 = 0.017$)
\end{enumerate}
In contrast, the remaining dimensions contribute very little: \comprehensibility ($\upomega^2 = 0.008$), \completeness ($\upomega^2 = 0.001$), \precision ($\upomega^2 = 0$), and \confidence ($\upomega^2 = 0$).

A second ANOVA model is used to examine interactions between dimensions. The results show that all interactions have limited explanatory power ($\upomega^2 \leq 0.04$), which supports the interpretation that the dimensions are largely orthogonal and capture different aspects of a statement’s truthfulness. However, the presence of interaction effects also suggests that workers take all dimensions into account when forming their judgments. This implies that none of the dimensions is redundant. Investigating whether additional dimensions could further enrich the evaluation framework is left for future work.

To further investigate the relationships and independence among dimensions, both individual and aggregated judgments are used to construct a statement $\times$ judgment matrix. A Principal Components Analysis (PCA) \index{Principal Components Analysis} is then performed on this matrix to identify orthogonal bases that explain the maximum variance in the data. In the resulting transformed space, the two principal components that account for the majority of the variance are considered.

\begin{figure}[tbp]
  \centering
  \begin{subfigure}{.49\linewidth}
    \centering
    \includegraphics[width=.95\linewidth]{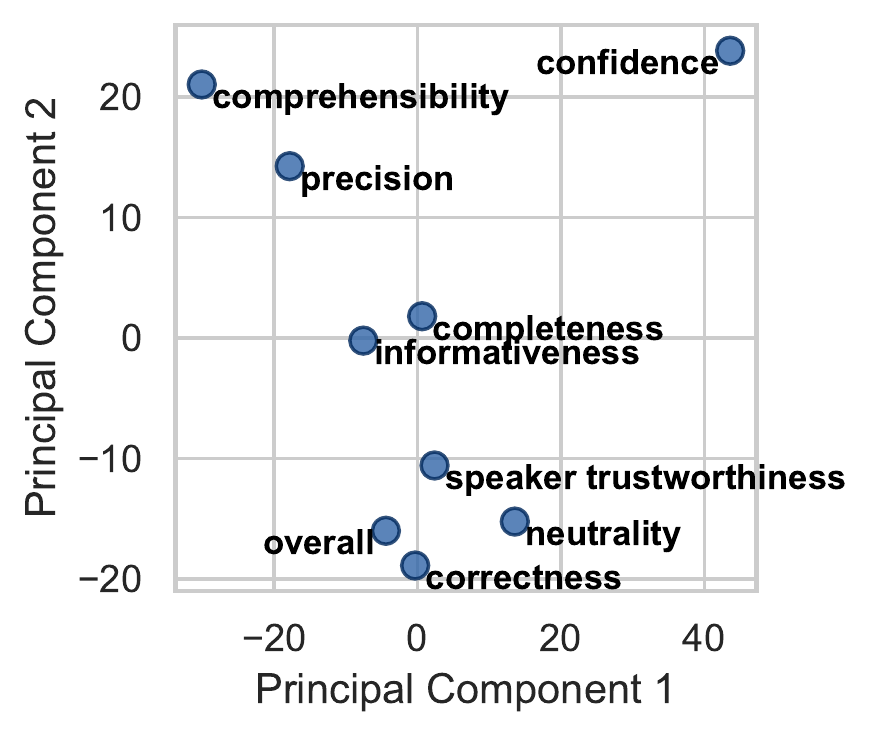}
    \caption{Individual judgments.}
    \label{cap:paper_ipm-sec:results-subsec:truthfulness-dimensions-fig:clustering_raw}
  \end{subfigure}
  \hfill
  \begin{subfigure}{.49\linewidth}
    \centering
    \includegraphics[width=.95\linewidth]{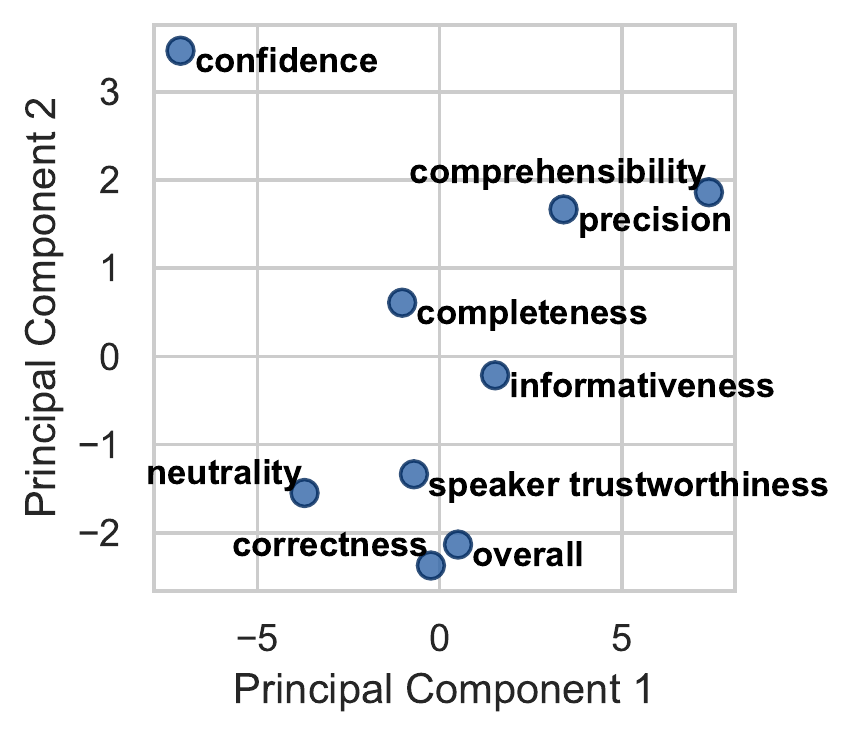}
    \caption{Aggregated judgments.}
    \label{cap:paper_ipm-sec:results-subsec:truthfulness-dimensions-fig:clustering_agg}
  \end{subfigure}
\caption{Principal component analysis (PCA) of the statements $\times$ judgments matrix. The plots show the projection of each truthfulness dimension onto the first two principal components. The left plot is based on individual judgments, and the right plot on aggregated judgments.}
  \label{cap:paper_ipm-sec:results-subsec:truthfulness-dimensions-fig:clustering}
\end{figure}

Figure~\ref{cap:paper_ipm-sec:results-subsec:truthfulness-dimensions-fig:clustering} presents the outcomes of the PCA, applied separately to individual (Figure~\ref{cap:paper_ipm-sec:results-subsec:truthfulness-dimensions-fig:clustering_raw}) and aggregated (Figure~\ref{cap:paper_ipm-sec:results-subsec:truthfulness-dimensions-fig:clustering_agg}) judgments. The analysis reveals that the dimensions most similar to \overalltruthfulness are \correctness, \speakertrustworthiness, and, to a lesser extent, \neutrality. This is particularly evident when observing the proximity of these dimensions to \overalltruthfulness in the PCA plots. The pattern is consistent for both individual and aggregated data.

This finding aligns with intuition: when workers assess the \overalltruthfulness of a statement, they likely rely on dimensions that are conceptually close to it. In contrast, dimensions such as \confidence, \comprehensibility, and \precision appear more isolated and show weak associations with \overalltruthfulness. These dimensions are also positioned farthest from \overalltruthfulness in the PCA space, reinforcing the idea that they capture different aspects of truthfulness. Future research should explore whether similar patterns hold for expert judgments. In summary, the PCA analysis confirms that each dimension contributes unique information and helps uncover meaningful relationships among them.

Since the individual dimensions measure different aspects of truthfulness, a natural hypothesis is that combining judgments across these dimensions could yield a better approximation of the ground truth than relying only on the \overalltruthfulness dimension. In other words, a weighted combination of individual dimensions might improve the agreement between crowd and expert judgments.

\begin{figure}[tbp]
  \centering
  \begin{subfigure}{.49\linewidth}
    \centering
    \includegraphics[width=.95\linewidth]{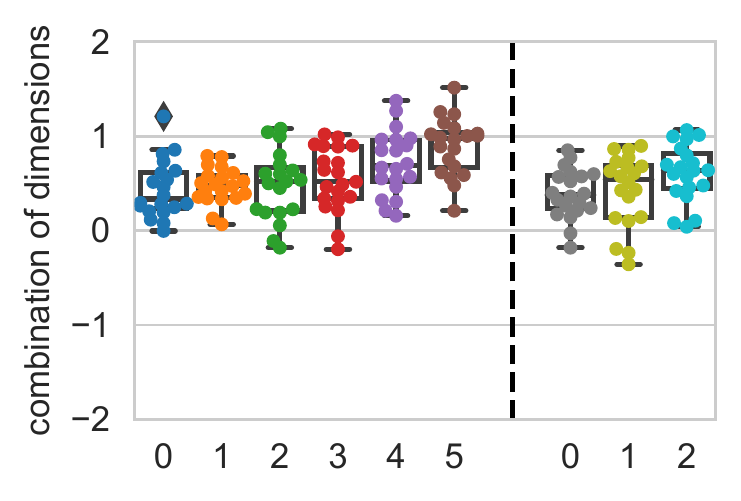}
    \caption{Combination using $\upomega^2$ values.}
    \label{cap:paper_ipm-sec:results-subsec:truthfulness-dimensions-fig:scales-comparison_omega}
  \end{subfigure}
  \hfill
  \begin{subfigure}{.49\linewidth}
    \centering
    \includegraphics[width=.95\linewidth]{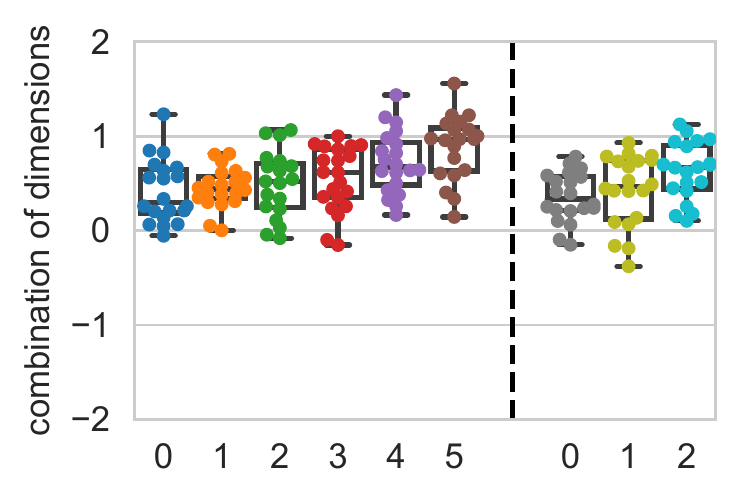}
    \caption{Combination using CRT scores.}
    \label{cap:paper_ipm-sec:results-subsec:truthfulness-dimensions-fig:scales-comparison-combination_crt}
  \end{subfigure}
  \caption{Truthfulness dimensions first aggregated using the mean function then combined. Compare with Figure~\ref{cap:paper_ipm2021-sec:results-subsec:judgment-reliability-subsec:agreement-external-fig:scale-comparison}.}
  \label{cap:paper_ipm-sec:results-subsec:truthfulness-dimensions-fig:scales-comparison-combination}
\end{figure}

To test this hypothesis, judgments for each truthfulness dimension are combined and used to predict the ground truth categories for both \politifact and \abc. This is done in an idealized scenario where ground truth values are assumed to be known for model estimation. An \index{ANOVA}ANOVA analysis is conducted to estimate the effect size of each dimension using the $\upomega^2$ \index{$\upomega^2$} index. Based on these values, the 10 judgments collected per statement are aggregated using a weighted mean, where weights are proportional to the respective $\upomega^2$ values. The resulting correlations with the ground truth are shown in Figure~\ref{cap:paper_ipm-sec:results-subsec:truthfulness-dimensions-fig:scales-comparison_omega}. While the combined judgments still show increasing median values with increasing truthfulness levels, the results do not outperform those in Figure~\ref{cap:paper_ipm2021-sec:results-subsec:judgment-reliability-subsec:agreement-external-fig:scale-comparison_overall}, where only the \overalltruthfulness judgments are used.

A second approach uses workers' \index{Cognitive!Reflection Test}CRT performance to weight their contributions. Judgments are first aggregated using a weighted mean, where weights correspond to each worker’s proportion of correct CRT answers normalized in the $[0.5,1]$ range. Then, dimensions are combined using a weighted mean based on the previously computed $\upomega^2$ values. The results are reported in Figure~\ref{cap:paper_ipm-sec:results-subsec:truthfulness-dimensions-fig:scales-comparison-combination_crt}. Again, the outcome does not show noticeable improvement over the previous strategies.

To better understand these results, a further \index{ANOVA}ANOVA analysis is performed. Two models are compared: one regressing the ground truth values on all dimensions, and the other on the \overalltruthfulness dimension alone. The residuals from both models are nearly identical, suggesting that the additional dimensions do not significantly improve the prediction. Moreover, the $\upomega^2$ index for the latter model is quite low (equal to $0.02$), confirming that \overalltruthfulness alone does not strongly predict the ground truth, and that simple combinations of the available dimensions are not sufficient either.

This analysis aligns with prior work on the decomposition of quality signals in system evaluation \cite{ferro2019using, ferro2016general, ferro2018toward, roitero2020leveraging, zampieri2019topic}. Overall, the findings indicate that linear combinations of the collected dimensions are not effective in improving agreement with expert judgments.

Future work could explore more advanced strategies. These include hierarchical models, which may require redesigning the experimental setup, or non-linear combinations of dimensions. External data such as the justification URLs provided by workers could also be integrated. Additionally, requesting further input from workers, such as confidence scores or textual justifications for each dimension, may enhance the quality of the collected data. However, this would require careful design to avoid increasing the cognitive load on participants.

\subsection{\ref{cap:paper_ipm2021-sec:research-questions_3}: Worker Behavior}

\label{cap:paper_ipm-sec:results-subsec:worker-behavior}

An attempt is made to consider worker behavior as a proxy for worker quality, in light of the inconclusive results from the combination of dimensions. The goal is to improve the correlation between collected judgments and ground truth by giving more weight to higher-quality workers. To estimate worker quality, the responses to the \index{Cognitive!Reflection Test}CRT are used. Individual judgments are aggregated using a weighted mean, where weights correspond to the normalized CRT scores. Each worker’s score is computed as the number of correct answers (out of 3) to the CRT questionnaire, normalized to the $[0.5, 1]$ range.

\begin{figure}[tbp]
  \centering
  \begin{subfigure}{.49\linewidth}
    \centering
    \includegraphics[width=.95\linewidth]{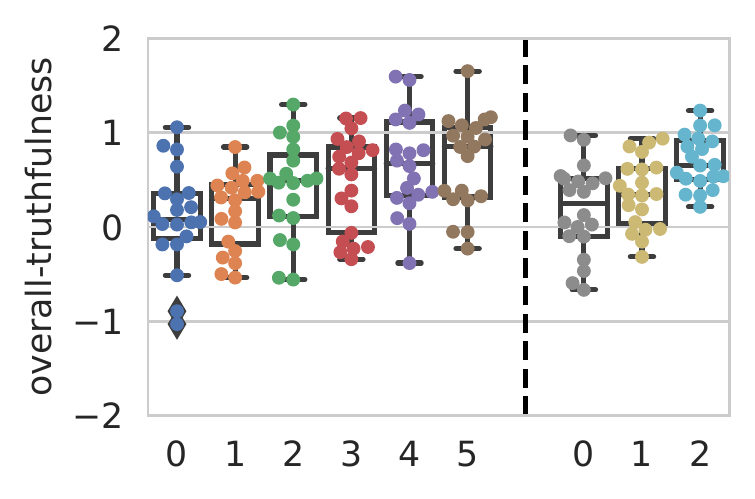}
    \caption{Original \politifact and \abc categories.}
    \label{cap:paper_ipm-sec:results-subsec:worker-behavior-fig:agg-crt_crt}
  \end{subfigure}
  \hfill
  \begin{subfigure}{.49\linewidth}
    \centering
    \includegraphics[width=.95\linewidth]{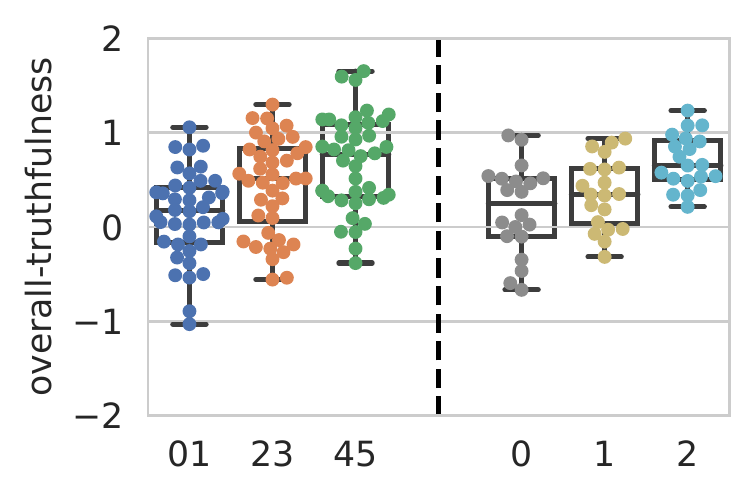}
    \caption{\politifact categories grouped into three different bins.}
    \label{cap:paper_ipm-sec:results-subsec:worker-behavior-fig:agg-crt_pol}
  \end{subfigure}
   \caption{Weighted mean aggregation of \overalltruthfulness judgments based on workers’ CRT scores, using the original \politifact and \abc categories (left plot) and a the \politifact categories grouped into three bins. Compare with Figure~\ref{cap:paper_ipm2021-sec:results-subsec:judgment-reliability-subsec:agreement-external-fig:scale-comparison} and Figure~\ref{cap:paper_ipm2021-sec:results-subsec:judgment-reliability-subsec:agreement-external-fig:scale-comparison-bin3}.}
  \label{cap:paper_ipm-sec:results-subsec:worker-behavior-fig:agg-crt}
\end{figure}

Figure~\ref{cap:paper_ipm-sec:results-subsec:worker-behavior-fig:agg-crt_crt} shows the correlation between the aggregated \overalltruthfulness values and the ground truth categories of \politifact and \abc. Figure~\ref{cap:paper_ipm-sec:results-subsec:worker-behavior-fig:agg-crt_pol} reports the same analysis after grouping the categories into 3 bins. The resulting trends closely resemble those in the top-left plots of Figure~\ref{cap:paper_ipm2021-sec:results-subsec:judgment-reliability-subsec:agreement-external-fig:scale-comparison} and Figure~\ref{cap:paper_ipm2021-sec:results-subsec:judgment-reliability-subsec:agreement-external-fig:scale-comparison-bin3}, suggesting that this approach does not significantly improve the correlation with the ground truth. Future work will explore more complex models of worker behavior and their interaction with aggregation strategies.

It must also be remarked that, when considering the individual (i.e., non-aggregated) judgments for each statement without gold questions, most workers tend to use a variety of labels across dimensions. Each worker provides 8 judgments by selecting from a five-point scale, without considering self-reported confidence. Only 12\% of workers used the same label for all dimensions, while 29\% used two distinct labels, 39\% used three, 18\% used four, and 2\% used all five. This indicates that the majority of workers make use of most of the judgment scale. These findings further support the independence of the dimensions (see Section~\ref{cap:paper_ipm2021-sec:research-questions_2} and Section~\ref{cap:paper_ipm-sec:results-subsec:truthfulness-dimensions}), suggesting that each dimension captures a distinct aspect of truthfulness.

\subsection{\ref{cap:paper_ipm2021-sec:research-questions_4}: Dimension Informativeness}

\label{cap:paper_ipm-sec:results-subsec:informativeness}

To assess the added value of collecting multidimensional judgments, this section examines whether some of the assessed dimensions can be approximated through computational methods. The focus is on their informativeness, which should not be confused with the truthfulness dimension labeled \informativeness used to judge the content of the statements. The first step involves testing whether certain dimensions can be synthesized computationally. Two dimensions for which computational proxies are identified are \comprehensibility and \correctness.

To approximate \comprehensibility, standard readability measures are used. These metrics quantify how easily a text can be understood and are therefore expected to correlate with perceived comprehensibility. For each statement, ten widely adopted readability scores are computed:
\begin{itemize}[label=--]
  \item Flesch-Kincaid Reading Ease~\cite{fleschYard} \index{Flesch-Kincaid!Reading Ease}
  \item Flesch-Kincaid Grade Level~\cite{fleschYard} \index{Flesch-Kincaid!Grade Level}
  \item Automated Readability Index~\cite{Smith1967-js} \index{Automated Readability Index}
  \item Gunning Fog Index~\cite{kincaid1975derivation} \index{Gunning Fog Index}
  \item Dale-Chall~\cite{dale1948formula} \index{Dale-Chall}
  \item Simple Measure of Gobbledygook (SMOG)~\cite{mclaughlin1969smog} \index{Simple Measure of Gobbledygook}
  \item Coleman-Liau Index~\cite{coleman1975readability} \index{Coleman-Liau Index}
  \item Forcast~\cite{caylor1973methodologies} \index{Forcast}
  \item Lesbarhets Index and Rate Index (LIX, RIX)~\cite{biornsson1968lasbarhet} \index{Lesbarhets!Index} \index{Lesbarhets!Rate Index}
\end{itemize}

All these scores show a low correlation with the crowdsourced \comprehensibility scores, with the highest \index{$\uprho$} $\uprho = 0.19$ observed for RIX. This indicates that the information captured by the workers through \comprehensibility judgments is not well approximated by automated readability measures and therefore remains a valuable dimension to crowdsource. As for \correctness, its scores are compared to the statement \polarity values obtained using the \texttt{Textblob} \index{Textblob} Python library.\footnote{\url{https://textblob.readthedocs.io/en/dev/}.} However, since \polarity reflects the emotional emphasis of the statement rather than its factual accuracy, it correlates only weakly with the \correctness scores (\index{$\uprho$} $\uprho = 0.13$).

Furthemore, the contribution of each judgment dimension to understanding the motivations behind the overall judgment~\cite{lawrence-reed-2019-argument} is investigated as follows. For \abc statements, the ground truth provides also an assessment rationale (e.g., ``cherry picking''). The Word Mover Distance \index{Word Mover Distance} ($\mathrm{wmd}$) \cite{kusner} is computed between each rationale and the name of each dimension, and its eventual correlation with the scores of that dimension is checked. Consider the case where there are two statements, $\text{statement}_i$ and $\text{statement}_j$, their \precision scores are 2 and 1 respectively, and their ground truth rationales are $\mathit{exaggeration}$ and $\mathit{wrong}$. In such a situation, the correlation is computed between the two scores (i.e., 2 and 1) and the semantic similarity of the word pairs (rationale, dimension name):
\begin{equation}
\mathrm{corr}\left((1,2), (\mathrm{wmd}(\mathit{exaggeration}, \mathit{precision}),\mathrm{wmd}(\mathit{wrong},\mathit{precision}))\right)
\label{cap:paper_ipm-sec:results-subsec:informativeness-eq:skeleton-total-cost}
\end{equation}
\myequations{Word Mover Distance computation between dimensions scores and ground truth rationales.}

The scores show a weak correlation with the semantic distance between the labels and the corresponding dimension name, with a peak at 0.3 Pearson's \index{$\uprho$}$\uprho$ correlation for \informativeness. However, combinations of similarity scores and metric scores show a higher correlation, such as:

\begin{itemize}[label=--]
\item \speakertrustworthiness vs. \completeness, similarity 0.30.
\item \overalltruthfulness vs. \informativeness, similarity 0.38.
\end{itemize}

These preliminary insights indicate that the dimension scores can help identify the motivation behind the overall assessment of a statement. The combinations of similarities and scores will be further investigated in the future.

\subsection{\ref{cap:paper_ipm2021-sec:research-questions_5}: Learning Truthfulness From Multidimensional Judgments}
\label{cap:paper_ipm-sec:results-subsec:machine-learning}

A machine learning-based approach is proposed to analyze the usefulness of multidimensional judgments and worker behavior in predicting expert judgments for both \politifact and \abc. In particular, two different strategies are explored. First, several supervised methods are evaluated in their ability to predict the exact truthfulness judgments provided by experts (Section~\ref{cap:paper_ipm-sec:results-subsec:machine-learning-subsec:supervised}). Second, unsupervised and hybrid approaches are used to estimate truthfulness judgments that are semantically close to the ground truth (Section~\ref{cap:paper_ipm-sec:results-subsec:machine-learning-subsec:unsupervised}).

\subsubsection{Supervised Approach}
\label{cap:paper_ipm-sec:results-subsec:machine-learning-subsec:supervised}

The aim of the supervised approach is to predict \politifact and \abc judgments. For \abc, both the simplified three-level scale and the original verdicts are considered, with the latter comprising 30 distinct labels in the sample used (Section~\ref{cap:paper_ipm2021-sec:exp-setup-subsec:crowdsourcing-task}). This original scale was adopted by experts during truthfulness judgment and is semantically more informative than the simplified version.

The following features are considered, and computed for each judgment. The one-hot-encoding of the worker identifiers in order to identify which worker provided the judgments, followed by the worker judgments on all the dimensions, and the 300-dimensional embedding of the string obtained from the concatenation of the query issued by the worker, and the title, snippet, and domain of the URL selected. The \texttt{SISTER (SImple SenTence EmbeddeR)}\footnote{\url{https://github.com/tofunlp/sister}} \index{SImple SenTence EmbeddeR} implementation is used to such an end. The rationale behind this set of embeddings is trying to capture the semantic relationship between the expert classification and the piece of information used by the worker to justify its judgment. After computing the features, the dataset is divided into training and test sets. To avoid any possible bias or overfitting the effectiveness metrics are computed over 3 folds obtained using stratified sampling. 
The following baselines are considered. The first (i.e., \lq\lq Most Frequent\rq\rq{}) predicts always the most frequent class present in the training set. The second (i.e., \lq\lq Weighted Sampling\rq\rq{}) predicts, for each instance in the test set, a weighted random choice among the classes present in the training set, where the weights are the frequencies of each class. The process for the second baseline is repeated \num{1000} times for each fold. Finally, the third baseline (i.e., \lq\lq Random Choice\rq\rq{}) simply returns a random class. Apart from the three baselines, the following supervised classification algorithms are used: 
\begin{itemize}[label=--]
\item Random Forest~\cite{Breiman2001}. \index{Random Forest}
\item Logistic Regression~\cite{kleinbaum2002logistic}. \index{Logistic Regression}
\item AdaBoost~\cite{Schapire2013}. \index{AdaBoost}
\item Support Vector Machine (SVM)~\cite{Shmilovici2005}. \index{Support Vector Machine}
\item Na\"ive Bayes~\cite{Webb2016}. \index{Na\"ive Bayes}
\end{itemize}
The \texttt{sklearn} \index{sklearn} implementation\footnote{\url{https://scikit-learn.org/stable/supervised_learning.html}} of the algorithms is used. The parameters used to train the algorithms, reported to allow reproducibility, can be found in the repository containing the dataset released.

The following features are considered and computed for each judgment. First, a one-hot encoding of worker identifiers is included to capture which worker provided each judgment. This is followed by the worker's judgments on all dimensions, and a 300-dimensional embedding of the concatenation of the query issued by the worker, along with the title, snippet, and domain of the selected URL. To compute these embeddings, the \texttt{SISTER (SImple SenTence EmbeddeR)} implementation\footnote{\url{https://github.com/tofunlp/sister}} \index{SImple SenTence EmbeddeR} is used. The rationale behind this representation is to capture the semantic relationship between the expert classification and the information selected by the worker as justification.
Once all features are computed, the dataset is split into training and test sets. To mitigate bias and prevent overfitting, evaluation metrics are computed over three folds, using stratified sampling.
Three baselines are considered:
\begin{itemize}[label=--]
  \item \lq\lq Most Frequent\rq\rq{}: always predicts the most frequent class in the training set.
  \item \lq\lq Weighted Sampling\rq\rq{}: for each test instance, randomly selects a class based on training set frequencies; this process is repeated \num{1000} times per fold.
  \item \lq\lq Random Choice\rq\rq{}: randomly selects a class uniformly.
\end{itemize}
In addition to the baselines, the following supervised classifiers are evaluated:
\begin{itemize}[label=--]
  \item Random Forest~\cite{Breiman2001} \index{Random Forest}
  \item Logistic Regression~\cite{kleinbaum2002logistic} \index{Logistic Regression}
  \item AdaBoost~\cite{Schapire2013} \index{AdaBoost}
  \item Support Vector Machine (SVM)~\cite{Shmilovici2005} \index{Support Vector Machine}
  \item Na\"ive Bayes~\cite{Webb2016} \index{Na\"ive Bayes}
\end{itemize}

All algorithms are implemented using the \texttt{sklearn} library\footnote{\url{https://scikit-learn.org/stable/supervised_learning.html}} \index{sklearn}. The specific training parameters used for each model are documented in the repository accompanying the released dataset to ensure reproducibility \cite{Soprano2023thesis}.

Table~\ref{cap:paper_ipm-sec:results-subsec:machine-learning-subsec:supervised-tab:effectiveness-abc-prediction} reports the effectiveness scores obtained when predicting the \politifact and \abc verdicts. To address class imbalance, the weighted average of \index{Precision}Precision, \index{Recall}Recall, and \index{F1}F1 is reported. This means that the scores for all classes are aggregated by weighting each class by its frequency.

Among all the methods, the Random Forest algorithm achieves the best performance across all datasets, outperforming both the random baselines and the other classification algorithms. To understand the reason for this performance gap, the importance of the features\footnote{\url{https://scikit-learn.org/stable/auto_examples/ensemble/plot_forest_importances.html}} used by the Random Forest is analyzed. The algorithm assigns similar importance to all dimensions of the embedding vector, which turn out to be the most informative features. The remaining features, namely the one-hot encoding of worker identifiers and the worker judgments, have lower importance than the embedding but still contribute meaningfully to the model's predictions. When either the worker identifiers or the judgment values are removed, performance metrics decrease. This suggests that Random Forest successfully leverages all input signals to learn a predictive model that generalizes well.

This result highlights the usefulness of integrating multiple types of information from workers, including their search sessions, for predicting expert verdicts. The finding holds even in the more challenging 30-class classification scenario corresponding to the original and semantically richer \abc verdicts.

The statistical significance of the metric scores is also assessed by comparing them with those of the best baseline. To this end, the \index{Wilcoxon Signed-Rank Test}Wilcoxon Signed-Rank Test~\cite{woolson2008test} is employed, as it is a non-parametric test suitable for paired data. To account for multiple comparisons, the \index{Bonferroni Correction}Bonferroni correction~\cite{Bland170, Sedgwicke509} is applied. None of the comparisons lead to statistically significant differences, with all \index{$p$}$p$-values greater than 0.05. This outcome is likely due to the small number of data points used in the test, which is three, since the data is split into 3 folds.

\begin{figure}[tbp]
  \centering
  \includegraphics[width=\linewidth]{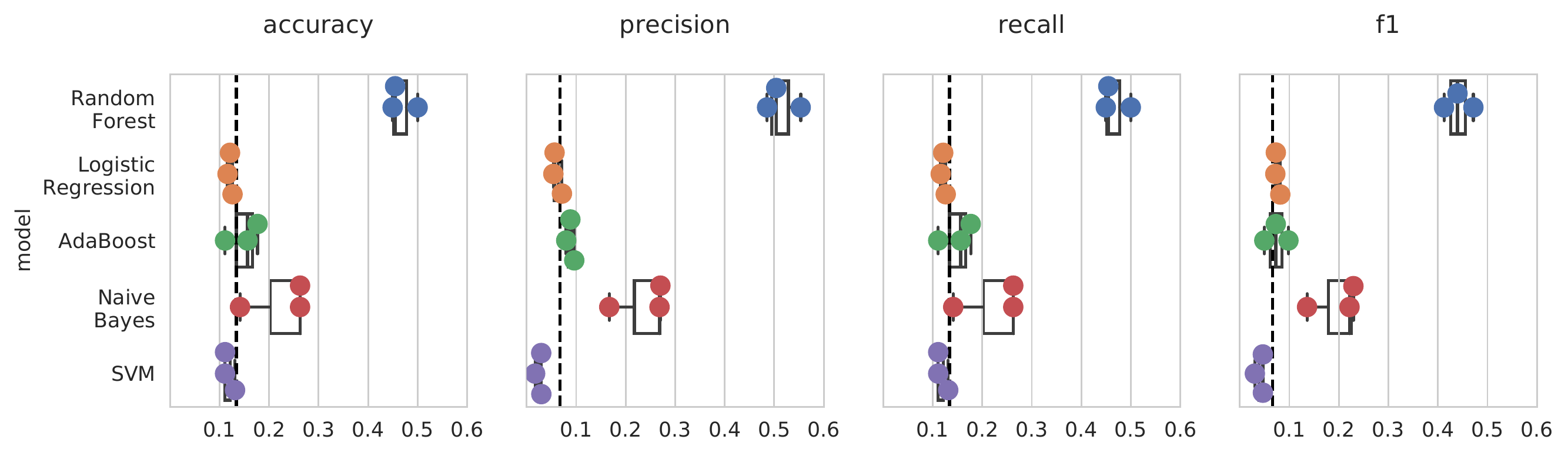}
  \caption{Effectiveness metrics across the 3 folds for the \abc 30-level case. The dashed line indicates the performance of the best baseline.}
  \label{cap:paper_ipm-sec:results-subsec:machine-learning-subsec:supervised-fig:stat-sign}
\end{figure}

As an additional analysis, the metric scores for each fold are plotted for all models, with the best baseline’s score indicated by a dashed line (note that the baseline always yields the same score across folds). The results are shown in Figure~\ref{cap:paper_ipm-sec:results-subsec:machine-learning-subsec:supervised-fig:stat-sign}. While statistical significance is not established, the plots suggest that the top-performing algorithms consistently outperform the best baseline. A similar pattern is observed for both the \politifact 6-level and \abc 3-level settings (not shown).

The bootstrap technique is employed to compute the 95\% confidence intervals for the most effective algorithm, namely Random Forest. A total of \num{100,000} stratified samples are drawn, and the 2.5-th and 97.5-th percentiles are calculated~\cite{bland2015statistics, linnet2000nonparametric} to estimate the range that covers the true mean statistic with 95\% likelihood. The results are reported in Table~\ref{cap:paper_ipm-sec:results-subsec:machine-learning-subsec:supervised-tab:effectiveness-abc-prediction}. Even the lower bound of the interval confirms that Random Forest significantly outperforms the best baseline.

As a final analysis, the performance of the machine learning algorithms is examined under different sets of judgment dimensions. This is motivated by the goal of assessing the utility of adopting a multidimensional scale. The task remains to predict the \politifact and \abc expert judgments. Specifically, the same algorithms listed in Table~\ref{cap:paper_ipm-sec:results-subsec:machine-learning-subsec:supervised-tab:effectiveness-abc-prediction} are trained using the following three feature sets:
\begin{enumerate}
\item All dimensions except \overalltruthfulness.
\item Only the \overalltruthfulness dimension.
\item All dimensions including \overalltruthfulness.
\end{enumerate}

Although detailed results are omitted, they are nearly identical to those shown in Table~\ref{cap:paper_ipm-sec:results-subsec:machine-learning-subsec:supervised-tab:effectiveness-abc-prediction}, with only minor fluctuations. In most cases, training with set (1), which excludes \overalltruthfulness, leads to slightly better effectiveness metrics than using set (3), which includes it. Both of these configurations outperform set (2), which uses only \overalltruthfulness. As before, the statistical significance of the differences is evaluated using the Wilcoxon signed-rank test with Bonferroni correction for multiple comparisons. None of the differences are statistically significant.

In summary, these findings suggest that using all individual dimensions leads to the best (albeit not statistically significant) predictive performance. Moreover, the inclusion of \overalltruthfulness as a feature does not lead to measurable improvement and is in fact outperformed by the combination of the remaining dimensions.

\begin{table}[tbp]
\caption{Effectiveness metrics for predicting expert judgments. Baselines are listed above the dashed line.}
\label{cap:paper_ipm-sec:results-subsec:machine-learning-subsec:supervised-tab:effectiveness-abc-prediction}
\centering
\adjustbox{max width=0.95\textwidth}{%
\begin{tabular}{l cccc}
\toprule
\textbf{Algorithm} & \textbf{Accuracy} & \textbf{Precision (W)} & \textbf{Recall (W)} & \textbf{F1 (W)} \\
\midrule
\multicolumn{5}{c}{\textbf{\politifact 6 Levels}} \\
\midrule
Random Choice                     & .167 & .167 & .167 & .167 \\
\hdashline
Random Forest                     & \textbf{.556} & \textbf{.561} & \textbf{.556} & \textbf{.554} \\
Random Forest (bootstrap CI)     & $[.477,.569]$ & $[.482,.574]$ & $[.477,.569]$ & $[.476,.568]$ \\
Logistic Regression               & .391 & .417 & .392 & .392 \\
AdaBoost                          & .327 & .340 & .327 & .327 \\
Naive Bayes                       & .165 & .185 & .165 & .064 \\
SVM                               & .225 & .213 & .226 & .207 \\
\midrule
\multicolumn{5}{c}{\textbf{\abc 3 Levels (Simplified)}} \\
\midrule
Random Choice                     & .333 & .333 & .333 & .333 \\
\hdashline
Random Forest                     & \textbf{.667} & \textbf{.670} & \textbf{.667} & \textbf{.665} \\
Random Forest (bootstrap CI)     & $[.594,.716]$ & $[.595,.720]$ & $[.594,.716]$ & $[.592,.715]$ \\
Logistic Regression               & .557 & .563 & .557 & .555 \\
AdaBoost                          & .560 & .562 & .560 & .559 \\
Naive Bayes                       & .579 & .584 & .579 & .576 \\
SVM                               & .392 & .391 & .392 & .379 \\
\midrule
\multicolumn{5}{c}{\textbf{\abc 30 Levels (Original)}} \\
\midrule
Random Choice                     & .033 & .033 & .033 & .033 \\
Most Frequent                     & .134 & .018 & .134 & .032 \\
Weighted Sampling                 & .067 & .067 & .067 & .066 \\
\hdashline
Random Forest                     & \textbf{.518} & \textbf{.562} & \textbf{.518} & \textbf{.491} \\
Random Forest (bootstrap CI)     & $[.426,.538]$ & $[.460,.605]$ & $[.426,.538]$ & $[.398,.514]$ \\
Logistic Regression               & .195 & .151 & .195 & .143 \\
AdaBoost                          & .148 & .088 & .148 & .073 \\
Naive Bayes                       & .203 & .221 & .203 & .181 \\
SVM                               & .154 & .052 & .154 & .075 \\
\bottomrule
\end{tabular}
}
\end{table}

\subsubsection{Unsupervised Approach}

\label{cap:paper_ipm-sec:results-subsec:machine-learning-subsec:unsupervised}

In addition to the supervised approach described in Section~\ref{cap:paper_ipm-sec:results-subsec:machine-learning-subsec:supervised}, this section evaluates the use of unsupervised methods for truthfulness prediction. Considering both supervised and unsupervised strategies offers a more complete view of the potential effectiveness of different approaches to predict expert judgments.

The goal here is not to predict the exact ground truth label, but rather to estimate a judgment that is semantically close to the ground truth and aligned in terms of \polarity. The focus is specifically on the \abc verdicts, which are semantically rich and nuanced. This analysis serves to explore the alignment between expert and crowd judgments. In particular, it aims to determine whether the weighted embeddings derived only from workers' multidimensional assessments are consistent with the evaluations provided by experts.

To this end, predictions are evaluated using two metrics: Word Mover’s semantic distance and sentiment difference. Sentiment scores are computed with \index{flair}\texttt{flair}~\cite{akbik-etal-2019-flair}.\footnote{\url{https://github.com/flairNLP/flair}} Sentiment values range between -1 and +1. While semantic similarity captures whether the rationale behind judgments is conceptually similar, sentiment difference reflects whether the polarities align. For instance, the terms \textit{comprehensible} and \textit{accurate} may have high semantic distance but relatively low sentiment difference, whereas \textit{comprehensible} and \textit{incomprehensible} have low semantic similarity but high polarity contrast.

The results are compared against three reference cases derived from the ground truth: worst, best, and average combinations of verdicts. Randomly selecting a ground truth verdict for each statement leads to an average semantic distance of 3.40, with 2.48 in the best case and 4.41 in the worst. Regarding sentiment difference, the worst possible value is 1.97, while the best (excluding exact matches) is 0.02. The average sentiment difference in the random scenario is 1.00. In the analysis, each statement is considered using the average of the corresponding worker judgments. The two strategies employed are detailed below.

\myparagraph{Weighted Average Word Embeddings} This approach assumes that the quality dimensions are positively connoted: for instance, when a worker assigns a +2 score to \comprehensibility, the corresponding overall verdict is taken to imply that the statement is comprehensible. To operationalize this, the word embedding of each dimension name is computed and weighted according to its score. The weighted embeddings are then averaged to produce a vector representation of the expected verdict. The label whose embedding is closest to this average vector is retrieved from the embedding dictionary. The resulting labels yield an average semantic distance from the ground truth of 4.14 and an average sentiment difference of 1.31, showing no improvement over random selection. This outcome is largely due to the method searching the entire embedding dictionary, whereas the ground truth judgments fall within a more specific semantic space related to quality assessment.

Although averaging embeddings introduces some information loss, this loss appears limited, since all the dimension names share a similar semantic context. To further examine this aspect, Figure~\ref{cap:paper_ipm-sec:results-subsec:machine-learning-subsec:unsupervised-fig:embed} shows the embeddings for each dimension and compares them to their average. The plots are produced using \index{t-Distributed Stochastic Neighbor Embedding} t-Distributed Stochastic Neighbor Embedding (t-SNE)~\cite{JMLR:v9:vandermaaten08a} to obtain a meaningful two-dimensional representation. Each plot includes an ellipse denoting the 95\% confidence interval for the set of embeddings. Each set can be interpreted as a sample from the distribution of judgments collected about a given quality dimension, weighted by the embedding of its name. The considerable overlap between each distribution and its average embedding suggests that the overall information loss is limited.

\myparagraph{Linear Regression} This supervised approach builds on the weighted average word embeddings described above. For each statement, an average word embedding of the judgments is computed, while the embedding of the corresponding ground truth verdict is also derived. A linear regression model is then trained to map between the two. A 3-fold cross-validation is used. For each prediction, the closest term in the embedding dictionary is retrieved. The resulting average semantic distance from the ground truth is 3.38, and the average sentiment difference is 0.41, outperforming the random selection baseline. These results indicate that the link between worker assessments and expert judgments is not as direct as previously assumed, but even a simple linear model can partially capture it. Future work will explore more sophisticated models and consider incorporating worker profiles.

\begin{figure}[tbp]
  \centering
  \begin{subfigure}{\linewidth}
    \centering
    \includegraphics[width=0.55\linewidth]{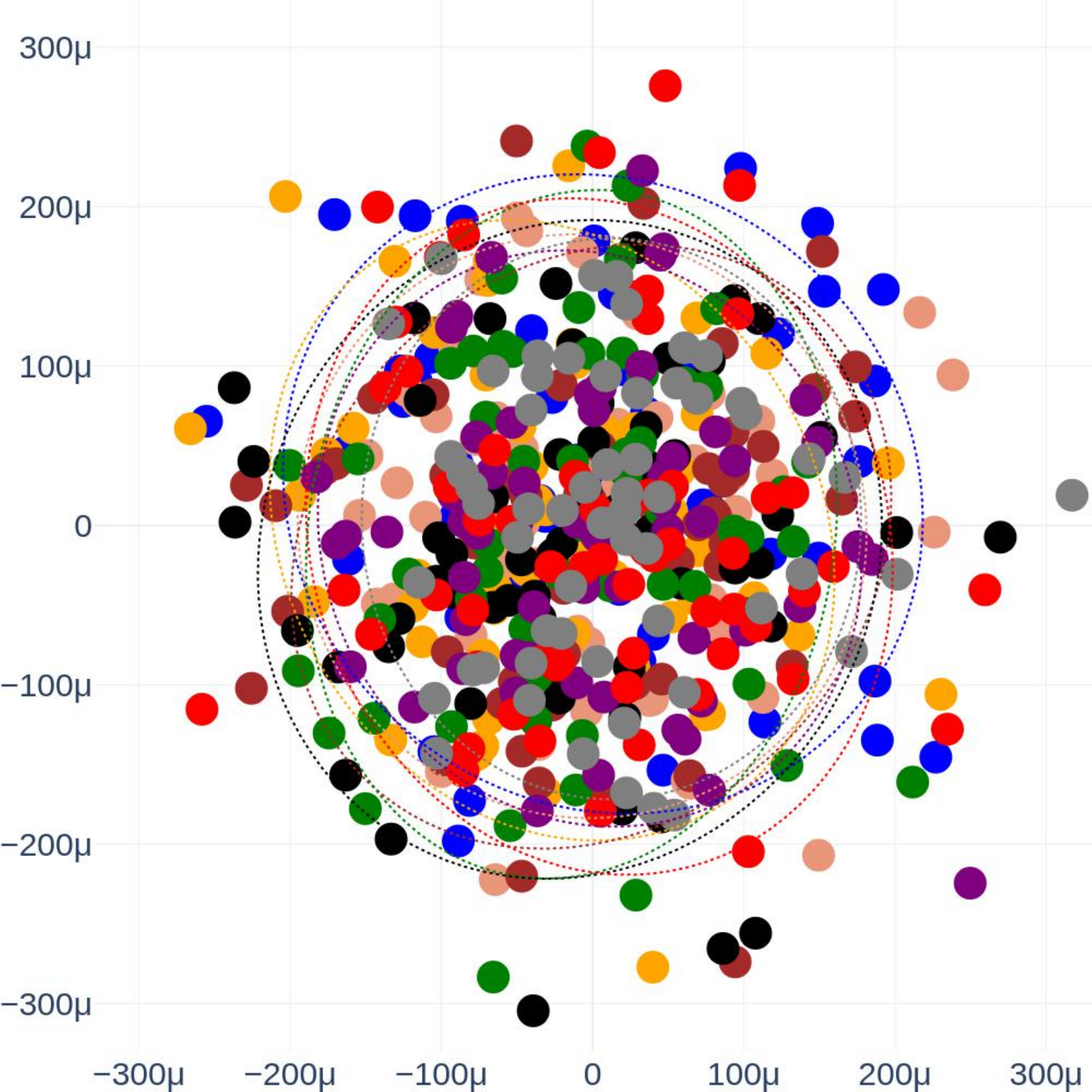}
    \caption{The whole set of dimensions.}
    \label{cap:paper_ipm-sec:results-subsec:machine-learning-subsec:unsupervised-fig:embed_all}
  \end{subfigure}

  \begin{subfigure}{0.49\linewidth}
    \centering
    \includegraphics[width=.9\linewidth]{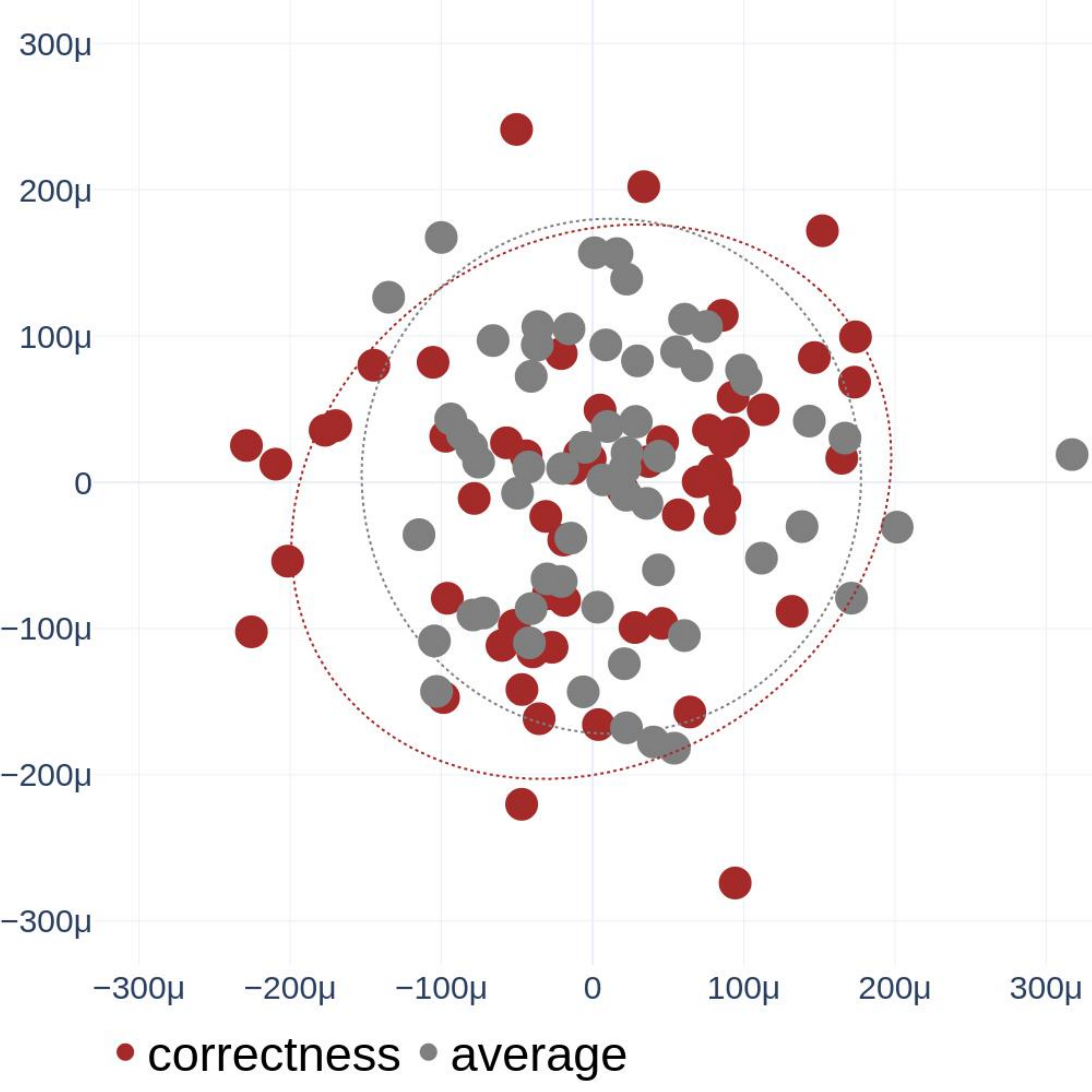}
    \caption{\correctness}
    \label{cap:paper_ipm-sec:results-subsec:machine-learning-subsec:unsupervised-fig:embed_correctness}
  \end{subfigure}
  \hfill
  \begin{subfigure}{0.49\linewidth}
    \centering
    \includegraphics[width=.9\linewidth]{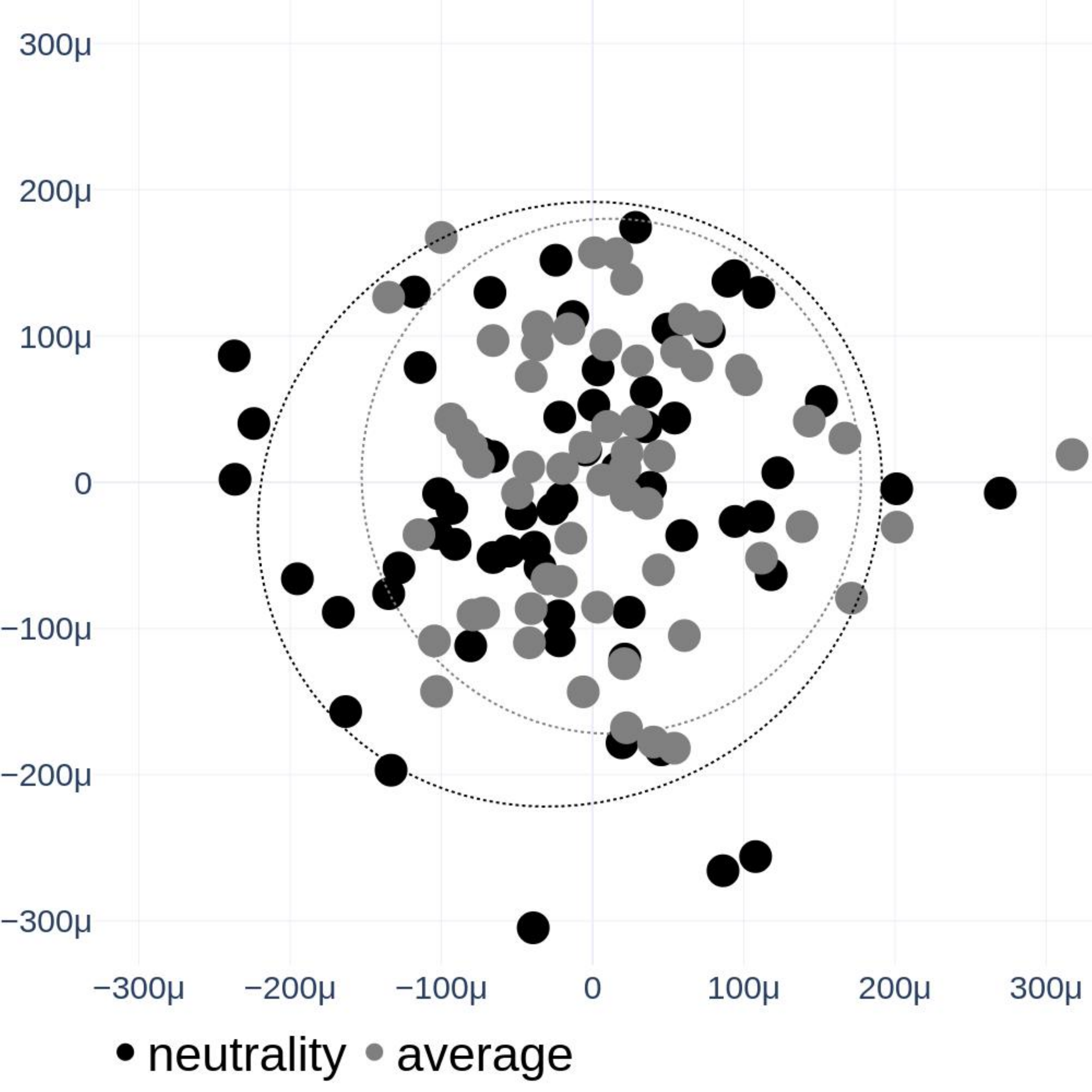}
    \caption{\neutrality}
    \label{cap:paper_ipm-sec:results-subsec:machine-learning-subsec:unsupervised-fig:embed_neutrality}
  \end{subfigure}

  \begin{subfigure}{0.49\linewidth}
    \centering
    \includegraphics[width=.9\linewidth]{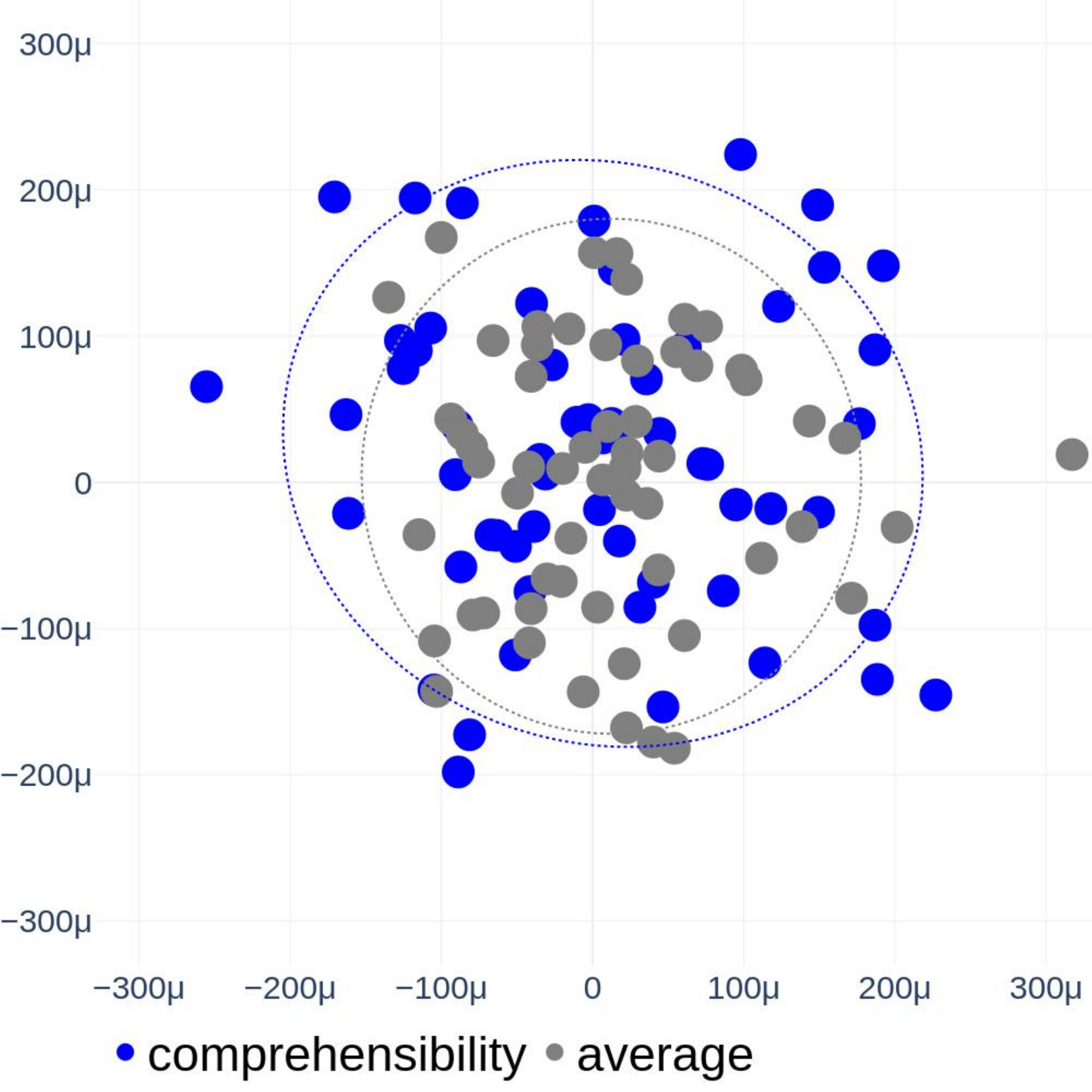}
    \caption{\comprehensibility}
    \label{cap:paper_ipm-sec:results-subsec:machine-learning-subsec:unsupervised-fig:comprehensibility}
  \end{subfigure}
  \hfill
  \begin{subfigure}{0.49\linewidth}
    \centering
    \includegraphics[width=.9\linewidth]{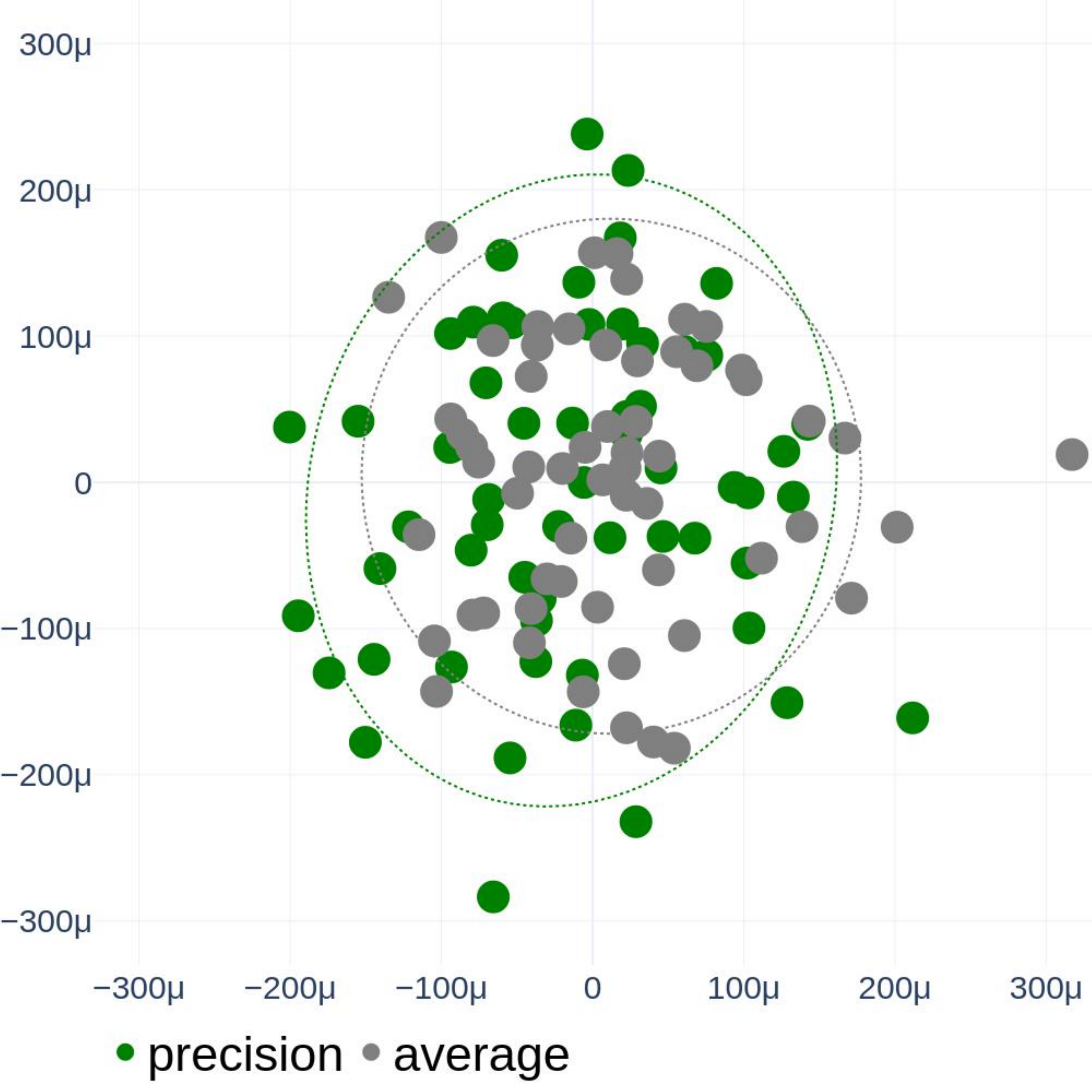}
    \caption{\precision}
    \label{cap:paper_ipm-sec:results-subsec:machine-learning-subsec:unsupervised-fig:embed_precision}
  \end{subfigure}
\end{figure}

\begin{figure}[tbp]\ContinuedFloat
  \centering
  \begin{subfigure}{0.49\linewidth}
    \centering
    \includegraphics[width=.9\linewidth]{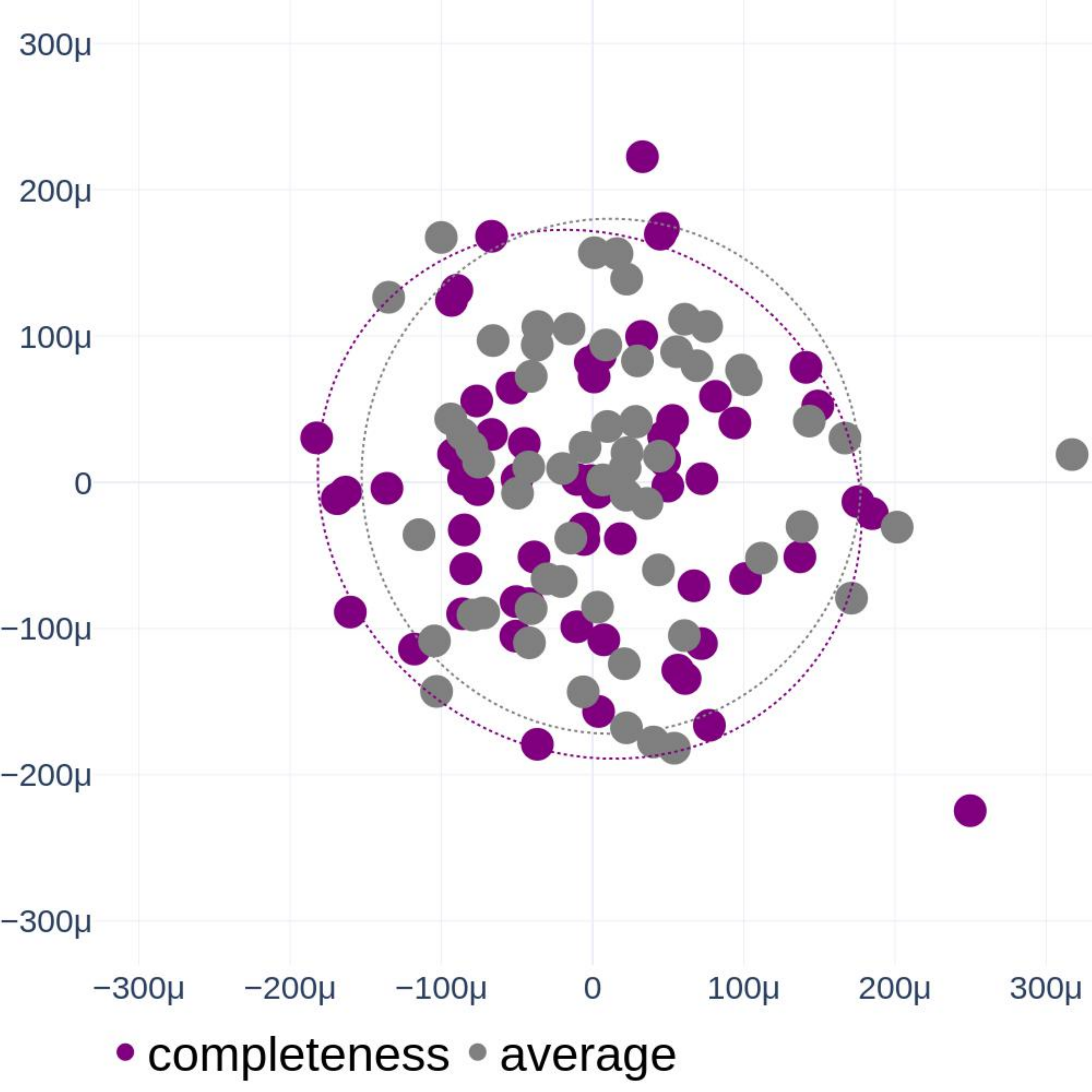}
    \caption{\completeness}
    \label{cap:paper_ipm-sec:results-subsec:machine-learning-subsec:unsupervised-fig:embed_completeness}
  \end{subfigure}
  \hfill
  \begin{subfigure}{0.49\linewidth}
    \centering
    \includegraphics[width=.9\linewidth]{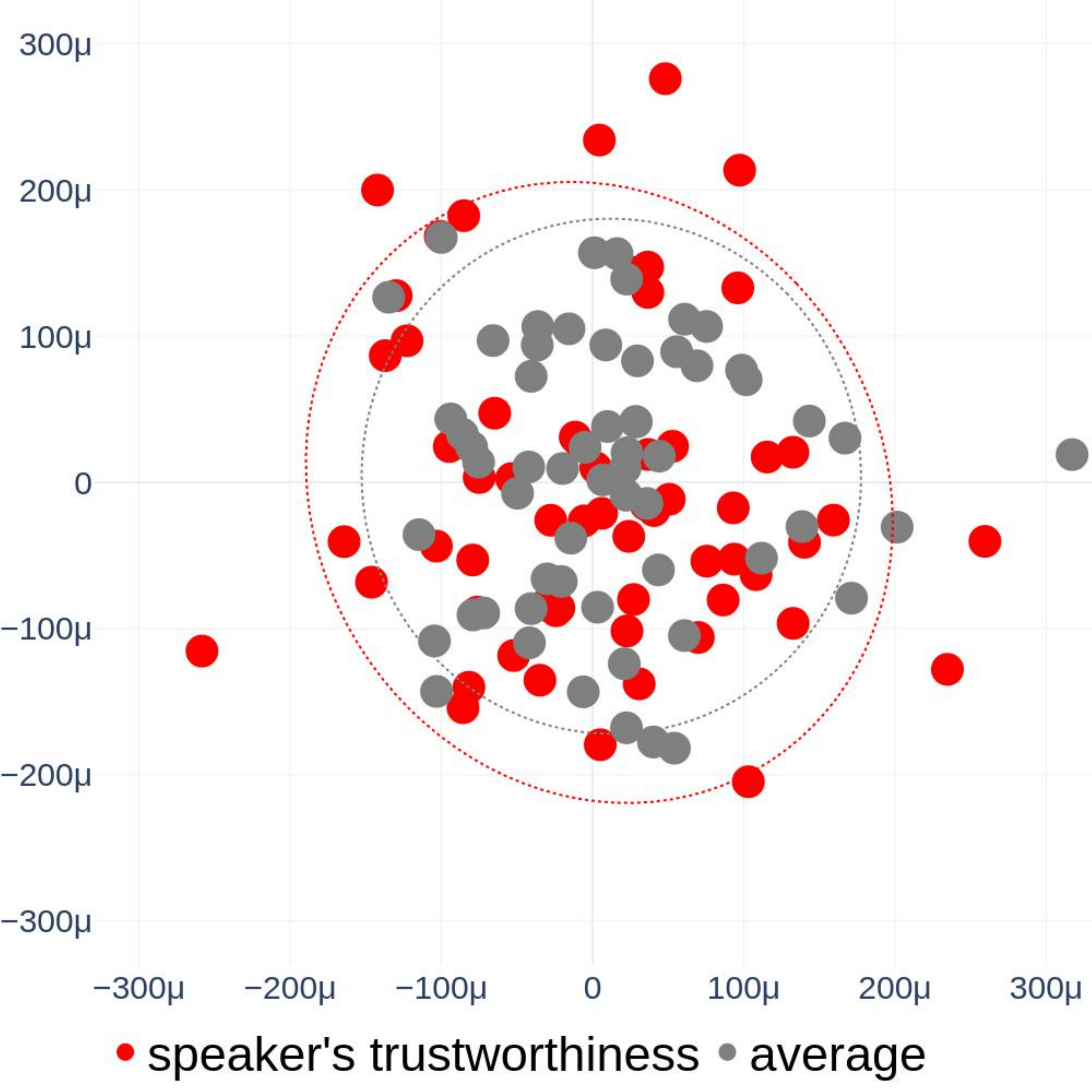}
    \caption{\speakertrustworthiness}
    \label{cap:paper_ipm-sec:results-subsec:machine-learning-subsec:unsupervised-fig:embed_trustworthiness}
  \end{subfigure}

  \begin{subfigure}{0.49\linewidth}
    \centering
    \includegraphics[width=.9\linewidth]{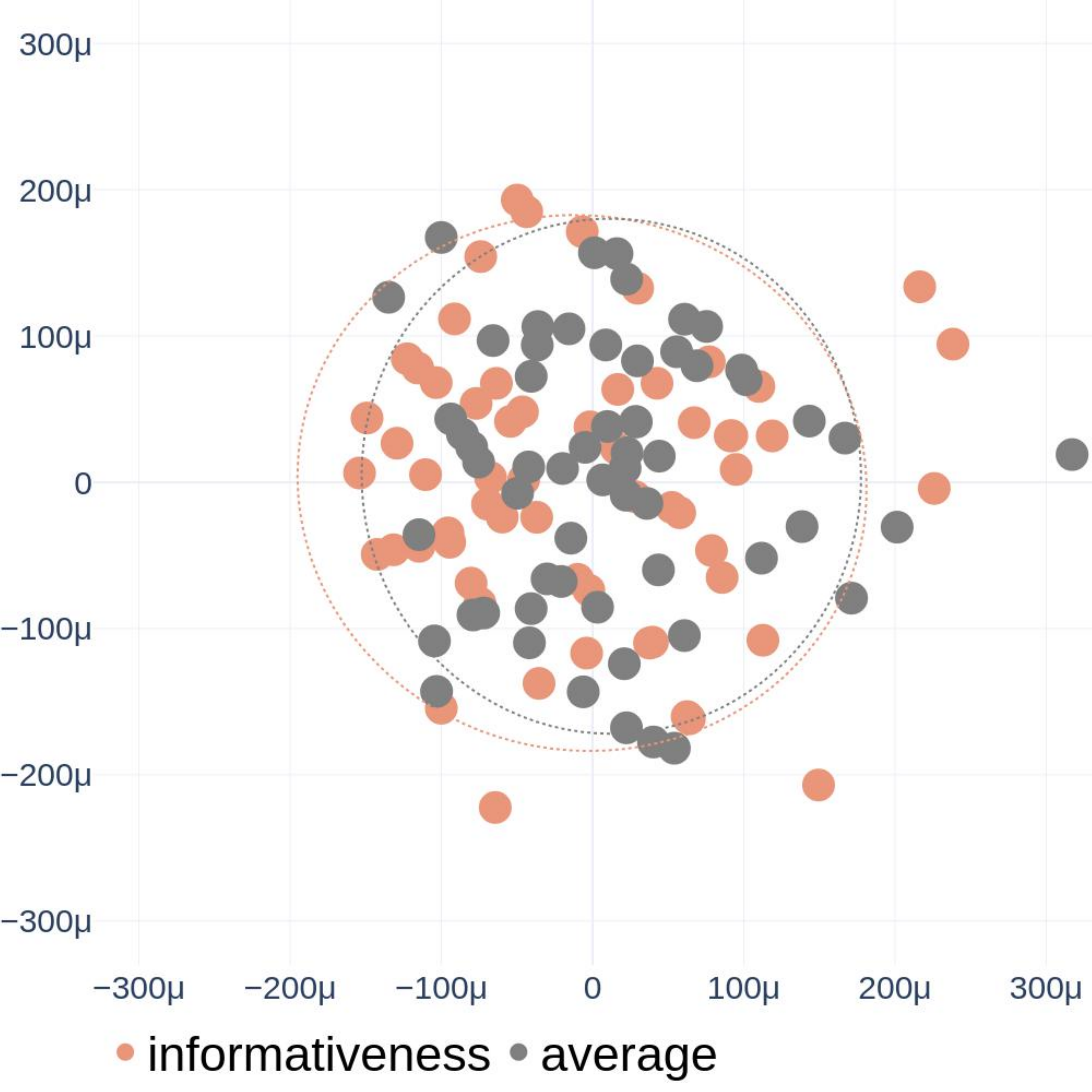}
    \caption{\informativeness}
    \label{cap:paper_ipm-sec:results-subsec:machine-learning-subsec:unsupervised-fig:embed_informativeness}
  \end{subfigure}
  \hfill
  \begin{subfigure}{0.49\linewidth}
    \centering
    \includegraphics[width=.9\linewidth]{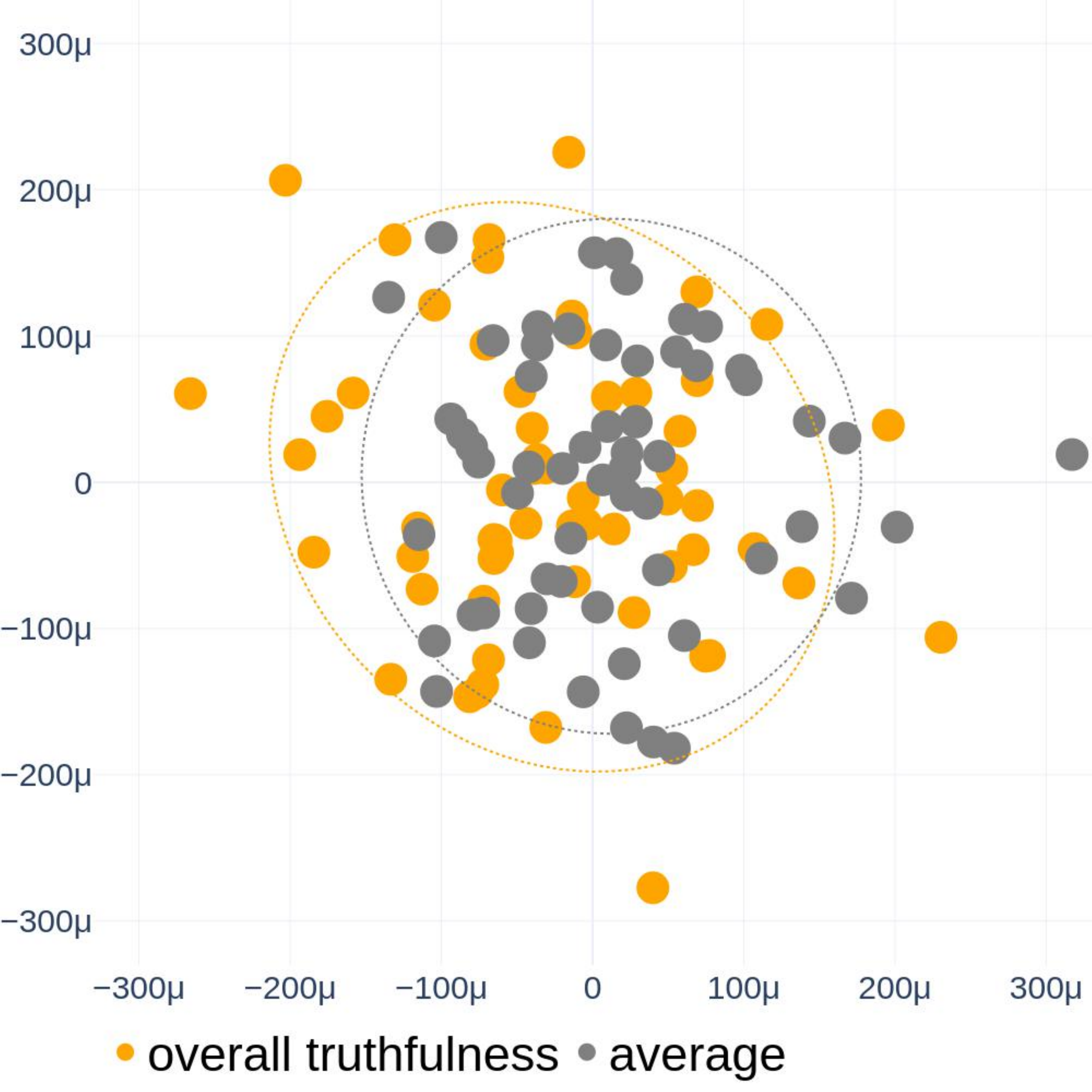}
    \caption{\overalltruthfulness}
    \label{cap:paper_ipm-sec:results-subsec:machine-learning-subsec:unsupervised-fig:embed_overall}
  \end{subfigure}

  \caption{Visualization of the embedding space. Each dimension is compared to the average embedding. In the first plot, colors match the legend used in the individual dimension plots.}
  \label{cap:paper_ipm-sec:results-subsec:machine-learning-subsec:unsupervised-fig:embed}
\end{figure}

\section{Summary}

\label{cap:paper_ipm2021-sec:discussion}

This chapter presents a study on the impact of collecting truthfulness judgments via crowdsourcing using multiple dimensions of truthfulness instead of a single one. This multidimensional approach enhances the explainability of the collected labels and enables additional quality control, as the increased input from crowd workers allows for cross-correlation. The answers to the research questions are summarized below.

\myparagraph{\ref{cap:paper_ipm2021-sec:research-questions_1}} Extensive evidence is provided that the truthfulness judgments collected from crowd workers across the seven dimensions of truthfulness are sound and reliable. The analysis of internal agreement among workers does not reveal issues with any individual dimension. Agreement with the expert-provided ground truth is good when measuring the same construct (i.e., \overalltruthfulness), and reasonable for the individual dimensions, with deviations that can be explained by the specific meaning of each dimension.

\myparagraph{\ref{cap:paper_ipm2021-sec:research-questions_2}} Multiple analyses demonstrate that the seven dimensions are independent, non-redundant, and capture distinct aspects of truthfulness. However, this independence could not be successfully exploited to combine individual dimension assessments into a composite score that leads to higher agreement with the ground truth.

\myparagraph{\ref{cap:paper_ipm2021-sec:research-questions_3}} Clear differences emerge in the behavior of individual crowd workers. Nevertheless, these behavioral signals could not be effectively leveraged to enhance the correlation between aggregated crowd judgments and expert labels.

\myparagraph{\ref{cap:paper_ipm2021-sec:research-questions_4}} The analyses on the informativeness of the various dimensions indicate that crowd data are not trivially replicable through automatic methods. Furthermore, the individual dimensions provide useful insights into the rationale behind each worker’s judgment.

\myparagraph{\ref{cap:paper_ipm2021-sec:research-questions_5}} Signals derived from workers, particularly their multidimensional judgments and search session data, can be effectively used to predict expert verdicts, both for \politifact and \abc.

\myparagraph{}
The next chapter presents a review of the cognitive biases that may arise during the fact-checking process, following the \prisma guidelines to ensure transparent and structured reporting. It provides a comprehensive list of these biases, categorizes them according to a defined scheme, and outlines a set of countermeasures to mitigate their effects.

\chapter{Cognitive Biases In Fact-Checking And Their Countermeasures: A Review}

\label{cap:paper_ipm2023_bias}

This chapter is based on the article published in the \lq\lq Information Processing \& Management\rq\rq{} journal~\cite{SOPRANO2024103672}. Section~\ref{cap:related_work-sec:crowdsourcing-truthfulness} and Section~\ref{cap:related_work-sec:worker-bias} provide the relevant related work. The remainder of the chapter is structured as follows: Section~\ref{cap:paper_ipm2023_bias-sec:research_questions} introduces the research questions, Section~\ref{cap:paper_ipm2023_bias-sec:methodology} outlines the methodology, and Section~\ref{cap:paper_ipm2023_bias-sec:results} presents the results. Finally, Section~\ref{cap:paper_ipm2023_bias-sec:discussion} summarizes the main findings and concludes the chapter.

\section{Research Questions}

\label{cap:paper_ipm2023_bias-sec:research_questions}

In addressing the challenges posed by cognitive biases that may impact the fact-checking activity, various studies have focused on different aspects of fact-checking. There is still a gap in comprehensively reviewing how cognitive biases specifically influence the fact-checking process. 

Existing literature reviews have primarily focused on technical and procedural aspects of fact-checking \cite{10.1145/3395046}, addressing cognitive biases only partially. For instance, some propose generative mechanisms for subsets of cognitive biases \cite{doi:10.1177/17456916221148147}, exclusively address partisanship-related biases \cite{doi:10.1080/10584609.2019.1668894}, review only those considered most important by the authors \cite{RUFFO2023100531}, or completely avoid this aspect \cite{WANG2019112552}. Some reviews narrow their focus to the context of political conversations only \cite{tucker2018social}, while other recent reviews addressing cognitive biases are contributions to other fields of study \cite{10.1093/bjs/znad004, doi:10.1080/09638237.2020.1766000, Eberhard2023, doi:10.1177/01655515211001777}. Existing reviews, in sum, often overlook the comprehensive set of cognitive biases that can significantly impact the outcomes of fact-checking activities. Thus, there is the need for a review aimed at filling these gaps. Such a review is necessary because cognitive biases are inherent in human judgments and, as a consequence, in the datasets used for training machine learning models for the fact-checking domain (Section~\ref{cap:related_work-sec:worker-bias}).

In summary, existing reviews often overlook the comprehensive set of cognitive biases that can significantly impact the outcomes of fact-checking activities. Therefore, there is a need for a review aimed at filling these gaps, also considering that cognitive biases are inherent in human judgments and, consequently, in the datasets used for training machine learning models in the fact-checking domain (Section~\ref{cap:related_work-sec:worker-bias}). These biases can subtly but profoundly influence the effectiveness and reliability of fact-checking processes. By identifying and understanding these biases, more robust and unbiased fact-checking methodologies can be developed, whether they are human-based or automatic. This would not only improve the accuracy of fact-checking, but also contribute to the broader discussion on the reliability and trustworthiness of information.

This chapter presents a review with the primary motivation of providing a comprehensive and systematic investigation into the cognitive biases that may manifest during the fact-checking process (Section~\ref{cap:intro-sec:fact-checking}), compromising its effectiveness in a real-world scenario. The review serves a fourfold purpose:
\begin{enumerate}
	\item To systematically identify the cognitive biases that are relevant to the fact-checking process.
	\item To provide a categorization of these biases and real-world examples to illustrate their impact on fact-checking.
	\item To propose potential countermeasures that can help mitigate the risk of cognitive biases manifesting in a fact-checking context.
	\item To provide the constituting blocks of a bias-aware fact-checking pipeline that helps to minimize such a risk. 
\end{enumerate} 

In more detail, the Preferred Reporting Items for Systematic Reviews and Meta-Analyses (\prisma) guidelines~\cite{moher2009preferred, Pagen71} are followed to support the transparent and structured reporting of the biases identified. Specifically, the focus is on retrieving a single formulation for each considered bias and, when possible, a single reference that frames it within a fact-checking-related scenario. The goal is not to compile an exhaustive list of references, but rather to construct a comprehensive list of cognitive biases relevant to fact-checking.

This review offers a novel perspective that complements and extends the existing literature but is not intended to be conclusive. It should be considered a starting point for further research in this area. The analysis provides insights and practical guidance for researchers, practitioners, and policymakers working on fact-checking and information quality. To the best of current knowledge, this is the first work to present a comprehensive set of countermeasures aimed at mitigating the manifestation of cognitive biases in the fact-checking process. The following research questions are investigated:

\begin{enumerate}[start=21, leftmargin=2.92em, label=RQ\arabic*]
\item \label{cap:paper_ipm2023_bias-sec:research-questions_1} Which cognitive biases may manifest during the fact-checking process?
\item \label{cap:paper_ipm2023_bias-sec:research-questions_2} Can the cognitive biases emerging during fact-checking be systematically categorized using an established classification scheme?
\item \label{cap:paper_ipm2023_bias-sec:research-questions_3} What countermeasures can be employed to prevent the manifestation of cognitive biases in fact-checking?
\item \label{cap:paper_ipm2023_bias-sec:research-questions_4} Can the key components of a bias-aware fact-checking pipeline be defined to minimize the risk of bias manifestation?
\end{enumerate}

\section{Methodology}

\label{cap:paper_ipm2023_bias-sec:methodology}

The \prisma guidelines, a widely adopted standard for reporting systematic reviews and meta-analyses, are introduced. Their application to the identification of literature on cognitive biases is then described (Section~\ref{cap:paper_ipm2023_bias-sec:methodology-subsec:prisma}). Specifically, multiple information sources are consulted to retrieve studies addressing one or more examples of cognitive bias, following a predefined search strategy. Data are extracted from the identified studies according to predefined eligibility criteria (Section~\ref{cap:paper_ipm2023_bias-sec:methodology-subsec:eligibility}). Finally, the resulting list of cognitive biases is filtered through a selection process designed to retain only those that may manifest during the fact-checking process (Section~\ref{cap:paper_ipm2023_bias-sec:methodology-subsec:selection-process}).

\subsection{The PRISMA Guidelines}

\label{cap:paper_ipm2023_bias-sec:methodology-subsec:prisma}

Preferred Reporting Items for Systematic Reviews and Meta-Analyses (\prisma) is an evidence-based minimum set of items for reporting on systematic reviews and meta-analyses. \citet{moher2009preferred} originally proposed the approach in 2009 as a reformulation of the QUORUM\index{QUORUM} guidelines \cite{MOHER19991896}. In 2020, \citet{Pagen71} proposed an updated version known as \lq\lq The PRISMA 2020 Statement\rq\rq. In the following, the acronym \prisma refers to this formulation.

\prisma is a transparent approach that has been widely adopted in various research fields. It aims to assist researchers in conducting high-quality systematic reviews and meta-analyses. Its clear and structured framework facilitates the identification, assessment, and synthesis of relevant data, ensuring that the review process is rigorous, replicable, and unbiased. At its core, \prisma consists of a checklist\footnote{\url{http://prisma-statement.org/PRISMAStatement/Checklist}} and a flow diagram,\footnote{\url{http://prisma-statement.org/PRISMAStatement/FlowDiagram}} both publicly available.

\begin{figure}[tpb]
   \centering
   \includegraphics[width=\linewidth]{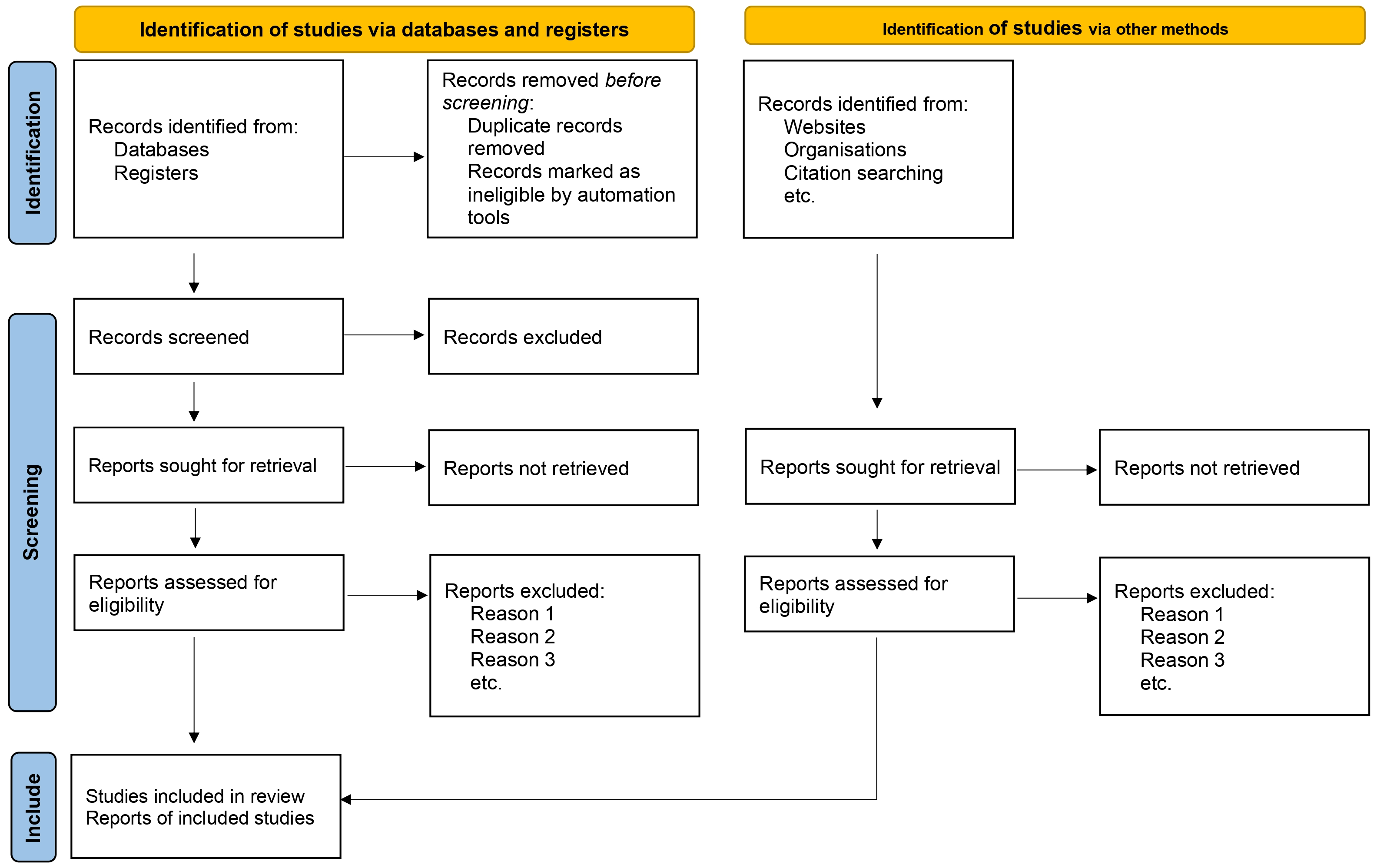}
 \caption{The \prisma flow diagram for new systematic reviews which included searches of databases, registers and other sources. Adapted from \citet{Pagen71}.}
   \label{cap:paper_ipm2023_bias-sec:methodology-subsec:prisma-fig:prisma-flow}
 \end{figure}

The \prisma checklist is composed of 27 items addressing the introduction, methods, results, and discussion sections of a systematic review report, summarized in Table~\ref{cap:paper_ipm2023_bias-sec:methodology-subsec:prisma-tab:prisma-checklist}. Items 10, 13, 16, 20, 23, and 24 are further split into sub-items (not shown in the table). On the other hand, the flow diagram depicts the flow of information through the different phases of a systematic review. It outlines the number of records identified, included, and excluded, along with the rationale for exclusions. The diagram is available in two forms, depending on whether the review is a new contribution or an updated version of an existing one. For the proposed review, being a new contribution, reliance is placed on the former version, as depicted in Figure~\ref{cap:paper_ipm2023_bias-sec:methodology-subsec:prisma-fig:prisma-flow}. If the figure is compared with Table~\ref{cap:paper_ipm2023_bias-sec:methodology-subsec:prisma-tab:prisma-checklist}, further details, mostly pertaining to items 5 to 10, and 16 of the checklist, are detailed in the diagram.

\begin{table}[tbp]
\centering
    \caption{The 27 items of the PRISMA checklist. Adapted from \citet{Pagen71}.}
    \label{cap:paper_ipm2023_bias-sec:methodology-subsec:prisma-tab:prisma-checklist}
\begin{tabular}{cl}
        \toprule
\textbf{Item \#} & \textbf{Section / Topic} \\
\midrule
1 & Title \\
\midrule
2 & Abstract \\
\midrule
3 & Rationale \\
4 & Objectives \\
\midrule
5 &  Eligibility Criteria\\
6 &  Information Sources \\
7 &  Search Strategy \\
8 &  Selection Process\\
9 &  Data Collection Process\\
10 & Data Items \\
11 & Study Risk Of Bias Assessment \\
12 & Effect Measures \\
13 & Synthesis Methods \\
14 & Reporting Bias Assessment \\
15 & Certainty Assessment \\
\midrule
16 & Study Selections \\
17 & Study Characteristics \\
18 & Risk Of Bias In Studies \\
19 & Results Of Individual Studies \\
20 & Results Of Syntheses \\
21 & Reporting Biases \\
22 & Certainty Of Evidences \\
\midrule
23 & Discussion \\
\midrule
24 & Registration And Protocol \\
25 & Support \\
26 & Competing Interests \\
27 & Availability Of Data \\
\bottomrule
\end{tabular}
\end{table}

The use of the checklist and the diagram should adhere to the guidelines provided in the \lq\lq Explanation and Elaboration Document\rq\rq{}~\cite{Pagen160}. This document aims to enhance the use, understanding, and dissemination of a review conducted using \prisma. Additionally, several extensions\footnote{\url{http://prisma-statement.org/Extensions/}} have been developed to facilitate the reporting of different types or aspects of systematic reviews. The \lq\lq PRISMA 2020 Extension For Abstracts\rq\rq{}, published together with the overall statement~\cite{Pagen71}, is a 12-item checklist that gives researchers a framework for condensing their systematic reviews into abstracts for journals or conferences. To summarize, the review proposed is conducted by relying on four main \prisma elements: 
 \begin{enumerate}
 	\item The \prisma 2020 Checklist.
 	\item The \prisma 2020 Flow Diagram for new systematic reviews which include searches of databases, registers and other sources.
 	\item The \prisma 2020 Extension For Abstracts.
 	\item The \prisma 2020 Explanation and Elaboration Document.
 \end{enumerate}
All these resources are part of the overall \lq\lq PRISMA 2020 Statement\rq\rq{}~\cite{Pagen71, Pagen160}, and are used to explore various sources in order to identify literature on cognitive biases through a predefined search strategy. During data collection, biases are extracted based on eligibility criteria. Subsequently, the list of cognitive biases is filtered to focus on those relevant to fact-checking.

Given the goal of identifying and extracting cognitive biases that might manifest in the fact-checking process from the entire set described in the literature, rather than seeking all research papers that address cognitive biases to some extent (Section~\ref{cap:paper_ipm2023_bias-sec:research_questions}), it's not necessary to address all the items provided by the \prisma checklist. Nevertheless, the importance of adhering to the original guidelines has to be recognized. Thus, the checklist items (Table~\ref{cap:paper_ipm2023_bias-sec:methodology-subsec:prisma-tab:prisma-checklist}) that have not been addressed in the review, either because they are not relevant for the goal of finding biases or simply because they are not needed, are initially detailed. In more detail, there is no need to: compute effect measures (Item 12), study heterogeneity and robustness of the synthesized results (Item 13), present assessments of risk of bias for each included study (Item 18), perform statistical analyses (Item 20), present assessments of risk of bias due to missing results for each synthesis (Item 21), present assessments of certainty in the body of evidence for each outcome (Item 22), provide particular registration information about the review (Item 24).

Most of the work performed to adopt \prisma concerns the alterations of the inclusion and exclusion criteria typically applied in a literature review and the collection and selection process of the cognitive biases presented in Section~\ref{cap:paper_ipm2023_bias-sec:methodology-subsec:eligibility} and Section~\ref{cap:paper_ipm2023_bias-sec:methodology-subsec:selection-process}; that is, how the items 5--10 and 13 reported in Table~\ref{cap:paper_ipm2023_bias-sec:methodology-subsec:prisma-tab:prisma-checklist} are approached, and how the processes described by the flow diagram shown in Figure~\ref{cap:paper_ipm2023_bias-sec:methodology-subsec:prisma-fig:prisma-flow} are performed. This tailored adaptation allows for adhering to \prisma's structured approach, incorporating predefined eligibility criteria, search strategies, and data extraction. This approach mitigates the risk of errors in the review process, even though it leads to a slightly different final outcome. In Section~\ref{cap:paper_ipm2023_bias-sec:methodology-subsec:eligibility}, the eligibility criteria, information sources, and search strategy are outlined, while Section~\ref{cap:paper_ipm2023_bias-sec:methodology-subsec:selection-process} provides details on the data collection and selection processes. The complete \lq\lq PRISMA Abstract Checklist\rq\rq and \lq\lq PRISMA Checklist\rq\rq are provided in Appendix~\ref{cap:paper_ipm2023_bias-appendix:prisma}.

\subsection{Eligibility Criteria, Information Sources, And Search Strategy}

\label{cap:paper_ipm2023_bias-sec:methodology-subsec:eligibility}

The process of building the list of cognitive biases that may manifest during the fact-checking process was initiated by defining three eligibility criteria for including or excluding a given work: 
\begin{enumerate}
    \item \label{cap:paper_ipm2023_bias-sec:methodology-subsec:eligibility-c1} Is a given bias described in a peer-reviewed literature work?
    \item \label{cap:paper_ipm2023_bias-sec:methodology-subsec:eligibility-c2} Does the bias have a clear definition? Are its causes and domains of application explained?
    \item \label{cap:paper_ipm2023_bias-sec:methodology-subsec:eligibility-c3} Can we frame a fact-checking related scenario which involves the bias, eventually supported by existing literature?
\end{enumerate}

In line with \prisma statement guidelines that emphasize the importance of achieving a balance between precision and recall based on review goals, eligibility criteria were defined to identify all cognitive biases with the outlined characteristics. Biases were excluded if they lacked a clear definition, were not well-established, or were considered irrelevant to fact-checking.

In terms of information sources, over 200 biases are documented on publicly available web pages. Wikipedia, for instance, catalogs 227 biases,\footnote{\url{https://en.wikipedia.org/wiki/List_of_cognitive_biases}} while The Decision Lab, an applied research firm, offers a list of 103 biases.\footnote{\url{https://thedecisionlab.com/biases/}} Additionally, various researchers have compiled their lists of cognitive biases; for example, \citet{dimara2018task} identify 154, and \citet{hilbert2012toward} lists 9. The search strategy involved exploring the literature retrieved by performing manual searches using each bias name as a query. The databases utilized include \texttt{Google Scholar},\footnote{\url{https://scholar.google.com/}}\index{Google!Scholar} 
\texttt{Scopus},\footnote{\url{https://www.scopus.com/}}\index{Scopus} 
\texttt{PubMed},\index{PubMed}\footnote{\url{https://pubmed.ncbi.nlm.nih.gov/}} 
\texttt{ACM Digital Library},\footnote{\url{https://dl.acm.org/}}\index{ACM Digital Library}
\texttt{Wiley Online Library},\footnote{\url{https://onlinelibrary.wiley.com/}}\index{Wiley Online Library}
\texttt{ACL Anthology},\footnote{\url{https://aclanthology.org/}}\index{ACL Anthology} and 
\texttt{DBLP}.\footnote{\url{https://dblp.org/}}\index{DBLP} The search terms comprised various combinations of keywords, including \spverb|cognitive bias|, \spverb|heuristics|, \spverb|systematic error|, \spverb|judgment|, and \spverb|decision-making|.

\subsection{Data Collection And Selection Process}

\label{cap:paper_ipm2023_bias-sec:methodology-subsec:selection-process}

\begin{figure}[tpb]
   \centering
   \includegraphics[width=.8\linewidth]{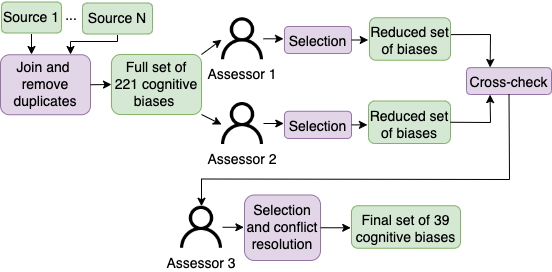}
 \caption{Data collection and selection process of the review.}
   \label{cap:paper_ipm2023_bias-sec:methodology-subsec:selection-process-fig:selection-draw}
 \end{figure}

The complete methodology for data collection and selection to which this review adheres is derived from the diagram shown in Figure~\ref{cap:paper_ipm2023_bias-sec:methodology-subsec:prisma-fig:prisma-flow} and presented in Figure~\ref{cap:paper_ipm2023_bias-sec:methodology-subsec:selection-process-fig:selection-draw}. The lists obtained from the information sources are consolidated by removing duplicates and conducting disambiguation for each bias, resulting in the final amount of \numbias cognitive biases. Given that a standard conceptualization or classification of biases is a debated issue \cite{gigerenzer2008bounded, hilbert2012toward}, and with the objective of maximizing the number of identified cognitive biases, two biases are included even if their differences are subtle. 

To select the cognitive biases that might manifest during the fact-checking process, each of the \numbias cognitive biases found in the literature and compiled in the final list underwent an analysis. This analysis, conducted individually for each bias, focused on their definitions, causes, and domains of application. In other words, each bias was evaluated against the eligibility criteria outlined in Section~\ref{cap:paper_ipm2023_bias-sec:methodology-subsec:eligibility}.
  
The selection process was conducted in the following manner: two authors of the review (referred to as \emph{Assessor 1}\index{Assessor!1} and \emph{Assessor 2}\index{Assessor!2}) individually and independently examined the entire set of \numbias biases. Both assessors, consequently, reviewed the comprehensive list of cognitive biases identified in the literature, examining each bias definition along with a collection of practical examples for each. Additionally, they provided justification for the inclusion or exclusion of specific biases by giving an example of their manifestation in a fact-checking scenario. Then, \emph{Assessor 1}\index{Assessor!1} and \emph{Assessor 2}\index{Assessor!2} compared their respective lists, discussed any conflicts that arose, and reached a consensus. In order to maximize recall, they chose to include a bias in the list even if its likelihood of manifesting was relatively low. After such a step, a third author (referred to as \emph{Assessor 3}\index{Assessor!3}) reviewed the finalized, conflict-free list of biases to ensure its comprehensiveness and consistency. While the selection process inherently involves some degree of subjectivity, the incorporation of discussion points, redundancy, and cross-checks establishes a robust and reliable methodology for identifying pertinent cognitive biases in the context of fact-checking.

The detailed process resulted in a list of \numbiasused cognitive biases that may manifest during fact-checking. It is important to note that the proposed list of cognitive biases should not be considered final but rather should be updated as new evidence of the effects of specific cognitive biases is published in the literature.

\section{Results}

\label{cap:paper_ipm2023_bias-sec:results}

Section~\ref{cap:paper_ipm2023_bias-sec:results-subsec:list} presents the list of \numbiasused cognitive biases selected and investigated. Their categorization is described in Section~\ref{cap:paper_ipm2023_bias-sec:results-subsec:categorization}. A list of \numcountermeasures countermeasures is proposed in Section~\ref{cap:paper_ipm2023_bias-sec:results-subsec:countermeasures}, and Section~\ref{cap:paper_ipm2023_bias-sec:results-sect:ideal-pipeline} introduces the bias-aware assessment pipeline to be applied in the context of fact-checking.

\subsection{\ref{cap:paper_ipm2023_bias-sec:research-questions_1}: List Of Cognitive Biases}

\label{cap:paper_ipm2023_bias-sec:results-subsec:list}

The \numbiasused cognitive biases, identified through the process outlined in Section~\ref{cap:paper_ipm2023_bias-sec:methodology}, are presented in alphabetical order. For each bias, a reference to the literature proposing a psychological explanation is provided, along with a brief description. Additionally, a scenario related to fact-checking, where such bias can manifest, is presented. Subsequently, when available, a reference related to fact-checking is provided to support this framing. The list of \numbiasused cognitive biases selected is presented below, while the complete list of the \numbias cognitive biases found is reported in Appendix~\ref{cap:paper_ipm2023_bias-appendix:biases}.

\begin{enumerate}[label=B\arabic*., leftmargin=*, font=\sffamily]
\item \mybiasname{cap:paper_ipm2023_bias-bias:affect-heuristic}{Affect Heuristic}~\cite{SLOVIC20071333}. To often rely on emotions, rather than concrete information, when making decisions. This allows one to conclude quickly and easily, but can also distort the reasoning and lead to making suboptimal choices. This bias can manifest when the assessor likes, for example, the speaker of an information item.

\item \mybiasnamecustomindex{cap:paper_ipm2023_bias-bias:anchoring-bias}{Anchoring Effect}{Anchoring Effect}~\cite{ni2019anchoring}. To rely too much on an information item (typically the first one acquired) when making a decision. This bias can occur when the assessor inspects more than one source of information when assessing the truthfulness of an information item~\cite{doi:10.1177/10776990211021800}. 

\item \mybiasname{cap:paper_ipm2023_bias-bias:attentional-bias}{Attentional Bias}~\cite{bar2007threat}. To misperceive because of recurring thoughts. This effect may occur due to the overwhelming amount of certain topics on news media over time, for example for an assessor who is asked to evaluate the truthfulness of \covid related information items \cite{Lee2023}.

\item \mybiasname{cap:paper_ipm2023_bias-bias:authority-bias}{Authority Bias} (also called \textsf{Halo Effect}\index{Halo Effect})~\cite{ries2006understanding}. To attribute higher accuracy to the opinion of an authority figure (unrelated to its content) and be more influenced by that opinion. This bias can manifest when the assessor is shown the speaker/organization making the information item~\cite{10.1093/cje/beac071}. 
     
\item \mybiasname{cap:paper_ipm2023_bias-bias:automation-bias}{Automation Bias}~\cite{cummings2017automation}. To rely on automated systems which might override correct decisions made by a human assessor. This bias can occur when the assessor is presented with the outcome of an automated system that is designed to help him/her to make an informed decision on a given information item.

\item \mybiasnamecustomindex{cap:paper_ipm2023_bias-bias:availability-cascade}{Availability Cascade}{Availability!Cascade}~\cite{kuran1998availability}. To attribute a higher plausibility to a belief just because it is public and more \lq\lq available\rq\rq{}. This bias might occur when the information item presented to the assessor contains popular beliefs or popular facts~\cite{effectiviology-availability}.

\item \mybiasnamecustomindex{cap:paper_ipm2023_bias-bias:availability-heuristic}{Availability Heuristic}{Availability!Heuristic}~\cite{groome2016introduction}: to overestimate the likelihood of events that are recent in the memory. This bias can occur when the assessors are evaluating recent information items~\cite{Hayibor2009}.
    
\item \mybiasname{cap:paper_ipm2023_bias-bias:backfire-effect}{Backfire Effect}~\cite{wood2019elusive}. To increase one's own original belief when presented with opposed evidence. This bias, which is based on a typical human reaction, can in principle always occur in fact-checking~\cite{SWIRETHOMPSON2020286}. 

\item \mybiasname{cap:paper_ipm2023_bias-bias:bandwagon-effect}{Bandwagon Effect}~\cite{kiss2014identifying}. To do (or believe) things because many other people do (or believe) the same. This bias manifests for example when an assessor is asked to evaluate an information item related to recent or debated topics, for which the media coverage is high.
    
\item \mybiasname{cap:paper_ipm2023_bias-bias:barnum-effect}{Barnum Effect} (also called \textsf{Forer Effect}\index{Forer Effect})~\cite{fichten1983popular}. To fill the gaps in vague information by including personal experiences or information. This bias can in principle always occur in fact-checking~\cite{ESCOLAGASCON2023111893}.
    
\item \mybiasname{cap:paper_ipm2023_bias-bias:base-rate-fallacy}{Base Rate Fallacy}~\cite{welsh2012seeing}. To focus on specific parts of information which support an information item and ignore the general information. This bias is related to the fact that the assessors are asked to report the piece of text or sources of information motivating their assessment.
    
 \item \mybiasnamecustomindex{cap:paper_ipm2023_bias-bias:belief-bias}{Belief Bias}{Belief!Bias}~\cite{leighton2004nature}. To attribute too much logical strength to someone's argument because of the validity of the conclusion. This bias is most likely to occur when the assessors are asked to evaluate factual information items~\cite{doi:10.1073/pnas.2104235118}.

\item \mybiasname{cap:paper_ipm2023_bias-bias:choice-supportive-bias}{Choice-Supportive Bias}~\cite{kafaee2021choice}. To remember one's own choices as better than they actually were. This bias might occur when an assessor is asked to perform a task more than one time, or when they is asked to revise their judgment; it might prevent assessors from revising their initially submitted score~\cite{10.3389/fpsyg.2017.02062}.
    
 \item \mybiasname{cap:paper_ipm2023_bias-bias:compassion-fade}{Compassion Fade}~\cite{leighton2004nature}. To act more compassionately towards a small group of victims. This bias can occur for example when the information item to be evaluated is related to minorities, or tragic events~\cite{doi:10.1177/1461444818760819}.
    
\item \mybiasnamecustomindex{cap:paper_ipm2023_bias-bias:confirmation-bias}{Confirmation Bias}{Confirmation Bias}~\cite{nickerson1998confirmation}. To focus on or to search for the information item which confirms prior beliefs. This bias can in principle always occur in fact-checking, for example if the assessor receives a true information that contradicts their prior beliefs, or if the assessor is asked to provide supporting evidence for their evaluation of such an information item.

\item \mybiasname{cap:paper_ipm2023_bias-bias:conjuction-fallacy}{Conjunction Fallacy} (also called \textsf{Linda Problem}\index{Linda Problem})~\cite{tversky1983extensional}. To assume that a conjunct event is more probable than a constituent event. \citet{tversky1983extensional} presented this effect with a well-known example. They propose the following description: \lq\lq Linda is 31 years old, single, outspoken, and very bright. She majored in philosophy. As a student, she was deeply concerned with issues of discrimination and social justice, and also participated in antinuclear demonstrations.\rq\rq{} asking afterwards which of the following alternative is more probable: 
	\begin{enumerate}[label=(\alph*)]
		\item Linda is a bank teller (\emph{constituent event}\index{Constituent Event}).
		\item Linda is a bank teller and is active in the feminist movement (\emph{conjunct event}\index{Conjunct Event}).
	\end{enumerate}
	Participants' intuitive methods for assessing probability often resulted in them concluding that the latter option was more likely than the former. Thus, this bias can potentially arise when an information item attributes simultaneous occurrences to a specific causal event, a pattern often found in conspiracy theories related to \covid~\cite{10.1002/acp.3860}.
	
\item \mybiasname{cap:paper_ipm2023_bias-bias:conservatism-bias}{Conservatism Bias}~\cite{luo1conservatism}. To revise one's belief insufficiently when presented with new evidence. Note that this bias is different from \mycitebiasnoindex{Con\-fir\-ma\-tion Bias}{\ref{cap:paper_ipm2023_bias-bias:confirmation-bias}}{Confirmation Bias}: the former deals with the revision of a belief, while the latter deals with new information. This bias, like \mycitebiasnoindex{Con\-fir\-ma\-tion Bias}{\ref{cap:paper_ipm2023_bias-bias:confirmation-bias}}{Confirmation Bias}, can occur when the assessor is asked to provide supporting evidence for their evaluation of such an information item.
    
\item \mybiasname{cap:paper_ipm2023_bias-bias:consistency-bias}{Consistency Bias}~\cite{clark2007stereotypes}. To attribute past events as resembling present behavior. This bias might occur when the assessor has evaluated an information item in the past and is asked to assess another information item coming from the same speaker and/or party.

\item \mybiasname{cap:paper_ipm2023_bias-bias:courtesy-bias}{Courtesy Bias}~\cite{jones1993courtesy}. To give a socially accepted answer to avoid offending anyone. This bias is often influenced by the assessor's personal experience and background, as well as the specific context in which they is providing their answer.
    
\item \mybiasname{cap:paper_ipm2023_bias-bias:declinism}{Declinism}~\cite{etchells2015declinism}. To see the past with a positive connotation and the future with a negative one. This bias is related to the temporal part of the information items that the assessor is evaluating~\cite{doi:10.1080/09636412.2022.2133626}.
    
\item \mybiasname{cap:paper_ipm2023_bias-bias:dunning-kruger-effect}{Dunning-Kruger Effect}~\cite{dunning2011dunning}. To overestimate oneself competence due to a lack of knowledge and skill in a certain area. This bias can occur typically to non-expert individuals, e.g., when an assessor is not trained and is overconfident about a given subject or certain topics. It is more likely to manifest with non-expert assessors in general, such as crowd workers, than expert fact-checkers like journalists.
 
 \item \mybiasname{cap:paper_ipm2023_bias-bias:framing-effect}{Framing Effect}~\cite{malenka1993framing}. To draw different conclusions from logically equivalent information items based on the context, the alternatives, and the presentation method. This bias is likely to manifest, for example, when negative and positive equivalents of information items to assess refer to two counterparts of a property that have respectively negative and positive connotations, and its presentation is in terms of the share that belongs to either one or the other of these counterparts \cite{doi:10.1177/10776990221117117}. For example, consider a scenario where a natural disaster claims the lives of 400 out of 1000 people, leaving 600 survivors. An information item could describe the outcome by stating either that:
  	 \begin{enumerate}[label=(\alph*)]
  	 	\item \lq\lq 60\% of people survived the disaster\rq\rq.
  	 	\item \lq\lq 40\% of people did not survive the disaster\rq\rq.
  	 \end{enumerate}
  	 That is, by framing the fact either positively or negatively.
  	 
\item \mybiasname{cap:paper_ipm2023_bias-bias:fundamental-attribution-error}{Fundamental Attribution Error}~\cite{harvey1981fundamental}. To under-emphasize situational and environmental factors for the behavior of an actor while over-emphasizing dispositional or personality factors. This bias is likely to manifest while fact-checking information items made by politically aligned speakers or news outlets, as in the case of a politician who says that people are poor because they are lazy. Furthermore, it is more likely to affect younger or older human assessors, since age differences influence its manifestation~\cite{10.1093/geronb/57.4.P312}.
     
\item \mybiasnamecustomindex{cap:paper_ipm2023_bias-bias:google-effect}{Google Effect}{Google!Effect}~\cite{brabazon2006google}. To forget information that can be found readily online by using search engines. This bias can manifest when a worker is required to use a search engine to find evidence, and/or when tasked to assess an information item at different time spans. For example, an assessor forgetting part of the information item right after reading it, because they knows that it is easily retrievable again if needed by querying a search engine~\cite{10.1145/3201064.3201095}.

\item \mybiasname{cap:paper_ipm2023_bias-bias:hindsight-bias}{Hindsight Bias} (also called \textsf{\lq\lq I-knew-it-all-along\rq\rq{} Effect} \index{I-knew-it-all-along Effect})~\cite{roese2012hindsight}. To see past events as being predictable at the time those events happened. Since it may cause distortions of memories of what was known before an event, this bias may manifest when an assessor is required to evaluate an event after some time or when is asked to evaluate the same information item multiple times at different time spans~\cite{Hom2022}.
     
\item \mybiasname{cap:paper_ipm2023_bias-bias:hostile-attribution-bias}{Hostile Attribution Bias}~\cite{pornari2010peer}. To interpret someone's behavior as hostile even if it is not. This bias can occur for assessors who have experienced discrimination from authority figures or the dominant social group. For example, a speaker from an under-represented ethnicity remarks oneself that they perceive something as offensive, thus fueling hostility~\cite{10.1002/ab.21655}.

\item \mybiasname{cap:paper_ipm2023_bias-bias:illusion-of-validity}{Illusion of Validity}~\cite{einhorn1978confidence}. To overestimate someone's judgment when the available information is consistent. This bias can occur for example when an assessor works with a set of previously true information items from a specific person and predicts that the subsequent set of information items will have the same outcome from the same person. 

\item \mybiasnamecustomindex{cap:paper_ipm2023_bias-bias:illusory-correlation}{Illusory Correlation}{Illusory!Correlation}~\cite{hamilton1976illusory}. To perceive the correlation between non-correlated events. This bias can manifest when an assessor works on multiple information items in a single task and may perceive nonexistent patterns between the items.

\item \mybiasnamecustomindex{cap:paper_ipm2023_bias-bias:illusory-truth-effect}{Illusory Truth Effect}{Illusory!Truth Effect}~\cite{newman2020truthiness}. To perceive an information item as true if it is easier to process or it has been stated multiple times. This bias can manifest for example when using straightforward or naive gold questions in a task to check for malicious assessors~\cite{BRASHIER2020104054}.

\item \mybiasname{cap:paper_ipm2023_bias-bias:ingroup-bias}{Ingroup Bias}~\cite{mullen1992ingroup}. To favor people belonging to one's own group. This bias can manifest for example when the assessors are required to work on information items related to their own political party, city, etc~\cite{10.1111/jcom.12284}.
    
\item \mybiasname{cap:paper_ipm2023_bias-bias:just-world-hypotesis}{Just-World Hypothesis}~\cite{lerner1978just}. To believe that the world is just. This bias can happen for example when the assessor is working with information items related to major political institutions, as people tend to assign to them higher scores in belief~\cite{rubin1975believes}.

\item \mybiasname{cap:paper_ipm2023_bias-bias:optimism-bias}{Optimism Bias} (also called \textsf{Optimistic Bias})\index{Optimistic Bias}~\cite{sharot2011optimism}. To be over-optimistic, underestimating the probability of undesirable outcomes and overestimating favorable and pleasing outcomes. This bias can occur for example when the information item provides some kind of assessment of the risk of an event manifesting, such as the likelihood of getting infected by \covid~\cite{g11030039}.
    
\item \mybiasnamecustomindex{cap:paper_ipm2023_bias-bias:ostrich-effect}{Ostrich Effect}{Ostrich!Effect} (also called \index{Ostrich!Problem}\textsf{Ostrich Problem}) \cite{karlsson2009ostrich}. To avoid potentially negative but useful information, such as feedback on progress, to avoid psychological discomfort. This bias can occur when an assessor avoids evaluating, explicitly or not, an information item which has a negative connotation according to their own beliefs.

\item \mybiasname{cap:paper_ipm2023_bias-bias:outocome-bias}{Outcome Bias}~\cite{baron1988outcome}. To judge a decision by its eventual outcome instead on the basis of the quality of the decision at the time it was made. This bias can manifest when the information item under consideration is related to a past event~\cite{robson2019outcome}.

\item \mybiasname{cap:paper_ipm2023_bias-bias:overconfidence-effect}{Overconfidence Effect}~\cite{dunning1990overconfidence}. To be too confident in one's own answers. This effect can manifest when the assessor is an expert in the field, as for example an expert journalist who performs fact-checking related to their writing or a medical specialist who assesses health-related information items.

\item \mybiasname{cap:paper_ipm2023_bias-bias:proportionality-bias}{Proportionality Bias} (also called \index{Major Event/Major Cause Heuristic}\textsf{Major Event/Major Cause Heuristic})~\cite{leman2007major}. To assume that big events have big causes. This innate human tendency can also explain why some individuals accept conspiracy theories. This bias can occur when the factual information being assessed deals with the causes and effects of a particular event~\cite{https://doi.org/10.1002/acp.3998}.
    
\item \mybiasname{cap:paper_ipm2023_bias-bias:salience-bias}{Salience Bias}~\cite{mullen1992ingroup}. To focus on items that are more prominent or emotionally striking and ignore those that are unremarkable, even though this difference is irrelevant by objective standards. For example, an information item detailing the numerous deaths of infants will receive more attention than an information item detailing a less emotionally striking fact. If those two facts are presented in the same information item, the score assessed for the prominent fact might drive the overall assessment of the whole information item.

\item \mybiasname{cap:paper_ipm2023_bias-bias:stereotypical-bias}{Stereotypical Bias}~\cite{heilman2012gender}. To discriminate against a personal trait (e.g., gender). Like \mycitebias{Ingroup Bias}{\ref{cap:paper_ipm2023_bias-bias:ingroup-bias}}, this bias can happen when the assessor, especially a crowd worker, identifies themselves with the group related to the information item they is assessing.
    
\item \mybiasname{cap:paper_ipm2023_bias-bias:telescoping-effect}{Telescoping Effect}~\cite{thompson1988telescoping}. To displace recent events backward in time and remote events forward in time, so that recent events appear more remote, and remote events, more recent. This bias might occur when the information item presented contains temporal references.

\end{enumerate}

\subsection{\ref{cap:paper_ipm2023_bias-sec:research-questions_2}: Categorization Of Cognitive Biases}

\label{cap:paper_ipm2023_bias-sec:results-subsec:categorization}

Considering at the same time all the \numbiasused biases that might manifest while performing fact-checking can be challenging for researchers and practitioners that aim studying their manifestation and/or impact in fact-checking settings. Thus, providing a second level of aggregation might support laying out the problem and facilitate further analysis, for instance by considering the type of fact-checking related task. To this end, the \numbiasused biases are further categorized from a task-based perspective, utilizing the classification scheme proposed by \citet{dimara2018task}. This scheme has been employed to address cognitive biases specifically affecting information visualization tasks. This scheme allows for aggregating the initial list based on the psychological explanations of why biases might occur in a fact-checking related context, as reported in Section~\ref{cap:paper_ipm2023_bias-sec:results-subsec:list}.

\citeauthor{dimara2018task} scheme for classifying cognitive biases works as follows. They first identify user tasks that might involve cognitive biases, thus generating a set of 7 task categories~\cite[Section~3.4]{dimara2018task} using open card-sorting analysis~\cite{10.5555/2835577.2835578}. The user task categories found are summarized in Table~\ref{cap:paper_ipm2023_bias-sec:results-subsec:categorization-tab:task-types}. Then, they focus on the relevance of cognitive biases with information visualization aspects. The initial classification of cognitive biases according to the user task included a fairly large number of biases for each category. Thus, they further refined the overall scheme by proposing a set of 5 sub-categories called \textsf{flavors}~\cite[Section~3.4]{dimara2018task} that focus on other types of similarities across cognitive biases, related with how each bias affects human cognition. Such flavors are summarized in Table~\ref{cap:paper_ipm2023_bias-sec:results-subsec:categorization-tab:task-flavors}.

\begin{table}[tbp]
\centering
\caption{User tasks that may involve cognitive biases, as proposed by~\citet{dimara2018task}.}
\label{cap:paper_ipm2023_bias-sec:results-subsec:categorization-tab:task-types}
\begin{tabular}{p{3.5cm}p{9cm}}
\toprule
\textbf{Task} & \textbf{Description} \\
\midrule
\textsf{Causal Attribution} & Tasks involving an assessment of causality. \\
\midrule
\textsf{Decision} & Tasks involving the selection of one over several alternative options. \\
\midrule
\textsf{Estimation} & Tasks where people are asked to assess the value of a quantity. \\
\midrule
\textsf{Hypothesis Assessment} & Tasks involving an investigation of whether one or more hypotheses are true or false. \\
\midrule
\textsf{Opinion Reporting} & Tasks where people are asked to answer questions regarding their beliefs or opinions on political, moral, or social issues. \\
\midrule
\textsf{Recall} & Tasks where people are asked to recall or recognize previous material. \\
\midrule
\textsf{Other} & Tasks which are not included in one of the previous categories.\\
\bottomrule
\end{tabular}
\end{table}

\begin{table}[tbp]
\centering
\caption{Phenomena that affect human cognition, as proposed by~\citet{dimara2018task}.}
\label{cap:paper_ipm2023_bias-sec:results-subsec:categorization-tab:task-flavors}
\begin{tabular}{p{2.5cm}p{10cm}}
\toprule
\textbf{Flavor} & \textbf{Description} \\
\midrule
\textsf{Association} & Cognition is biased by associative connections between information items. \\
\midrule
\textsf{Baseline} & Cognition is biased by comparison with (what is perceived as) a baseline. \\
\midrule
\textsf{Inertia} & Cognition is biased by the prospect of changing the current state. \\
\midrule
\textsf{Outcome} & Cognition is biased by how well something fits an expected or desired outcome. \\
\midrule
\textsf{Self-Perspective} & Cognition is biased by a self-oriented viewpoint. \\ 
\bottomrule
\end{tabular}
\end{table}

The set of \numbiasused that might manifest in the fact-checking activity is thus categorized by assessing their potential impact on different types of tasks and how they influence human cognition. This categorization involves assigning biases to specific task/flavor combinations according to the scheme by \citet[Table~2]{dimara2018task}. The categorization process involved evaluating each bias identified in the initial list against the seven task categories and five flavors. This was accomplished by initially determining the most likely fact-checking task that each bias could influence (Table~\ref{cap:paper_ipm2023_bias-sec:results-subsec:categorization-tab:task-types}). Following this, an analysis was conducted to examine how each bias affects human cognition, aligning them with one of the flavors (Table~\ref{cap:paper_ipm2023_bias-sec:results-subsec:categorization-tab:task-flavors}). This dual-level categorization facilitates a detailed understanding of how each bias could potentially manifest in various aspects of fact-checking. Within the list of \numbiasused cognitive biases, 35 biases are classified using the same task/flavor combination in both the classification considered by \citeauthor{dimara2018task} and the one conducted in this review, although the two classifications address cognitive biases in different contexts. Furthermore, four biases are not considered by \citeauthor{dimara2018task} in their classification. These biases are:  
\begin{itemize}
	\item[--] \mycitebias{Consistency Bias}{\ref{cap:paper_ipm2023_bias-bias:consistency-bias}}.
	\item[--] \mycitebias{Courtesy Bias}{\ref{cap:paper_ipm2023_bias-bias:courtesy-bias}}.
	\item[--] \mycitebias{Proportionality Bias}{\ref{cap:paper_ipm2023_bias-bias:proportionality-bias}}.
	\item[--] \mycitebias{Salience Bias}{\ref{cap:paper_ipm2023_bias-bias:salience-bias}}.
\end{itemize}
An in-depth analysis has been conducted to appropriately assign these biases to both a task category and a flavor. This involved assessing the nature and implications of each bias and determining the most relevant task and flavor based on their characteristics and impact on cognitive processes during fact-checking.

The task/flavor classification outlined in Table~\ref{cap:paper_ipm2023_bias-sec:results-subsec:categorization-tab:bias-taxonomy-alt} offers a structured and detailed approach to understanding the multifaceted ways in which cognitive biases can impact the fact-checking process. It has to be acknowledged that, unlike the selection of the \numbiasused biases, where a structured \prisma-based approach was feasible, this classification is inherently more subjective, as it relies on agreement between evaluators. By mapping each bias to specific fact-checking tasks and cognitive influences, a comprehensive framework is provided. This framework aims to assist researchers and practitioners in identifying and addressing potential biases in their work.

To provide an interpretation of the categorization scheme based on tasks and flavors in a fact-checking context, consider the example of a \textsf{Decision} task as defined by~\citet{dimara2018task} (Table~\ref{cap:paper_ipm2023_bias-sec:results-subsec:categorization-tab:task-types}). In this scenario, an assessor is asked to choose which information item is more truthful among two different alternatives. Further hypothesized is that the two information items are made by politicians, where one of them belongs to the governing party. In such a case, one could argue that the trustworthiness of the speaker might be linked to the assessor being affected by the \mycitebias{Authority Bias}{\ref{cap:paper_ipm2023_bias-bias:authority-bias}}, as the assessor might believe that being part of the governing party implies higher trustworthiness for the speaker. Moreover, since reasoning (or, as~\citeauthor{dimara2018task} refer to it, cognition) is biased by an associative connection between the two pieces of information, the underlying flavor can be concluded to be \textsf{Association}.

\begin{table}[htbp]
\centering
\caption{Categorization of cognitive biases, adapted from \citet{dimara2018task}.}
\label{cap:paper_ipm2023_bias-sec:results-subsec:categorization-tab:bias-taxonomy-alt}
\begin{tabular}{@{}P{2.1cm}@{~}P{2.9cm}@{~}P{2.9cm}@{~}P{2.9cm}@{~}P{2.9cm}@{~}P{2.9cm}@{}} 
    \toprule
    &\textbf{Association} & \textbf{Baseline} & \textbf{Inertia} & \textbf{Outcome} & \textbf{Self-Perspective} \\
\midrule
\textbf{Causal Attribution} 
& -- & -- & -- &
\mycitebias{Hostile Attribution Bias}{\ref{cap:paper_ipm2023_bias-bias:hostile-attribution-bias}}
\mycitebiashyphennoindex{Just}{World Hypothesis}{\ref{cap:paper_ipm2023_bias-bias:just-world-hypotesis}}{Just-World Hypothesis}
&
\mycitebiasnoindex{Fun\-da\-men\-tal Attribution Error}{\ref{cap:paper_ipm2023_bias-bias:fundamental-attribution-error}}{Fundamental Attribution Error}
\mycitebias{Ingroup Bias}{\ref{cap:paper_ipm2023_bias-bias:ingroup-bias}}
\\
\midrule
\textbf{Decision} 
&
\mycitebias{Authority Bias}{\ref{cap:paper_ipm2023_bias-bias:authority-bias}}
\mycitebias{Automation Bias}{\ref{cap:paper_ipm2023_bias-bias:automation-bias}}

\mycitebias{Framing Effect}{\ref{cap:paper_ipm2023_bias-bias:framing-effect}}
& -- & -- & -- & --
\\
\midrule
\textbf{Estimation} 
&
\mycitebiasnoindex{A\-vail\-a\-bil\-i\-ty Heuristic}{\ref{cap:paper_ipm2023_bias-bias:availability-heuristic}}{Availability!Heuristic}
\mycitebias{Conjunction Fallacy}{\ref{cap:paper_ipm2023_bias-bias:conjuction-fallacy}}
&
\mycitebiasnoindex{Anchoring Ef\-fect\xspace\xspace\xspace\xspace\xspace\xspace\xspace\xspace\xspace\xspace}{\ref{cap:paper_ipm2023_bias-bias:anchoring-bias}}{Anchoring Effect}
 \mycitebiasnoindex{Base Rate Fal\-la\-cy\xspace\xspace\xspace\xspace}{\ref{cap:paper_ipm2023_bias-bias:base-rate-fallacy}}{Base Rate Fallacy}
 \mycitebiasnoindex{Compassion Fade\xspace\xspace\xspace\xspace}{\ref{cap:paper_ipm2023_bias-bias:compassion-fade}}{Compassion Fade}
 \mycitebiashyphennoindex{Dunning}{Kru\-ger Effect}{\ref{cap:paper_ipm2023_bias-bias:dunning-kruger-effect}}{Dunning-Kruger Effect}
 \mycitebiasnoindex{O\-ver\-con\-fi\-dence Effect}{\ref{cap:paper_ipm2023_bias-bias:overconfidence-effect}}{Overconfidence Effect}
&
\mycitebiasnoindex{Con\-ser\-vatism Bias}{\ref{cap:paper_ipm2023_bias-bias:conservatism-bias}}{Conservatism Bias}
&
\mycitebias{Illusion of Validity}{\ref{cap:paper_ipm2023_bias-bias:illusion-of-validity}}
\mycitebias{Outcome Bias}{\ref{cap:paper_ipm2023_bias-bias:outocome-bias}}
&
\mycitebias{Optimism Bias}{\ref{cap:paper_ipm2023_bias-bias:optimism-bias}}
\mycitebias{Salience Bias}{\ref{cap:paper_ipm2023_bias-bias:salience-bias}}
\\
\midrule
\textbf{Hypothesis Assessment} 
&
 \mycitebiasnoindex{Availability Cascade}{\ref{cap:paper_ipm2023_bias-bias:availability-cascade}}{Availability!Cascade}
 \mycitebiasnoindex{Illusory Truth Effect}{\ref{cap:paper_ipm2023_bias-bias:illusory-truth-effect}}{Illusory!Truth Effect}
& -- & -- &
\mycitebias{Barnum Effect}{\ref{cap:paper_ipm2023_bias-bias:barnum-effect}}

\mycitebiasnoindex{Belief Bias}{\ref{cap:paper_ipm2023_bias-bias:belief-bias}}{Belief!Bias}

\mycitebiasnoindex{Con\-fir\-ma\-tion Bias}{\ref{cap:paper_ipm2023_bias-bias:confirmation-bias}}{Confirmation Bias}

\mycitebiasnoindex{Illusory Correlation}{\ref{cap:paper_ipm2023_bias-bias:illusory-correlation}}{Illusory!Correlation}
& --
\\
\midrule
\textbf{Opinion Reporting} 
& -- & 
\mycitebiasnoindex{Pro\-por\-tion\-al\-i\-ty Bias}{\ref{cap:paper_ipm2023_bias-bias:proportionality-bias}}{Proportionality Bias}
&
\mycitebias{Backfire Effect}{\ref{cap:paper_ipm2023_bias-bias:backfire-effect}}
&
\mycitebias{Bandwagon Effect}{\ref{cap:paper_ipm2023_bias-bias:bandwagon-effect}}
\mycitebiasnoindex{Ster\-e\-o\-typ\-i\-cal Bias}{\ref{cap:paper_ipm2023_bias-bias:stereotypical-bias}}{Stereotypical Bias}
&
\mycitebias{Courtesy Bias}{\ref{cap:paper_ipm2023_bias-bias:courtesy-bias}}
\\
\midrule
\textbf{Recall} 
&
\mycitebiasnoindex{Google Effect}{\ref{cap:paper_ipm2023_bias-bias:google-effect}}{Google!Effect}
\mycitebias{Telescoping Effect}{\ref{cap:paper_ipm2023_bias-bias:telescoping-effect}}
& -- &
\mycitebias{Con\-sist\-en\-cy Bias}{\ref{cap:paper_ipm2023_bias-bias:consistency-bias}}
&
\mycitebiashyphennoindex{Choice}{Supportive Bias}{\ref{cap:paper_ipm2023_bias-bias:choice-supportive-bias}}{Choice-Supportive Bias}
\mycitebias{Declinism}{\ref{cap:paper_ipm2023_bias-bias:declinism}}

\mycitebias{Hindsight Bias}{\ref{cap:paper_ipm2023_bias-bias:hindsight-bias}}
& -- 
\\
\midrule
\textbf{Other} 
&
\mycitebias{Attentional Bias}{\ref{cap:paper_ipm2023_bias-bias:attentional-bias}}
& -- & -- &
\mycitebiasnoindex{Ostrich Effect}{\ref{cap:paper_ipm2023_bias-bias:ostrich-effect}}{Ostrich!Effect}
& --
\\
\bottomrule
\end{tabular}
\end{table}

\subsection{\ref{cap:paper_ipm2023_bias-sec:research-questions_3}: List Of Countermeasures}

\label{cap:paper_ipm2023_bias-sec:results-subsec:countermeasures}

The literature allows for specifying \numcountermeasures countermeasures that can be employed in a fact-checking context to help prevent manifesting the cognitive biases outlined in Table \ref{cap:paper_ipm2023_bias-sec:results-subsec:categorization-tab:bias-taxonomy-alt}. Each countermeasure is detailed in the following (\textsf{C1}--\textsf{C\numcountermeasures}). 

The countermeasure selection process involves two steps. First, the literature is inspected to identify works addressing specific biases, and second, countermeasures applicable when assessing the truthfulness of an information item are chosen. This approach details how to remove individual biases; however, it should be noted that the removal of one bias as a result of the application of a countermeasure might lead to the manifestation of another one. For instance, \citet{Park_Park_Kang_2021} show that unexpected biases often arise in a fact-checking scenario. Therefore, there might not exist a systematic way to safely remove all the possible sources of bias altogether. Hence, researchers and practitioners should aim at finding a good compromise between the possibility of bias manifestation and the specific experimental setting. The \numcountermeasures identified countermeasures are listed below in alphabetical order, and for each countermeasure, the relevant literature examined is cited.

\begin{enumerate}[label=C\arabic*., leftmargin=*, font=\itshape]

\item \mycountermeasurename{cap:paper_ipm2023_bias-count:custom-search-engine}{Custom Search Engine}. 
Researchers and practitioners should be extremely careful with the system supplied to the assessors to help them retrieving some kind of supporting evidence, since such a system can be biased~\cite{diaz2008through,2005measuring, otterbacher2018investigating, wilkie2014best}. Researchers should employ a custom and controllable search engine when asking the assessors to evaluate an information item. The assessors might be influenced by the score assigned to the news by a news agency or an online website for the very same information item. Thus, the researcher may tune the search engine parameters to limit the bias that each assessor encounters during a fact-checking activity due to the result source.

\item \mycountermeasurename{cap:paper_ipm2023_bias-count:inform-decision-makers}{Inform Assessors}.
Researchers should always inform assessors about the presence of any kind of automatic (e.g., AI-based) system designed to provide support during the assessment activity, for example by asking them for confirmation or rejection while using such systems, thus limiting \mycitebiasnoindex{Au\-to\-ma\-tion Bias}{\ref{cap:paper_ipm2023_bias-bias:automation-bias}}{Automation Bias}~\cite{10.1136/amiajnl-2011-000089, 10.3389/fpsyg.2023.1118723}. This includes, for example, the presence of a search engine helping them finding some kind of evidence~\cite{draws2022bias, roitero2020crowd, roitero2020covid, SOPRANO2021102710}.

\item \mycountermeasurename{cap:paper_ipm2023_bias-count:discuss}{Discussion}.
 Researchers should allow a synchronous discussion among assessors when possible. In fact, when evaluating the truthfulness of an information item each individual is more prone to accept information items that are consistent with their set of beliefs \cite{la2020crowdsourcing, lewandowsky2012misinformation, roitero2020crowd}. \citet{discussion} and \citet{szpara2005national}, indeed, proved the effectiveness of conducting a synchronous discussion between different assessors to reduce their own bias. \citet{doi:10.1080/01421590220125321} and \citet{doi:10.1080/02602938.2017.1370533} show how discussion among assessors improves the overall assessment quality.
 
 \item \mycountermeasurename{cap:paper_ipm2023_bias-count:good-mood}{Engagement}.
    It is important to put the assessors in a good mood when performing a fact-checking task. \citet{cheng2010debiasing} show that engaged assessors are less likely to experience both:
    \begin{itemize}
        \item[--] \mycitebias{Framing Effect}{\ref{cap:paper_ipm2023_bias-bias:framing-effect}}.
        \item[--] \mycitebiasnoindex{Illusory Correlation}{\ref{cap:paper_ipm2023_bias-bias:illusory-correlation}}{Illusory!Correlation}.
    \end{itemize}
    Moreover, \citet{furnham2011literature} show that if assessors are engaged they are less likely to experience:
    \begin{itemize}
        \item[--] \mycitebiasnoindex{Anchoring Effect}{\ref{cap:paper_ipm2023_bias-bias:anchoring-bias}}{Anchoring Effect}.
        \item[--] \mycitebiasnoindex{Ostrich Effect}{\ref{cap:paper_ipm2023_bias-bias:ostrich-effect}}{Ostrich!Effect}.
    \end{itemize}
    
 \item \mycountermeasurename{cap:paper_ipm2023_bias-count:instructions}{Instructions}.
 Another important aspect to consider consists of formulating an adequate set of instructions. \citet{GILLIER201835} have shown that a set of instructions helps assessors in coming up with new ideas when performing a crowdsourcing task. Even though \citet{10.1145/3078714.3078715} explain that assessors can perform a task even if they have a sub-optimal understanding of the work requested, task instructions clarity should be taken into account. Furthermore, the assessors should be encouraged explicitly to be skeptical about the information that they are evaluating \cite{lewandowsky2012misinformation}. Indeed, \citet{preexposure1} and \citet{preexposure2} prove that pre-exposure warning (i.e., telling explicitly a person that they could be exposed to something) reduces the overall impact on the person itself. Thus, showing a set of assessment instructions can be seen as a pre-exposure warning against the impact of misinformation on the assessor. 
 
\item \mycountermeasurename{cap:paper_ipm2023_bias-count:supporting-evidence}{Require Evidence}.    
Requiring the assessors to provide supporting evidence for their judgments is another effective countermeasure with several advantages. It encourages the assessor to focus on verifiable facts. \citet{lewandowsky2012misinformation} explain that such a countermeasure increases the perceived familiarity with the information item, reinforcing the assessor perceived trustworthiness of the information item itself. They also show that reporting a small set of facts as evidence has the effect of discouraging possible criticisms by other assessors, thus reinforcing the assessment provided. \citet{10.2307/40213302} observes such a phenomenon in public debates. Furthermore, asking the assessors to come up with arguments to support their assessment has proven to reduce:  
    \begin{itemize}
        \item[--] \mycitebiasnoindex{Anchoring Effect}{\ref{cap:paper_ipm2023_bias-bias:anchoring-bias}}{Anchoring Effect}, as shown by \citet{mussweiler2000overcoming}.
        \item[--] \mycitebias{Base Rate Fallacy}{\ref{cap:paper_ipm2023_bias-bias:base-rate-fallacy}}, as shown by \citet{kahneman1973psychology}.
        \item[--] \mycitebias{Framing Effect}{\ref{cap:paper_ipm2023_bias-bias:framing-effect}}, as shown by \citet{kim2005framing, cheng2010debiasing}.
        \item[--] \mycitebias{Illusion of Validity}{\ref{cap:paper_ipm2023_bias-bias:illusion-of-validity}}, as shown by \citet{kahneman1973psychology}.
        \item[--] \mycitebiasnoindex{Illusory Correlation}{\ref{cap:paper_ipm2023_bias-bias:illusory-correlation}}{Illusory!Correlation}, as shown by \citet{matute2011illusions}.
    \end{itemize}
    However, requesting for evidence may be a source of bias itself. \citet{luo1conservatism} and \citet{wood2019elusive} show, indeed, that such a request can lead to the manifestation of, respectively:
    \begin{itemize}
        \item[--] \mycitebias{Backfire Effect}{\ref{cap:paper_ipm2023_bias-bias:backfire-effect}}.
        \item[--] \mycitebias{Conservatism Bias} {\ref{cap:paper_ipm2023_bias-bias:conservatism-bias}}.
    \end{itemize}
   Thus, the requester of the fact-checking activity should address this matter carefully.
   
\item \mycountermeasurename{cap:paper_ipm2023_bias-count:randomized-experimental-design}{Randomized or Constrained Experimental Design}. Using a Randomized or Constrained Experimental Design is helpful in reducing biases. Indeed, different assessors should evaluate different information items. Moreover, each set of items should be evaluated according to a different order, and the assignment of an information item to a given assessor should be such that the item overlap between every two assessors is minimum. If such a constraint can not be satisfied, a randomization process should minimize the chances of overlap between items and assessors~\cite{CESCHIA2022105995, 10.1145/3494522}.

\item \mycountermeasurename{cap:paper_ipm2023_bias-count:more-one-assessor}{Redundancy and Diversity}.
    Redundancy should be employed when asking more than one assessor to fact-check a set of information items. Indeed, the same information item should be evaluated by different assessors. Each item can thus be characterized by a final score, that should be computed by aggregating the individual scores provided by each assessor. In this way, the individual bias of each assessor is mitigated by the remaining assessors. If the population of assessors is diverse enough, one can ideally expect less bias from the fact-checking activity. The population of assessors should thus be as variegated as possible, in terms of both background and experience~\cite{difallah2018demographics}.
    
 \item \mycountermeasurename{cap:paper_ipm2023_bias-count:revise-scores}{Revision}.
 Asking the assessors to revise and/or double-check their answers or even provide them with alternative labels is a useful countermeasure to reduce many biases. In more detail, \citet{cheng2010debiasing}, \citet{kahneman2011thinking}, \citet{kahneman1973psychology}, \citet{kim2005framing}, \citet{mussweiler2000overcoming}, and \citet{effectiviology-bandwagon} show that assessment revision helps reducing:
    \begin{itemize}
        \item[--] \mycitebiasnoindex{Anchoring Effect}{\ref{cap:paper_ipm2023_bias-bias:anchoring-bias}}{Anchoring Effect}.
        \item[--] \mycitebiasnoindex{Availability Heuristic}{\ref{cap:paper_ipm2023_bias-bias:availability-heuristic}}{Availability!Heuristic}.
        \item[--] \mycitebias{Bandwagon Effect}{\ref{cap:paper_ipm2023_bias-bias:bandwagon-effect}}.
        \item[--] \mycitebias{Base Rate Fallacy}{\ref{cap:paper_ipm2023_bias-bias:base-rate-fallacy}}.
        \item[--] \mycitebias{Framing Effect}{\ref{cap:paper_ipm2023_bias-bias:framing-effect}}.
    \end{itemize}
    Furthermore, \citet{bollinger2011calorie}, \citet{cooper2014training}, \citet{hettiachchi2021challenge}, and \citet{mussweiler2000overcoming} show that providing feedback to assessors while performing a given task is useful to reduce biases such as:
    \begin{itemize}
    \item [--]\mycitebiasnoindex{Anchoring Effect}{\ref{cap:paper_ipm2023_bias-bias:anchoring-bias}}{Anchoring Effect}.
    \item [--]\mycitebias{Attentional Bias}{\ref{cap:paper_ipm2023_bias-bias:attentional-bias}}.
    \item [--]\mycitebias{Salience Bias}{\ref{cap:paper_ipm2023_bias-bias:salience-bias}}.
    \end{itemize}
    
\item \mycountermeasurename{cap:paper_ipm2023_bias-count:time-allocated}{Time}.
    Researchers should be careful when setting the time available for each assessor to fact-check a given information item. An adequate amount of time should be left to the assessor. There are  advantages and disadvantages of granting the assessor with a small or large amount of time. For instance, one may assume that providing the assessor with more time will encourage careful consideration of the decision, thus helping to avoid potential \mycitebiasnoindex{Anchoring Effect}{\ref{cap:paper_ipm2023_bias-bias:anchoring-bias}}{Anchoring Effect}. However, \citet{furnham2011literature} show that overthinking might actually increase such a bias. On the other hand,~\citet{effectiviology-bandwagon} shows that assessors left with an adequate amount of time experienced a reduction of the \mycitebias{Bandwagon Effect}{\ref{cap:paper_ipm2023_bias-bias:bandwagon-effect}}.
    
\item \mycountermeasurename{cap:paper_ipm2023_bias-count:training}{Training}. 
\citet{dugan1988effects}, \citet{kazdin1977artifact}, \citet{lievens2001assessor}, \citet{pell2008assessor}, and \citet{szpara2005national} show that training an assessor increases accuracy and reduces the chances for bias to manifest. Thus, assessors training is a useful countermeasure against biases within any context.

 \end{enumerate}

\subsection{\ref{cap:paper_ipm2023_bias-sec:research-questions_4}: Towards A Bias-Aware Judgment Pipeline For Fact-Checking}

\label{cap:paper_ipm2023_bias-sec:results-sect:ideal-pipeline}

Section~\ref{cap:paper_ipm2023_bias-sec:results-subsec:countermeasures} presents \numcountermeasures countermeasures aimed at mitigating the risk of manifestation of the \numbiasused cognitive biases identified in Section~\ref{cap:paper_ipm2023_bias-sec:results-subsec:list} and categorized in Section~\ref{cap:paper_ipm2023_bias-sec:results-subsec:categorization}. These countermeasures form the basis for defining the core components of a fact-checking pipeline designed to minimize cognitive bias, as outlined in Table~\ref{cap:paper_ipm2023_bias-sec:results-sect:ideal-pipeline-tab:ideal-pipeline}. The table summarizes, for each phase of the fact-checking process, the relevant countermeasures, a brief description of each, and the associated biases.

More specifically, the first column of the table details the task phase where the specific countermeasure can be applied: before the task (i.e., pre-task), when the assessor is performing the task (i.e., during the task), or after the task when the assessment has been made (i.e., post-task). Furthermore, a set of countermeasures that are not bound to a specific task purpose (i.e., general purpose) is also listed. The remaining columns detail the countermeasures that can be adopted within the corresponding phase, and for each of them, a brief description together with the set of biases that they reduce.

As an example, the first row of the table deals with the adoption of the countermeasure \mycitecountermeasure{Randomized or Constrained Experimental Design}{\ref{cap:paper_ipm2023_bias-count:randomized-experimental-design}}, i.e., randomizing the process that assigns assessors and information items to enforce diversity and randomness in the pairing. This is a general-purpose countermeasure since it can be applied in different task phases (e.g., when designing the task offline or dynamically when a new assessor is assigned to a new information item). It allows for the mitigation or removal of:
\begin{itemize}
	\item[--] \mycitebiasnoindex{Anchoring Effect}{\ref{cap:paper_ipm2023_bias-bias:anchoring-bias}}{Anchoring Effect}, as the assessor is less likely to rely on a specific information item, given that they inspect more than one with different characteristics.
	\item[--] \mycitebias{Bandwagon Effect}{\ref{cap:paper_ipm2023_bias-bias:bandwagon-effect}}, as the assessor is less likely to be presented with a set of items all related to their personal beliefs or to debated topics with high coverage.
\end{itemize}

In light of the detailed review of current fact-checking practices by prominent organizations like \politifact, \abc, and \factcheckorg presented in Section~\ref{cap:intro-sec:fact-checking}, one can now discuss how the proposal of a bias-aware pipeline can be considered as an enhancement to these existing practices. The key distinction lies in the systematic integration of cognitive bias countermeasures at various stages of the fact-checking process, which is not explicitly addressed in the conventional practices followed by organizations.

The proposed pipeline, as outlined in Table~\ref{cap:paper_ipm2023_bias-sec:results-sect:ideal-pipeline-tab:ideal-pipeline}, includes specific countermeasures targeting known cognitive biases. For instance, in \politifact's process, consensus among editors and reporters for the final rating can be influenced by biases like the \mycitebiasnoindex{Anchoring Effect}{\ref{cap:paper_ipm2023_bias-bias:anchoring-bias}}{Anchoring Effect} or the \mycitebias{Bandwagon Effect}{\ref{cap:paper_ipm2023_bias-bias:bandwagon-effect}}. The proposed pipeline suggests employing randomized pairing of assessors and information items and synchronous discussion between assessors to mitigate these biases. Similarly, in \abc's process, involving a collaborative review where a team decides the final verdict, the proposed countermeasures, such as requiring evidence to support assessments and revising the assessments to encourage a systematic evaluation of the evidence that supports the assessments, could benefit to counter biases such as \mycitebias{Framing Effect}{\ref{cap:paper_ipm2023_bias-bias:framing-effect}} or \mycitebiasnoindex{Illusory Correlation}{\ref{cap:paper_ipm2023_bias-bias:illusory-correlation}}{Illusory!Correlation}. \factcheckorg’s approach, emphasizing evidence retrieval and a review team, aligns with the recommendation of using a custom search engine and informing assessors about AI-based support systems to reduce the impact of biases like \mycitebias{Automation Bias}{\ref{cap:paper_ipm2023_bias-bias:automation-bias}}.

In summary, while the existing practices of these organizations adhere to high standards of transparency, accuracy, and thoroughness, the integration of a bias-aware pipeline could further enhance the objectivity and reliability of the fact-checking process by explicitly addressing and mitigating the influence of cognitive biases. This would not only strengthen the trustworthiness of the fact-checking results but also align with the evolving needs of a complex information landscape where biases can significantly impact the interpretation and verification of information. In the future, the plan is to engage directly with organizations like \politifact, \abc, and \factcheckorg to discuss the practical testing and implementation of such a bias-aware pipeline in their fact-checking processes.

\begin{table}[tbp]
   \centering
    \caption{Constituting elements of a bias-aware judgment pipeline.}
    \label{cap:paper_ipm2023_bias-sec:results-sect:ideal-pipeline-tab:ideal-pipeline}
\begin{tabular}{P{2.1cm}@{\phantom{a}}P{4.8cm}P{5cm}P{4.1cm}}
\toprule
\textbf{Task Phase} & \textbf{Countermeasure} & \textbf{Brief Description} & \textbf{Biases Involved} \\
\midrule

\multirow{3}{=}{\textbf{General Purpose}} 
& \mycitecountermeasure{Randomized or Con\-strained Experimental Design}{\ref{cap:paper_ipm2023_bias-count:randomized-experimental-design}}
& Employ a randomization process when pairing assessors and information items 
& Bias in General
\\
\addlinespace

& \mycitecountermeasure{Redundancy and Diversity}{\ref{cap:paper_ipm2023_bias-count:more-one-assessor}} 
& Use more than one assessor for each information item, and a variegated pool of assessors
& Bias in General 
\\
\addlinespace

& \mycitecountermeasure{Time}{\ref{cap:paper_ipm2023_bias-count:time-allocated}} 
& Allocate an adequate amount of time for the assessors to perform the task 
& \mycitebiasnoindex{Anchoring Effect}{\ref{cap:paper_ipm2023_bias-bias:anchoring-bias}}{Anchoring Effect}
  \mycitebias{Bandwagon Effect}{\ref{cap:paper_ipm2023_bias-bias:bandwagon-effect}}
\\

\midrule
\multirow{3}{=}{\textbf{Pre-Task}} 

& \mycitecountermeasure{Engagement}{\ref{cap:paper_ipm2023_bias-count:good-mood}}
& Put the assessors in a good mood and keep them engaged 
& \mycitebiasnoindex{Anchoring Effect}{\ref{cap:paper_ipm2023_bias-bias:anchoring-bias}}{Anchoring Effect}
  \mycitebias{Framing Effect}{\ref{cap:paper_ipm2023_bias-bias:framing-effect}}
  \mycitebiasnoindex{Illusory Correlation}{\ref{cap:paper_ipm2023_bias-bias:illusory-correlation}}{Illusory!Correlation}
  \mycitebiasnoindex{Ostrich Effect}{\ref{cap:paper_ipm2023_bias-bias:ostrich-effect}}{Ostrich!Effect}
\\
\addlinespace

& \mycitecountermeasure{Instructions}{\ref{cap:paper_ipm2023_bias-count:instructions}}
& Prepare a clear set of instructions to the assessors before the task
& \mycitebiashyphennoindex{Dunning}{Kruger Effect\xspace\xspace\xspace\xspace\xspace\xspace\xspace\xspace\xspace\xspace}{\ref{cap:paper_ipm2023_bias-bias:dunning-kruger-effect}}{Dunning-Kruger Effect}
  \mycitebias{Overconfidence Effect}{\ref{cap:paper_ipm2023_bias-bias:overconfidence-effect}}
\\
\addlinespace

& \mycitecountermeasure{Training}{\ref{cap:paper_ipm2023_bias-count:training}}
& Train the Assessors before the task 
& Bias in General
\\

\midrule
\multirow{6}{=}{\textbf{During the Task}} 

& \mycitecountermeasure{Custom Search Engine}{\ref{cap:paper_ipm2023_bias-count:custom-search-engine}}
& Deploy a custom search engine 
& Bias in General
\\

& \mycitecountermeasure{Inform Assessors}{\ref{cap:paper_ipm2023_bias-count:inform-decision-makers}}
& Inform the assessors about AI-Based support systems  
& \mycitebias{Automation Bias}{\ref{cap:paper_ipm2023_bias-bias:automation-bias}}
\\
\addlinespace

& \mycitecountermeasure{Discussion}{\ref{cap:paper_ipm2023_bias-count:discuss}}
& Synchronous discussion between assessors
& Bias in General
\\
\addlinespace

& \mycitecountermeasure{Require Evidence}{\ref{cap:paper_ipm2023_bias-count:supporting-evidence}}
& Ask the assessors to provide supporting evidence 
& \mycitebiasnoindex{Anchoring Effect\xspace\xspace\xspace}{\ref{cap:paper_ipm2023_bias-bias:anchoring-bias}}{Anchoring Effect}
  \mycitebiasnoindex{Backfire Effect\xspace\xspace\xspace}{\ref{cap:paper_ipm2023_bias-bias:backfire-effect}}{Backfire Effect}
  \mycitebias{Base Rate Fallacy}{\ref{cap:paper_ipm2023_bias-bias:base-rate-fallacy}}
  \mycitebiasnoindex{Conservatism Bias}{\ref{cap:paper_ipm2023_bias-bias:conservatism-bias}}{Conservatism Bias}
  \mycitebias{Framing Effect}{\ref{cap:paper_ipm2023_bias-bias:framing-effect}}
  \mycitebias{Illusion of Validity}{\ref{cap:paper_ipm2023_bias-bias:illusion-of-validity}}
  \mycitebiasnoindex{Illusory Correlation}{\ref{cap:paper_ipm2023_bias-bias:illusory-correlation}}{Illusory!Correlation}
\\
\addlinespace

& \mycitecountermeasure{Revision}{\ref{cap:paper_ipm2023_bias-count:revise-scores}}
& Ask the assessors to revise the assessments & 
  \mycitebiasnoindex{Anchoring Effect\xspace\xspace\xspace}{\ref{cap:paper_ipm2023_bias-bias:anchoring-bias}}{Anchoring Effect}
  \mycitebias{Attentional Bias}{\ref{cap:paper_ipm2023_bias-bias:attentional-bias}}
  \mycitebiasnoindex{Availability Heuristic}{\ref{cap:paper_ipm2023_bias-bias:availability-heuristic}}{Availability!Heuristic}
  \mycitebias{Bandwagon Effect}{\ref{cap:paper_ipm2023_bias-bias:bandwagon-effect}}
  \mycitebias{Base Rate Fallacy}{\ref{cap:paper_ipm2023_bias-bias:base-rate-fallacy}}  
  \mycitebias{Framing Effect}{\ref{cap:paper_ipm2023_bias-bias:framing-effect}} 
  \mycitebias{Salience Bias}{\ref{cap:paper_ipm2023_bias-bias:salience-bias}}
\\

\midrule
\textbf{Post-Task}
&\mycitecountermeasure{Redundancy and Diversity}{\ref{cap:paper_ipm2023_bias-count:more-one-assessor}} & Aggregate the final scores 
& Bias in General \\

\bottomrule
\end{tabular}
\end{table}

\section{Summary}

\label{cap:paper_ipm2023_bias-sec:discussion}

The review presented in this chapter addresses the issue of cognitive biases that may emerge during fact-checking tasks. The study synthesizes existing literature and follows the \prisma guidelines to ensure transparency and completeness in reporting. This constitutes the first comprehensive attempt to examine and address cognitive biases across the various stages of the fact-checking process. The answers to the research questions are summarized as follows.

\myparagraph{\ref{cap:paper_ipm2023_bias-sec:research-questions_1}} A subset of \numbiasused out of \numbias cognitive biases has been identified as potentially manifesting during fact-checking. Each bias is discussed in detail, and real-world examples are provided for \numbiaswithscenario out of these \numbiasused biases.

\myparagraph{\ref{cap:paper_ipm2023_bias-sec:research-questions_2}} The \numbiasused biases are categorized using a classification scheme drawn from the literature, which has been extended to include four additional biases not previously considered.

\myparagraph{\ref{cap:paper_ipm2023_bias-sec:research-questions_3}} A total of \numcountermeasures countermeasures are proposed to mitigate the influence of the \numbiasused cognitive biases throughout the fact-checking process.

\myparagraph{\ref{cap:paper_ipm2023_bias-sec:research-questions_4}} The key components of a bias-aware fact-checking pipeline are outlined. Each component is associated with one or more countermeasures designed to reduce the risk of bias manifestation.

\myparagraph{}
The next chapter presents an exploratory analysis of the data collected during the experiments described in Chapter~\ref{cap:paper_ipm-sec:results-subsec:informativeness-eq:skeleton-total-cost}. The goal is to identify cognitive biases that may actually arise when non-expert human judges engage in fact-checking. To support this analysis, a new set of crowdsourced truthfulness judgments is collected and used to validate the derived hypotheses.

\chapter{The Effect Of Crowd Worker Biases In Fact-Checking Tasks}

\label{cap:paper_facct2022}

This chapter is based on the article published at the 2022 ACM Conference on Fairness, Accountability, and Transparency~\cite{draws2022bias}. Section~\ref{cap:related_work-sec:crowdsourcing-truthfulness} and Section~\ref{cap:related_work-sec:worker-bias} provide an overview of the relevant related work. Section~\ref{cap:paper_facct2022-sec:research-questions} introduces the research questions. Section~\ref{cap:paper_facct2022-sec:exp-study} presents the exploratory study, while Section~\ref{cap:paper_facct2022-sec:exp-setup} describes the experimental setup of a novel crowdsourcing study. Section~\ref{cap:paper_facct2022-sec:results} reports both a descriptive analysis and the main results. Finally, Section~\ref{cap:paper_facct2022-sec:discussion} summarizes the main findings and concludes the chapter.

\section{Research Questions}

\label{cap:paper_facct2022-sec:research-questions}

This chapter investigates which systematic biases may decrease data quality for crowdsourced truthfulness judgments. Initially, an exploratory study is conducted on an earlier collected data set containing crowdsourced truthfulness judgments for political statements (Section~\ref{cap:paper_ipm2021-sec:exp-setup}). These data also contain information on the political leaning of statements as well as individual worker characteristics (e.g., workers' level of education and political leaning). The findings from these exploratory analyses are used to formulate specific hypotheses concerning which individual characteristics of statements or workers and what cognitive worker biases (Section~\ref{cap:paper_ipm2023_bias-sec:results-subsec:list}) may affect the accuracy of crowd workers' truthfulness judgments. To test these hypotheses, a new, preregistered crowdsourcing study is conducted. The following research questions are investigated:

\begin{enumerate}[start=25, leftmargin=2.92em, label=RQ\arabic*]
\item \label{cap:paper_facct2022-sec:research-questions_1} Which individual characteristics of crowd workers and statements contribute to systematic biases in workers' truthfulness judgments?
\item \label{cap:paper_facct2022-sec:research-questions_2} Which cognitive biases can influence crowd workers' truthfulness judgments?
\item \label{cap:paper_facct2022-sec:research-questions_3} Are different truthfulness dimensions affected by distinct types of bias?
\end{enumerate}

\section{Exploratory Study}

\label{cap:paper_facct2022-sec:exp-study}

This exploratory study builds on the dataset originally collected to evaluate multiple dimensions of truthfulness. The experimental setup, sample, and task design used for data collection are those described in Section~\ref{cap:paper_ipm2021-sec:exp-setup}. This section focuses specifically on the exploratory analyses and the resulting hypotheses. Section~\ref{cap:paper_facct2022-sec:exp-study-subsec:preprocessing} outlines the preprocessing steps applied to the previously collected data. Section~\ref{cap:paper_facct2022-sec:exp-study-subsec:exploratory} presents the exploratory analyses. Finally, Section~\ref{cap:paper_facct2022-sec:exp-study-subsec:hypotheses} introduces the seven hypotheses derived from the study.

\subsection{Data Preprocessing}

\label{cap:paper_facct2022-sec:exp-study-subsec:preprocessing}

Several preprocessing steps are performed on the data described in Section~\ref{cap:paper_ipm2021-sec:exp-setup} to make them suitable for the analysis. First, the judgment scales are transformed and mapped to a common set of labels (Section~\ref{cap:paper_facct2022-sec:exp-study-subsec:preprocessing-subsec:transformations}). Then, three metrics are computed to evaluate potential annotation bias (Section~\ref{cap:paper_facct2022-sec:exp-study-subsec:preprocessing-subsec:annotation-bias}). Finally, three additional metrics are calculated to assess worker-related bias (Section~\ref{cap:paper_facct2022-sec:exp-study-subsec:preprocessing-subsec:worker-bias}).

\subsubsection{Scale Transformations}

\label{cap:paper_facct2022-sec:exp-study-subsec:preprocessing-subsec:transformations}

Each statement in the dataset described in Section~\ref{cap:paper_ipm2021-sec:exp-setup} includes a truthfulness judgment from either \politifact (Section~\ref{cap:dataset-sec:politifact}) or \abc (Section~\ref{cap:dataset-sec:abc}), along with corresponding judgments from crowd workers. However, these sources use different ordinal judgment scales: \politifact employs a six-level scale, \abc a three-level scale, and crowd workers a five-level \index{Likert}Likert scale. To enable comparison across these different sources, the scales must be aligned.

Assuming that the \politifact, \abc, and \index{Likert}Likert scales are linear and equally spaced, a common assumption also adopted in previous experiments (see Section~\ref{cap:paper_sigir2020-sec:exp-setup-subsec:scales}), the \politifact and \index{Likert}Likert judgments are converted to the three-level scale used by \abc. This unified scale comprises three labels: \negativegt ($-1$), \neutralgt ($0$), and \positivegt ($1$). Note that two of these labels overlap with those used by \abc; however, they should be interpreted here as elements of a shared and newly defined set used for mapping purposes. This conversion is necessary due to intra-dataset inconsistencies in both \politifact and \abc, as discussed in their respective sections. The transformation procedure is as follows:
\begin{itemize}
\item \politifact: 
   	\begin{itemize}
   	 \item \politifactzero and \politifactone mapped into \negativegt ($-1$).
   	 \item \politifacttwo and \politifactthree mapped into \neutralgt ($0$).
   	 \item \politifactfour and \politifactfive mapped into \positivegt ($1$).
   	\end{itemize}
\item \abc: 
   \begin{itemize}
   \item \abczero and \abctwo maintain their original semantic meaning.
   \item \abcone mapped into \neutralgt ($0$).
   \end{itemize}
\item \index{Likert}Likert scale: 
   \begin{itemize}
    \item $-2$ and $-1$ mapped into \negativegt ($-1$). 
    \item $0$ mapped into \neutralgt ($0$).
    \item $+1$ and $+2$ mapped into \positivegt ($1$).
 \end{itemize}
\end{itemize}

\subsubsection{Judgment Bias Metrics}
\label{cap:paper_facct2022-sec:exp-study-subsec:preprocessing-subsec:annotation-bias}

Three metrics are computed to quantify and evaluate judgment bias. Both \emph{external} errors (i.e., comparing crowd judgments with the ground truth) and \emph{internal} errors (i.e., comparing a crowd judgment with other crowd judgments on the same items) are considered. The metrics are defined as follows:
\begin{itemize}[label=--]
  \item \texttt{External Error} \exterr: the difference between a worker's \overalltruthfulness judgment (Section~\ref{cap:paper_ipm2021-sec:exp-setup-subsec:seven-dim}) and the corresponding item's expert-assessed ground truth judgment. This metric indicates whether a crowd worker overestimates or underestimates the \overalltruthfulness of a statement.
  \item \texttt{External Absolute Error} \extabserr: the absolute difference between a crowd worker's \overalltruthfulness judgment and the item's ground truth label. Unlike \exterr, this metric captures the \textit{magnitude} of bias without direction.
  \item \texttt{Internal Error} \interr: the difference between a worker's judgment and the average judgment of other crowd workers for the same statement. This metric is computed nine times overall: once for \overalltruthfulness, once for \confidence, and once for each of the seven truthfulness dimensions (Section~\ref{cap:paper_ipm2021-sec:exp-setup-subsec:seven-dim}).
\end{itemize}

The value ranges for \exterr and \interr are $[-2, 2]$, while \extabserr ranges in $[0, 2]$. For example, if the ground truth label is positive ($1$) and a worker's judgment is negative ($-1$), then \exterr equals $-2$, and \extabserr equals $2$. These metrics enable both directional and magnitude-based analyses of individual judgments, offering a detailed view of disagreement and potential bias in crowd-provided data.

\subsubsection{Worker Bias Metrics}
\label{cap:paper_facct2022-sec:exp-study-subsec:preprocessing-subsec:worker-bias}

Aggregate bias metrics are computed to evaluate each worker's individual degree of bias, based on the annotation bias metrics introduced in Section~\ref{cap:paper_facct2022-sec:exp-study-subsec:preprocessing-subsec:annotation-bias}. Specifically, the following values are calculated for each worker:
\begin{itemize}[label=--]
  \item Mean external error \extmeanerr
  \item Mean external absolute error \extmeanabserr
  \item Mean internal error \intmeanerr computed separately for \overalltruthfulness, \confidence, and each of the seven truthfulness dimensions (Section~\ref{cap:paper_ipm2021-sec:exp-setup-subsec:seven-dim})
\end{itemize}
These 11 worker-specific metrics serve as dependent variables in the exploratory study.

\subsection{Exploratory Analyses}
\label{cap:paper_facct2022-sec:exp-study-subsec:exploratory}

A series of exploratory analyses is conducted on the dataset described in Section~\ref{cap:paper_ipm2021-sec:exp-setup} to identify potential systematic biases in crowd workers' truthfulness judgments. Specifically, various worker-related attributes (e.g., political views, average time per judgment) are used as independent variables, while the aggregate worker bias metrics introduced in Section~\ref{cap:paper_facct2022-sec:exp-study-subsec:preprocessing-subsec:worker-bias} serve as dependent variables. The sample is relatively balanced in terms of demographics (e.g., age group, income) and political orientation (e.g., conservative vs. liberal). 

It is important to note that the results presented in this subsection (e.g., \textit{p}-values from hypothesis tests) are exploratory in nature. The analyses are intended solely to generate concrete hypotheses, which will be tested on new data in Section~\ref{cap:paper_facct2022-sec:exp-study-subsec:hypotheses}.

\subsubsection{Exploring Workers' \extmeanerr}
\label{cap:paper_facct2022-sec:exp-study-subsec:exploratory-subsec:worker-ext-mean-err}

The exploratory analysis begins by computing workers' \extmeanerr, which corresponds to the average difference between a crowd worker's judgment and the corresponding ground truth label. On average, workers tend to overestimate truthfulness (\extmeanerr = $0.32$, $\upsigma = 0.42$, $t = 10.93$, $p < 0.001$; one-sample $t$-test against zero).

Linear regression models~\cite{xiaogang2012reg} and ANOVA tests (including Tukey-adjusted post-hoc comparisons~\cite{olejnikAnova}) reveal that workers identifying as \veryconservative and/or \republican tend to overestimate truthfulness significantly more than other groups. Specifically, \veryconservative workers show significantly higher \extmeanerr scores compared to other political orientations ($p = [0.006, 0.050]$), and \republican workers compared to other party affiliations ($p = [0.012, 0.089]$). Additionally, workers agreeing with the southern border statement (see Appendix~\ref{cap:paper_facct2022-appendix:quest-crt}) also show significantly higher overestimation than those who disagreed ($p = 0.004$), although this effect largely overlaps with party affiliation, as 78\% of such workers identified as \republican.

Further analysis suggests that overestimation may be particularly strong when workers assess statements aligned with their political beliefs. This is especially evident among \republican respondents (Figure~\ref{cap:paper_facct2022-sec:exp-study-subsec:exploratory-subsec:worker-ext-mean-err-fig:polviews-diff_crt-crt}). Paradoxically, this trend leads to more accurate judgments on statements affiliated with opposing views, given the general bias toward overestimation. This behavior may be attributed to cognitive biases such as the \mycitebias{Affect Heuristic}{\ref{cap:paper_ipm2023_bias-bias:affect-heuristic}} (i.e., overestimating truthfulness when the speaker is perceived favorably) and the \mycitebias{Confirmation Bias}{\ref{cap:paper_ipm2023_bias-bias:confirmation-bias}} (i.e., reinforcing one's preexisting beliefs).

\subsubsection{Exploring Workers' \extmeanabserr}
\label{cap:paper_facct2022-sec:exp-study-subsec:exploratory-subsec:worker-ext-mean-abs-err}

\begin{figure}[tbp]
  \centering
  \begin{subfigure}{.8\linewidth}
    \centering
    \includegraphics[width=.75\linewidth]{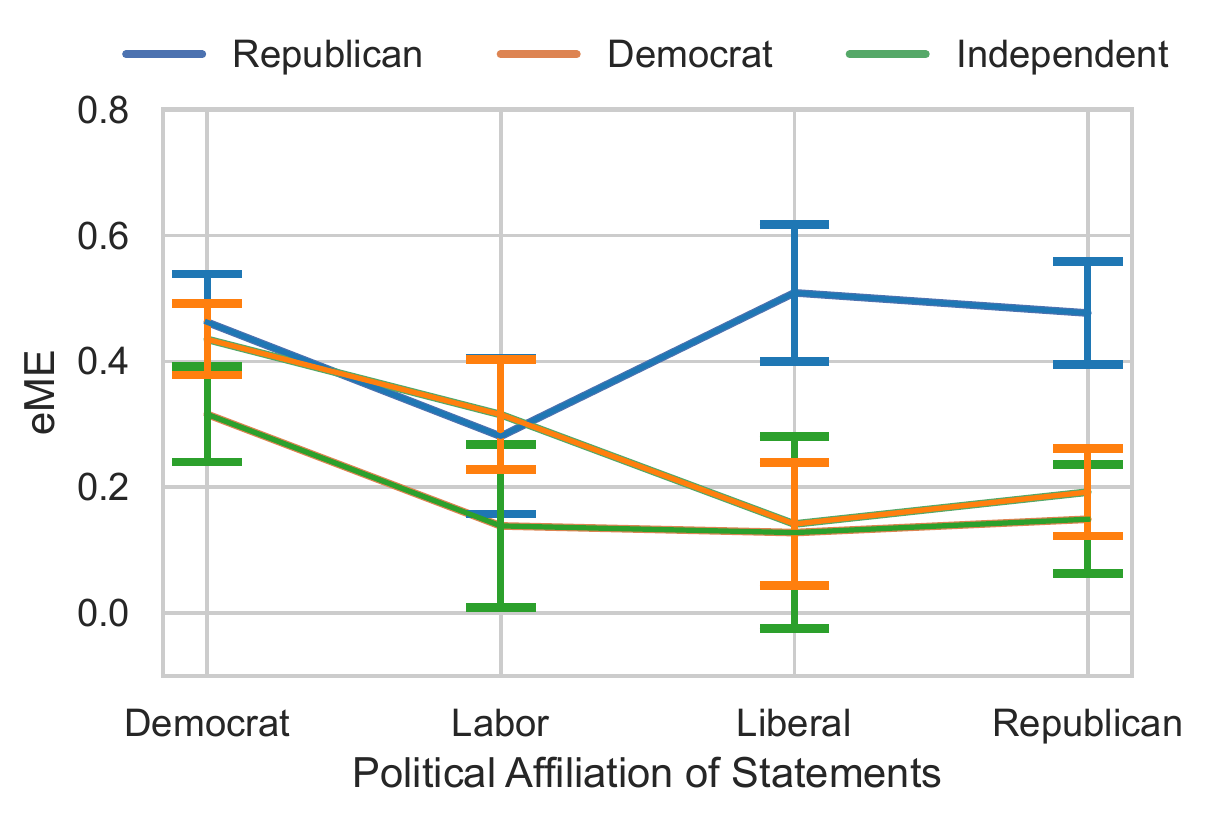}
    \caption{Mean \extmeanerr by political affiliation of statements and workers.}
    \label{cap:paper_facct2022-sec:exp-study-subsec:exploratory-subsec:worker-ext-mean-err-fig:polviews-diff_crt-crt}
  \end{subfigure}
  
  \begin{subfigure}{.8\linewidth}
    \centering
    \includegraphics[width=.75\linewidth]{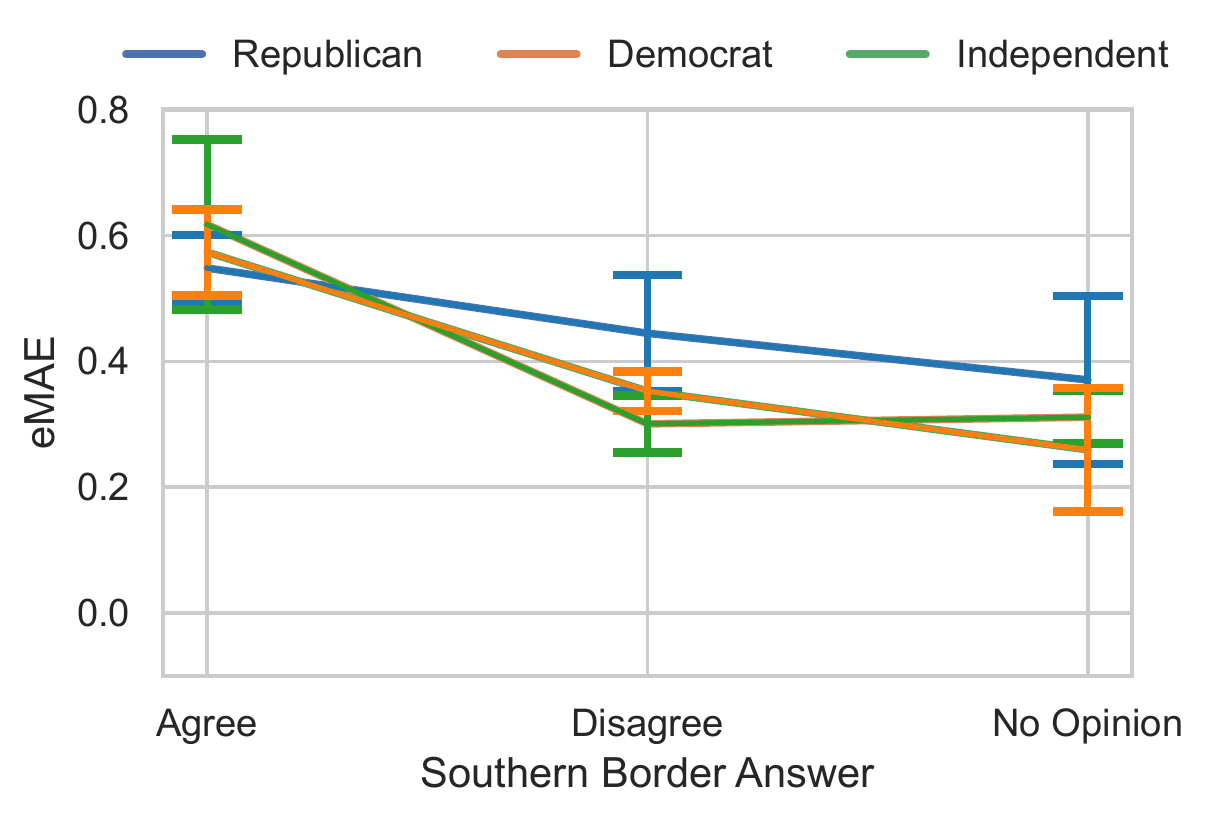}
    \caption{Mean \extmeanabserr by southern border response and political affiliation of workers.}
    \label{cap:paper_facct2022-sec:exp-study-subsec:exploratory-subsec:worker-ext-mean-err-fig:polviews-diff_crt-pol}
  \end{subfigure}
  
  \caption{Systematic differences in crowd worker bias metrics (\extmeanerr and \extmeanabserr) across political affiliations and issue-related responses. Four workers who selected affiliations other than \democratic, \independent, or \republican are excluded.}
  \label{cap:paper_facct2022-sec:exp-study-subsec:exploratory-subsec:worker-ext-mean-err-fig:polviews-diff}
\end{figure}

The \extmeanabserr, corresponding to the mean absolute difference between a crowd worker's judgment and the respective item's ground truth label, is also considered. The average \extmeanabserr in the dataset is $0.42$ ($\upsigma = 0.31$), indicating that workers were, on average, somewhat biased in their assessments (i.e., \extabserr ranged from $0$ to $1.11$). Consistent with the previous analysis, workers identifying as \veryconservative (Tukey-adjusted $p = [0.012, 0.200]$), \republican ($p = [0.031, 0.129]$), or agreeing with the southern border statement ($p < 0.001$) exhibit significantly higher \extmeanabserr, as shown in Figure~\ref{cap:paper_facct2022-sec:exp-study-subsec:exploratory-subsec:worker-ext-mean-err-fig:polviews-diff}.

In general, the worker groups exhibiting higher bias also tend to complete the task more quickly. While no significant effect of \cognitivereflection on \extmeanabserr is found when accounting for all independent variables simultaneously, workers with lower \cognitivereflection scores also tend to provide faster responses. This suggests that cognitive reasoning ability may partially account for the observed bias, although the effect appears too small to detect in this exploratory setting.

Another explanation may be workers' \beliefscience (Appendix~\ref{cap:paper_facct2022-appendix:quest-crt-sec:belief}). Indeed, 78\% of the \disagree responses regarding additional environmental regulations comes from \veryconservative workers. Given the well-established scientific consensus on environmental issues, a lack of trust in scientific findings may lead some workers to distrust statements supported by scientific evidence. Although the number of such \disagree responses is too small to draw direct conclusions, \beliefscience may represent an underlying factor influencing the quality of truthfulness judgments.

Finally, the analyses reveal a positive correlation between workers' average self-reported \meanconfidence and their \extmeanabserr ($\upbeta = 0.14$, $p < 0.001$). This relationship may reflect the influence of the \mycitebias{Overconfidence Effect}{\ref{cap:paper_ipm2023_bias-bias:overconfidence-effect}} (Section~\ref{cap:paper_ipm2023_bias-sec:results-subsec:list}), whereby overly confident workers provide less accurate judgments.

\subsubsection{Exploring Worker's \intmeanerr}
\label{cap:paper_facct2022-sec:exp-study-subsec:exploratory-subsec:worker-int-mean-err}

Finally, the \intmeanerr is investigated, which corresponds to the mean difference between a worker's judgment and the average judgment of other workers on the same statements.

Workers with some postgraduate or professional schooling (but no completed postgraduate degree) tend to report higher \confidence in their judgments compared to most other education groups \index{$p$}(Tukey-adjusted $p = [<0.001, 0.018]$). Similarly, the more a worker identifies as \conservative, the higher their self-reported \confidence relative to others who annotated the same items. In general, higher \confidence values are observed among worker groups that also show greater bias. This points to a potential \mycitebias{Overconfidence Effect}{\ref{cap:paper_ipm2023_bias-bias:overconfidence-effect}} and suggests that \confidence may act as a proxy for explaining political skewness in truthfulness judgments.

Among all the \intmeanerr dimensions, the strongest predictor of \extmeanerr is the \correctness dimension ($\upbeta = 0.51$, $p < 0.001$). This indicates that workers who rate \correctness higher than their peers tend to also overestimate \overalltruthfulness. The result reinforces earlier findings (Section~\ref{cap:paper_ipm-sec:results-subsec:truthfulness-dimensions}) that workers often consider \correctness commensurable with \overalltruthfulness.

In addition, workers identifying as \democratic or \republican tend to give higher judgments across most dimensions compared to those identifying as \independent or other. Due to the general trend of overestimation, this leads the latter group to provide more accurate truthfulness judgments. These differences, though small, suggest that workers with stronger \emph{trust in politics}, as inferred from their party identification, may display higher overall bias.

Further evidence supporting this hypothesis comes from responses to the southern border question: workers who answered \lq\lq no opinion\rq\rq{} consistently rated \speakertrustworthiness lower than other groups (Figure~\ref{cap:paper_facct2022-sec:exp-study-subsec:exploratory-subsec:worker-ext-mean-err-fig:polviews-diff}). Lastly, among the nine \intmeanerr dimensions, \speakertrustworthiness is the strongest predictor of \extmeanabserr ($\upbeta = 0.16$, $p = 0.040$). This again suggests the influence of the \mycitebias{Affect Heuristic}{\ref{cap:paper_ipm2023_bias-bias:affect-heuristic}}—workers may overestimate truthfulness when they view a statement's speaker more favorably.

\subsection{Hypotheses For The Novel Data Collection}

\label{cap:paper_facct2022-sec:exp-study-subsec:hypotheses}

Seven hypotheses are derived from the exploratory analyses and are tested on a newly collected dataset. These hypotheses are categorized according to whether they refer to general worker characteristics (Section~\ref{cap:paper_facct2022-sec:exp-study-subsec:hypotheses-subsec:worker-traits}) or to task-specific cognitive biases (Section~\ref{cap:paper_facct2022-sec:exp-study-subsec:hypotheses-subsec:cognitive-biases}).

\subsubsection{\ref{cap:paper_facct2022-sec:research-questions_1}: General Worker Traits}

\label{cap:paper_facct2022-sec:exp-study-subsec:hypotheses-subsec:worker-traits}

The following three (3) hypotheses concern expectations about which worker groups may be more prone to biased judgments.

\begin{enumerate}[leftmargin=*]
	\item \hyphoonea:
	\begin{itemize}[leftmargin=*]
	\item[--] \emph{Formulation}: workers with stronger \trustpolitics tend to be less accurate in judging the \overalltruthfulness of statements.
	\item[--] \emph{Rationale}: in the exploratory study, workers identifying as \democratic or \republican (i.e., affiliated with traditional political parties) exhibited lower judgment accuracy than others. A possible explanation is that stronger \trustpolitics—defined as the belief that politicians and governmental institutions are trustworthy and act with good intentions—may increase party identification. This, in turn, may reduce skepticism toward political statements, leading to an overestimation of their truthfulness.
	\end{itemize}

	\item \hyphooneb:
	\begin{itemize}[leftmargin=*]
	\item[--] \emph{Formulation}: workers with stronger \beliefscience are more accurate in judging the \overalltruthfulness of statements.
    \item[--] \emph{Rationale}: in the exploratory study, workers who responded \disagree to the environmental regulations question tended to show greater bias. A possible underlying variable is \beliefscience, defined as the belief that scientific evidence is trustworthy and essential to societal development. Low trust in science may lead workers to distrust statements referring to scientific findings (e.g., on climate change), thus hindering their ability to judge truthfulness accurately.
	\end{itemize}

	\item \hyphoonec:
	\begin{itemize}[leftmargin=*]
	\item[--] \emph{Formulation}: workers with higher \cognitivereflection are more accurate in judging the \overalltruthfulness of statements.
    \item[--] \emph{Rationale}: the exploratory study shows that workers with lower cognitive reflection tend to complete the task more quickly, which is generally associated with greater bias. Although no direct association between \cognitivereflection and bias was observed, a weak relationship may exist but be difficult to detect, especially given that many participants are likely familiar with the \index{Cognitive!Reflection Test}CRT~\cite{haigh2016has}.
    \end{itemize}
\end{enumerate}

\subsubsection{\ref{cap:paper_facct2022-sec:research-questions_2}: Cognitive Biases}

\label{cap:paper_facct2022-sec:exp-study-subsec:hypotheses-subsec:cognitive-biases}

The following four (4) hypotheses concern cognitive biases that may affect crowd workers' truthfulness judgments.

\begin{enumerate}[leftmargin=*]
	\item \hyphotwoa:
	\begin{itemize}[leftmargin=*]
	\item[--] \emph{Formulation}: workers generally tend to overestimate the truthfulness of statements.
	\item[--] \emph{Rationale}: in the exploratory study, workers systematically overestimated truthfulness. A similar tendency is expected to be observed in the novel dataset.
	\end{itemize}

	\item \hyphotwob:
	\begin{itemize}[leftmargin=*]
	\item[--] \emph{Formulation}: workers' tendency to over- or underestimate the \overalltruthfulness of a statement is influenced by how much they like the statement speaker.
	\item[--] \emph{Rationale}: the exploratory analyses reveal several associations consistent with the \mycitebias{Affect Heuristic}{\ref{cap:paper_ipm2023_bias-bias:affect-heuristic}} (see Section~\ref{cap:paper_ipm2023_bias-sec:results-subsec:list}), whereby judgments are affected by the perceived likability of the source.
	\end{itemize}

	\item \hyphotwoc:
	\begin{itemize}[leftmargin=*]
	\item[--] \emph{Formulation}: workers' tendency to over- or underestimate the \overalltruthfulness of a statement is influenced by the extent to which they personally support the goal expressed in the statement.
	\item[--] \emph{Rationale}: certain patterns observed in the exploratory study suggest the presence of \mycitebias{Confirmation Bias}{\ref{cap:paper_ipm2023_bias-bias:confirmation-bias}} (see Section~\ref{cap:paper_ipm2023_bias-sec:results-subsec:list}), in which judgments align with the worker’s prior beliefs or attitudes.
	\end{itemize}

	\item \hyphotwod:
	\begin{itemize}[leftmargin=*]
	\item[--] \emph{Formulation}: workers with higher \meanconfidence in their ability to judge truthfulness exhibit greater bias.
	\item[--] \emph{Rationale}: in the exploratory study, higher self-reported confidence correlates with greater judgment bias. This supports the potential presence of the \mycitebias{Overconfidence Effect}{\ref{cap:paper_ipm2023_bias-bias:overconfidence-effect}}, which is expected to be confirmed in the novel data.
	\end{itemize}
\end{enumerate}

\section{Experimental Setting}

\label{cap:paper_facct2022-sec:exp-setup}

A novel crowdsourcing experiment is conducted to test the hypotheses outlined in Section~\ref{cap:paper_facct2022-sec:exp-study-subsec:hypotheses}. The hypotheses, research design, and data analysis plan were preregistered before the data collection.\footnote{The preregistration is available at \url{https://osf.io/5jyu4}.} For clarity, some variable definitions from earlier sections are repeated here.

\subsection{Exploratory, Dependent, and Independent Variables}

\label{cap:paper_facct2022-sec:exp-setup-subsec:variables}

A total of three groups of variables are recorded and analyzed in the study: exploratory, dependent, and independent. Exploratory variables are used to characterize the dataset but are not employed to test conclusive hypotheses. Dependent and independent variables are used to test the hypotheses introduced in Section~\ref{cap:paper_facct2022-sec:exp-study-subsec:hypotheses}.

The nine (9) \emph{exploratory variables} are listed below. They are labeled as ``exploratory'' because they are analyzed to describe and understand the data but not used in hypothesis testing. The first seven originate from the initial demographic questionnaire (Appendix~\ref{cap:paper_sigir2020-appendix:quest-crt-sec:initial}), while the last two are computed from workers' task annotations:

\begin{itemize}[label=--]
  \item \texttt{Age Group} (categorical): from the demographic questionnaire.
  \item \texttt{Gender} (categorical): from the demographic questionnaire.
  \item \texttt{Level of Education} (ordinal categorical): from the demographic questionnaire.
  \item \texttt{Yearly Income (Before Taxes)} (ordinal categorical): from the demographic questionnaire
  \item \texttt{Political Views} (categorical): from the demographic questionnaire.
  \item \texttt{Opinion About US Southern Border Wall} (binary categorical): from the demographic questionnaire.
  \item \texttt{Opinion About US Environmental Regulation} (binary categorical): from the demographic questionnaire.
  \item \interr\ (continuous): difference between a worker’s judgment and the average judgment of other crowd workers for the same statement.
  \item \intmeanerr\ (continuous): mean of \interr, aggregated per worker.
\end{itemize}

The three (3) \emph{dependent variables} quantify worker-level bias and are used as outcome variables in the analyses:

\begin{itemize}[label=--]
  \item \exterr\ (continuous): difference between a worker’s \overalltruthfulness\ judgment and the expert-assessed ground truth of the statement.
  \item \extmeanerr\ (continuous): mean of \exterr\ for each worker (Section~\ref{cap:paper_facct2022-sec:exp-study-subsec:exploratory-subsec:worker-ext-mean-err}).
  \item \extmeanabserr\ (continuous): mean of the absolute value of \exterr\ for each worker (Section~\ref{cap:paper_facct2022-sec:exp-study-subsec:exploratory-subsec:worker-ext-mean-abs-err}).
\end{itemize}

The seven (7) \emph{independent variables} are used to explain the variation in the dependent variables. They are derived from workers’ responses to additional questionnaires or inferred from task interaction:

\begin{itemize}[label=--]
  \item \trustpolitics\ (continuous, $[-2,2]$): measured via the CTGO questionnaire as the average response score.
  \item \beliefscience\ (continuous, $[-2,2]$): measured via the BISS questionnaire as the average response score.
  \item \cognitivereflection\ (ordinal, $[0,4]$): number of correct answers in the Cognitive Reflection Test (CRT); time spent on CRT also recorded.
  \item \politicalpartyaffiliation\ (categorical): whether workers self-identify as Republican, Democrat, Independent, or Other (Initial Questionnaire, Q5).
  \item \affectspeaker\ (ordinal, $[-3,3]$): degree to which workers like each statement’s speaker (5-point Likert scale with “I don’t know” option).
  \item \meanconfidence\ (ordinal, $[-2,2]$): average of workers' self-reported confidence across all their judgments.
  \item \statementsupport\ (categorical): inferred approximation of support for the statement goal, based on workers’ political orientation.
\end{itemize}

\subsection{Crowdsourcing Task}

\label{cap:paper_facct2022-sec:exp-setup-subsec:crowdsourcing-task}

The crowdsourcing task allows recording 9 descriptive and exploratory variables, 3 dependent variables, and 7 independent variables (Section~\ref{cap:paper_facct2022-sec:exp-setup-subsec:variables}). The experimental design follows the same structure used to judge the multiple dimensions of truthfulness (Section~\ref{cap:paper_ipm2021-sec:exp-setup}). Specifically, the same interface and set of \index{HIT}HITs are used to keep the new task as similar as possible. However, the variables defined for the current study require modifications to the original task design in order to be properly recorded and addressed. To this end, the following adjustments are implemented:
\begin{enumerate}
	\item The generalized version of the \lq\lq Citizen Trust in Government Organizations\rq\rq{} questionnaire~\cite{grimmelikhuijsen2017validating} (\index{CTGO}CTGO, Appendix~\ref{cap:paper_facct2022-appendix:quest-crt-sec:trust}) is added to the task to measure workers' \trustpolitics.
	\item The \lq\lq Belief in Science Scale\rq\rq{}~\cite{dagnall2019evaluation} (\index{BISS}BISS, Appendix~\ref{cap:paper_facct2022-appendix:quest-crt-sec:belief}) is added to capture workers' \beliefscience.
	\item A single five-point \index{Likert}Likert scale is introduced to assess the degree to which workers like the speaker of each statement (\affectspeaker). An additional option is provided to allow workers to indicate that they do not know the speaker.
\end{enumerate}
These two additional questionnaires are presented in the task immediately after the original demographic questionnaire (Appendix~\ref{cap:paper_sigir2020-appendix:quest-crt-sec:initial}).

The crowdsourcing task aims to collect data from at least $255$ crowd workers. This required sample size is determined through a power analysis for a Between-Subjects \index{ANOVA}ANOVA (fixed effects, special, main effects, and interactions; see Section~\ref{cap:paper_facct2022-sec:results-subsec:hypothesis-tests}) using the \index{G*Power}\texttt{G*Power} software~\cite{gpower}. The analysis specifies a small effect size of \index{$F$}$F = 0.10$, based on the findings of the exploratory study, along with a significance threshold of \index{$\upalpha$}$\upalpha = 0.05 / 7 = 0.007$ (to account for multiple hypothesis testing), and statistical power of \index{$\upbeta$}$(1 - \upbeta) = 0.8$.
The analysis accounts for three between-subjects groups (\republican, \democratic, and \independent/else) and four within-subjects groups (\republican, \democratic, \liberal, and \labor). The required sample size is computed individually for each hypothesis based on its respective degrees of freedom.

To meet these requirements, the task publishes $200$ \index{HIT}HITs on the \mturk platform, covering the $180$ statements defined in Section~\ref{cap:paper_ipm2021-sec:exp-setup} and described in detail in Section~\ref{cap:paper_sigir2020-sec:exp-setup}. A total of $2200$ individual judgments is collected. Only crowd workers based in the United States are eligible to participate. Each worker receives a compensation of \$2 USD for completing the task, calculated to align with the estimated minimum time required and the United States minimum wage of \$7.25 USD per hour.

\section{Descriptive Analysis}

\label{cap:paper_facct2022-sec:desc-stat}

Initially, Section~\ref{cap:paper_facct2022-sec:desc-stat-subsec:worker-demographics} presents demographic information about the participating crowd workers. Finally, Section~\ref{cap:paper_facct2022-sec:desc-stat-subsec:task-abandonment} reports on the task abandonment rate.

\subsection{Worker Demographics}
\label{cap:paper_facct2022-sec:desc-stat-subsec:worker-demographics}

Overall, \num{302} workers completed the crowdsourcing task. Nearly 36\% of them are between 26 and 35 years old, while 34\% are between 35 and 50 years old. The majority (52\%) hold a college or bachelor's degree. Regarding total income before taxes, 25\% of workers earn between \$50k and less than \$75k, while 19\% earn between \$75k and less than \$100k. 
Considering political views, 27\% of workers identify as \moderate, 27\% as \conservative, and 26\% as \liberal. In terms of party affiliation, 53\% consider themselves \democratic, 27\% \republican, and 17\% \independent. A majority (53\%) agree with building a wall on the US southern border, while 25\% disagree. Finally, a large majority (84\%) believe that the government should increase environmental regulations to prevent climate change, with only 9\% disagreeing.

In general, the sample is well balanced across most categories and is comparable to the one described in Section~\ref{cap:paper_ipm2021-sec:exp-setup-subsec:task-abandonment}, except for the reversed trend on the southern border question, where most workers in that earlier setting disagreed with building a wall.

\subsection{Task Abandonment}
\label{cap:paper_facct2022-sec:desc-stat-subsec:task-abandonment}

The abandonment rate of the crowdsourcing task is measured using the definition provided by \citet{8873609}, namely the number of workers who voluntarily terminated the task before completion. Overall, \num{2,742} workers participated. Of these, 302 (11\%) completed the task, while 2065 (75\%) voluntarily abandoned it. Additionally, 375 workers (14\%) failed at least one quality check at the end of the task. Each worker was allowed up to 10 attempts to complete the task.

\begin{figure}[tbp]
  \centering
  \begin{subfigure}{.49\linewidth}
    \centering
    \includegraphics[width=\linewidth]{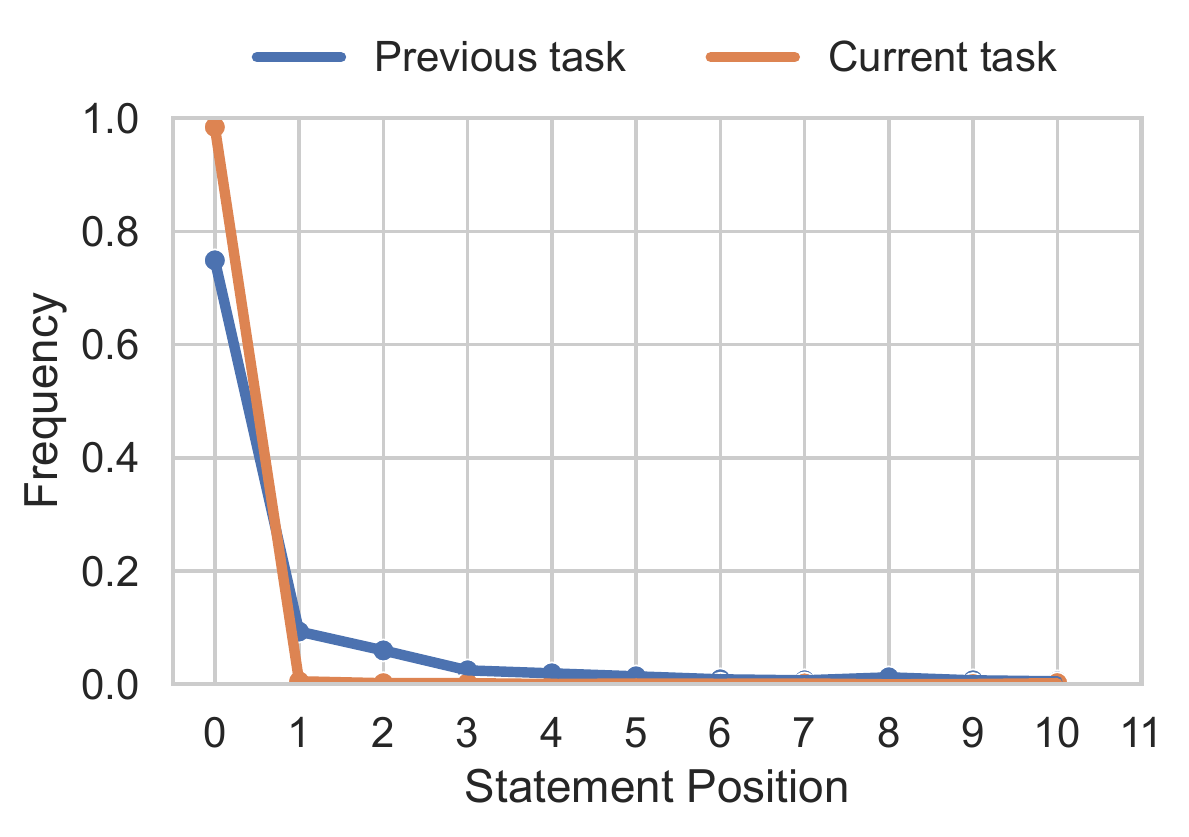}
    \caption{Distribution of task abandonment.}
    \label{cap:paper_facct2022-sec:desc-stat-subsec:task-abandonment-distribution_aband}
  \end{subfigure}
  \hfill
  \begin{subfigure}{.49\linewidth}
    \centering
    \includegraphics[width=\linewidth]{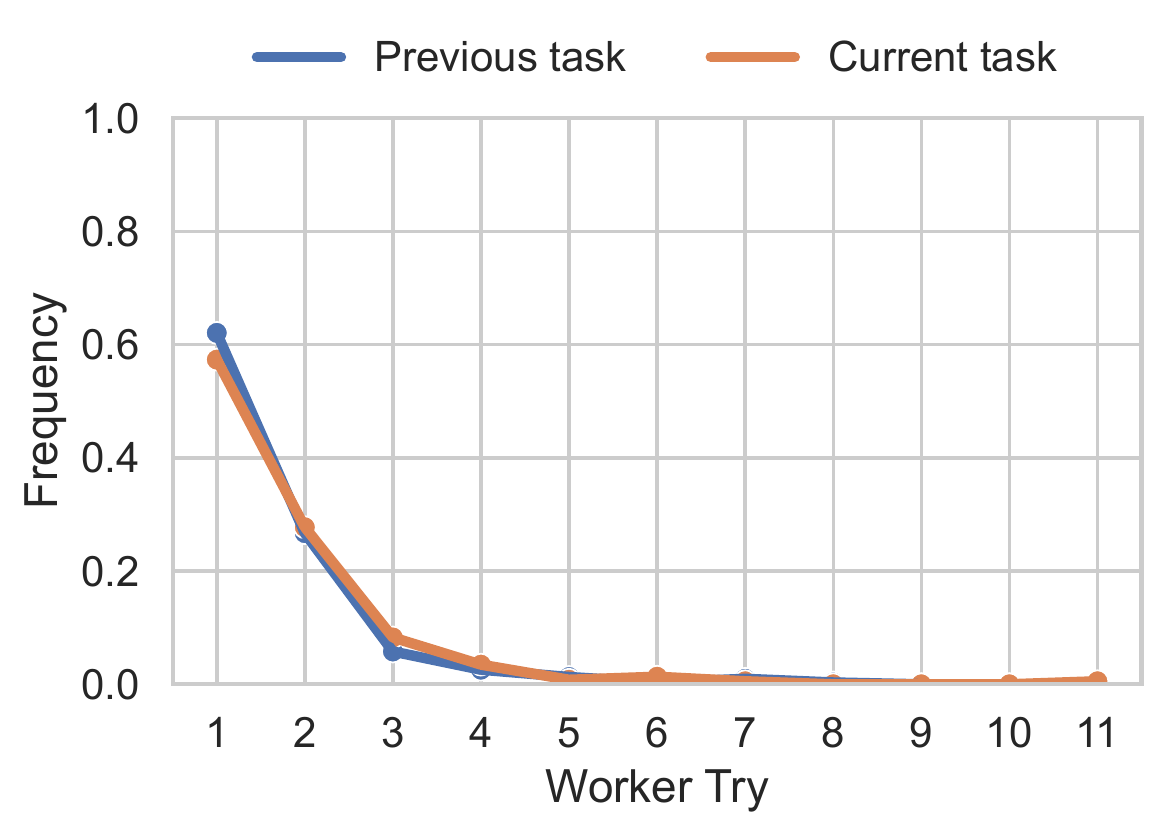}
    \caption{Distribution of task failures.}
    \label{cap:paper_facct2022-sec:desc-stat-subsec:task-abandonment-distribution_fail}
  \end{subfigure}
  
  \caption{Comparison of abandonment and failure rates. Orange: current task (Section~\ref{cap:paper_facct2022-sec:exp-setup-subsec:crowdsourcing-task}); blue: previous task (Section~\ref{cap:paper_ipm2021-sec:exp-setup-subsec:crowdsourcing-task}).}
  \label{cap:paper_facct2022-sec:desc-stat-subsec:task-abandonment-distribution}
\end{figure}

Figure~\ref{cap:paper_facct2022-sec:desc-stat-subsec:task-abandonment-distribution} provides a comparison of the distributions of abandonment and task failure. Specifically, Figure~\ref{cap:paper_facct2022-sec:desc-stat-subsec:task-abandonment-distribution_aband} shows how many workers abandoned the task after annotating a given number of statements. The vast majority (98\%) did so after the first statement. Abandonment after the first statement is negligible. The abandonment rate is 18\% higher than that observed in the task described in Section~\ref{cap:paper_ipm2021-sec:exp-setup-subsec:crowdsourcing-task}, which featured fewer questions. The increase may be explained by the added burden of two additional questionnaires and one more evaluation dimension in the present task (Section~\ref{cap:paper_facct2022-sec:exp-setup-subsec:crowdsourcing-task}), possibly leading more workers to feel overwhelmed or lose interest earlier. In the previous task, some workers continued up to the fourth statement before abandoning. Nonetheless, the overall trend remains: most workers leave the task after the first annotation.

Figure~\ref{cap:paper_facct2022-sec:desc-stat-subsec:task-abandonment-distribution_fail} shows how many workers failed at least one quality check after submitting their work. Most of those who failed attempted the task only once (216, 58\%), with 103 (27\%) attempting it a second time. The remaining 15\% tried three to ten times. The failure rate dropped from 18\% to 14\% compared to the earlier task, indicating that those who submitted their work were less likely to fail. However, the overall failure distribution is consistent with that of previous tasks.

\subsection{Agreement}

\label{cap:paper_facct2022-sec:desc-stat-subsec:agreement}

Internal agreement among workers is computed using Krippendorff's \index{$\upalpha$}$\upalpha$ \cite{krippendorff2011computing} at the unit level. The choice of this metric is motivated in Section~\ref{cap:paper_sigir2020-sec:results-subsec:agreement} and Section~\ref{cap:paper_ipm-sec:results-subsec:judgment-reliability}. Overall, there is a low level of agreement among workers for each considered truthfulness dimension, consistent with previous tasks (see Section~\ref{cap:paper_sigir2020-sec:results-subsec:agreement} and Section~\ref{cap:paper_ipm-sec:results-subsec:judgment-reliability}).

\begin{figure}[tbp]
  \centering
  \begin{subfigure}{.49\linewidth}
    \centering
    \includegraphics[width=.95\linewidth]{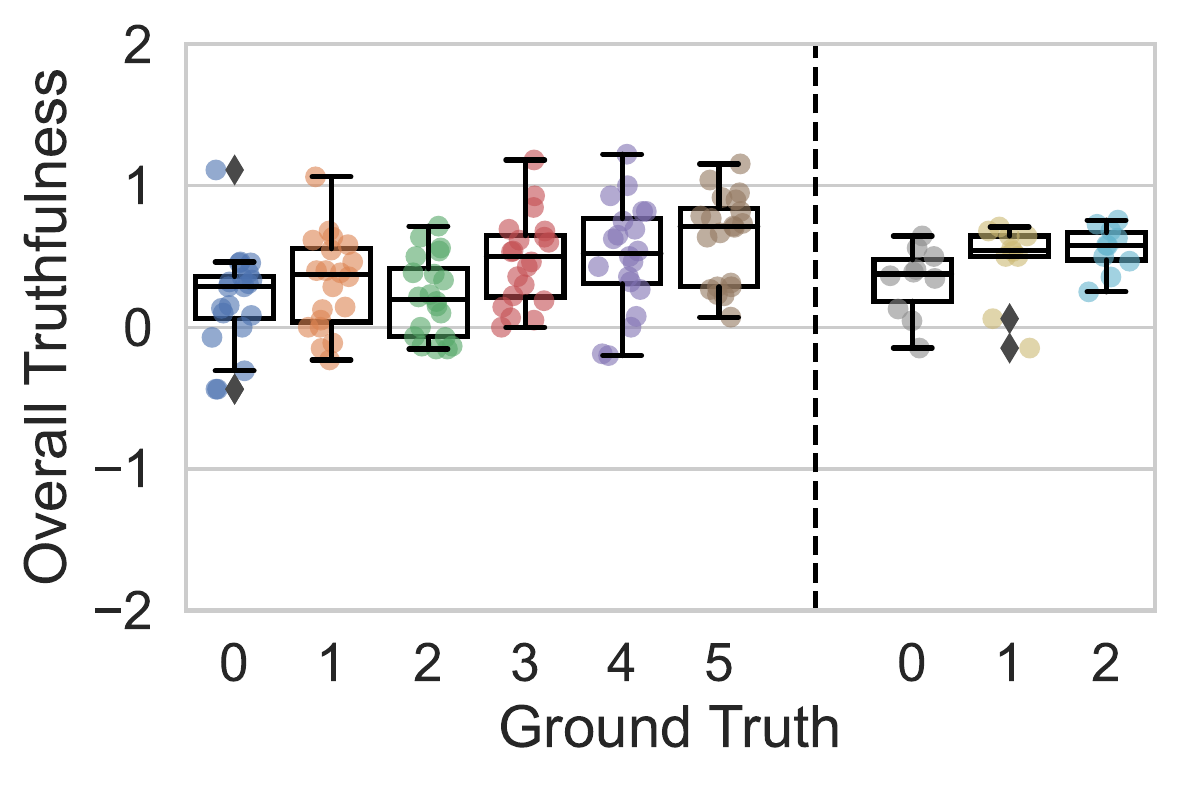}
    \caption{\overalltruthfulness}
    \label{cap:paper_facct2022-sec:desc-stat-subsec:agreement-fig:agreement-ext_overall}
  \end{subfigure}
  \hfill
  \begin{subfigure}{.49\linewidth}
    \centering
    \includegraphics[width=.95\linewidth]{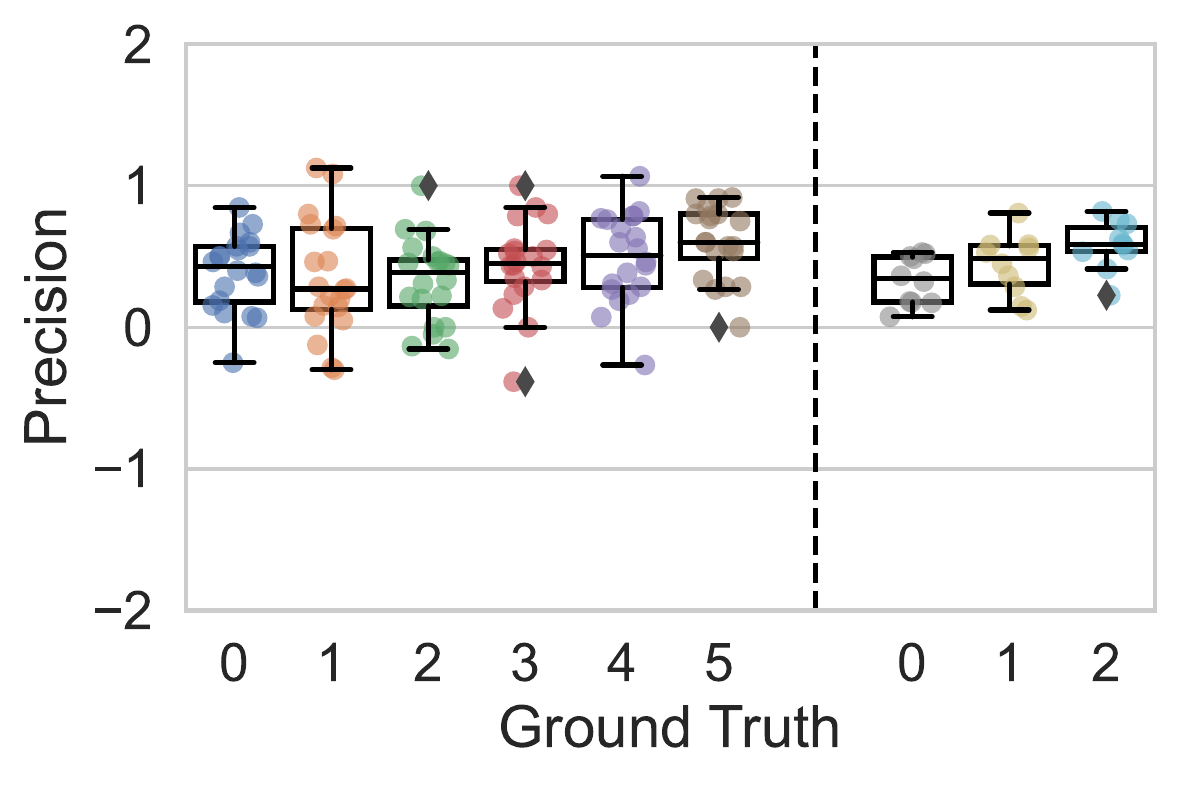}
    \caption{\precision}
    \label{cap:paper_facct2022-sec:desc-stat-subsec:agreement-fig:agreement-ext_precision}
  \end{subfigure}
  \begin{subfigure}{.49\linewidth}
    \centering
    \includegraphics[width=.95\linewidth]{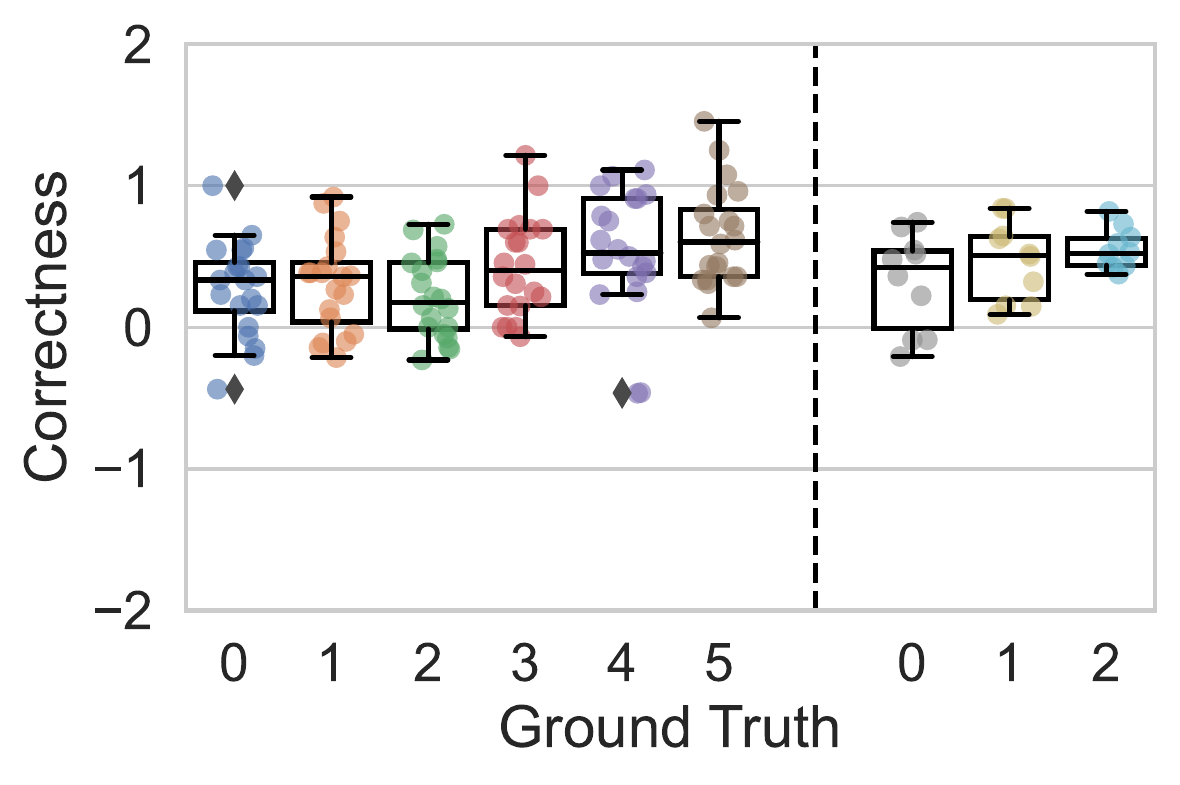}
    \caption{\correctness}
    \label{cap:paper_facct2022-sec:desc-stat-subsec:agreement-fig:agreement-ext_correctness}
  \end{subfigure}
 
  \caption{Agreement between workers’ judgments and ground truth for three dimensions, with a breakdown by \politifact and \abc labels. Compare with Figure~\ref{cap:paper_ipm2021-sec:results-subsec:judgment-reliability-subsec:agreement-external-fig:scale-comparison}.}
  \label{cap:paper_facct2022-sec:desc-stat-subsec:agreement-fig:agreement-ext}
\end{figure}

External agreement between workers’ aggregated scores for \overalltruthfulness and the corresponding expert scores is also measured. It should be noted that the judgment scales used by the experts and workers differ. The experts used six- (\politifact) or three-level scales (\abc), while the workers used a five-level scale. Figure~\ref{cap:paper_facct2022-sec:desc-stat-subsec:agreement-fig:agreement-ext} shows a box plot for each ground truth label. \politifact is displayed to the left of the dotted line, with labels ranging from 0 (\politifactzero) to 5 (\politifactfive). \abc is displayed to the right of the dotted line, with labels ranging from 0 (\abczero) to 2 (\abctwo). The figure can be directly compared with Figure~\ref{cap:paper_ipm2021-sec:results-subsec:judgment-reliability-subsec:agreement-external-fig:scale-comparison}.

Workers tend to provide judgments with higher mean values when moving from left to right (i.e., when considering higher ground truth values), in agreement with the expert assessments. Although Figure~\ref{cap:paper_facct2022-sec:desc-stat-subsec:agreement-fig:agreement-ext} shows only the \overalltruthfulness, \precision, and \correctness dimensions as examples, the patterns for the remaining dimensions are similar. \overalltruthfulness directly correlates with the ground truth, whereas the other dimensions capture orthogonal and independent information not directly measured by the experts. The interquartile range is narrower for \abc statements compared to the analysis in Section~\ref{cap:paper_ipm2021-sec:results-subsec:judgment-reliability-subsec:agreement-external}.

\section{Results}

\label{cap:paper_facct2022-sec:results}

Section~\ref{cap:paper_facct2022-sec:results-subsec:hypothesis-tests} discusses the outcomes of the hypothesis tests introduced in Section~\ref{cap:paper_facct2022-sec:exp-study-subsec:hypotheses}, while Section~\ref{cap:paper_facct2022-sec:results-subsec:exp-analyses} identifies trends that emerged from the exploratory analyses.

\subsection{Hypothesis Tests}

\label{cap:paper_facct2022-sec:results-subsec:hypothesis-tests}

The hypothesis testing involves conducting various statistical analyses. Table~\ref{cap:paper_facct2022-sec:results-subsec:hypothesis-tests-tab:summary} summarizes the overall aim of each hypothesis, the analysis performed to assess it, and the values obtained.

The multiple linear regression shows no evidence of a relationship between \extmeanabserr and \trustpolitics (\hyphooneashort) or \cognitivereflection (\hyphoonecshort). However, both \beliefscience (\hyphoonebshort) and \meanconfidence (\hyphotwodshort) are significant predictors of \extmeanabserr. Workers with stronger \beliefscience and those with greater \meanconfidence appear more biased in their truthfulness judgments compared to others, which is partly unexpected.

Workers generally overestimated truthfulness, as their mean \extmeanerr ($0.33$, \index{$\upsigma$}$\upsigma = 0.46$) was significantly greater than 0, according to the one-sample \textit{t}-test performed (\hyphotwoashort). Correlation analysis shows a significant positive relationship between \affectspeaker and \exterr: the more workers liked the statement speaker, the more they overestimated truthfulness. Conversely, the more they disliked the speaker, the more they underestimated it (\hyphotwobshort).

The \index{ANOVA}ANOVA, conducted with the statement's affiliated party and the worker's affiliated party as independent variables and \exterr as the dependent variable (\hyphotwocshort), reveals no evidence of an interaction effect between the two. This result suggests that no conclusion can be drawn about whether workers over- or underestimate truthfulness to a different degree depending on whether the statement aligns with their preferred political party or orientation. In summary, there is evidence supporting some of the hypotheses, suggesting that:
\begin{itemize}[label=--]
\item Workers with stronger beliefs in science are more accurate in their truthfulness judgments (\hyphoonebshort).
\item Workers with greater confidence are more biased in their truthfulness judgments (\hyphotwodshort).
\item Workers generally overestimate truthfulness (\hyphotwoashort).
\item Workers' truthfulness judgments are affected by the extent to which they like the statement speaker (\hyphotwobshort).
\end{itemize}
There is also evidence of a relationship between \beliefscience and bias in truthfulness judgments. However, the results show that workers with stronger \beliefscience are actually more biased than others, contrary to the prediction of (\hyphoonebshort).

\begin{table}[tbp]
\centering
\caption{Statistical analyses performed to assess the hypotheses.}
\label{cap:paper_facct2022-sec:results-subsec:hypothesis-tests-tab:summary}
\begin{tabular}{p{0.5cm}p{6.7cm}p{4.9cm}p{3.2cm}}
\toprule
\textbf{ID} & \textbf{Interpretation} & \textbf{Analysis} & \textbf{Values} \\
\midrule
\hyphooneashort & \trustpolitics as a predictor of \extmeanabserr & Multiple linear regression & \index{$\upbeta$}$\upbeta = -0.04$, \index{$p$}$p = 0.020$ \\
\midrule
\hyphoonebshort & \beliefscience as a predictor of \extmeanabserr & Multiple linear regression & \index{$\upbeta$}$\upbeta = 0.07$, \index{$p$}$p = 0.003$ \\
\midrule
\hyphoonecshort & \cognitivereflection as a predictor of \extmeanabserr & Multiple linear regression & \index{$\upbeta$}$\upbeta = 0.02$, \index{$p$}$p = 0.152$ \\
\midrule
\hyphotwoashort & Difference between \extmeanerr and 0 & One-sample \textit{t}-test & \index{$t$}$t = 12.18$, \index{$p$}$p < 0.001$ \\
\midrule
\hyphotwobshort & Relationship between \affectspeaker and \exterr & Spearman correlation analysis & \index{$\uprho$}$\uprho = 0.25$, \index{$p$}$p < 0.001$ \\
\midrule
\hyphotwocshort & Interaction effect between political affiliations on \exterr & Factorial mixed \index{ANOVA}ANOVA: \exterr as DV; workers’ affiliation (between-subjects), statement’s affiliation (within-subjects) & \index{$F$}$F = 1.59$, \index{$p$}$p = 0.112$ \\
\midrule
\hyphotwodshort & \meanconfidence as a predictor of \extmeanabserr & Multiple linear regression & \index{$\upbeta$}$\upbeta = 0.06$, \index{$p$}$p \leq 0.001$ \\
\bottomrule
\end{tabular}
\end{table}

\index{$\upbeta$} \index{$\uprho$} \index{$F$}

\subsection{Exploratory Analyses}

\label{cap:paper_facct2022-sec:results-subsec:exp-analyses}

Section~\ref{cap:paper_facct2022-sec:results-subsec:exp-analyses-subsec:remarks} provides initial remarks on the analyses performed. Section~\ref{cap:paper_facct2022-sec:results-subsec:exp-analyses-subsec:pol-aff} examines the worker characteristics associated with systematic biases in crowdsourced truthfulness judgments. Section~\ref{cap:paper_facct2022-sec:results-subsec:exp-analyses-subsec:pred-ext-mean-err} addresses the cognitive biases that manifest during the crowdsourcing task. Finally, Section~\ref{cap:paper_facct2022-sec:results-subsec:exp-analyses-subsec:dimensions} investigates how these biases affect the individual dimensions of truthfulness.

\subsubsection{Initial Remarks}

\label{cap:paper_facct2022-sec:results-subsec:exp-analyses-subsec:remarks}

In addition to the descriptive analyses (Section~\ref{cap:paper_facct2022-sec:desc-stat}) and hypothesis tests (Section~\ref{cap:paper_facct2022-sec:exp-study-subsec:hypotheses}), further exploratory analyses, unregistered in the preregistration, are conducted on the collected data. These analyses aim to refine the interpretation of the hypothesis test results and to uncover potentially relevant trends that were not addressed in the planned analyses.

Accordingly, the results should be interpreted in relation to the initial exploratory study (Section~\ref{cap:paper_facct2022-sec:exp-study}), the experimental variables (Section~\ref{cap:paper_facct2022-sec:exp-setup-subsec:variables}), and the outcomes of the hypothesis tests (Section~\ref{cap:paper_facct2022-sec:results-subsec:hypothesis-tests}).

\subsubsection{\ref{cap:paper_facct2022-sec:research-questions_1}: The Role of Workers’ and Statements’ Political Affiliations}

\label{cap:paper_facct2022-sec:results-subsec:exp-analyses-subsec:pol-aff}

The \index{ANOVA}ANOVA analysis reveals no evidence of an interaction effect between workers’ and statements’ political affiliations in predicting \exterr (\hyphotwocshort). This suggests that workers do not systematically overestimate or underestimate truthfulness based on alignment with the political party associated with the statement. The same model also shows no evidence of a main effect of workers’ political affiliation on \exterr ($F = 1.43$, \index{$p$}$p = 0.232$), indicating that workers’ political affiliation may not influence truthfulness judgments in this context.

However, a significant main effect is observed for statements’ political affiliation ($F = 10.55$, \index{$p$}$p < 0.001$). Comparisons across different affiliations show that workers significantly overestimate the truthfulness of statements associated with the Australian \labor Party (mean \exterr = $0.51$, Tukey-adjusted \index{$p$}$p = [<0.001, 0.018]$), and significantly underestimate statements affiliated with the Australian \liberal Party (mean \exterr = $0.08$, Tukey-adjusted \index{$p$}$p = [<0.001, 0.014]$). Statements affiliated with the \republican and \democratic parties receive similar ratings on average.

These findings suggest that the political affiliation of the statement plays a role in predicting bias in crowd workers’ truthfulness judgments—even, or perhaps especially, when those parties are unfamiliar to the workers. This is particularly relevant given that all participants in the crowdsourcing task are US-based (Section~\ref{cap:paper_facct2022-sec:exp-setup-subsec:crowdsourcing-task}).

\subsubsection{\ref{cap:paper_facct2022-sec:research-questions_2}: Predicting \extmeanabserr}

\label{cap:paper_facct2022-sec:results-subsec:exp-analyses-subsec:pred-ext-mean-err}

\begin{figure}[tbp]
  \centering
  \includegraphics[width=\linewidth]{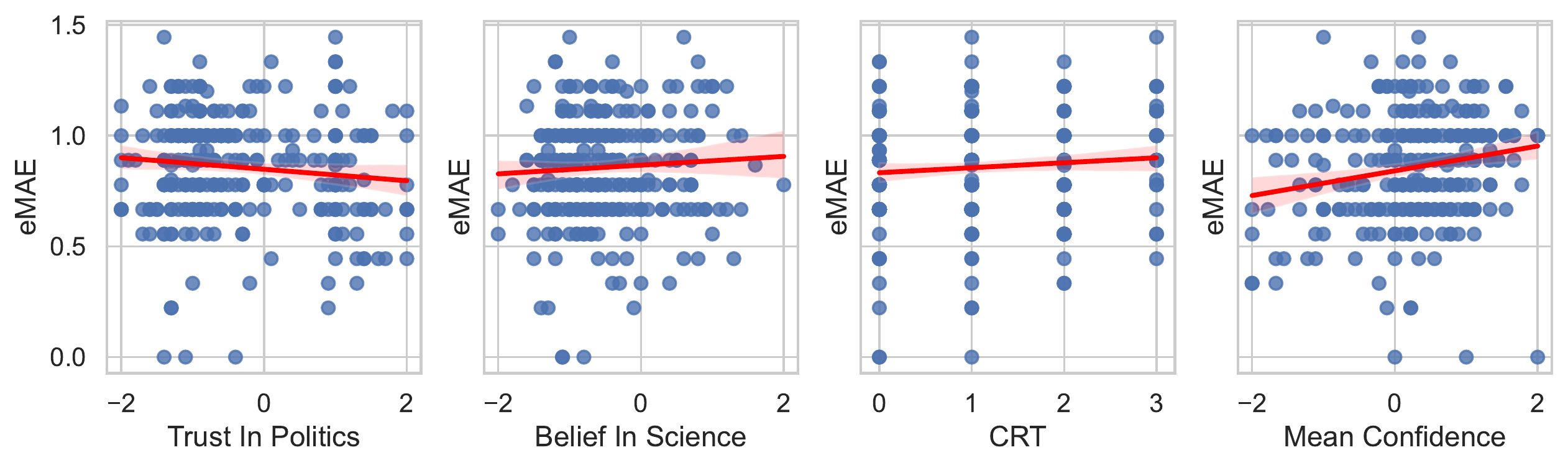}
  \caption{Scatter plots showing the relationships between workers’ \extmeanabserr and their (from left to right) \trustpolitics (\hyphooneashort), \beliefscience (\hyphoonebshort), \cognitivereflection (\hyphoonecshort), and \meanconfidence (\hyphotwodshort).}
  \label{cap:paper_facct2022-sec:results-subsec:exp-analyses-subsec:pred-ext-mean-err-fig:hyp_tests}
\end{figure}

The multiple linear regression identifies workers’ \beliefscience and \meanconfidence as significant predictors of \extmeanabserr (\hyphoonebshort and \hyphotwodshort). Interestingly, individual Spearman correlation analyses show that only \meanconfidence is substantially correlated with \extmeanabserr (\index{$\uprho$}$\uprho = 0.20$, \index{$p$}$p < 0.001$), whereas \beliefscience is not (see also Figure~\ref{cap:paper_facct2022-sec:results-subsec:exp-analyses-subsec:pred-ext-mean-err-fig:hyp_tests}).

This suggests that \beliefscience becomes a relevant predictor of \extmeanabserr only when considered in conjunction with \trustpolitics and/or \cognitivereflection, as in the multiple regression model. These two variables may therefore still play an important role in predicting workers’ \extmeanabserr, even though direct evidence for their individual effects was not found.

\subsubsection{\ref{cap:paper_facct2022-sec:research-questions_3}: Looking At Individual Truthfulness Dimensions}

\label{cap:paper_facct2022-sec:results-subsec:exp-analyses-subsec:dimensions}

In addition to an overall tendency toward overestimation of truthfulness, the hypothesis tests (Section~\ref{cap:paper_facct2022-sec:results-subsec:hypothesis-tests}) indicate that workers’ \beliefscience, \meanconfidence, and their affective stance toward the statement speaker (\affectspeaker) may be related to bias in truthfulness judgments. Therefore, the individual truthfulness dimensions are analyzed to determine which specific dimensions are affected by these biases, in order to gain further insight into their nature.

The best \intmeanerr predictors of \extmeanabserr are \neutrality and \comprehensibility. Workers exhibit more bias when they assign higher scores to \neutrality (\index{$\upbeta$}$\upbeta = 0.10$, \index{$p$}$p = 0.001$) or lower scores to \comprehensibility (\index{$\upbeta$}$\upbeta = -0.08$, \index{$p$}$p = 0.013$). Additionally, workers’ \beliefscience is not a significant predictor for any truthfulness dimension except \overalltruthfulness, whereas \meanconfidence is a significant predictor for all \intmeanerr measures.

Other noteworthy relationships include a negative association between workers’ \trustpolitics and \neutrality scores (\index{$\upbeta$}$\upbeta = -0.09$, \index{$p$}$p = 0.028$), and a positive association between \cognitivereflection and \comprehensibility (\index{$\upbeta$}$\upbeta = 0.08$, \index{$p$}$p = 0.027$). Finally, \affectspeaker is positively related to all considered truthfulness dimensions.

\section{Summary}

\label{cap:paper_facct2022-sec:discussion}

This chapter investigates the impact of worker biases in crowdsourced fact-checking by addressing three research questions. To support the analysis, an initial exploratory study is conducted to formulate several hypotheses, based on the dataset described in Section~\ref{cap:paper_ipm2021-sec:exp-setup}. These hypotheses are then tested in a novel crowdsourcing experiment. The answers to the research questions can be summarized as follows.

\myparagraph{\ref{cap:paper_facct2022-sec:research-questions_1}}
No evidence was found for any influence of workers’ \trustpolitics (\hyphooneashort) or \cognitivereflection (\hyphoonecshort) on truthfulness judgments. However, the results suggest a relationship with workers’ degree of \beliefscience (\hyphoonebshort). Contrary to expectations, those reporting a stronger belief in science tend to be less accurate in their truthfulness assessments.

\myparagraph{\ref{cap:paper_facct2022-sec:research-questions_2}}
Although no interaction was found between workers’ and statements’ political affiliations (\hyphotwocshort), indicating no support for \mycitebias{Confirmation Bias}{\ref{cap:paper_ipm2023_bias-bias:confirmation-bias}}, workers generally overestimate truthfulness (\hyphotwoashort). The findings also support the presence of the \mycitebias{Affect Heuristic}{\ref{cap:paper_ipm2023_bias-bias:affect-heuristic}}: the more workers like the speaker of a statement (\affectspeaker), the more they overestimate its truthfulness, and vice versa (\hyphotwobshort). Additionally, there is evidence for the \mycitebias{Overconfidence Effect}{\ref{cap:paper_ipm2023_bias-bias:overconfidence-effect}}, as workers with higher self-reported confidence in their ability to judge truthfulness (\meanconfidence) tend to be less accurate (\hyphotwodshort).

\myparagraph{\ref{cap:paper_facct2022-sec:research-questions_3}}
Exploratory analyses show that more biased workers tend to assign higher scores to \neutrality and lower scores to \comprehensibility. Furthermore, workers’ \trustpolitics is negatively associated with their \neutrality ratings.

\myparagraph{} The next chapter presents a machine learning-based architecture designed to jointly predict the truthfulness of information items and generate human-readable explanations. The underlying models are validated and calibrated, and an extensive human evaluation is conducted to assess the impact of the generated explanations.

\chapter{A Neural Model To Jointly Predict and Explain Truthfulness}

\label{cap:paper_jdiq2022}

This chapter is based on the article published in the \lq\lq Journal of Data and Information Quality\rq\rq{}~\cite{brand2022neural}, and extends the earlier version presented at the 2021 Truth and Trust Online Conference~\cite{brand2021jointly}. Section~\ref{cap:related_work-sec:crowdsourcing-truthfulness} and Section~\ref{cap:related_work-sec:afc-models} review the relevant related work. Section~\ref{cap:paper_jdiq2022-sec:research_questions} introduces the research questions, while Section~\ref{cap:paper_jdiq2022-sec:model-definition} presents the proposed architecture. Section~\ref{cap:paper_jdiq2022-sec:exp-setup} outlines the experimental setup used to validate the model. Section~\ref{cap:paper_jdiq2022-sec:results} summarizes the main findings and concludes the chapter.

\section{Research Questions}

\label{cap:paper_jdiq2022-sec:research_questions}

This chapter proposes and experimentally evaluates a system that jointly performs truthfulness prediction and explanation generation within the same model. This approach is novel compared to traditional post-hoc explainability methods, which are applied on top of existing machine learning models. By integrating explanation generation into the model itself, the resulting explanations more directly reflect the decision-making process of the truthfulness prediction component.
In addition, the results demonstrate that large transformer models are sufficiently flexible to perform both tasks simultaneously, generating explanations without compromising performance on the original task. This dual capability enables human end users to better interpret the model’s outputs, fostering a more trustworthy interaction between users and deep learning models.

The overarching goal is to support wider adoption of automated fact-checking systems by developing a model capable of both evaluating the truthfulness of a statement and generating a human-interpretable explanation for its prediction. The following research questions are investigated:

\begin{enumerate}[start=28, leftmargin=2.9em, label=RQ\arabic*]
    \item \label{cap:paper_jdiq2022-sec:research-questions_1} How can a deep learning model be designed to classify information truthfulness while simultaneously generating a natural language explanation that supports its classification decision?
    \item \label{cap:paper_jdiq2022-sec:research-questions_2} Can such a model achieve both accurate classification decisions and high-quality natural language explanations?
    \item \label{cap:paper_jdiq2022-sec:research-questions_3} Are machine-generated explanations useful for helping humans better judge information truthfulness?
    \item \label{cap:paper_jdiq2022-sec:research-questions_4} Can the deep learning model be calibrated to produce reliable confidence scores for its truthfulness predictions?
\end{enumerate}

\section{\ref{cap:paper_jdiq2022-sec:research-questions_1}: \ebart Definition}

\label{cap:paper_jdiq2022-sec:model-definition}

Many of the systems in the reviewed literature use separate \transformer models for truthfulness prediction and explanation generation. In contrast, the architecture proposed in this chapter, called \ebart, jointly outputs both a truthfulness prediction and a human-readable, abstractive explanation.

\begin{figure}[tpb]
\centering
\includegraphics[width=.9\linewidth]{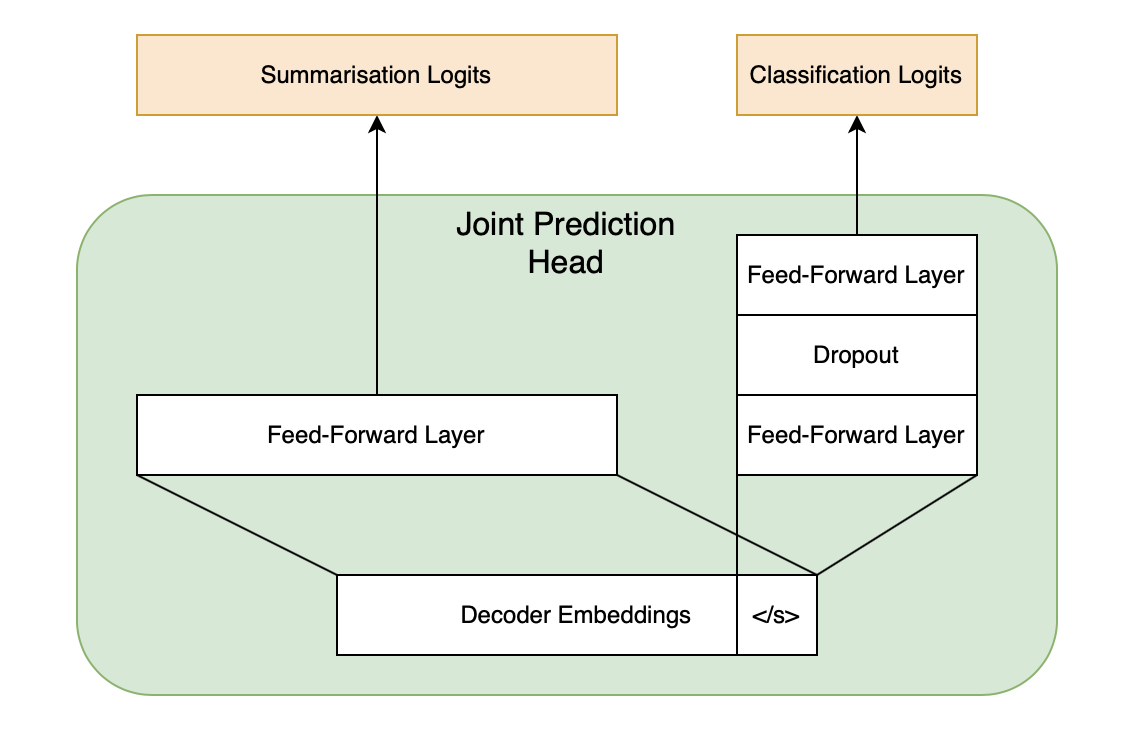}
\caption{The \jointpredictionhead of the \ebart architecture.}
\label{cap:paper_jdiq2022-sec:model-definition-fig:jointhead}
\end{figure}

Adapting the \bartlarge encoder-decoder model to this downstream task required the development of a \jointpredictionhead, shown in Figure~\ref{cap:paper_jdiq2022-sec:model-definition-fig:jointhead}. This component sits atop the \bart model \cite{lewis2019bart} and transforms the transformer hidden states into the desired output format. Both the base \bart model and the \jointpredictionhead can be fine-tuned together, allowing the pre-trained \bart weights to be adapted to the joint prediction task.

\begin{figure}[tpb]
    \centering
    \includegraphics[width=.9\linewidth]{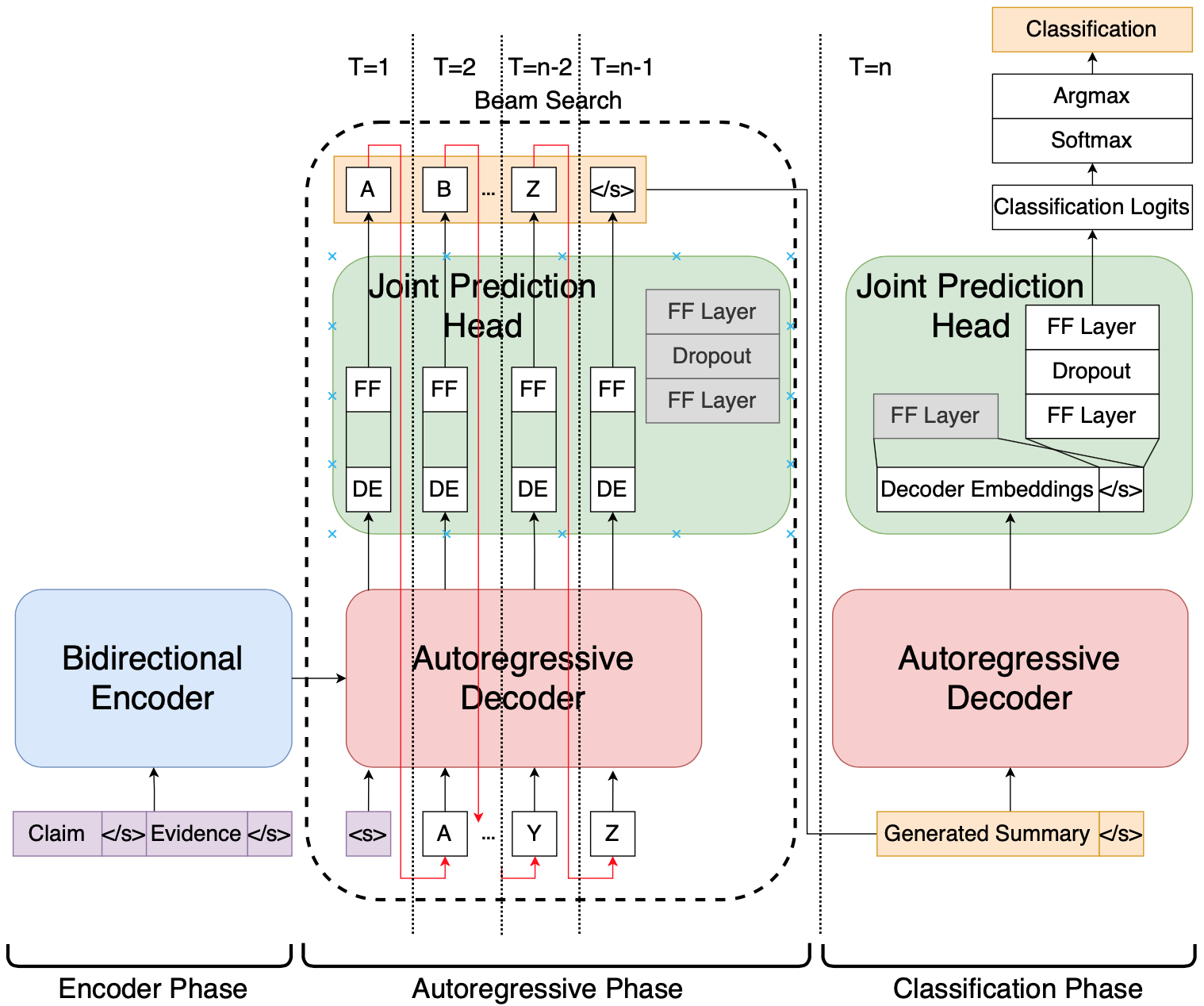}
    \caption{The inference process of the \ebart architecture.}
    \label{cap:paper_jdiq2022-sec:model-definition-fig:inference}
\end{figure}

The \jointpredictionhead is also highlighted in green in Figure~\ref{cap:paper_jdiq2022-sec:model-definition-fig:inference}. This head takes as input the final decoder hidden state embeddings. All embeddings are passed to a single feed-forward layer, which produces a sequence of logits forming the basis of the predicted explanation.

To facilitate classification, the hidden state embeddings corresponding to the final sequence separator token (\spverb|</s>| in \bart) are extracted and passed to a small feed-forward network that shapes the output to match the number of target classes. The resulting logits are then passed through a final softmax layer to obtain class probabilities. Unlike \bert \cite{devlin2018bert}, which uses the embedding corresponding to the \spverb|[CLS]| token prepended to the input for classification, \bart uses the final sequence separator token instead. This design reflects the fact that the decoder can only attend to tokens on its left, meaning classification is conditioned on the full generated sequence.

The summarization component of the \jointpredictionhead consists of a single feed-forward layer. Its input dimension is equal to the decoder embedding size (768), and its output dimension corresponds to the model’s vocabulary size. During greedy generation, the head uses the argmax of the raw logits at each step.

It is useful to distinguish between training and inference, as they differ slightly due to the auto-regressive nature of the \bart decoder. Figure~\ref{cap:paper_jdiq2022-sec:model-definition-fig:inference} shows the inference process. Inference begins by running the encoder on the tokenized input to generate the encoder hidden states, as in the training phase. The decoder is then initialized with the start sequence token \index{<s>}(\texttt{<s>} in \bart), which differs from the training setup. It proceeds to generate logits auto-regressively, guided by beam search.
In the final phase of inference, the decoder is run with the entire generated sequence as input. At this point, the \jointpredictionhead extracts the embeddings corresponding to the token immediately before the final sequence separator token. This step is performed to mirror the training process. The extracted embeddings are passed to the classification component of the \jointpredictionhead, followed by a softmax layer, which produces the final classification.

\begin{figure}[tpb]
    \centering
    \includegraphics[width=.85\linewidth]{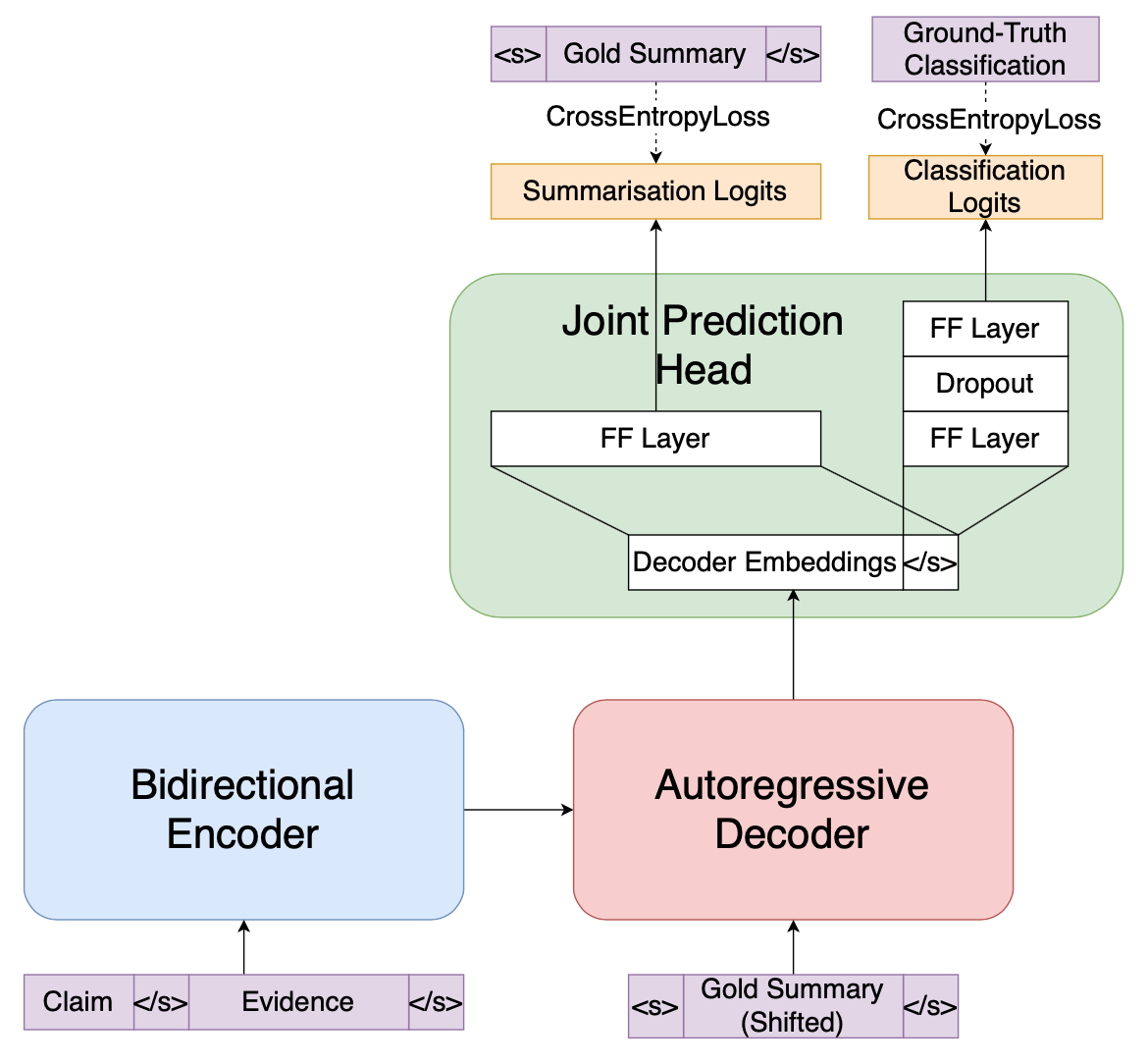}
    \caption{The training configuration of the \ebart architecture.}
    \label{cap:paper_jdiq2022-sec:model-definition-fig:training}
\end{figure}

Figure~\ref{cap:paper_jdiq2022-sec:model-definition-fig:training} shows the training process. During training, the encoder generates hidden states from the tokenized input, which are then passed to the decoder. The tokenized gold summary is fed into both the input and the summarization output of the decoder, with the input shifted right by one token. This setup conditions the decoder to predict the next token given the current one. At the same time, the classification labels are provided to the classification output of the \jointpredictionhead. The model weights for the two main \lq\lq pathways\rq\rq{} in the \jointpredictionhead, namely the generation path and the classification path, are not shared. However, all layers before the head share weights between the classification and generation tasks, as in the base \bart model.

The overall loss is computed as a weighted sum, using parameters \index{$\upalpha$}$\upalpha$ and $\left(1 - \upalpha\right)$, of two \index{Cross Entropy Loss}\texttt{Cross Entropy Loss} terms: one between the summarization logits and the gold summary, and one between the classification logits and the ground truth label. The first goal is to train the model to reconstruct the original summary text after it has been corrupted by a noising function. The second goal is to adjust the model weights so that the final classification logits align with the target class representation. In this way, the loss is optimized jointly for both the generation and classification tasks.

The code, training procedure, and regularization parameters from the original paper \cite{lewis2019bart} and its associated repository\footnote{\url{https://github.com/pytorch/fairseq/blob/main/examples/bart/README.md}} are adopted for this work. Additional details are provided by \citet[Section~5]{lewis2019bart}.
Specifically, the pre-trained \bart weights are used as a starting point, and further training is carried out jointly on both the \bart model and the custom \jointpredictionhead. While the \ebart architecture is heavily based on \bart, it introduces important modifications. In particular, \ebart does not simply add a classification layer on top of the original model. Instead, it incorporates the \jointpredictionhead (shown in Figure~\ref{cap:paper_jdiq2022-sec:model-definition-fig:inference} and Figure~\ref{cap:paper_jdiq2022-sec:model-definition-fig:training}), which enables joint modeling of both classification and generation. This design allows the model to simultaneously predict the truthfulness of a statement and generate a corresponding explanation.

\section{Experimental Setting}

\label{cap:paper_jdiq2022-sec:exp-setup}

Three datasets are used: \fever (Section~\ref{cap:dataset-sec:fever}), \efever (Section~\ref{cap:dataset-sec:e-fever}), and \esnli (Section~\ref{cap:dataset-sec:e-snli}). Section~\ref{cap:paper_jdiq2022-sec:exp-setup-subsec:training} describes the training methodology used for the \ebart architecture, while Section~\ref{cap:paper_jdiq2022-sec:exp-setup-subsec:evaluation} presents the evaluation approach.
 
\subsection{Training Methodology}

\label{cap:paper_jdiq2022-sec:exp-setup-subsec:training}

Two different versions of the model are trained to evaluate performance on the \fever and \efever datasets. The first version, \ebartsmall, is trained on the subset of the \efever training set that excludes examples with \nulltext explanations (Section~\ref{cap:dataset-sec:e-fever}), resulting in \num{40,702} examples. To preprocess the data, the \index{+}\spverb|+| character used to separate page titles from evidence fragments is removed. Model inputs are tokenised and formatted as \index{<s>}\spverb|<s>claim</s>evidence</s>|. Truthfulness labels are converted to numerical values, and explanations are tokenised similarly. This processed dataset is then used to fine-tune the \bartlarge model with a \jointpredictionhead for 3 epochs.

The second version, \ebartfull, follows the same training procedure but includes the entire \efever training set, incorporating also the examples with \nulltext explanations, for a total of \num{50,000} examples. Training is performed on two platforms: \index{Google!Colab}Google Colab with an \index{NVIDIA!T4 GPU}NVIDIA T4 GPU (16 GB memory), and \index{Microsoft Azure}Microsoft Azure with a 12 GB \index{NVIDIA!Tesla K80 GPU}NVIDIA Tesla K80 GPU. The \bart-based models have approximately \num{375,000000} parameters and require around 5 hours to fine-tune with the \jointpredictionhead for 3 epochs on the NVIDIA T4 GPU using the \efever dataset.

\subsection{Evaluation Methodology}

\label{cap:paper_jdiq2022-sec:exp-setup-subsec:evaluation}

The development split of the \efever dataset is prepared in the same way as the training split, resulting in two versions: \efeversmall and \efeverfull, which respectively exclude and include examples with \nulltext explanations. When evaluating truthfulness prediction accuracy, it has been observed that including the \notenoughinfo class may lead to an underestimation of the model's actual classification performance. Table~\ref{cap:paper_jdiq2022-sec:exp-setup-subsec:evaluation-tab:example-evauluation} shows an example whose ground truth label is \notenoughinfo. However, manual inspection reveals that both the explanation and evidence indicate the claim is in fact refuted—a prediction correctly made by \ebart.

For this reason, two sets of results are reported: one that includes all examples and one that excludes those labeled as \notenoughinfo in \efever. Section~\ref{cap:paper_jdiq2022-sec:results} presents the experiments conducted to evaluate the proposed approach.

\begin{table}[tpb]
    \centering
    \caption{Example where the ground truth label is \notenoughinfo, and the one predicted by \ebart is \refutes.}
    \label{cap:paper_jdiq2022-sec:exp-setup-subsec:evaluation-tab:example-evauluation}
    \begin{tabular}{p{3.5cm}p{5.6cm}p{3.8cm}}
    \toprule
    \textbf{Statement} & \textbf{Evidence} & \textbf{Generated Explanation} \\
    \midrule
    Marnie was directed by someone who was \lq\lq The Master of Nothing\rq\rq{}. &
    Alfred Hitchcock Sir Alfred Joseph Hitchcock (13 August 1899–29 April 1980) was an English film director and producer, at times referred to as \lq\lq The Master of Suspense\rq\rq{}. Marnie (film) Marnie is a 1964 American psychological thriller film directed by Alfred Hitchcock. &
    Marnie was directed by Alfred Hitchcock, who was \lq\lq The Master of Suspense\rq\rq{}. \\
    \bottomrule
    \end{tabular}
\end{table}

\section{Results}

\label{cap:paper_jdiq2022-sec:results}

Section~\ref{cap:paper_jdiq2022-sec:results-subsec:evaluation} reports performance metrics and comparisons for \ebart and the joint prediction model. Section~\ref{cap:paper_jdiq2022-sec:results-subsec:crowdsourcing-task} examines the impact of model-generated explanations as assessed by crowd workers. Finally, Section~\ref{cap:paper_jdiq2022-sec:results-subsec:calibration-generation} addresses model calibration, focusing on the reliability of the predicted confidence scores.

\subsection{\ref{cap:paper_jdiq2022-sec:research-questions_2}: \ebart Evaluation And Validation}

\label{cap:paper_jdiq2022-sec:results-subsec:evaluation}

The effectiveness of the \ebart model is evaluated on the original \fever dataset in Section~\ref{cap:paper_jdiq2022-sec:results-subsec:evaluation-subsec:fever}. Section~\ref{cap:paper_jdiq2022-sec:results-subsec:evaluation-subsec:efever} addresses the \efever dataset, while Section~\ref{cap:paper_jdiq2022-sec:results-subsec:evaluation-subsec:esnli} focuses on \esnli. Sections~\ref{cap:paper_jdiq2022-sec:results-subsec:evaluation-subsec:validation-exp-1} and \ref{cap:paper_jdiq2022-sec:results-subsec:evaluation-subsec:validation-exp-2} present two experiments designed to validate the use of joint models.

\subsubsection{Evaluation: Original \fever}

\label{cap:paper_jdiq2022-sec:results-subsec:evaluation-subsec:fever}

The classification performance of \ebart on the original \fever development set is reported to enable comparison with existing models. The \domlin system~\cite{stammbach2019team} is used only for evidence retrieval (its truthfulness predictions are discarded) and provides evidence for \num{17,000} out of the \num{20,000} examples in the development set. The \ebart model generates truthfulness predictions for these 17k examples, while the remaining ones are labeled as \notenoughinfo, following the procedure described by \citet{stammbach2019team}. Results are reported on the development set rather than the test set, since ground-truth labels for the test set are not publicly available.

\ebartsmall and \ebartfull achieve label accuracies of \num{75.0} and \num{75.1}, respectively, on the \fever dataset, outperforming previously published methods. For comparison, reported accuracies of other models on this dataset include:

\begin{itemize}[label=--]
\item \ukpathene: 68.5\%, by \citet{hanselowski2018ukp}.
\item \unc: 69.6\%, by \citet{nie2019combining}.
\item \bertbased: 74.6\%, by \citet{soleimani2020bert}.
\item \domlin: 72.1\%, by \citet{stammbach2019team}.
\item \uclmr: 69.7\%, by \citet{yoneda2018ucl}.
\end{itemize}

\ebart compares favorably to prior work, despite being trained on the \efever dataset, which contains \num{95,000} fewer examples than \fever (the dataset used to train the other models). This performance gain is hypothesized to result from the use of \bart as the base model and from requiring the model to attend more closely to the most relevant evidence when generating explanations. The most informative comparison is between \ebart and \domlin, since both use the same evidence retrieval mechanism, allowing the contribution of \ebart over standard truthfulness classifiers to be isolated.

\subsubsection{Evaluation: \efever}

\label{cap:paper_jdiq2022-sec:results-subsec:evaluation-subsec:efever}

Table~\ref{cap:paper_jdiq2022-sec:results-subsec:evaluation-subsec:efever-tab:performance-efever} reports the results obtained on the \efever development set. Since no other results have been published for this dataset, a comprehensive snapshot of \ebart's performance is provided. As expected, both models perform better on \efeversmall, which contains fewer inconclusive examples. More surprisingly, \ebart achieves consistent performance regardless of whether it is trained on \efeversmall or \efeverfull. This suggests that \ebart is robust even in scenarios where evidence is sparse.

The \rouge{} metrics~\cite{lin-2004-rouge} are used to evaluate the overlap between the generated explanations and those in the \efever dataset. However, these scores are not necessarily indicative of explanation quality. For instance, an explanation generated by \texttt{GPT-3}\index{GPT-3} may include additional information not present in the reference. Whether such elaboration leads to a better explanation than a more concise one depends on the intended use case and is ultimately subjective.

\begin{table}[tbp]
    \centering
    \caption{Effectiveness of \ebart on the \efever dataset.}
    \label{cap:paper_jdiq2022-sec:results-subsec:evaluation-subsec:efever-tab:performance-efever}
    \begin{tabular}{ll*{6}{c}}
    \toprule
    \textbf{Model} & \textbf{Dataset} &
    \begin{tabular}[c]{@{}c@{}}\textbf{Accuracy}\\\textbf{(No N.E.I.)}\end{tabular} &
    \begin{tabular}[c]{@{}c@{}}\textbf{Accuracy}\\\textbf{(Full)}\end{tabular} &
    \begin{tabular}[c]{@{}c@{}}\rouge\\\textbf{1}\end{tabular} &
    \begin{tabular}[c]{@{}c@{}}\rouge\\\textbf{2}\end{tabular} &
    \begin{tabular}[c]{@{}c@{}}\rouge\\\textbf{L}\end{tabular} &
    \begin{tabular}[c]{@{}c@{}}\rouge\\\textbf{Sum}\end{tabular} \\
    \midrule
    \ebartsmall & \efeversmall & \num{87.2} & \num{78.2} & \num{73.581} & \num{64.365} & \num{71.434} & \num{71.585} \\
    \ebartsmall & \efeverfull  & \num{85.4} & \num{77.1} & \num{59.447} & \num{50.177} & \num{57.697} & \num{57.782} \\
    \ebartfull  & \efeversmall & \num{87.1} & \num{78.1} & \num{64.530} & \num{55.283} & \num{62.691} & \num{62.820} \\
    \ebartfull  & \efeverfull  & \num{85.2} & \num{77.2} & \num{65.511} & \num{57.598} & \num{64.071} & \num{64.144} \\
    \bottomrule
    \end{tabular}
\end{table}

Table~\ref{cap:paper_jdiq2022-sec:exp-setup-subsec:evaluation-tab:example-evauluation}, Table~\ref{cap:paper_jdiq2022-sec:results-subsec:evaluation-subsec:efever-tab:example-1}, and Table~\ref{cap:paper_jdiq2022-sec:results-subsec:evaluation-subsec:efever-tab:example-2} present examples from the development set. Table~\ref{cap:paper_jdiq2022-sec:results-subsec:evaluation-subsec:efever-tab:example-1} illustrates that the model is able to express uncertainty when no relevant evidence is found. In the generated explanation, the phrase \lq\lq however does not say whether...\rq\rq{} conveys in a human-readable way that the evidence retrieval mechanism did not provide sufficient support for a confident prediction.

Table~\ref{cap:paper_jdiq2022-sec:results-subsec:evaluation-subsec:efever-tab:example-2} shows an example where the model correctly identifies a contradiction between the claim that Janet Leigh was incapable of writing and the evidence, which states that she authored four books. The model produces a coherent explanation that highlights this inconsistency, extracting relevant facts to support the justification. Importantly, it also assigns the correct label to the statement, in alignment with the generated explanation.

\begin{table}[tpb]
    \centering
    \caption{Statement where the ground truth label is \supports and the prediction by \ebart is also \supports.}
    \label{cap:paper_jdiq2022-sec:results-subsec:evaluation-subsec:efever-tab:example-1}
    \begin{tabular}{p{3.2cm}p{5.6cm}p{6.6cm}}
    \toprule
    \textbf{Statement} & \textbf{Evidence} & \textbf{Generated Explanation} \\
    \midrule 
    CBS is the network that aired The Millers & The Millers The multi-camera series aired from October 3, 2013 to July 18, 2015 and ran 34 episodes over two seasons on CBS. CBS announced the cancellation of The Millers on November 14, 2014, four episodes into the show's second season. & The Millers aired on CBS, however it does not say whether it was the network that aired it. \\
    \bottomrule
    \end{tabular}
\end{table}

\begin{table}[tpb]
    \centering
    \caption{Statement where the ground truth label is \refutes and the prediction by \ebartsmall is also \refutes.}
    \label{cap:paper_jdiq2022-sec:results-subsec:evaluation-subsec:efever-tab:example-2}
    \begin{tabular}{p{3.2cm}p{5.6cm}p{2.9cm}p{3.7cm}}
    \toprule
    \textbf{Statement} & \textbf{Evidence} & \textbf{Gold Explanation} & \textbf{Generated Explanation} \\
    \midrule 
    Janet Leigh was incapable of writing. & Janet Leigh Janet Leigh (born Jeanette Helen Morrison; July 6, 1927-October 3, 2004) was an American actress, singer, dancer and author. Janet Leigh She also wrote four books between 1984 and 2002, including two novels. & Janet Leigh was capable of writing. & Janet Leigh wrote four books between 1984 and 2002, including two novels. \\
    \bottomrule
    \end{tabular}
\end{table}

\subsubsection{Evaluation: \esnli}
\label{cap:paper_jdiq2022-sec:results-subsec:evaluation-subsec:esnli}

The \esnli task presents a challenge similar to that of \efever, in which the entailment between two sentences must be predicted and an explanation generated. This setup is conceptually close to evaluating the truthfulness of a statement with respect to supporting evidence. A modified version of the \ebart architecture is trained specifically on the \esnli dataset. The data is prepared by enumerating the labels, removing noisy entries, and tokenizing the summaries. The premise and hypothesis sentences are concatenated and tokenized in the same manner as the statement and evidence used in the \efever evaluation.

\ebart achieves a label accuracy of \num{90.1} and a \bleu{} score of \num{32.70}~\cite{10.3115/1073083.1073135}. The model originally proposed alongside the \esnli dataset, \index{e-INFERSENT}\texttt{e-INFERSENT}, achieves an accuracy of 84.0 and a \bleu{} score of 22.4~\cite{camburu2018snli}. As in \citeauthor{camburu2018snli}, the first two gold explanations are used as references to compute the \bleu{} score for explainable models.

The following are the best-performing models published in the literature,\footnote{\url{https://nlp.stanford.edu/projects/snli/}} which, however, do not provide explanations. They are included here for additional comparison:

\begin{itemize}[label=--]
\item \dcrcoan: 90.1\%, by \citet{kim2019semantic}.
\item \mtdnn: 91.6\%, by \citet{liu2019multi}.
\item \camtl: 92.1\%, by \citet{pilault2020conditionally}.
\item \lmtrans: 89.9\%, by \citet{radford2018improving}.
\item \sjrc: 91.3\%, by \citet{zhang2018explicit}.
\item \sembert: 91.9\%, by \citet{zhang2020semantics}.
\end{itemize}

The performance of \ebart compares favorably with the state-of-the-art on this distinct task, reinforcing the findings from the \fever evaluation and further confirming that \ebart is flexible enough to generate explanations without compromising its performance on the main task. Table~\ref{cap:paper_jdiq2022-sec:results-subsec:evaluation-subsec:esnli-tab:example-1} and Table~\ref{cap:paper_jdiq2022-sec:results-subsec:evaluation-subsec:esnli-tab:example-2} provide examples that complement the \bleu{} metric through manual verification of explanation quality.

The explanation shown in Table~\ref{cap:paper_jdiq2022-sec:results-subsec:evaluation-subsec:esnli-tab:example-1} illustrates \ebart's ability to focus on the most relevant details of the input. \index{Sentence!1}Sentence 1 includes superfluous information about an additional person in a wool hat, whereas \index{Sentence!2}Sentence 2 does not mention this. The generated explanation correctly reflects this, discussing only the differences involving the tan girl while omitting the irrelevant detail.
Similarly, Table~\ref{cap:paper_jdiq2022-sec:results-subsec:evaluation-subsec:esnli-tab:example-2} presents a case in which the model identifies a subtle discrepancy between the input sentences that undermines a full entailment relation. The fact that \index{Sentence!2}Sentence 2 makes a stronger claim about the family going to the beach suggests that \index{Sentence!1}Sentence 1 does not fully represent the situation. The \ebart architecture captures this nuance and correctly labels the pair as \neutral. The generated explanation reflects this distinction, highlighting that \index{Sentence!2}Sentence 2 introduces important details omitted in \index{Sentence!1}Sentence~1.

\begin{table}[tbp]
    \centering
    \caption{Example where the ground truth label is \entailment and the prediction by \ebart is also \entailment.}
    \label{cap:paper_jdiq2022-sec:results-subsec:evaluation-subsec:esnli-tab:example-1}
    \begin{tabular}{p{4.2cm}p{4.2cm}p{4.2cm}}
    \toprule
    \textbf{Sentence 1} & \textbf{Sentence 2} & \textbf{Generated Explanation} \\
    \midrule
    One tan girl with a wool hat is running and leaning over an object, while another person in a wool hat is sitting on the ground. &
    A tan girl runs leans over an object &
    One tan girl with a wool hat is running and leaning over an object is a rephrasing of a tan girl runs leans over an object. \\
    \bottomrule
    \end{tabular}
\end{table}

\begin{table}[tpb]
    \centering
    \caption{Example where the ground truth label is \neutral and the prediction by \ebart is also \neutral.}
    \label{cap:paper_jdiq2022-sec:results-subsec:evaluation-subsec:esnli-tab:example-2}
    \begin{tabular}{p{4.2cm}p{4.2cm}p{4.2cm}}
    \toprule
    \textbf{Sentence 1} & \textbf{Sentence 2} & \textbf{Generated Explanation} \\
    \midrule
    A young family enjoys feeling ocean waves lap at their feet. &
    A young man and woman take their child to the beach for the first time. &
    Just because a young family enjoys feeling ocean waves lap at their feet does not mean that they take their child to the beach for the first time. \\
    \bottomrule
    \end{tabular}
\end{table}

\subsubsection{Validation: Experiment 1}

\label{cap:paper_jdiq2022-sec:results-subsec:evaluation-subsec:validation-exp-1}

The performance of \ebart is compared to a pipeline that performs truthfulness prediction and explanation generation using two independent models, in order to assess the capabilities of the underlying components and to further address \ref{cap:paper_jdiq2022-sec:research-questions_2}. The architecture and training methodology were kept as consistent as possible to ensure comparability between the two approaches. 

The separate pipeline, named \bartseparate, uses a \bart-based sequence classifier and a separate \bart-based model for explanation generation. Both \ebart and \bartseparate are initialized with the same pre-trained weights and trained and evaluated on \efeversmall. In \bart, the statement and evidence are provided as input to both the encoder and decoder. In contrast, \ebart uses the much shorter gold summary as input to the decoder. As a result, due to memory constraints, inputs are truncated to a maximum length of 256 tokens, which affects only 4.56\% of the examples.

To address convergence issues, a virtual batch size of 32 is used when training the classifier (batch size four with eight gradient accumulation steps). The sequence generator is trained with a batch size of two and two gradient accumulation steps, also due to hardware memory limitations. In comparison, the joint model is trained with a batch size of four and no gradient accumulation.

Before presenting the results, it is important to highlight the trade-off between effectiveness and resource requirements when training a single joint model versus two separate models. Training separate models entails either doubling the required computational resources (if trained in parallel, since each model must be trained independently on a GPU), or doubling the training time if executed sequentially. In contrast, a joint model must manage more parameters, but it can be trained on both tasks simultaneously.

Table~\ref{cap:paper_jdiq2022-sec:results-subsec:evaluation-subsec:validation-exp-1-tab:joint-v-separate} shows that the prediction performance of \ebart and \bartseparate is nearly identical, with the latter being marginally more effective. Manual inspection of the generated explanations suggests that both models produce outputs of comparable quality in terms of expressiveness and cohesiveness. This result supports the findings from the evaluations on \efever and \esnli: \ebart is capable of jointly generating explanations without compromising performance on the primary classification task.

\begin{table}[tbp]
    \centering
    \caption{Effectiveness of \ebart and \bartseparate on the \efeversmall dataset.}
    \label{cap:paper_jdiq2022-sec:results-subsec:evaluation-subsec:validation-exp-1-tab:joint-v-separate}
    \begin{tabular}{lcccccc}
    \toprule
    \textbf{Model} & 
    \begin{tabular}[c]{@{}c@{}}\textbf{Accuracy}\\\textbf{(No N.E.I.)}\end{tabular} & 
    \begin{tabular}[c]{@{}c@{}}\textbf{Accuracy}\\\textbf{(Full)}\end{tabular} & 
    \begin{tabular}[c]{@{}c@{}}\rouge\\\textbf{1}\end{tabular} & 
    \begin{tabular}[c]{@{}c@{}}\rouge\\\textbf{2}\end{tabular} & 
    \begin{tabular}[c]{@{}c@{}}\rouge\\\textbf{L}\end{tabular} & 
    \begin{tabular}[c]{@{}c@{}}\rouge\\\textbf{Sum}\end{tabular} \\
    \midrule
    \ebart & \num{87.2} & \num{78.2} & \num{73.581} & \num{64.365} & \num{71.434} & \num{71.585} \\
    \bartseparate & \num{88.1} & \num{78.9} & \num{73.070} & \num{63.634} & \num{71.005} & \num{71.136} \\
    \bottomrule
    \end{tabular}
\end{table}

\subsubsection{Validation: Experiment 2}

\label{cap:paper_jdiq2022-sec:results-subsec:evaluation-subsec:validation-exp-2}

This experiment investigates whether the internal consistency between the predicted truthfulness and the corresponding explanation differs between \ebart and \bartseparate. To this end, the same models from Experiment 1 (Section~\ref{cap:paper_jdiq2022-sec:results-subsec:evaluation-subsec:validation-exp-1}) are reused, with the addition of a \lq\lq judge\rq\rq{} model trained to predict the truthfulness of a statement based solely on its explanation.
The judge model is a \bart-based sequence classifier trained using the ground-truth truthfulness labels and gold explanations from the \efeversmall dataset. Since it is trained independently, its parameters are not influenced by either \ebart or \bartseparate.

In this experiment, the statements from the development set are paired with the explanations generated by \ebart. These pairs are passed to the judge model, which predicts a truthfulness label. This prediction is then compared to the original label predicted by \ebart, and the agreement between the two is used to compute accuracy. The same procedure is applied to the outputs of \bartseparate. 

The results of this consistency evaluation are shown in Table~\ref{cap:paper_jdiq2022-sec:results-subsec:evaluation-subsec:validation-exp-2-tab:joint-v-separate}, and indicate that \ebart achieves higher consistency accuracy, as determined by the judge model. This suggests that the truthfulness prediction and explanation produced by \ebart are more closely aligned than those generated by \bartseparate. These findings provide evidence that joint models may offer greater inherent interpretability compared to pipelines that generate explanations in a post-hoc manner. While this does not constitute definitive proof, it supports the hypothesis that consistency can be improved through joint prediction and explanation.

\begin{table}[tbp]
    \centering
    \caption{Internal consistency of \ebart and \bartseparate on the \efeversmall dataset.}
    \label{cap:paper_jdiq2022-sec:results-subsec:evaluation-subsec:validation-exp-2-tab:joint-v-separate}
    \begin{tabular}{lcc}
    \toprule
     \textbf{Model} & \textbf{Accuracy (No N.E.I.)} & \textbf{Accuracy (Full)} \\
    \midrule
    \ebart & \num{91.8} & \num{86.8} \\
    \bartseparate & \num{90.4} & \num{85.8} \\
    \bottomrule
    \end{tabular}
\end{table}

\subsection{\ref{cap:paper_jdiq2022-sec:research-questions_3}: Testing The Impact Of The Explanations Generated}

\label{cap:paper_jdiq2022-sec:results-subsec:crowdsourcing-task}

The usefulness of the explanations generated by the \ebart model is evaluated through a set of experiments involving crowd workers. In these experiments, human annotators perform truthfulness classification tasks under different conditions. Section~\ref{cap:paper_jdiq2022-sec:results-subsec:crowdsourcing-task-subsec:design} outlines the experimental design, Section~\ref{cap:paper_jdiq2022-sec:results-subsec:crowdsourcing-task-subsec:agreement-ext} reports the external agreement scores, Section~\ref{cap:paper_jdiq2022-sec:results-subsec:crowdsourcing-task-subsec:agreement-int} covers internal agreement, and Section~\ref{cap:paper_jdiq2022-sec:results-subsec:crowdsourcing-task-subsec:summary} summarizes the main findings.

\subsubsection{Crowdsourcing Task}

\label{cap:paper_jdiq2022-sec:results-subsec:crowdsourcing-task-subsec:design}

The crowdsourcing task is published on the \mturk platform. Four versions of the same task are designed to assess the impact of machine-generated explanations on truthfulness judgments. In each version, workers are shown a statement from the \fever dataset and asked to provide a truthfulness judgment using a two-level judgment scale, along with a written justification. Requiring justifications has been shown to improve annotation quality~\cite{kutlu2020annotator}. The available labels are \bartfalse and \barttrue.

Each worker evaluates four statements—two labeled as \refutes and two as \supports in the ground truth—and each statement is judged by ten distinct workers. To mitigate bias, a randomization process is applied when assigning statements to workers (i.e., in the \index{HIT}HITs published on the platform). The same assignments are maintained across all task versions for consistency, except in one specific case described later. Workers are restricted to completing only one task version. To ensure data quality and discourage adversarial behavior, a minimum of two seconds is required per judgment. The four task versions are:
\begin{enumerate}
	\item \taskone: workers see only the statement from the \fever dataset and provide a truthfulness judgment and justification.
	\item \tasktwo: workers see the statement and the explanation generated by the \ebart architecture, then provide a judgment and justification.
	\item \taskthree: workers see the statement and the ground truth explanation, then provide a judgment and justification.
	\item \taskfour: workers see the statement, the ground truth explanation, and the \ebart-generated explanation. They provide a judgment, a justification, and indicate which explanation they find more informative.
\end{enumerate}
A manual inspection of the dataset revealed that, in some \index{HIT}HITs, the ground truth explanation and the explanation generated by \ebart were identical. To avoid this overlap, the statements were re-sampled to ensure that the two explanations differed by at least one character. 

The experimental setup described above enables both implicit and explicit comparisons. Comparing \taskone (no explanation) with \tasktwo (\ebart explanation) and \taskthree (ground truth explanation) allows assessing the effect of providing an explanation. Furthermore, an implicit comparison between the two types of explanations, namely the ones generated by \ebart and the ground truth, is also possible. \taskfour (explanation preferred) includes an explicit comparison, allowing workers to indicate which explanation they find more informative.

\subsubsection{External Agreement}

\label{cap:paper_jdiq2022-sec:results-subsec:crowdsourcing-task-subsec:agreement-ext}

\begin{figure}[tbp]
  \centering
  \begin{subfigure}{.49\linewidth}
    \centering
    \textbf{\qquad Raw Judgments}
    \includegraphics[width=.9\linewidth]{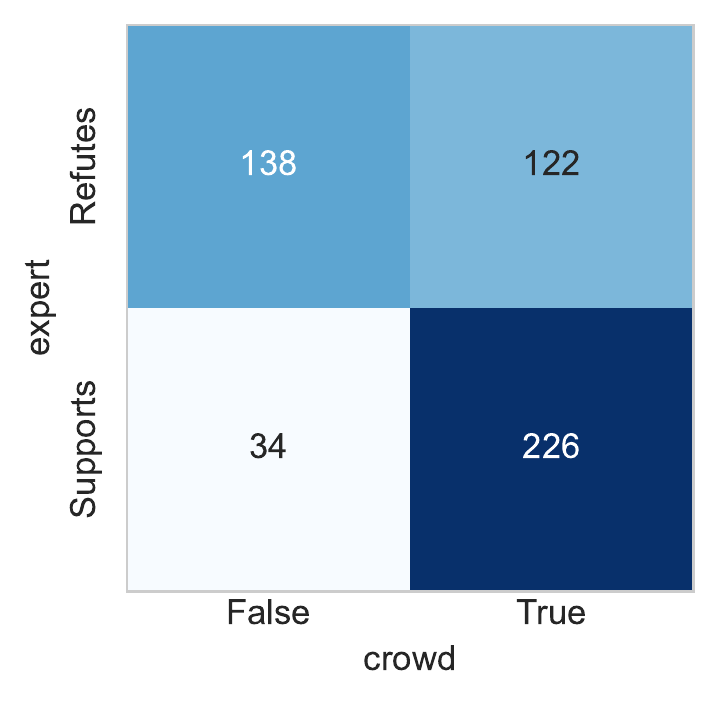}
    \caption{\taskone.}
    \label{cap:paper_jdiq2022-sec:results-subsec:crowdsourcing-task-subsec:agreement-ext-fig:agreement-ext_t1-raw}
  \end{subfigure}
  \begin{subfigure}{.49\linewidth}
    \centering
    \textbf{\qquad Aggregated Judgments}
    \includegraphics[width=.9\linewidth]{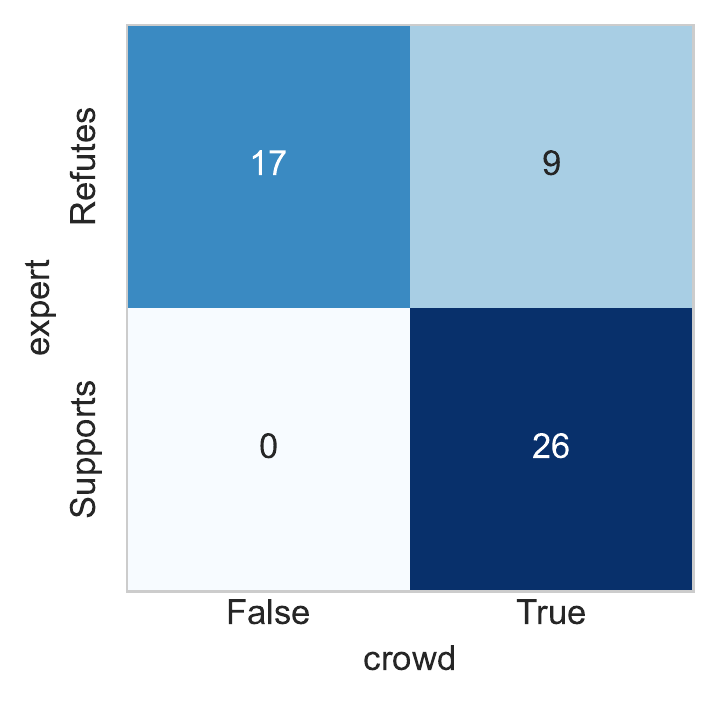}
    \caption{\taskone.}
    \label{cap:paper_jdiq2022-sec:results-subsec:crowdsourcing-task-subsec:agreement-ext-fig:agreement-ext_t1-agg}
  \end{subfigure}
  \begin{subfigure}{.49\linewidth}
    \centering
    \includegraphics[width=.9\linewidth]{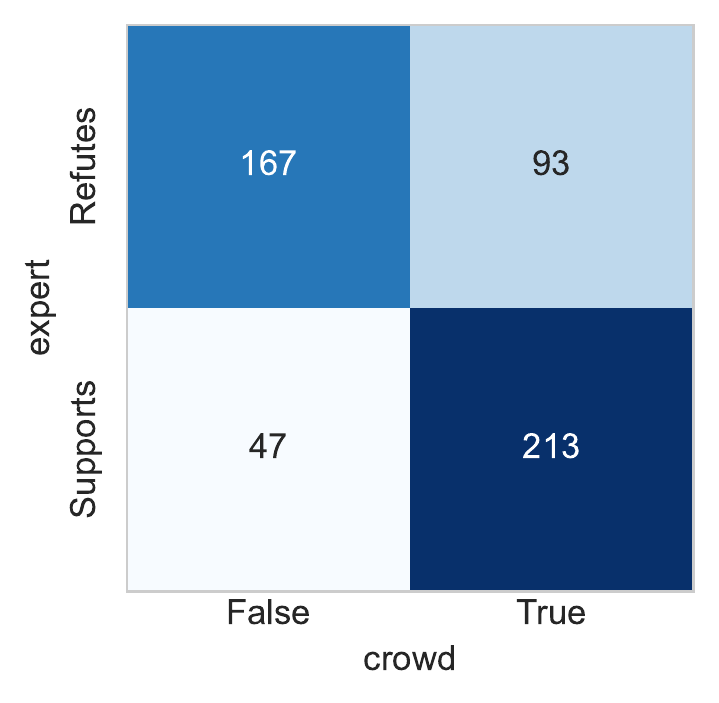}
    \caption{\tasktwo.}
    \label{cap:paper_jdiq2022-sec:results-subsec:crowdsourcing-task-subsec:agreement-ext-fig:agreement-ext_t2-raw}
  \end{subfigure}
  \begin{subfigure}{.49\linewidth}
    \centering
    \includegraphics[width=.9\linewidth]{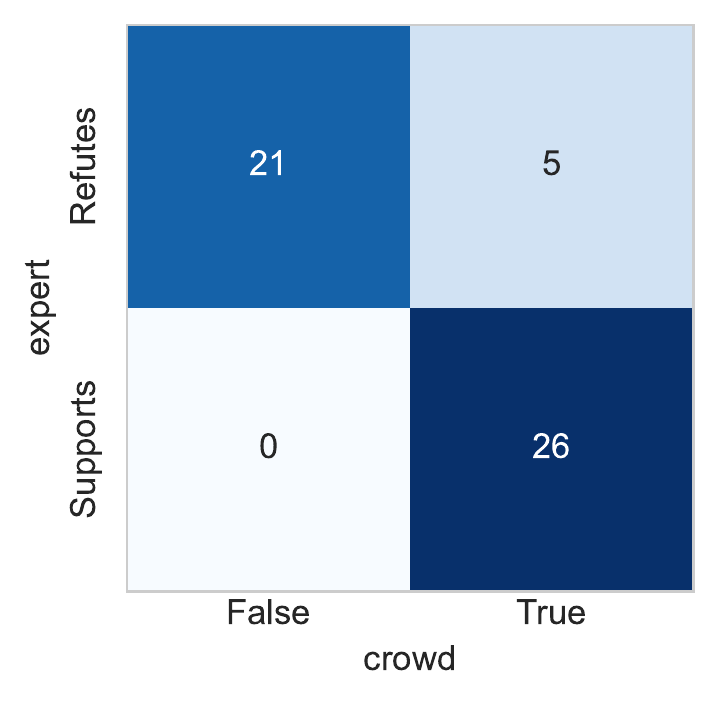}
    \caption{\tasktwo.}
    \label{cap:paper_jdiq2022-sec:results-subsec:crowdsourcing-task-subsec:agreement-ext-ext_t2-agg}
  \end{subfigure}
  \begin{subfigure}{.49\linewidth}
    \centering
    \includegraphics[width=.9\linewidth]{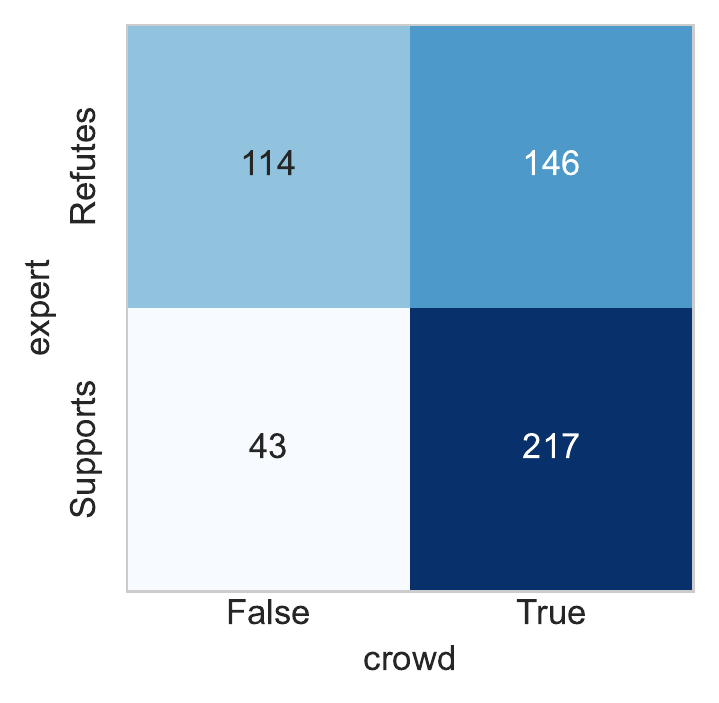}
    \caption{\taskthree.}
    \label{cap:paper_jdiq2022-sec:results-subsec:crowdsourcing-task-subsec:agreement-ext-ext_t3-raw}
  \end{subfigure}
  \begin{subfigure}{.49\linewidth}
    \centering
    \includegraphics[width=.9\linewidth]{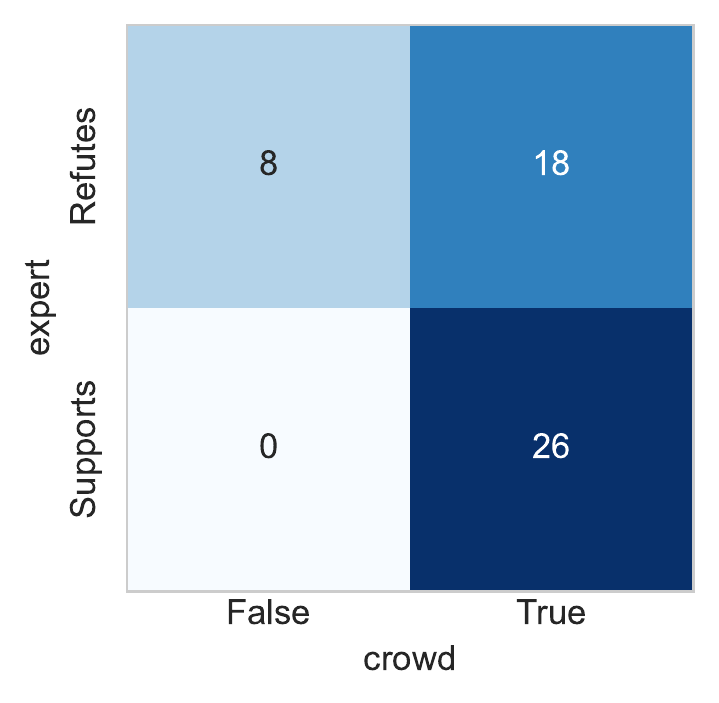}
    \caption{\taskthree.}
    \label{cap:paper_jdiq2022-sec:results-subsec:crowdsourcing-task-subsec:agreement-ext-fig:agreement-ext_t3-agg}
  \end{subfigure}
  \caption{External agreement between ground truth and crowd for raw (first column) and aggregated (second column) truthfulness judgments. Each cell represents either the count of judgments (first column) or the count of statements (second column). Correctly classified statements lie on the main diagonal.}
  \label{cap:paper_jdiq2022-sec:results-subsec:crowdsourcing-task-subsec:agreement-ext-fig:agreement-ext}
\end{figure}

\begin{figure}[tbp]
  \centering
  \begin{subfigure}{.49\linewidth}
    \centering
    \textbf{\qquad Raw Agreement}
    \includegraphics[width=.9\linewidth]{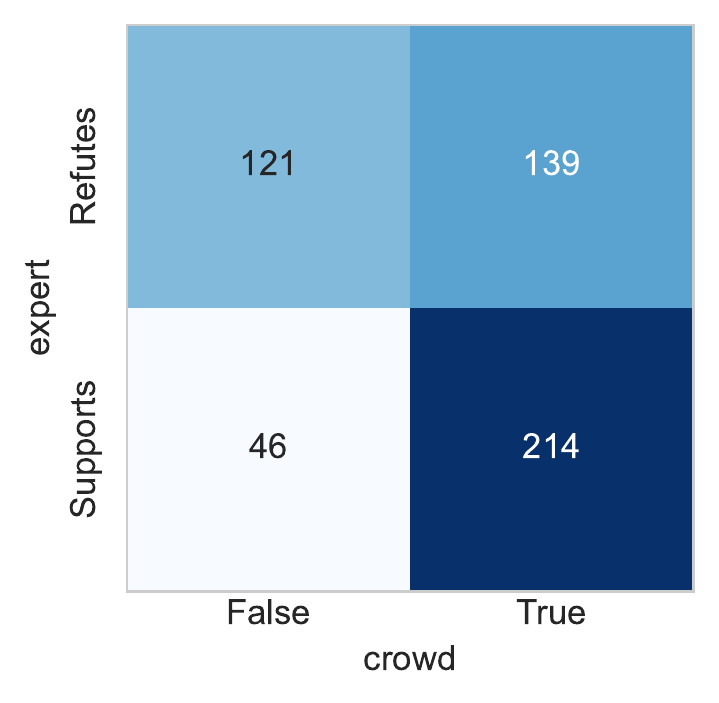}
    \caption{\taskfour.}
    \label{cap:paper_jdiq2022-sec:results-subsec:crowdsourcing-task-subsec:agreement-ext-fig:preference-ext_t4-raw}
  \end{subfigure}
  \begin{subfigure}{.49\linewidth}
    \centering
    \textbf{\qquad Aggregated Agreement}
    \includegraphics[width=.9\linewidth]{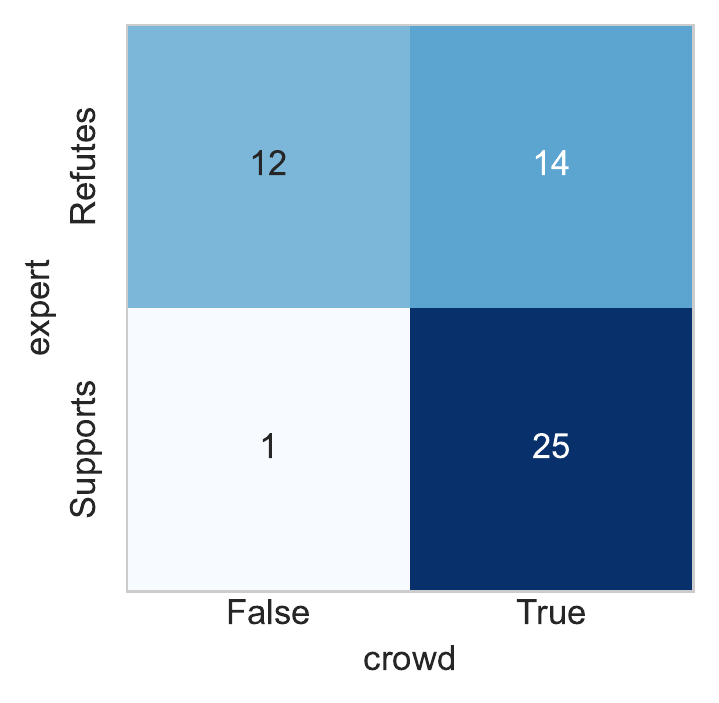}
    \caption{\taskfour.}
    \label{cap:paper_jdiq2022-sec:results-subsec:crowdsourcing-task-subsec:agreement-ext-fig:preference-ext_t4-agg}
  \end{subfigure}
  \begin{subfigure}{.49\linewidth}
    \centering
    \textbf{\qquad Explanation Preferred}
    \includegraphics[width=.9\linewidth]{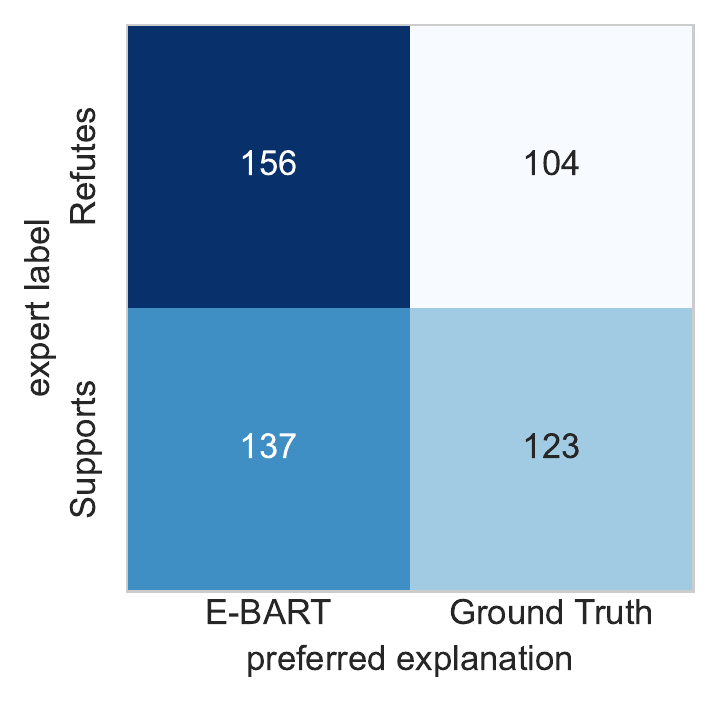}
    \caption{\taskfour.}
    \label{cap:paper_jdiq2022-sec:results-subsec:crowdsourcing-task-subsec:agreement-ext-ext_t4-pref}
  \end{subfigure}
  \caption{External agreement between ground truth and crowd for raw (left chart) and aggregated (center chart) truthfulness judgments. The right chart shows worker preferences between explanations. Each cell represents either the count of judgments (left and right charts) or the number of statements (center chart).}
  \label{cap:paper_jdiq2022-sec:results-subsec:crowdsourcing-task-subsec:agreement-ext-fig:preference-ext}
\end{figure}

Figure~\ref{cap:paper_jdiq2022-sec:results-subsec:crowdsourcing-task-subsec:agreement-ext-fig:agreement-ext} and Figure~\ref{cap:paper_jdiq2022-sec:results-subsec:crowdsourcing-task-subsec:agreement-ext-fig:preference-ext} show the external agreement between the ground truth and the crowd, considering both individual judgments and those aggregated across the ten workers assigned to each statement. Aggregation is performed using majority vote. The corresponding accuracy scores for the different task versions, based on pilot data, are as follows:
\begin{itemize}
  \item[--] \taskone: $0.70$ for raw judgments and $0.83$ for aggregated judgments.
  \item[--] \tasktwo: $0.73$ for raw judgments and $0.90$ for aggregated judgments.
  \item[--] \taskthree: $0.64$ for raw judgments and $0.65$ for aggregated judgments.
  \item[--] \taskfour: $0.64$ for raw judgments and $0.71$ for aggregated judgments.
\end{itemize}

A non-parametric \index{ANOVA}ANOVA with post-hoc test\footnote{\url{https://scikit-posthocs.readthedocs.io/en/latest/}} is run to account for statistical significance. A non-parametric \index{Kruskal-Wallis H Test}Kruskal-Wallis H test (one-way non-parametric \index{ANOVA}ANOVA) is used to test whether the samples originate from the same distribution, given that normality assumptions are violated. The null hypothesis—that the population medians of all considered tasks are equal—can be rejected since a p-value \index{$p$}$p<0.05$ is obtained. A post-hoc test must then be run to identify which tasks differ in their medians. To this end, the \index{Conover Test}Conover test \cite{doi:10.1080/00949655.2018.1438437, doi:10.1080/00401706.1981.10487680} is employed. The results indicate that the task pairs for which the null hypothesis can be rejected (i.e., those with statistically significant differences) are:
\begin{itemize}
  \item[--] \taskone\ vs. \tasktwo\ (\index{$p$}$p<0.05$).
  \item[--] \tasktwo\ vs. \taskthree\ (\index{$p$}$p<0.01$).
  \item[--] \taskthree\ vs. \taskfour\ (\index{$p$}$p<0.01$), for both raw and aggregated judgments.
\end{itemize}

\subsubsection{Internal Agreement}

\label{cap:paper_jdiq2022-sec:results-subsec:crowdsourcing-task-subsec:agreement-int}

The internal agreement among workers is measured using Krippendorff's \index{$\upalpha$}$\upalpha$ \cite{krippendorff2011computing, zapf2016measuring}, following the methodology adopted in Section~\ref{cap:paper_sigir2020-sec:results-subsec:agreement-internal}, Section~\ref{cap:paper_pauc2021-sec:results-subsec:crowd-accuracy-subsec:int-agreement}, and Section~\ref{cap:paper_ipm-sec:results-subsec:judgment-reliability-subsec:agreement-internal}. The \index{$\upalpha$}$\upalpha$ score is computed on individual judgments, with the following agreement values observed across the four tasks:
\begin{itemize}[label=--]
  \item $0.19$ for \taskone
  \item $0.24$ for \tasktwo
  \item $0.10$ for \taskthree
  \item $0.10$ for \taskfour
\end{itemize}
These values suggest a consistently low level of agreement among workers across all task versions.

\subsubsection{Summary}

\label{cap:paper_jdiq2022-sec:results-subsec:crowdsourcing-task-subsec:summary}

The results lead to the following observations. Providing the \ebart explanation (\tasktwo) improves performance compared to showing no explanation (\taskone), whereas this improvement is not observed when the ground truth explanation is shown instead (\taskthree). The implicit comparison between \tasktwo and \taskthree indicates that crowd workers are more accurate when shown the \ebart explanation rather than the ground truth. This outcome is further supported by the explicit preference judgments collected in \taskfour (Figure~\ref{cap:paper_jdiq2022-sec:results-subsec:crowdsourcing-task-subsec:agreement-ext-ext_t4-pref}).

Moreover, showing the \ebart explanation in \tasktwo reduces the number of false positives (i.e., false statements that are mistakenly perceived as true) from 122 to 93. This reduction does not occur when the ground truth explanation is shown. Therefore, explanations generated by \ebart appear to make crowd workers more skeptical toward the truthfulness of statements (see also Table~\ref{cap:paper_jdiq2022-sec:results-subsec:evaluation-subsec:efever-tab:example-1} for an example). This trend does not hold for false negatives (i.e., true statements that are mistakenly perceived as false).

In the context of misinformation, false positives are typically more harmful than false negatives, meaning it is preferable to be overly skeptical than to fail to identify a false statement. Finally, looking at the accuracy scores, a simple majority aggregation of crowd judgments under the conditions of \tasktwo achieves 90\% non-expert label accuracy, offering promising evidence for the viability of crowdsourced truthfulness assessments (Section~\ref{cap:paper_sigir2020-sec:discussion}).

\subsection{\ref{cap:paper_jdiq2022-sec:research-questions_4}: Network Calibration And Generation Of Confidence Scores}

\label{cap:paper_jdiq2022-sec:results-subsec:calibration-generation}

Producing confidence scores alongside the truthfulness classification would improve the transparency and interpretability of the \ebart predictions. These scores reflect how confident the model is in its output and may depend on various factors, such as the quality and quantity of the supporting evidence or the similarity of the input statement to the training data. \ebart generates predictions using a softmax layer applied to a set of logits, with dimensions corresponding to the number of target classes. The softmax output yields a probability-like score for each class. However, this score is not guaranteed to be well-calibrated—that is, it may not accurately reflect the true likelihood of correctness \cite{10.5555/3305381.3305518}. Although calibration issues are common in modern deep neural networks, it is not yet clear whether they systematically affect transformer-based models. This motivates an investigation into the calibration properties of the \ebart model.

A number of post-processing techniques have been developed to correct calibration errors, making the output of the final softmax layer interpretable as a confidence score. Notable methods include Bayesian Binning into Quantiles \cite{10.5555/2888116.2888120}, Platt Scaling \cite{platt1999probabilistic}, and Histogram Binning \cite{10.5555/645530.655658}. A tutorial on calibration measures is provided by \citet{10.1093/jamia/ocz228}. Among these methods, Temperature Scaling \cite{10.5555/3305381.3305518} has proven particularly effective for various neural network architectures, including multi-class classifiers, and is relatively straightforward to implement. This approach introduces a single parameter—the temperature ($T>0$)—and uses it to compute confidence predictions from the model's logits $z_i$, as shown in Equation~\ref{cap:paper_jdiq2022-sec:results-subsec:calibration-generation-eq:temp-scaling}, where $\upsigma_{SM}$ denotes the softmax function:

\begin{equation}
\hat{q_i} = \max_{k} \sigma_{SM}( z_i / T )^k
\end{equation}
\label{cap:paper_jdiq2022-sec:results-subsec:calibration-generation-eq:temp-scaling}
\myequations{Confidence prediction using Temperature Scaling \cite{10.5555/3305381.3305518}.}

Applying Temperature Scaling does not alter the model's classification outputs and therefore does not affect its accuracy \cite{10.5555/3305381.3305518}. The temperature parameter must be tuned on a validation set rather than the training set. For this purpose, the first 9,999 examples from the \efever development set are used to train the temperature parameter, while the remaining instances are used for validation, since the dataset does not include a separate test split.

The temperature parameter is inserted after the final fully connected layer of the original model, with all other model parameters kept frozen. Training is performed using the LBFGS optimizer \cite{Liu1989} for 10,000 iterations. It is important to note that the \ebart model is executed in auto-regressive (inference) mode to generate the logits used in the calibration step. Finally, the model is tested on the held-out validation data by applying the learned temperature parameter prior to the final softmax computation.

\begin{figure}[tbp]
  \centering
  \begin{subfigure}{.49\linewidth}
    \centering
    \includegraphics[width=.9\linewidth]{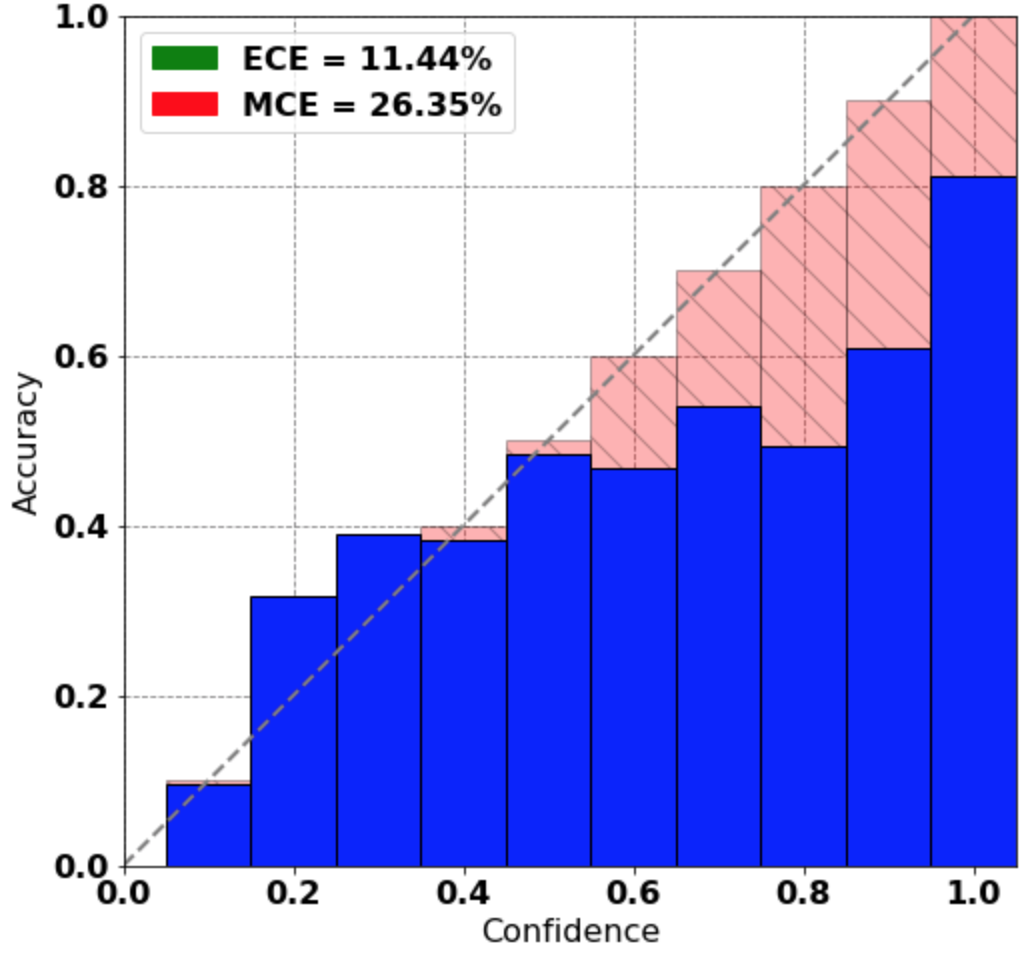}
    \caption{Before calibration.}
    \label{cap:paper_jdiq2022-sec:results-subsec:calibration-generation-fig:calibration_before}
  \end{subfigure}
  \begin{subfigure}{.49\linewidth}
    \centering
    \includegraphics[width=.9\linewidth]{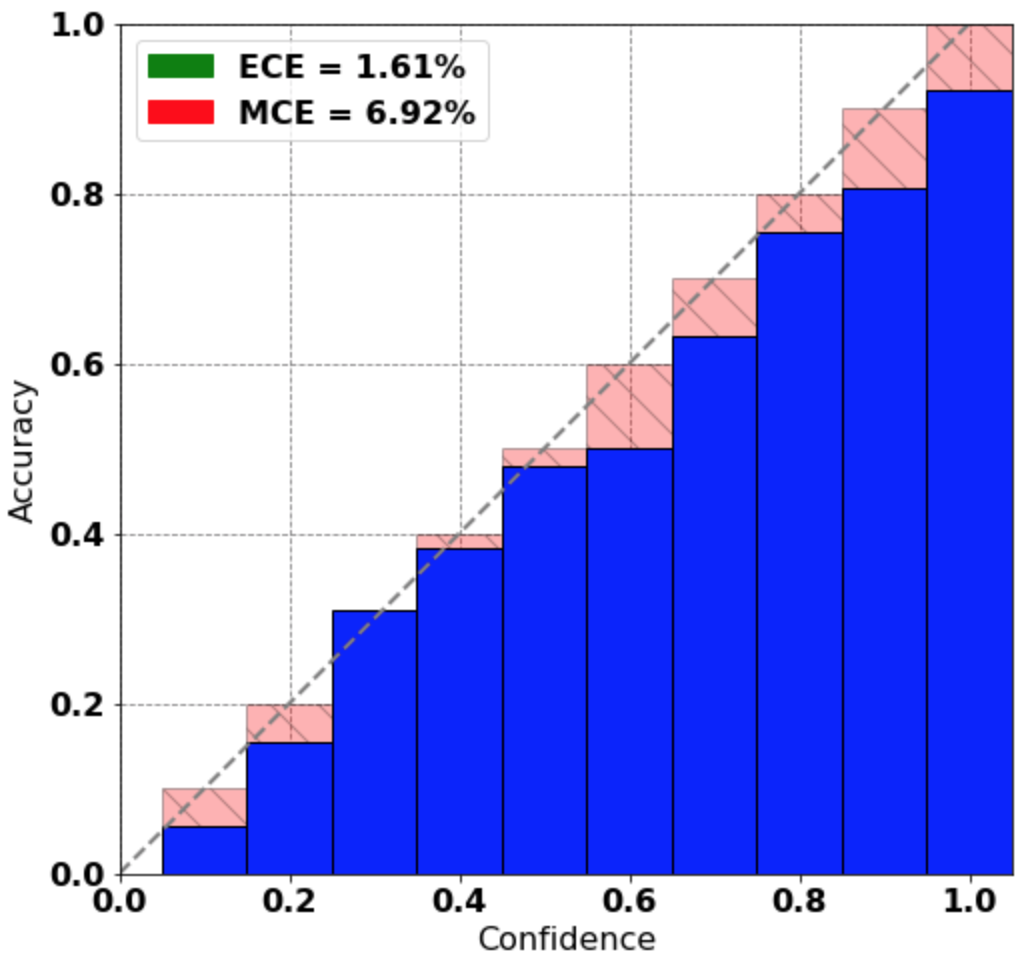}
    \caption{After calibration.}
    \label{cap:paper_jdiq2022-sec:results-subsec:calibration-generation-fig:calibration_after}
  \end{subfigure}
  \caption{Reliability diagrams for the \ebartfull model before and after calibration. Calibration is performed using Temperature Scaling~\cite{10.5555/3305381.3305518} to align predicted confidence with observed accuracy.}
  \label{cap:paper_jdiq2022-sec:results-subsec:calibration-generation-fig:calibration}
\end{figure}

A reliability diagram is produced for the model both before and after calibration to assess the quality of the confidence scores. Additionally, the Expected Calibration Error (ECE)\cite{10.5555/2888116.2888120} \index{Expected Calibration Error} and the Maximum Calibration Error (MCE)\cite{8627976, 10.5555/3305381.3305518} \index{Maximum Calibration Error} are computed using ten bins. Figure~\ref{cap:paper_jdiq2022-sec:results-subsec:calibration-generation-fig:calibration} presents the results for the \ebartfull model under both conditions. The dotted 45-degree line in the reliability diagrams represents perfect calibration; deviations from this line indicate discrepancies between the predicted confidence and actual accuracy.

The diagrams reveal that the original \ebart model is poorly calibrated, with an ECE of 11.44\% and an MCE of 26.35\%. After applying temperature scaling, the model’s calibration improves considerably. The ECE drops to 1.61\%, and the MCE decreases to 6.92\%. This improvement means the softmax scores can be more reliably interpreted as calibrated confidence values.

Table~\ref{cap:paper_jdiq2022-sec:results-subsec:calibration-generation-tab:confidence-1} and Table~\ref{cap:paper_jdiq2022-sec:results-subsec:calibration-generation-tab:confidence-2} show how calibration affects the model’s output. The first example shows a case where \ebart predicts the correct truthfulness label with high confidence. In contrast, the second example highlights a failure case: the ground truth label is refuted, but both the model and the ground truth explanation fail to clearly convey this. \ebart incorrectly predicts supported and, crucially, reports low confidence in this decision. Prior to calibration, the only clue to the model’s uncertainty would have been the limited quality of the explanation. With calibrated confidence scores, there is now a quantitative indicator that the model's output should be treated with caution.

\begin{table}[tpb]
    \centering
    \caption{Statement where the ground truth label is \supports and the one predicted by \ebart is \supports. Calibrated confidence score: $0.851$; original (uncalibrated) confidence: $0.987$ (\ebartfull).}
    \label{cap:paper_jdiq2022-sec:results-subsec:calibration-generation-tab:confidence-1}
    \begin{tabular}{p{3.1cm}p{5.2cm}p{2.9cm}p{3.8cm}}
    \toprule
     \textbf{Statement} & \textbf{Evidence} & \textbf{Explanation Gold} & \textbf{Explanation Generated} \\
    \midrule
    Ekta Kapoor worked on an Indian soap opera that premiered in 2000. & 
    Kyunki Saas Bhi Kabhi Bahu Thi (Because a mother-in-law was once a daughter-in-law, too) is an Indian soap opera that premiered on 3 July 2000 on Star Plus. Ekta Kapoor. Some soap operas she had worked on include [...]. Pavitra Rishta (Sacred Ties) is a 2009 Indian soap opera produced by Ekta Kapoor of Balaji Telefilms, that aired on Zee TV. Kasautii Zindagii Kay (The criterion of life), often abbreviated as KZK, is an Indian soap opera created by Ekta Kapoor's Balaji Telefilms for the channel STAR Plus. &
    Ekta Kapoor worked on the Indian soap opera Kyunki Saas Bhi Kabhi Bahu Thi, which premiered in 2000. &
    Ekta Kapoor worked on the soap opera Kyunki Saas Bhi Kabhi Bahu Thi, which premiered in 2000. \\
    \bottomrule
    \end{tabular}
\end{table}

\begin{table}[tpb]
    \centering
    \caption{Statement where the ground truth label is \refutes and the one predicted by \ebart is \supports. Calibrated confidence score: $0.432$; original (uncalibrated) confidence: $0.550$ (\ebartfull).}
    \label{cap:paper_jdiq2022-sec:results-subsec:calibration-generation-tab:confidence-2}
    \begin{tabular}{p{3cm}p{5.2cm}p{2.9cm}p{3.8cm}}
    \toprule
     \textbf{Statement} & \textbf{Evidence} & \textbf{Explanation Gold} & \textbf{Explanation Generated} \\
    \midrule
    Henry III assumed the throne when he was 2 years old. & 
    Henry III of France  Henry III (19 September 1551 – 2 August 1589; born Alexandre Édouard de France, Henryk Walezy, Henrikas Valua) was a monarch of the House of Valois who was elected the monarch of the Polish-Lithuanian Commonwealth from 1573 to 1575 and ruled as King of France from 1574 until his death. & 
    Henry III was elected the monarch of the Polish-Lithuanian Commonwealth when he was 2 years old. & 
    Henry III was born on 19 September 1551 and died on 2 August 1589. \\
    \bottomrule
    \end{tabular}
\end{table}

\subsection{Summary}

\label{cap:paper_jdiq2022-sec:discussion}

This chapter investigates the potential of Automated Fact-Checking (AFC) models capable of jointly producing a truthfulness prediction and a human-readable explanation. To this end, the \ebart architecture is proposed and evaluated on the \efever and \esnli datasets. The experimental results show that \ebart achieves performance comparable to state-of-the-art models, while also generating coherent and relevant explanations. The underlying hypothesis is that combining truthfulness prediction with explanation enhances the transparency of AFC systems and fosters greater trust in their outputs. The answers to the research questions are summarized as follows.

\myparagraph{\ref{cap:paper_jdiq2022-sec:research-questions_1}} The \ebart architecture is introduced by developing a \jointpredictionhead positioned atop the \bart model. This component manipulates the transformer's hidden states, enabling the model to classify the truthfulness of a statement and generate an explanation simultaneously.

\myparagraph{\ref{cap:paper_jdiq2022-sec:research-questions_2}} The evaluation of \ebart shows that the architecture is competitive with the state-of-the-art on the \efever and \esnli tasks. Moreover, the generation of explanations does not significantly hinder the model's ability to perform its primary task of truthfulness prediction. When generated jointly, truthfulness predictions and explanations exhibit greater internal coherence than when produced separately.

\myparagraph{\ref{cap:paper_jdiq2022-sec:research-questions_3}} The extensive human evaluation conducted through a crowdsourcing task demonstrates that the explanations generated by \ebart generally improve users' ability to identify misinformation and promote a more skeptical attitude toward claims. These explanations are also competitive with the ground truth explanations in terms of utility.

\myparagraph{\ref{cap:paper_jdiq2022-sec:research-questions_4}} The \ebart architecture is calibrated using the Temperature Scaling technique to produce reliable confidence scores for its truthfulness predictions. The results confirm that the model was not well calibrated prior to this post-processing step.

\myparagraph{}
The next chapter concludes this thesis. It begins by summarizing the contributions presented throughout the work. Then, for each meta-research question, it outlines the practical implications, discusses current limitations of the proposed approaches, suggests directions for future work, and presents the final conclusions.

\chapter{Discussion}

\label{cap:conclusions}

This chapter provides a comprehensive reflection on the findings and contributions of the thesis. Section~\ref{cap:conclusions-sec:contributions} summarizes the main outcomes of the experimental work. Section~\ref{cap:conclusions-sec:practical} discusses their practical implications, while Section~\ref{cap:conclusions-sec:limitations} outlines the limitations encountered. Section~\ref{cap:conclusions-sec:future-work} proposes directions for future research. Each of these sections addresses the meta-research questions introduced in Section~\ref{cap:research-questions}. Section~\ref{cap:conclusions-sec:conclusions} provides the overall conclusions of the thesis, and Section~\ref{cap:conclusion-sec:acks} closes with acknowledgements.

\section{Contributions}

\label{cap:conclusions-sec:contributions}

The main contributions developed throughout the thesis are grouped according to the meta-research questions. 

Section~\ref{cap:conclusions-sec:contributions-subsec:mrq-1} presents the results related to the assessment of online (mis)information. Section~\ref{cap:conclusions-sec:contributions-subsec:mrq-2} discusses the characterization of cognitive biases and their impact on fact-checking. Finally, Section~\ref{cap:conclusions-sec:contributions-subsec:mrq-3} outlines the contributions concerning the prediction of truthfulness and the generation of explanations.

\subsection{\mrqone}

\label{cap:conclusions-sec:contributions-subsec:mrq-1}

This thesis improves the understanding of how non-expert crowd workers can contribute to the assessment of online (mis)information, with a focus on truthfulness assessment and longitudinal dynamics in crowdsourcing settings.

The findings demonstrate that the choice of judgment scale used to collect truthfulness labels does not significantly affect their quality. When workers' responses are appropriately aggregated and the levels of truthfulness are merged, the resulting crowdsourced data correlates well with expert fact-checker assessments, despite generally low inter-worker agreement. The background of crowd workers influences the judgments they provide (\ref{cap:paper_sigir2020-sec:research-questions_1}--\ref{cap:paper_sigir2020-sec:research-questions_4}).

Crowd workers can effectively detect and classify recent online (mis)information, such as content related to the \covid pandemic. Both crowdsourced and expert assessments can be transformed and aggregated to improve overall quality. The accuracy of judgments is influenced by workers’ backgrounds, the sources of information they rely on, and their behavioral patterns. The longitudinal study confirms that the passage of time has a significant impact on judgment quality, affecting both novice and experienced annotators. A detailed failure analysis of the statements misjudged by crowd workers is provided (\ref{cap:paper_pauc2021-sec:research-questions_1}--\ref{cap:paper_pauc2021-sec:research-questions_8}).

This thesis also presents the first large-scale survey aimed at understanding the barriers to conducting longitudinal studies on crowdsourcing platforms. The study spans three major commercial platforms and combines both quantitative and qualitative analyses. Based on this investigation, a set of \numrecommendations recommendations is proposed for researchers and practitioners, along with \numpractices best practices for commercial crowdsourcing platforms, to better support longitudinal research (\ref{cap:paper_tsc2024-sec:research-questions_1}--\ref{cap:paper_tsc2024-sec:research-questions_3}).

Truthfulness judgments provided by crowd workers across seven dimensions are shown to be sound and reliable. Agreement with expert judgments is high when the same dimension is considered, and remains reasonable across different dimensions, with differences reflecting the specific semantics of each category. Analyses confirm that the seven dimensions are independent, non-redundant, and capture complementary facets of truthfulness. Additionally, informativeness analyses show that these dimensions help interpret the rationale behind workers’ judgments. Signals derived from the workers, in particular their labels and search behaviors, are effective in predicting expert verdicts (\ref{cap:paper_ipm2021-sec:research-questions_1}--\ref{cap:paper_ipm2021-sec:research-questions_5}).

\subsection{\mrqtwo}

\label{cap:conclusions-sec:contributions-subsec:mrq-2}

This thesis contributes to a deeper understanding of how cognitive biases influence fact-checking activities, particularly in crowdsourced settings.

A systematic review of cognitive biases in the misinformation literature is conducted to identify those most relevant to fact-checking. Using the \index{PRISMA}PRISMA methodology, a curated subset of \numbiasused out of \numbias cognitive biases is selected and described. A classification scheme is introduced, together with a set of countermeasures aimed at mitigating their effects. Based on these findings, a bias-aware fact-checking pipeline is proposed, with each countermeasure conceptually mapped to a specific component of the process (\ref{cap:paper_ipm2023_bias-sec:research-questions_1}--\ref{cap:paper_ipm2023_bias-sec:research-questions_4}).

The effort to uncover evidence of bias led to informative results. Experimental findings indicate that workers’ trust in politics and cognitive reflection abilities do not significantly influence truthfulness judgments. However, contrary to initial expectations, a stronger belief in science is associated with greater accuracy in assessments (\ref{cap:paper_facct2022-sec:research-questions_1}--\ref{cap:paper_facct2022-sec:research-questions_3}).

\subsection{\mrqthree}

\label{cap:conclusions-sec:contributions-subsec:mrq-3}

This thesis advances automated fact-checking by proposing and evaluating a model that improves both prediction quality and explanation generation. The \ebart architecture is introduced to jointly produce truthfulness classifications and human-readable explanations (\ref{cap:paper_jdiq2022-sec:research-questions_1}). Experimental results show that \ebart performs comparably to state-of-the-art systems while delivering more coherent and informative explanations.

Human evaluation confirms that these explanations enhance users’ ability to detect misinformation and increase their skepticism toward online claims. In addition, \ebart is calibrated using temperature scaling to provide more reliable confidence estimates for its predictions (\ref{cap:paper_jdiq2022-sec:research-questions_2}--\ref{cap:paper_jdiq2022-sec:research-questions_4}).

\section{Practical Implications}

\label{cap:conclusions-sec:practical}

The findings presented in this thesis have several practical implications for both researchers and practitioners working on misinformation assessment and automated fact-checking. 

Section~\ref{cap:conclusions-sec:practical-subsec:mrq-1} highlights the insights related to the assessment of online (mis)information. Section~\ref{cap:conclusions-sec:practical-subsec:mrq-2} discusses how the characterization of cognitive biases can inform fact-checking processes. Section~\ref{cap:conclusions-sec:practical-subsec:mrq-3} outlines how the proposed model for joint prediction and explanation can be applied in practice. 

To show how these insights transfer to a new domain, Section~\ref{cap:paper_is2022} presents a concrete application to product-review quality assessment. This case study shows that the multidimensional truthfulness developed for fact-checking also improve the detection of high-quality opinions in a review quality judgment setting.

\subsection{\mrqone}

\label{cap:conclusions-sec:practical-subsec:mrq-1}

The main practical implications of experiments aimed at understanding the ability of human assessors to address misinformation are summarised below. These include, for instance, the study of the effects of judgment scales and worker background described in Chapter~\ref{cap:paper_sigir2020}, as well as the multidimensional notion of truthfulness proposed in Chapter~\ref{cap:paper_ipm2021}.

A series of experiments examined how accurately crowd workers can detect and categorise online (mis)information. Results show that non-expert assessors are able to judge political statements, as well as more recent items related to the \covid pandemic, and they can apply a multidimensional truthfulness scale with reasonable reliability. However, agreement among workers remains modest, even when multiple dimensions are considered. For this reason, researchers should prioritise experienced contributors when high overlap with expert labels is required.

Background questionnaires offer little value as a proxy for worker quality. Neither political affiliation nor self-reported confidence improves the accuracy of aggregated labels. Relying on such attributes may introduce unnecessary selection bias. Instead, judgement quality is influenced by statement features and the availability of supporting evidence. Statements that lack verifiable information are harder to rate, and assessors sometimes focus on a single fragment of text or source. Requesting a short textual justification and providing a custom search interface encourage workers to consult multiple sources and document their reasoning.

Label distributions tend to be skewed toward positive Likert values, such as \agree and \completelyagree. Categories at the negative end of the judgment scale like \politifactpantsfire and \politifactfalse are often confused, suggesting that fine-grained distinctions in this region may not be meaningful for non-experts. Aggregating individual labels using the arithmetic mean improves agreement with expert judgements, whereas naïve combinations of the seven dimensions do not. The dimensions themselves are statistically independent and capture information that differs from automatically computed measures. They can therefore be reused in future crowdsourcing tasks without strong collinearity concerns.

Temporal effects also play a role. Data collected in batches that are closer in time tend to be more consistent, and returning workers usually outperform novices. Presenting the same item in different positions helps mitigate order effects, and workers become faster as they progress through a task, reflecting a learning curve. Finally, quality-control checks embedded in the task design remain essential for obtaining useful labels.

\subsection{\mrqtwo}
\label{cap:conclusions-sec:practical-subsec:mrq-2}

The characterisation of cognitive biases presented in Chapter~\ref{cap:paper_ipm2023_bias}, together with the proposed bias-aware assessment pipeline for fact-checking and the investigation described in Chapter~\ref{cap:paper_facct2022}, leads to a number of practical lessons for crowdsourced truthfulness assessment and related tasks.

First, the taxonomy of biases clarifies which cognitive errors are likely to manifest when workers judge misinformation. This knowledge can inform professional fact-checking workflows \cite{ceci2020psychology} and guide artificial intelligence practitioners in building models that remain robust when training data are biased. Although crowd workers are generally reliable, personal attributes such as \beliefscience or biases such as the \mycitebias{Affect Heuristic}{\ref{cap:paper_ipm2023_bias-bias:affect-heuristic}} and the \mycitebias{Overconfidence Effect}{\ref{cap:paper_ipm2023_bias-bias:overconfidence-effect}} may reduce accuracy. Researchers and practitioners are therefore encouraged to assess, document, and where possible mitigate these biases, for example by following checklists and task-design guidelines \cite{draws2021ChecklistCombatCognitive, eickhoff2018cognitive, hube2019understanding}. Table~\ref{cap:paper_ipm2023_bias-sec:results-sect:ideal-pipeline-tab:ideal-pipeline} summarises concrete countermeasures.

Several design choices have proved effective. Limiting the information shown to workers, for instance by means of a custom search engine (\mycitecountermeasure{Custom Search Engine}{\ref{cap:paper_ipm2023_bias-count:custom-search-engine}}), helps them focus on evidence that is truly relevant. Allowing open discussion among assessors (\mycitecountermeasure{Discussion}{\ref{cap:paper_ipm2023_bias-count:discuss}}) can surface extreme individual views that might otherwise go unnoticed, while inviting workers to revise their initial labels (\mycitecountermeasure{Revision}{\ref{cap:paper_ipm2023_bias-count:revise-scores}}) reduces the \mycitebiasnoindex{Anchoring Effect}{\ref{cap:paper_ipm2023_bias-bias:anchoring-bias}}{Anchoring Effect}. Clear, well-tested instructions (\mycitecountermeasure{Instructions}{\ref{cap:paper_ipm2023_bias-count:instructions}}) are essential to avoid unwanted priming.

Whenever feasible, requesters should hide cues such as speaker identity or political affiliation, as these tend to trigger the Affect Heuristic. Measuring constructs like \beliefscience \cite{dagnall2019evaluation} provides a basis for detecting systematic bias, and workers with moderate levels of political affiliation, belief in science, and self-confidence often produce higher-quality judgments. Conversely, unnecessary instruments add cognitive load: the Cognitive Reflection Test, for example, shows no link to truthfulness-judgment quality and can distract workers from the main task. Finally, labels produced by highly self-confident workers should be adjusted with care, since such workers tend to exhibit greater bias on average.

\subsection{\mrqthree}
\label{cap:conclusions-sec:practical-subsec:mrq-3}

Experiments in Chapter~\ref{cap:paper_jdiq2022} demonstrate that truthfulness labels collected from crowd workers can be used to predict expert judgements in a supervised setting. When only structured features are available, a Random Forest classifier achieves the highest effectiveness among the conventional algorithms evaluated, outperforming all baselines.

The \ebart architecture offers additional advantages in practical deployments. By generating a truthfulness label and a human-readable explanation in a single forward pass, \ebart eliminates the need for separate explainability modules and enhances transparency for end users. Fine-tuning the pre-trained weights together with the \jointpredictionhead leads to competitive accuracy while preserving fluency and relevance in the generated rationales. Human evaluation shows that these explanations help reviewers verify predictions more efficiently and foster greater trust in automated outputs.

Proper calibration of model confidence scores, detailed in Section~\ref{cap:paper_jdiq2022-sec:results-subsec:calibration-generation}, further improves reliability. Temperature scaling reduces the expected calibration error by more than 50\%, producing probability estimates that align closely with observed accuracy. Practitioners can set risk-adjusted decision thresholds, route low-confidence cases to human moderators, and maintain consistent precision targets across deployments.

\subsection{Multidimensional Reviews Quality Judgment}

\label{cap:paper_is2022}

This section summarises a study published in the \lq\lq Information Systems\rq\rq{} journal~\cite{CEOLIN2022102107}, which extends earlier work presented at the 20th International Conference on Web Engineering~\cite{10.1007/978-3-030-74296-6_6}. Background on argument mining relevant to this study appears in Section~\ref{cap:related_work-sec:argument-mining}. The work demonstrates how the multidimensional truthfulness framework introduced in Chapter~\ref{cap:paper_ipm2021} can be adapted to assess the quality of online product reviews.

Section~\ref{cap:paper_is2022-sec:motivation-background} introduces the motivation and background for adapting multidimensional truthfulness assessment to the product reviews setting. Section~\ref{cap:paper_is2022-subsec:framework} presents the argumentation-based framework used to model and evaluate the quality of reviews. Section~\ref{cap:paper_is2022-subsec:crowdsourcing-task} describes the construction of the dataset and the design of the crowdsourcing task. Section~\ref{cap:paper_is2022-subsec:evaluation} discusses the results obtained and summarises the main findings.

\subsubsection{Motivation And Background}
\label{cap:paper_is2022-sec:motivation-background}

Online reviews can be a valuable source of information, as they allow users to benefit from the experiences of others who have expressed their opinions about a product to purchase, a room to book, and so on. These opinions are useful only if high-quality reviews can be identified and those of low quality can be dismissed (e.g., due to bias, incompleteness, irrelevance). Over the past years, research has characterised the trustworthiness of reviews in several ways, including user reputation and quality assessment. However, while reviews focus on specific products or services, they often express multifaceted views about the target. To assess the quality and trustworthiness of a review, it is important to understand the arguments it provides, their strengths, and which aspects of the product are presented as positive or negative evidence. In other words, reviews serve as a way for users to express their opinions on a given product or service and can be represented as rating–description pairs. This structure includes a rating (often on a $1$–$5$ \index{Likert}Likert scale) for the product’s quality, enriched with a textual description that motivates the score. The textual description can provide one or more arguments in support of the given \index{Likert}Likert scale rating.

\subsubsection{Argumentation-Based Framework}
\label{cap:paper_is2022-subsec:framework}

Research on the assessment of the quality and credibility of product reviews has mostly focused on linguistic aspects, such as readability and linguistic errors \cite{ghose, korfiatis, Ocampo, wu}. \citet{wathen}, on the other hand, proposes an approach that examines credibility factors. Lastly, \citet{wyner} explores a connection between natural language processing and argumentation reasoning. While their method classifies various tokens more thoroughly as different kinds of arguments, it does so in a semi-supervised fashion. This suggests that argumentation reasoning could be employed to analyse these textual descriptions.

\citet{lawrence-reed-2019-argument} and \citet{10.1145/2850417} provide extensive surveys of argument mining methods. Exploring the applicability and adaptability of these methods is worthwhile, although the setting considered here presents a difference: the arguments to be mined are extracted from short review descriptions accompanied by a numerical rating. This feature distinguishes the task from most cases covered in existing surveys.

\citet{wagemans} propose a classification of argument types, known as \index{Periodic Table Of Arguments} the \lq\lq Periodic Table of Arguments\rq\rq{},\footnote{\url{https://periodic-table-of-arguments.org/}} which may be operationalised in future work. Identifying and reasoning about argument types is important for understanding the strength and relevance of arguments in debates. \citet{ijcai2021-600} survey methods for determining the strength of argumentation-based approaches, with a focus on their use in explainability. However, these methods typically deal with semi-structured data and may require adaptation for short, structured items like reviews, which combine numerical ratings with textual explanations. This structure is also relevant in fact-checking, where truthfulness can be interpreted as a numerical score.

\citet{stab-etal-2018-cross} investigate argument mining across heterogeneous sources. \citet{michel:hal-02939363} integrate natural language processing and knowledge graphs to support argument reasoning. Both contributions offer valuable techniques that can enhance the argument mining process. Similarly, \citet{DBLP:conf/atal/CocarascuRT19} propose a relevant approach using argumentation reasoning to aggregate and explain movie reviews—an idea that could be adapted for product reviews. Along the same lines, \citet{BRIGUEZ20146467} apply argumentation reasoning in the context of movie recommendation. Finally, \citet{Moser2020UseOC} introduce a method for visualising argumentation graphs to support explanation and exploration. While developed for biomedical applications, their graphical model could be adapted for simpler argumentation settings such as product reviews.

Formal argumentation implements argumentation theory, the interdisciplinary study of how conclusions can be drawn from premises through logical reasoning. In this formal setting, arguments are the atomic units of analysis, and the theory aims to analyse the complex graph formed by the relationships between them, determining whether an argument attacks or supports another, and which arguments ultimately prevail. 

In the proposed setting, review descriptions are analysed to identify arguments that support the corresponding scores. A value-based semantics extending the model of \citet{Caminada} is adopted to describe the conflict and support dynamics among tokens within a set of product reviews. The approach falls within the growing family of weighted argumentation frameworks that extend Dung's standard model. It introduces an ordering of weighted attacks, differing in some aspects from previously proposed frameworks. These dynamics are formalised within a graph structure, where nodes represent arguments (i.e., reviews), and edges represent attack relations between them. Reviews are included in the same graph if they refer to at least one common feature of the product under evaluation. Edges encode the attack relationship between reviews that assign different scores to the shared feature. The direction of the attack is determined by the relevance of the tokens and the score values of the reviews. The semantics of the resulting graph is defined by a standard labelling function from formal argumentation theory:
\begin{itemize}
  \item[--] A review is labelled \texttt{in} \index{In} if all its attackers are labelled \texttt{out}.
  \item[--] A review is labelled \texttt{out} \index{Out} if at least one of its attackers is labelled \texttt{in}.
  \item[--] A review is labelled \texttt{undec} \index{Undec} if not all its attackers are \texttt{out} and none is \texttt{in}.
\end{itemize}

\begin{figure}[tpb]
    \centering
    \includegraphics[width=0.6\linewidth]{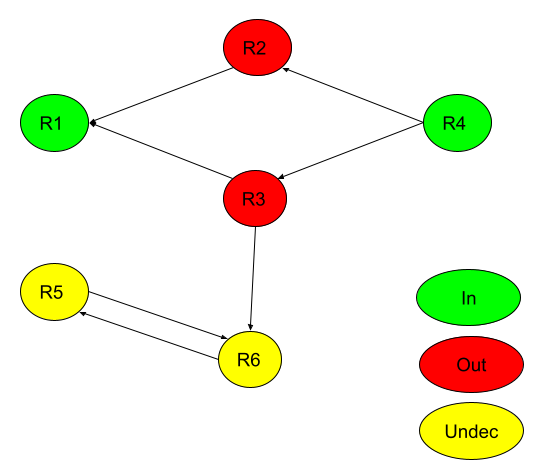}
    \caption{Example of review labelling based on the adopted argumentation-theoretic semantics.}
    \label{cap:paper_is2022-fig:argumentation_scheme}
\end{figure}

Figure~\ref{cap:paper_is2022-fig:argumentation_scheme} shows this semantics. R1 \index{R!1} and R4 \index{R!4} are labelled \texttt{in} because either all their attackers are \texttt{out} (R1) or they have no attackers (R4). R2 \index{R!2} and R3 \index{R!3} are labelled \texttt{out} because their attacker is \texttt{in}. R5 \index{R!5} and R6 \index{R!6} are labelled \texttt{undec} because their attackers are either \texttt{undec} or \texttt{out}. 

The semantics relies on a scoring system for tokens extracted from reviews via natural language processing, enabling their translation into a graph construction algorithm. A full technical description of the pipeline and formal semantics is beyond the scope of this section. However, the implementation is publicly available to the research community.\footnote{\url{https://github.com/davideceolin/FAReviews}}

\subsubsection{Crowdsourcing Task}
\label{cap:paper_is2022-subsec:crowdsourcing-task}

The model is evaluated on the Amazon Review Dataset~\cite{ni}, specifically on the Amazon Fashion 5-core subset, which contains \num{3,176} unique reviews by 406 users covering 31 products. One goal of the analysis using crowdsourcing is to obtain a stratified sample with an equal number of reviews for each star rating. However, the 5-core subset does not provide a sufficient number of reviews to support this goal. The 5-core subset is a filtered version of the full Amazon Review Dataset, where each user and item must have at least five reviews. However, the publicly available 5-core files include only a subset of user/item pairs that satisfy this condition. To enable a more complete evaluation, the full 5-core dataset is reconstructed.

The complete Amazon Review Dataset comprises \num{883,636} reviews (\num{871,502} after duplicate removal) written by \num{746,352} users about \num{185,241} products. To rebuild a balanced 5-core dataset, five reviews are sampled for each product ASIN\index{ASIN} code, each written by a different user. Similarly, each user is associated with five reviews written for five different products. This ensures that all user/product pairs are unique and that both users and products have exactly five reviews, thus adhering to the 5-core assumption.

The resulting reconstructed 5-core dataset contains 148,588 reviews across 29,958 products, which corresponds to 16.81\% of the original Amazon Review Dataset.\index{Amazon!Review Dataset} The sample used in the earlier version of this study~\cite{10.1007/978-3-030-74296-6_6} is extended by sampling reviews from the full 5-core dataset to balance the number of reviews across star ratings and upvote counts. The final stratified and augmented sample comprises 670 reviews, representing 0.45\% of the original dataset.

A crowdsourcing task is deployed to collect 670 reviews by randomly selecting a product and then drawing one of its reviews, continuing this process until a balanced number of reviews is obtained for each review score. This results in 134 reviews for each score value. Each review is evaluated by five workers, and each worker assesses ten reviews. Workers are located in the United States, and the tasks are compensated with \$0.90 each. The task is conducted through the \mturk platform.

Each worker is presented with the corresponding product description as provided in the Amazon dataset, followed by a review. The worker is then asked to evaluate the review using a 5-point \index{Likert}Likert scale (from \completelydisagree to \completelyagree) across seven quality dimensions (Appendix~\ref{cap:paper_is2022-appendix:mult-scale-rev}), based on prior work on multidimensional quality assessment described in Chapter~\ref{cap:paper_ipm2021}. Due to the subjective nature of reviews, determining the truthfulness of the information is particularly challenging, as it involves judging the authenticity of the review itself. For this reason, the quality dimensions are slightly adapted to emphasise more subjective aspects such as reliability.

\subsubsection{Evaluation And Findings}
\label{cap:paper_is2022-subsec:evaluation}

The collected data indicate that the classification produced by the argumentation framework is correlated with review quality, particularly with respect to \comprehensibility and \completeness. Among the readability measures tested, ARI~\cite{Smith1967-js} \index{ARI} shows the strongest correlation with argumentation-based judgments. However, readability scores alone are not sufficient to identify reviews with higher \overalltruthfulness, as all readability metrics exhibit weak correlation with \comprehensibility. These findings support the effectiveness of the argumentation-based approach and underscore the importance of applying logical reasoning over ranked arguments to produce labels that align with \overalltruthfulness and \comprehensibility assessments.

To summarise, the crowdsourcing task using the multidimensional scale originally designed for truthfulness assessment confirms the potential of argumentation graphs to provide useful explanations for product reviews. The argumentation-based framework represents an initial step towards a reliable and transparent method for evaluating the quality of online opinions.

Nonetheless, several directions for future work remain open. Currently, all tokens in a review are na\"{i}vely treated as potential indicators of arguments, with their importance weighted accordingly. This highlights the need for refinement in the argument mining and classification steps. The approaches outlined in the following sections may offer valuable insights for improving the analysis of product reviews and for supporting fact-checking tasks involving statements made by public figures, even when such methods are not explicitly designed for this type of data.

\section{Limitations}

\label{cap:conclusions-sec:limitations}

Having outlined the practical implications of the thesis, it is important to acknowledge the limitations that qualify them. Section~\ref{cap:conclusions-sec:limitations-subsec:mrq-1} discusses limitations related to the assessment of online (mis)information. Section~\ref{cap:conclusions-sec:limitations-subsec:mrq-2} outlines issues concerning the characterization of the proposed cognitive biases and their impact on the fact-checking process. No substantial limitations were observed in the model used to predict the truthfulness of information items and to generate explanations, owing to its consistent performance across evaluation settings.

\subsection{\mrqone}
\label{cap:conclusions-sec:limitations-subsec:mrq-1}

An important limitation concerns the structure of the ground truth labels provided by \politifact. According to publicly available information on the \politifact\ assessment process \cite{doi:10.1177/2053168018786848, politifactprinciples, politifactbadge} (Section~\ref{cap:intro-sec:fact-checking}), only the final label assigned by experts is available. Each statement is judged by three editors and a reporter, who reach a consensus for the final judgment. Access to the unaggregated individual judgments would enable, for example, a more detailed analysis of disagreement between assessors and the aggregated ground truth. However, these individual judgments are not publicly released for the \politifact\ dataset.

Some of the studies conducted (e.g., those presented in Chapter~\ref{cap:paper_pauc2021}, Chapter~\ref{cap:paper_ipm2021}, and Chapter~\ref{cap:paper_facct2022}) rely exclusively on statements sampled from the \politifact dataset. To generalize the findings, comparisons with multiple datasets are necessary. Furthermore, having each statement evaluated by only 5 or 10 workers does not guarantee strong statistical power; a larger sample of annotators may be required to draw more robust conclusions and reduce the standard deviation of the results. The longitudinal study on the effect of information recency also showed a relatively low number of returning participants. Using a different platform, such as \prolific\index{Prolific}, may improve worker retention.

No ground truth exists for each of the seven dimensions of truthfulness (Section~\ref{cap:paper_ipm2021-sec:exp-setup-subsec:seven-dim}). If such information were available, a more direct comparison between expert and worker judgments could be performed. However, this would present several challenges. The subjectivity of information quality dimensions varies; for instance, \comprehensibility\ is generally more subjective than \precision. In addition, comparisons with an expert-provided ground truth could be misleading, since observed differences may reflect individual interpretations of the dimensions rather than systematic disagreement.

As noted in Section~\ref{cap:paper_ipm2021-sec:exp-setup-subsec:crowdsourcing-task}, workers provide responses on a five-level \index{Likert}Likert scale, ranging from \completelydisagree\ to \completelyagree. The scale is ordinal, but the categories are not strictly equidistant, as each label has a distinct semantic meaning. Consequently, the perceived distance between successive categories can vary across workers. For instance, the gap between \completelydisagree and \neitheraord\ may not be twice that between \completelydisagree and \disagree. Under these conditions, using the mean as an aggregation function may be inappropriate, since it assumes equal spacing between categories. This limitation can be mitigated only if adjacent categories are perceived as approximately equidistant, which is plausible due to the inclusion of numeric labels. Alternative aggregation functions, such as the median or majority vote, also have limitations, as they discard potentially informative variation in the responses. A more detailed discussion of mean-based aggregation in crowdsourced truthfulness assessment appears in Section~\ref{cap:paper_sigir2020-sec:exp-setup-subsec:scales}.

A further limitation of the work described in Chapter~\ref{cap:paper_ipm2021} concerns the combination of multiple dimensions to improve agreement between worker and expert labels. As discussed in Section~\ref{cap:paper_ipm-sec:results-subsec:truthfulness-dimensions}, attempts were made to combine individual dimensions in ways that could enhance agreement between crowd and expert judgments. However, additional analyses did not reveal any improvement in internal agreement among assessors or in external agreement between the crowd and expert labels.

Another limitation concerns the possible presence of malicious or inattentive workers. Quality control mechanisms were implemented in every task to ensure data reliability as detailed, for instance, in Section~\ref{cap:paper_ipm2021-sec:exp-setup-subsec:crowdsourcing-task}. Workers who failed the checks were prevented from submitting their responses, whereas those who passed were assumed to have completed the task in good faith. This assumption is supported by post-hoc analyses of response distributions, time on task, and behavioural patterns, none of which revealed suspicious or outlier behaviour. It is therefore reasonable to regard the submitted data as high quality. A further limitation may lie in the number and sampling of statements used in the experiments; however, manual inspection detected no noticeable bias in language level, terminology, length, or other observable features. 

In addition, task-abandonment statistics show that the abandonment rate decreases steadily as workers progress through a task (Section~\ref{cap:paper_ipm2021-sec:exp-setup-subsec:task-abandonment}). Most workers leave immediately after reading the instructions, followed by those who complete only a single judgment. This pattern suggests that workers prefer to quit if the task is not appealing rather than attempt it maliciously. Similar trends were observed in other studies (Section~\ref{cap:paper_sigir2020-sec:desc-stat-subsec:task-abandonment}, Section~\ref{cap:paper_pauc2021-sec:exp-setup-subsec:task-abandonment}, and Section~\ref{cap:paper_ipm2021-sec:exp-setup-subsec:task-abandonment}). Overall, no anomalous behaviour was detected, although worker behaviour should continue to be monitored in future work.

Considering the study presented in Chapter~\ref{cap:paper_tsc2024}, there are additional limitations to acknowledge, the first concerning the recruitment of workers with previous experience in longitudinal studies. Depending on the platform, participants were either selected based on recorded experience (\mturk and \prolific) or asked about their experience during recruitment (\toloka), as detailed in Section~\ref{cap:paper_tsc2024-sec:exp-setup-subsec:survey-design}. However, recruiting experienced workers alone does not guarantee a full understanding of worker behaviour in longitudinal settings. The sample gathered across the three platforms was relatively small: although several survey items produced statistically significant results, it may not reflect the heterogeneity of the wider worker population. Moreover, the absence of behavioural logs limits the ability to verify whether applying the survey findings would lead to the intended outcomes in practice.

The survey design had two further weaknesses. First, some questions would have been more informative if a Likert scale had been used instead of binary responses (see Section~\ref{cap:paper_tsc2024-sec:results-subsec:rq1-analysis-p1-question:previous-experiences}). Second, certain items employed single-choice radio buttons where multiple-choice answers would have been more appropriate, which may have introduced response bias (see Section~\ref{cap:paper_tsc2024-sec:results-subsec:rq1-analysis-p1-question:participation-incentives-previous}).

Finally, demographic and background factors can affect participation in longitudinal studies. Younger workers, for example, may have more available time and may respond differently from older workers. While \prolific and \toloka provide demographic information, \mturk does not. Future work could address this limitation by including demographic questions in the survey or by applying platform-specific screening criteria. Recruiting participants from diverse age groups is feasible on all three platforms, although other characteristics may be more difficult to control.

\subsection{\mrqtwo}
\label{cap:conclusions-sec:limitations-subsec:mrq-2}

Beyond those related to truthfulness assessment, additional limitations emerge from the study of cognitive biases that may affect human assessors during fact-checking, as presented in Chapter~\ref{cap:paper_ipm2023_bias}.

First, subjectivity is inherent in both the identification and classification of the \numbias cognitive biases. These processes rely on the authors’ interpretation of each bias, and subjectivity also influences the fact-checking scenarios devised to illustrate them. For \numbiaswithscenario of the \numbiasused biases, the scenarios were grounded in existing literature, whereas no fact-checking-related references were found for the remaining \numbiaswithnoscenario; the corresponding examples therefore reflect the authors’ judgement. Because these steps involve subjective decisions, other researchers might identify or classify the biases differently. 

Although the \prisma-guided methodology described in Section~\ref{cap:paper_ipm2023_bias-sec:methodology} aims to produce a comprehensive list, some biases may have been overlooked given the breadth of the literature. The resulting list is therefore unlikely to be exhaustive, and the findings may not generalise to all fact-checking contexts or populations. Assessing the effectiveness and general applicability of the countermeasures described in Section~\ref{cap:paper_ipm2023_bias-sec:results-subsec:countermeasures} is also challenging, as their impact may vary depending on context, individual differences, and the nature of the misinformation.

Moreover, the review treats each cognitive bias as an independent factor, although biases may interact in complex ways that amplify or attenuate their effects. Future research should examine these interactions experimentally, assess their cumulative influence on assessor performance, and test the proposed countermeasures in practical settings. Finally, the catalogue of biases in Section~\ref{cap:paper_ipm2023_bias-sec:results-subsec:list} reflects the current state of knowledge. As cognitive psychology evolves, new biases may emerge or existing ones may be refined. Periodic updates will be necessary to ensure that the analysis remains aligned with the most current evidence.

\section{Future Directions}
\label{cap:conclusions-sec:future-work}

Building on the limitations outlined above, this thesis advances the goal of mitigating misinformation, potentially in real time, and identifies several directions for future investigation. In particular, a hybrid strategy that combines automatic machine-learning classifiers, crowdsourced judgments, and a limited pool of expert assessors is expected to lead to more effective solutions.

Section~\ref{cap:conclusions-sec:future-work-subsec:mrq-1} outlines research lines that could improve online (mis)information assessment. Section~\ref{cap:conclusions-sec:future-work-subsec:mrq-2} proposes experiments on the impact of cognitive biases. Section~\ref{cap:conclusions-sec:future-work-subsec:mrq-3} discusses future work on predicting truthfulness and generating explanations. Finally, Section~\ref{cap:paper_tois2023} introduces ongoing research that is expected to converge with the topics addressed in this thesis.

\subsection{\mrqone}
\label{cap:conclusions-sec:future-work-subsec:mrq-1}

The crowdsourcing-based approaches described in this thesis could be integrated with machine learning to support fact-checking experts in a human-in-the-loop process~\cite{demartini2020human}. This integration could involve extending existing information-access tools such as \index{FactCatch} \texttt{FactCatch}~\cite{nguyen2020factcatch} or \index{Watch 'n' Check} \texttt{Watch 'n' Check}~\cite{cerone2020wnc}. In the long term, a rating or flagging mechanism could be implemented within social media platforms to enable users to express judgments about the truthfulness of statements. This is a complex task that will require addressing ethical concerns, including potential misuse by polarised groups, the treatment of under-represented minorities, and manipulative behaviour resulting from numerical imbalances.

Additionally, a detailed investigation into the perceived distances between categories on truthfulness scales could inform more advanced methods for aggregating and interpreting crowdsourced judgments. The resulting resources could also help clarify how crowd agreement can assist experts in evaluating the check-worthiness of statements.

The studies described in Figure~\ref{cap:paper_sigir2020-sec:desc-stat-subsec:score-distribution-fig:scores_distributions_s3}, Figure~\ref{cap:paper_sigir2020-sec:desc-stat-subsec:score-distribution-fig:scores_distributions_s6}, and Figure~\ref{cap:paper_sigir2020-sec:desc-stat-subsec:score-distribution-fig:scores_distributions_s100} explore the use of different pointwise scales for collecting truthfulness judgments (Section~\ref{cap:paper_sigir2020-sec:exp-setup-subsec:scales}). Another promising approach worth investigating is magnitude estimation\cite{10.1111/j.1745-4557.1977.tb00942.x}, a psychophysical scaling technique in which observers assign numbers to stimuli based on perceived intensity. This method has seen applications in several domains, including Information Retrieval\cite{10.1145/3002172}.

The longitudinal study described in Chapter~\ref{cap:paper_pauc2021} should be replicated using statements verified by other fact-checking organizations, such as those indexed by the \texttt{Google Fact Check Explorer}\footnote{\url{https://toolbox.google.com/factcheck/explorer}}, to support the generalisability of the findings. The study should also be reproduced on other crowdsourcing platforms. For instance, \citet{Qarout2019} show that experiments replicated across different platforms can yield significantly different levels of data quality. Another avenue for future work is to perform a longitudinal study using the task design described in Section~\ref{cap:paper_ipm2021-sec:exp-setup-subsec:crowdsourcing-task}, and to compare its outcomes with those of the existing dataset.

In addition, one-on-one interviews with both crowd workers and requesters should be conducted to gain deeper insight into the benefits and challenges of participating in longitudinal studies. Intervention-based studies could also be employed to test features and experimental configurations aimed at improving worker retention and increasing satisfaction for both participants and requesters. These efforts would contribute to a more robust and sustainable framework, enabling both parties to engage effectively in longitudinal research.

The seven dimensions of truthfulness introduced in Chapter~\ref{cap:paper_ipm2021} generally show no correlation, with the only exception of \overalltruthfulness and \correctness. This suggests that the full set of dimensions can be reused to collect crowdsourced truthfulness labels. However, it also indicates that the current set may not be optimal. Future work should investigate the relationships and correlations between dimensions to identify a more effective and streamlined subset. The proposed multidimensional approach also enhances explainability when compared to systems or data collection strategies based on a single quality dimension. Judgments across multiple dimensions can reveal which specific facet(s) of a statement cause uncertainty or disagreement, thereby supporting more informed truthfulness assessments. In addition, expert-annotated data should be collected for all dimensions in future studies, as ground truth is currently available only for \overalltruthfulness. Advanced methods should also be explored to improve the combination of dimensions and thereby increase overall agreement.

Another promising direction for future work involves leveraging the geographical data of crowd workers. Access to the tasks described in this thesis was often, though not always, restricted to workers based in the United States. However, collecting or requesting such data raises additional challenges, particularly in terms of privacy and ethics. Workers may prefer not to be tracked, and any policy requiring geographical information must respect these preferences. Nevertheless, such data could offer valuable insights into potential correlations between worker performance and the geopolitical context of their country of residence.

Furthermore, the experimental settings adopted have enabled the collection of a substantial volume of data, which may support additional analyses and applications. These include, for instance, the examination of confidence scores, selected URLs, complex combinations of dimensions, and textual justifications. One possible line of investigation could explore the URLs submitted as evidence for each assessment, along with the content of the corresponding web pages, to construct a corpus of documents organised by truthfulness level.

Conversational information seeking~\cite{10.1145/3477495.3532678} involves interaction sequences between one or more users and an information system. These interactions are primarily based on natural language dialogue, although they may also include other modalities such as clicks, touch, or body gestures. The use of conversational agents such as smart speakers, virtual assistants, and chatbots is steadily increasing~\cite{bitterly_2019}, a trend that was further accelerated by the COVID-19 pandemic. For instance, individuals working from home have shown a higher likelihood of requesting news and updates through these agents~\cite{perez_2020}.

Enabling users to assess the truthfulness of information through conversational agents could offer several advantages. Performing such tasks via a smart speaker in a voice-only format may enhance user engagement and interest. A similar benefit may be expected in the case of chatbots, since many users are already familiar with such interfaces. A crowdsourcing experiment focused on information truthfulness could be conducted to evaluate whether this setting improves worker quality, behaviour, satisfaction, and engagement. Tools that support crowdsourcing through conversational interfaces already exist~\cite{10.1145/3320435.3320439, 10.1145/3406865.3418572} and could be effectively employed for this purpose.

People are rarely exposed to a single piece of information from one source at a time~\cite{doi:10.1126/sciadv.aay3539}. Moreover, individuals are often capable of evaluating multiple items in a short period. In contrast, the crowdsourcing experiments described in this thesis were conducted using a pointwise setup, in which each worker assessed one information item at a time, sequentially. An alternative approach would involve presenting workers with multiple items simultaneously, using a pairwise design~\cite{checco2016pairwise}. For instance, exposing workers to two or more statements could help assess the impact of comparative evaluation on response quality and behaviour. This concept could be extended further by asking workers to produce a ranked list of statements. Many machine learning systems already rely on pairwise or listwise approaches to learn how to rank items~\cite{DBLP:series/synthesis/2014Li}, which points to a methodological gap between how data is collected for training machine learning models and how labels are gathered through crowdsourcing.

Another promising research direction involves time constraints. Limiting the time available for judgment has been shown to improve data quality in some contexts~\cite{DBLP:conf/hcomp/MaddalenaBNDMD16}. It would be useful to explore whether this effect applies to truthfulness judgments. A dedicated crowdsourcing experiment could require workers to provide their assessments within a fixed time window. If successful, this strategy could help optimise task costs by identifying the optimal balance: allowing enough time to avoid pressure while discouraging multitasking. This is particularly relevant on commercial crowdsourcing platforms such as Prolific, where compensation is tied to the estimated time required to complete the task (Appendix~\ref{cap:paper_wsdm2022-sec:crowdsourcing_platforms-subsec:prolific}).

\subsection{\mrqtwo}
\label{cap:conclusions-sec:future-work-subsec:mrq-2}

Expanding on the findings related to cognitive biases, future work should also address broader socio-technical factors that influence the fact-checking process and the public’s perception of truth.

Modern computational propaganda and the configuration of social media platforms rely on communication techniques that not only misinform but also hinder critical thinking. This erosion weakens the public's ability to share a common understanding of social reality~\cite{waisbord2018truth}. It is therefore important to investigate the socio-technical features, platform metrics, and algorithmic settings that influence the content production pipeline, in order to strengthen communities’ resilience to the deterioration of the public sphere. 

Further cross-disciplinary research is needed to explore how theories from social science, psycholinguistics, and cognitive science can help explain the findings reported in this thesis. For example, the methodology proposed by~\citet{sethi2019fact} could be used to examine the emotional responses of crowd workers to misinformation.

The review of cognitive biases that may emerge during the fact-checking process, as presented in Chapter~\ref{cap:paper_ipm2023_bias}, provides a foundation for future research. The identified biases can inform the design of targeted experiments aimed at understanding and mitigating their effects. Future studies should also incorporate quantitative analysis alongside the existing qualitative approach. For example, a bibliometric analysis could be conducted to identify the most influential contributions within the referenced literature, offering insights into which cognitive biases most frequently affect human judgment in fact-checking. Additionally, statistical methods could be employed to quantify the strength and prevalence of these influences. Such investigations would contribute to a more comprehensive understanding of how cognitive biases interact with the fact-checking process.

For instance, Section~\ref{cap:paper_ipm2023_bias-sec:results-subsec:list} and Table~\ref{cap:paper_ipm2023_bias-sec:results-sect:ideal-pipeline-tab:ideal-pipeline} suggest that prompting assessors to revise their judgments (i.e., adopting \mycitecountermeasure{Revision}{\ref{cap:paper_ipm2023_bias-count:revise-scores}}) may help mitigate the \mycitebiasnoindex{Anchoring Effect}{\ref{cap:paper_ipm2023_bias-bias:anchoring-bias}}{Anchoring Effect}. Building on this, a between-subjects experiment for truthfulness classification could be designed, in which assessors are divided into two disjoint groups. Both groups would receive an initial piece of information before each assessment (i.e., the anchor); however, only the first group would be asked to revise their initial judgment before submission. By comparing the resulting annotations, researchers could empirically assess the degree of the \mycitebias{Anchoring Effect}{\ref{cap:paper_ipm2023_bias-bias:anchoring-bias}}. Based on the findings, practitioners might adopt the configuration that yields less biased annotations when collecting final truthfulness judgments.

More broadly, future research should explore experimental designs that evaluate the impact of cognitive biases on assessor performance and test the effectiveness of proposed countermeasures. It is also essential to investigate how biases interact and whether their combined effects further compromise the reliability of human assessments.

Additional studies should be carried out using the multidimensional truthfulness judgments described in Section~\ref{cap:paper_ipm2021-sec:results}. Some evidence suggests that specific dimensions are more susceptible to cognitive bias effects. Accordingly, strategies for correcting and de-biasing individual truthfulness dimensions should be tested. These efforts would enable the creation of non-biased datasets that could be used to train state-of-the-art deep learning algorithms for automatic misinformation assessment. Confidence scores produced by such algorithms could then be compared with workers’ self-reported confidence levels. Finally, these de-biasing approaches could be further refined by examining how disinformation targets specific subgroups in different ways~\cite{doi:10.1080/23738871.2018.1462395}.

\subsection{\mrqthree}

\label{cap:conclusions-sec:future-work-subsec:mrq-3}

Building on the findings related to the automatic prediction and explanation of truthfulness, future research should further investigate machine learning techniques to support scalable, data-driven fact-checking.

Several studies have proposed methods to analyze news attributes in order to assess their truthfulness~\cite{Horne_Adali_2017, rashkin-etal-2017-truth, DBLP:conf/ecir/ShresthaS21}. These approaches warrant further exploration to enhance the automation of the fact-checking pipeline. The machine learning techniques presented in Chapter~\ref{cap:paper_jdiq2022}, for instance, include an unsupervised strategy that uses static word embeddings to estimate the semantic proximity of labels and to predict truthfulness judgments. However, such embeddings may lose important contextual information due to averaging.

The query terms and textual justifications provided by high-quality workers could also be leveraged to train supervised models and to construct a curated set of effective fact-checking queries. Future work should therefore explore more advanced embedding methods, including contextualized embeddings, to improve predictive performance.

The \ebart architecture introduced in this thesis also presents opportunities for further investigation. Its joint modeling approach allows for extensions such as using saliency maps to analyze the influence of evidence passages on classification outcomes. Future research could also examine the relationship between the two components of the architecture, specifically the discriminative and the generative models, to better understand their individual contributions and how they interact.

\subsection{Statistical Power in Crowdsourcing}
\label{cap:paper_tois2023}

This section summarises a study published in the \lq\lq ACM Transactions on Information Systems\rq\rq{} journal~\cite{roitero2023setsize}. It investigates how to improve the statistical robustness of crowdsourcing experiments, particularly in the context of evaluation tasks such as those used in Information Retrieval (IR).

Section~\ref{cap:paper_tois2023-sec:motivation-background} outlines the motivation for using crowdsourcing in IR evaluation and its relevance to truthfulness assessment. Section~\ref{cap:paper_tois2023-sec:statistical-power} introduces a methodology to estimate the number of workers needed to ensure statistical power in annotation tasks.

\subsubsection{Motivation and Background}
\label{cap:paper_tois2023-sec:motivation-background}

Test collections provide a standard framework for assessing the effectiveness of IR systems. However, the most resource-intensive aspect of creating a test collection is obtaining relevance judgments, which requires substantial human effort and financial investment. To address this limitation, prior research has explored the use of crowdsourcing as a scalable and cost-effective alternative to traditional expert assessments~\cite{alonso:mizzaro:2012, webcrowd25k_second, Maddalena:2017:CRM:3026478.3002172, webcrowd25k_first, Roitero:2018:FRS:3209978.3210052, roitero2021effect, alonso2009can, yang2018pairwise}.

This line of work aligns with the broader objectives of this thesis by providing strategies to optimize the design of crowdsourcing experiments, ensuring that the collected data is both cost-efficient and statistically robust. These insights can inform future truthfulness assessment efforts by guiding the selection of task parameters that balance data quality, reliability, and scalability. In particular, principles such as configuring the number of judgments per item, selecting suitable workers, and setting appropriate agreement thresholds can help produce more reliable annotations. These methods represent a promising direction for scaling the collection of high-quality truthfulness judgments and contribute to the overall aim of enabling robust and interpretable fact-checking pipelines.

\subsubsection{Towards Statistically Robust Crowdsourcing}
\label{cap:paper_tois2023-sec:statistical-power}

A formal study of statistical power and significance in the setting of crowdsourcing experiments has received little attention~\cite{eickhoff2013increasing}, unlike in disciplines such as Information Retrieval. However, some researchers have considered statistical power in crowdsourcing, even though a comprehensive study detailing the effects and implications of experimental design is lacking. \citet{kittur2008crowdsourcing} point out the importance of good experimental design for obtaining reliable results from the crowd. \citet{5946971} propose a tool to conduct Mean Opinion Score~\cite{Streijl2016}\index{Mean Opinion Score} tests to evaluate signal processing methods using crowdsourcing, taking into account the statistical significance of the sample. \citet{behrend2011viability} compare the use of crowdsourcing platforms versus university students for survey data in behavioral research, with attention to statistical power. \citet{eickhoff2013increasing} address statistical significance to identify malicious workers and improve task robustness. \citet{landy2020crowdsourcing} compare the results of fifteen research teams on a common problem and investigate how design choices influence statistical significance.

The design of the crowdsourcing tasks aimed at collecting relevance judgments to build a so-called test collection is typically left to researchers and practitioners, who select and design the annotation task with a focus on the number of documents to be judged. In contrast, the number of workers assigned to judge each document is often set in a heuristic manner using simple rules of thumb. As a result, the annotated dataset produced by these crowdsourcing tasks often lacks guarantees that it satisfies predefined statistical requirements. These requirements include, for example, ensuring sufficient statistical power to distinguish each pair of documents in a statistically significant way. Meeting this requirement would mean that, if a document is judged to be more relevant than another, it is possible to conclude that the observed difference is not due to chance.

A methodology to estimate the number of crowd workers required to produce a test collection that achieves a given statistical power can be proposed. This methodology extends prior work based on the \index{t-test}t-test and one-way \index{ANOVA}ANOVA, and allows researchers and practitioners to estimate in advance the number of workers to employ in a crowdsourcing task in order to obtain an annotated dataset with a minimum level of statistical power guaranteed by design. The methodology is being experimentally evaluated using multiple publicly available datasets. The results show that it can provide reliable estimates of the number of workers required to distinguish documents in a statistically significant way, as well as the number of documents needed to distinguish workers in a statistically significant way.

At present, the methodology is being studied in the field of Information Retrieval systems evaluation. However, it is general and can be applied in other domains. In the context of misinformation assessment conducted using crowdsourcing-based approaches, each statement is evaluated by multiple human workers to filter out noise, malicious input, and cognitive biases that may affect their judgments. These judgments are typically aggregated to improve their quality, often using some form of aggregation function. Since numerous comparisons among statements are required, a methodology that can compute the required number of judges (in this case, crowd workers) and ensure that differences between statements are statistically significant would be extremely useful. 

Such a methodology, in other words, would provide a sound statistical foundation for concluding that \lq\lq statement X is perceived as more truthful than statement Y\rq\rq{}.

\section{Conclusions}
\label{cap:conclusions-sec:conclusions}

This thesis addresses the pressing challenge posed by the ever-increasing volume of (mis)\-in\-for\-ma\-ti\-on spreading online, exploring three main research directions. It shows that non-expert human judges are capable of objectively categorizing recent online misinformation, even when applying a multidimensional notion of truthfulness. The research also identifies a total of \numbiasused cognitive biases that may influence fact-checking activities, providing preliminary evidence of their presence in crowdsourced fact-checking experiments. Lastly, it introduces a machine-learning-based model, \ebart, which can predict the truthfulness of information while also generating valuable explanations for human users.

A collaborative process involving non-expert human judges, expert fact-checkers, and automatic fact-checking models could offer a scalable and decentralized hybrid mechanism to counter the growing spread of online misinformation. The characterization of cognitive biases that may emerge during fact-checking activities can support the development of more robust, aware, and bias-resistant pipelines for collecting reliable truthfulness judgments at scale. The ability to predict truthfulness and generate explanations simultaneously makes automatic fact-checking models more transparent and fosters greater user trust.

Ultimately, this thesis contributes to the design and development of systems to combat misinformation. The research not only advances the three core directions but also opens several avenues for future work. It aims to support a better understanding of how to harness crowd-powered human intelligence to build robust, trustworthy, explainable, and transparent systems aligned with the principles endorsed by fact-checking organizations.\footnote{\url{https://www.ifcncodeofprinciples.poynter.org/}} 

The complete dataset collected and analyzed is publicly available at: \url{https://doi.org/10.17605/OSF.IO/JR6VC}~\cite{Soprano2023thesis}.

\section{Acknowledgments}

\label{cap:conclusion-sec:acks}

The work described in this thesis is supported by several funding sources and collaborative initiatives.

Chapters~\ref{cap:paper_sigir2020}, \ref{cap:paper_pauc2021}, and \ref{cap:paper_ipm2021} are supported by the Australian Research Council Discovery Project DP190102141.\footnote{\url{http://purl.org/au-research/grants/arc/DP190102141}} Chapter~\ref{cap:paper_sigir2020} also received a Facebook Research Award,\footnote{\url{https://research.facebook.com/research-awards/the-online-safety-benchmark-request-for-proposals/}} while Chapter~\ref{cap:paper_pauc2021} is additionally supported by the HEaD project \lq\lq Higher Education and Development -- 1619942002 / 1420AFPLO1\rq\rq{} (Region Friuli--Venezia Giulia) and the MISTI \lq\lq MIT International Science and Technology Initiatives -- Seed Fund\rq\rq{} (MIT-FVG Seed Fund Project). Chapter~\ref{cap:paper_ipm2021} also benefits from data access facilitated by Devi Mallal at RMIT ABC Fact Check.

Chapter~\ref{cap:paper_tsc2024} is partially supported by the Next Generation EU PRIN 2022 project \lq\lq MoT--The Measure of Truth\rq\rq{} (20227F2ZN3_001, CUP G53D23002800006), the TU Delft AI Initiative, and the Delft Design@Scale AI Lab.

Chapter~\ref{cap:paper_ipm2023_bias} is supported by multiple sources, including the ARC DECRA Fellowship DE200100064, the ARC Centre of Excellence for Automated Decision-Making and Society (CE200100005), the ARC Training Centre for Information Resilience (IC200100022), the University of Udine’s Interdepartmental Project on Artificial Intelligence (2020--2025), and the Netherlands eScience Center project \lq\lq The Eye of the Beholder\rq\rq{} (027.020.G15). Additional support comes from the PRIN 2022 project \lq\lq MoT--The Measure of Truth\rq\rq{} (20227F2ZN3_001, CUP G53D23002800006). This research is also part of the AI, Media \& Democracy Lab\footnote{\url{https://www.aim4dem.nl/}} (NWO project NWA.1332.20.009).

Chapter~\ref{cap:paper_facct2022} and Section~\ref{cap:paper_is2022} are partially supported by The Credibility Coalition.\footnote{\url{https://credibilitycoalition.org/}} Section~\ref{cap:paper_is2022} also received funding from the \lq\lq Departments of Excellence 2018--2022\rq\rq{} project awarded to the Department of Philosophy of the University of Milan \lq\lq Piero Martinetti\rq\rq{}, and from the PRIN 2020 Grant 2020SSKZ7R\footnote{\url{https://sites.unimi.it/brio/about/}} funded by MIUR.

Damiano Spina is the recipient of an ARC DECRA Fellowship (DE200100064), an Associate Investigator at the ARC Centre of Excellence for Automated Decision-Making and Society (CE200100005), and a research collaborator at RMIT FactLab. Gianluca Demartini is a Chief Investigator at the ARC Training Centre for Information Resilience (IC200100022). 

The contributions of all crowd workers who participated in the studies are gratefully acknowledged. The opinions, findings, and conclusions presented in this thesis are those of the authors of each study and do not necessarily reflect the views of the sponsors.

\appendix

\chapter{Crowd\_Frame: Design and Deploy Crowdsourcing Tasks}


\label{cap:paper_wsdm2022}

This appendix is based on the article published at the Fifteenth ACM International Conference on Web Search and Data Mining~\cite{10.1145/3488560.3502182}. It presents \crowdframe, a software system designed to support the complete workflow of crowdsourcing activities. Section~\ref{cap:paper_wsdm2022-sec:crowdsourcing_platforms} outlines the design and deployment process of crowdsourcing tasks across three different platforms. Section~\ref{cap:paper_wsdm2022-sec:research_question} introduces the overarching objectives that the system is intended to address. Section~\ref{cap:paper_wsdm2022-sec:system-design} provides a detailed analysis of the system's four core components. Section~\ref{cap:paper_wsdm2022-sec:usage} explains how to configure and deploy a task, while Section~\ref{cap:paper_wsdm2022-sec:usage-sec:task-performing} describes the worker recruitment process. Section~\ref{cap:paper_wsdm2022-sec:usage-sec:task-result} details the structure of the output data. Section~\ref{cap:paper_wsdm2022-sec:conclusions} summarizes the current implementation status, and Section~\ref{cap:paper_wsdm2022-sec:future_work} outlines directions for future development.

\section{Crowdsourcing Platforms}
\label{cap:paper_wsdm2022-sec:crowdsourcing_platforms}

Understanding how crowdsourcing platforms support the design and deployment of tasks is important, especially since the process can be complex and prone to errors. This understanding helps highlight the improvements introduced by using \crowdframe. The following sections describe three platforms: \mturk (Section~\ref{cap:paper_wsdm2022-sec:crowdsourcing_platforms-subsec:mturk}), \toloka\index{Toloka} (Section~\ref{cap:paper_wsdm2022-sec:crowdsourcing_platforms-subsec:toloka}), and \prolific\index{Prolific} (Section~\ref{cap:paper_wsdm2022-sec:crowdsourcing_platforms-subsec:prolific}).

\subsection{Amazon Mechanical Turk}

\label{cap:paper_wsdm2022-sec:crowdsourcing_platforms-subsec:mturk}

\begin{figure}[tbp]
  \centering
  \includegraphics[width=\linewidth]{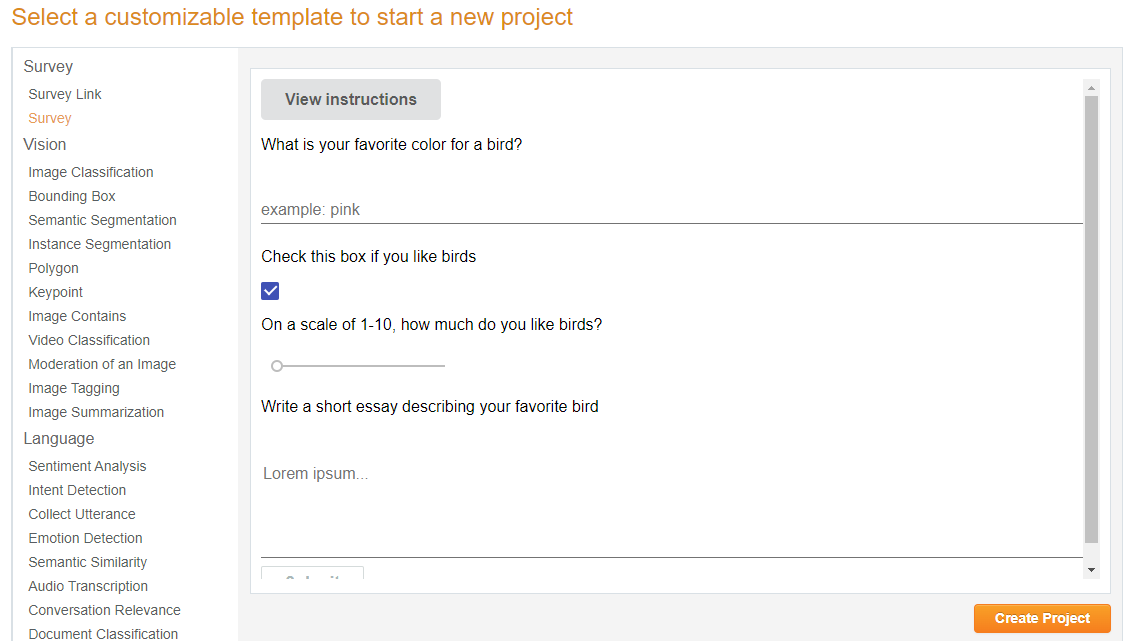}
  \caption{Selecting the task type on \mturk.}
  \label{cap:paper_wsdm2022-sec:crowdsourcing-platforms-fig:mturk-step-1}
\end{figure}

The task design workflow for a requester using Amazon Mechanical Turk involves three main phases. First, the requester must select the type of task to publish (Figure~\ref{cap:paper_wsdm2022-sec:crowdsourcing-platforms-fig:mturk-step-1}). The platform offers several customizable templates, although the requester may also opt to start from a blank template.

\begin{figure}[tbp]
  \centering
    \includegraphics[width=\linewidth]{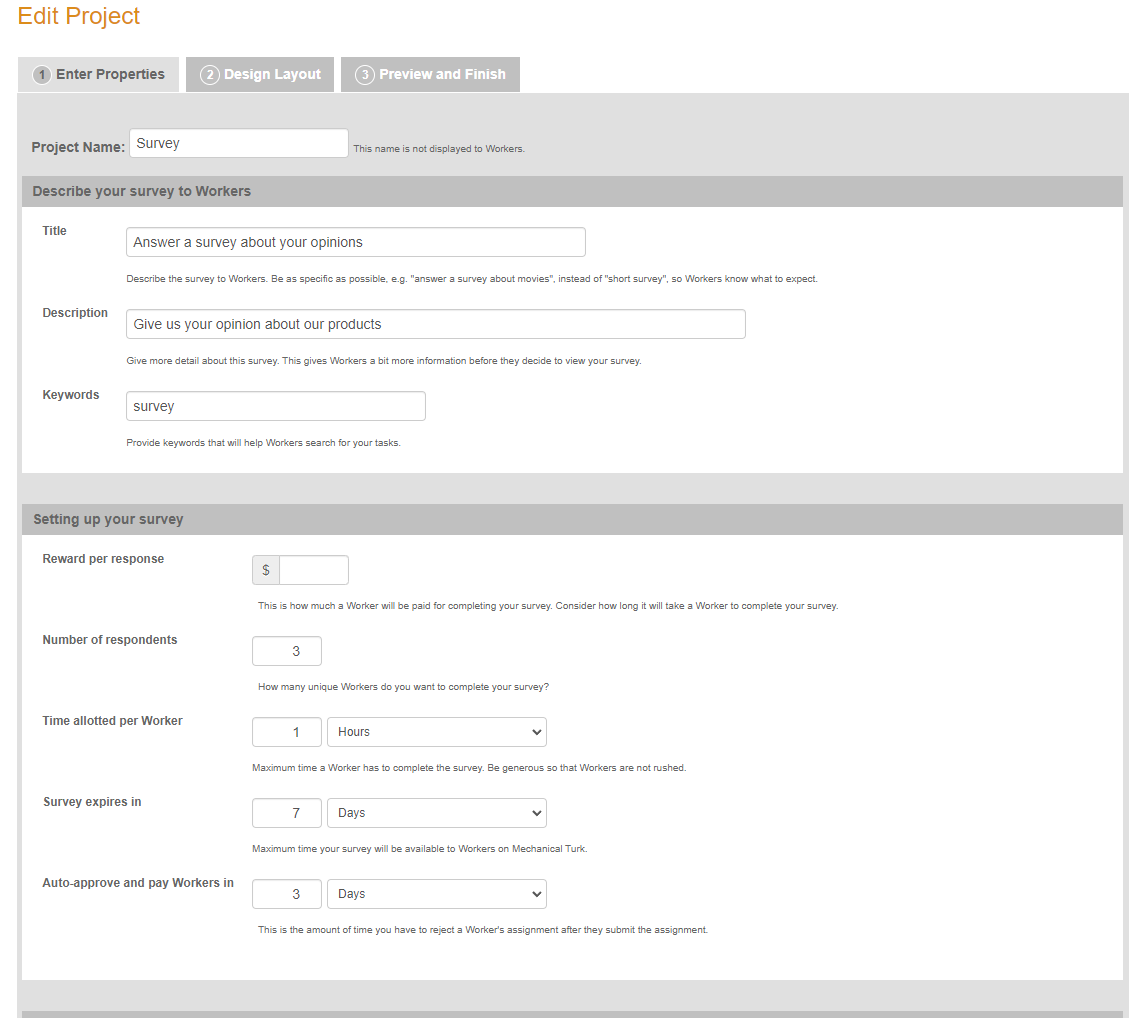}
\caption{Configuration of task parameters on \mturk.}
\label{cap:paper_wsdm2022-sec:crowdsourcing-platforms-fig:mturk-step-2}
\end{figure}

Then, the requester proceeds to design the overall task. First, they input the task name and description and set five configuration parameters (Figure~\ref{cap:paper_wsdm2022-sec:crowdsourcing-platforms-fig:mturk-step-2}). These parameters include the number of workers to recruit, the time allowed to complete the task, the reward amount in USD\$, the task expiration date, and the auto-approve threshold (in days) for workers' submissions. The requester can also define various criteria to filter the workers to be recruited. \mturk uses the term \lq\lq Qualification Type\rq\rq{} \index{Qualification Type} to refer to such criteria. Qualification types include the worker's country of residence, age, household income, and other attributes. It is also possible to define custom qualification types.\footnote{\url{https://docs.aws.amazon.com/AWSMechTurk/latest/AWSMechanicalTurkRequester/WorkWithCustomQualType.html}} 

After setting the parameters and criteria, the requester designs the task interface. This must be done using a set of custom markup tags known as \index{Crowd Elements} Crowd Elements.\footnote{\url{https://docs.aws.amazon.com/AWSMechTurk/latest/AWSMturkAPI/ApiReference_HTMLCustomElementsArticle.html}} Each task must include the \spverb|<crowd-form>| tag, which is the core element of the interface code. Listing~\ref{cap:paper_wsdm2022-sec:crowdsourcing-platforms-list:sample-task} shows the code for a sample task designed using Crowd Elements.

The requester can preview each \index{HIT}HIT of the task once the design is finalized. The task can then be published any number of times. \mturk uses the term \lq\lq batch\rq\rq{} to indicate a single set of workers recruited within a given crowdsourcing task (see also Section~\ref{cap:intro-sec:terminology}). The requester can recruit multiple batches of workers simultaneously. Recruiting a batch involves providing a special CSV file to specify the input and output data. The platform treats each column of this CSV file as a variable. Each row is assigned to one worker, and its values are used to initialize the task inputs and expected outputs. Therefore, the requester must supply a file with $n$ rows to recruit $n$ workers. Each worker accepts the \index{HIT}HIT initialized through this mechanism and completes the task. The requester then approves or rejects the submission. The final results are made available by the platform through a second \index{CSV}CSV file once all workers in the batch have completed their \index{HIT}HIT.

\begin{lstlisting}[style=htmlstyle, caption={Interface of a sample task on \mturk, built using \index{Crowd Elements}Crowd Elements.}, label={cap:paper_wsdm2022-sec:crowdsourcing-platforms-list:sample-task}]
<script src="https://assets.crowd.aws/crowd-html-elements.js"></script>
<crowd-form answer-format="flatten-objects">
  <crowd-instructions link-text="View instructions" link-type="button">
    <short-summary>
      <p>Provide a brief instruction here</p>
    </short-summary>
    <detailed-instructions>
      <h3>Provide more detailed instructions here</h3>
      <p>Include additional information</p>
    </detailed-instructions>
    <positive-example>
      <p>Provide an example of a good answer here</p>
      <p>Explain why it's a good answer</p>
    </positive-example>
    <negative-example>
      <p>Provide an example of a bad answer here</p>
      <p>Explain why it's a bad answer</p>
    </negative-example>
  </crowd-instructions>
  <div>
    <p>What is your favorite color for a bird?</p>
    <crowd-input name="favoriteColor" placeholder="example: pink" required>
    </crowd-input>
  </div>
  <div>
    <p>Check this box if you like birds</p>
    <crowd-checkbox name="likeBirds" checked="true" required>
    </crowd-checkbox>
  </div>
  <div>
    <p>On a scale of 1-10, how much do you like birds?</p>
    <crowd-slider name="howMuch" min="1" max="10" step="1" required>
    </crowd-slider>
  </div>
</crowd-form>
\end{lstlisting}

\subsection{Toloka}
\label{cap:paper_wsdm2022-sec:crowdsourcing_platforms-subsec:toloka}

\begin{figure}[tpb]
  \centering
  \includegraphics[width=\linewidth]{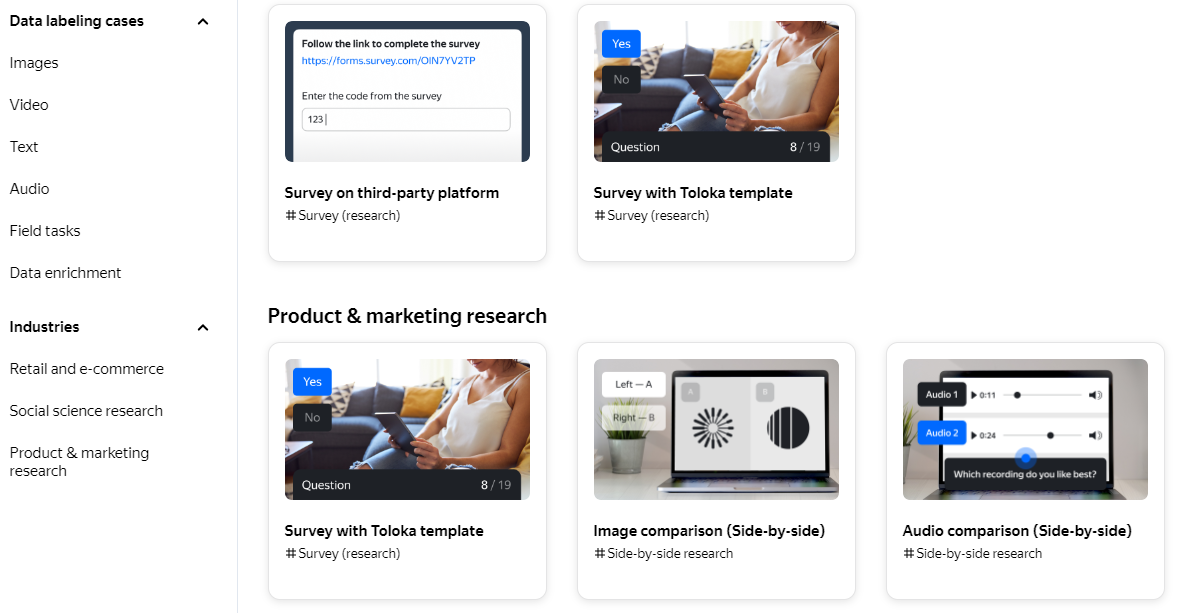}
  \caption{Selection of task type for a project on \toloka\index{Toloka}.}
  \label{cap:paper_wsdm2022-sec:crowdsourcing-platforms-fig:toloka-step-1}
\end{figure}

The task design workflow for a requester using \toloka involves three phases. Initially, the requester selects the type of task to be performed (Figure~\ref{cap:paper_wsdm2022-sec:crowdsourcing-platforms-fig:toloka-step-1}) and provides the task name and description, along with a private comment. This comment can be useful, for example, to distinguish between two projects with the same name. \toloka{} uses the term \lq\lq Project\rq\rq{} to refer to a crowdsourcing task.

\begin{figure}[tpb]
  \centering
  \includegraphics[width=.9\linewidth]{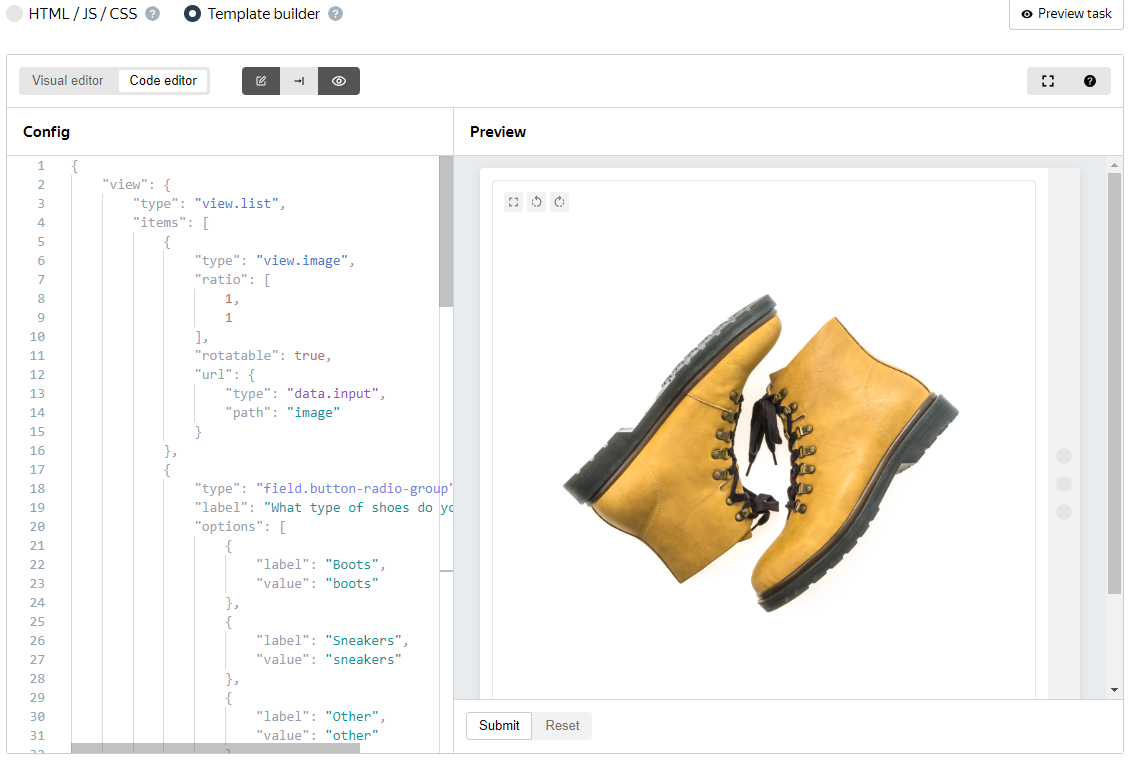}
  \caption{Interface design for a project on \toloka using the template builder.}
  \label{cap:paper_wsdm2022-sec:crowdsourcing-platforms-fig:toloka-step-2}
\end{figure}

\begin{figure}[tpb]
  \centering
  \includegraphics[width=0.7\linewidth]{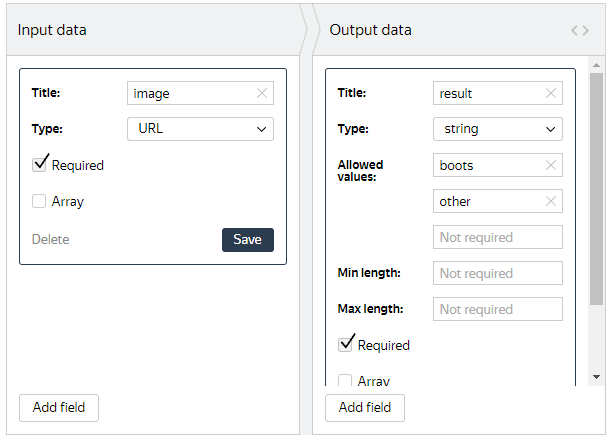}
  \caption{Specification of input and output data for a project on \toloka.}
  \label{cap:paper_wsdm2022-sec:crowdsourcing-platforms-fig:toloka-step-3}
\end{figure}

Then, the requester designs the task interface. They can use either standard HTML markup tags together with \index{JavaScript} JavaScript code and \index{CSS} CSS styling, or the \toloka\index{Toloka} template builder\footnote{\url{https://toloka.ai/docs/template-builder/index.html}} (Figure~\ref{cap:paper_wsdm2022-sec:crowdsourcing-platforms-fig:toloka-step-2}). The template builder is based on a set of predefined \index{JSON} JSON objects that are used to initialize the interface. The specification of input and output data\footnote{\url{https://toloka.ai/docs/guide/concepts/incoming.html}} must be declared in a separate section of the user interface (Figure~\ref{cap:paper_wsdm2022-sec:crowdsourcing-platforms-fig:toloka-step-3}). This data can be specified either manually, using dedicated JSON objects, or through the graphical interface. The declared input and output data are then referenced when building the task interface. Listing~\ref{cap:paper_wsdm2022-sec:crowdsourcing-platforms-list:toloka-template-builder} shows the JSON specification of the sample image classification task interface shown in Figure~\ref{cap:paper_wsdm2022-sec:crowdsourcing-platforms-fig:toloka-step-2}. Listing~\ref{cap:paper_wsdm2022-sec:crowdsourcing-platforms-list:toloka-input-spec} and Listing~\ref{cap:paper_wsdm2022-sec:crowdsourcing-platforms-list:toloka-output-spec} show the corresponding input and output JSON specifications. The requester then writes the instructions that the workers will read when accepting a \index{HIT}HIT. \toloka refers to workers as either \lq\lq Tolokers\rq\rq{} or \lq\lq Users\rq\rq{}.

\begin{figure}[tpb]
 \centering
\includegraphics[width=0.9\linewidth]{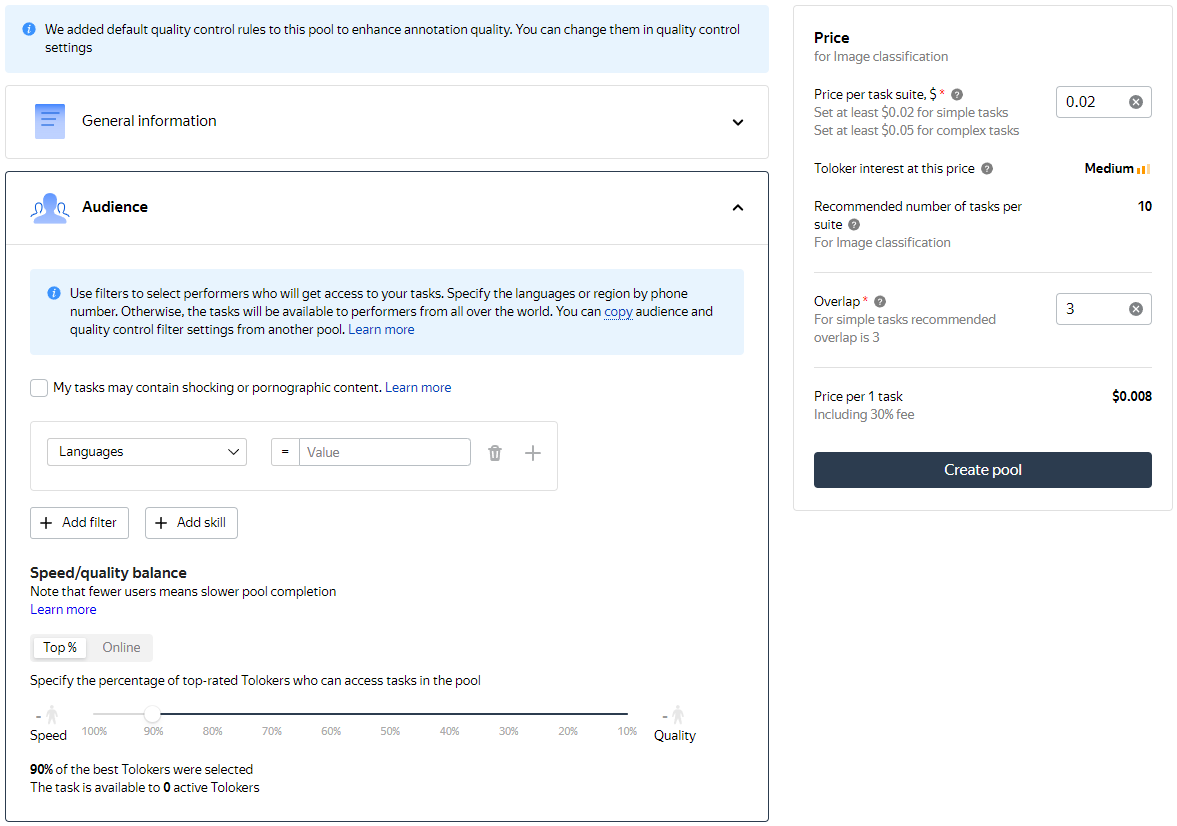}
\caption{Worker attributes configuration for a pool of a project on \toloka.}
\label{cap:paper_wsdm2022-sec:crowdsourcing-platforms-fig:toloka-step-4}
\end{figure}

The requester can define multiple pools of workers to recruit for a given task once its design is finalized. \toloka uses the word \lq\lq Pool\rq\rq{} to indicate groups of workers who share a predefined set of attributes. Specifically, the requester specifies a name and description for each pool, along with the set of required attributes. These attributes may include the language spoken, geographic region, operating system, and others. The requester also selects a speed/quality balance percentage, which determines the proportion of top-rated workers allowed to access the pool. A higher quality requirement typically results in slower pool completion times. Additionally, the requester can define quality control mechanisms and rules, and may require overlap for each \index{HIT}HIT published within the pool. Finally, the requester sets the reward for completing each \index{HIT}HIT. Figure~\ref{cap:paper_wsdm2022-sec:crowdsourcing-platforms-fig:toloka-step-4} shows part of the pool configuration interface in \toloka. Each pool can be started, paused, and stopped independently, and multiple pools can be active simultaneously.

\begin{lstlisting}[style=jsonstyle, caption={Image classification JSON configuration for the interface of a project on \toloka built using the template builder.}, label={cap:paper_wsdm2022-sec:crowdsourcing-platforms-list:toloka-template-builder}]
{
  "view": {
    "type": "view.list",
    "items": [
      {
        "type": "view.image",
        "ratio": [1,1],
        "rotatable": true,
        "url": { "type": "data.input", "path": "image" }
      },
      {
        "type": "field.button-radio-group",
        "label": "What type of shoes do you see?",
        "options": [ 
          { "label": "Boots", "value": "boots"}, 
          ...
        ],
        "data": { "type": "data.output", "path": "result"},
        "validation": { "type": "condition.required" }
      }
    ]
  },
  "plugins": [
    { 
      "type": "plugin.toloka", 
      "layout": { "kind": "scroll", "taskWidth": 600 } 
    },
    {
      "1": {
        "type": "action.set",
        "data": { "type": "data.output", "path": "result" },
        "payload": "boots"
      },
      "2": {
        ...
      },
      "type": "plugin.hotkeys"
    }
  ],
  "vars": {}
}
\end{lstlisting}

\begin{lstlisting}[style=jsonstyle, caption={Input data specification for a project on \toloka.}, label={cap:paper_wsdm2022-sec:crowdsourcing-platforms-list:toloka-input-spec}]
{
  "image": {
    "type": "url",
    "required": true
  }
}
\end{lstlisting}

\begin{lstlisting}[style=jsonstyle, caption={Output data specification for a project on \toloka.}, label={cap:paper_wsdm2022-sec:crowdsourcing-platforms-list:toloka-output-spec}]
{
  "result": {
    "type": "string",
    "allowed_values": [ "boots", "other" ],
    "required": true
  }
}
\end{lstlisting}

The requester must provide values for the input and output data defined during the design phase in order to start publishing the task on \toloka \index{Toloka} and recruit workers. The mechanism is similar to that of \mturk, described in Section~\ref{cap:paper_wsdm2022-sec:crowdsourcing_platforms-subsec:mturk}. A special file containing the input and output values must be provided. The file can be in XLSX, TSV, or JSON format.\index{XLSX} \index{CSV} \index{JSON} Each column (or attribute, in the case of a JSON file) is labeled using the prefixes \spverb|INPUT:| or \spverb|OUTPUT:| and then matched to the corresponding input or output field defined by the requester. Each row of the file is assigned to a single worker, and its values are used to initialize the data. In other words, the requester must provide a file with $n$ rows to recruit $n$ workers. Listing~\ref{cap:paper_wsdm2022-sec:crowdsourcing-platforms-list:toloka-xlsx-file} shows the content of a sample XLSX file used to recruit two workers within a pool for a task that includes a single input field named \spverb|image|. Each worker accepts the assigned \index{HIT}HIT and completes the task. The requester can then approve or reject the reward for each submitted \index{HIT}HIT. The final results are provided by \toloka as a second TSV file at the end of the pool.

\begin{lstlisting}[style=textstyle, caption={Input data initialization for two \index{HIT}HITs in a pool of a project on \toloka.}, label={cap:paper_wsdm2022-sec:crowdsourcing-platforms-list:toloka-xlsx-file}]
INPUT:image
https://labs-images-testing.s3.yandex.net/presets/for%20tb%20and%20dataset/leather-boots.jpg
https://labs-images-testing.s3.yandex.net/presets/for%20tb%20and%20dataset/pair-trainers.jpg
\end{lstlisting}

\subsection{Prolific}

\label{cap:paper_wsdm2022-sec:crowdsourcing_platforms-subsec:prolific}

\begin{figure}[tpb]
 \centering
\includegraphics[width=.9\linewidth]{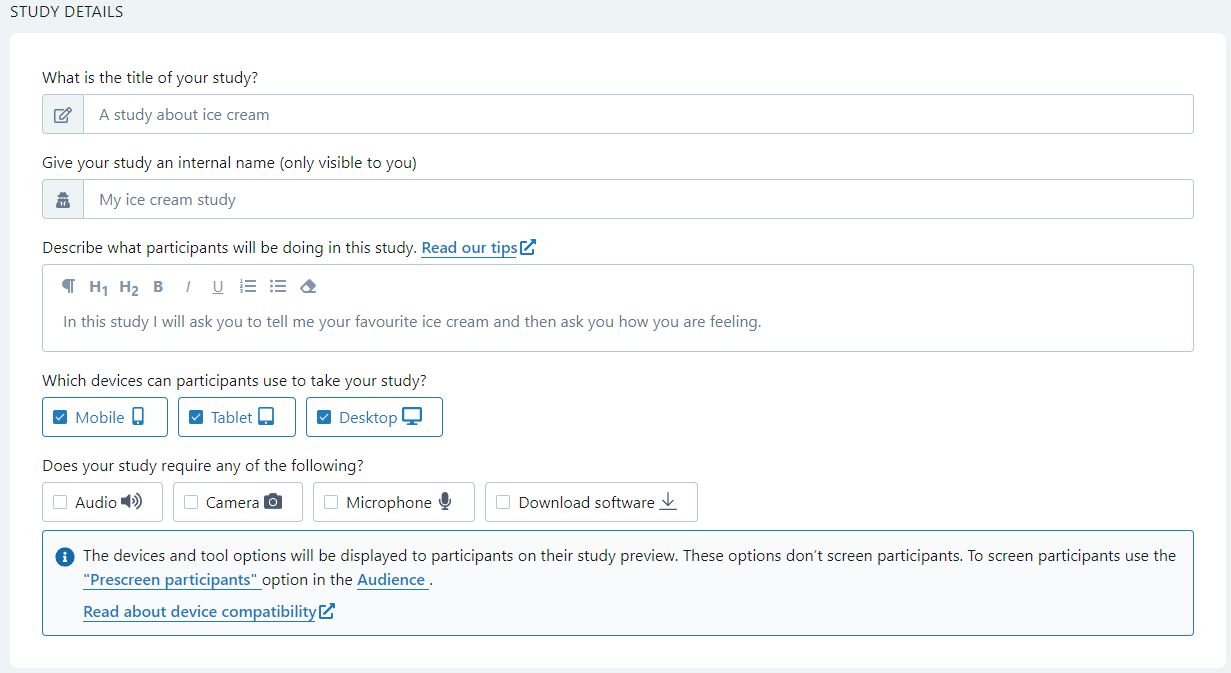}
\caption{Initial configuration parameters for a study on \prolific.}
\label{cap:paper_wsdm2022-sec:crowdsourcing-platforms-fig:prolific-step-1}
\end{figure}

The task deployment workflow for a requester using the \prolific platform involves four phases. Initially, the requester sets five parameters related to the task (Figure~\ref{cap:paper_wsdm2022-sec:crowdsourcing-platforms-fig:prolific-step-1}). \prolific uses the term \lq\lq Study\rq\rq{} to refer to a task, and \lq\lq Participant\rq\rq{} to refer to a worker. The preliminary parameters include the study name and description, as well as the devices that crowd workers can use to participate in the task. The requester can also indicate whether the study requires access to the worker’s microphone, camera, or audio, and whether participants need to install additional software to complete the task.

\begin{figure}[tpb]
 \centering
\includegraphics[width=.9\linewidth]{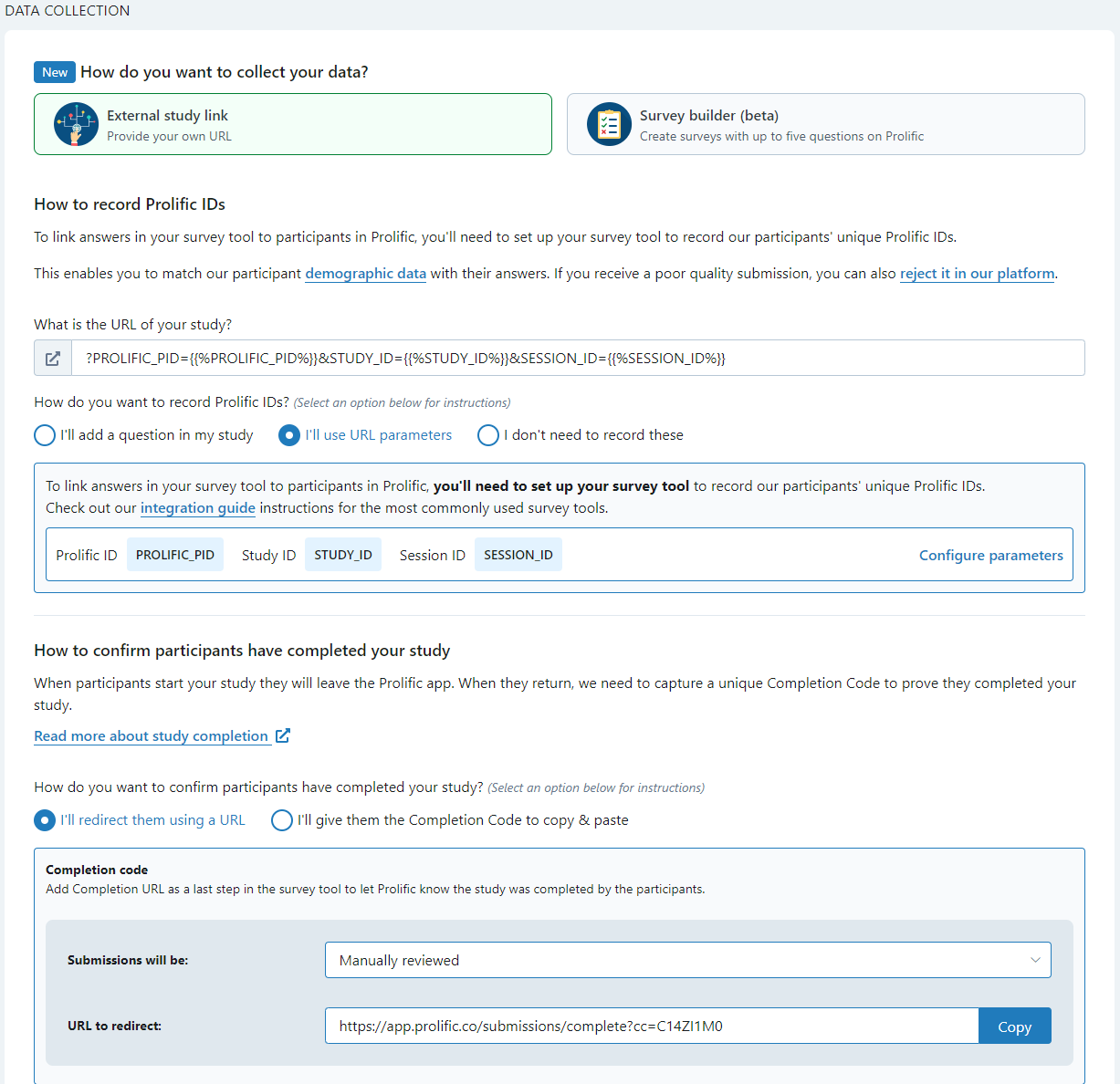}
\caption{Data collection configuration for a study on \prolific.}
\label{cap:paper_wsdm2022-sec:crowdsourcing-platforms-fig:prolific-step-2}
\end{figure}

The second phase of the task deployment workflow on \prolific requires the requester to specify the data collection modality (Figure~\ref{cap:paper_wsdm2022-sec:crowdsourcing-platforms-fig:prolific-step-2}). The most evident difference compared to \mturk (Section~\ref{cap:paper_wsdm2022-sec:crowdsourcing_platforms-subsec:mturk}) and \toloka (Section~\ref{cap:paper_wsdm2022-sec:crowdsourcing_platforms-subsec:toloka}) is that \prolific does not provide any built-in interface to design the task. Instead, the requester must rely on external tools to create and host the task, and then provide its URL to \prolific.

There are two approaches to assign an anonymous identifier to each worker for the external task. The first approach is to explicitly ask workers to input their \prolific identifiers. The second approach is to have \prolific automatically append each worker’s identifier to the external URL. This method also allows appending identifiers for the task and session. \prolific uses the term \lq\lq Session\rq\rq{} to refer to the current batch of recruited workers.
Additionally, the requester can choose between two methods for confirming task completion. One involves embedding a return URL in the external software interface, such as within a button, to redirect the worker back to \prolific. The other method consists of displaying a completion code that the worker copies and pastes manually into \prolific’s interface.

The third phase involves configuring the audience for the task. \prolific uses the term \lq\lq audience\rq\rq{} to refer to the workers to be recruited for the current task. The requester specifies how many workers should be recruited and from which country. They can also choose among three sampling modalities to select the workers. Additionally, various criteria can be applied to further filter the audience.

\begin{figure}[tbp]
 \centering
\includegraphics[width=0.9\linewidth]{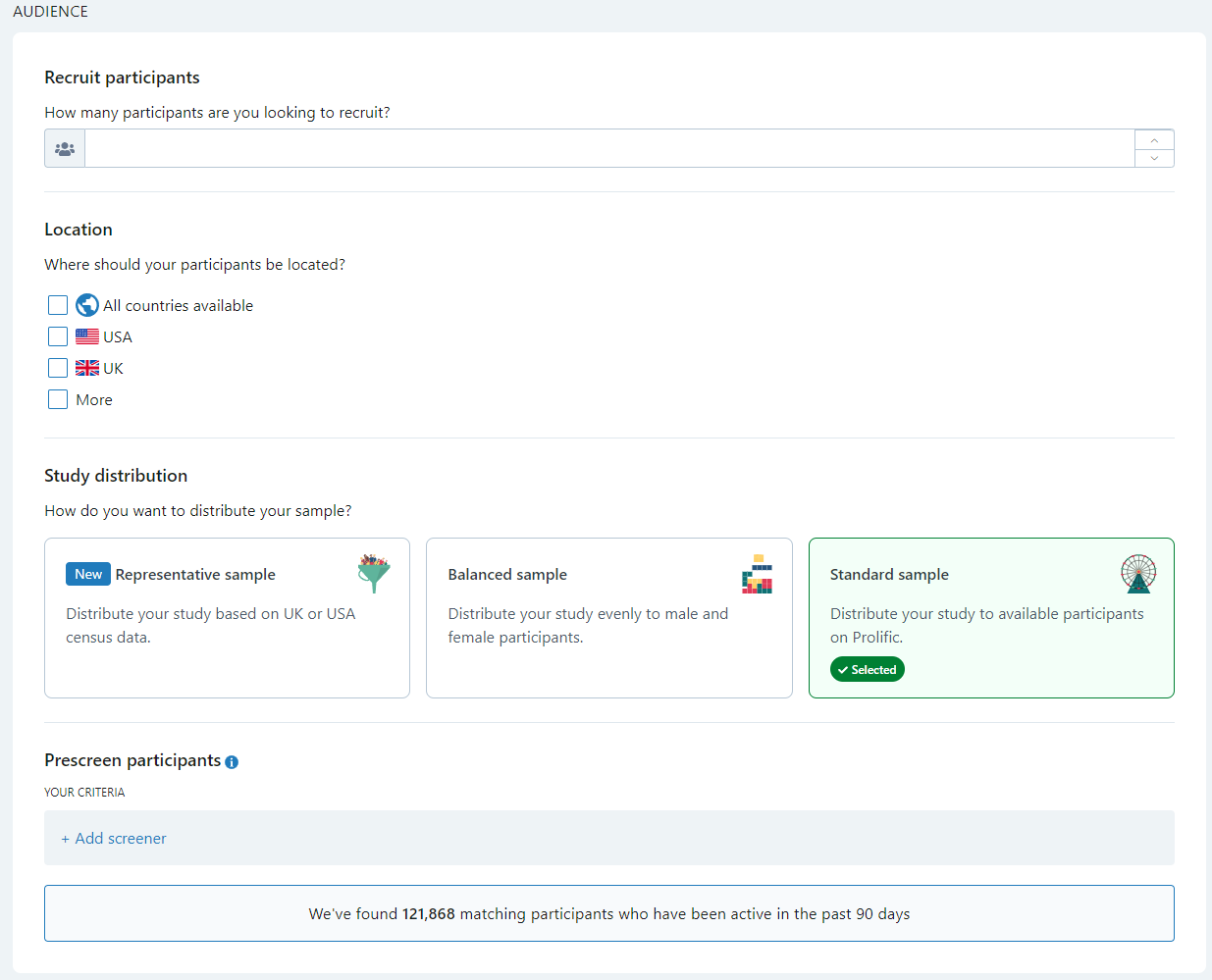}
\caption{Audience configuration for a study on \prolific.}
\label{cap:paper_wsdm2022-sec:crowdsourcing-platforms-fig:prolific-step-3}
\end{figure}

The fourth phase of the task deployment workflow on \prolific requires the requester to specify the estimated completion time and the final reward in pounds. The platform then calculates the corresponding hourly pay rate and notifies the requester whether the proposed amount is sufficient. While the platform does not prevent publication of the task if the reward is insufficient, it may pause the task if the actual median completion time exceeds the estimated time. In such cases, it may also suggest providing additional payments. Once the deployment is finalized, the requester can publish the task and wait for its completion. The status of each \index{HIT}HIT can be monitored in real time, and the payment can be approved or denied.

\begin{figure}[tbp]
 \centering
\includegraphics[width=0.9\linewidth]{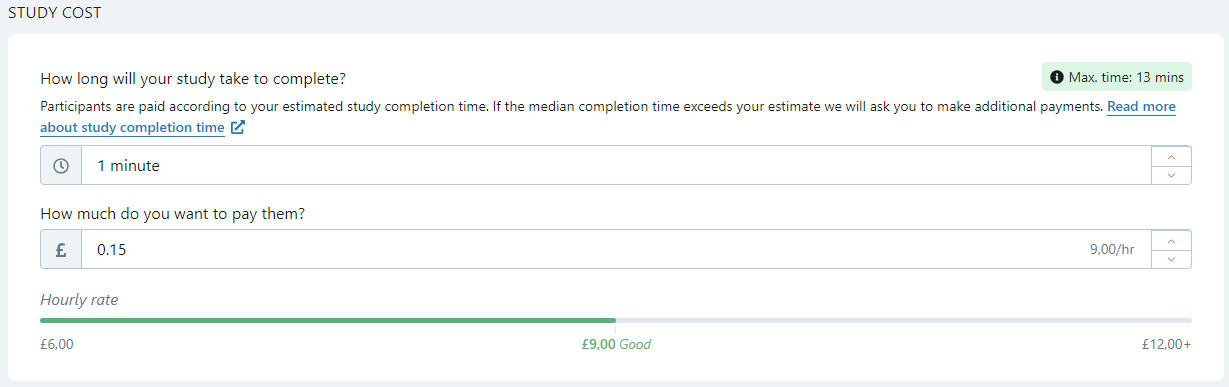}
\caption{Study cost configuration for a study on \prolific.}
\label{cap:paper_wsdm2022-sec:crowdsourcing-platforms-fig:prolific-step-4}
\end{figure}

\subsection{Discussion}

The design and deployment workflow of a crowdsourcing task is often cumbersome and unintuitive, as described in Section~\ref{cap:intro-sec:crowdsourcing-activity}. Consider \mturk: the task interface code must be written in a single box and combines a custom subset of HTML tags with \index{CSS} CSS and \index{JavaScript} JavaScript statements. This results in a mixture of business and presentation logic. Hidden form fields containing JSON objects are used to store data values. Additionally, a separate file must be built for each task batch. \citet{la2020crowdsourcing} deployed a misinformation assessment task directly on \mturk, collecting judgments for 120 politically related statements from 10 distinct crowd workers. They deployed 400 different \index{HIT}HITs. The task used a paginated structure built with \mturk's custom markup tags, and custom JavaScript had to be written to show or hide elements of the user interface, since only one \spverb|<crowd-form>| tag is allowed.

\toloka shares many of the same difficulties encountered when using \mturk, while adding further complexity. Quality control rules and pool initialization criteria are non-trivial to configure. The allocation mechanisms for \index{HIT}HITs can be confusing, as the platform groups them into overlapping task suites. On the other hand, the pool-based recruitment system allows for more fine-grained control over the worker selection process.

\prolific offers the most streamlined and accessible workflow among the three platforms. However, the lack of a built-in interface for task creation requires task requesters to rely entirely on external tools, which can be a significant barrier for those without the necessary technical background.

\section{Aims}

\label{cap:paper_wsdm2022-sec:research_question}

Several tools exist to assist task requesters when using crowdsourcing-based approaches (Section~\ref{cap:related_work-sec:crowdsourcing-tools}). However, none of these tools addresses the full set of difficulties outlined in Section~\ref{cap:paper_wsdm2022-sec:crowdsourcing_platforms}. One possible solution is to use crowdsourcing platforms exclusively to recruit the required workforce. The recruited workers access the task, which is deployed using external software on a separate platform. After completing the task, they return to the crowdsourcing platform to receive their reward. In this configuration, the task design and deployment processes are managed entirely by external software.

\citet{10.1145/3488560.3502182} introduced \crowdframe, a software system designed to simplify the creation and deployment of a wide range of crowdsourcing tasks, independent of the chosen platform. Tasks can include various sets of \index{HIT}HITs and are deployed in a customizable and controllable environment. The software is freely available for download\footnote{\url{https://github.com/Miccighel/Crowd_Frame}} and has already been used by researchers and practitioners in multiple studies~\cite{brand2021jointly, brand2022neural, CEOLIN2022102107, draws2022bias, roitero2020crowd, roitero2021crowd, soprano2023loyalty, SOPRANO2021102710}.

\section{System Design}

\label{cap:paper_wsdm2022-sec:system-design}

Section~\ref{cap:paper_wsdm2022-sec:system-design-subsec:general-arch} outlines the general architecture of \crowdframe. Section~\ref{cap:paper_wsdm2022-sec:system-design-subsec:generator} describes the \generator component, which allows requesters to configure tasks. Section~\ref{cap:paper_wsdm2022-sec:system-design-subsec:skeleton} presents the \skeleton component, which serves as the main interface for workers. Optionally, the skeleton may embed an instance of the \searchengine component (Section~\ref{cap:paper_wsdm2022-sec:system-design-subsec:search-engine}), which enables workers to retrieve supporting evidence during the task. Finally, Section~\ref{cap:paper_wsdm2022-sec:system-design-subsec:logger} describes the \logger component, which can be used to capture workers' behavior throughout the task.

\subsection{General Architecture}

\label{cap:paper_wsdm2022-sec:system-design-subsec:general-arch}

\begin{figure}[tpb]
  \centering
  \includegraphics[width=0.8\linewidth]{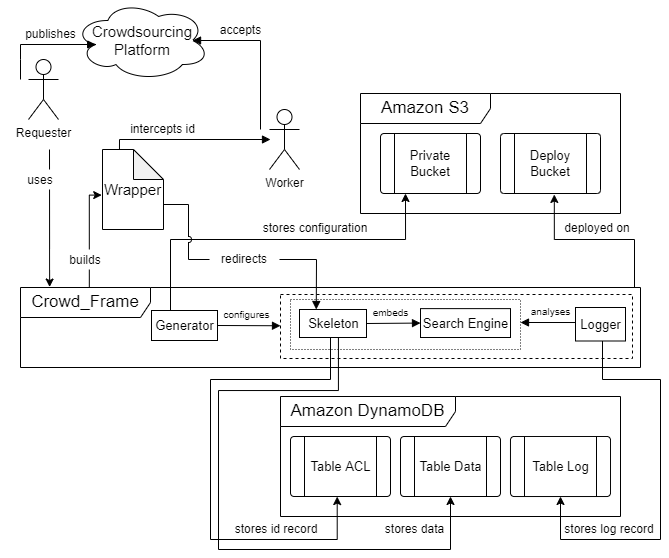}
  \caption{General architecture of \crowdframe.}
  \label{cap:paper_wsdm2022-sec:design-fig:architecture}
\end{figure}

\crowdframe is a client-side application developed using Angular,\footnote{\url{https://angular.io/}} an open-source framework for web development. Figure~\ref{cap:paper_wsdm2022-sec:design-fig:architecture} shows its architecture. The software is composed of four main components: \generator, \skeleton, \searchengine, and \logger. It relies on Amazon S3\footnote{\url{https://aws.amazon.com/it/s3/}}\index{Amazon!Web Services!S3} to store task configurations and source files. Amazon S3 is an object storage service designed to store and retrieve any type and amount of data. The software also uses Amazon DynamoDB,\footnote{\url{https://aws.amazon.com/it/dynamodb/}}\index{Amazon!Web Services!DynamoDB} a fully managed \index{NoSQL} NoSQL key-value and document database service. It is serverless and scales automatically based on workload.

A requester uses the \generator to instantiate and configure tasks through a simple user interface. The resulting configuration is uploaded to a private S3 bucket (i.e., a storage resource). The requester can then publish a set of \index{HIT}HITs on the selected crowdsourcing platform. A wrapper acts as a bridge between the platform and the workers. It redirects each recruited worker to the application deployed on a public S3 bucket. The worker interacts with the \skeleton component to perform the task. The \skeleton may embed a custom and configurable \searchengine. The \logger monitors and records the worker's behavior during the task. All data produced are stored in a \index{Amazon!Web Services!DynamoDB} DynamoDB table. Each component may also rely on external services to operate.

\subsection{Generator}

\label{cap:paper_wsdm2022-sec:system-design-subsec:generator}

The \generator component allows requesters to design and customize the configuration of a crowdsourcing task, as shown in Figure~\ref{cap:paper_wsdm2022-sec:design-fig:architecture}. 

\subsubsection{Use Cases}

\begin{figure}[tpb]
  \centering
    \includegraphics[width=\linewidth]{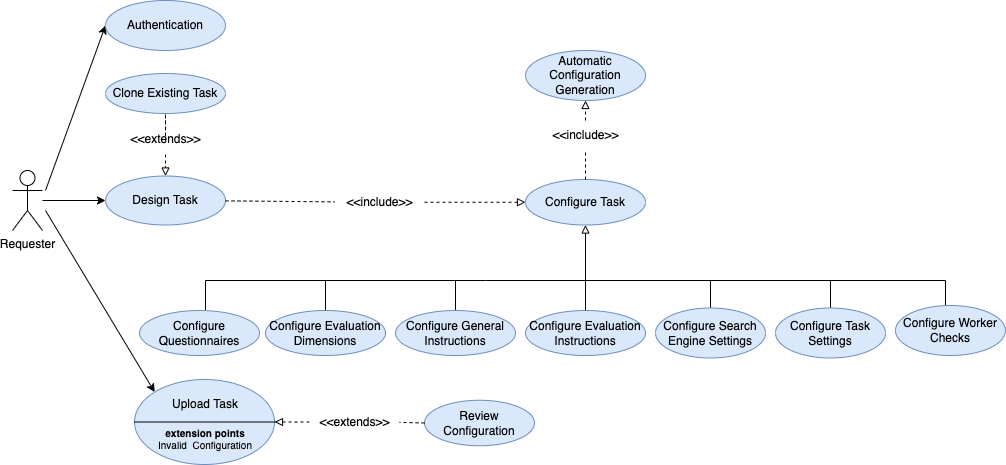}
\caption{Use case diagram for a requester designing and deploying a task using \crowdframe.}
\label{cap:paper_wsdm2022-sec:system-design-subsec:generator-subsec:use-case-fig:requester}
\end{figure}

The diagram in Figure~\ref{cap:paper_wsdm2022-sec:system-design-subsec:generator-subsec:use-case-fig:requester} provides a high-level overview of the interaction between a requester and the \generator component of \crowdframe. The requester authenticates to gain access to the component and proceeds to design or customize the crowdsourcing task. The component also allows cloning the configuration of a previously deployed task. The configuration process consists of seven steps, described in detail in Section~\ref{cap:paper_wsdm2022-sec:system-design-sec:generator-sec:architecture}. The system automatically generates and updates the task configuration at each step. Once satisfied, the requester finalizes the upload. If the configuration contains any errors, they must be resolved before uploading is permitted.

\subsubsection{Architecture}

\label{cap:paper_wsdm2022-sec:system-design-sec:generator-sec:architecture}

The \generator component is implemented as a form composed of eight distinct steps. In each step, the requester configures a specific aspect of the task. Table~\ref{cap:paper_wsdm2022-sec:system-design-sec:generator-tab:steps} provides a brief description of the purpose of each configuration step. T

he first step concerns questionnaires. It enables the creation of one or more questionnaires to be completed by workers either before or after the task, depending on the requester’s needs. Three types of questionnaires are available. The \texttt{Standard} \index{Standard} option supports basic questionnaires in which workers respond to questions using predefined choices, free-text answers, or numerical values. The \texttt{CRT} \index{Cognitive!Reflection Test} option allows the inclusion of Cognitive Reflection Tests~\cite{Frederick2005}, commonly used to assess workers' cognitive abilities. The \texttt{Likert} \index{Likert} option supports questionnaires where workers respond using a Likert scale \cite{likert}. The \generator can be easily extended to support additional questionnaire types.

The second step concerns the evaluation dimensions. It allows the requester to configure the aspects that workers must evaluate for each element of the assigned \index{HIT}HIT. A dimension may require expressing a judgment using some type of scale. \crowdframe supports all major rating scale types~\cite{1996-97594-000}, including categorical, interval, and ratio scales such as magnitude estimation~\cite{moskowitz1977magnitude}. Additional features can be enabled within each evaluation dimension. For example, the requester may require the worker to provide a textual justification or to search for a URL using the integrated custom search engine.

The third step concerns task instructions. The requester writes the instructions using a rich text editor. These instructions are shown to the worker before starting the task. Similarly, the fourth step concerns evaluation instructions, which are shown during task execution while the worker evaluates each element of the \index{HIT}HIT. These instructions explain how to assess each element along the configured dimensions.

The fifth step concerns the \searchengine. It allows the requester to select the preferred search provider and to define a list of domains to be filtered from the search results.

The sixth step concerns general task settings. The requester can set the maximum number of attempts allowed for each worker. This enables workers to revise their responses if they fail the quality checks at the end of the task and helps reduce task abandonment~\cite{han2019all, 8873609}. The requester can also configure the minimum time that each worker must spend on each \index{HIT}HIT element. This check is useful to prevent automated bots from completing the task.

Additionally, the requester can enable a countdown timer to limit the total time available for completing the assessment. \citet{DBLP:conf/hcomp/MaddalenaBNDMD16} show that such a constraint can improve worker quality and help optimize the cost of a crowdsourcing task. The requester can also enable an annotation interface, allowing workers to annotate and label texts, such as in the annotation of social media conversations~\cite{10.1145/2740908.2743052}. Previous batches of workers can be blocked from participating in the current task, which is particularly useful in longitudinal studies where the same task is repeated with new participants only~\cite{roitero2021crowd}. Finally, the requester must upload the \index{HIT}HITs for the task.

The seventh step involves additional checks on workers. The requester can manually specify a list of worker identifiers to be allowed or blocked for the current task. This provides fine-grained control over worker selection without blocking an entire batch. 

The eighth and final step displays a summary of the generated configuration. The requester can upload the configuration to the private S3 bucket once satisfied.

\begin{table}[tpb]
\centering
\caption{Configuration steps available in the \generator component of \crowdframe.}
\begin{tabular}{C{1.2cm}lp{7.7cm}}
  \toprule
   \textbf{Step \#} & \textbf{Name} & \textbf{Description}  \\
    \midrule
	1 & Questionnaires & Allows creating one or more questionnaires that workers will fill out before or after task execution. \\
	\midrule
	2 & Evaluation Dimensions & Allows configuring what the worker will assess for each element of the \index{HIT}HIT assigned. \\ 
	\midrule
	3 & Task Instructions & Allows configuring the instructions shown to each worker before starting the task. \\ 
	\midrule
	4 & Evaluation Instructions & Allows configuring the instructions shown to each worker while assessing each element of the \index{HIT}HIT assigned. \\
	\midrule
	5 & Search Engine & Allows configuring the desired search provider. It is also possible to specify a list of domains to filter from the search results. \\
	\midrule
	6 & Task Settings & Allows configuring general task settings, such as the maximum number of attempts for each worker, the use of an annotation interface, and more. \\
	\midrule
	7 & Worker Checks & Allows configuring additional filters and checks on the workers recruited. \\
	\midrule
	8 & Summary & Allows reviewing and uploading the final configuration. \\
	\bottomrule
\end{tabular}
\label{cap:paper_wsdm2022-sec:system-design-sec:generator-tab:steps}
\end{table}

\subsubsection{Case Study}

\label{cap:paper_wsdm2022-sec:case-study}

\citet{roitero2021crowd} use \crowdframe to deploy a misinformation assessment task aimed at evaluating whether crowd workers can identify and accurately classify online misinformation related to the COVID-19 pandemic, as described in Chapter~\ref{cap:paper_pauc2021}. Their task includes two questionnaires: one to collect background information and one to estimate workers' cognitive abilities. Each worker is shown a set of statements and asked to provide truthfulness assessments using a six-level scale. For each assessment, the worker must supply a supporting URL via a custom search engine and write a textual justification.

\begin{figure}[tbp]
  \centering
    \includegraphics[width=\linewidth]{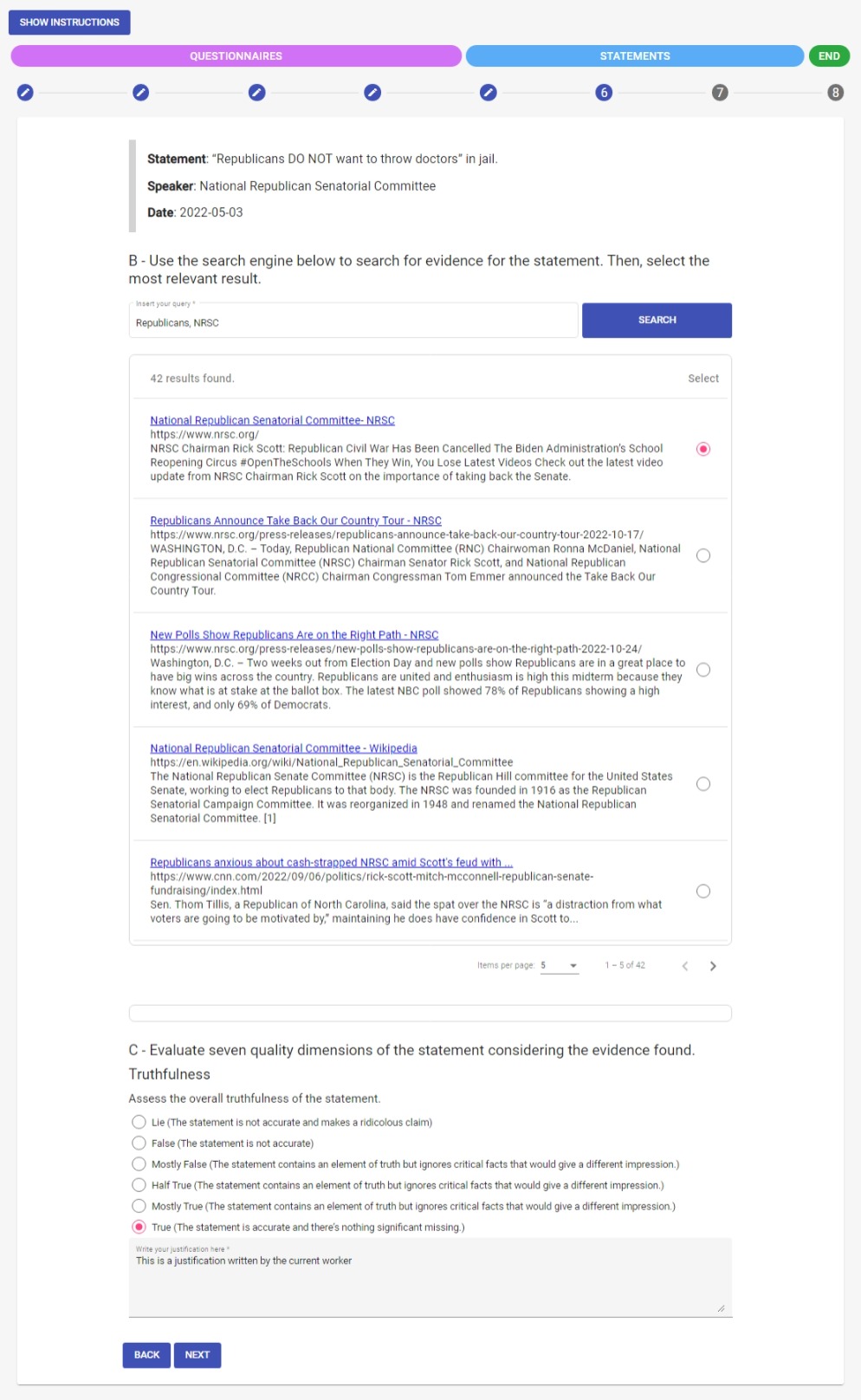}
\caption{Interface shown to workers in the misinformation assessment task deployed by \citet{roitero2021crowd} using \crowdframe.}
\label{cap:paper_wsdm2022-fig:screen}
\end{figure}

Figure~\ref{cap:paper_wsdm2022-fig:screen} shows the resulting interface. A requester can easily replicate the setup of \citet{roitero2021crowd} using \crowdframe. The configuration includes one standard questionnaire and three \index{Cognitive!Reflection Test}CRT questionnaires. A single evaluation dimension is defined, namely overall truthfulness, measured using a six-level categorical scale. Quality control is enforced on the ratings: workers must assign a higher truthfulness score to clearly true statements than to clearly false ones. Moreover, search results from three fact-checking websites are excluded. A sample configuration replicating this task is available in the \crowdframe repository.\footnote{\url{https://github.com/Miccighel/Crowd_Frame/tree/master/examples/misinformation_assessment}}

\subsection{Skeleton}

\label{cap:paper_wsdm2022-sec:system-design-subsec:skeleton}

The \skeleton component enables workers to perform the task after being recruited, as shown in Figure~\ref{cap:paper_wsdm2022-sec:design-fig:architecture}.

\subsubsection{Use Cases}

\begin{figure}[tpb]
  \centering
    \includegraphics[width=.9\linewidth]{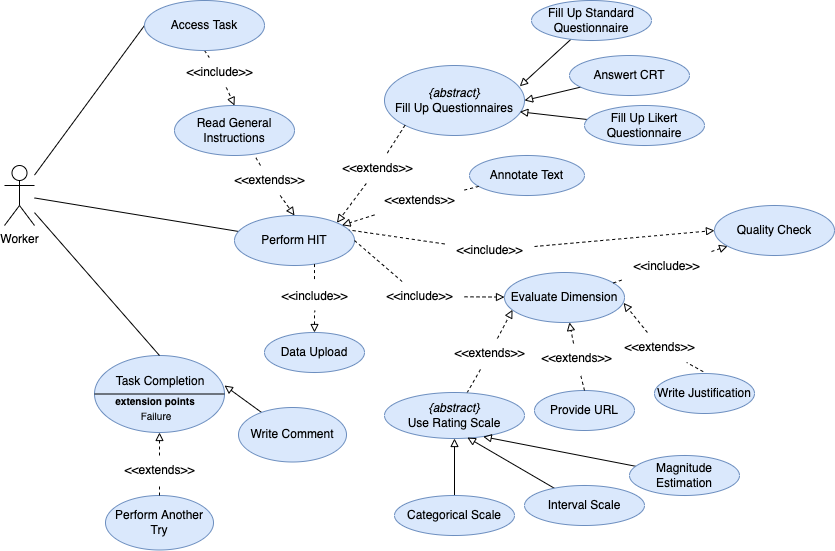}
\caption{Use case diagram of a worker performing a task deployed using \crowdframe.}
\label{cap:paper_wsdm2022-sec:system-design-subsec:generator-subsec:use-case-fig:worker}
\end{figure}

The diagram in Figure~\ref{cap:paper_wsdm2022-sec:system-design-subsec:generator-subsec:use-case-fig:worker} provides a high-level overview of the interaction between a worker and the \skeleton component of \crowdframe. After being recruited, the worker accesses the deployed task. They are initially required to read the general task instructions before starting work on the assigned \index{HIT}HIT. Once this step is completed, they can begin.

The task may involve completing one or more questionnaires, either at the beginning or the end. Three types of questionnaires can be configured, as described in Section~\ref{cap:paper_wsdm2022-sec:system-design-sec:generator-sec:architecture}. Workers are required to evaluate every dimension for each element of the \index{HIT}HIT. A dimension may require providing a rating, submitting a URL found using a search engine, or writing a justification. The types of rating scales used are implemented as detailed in Section~\ref{cap:paper_wsdm2022-sec:system-design-sec:generator-sec:architecture}.

Data is uploaded continuously throughout the task. Quality checks are applied either to user behavior during task execution or to the values submitted for each dimension. If these checks are not passed, the task may terminate without completion. In such cases, the worker may be allowed to retry. If the task concludes successfully, the worker has the option to leave a final comment for the requester.

\subsubsection{Architecture}

\begin{figure}[tpb]
  \centering
    \includegraphics[width=.9\linewidth]{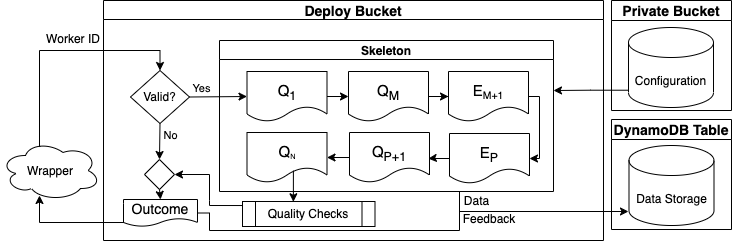}
\caption{Execution flow of the \skeleton component in \crowdframe.}
\label{cap:paper_wsdm2022-sec:skeleton-fig:skeleton}
\end{figure}

Figure~\ref{cap:paper_wsdm2022-sec:skeleton-fig:skeleton} illustrates the interaction between a crowd worker and the \skeleton component. The worker accesses the application deployed on the public bucket through the wrapper, with the identifier passed as a URL parameter. Upon reaching the current deployment, the identifier is stored in an access control list (i.e., the DynamoDB table shown in Figure~\ref{cap:paper_wsdm2022-sec:design-fig:architecture}). This enables the system to associate the worker with the data generated during task execution. Additionally, it allows tracking how much time the worker has left to complete the assigned task. If needed, the requester can restrict workers from accessing the deployed task multiple times.

The worker is initially shown the general instructions. The task is unlocked after an initial check on the identifier. The \skeleton component fetches the task configuration from the private bucket and instantiates the required layout. The worker then proceeds to perform the task configured by the requester.

The \skeleton creates a page for each of the $[0, M]$ questionnaires presented at the start of the task. It then initializes one page for each of the $[M+1, P]$ elements to be evaluated. Finally, it sets up a page for each of the $[P+1, N]$ questionnaires shown at the end of the task. The appearance of each page depends on the questionnaire type and the parameters of the corresponding dimension. Evaluation instructions are displayed on every page.

The worker can navigate freely between elements and complete the required tasks. Quality checks are triggered after the final element or questionnaire. If these checks are passed, an outcome page is shown to the worker, where they may also provide a final comment. Afterward, the worker is redirected to the crowdsourcing platform to receive their reward.

Throughout the entire execution flow, the \skeleton stores all generated data in the \index{Amazon!Web Services!DynamoDB}DynamoDB table.

\subsubsection{Wrapper}

Figure~\ref{cap:paper_wsdm2022-sec:design-fig:architecture} shows the presence of a wrapper between the workers recruited on a crowdsourcing platform and a task deployed using \crowdframe. This mechanism is used to identify each worker and assign a corresponding \index{HIT}HIT. It relies on the platform’s capability to accept input variables. The requester provides the crowdsourcing platform with a file containing input and output tokens. The input token is an alphanumeric string used to associate a worker with a specific \index{HIT}HIT, while the output token is used by the worker to confirm successful task completion.

\begin{figure}[tpb]
  \centering
    \includegraphics[width=.8\linewidth]{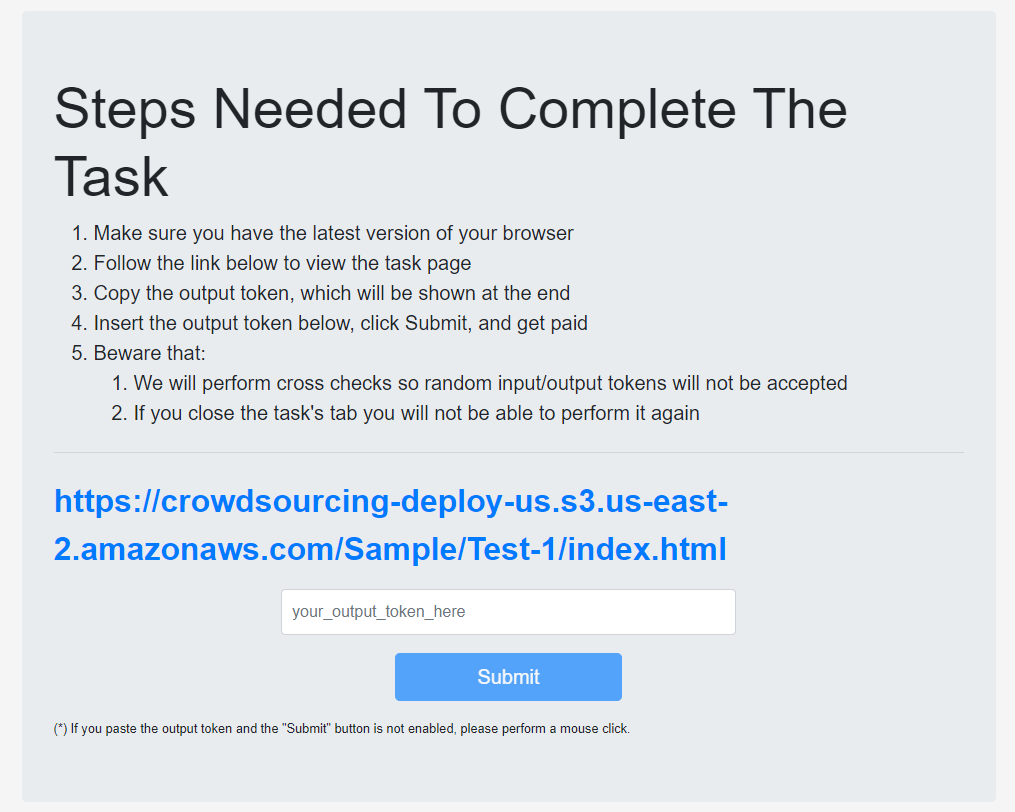}
\caption{Wrapper interface for a task configured in \crowdframe and deployed via \mturk for recruiting workers.}
  \label{cap:paper_wsdm2022-sec:wrapper-fig:interface}
\end{figure}

The wrapper fetches the platform-specific identifier of the worker. Figure~\ref{cap:paper_wsdm2022-sec:wrapper-fig:interface} shows the wrapper interface used to redirect a worker recruited on \mturk. The interface for a task deployed on \toloka is similar to the one shown in Figure~\ref{cap:paper_wsdm2022-sec:wrapper-fig:interface}. On the other hand, \prolific does not require a wrapper at all, as the requester must provide a direct URL to the externally deployed task.

The worker recruited on the crowdsourcing platform clicks the link shown by the wrapper and is redirected to the deployment on the public S3 bucket. They are then automatically and implicitly matched with an available input token that has not yet been allocated. \crowdframe implements a custom allocation scheme that ensures consistent assignment of \index{HIT}HITs.

Upon task completion, the worker is shown an output token. Depending on the platform, this token must be copied back into the wrapper interface. If the output token provided matches the originally assigned input token, the worker can be paid. Worker identifiers and input/output tokens are essential to ensure correct matching between workers and \index{HIT}HITs.

\subsubsection{Data Format}

\label{cap:paper_wsdm2022-sec:skeleton-subsec:data-format}

The data produced throughout the task is stored in \index{Amazon!Web Services!DynamoDB}DynamoDB tables, as shown in Figure~\ref{cap:paper_wsdm2022-sec:design-fig:architecture}. DynamoDB is a NoSQL \index{NoSQL} database service, which means that records inserted into tables can contain structured data. The \skeleton component uses two different tables. One of these stores the access control list for the deployed task. More specifically, the access control list includes a single record for each worker. The attributes in this table vary depending on the platform from which the worker originates.

Listing~\ref{cap:paper_wsdm2022-sec:system-design-subsec:skeleton-subsec:data-format-list:acl-record} shows a CSV \index{CSV} dump of a single record from the access control list table. The attributes include timestamps that indicate when the worker participated in, completed, or abandoned the task. The record also contains the identifiers and tokens associated with the assigned \index{HIT}HIT. Additional attributes capture whether the worker is eligible for payment, how many attempts they made, and other relevant metadata.

\begin{lstlisting}[style=csvstyle, caption={Format of the DynamoDB table storing the access control list for a task deployed using \crowdframe.}, label={cap:paper_wsdm2022-sec:system-design-subsec:skeleton-subsec:data-format-list:acl-record}]
"identifier","access_counter","batch_name","folder","generated","in_progress","ip_address","ip_source","paid","platform","status_code","task_name","time_arrival","time_completion","time_expiration","time_expiration_nearest","time_expired","time_removal","token_input","token_output","try_current","try_left","unit_id","user_agent","user_agent_source"
"<<anonymized>>","1","Batch-9-Toloka","<<anonymized>>","true","false","<<anonymized>>","cf","true","toloka","200","<<anonymized>>","Fri, 17 Mar 2023 09:25:36 GMT","Fri, 17 Mar 2023 09:30:04 GMT","Fri, 17 Mar 2023 10:25:36 GMT","Fri, 17 Mar 2023 10:25:54 GMT","false",","PKPBSINUKHS","BHPCIBOKBDL","1","10","unit_44","Mozilla/5.0 (Linux; Android 11; TECNO KG7h) AppleWebKit/537.36 (KHTML, like Gecko) Chrome/108.0.0.0 Mobile Safari/537.36","cf"
\end{lstlisting}

The second table stores the data produced by the worker. It includes multiple records per worker, depending on how they perform the task. The structure of the table remains consistent across all entries.

The \spverb|sequence| attribute combines the worker and \index{HIT}HIT identifiers with the index of the element being evaluated and a sequence number. This combination uniquely identifies each piece of stored data. The \spverb|access| attribute indicates how many times the worker accessed a specific element of the \index{HIT}HIT. The \spverb|action| attribute specifies whether the worker is progressing through the assigned \index{HIT}HIT or revisiting a previously evaluated element.

The \spverb|element| attribute describes the type of data stored. The \spverb|index| attribute refers to the current element of the \index{HIT}HIT. The \spverb|sequence_number| attribute identifies the order of data entries for the current worker. The \spverb|time| attribute records the data upload timestamp. The \spverb|try| attribute indicates the attempt number made by the worker. Finally, the \spverb|data| attribute contains the actual data produced during task execution.
Listing~\ref{cap:paper_wsdm2022-sec:system-design-subsec:skeleton-subsec:data-format-list:data-record} shows the CSV \index{CSV} dump of a single record from the data table of a task.

\begin{lstlisting}[style=jsonstyle, caption={Format of the DynamoDB table storing the data produced by a worker during a task deployed using \crowdframe.}, label={cap:paper_wsdm2022-sec:system-design-subsec:skeleton-subsec:data-format-list:data-record}]
"identifier","sequence","access","action","data","element","index","sequence_number","time","try"
"<<anonymized>>","<<anonymized>>-unit_118-1-10","1","Next",{...},"document","6","10","Tue, 23 Aug 2022 11:49
\end{lstlisting}

Each piece of data is stored as a JSON \index{JSON} object. It may contain values for a set of evaluation dimensions, the outcome of quality checks, questionnaire responses, and other relevant information. Listing~\ref{cap:paper_wsdm2022-sec:system-design-subsec:skeleton-subsec:data-format-list:data-payload} shows a JSON payload stored in the DynamoDB table for a single element of a \index{HIT}HIT assigned to a worker. In this example, the worker evaluates nine dimensions, one of which requires submitting a URL. The payload therefore includes both the search queries issued and the corresponding responses retrieved.

\begin{lstlisting}[style=jsonstyle, caption={JSON data stored for an element of a \index{HIT}HIT evaluated by a worker during a task deployed using \crowdframe.}, label={cap:paper_wsdm2022-sec:system-design-subsec:skeleton-subsec:data-format-list:data-payload}]
{
  "info": {
    "action": "Next",
    "access": 1,
    "try": 1,
    "index": 6,
    "sequence": 10,
    "element": "document"
  },
  "answers": { ... },
  "notes": [ ... ],
  "dimensions_selected": {
    "data": [ ... ],
    "amount": 9
  },
  "queries": {
    "data": [ ... ],
    "amount": 1
  },
  "timestamps_start": [ 1661254995.22 ],
  "timestamps_end": [ 1661255107.96 ],
  "timestamps_elapsed": 112.74000000953674,
  "countdowns_times_start": [],
  "countdowns_times_left": [],
  "countdowns_expired": [],
  "accesses": 1,
  "responses_retrieved": {
    "data": [ ... ],
    "amount": 15,
    "groups": 1
  },
  "responses_selected": {
    "data": [ ... ],
    "amount": 1
  }
}
\end{lstlisting}

\subsubsection{Cost Estimation}

\label{cap:paper_wsdm2022-sec:skeleton-subsec:cost-estimation}

The architecture implemented using Amazon Web Services (AWS) follows a pay-per-use model, with costs varying depending on request size, frequency, and other factors. The AWS Pricing Calculator\footnote{\url{https://calculator.aws/#/}} enables reasonably accurate cost predictions. 
All estimates reported below refer to the AWS Region \textsf{US-East-2 (Ohio)}.
To estimate the impact of the \skeleton component on the overall cost, the most recent variant of the task deployed by \citet{SOPRANO2021102710} (described in Chapter~\ref{cap:paper_ipm2021}) and \citet{draws2022bias} (described in Chapter~\ref{cap:paper_facct2022}) is considered. This variant involves 200 workers recruited from \prolific in a single batch.

The first service considered is \index{Amazon!Web Services!S3} \textsf{S3}. The pricing\footnote{\url{https://aws.amazon.com/s3/pricing/}} depends on several cost components, all determined by the selected storage class. \crowdframe adopts the \spverb|Standard| \index{Standard} class, which is generally recommended for typical storage needs without specific performance or access constraints.

A storage fee of \$0.023~USD per GB is applied for the first 50~TB per month across the region. In addition, a fee of \$0.005~USD is charged for every 1,000 HTTP requests of type PUT, COPY, POST, and LIST \index{PUT} \index{COPY} \index{POST} \index{LIST}, and \$0.004~USD for all other request types. Bandwidth usage also incurs charges: the first 100~GB of data transferred out of S3 to the internet is free per AWS account; beyond that, a fee of \$0.09~USD per GB applies for the first 10~TB per month. Data transfer into S3 from the internet is always free. Additional features not used by \crowdframe may add further costs, but are not considered in this estimation.

The amount of data stored in the S3 bucket corresponds to the configuration and source code files required for each batch of the task. Each time a worker accesses the task, a total of \(7 + 3 = 10\) files are transferred. The number of accesses per worker is recorded in the access control list. The configuration file has a size of approximately 0.98~MB, and the source code is approximately 5.38~MB. The total storage requirement is therefore \((5.38 + 0.98) = 0.00636\)~GB per month.

With 214 recorded task accesses, the system generates a total of \(214 \times (7 + 3) = 2{,}140\) HTTP requests. The volume of data transferred amounts to \(214 \times 0.00636 = 1.36\)~GB per month.
Equation~\ref{cap:paper_wsdm2022-sec:system-design-subsec:skeleton-subsec:cost-estimation-eq:s3-storage-cost} shows the detailed computation of the storage cost component.  
Equation~\ref{cap:paper_wsdm2022-sec:system-design-subsec:skeleton-subsec:cost-estimation-eq:s3-data-retrieval-cost} details the cost of the data retrieval component in terms of HTTP requests.  
Equation~\ref{cap:paper_wsdm2022-sec:system-design-subsec:skeleton-subsec:cost-estimation-eq:s3-data-transfer-cost} presents the data transfer cost component.  
Finally, Equation~\ref{cap:paper_wsdm2022-sec:system-design-subsec:skeleton-subsec:cost-estimation-eq:s3-cost} reports the total estimated cost.

\begin{equation}
\begin{split}
\textrm{Data Storage} & = \textrm{\$0.023 * Data Size (TB/Month)} \\
              & = \$0.023 * (0.00636/1024) \\
              & = \$0.023 * 0.00000621 = \$0.00000014\quad(\$0 \iff \textrm{free tier})\\
              & \quad\>\textrm{storage class: standard} \\  
              & \quad\>\textrm{threshold: 50 TB/Month}  \\
              & \quad\>\textrm{free tier: 5 GB/Month } 
\end{split}
\label{cap:paper_wsdm2022-sec:system-design-subsec:skeleton-subsec:cost-estimation-eq:s3-storage-cost}
\end{equation}
\myequations{Sample estimation of the storage cost for the use of S3 by the \skeleton component of \crowdframe.}

\begin{equation}
\begin{split}
\textrm{Data Retrieval} & = \$0.0000004  * \textrm{Requests Amount} \\
                & = \$0.0000004 * (\textrm{Worker Accesses} * (\text{Files Retrieved})) \\
                & = \$0.0000004 * (214 * (7+3)) \\
                & = \$0.0000004 * 2140 = \$0.000856\quad(\$0 \iff \textrm{free tier})\\
                & \quad\>\textrm{storage class: standard} \\  
                & \quad\>\textrm{threshold: 1000 requests}  \\
                & \quad\>\textrm{free tier: 2000 requests} 
\end{split}
\label{cap:paper_wsdm2022-sec:system-design-subsec:skeleton-subsec:cost-estimation-eq:s3-data-retrieval-cost}
\end{equation}
\myequations{Sample estimation of the data retrieval cost for the use of S3 by the \skeleton component of \crowdframe.}

\begin{equation}
\begin{split}
\textrm{Data Transfer} & = \$0.09 * \textrm{Data Transfer Size} \\
              & = \$0.09 * \lceil\textrm{Worker Accesses} * \textrm{Data Size (GB/Month)}\rceil \\
              & = \$0.09 * \lceil214 * 0.00636\rceil \\
              & =  \$0.09 * 1 = \$0.09\quad(\$0 \iff \textrm{free tier})\\
              & \quad\>\textrm{storage class: standard} \\  
              & \quad\>\textrm{threshold: 10 TB/Month}  \\
              & \quad\>\textrm{free tier: 100 GB/Month } 
\end{split}
\label{cap:paper_wsdm2022-sec:system-design-subsec:skeleton-subsec:cost-estimation-eq:s3-data-transfer-cost}
\end{equation}
\myequations{Sample estimation of the data transfer cost for the use of S3 by the \skeleton component of \crowdframe.}

\begin{equation}
\begin{split}
\textrm{S3} & = \textrm{Data Storage} + \textrm{Data Retrieval} + \textrm{Data Transfer}\\
              & = \$0.00000014 + \$0.000856 + \$0.09 = \$0.09085614
\end{split}
\label{cap:paper_wsdm2022-sec:system-design-subsec:skeleton-subsec:cost-estimation-eq:s3-cost}
\end{equation}
\myequations{Cost estimation for the use of Amazon S3 by the \skeleton component.}

The second service considered is \index{Amazon!Web Services!DynamoDB} \textsf{DynamoDB}. The service offers two capacity modes,\footnote{\url{https://aws.amazon.com/dynamodb/pricing/}} each with distinct billing options for processing read and write requests on tables. The \emph{on-demand capacity mode} charges based on the actual number of read and write operations performed by the application. In contrast, the \emph{provisioned capacity mode} charges based on the number of reads and writes that the application is expected to require in advance. \crowdframe adopts the on-demand mode, as the volume of read and write requests depends on the number of workers recruited.

A write request unit (WRU) \index{WRU} is the billing unit of a set of API calls used to write data to tables. A standard write request unit can store items up to 1~KB. Additional write request units are used if the item is larger than 1~KB. A transactional write requires two units.

A read request unit (RRU) \index{RRU} is the billing unit of API calls used to read data from tables. A strongly consistent read request of up to 4~KB requires one read request unit. Additional read request units are used if the item is larger than 4~KB. An eventually consistent read request requires one-half request unit, while a transactional read requires four read request units. The type of read request chosen has an impact on consistency.\footnote{\url{https://docs.aws.amazon.com/amazondynamodb/latest/developerguide/HowItWorks.ReadConsistency.html}} DynamoDB uses eventually consistent reads unless specified otherwise. \crowdframe sends read requests using the default parameters, thus relies on this type of read. The response might not reflect the result of a recently completed write operation when the read request is eventually consistent. In such a case, a second read request is needed to retrieve the most up-to-date data. On the other hand, the use of strongly consistent reads has some constraints and leads to higher throughput usage, thus being more expensive.

The fee applied for each write or read request unit depends on the storage class chosen for the tables when the on-demand capacity mode is used. DynamoDB offers a Standard table class and a Standard-infrequent access class. \crowdframe relies on the former class. The on-demand capacity mode charges \$1.25 for one million write request units and \$0.25 for one million read request units.

Data storage is another table-class-dependent cost component that must be considered. The first 25~GB/Month are free when using the Standard class, and a fee of \$0.25 is applied per GB/Month thereafter. There is no free quota when using the Standard-infrequent table class. The fee is thus \$0.10 per GB/Month. The service applies additional fees for the use of other features not required by \crowdframe.

The variant of the task deployed by \citet{SOPRANO2021102710} and \citet{draws2022bias} considered uses two of the three tables shown in Figure~\ref{cap:paper_wsdm2022-sec:design-fig:architecture}. The component uses only the access control list table and the data table. The log table is used by the \logger component. Its cost estimation process is described in Section~\ref{cap:paper_wsdm2022-sec:system-design-subsec:logger-subsubsec:cost-estimation}.
The data table has records of varying sizes. This depends on the type of data stored, as explained in Section~\ref{cap:paper_wsdm2022-sec:skeleton-subsec:data-format}. In the variant of the task considered, there are at minimum 13 records for each try of each worker, depending on how they behave during the task. These records are broken down into (at least) 8 document records, 3 questionnaire records, a single record containing the task setup and worker attributes, and a single record addressing quality checks. An additional record may appear if the worker provides a final comment to the requester.

Table~\ref{cap:paper_wsdm2022-sec:skeleton-subsec:cost-estimation-tab:skeleton-example} shows the parameters needed to estimate the cost for a single worker of the task considered. The data table for the whole 200 workers recruited for the task contains a total of \num{2,183} document records, 787 questionnaire records, 262 general data records, 256 quality checks records, and 41 comment records.

{
\setlength{\tabcolsep}{5pt}
\begin{table}[tpb]
\centering
\small
\caption{Sample cost estimation parameters for the use of the data table by the \skeleton component of \crowdframe.}
\begin{tabular}{l
                S[table-format=2.0]
                S[table-format=3.2]
                S[table-format=3.0]
                S[table-format=2.1]}
  \toprule
  \textbf{Record Type} & 
  \textbf{Average Amount} & 
  \textbf{Average Size (KB)} & 
  \textbf{Average WRUs} & 
  \textbf{Average RRUs} \\
  \midrule
  \spverb|document|      & 11 & 182.85 & 183 & 23   \\
  \spverb|questionnaire| &  4 &  14.48 &  15 &  2   \\
  \spverb|data|          &  1 &  38.86 &  39 &  5   \\
  \spverb|checks|        &  1 &   1.27 &   1 & 0.5  \\
  \spverb|comment|       &  1 &   0.04 &   1 & 0.5  \\
  \midrule
  Total                  & 18 & 237.50 & 239 & 31   \\
  \bottomrule
\end{tabular}
\label{cap:paper_wsdm2022-sec:skeleton-subsec:cost-estimation-tab:skeleton-example}
\end{table}
}

Equation~\ref{cap:paper_wsdm2022-sec:system-design-subsec:skeleton-subsec:cost-estimation-eq:sample-table-data-wru} shows the detailed computation of the write request units cost component of the data table. Equation~\ref{cap:paper_wsdm2022-sec:system-design-subsec:skeleton-subsec:cost-estimation-eq:sample-table-data-rru} shows the detailed computation of the read request units cost component of the data table. Equation~\ref{cap:paper_wsdm2022-sec:system-design-subsec:skeleton-subsec:cost-estimation-eq:sample-table-data-data-storage} shows the computation of the storage cost of the data written. Equation~\ref{cap:paper_wsdm2022-sec:system-design-subsec:skeleton-subsec:cost-estimation-eq:sample-table-data} summarizes the contribution of each cost component for the use of the data table by \crowdframe for the sample task considered.

\begin{equation}
\begin{split}
\textrm{WRUs} & = \$1.25 * \textrm{(Workers Amount} * \textrm{Avg. Message Number}) * \\
              & \quad\> \lceil\textrm{Avg. Payload Size (KB) / Unit Amount}\rceil\\
              & = \$0.00000125 * (200 * 18) * \\
              & \quad\> \lceil(182.85 + 14.48 + 38.86 + 1.27 + 0.04)/1\rceil \\
              & = \$0.00000125 * 3600 * (238\>\textrm{Avg. WRUs / Record})\\
              & = \$0.00000125 * 856800\>\textrm{Billable WRUs}  = \$1.07\\
              & \quad\>\>\textrm{threshold: 1 million WRUs}
\end{split}
\label{cap:paper_wsdm2022-sec:system-design-subsec:skeleton-subsec:cost-estimation-eq:sample-table-data-wru}
\end{equation}
\myequations{Sample estimation of the WRUs cost for the use of the data table by the \skeleton component.}

\begin{equation}
\begin{split}
\textrm{RRUs} & = \$0.25 * (0.5 * (\textrm{Workers Amount} * \textrm{Avg. Message Number}) * \\
              & \quad\> \lceil\textrm{Avg. Payload Size (KB) / Unit Amount}\rceil)\\
              & = \$0.00000025 * (0.5* (200 * 18) * \\
              & \quad\> \lceil182.85 + 14.48 + 38.86 + 1.27 + 0.04)/4\rceil) \\
              & = \$0.00000025 * (0.5 * (3600 * (60\>\textrm{Avg. RRUs / Record}))\\
              & = \$0.00000025 * 108000\>\textrm{Billable RRUs}  = \$0.03\\
              & \quad\>\>\textrm{threshold: 1 million RRUs}
\end{split}
\label{cap:paper_wsdm2022-sec:system-design-subsec:skeleton-subsec:cost-estimation-eq:sample-table-data-rru}
\end{equation}
\myequations{Sample estimation of the RRUs cost for the use of the data table by the \skeleton component.}

\begin{equation}
\begin{split}
\textrm{Data Storage} & = \$0.25 * \textrm{(Message Number * Avg. Payload Size (GB))} \\
              & = \$0.25 * (200 * ((182.85 + 14.48 + 38.86 + 1.27 + 0.04)/1024/1024))) \\
              & = \$0.25 * (200 * ((239/1024)/1024)) \\
              & = \$0.25 * 0.00022793\>\textrm{GB/Month} = \$0.00005698\quad(\$0 \iff \textrm{free tier})\\
              & \quad\>\>\textrm{threshold: 1 GB/Month} \\ 
              & \quad\>\>\textrm{note: on-demand capacity mode, standard table class} \\
              & \quad\>\>\textrm{free tier: 25 GB/Month } 
\end{split}
\label{cap:paper_wsdm2022-sec:system-design-subsec:skeleton-subsec:cost-estimation-eq:sample-table-data-data-storage}
\end{equation}
\myequations{Sample estimation of the data storage cost for the use of the data table by the \skeleton component of \crowdframe.}

\begin{equation}
\begin{split}
\textrm{Data Table} & = \textrm{WCUs} + \textrm{RCUs} + \textrm{Data Storage}\\
              & = \$1.07 + \$0.03 + \$0.00005698 = \$1.10005698
\end{split}
\label{cap:paper_wsdm2022-sec:system-design-subsec:skeleton-subsec:cost-estimation-eq:sample-table-data}
\end{equation}
\myequations{Sample overall estimation of the cost for the use of the data table by the \skeleton component of \crowdframe.}

The access control list table has an average record size of roughly 800~B. Such an item consumes 0.5~RRUs when read with eventual consistency and 1~WRU when written. A total of 200 workers are recruited, and each worker has a single record. The total size of the table is roughly 160~KB, thus 160~WRUs are used.

There may be a certain number of additional RRUs performed on the table. These RRUs occur when a worker does not complete the assigned task on time and the unit must be reallocated. In such a case, the table must be scanned to determine which unit to reallocate and to update the affected workers' records. Such a scenario does not occur when considering the aforementioned variant. Otherwise, the RRUs computation depends on how many records were already present in the access control list when each additional worker was recruited. 

Generally, the overall impact of the access control list table is negligible due to the small number of records it stores, the limited number of request units consumed, and its minimal storage size.

Equation~\ref{cap:paper_wsdm2022-sec:system-design-subsec:skeleton-subsec:cost-estimation-eq:sample-table-acl-wru} shows the detailed computation of the write request units cost component of the access control list table. Equation~\ref{cap:paper_wsdm2022-sec:system-design-subsec:skeleton-subsec:cost-estimation-eq:sample-table-acl-rru} shows the detailed computation of the read request units cost component of the access control list table. Equation~\ref{cap:paper_wsdm2022-sec:system-design-subsec:skeleton-subsec:cost-estimation-eq:sample-table-acl-data-storage} shows the computation of the storage cost of the records written. Equation~\ref{cap:paper_wsdm2022-sec:system-design-subsec:skeleton-subsec:cost-estimation-eq:sample-table-acl} further summarizes the contribution of each cost component for the use of the access control list table by \crowdframe for the sample task considered.

\begin{equation}
\begin{split}
\textrm{WRUs} & = \$1.25 * \textrm{(Workers Amount} * \textrm{Avg. Message Number}) * \\
              & \quad\> \lceil\textrm{Avg. Payload Size (KB) / Unit Amount}\rceil\\
              & = \$0.00000125 * (200 * 1) * \\
              & \quad\> \lceil(800/1024)/1\rceil \\
              & = \$0.00000125 * 3600 * (1\>\textrm{Avg. WRUs / Record})\\
              & = \$0.00000125 * 3600\>\textrm{Billable WRUs}  = \$0.045\\
              & \quad\>\>\textrm{threshold: 1 million WRUs}
\end{split}
\label{cap:paper_wsdm2022-sec:system-design-subsec:skeleton-subsec:cost-estimation-eq:sample-table-acl-wru}
\end{equation}
\myequations{Sample estimation of the WRUs cost for the use of the access control list table by the \skeleton component.}

\begin{equation}
\begin{split}
\textrm{RRUs} & = \$0.25 * (0.5 * (\textrm{Workers Amount} * \textrm{Avg. Message Number}) * \\
              & \quad\> \lceil\textrm{Avg. Payload Size (KB) / Unit Amount}\rceil)\\
              & = \$0.00000025 * (0.5* (200 * 1) * \\
              & \quad\> \lceil(800/1024)/4\rceil) \\
              & = \$0.00000025 * (0.5 * (3600 * (1 \>\textrm{Avg. RRUs / Record}))\\
              & = \$0.00000025 * 1800\>\textrm{Billable RRUs}  = \$0.00045\\
              & \quad\>\>\textrm{threshold: 1 million RRUs}
\end{split}
\label{cap:paper_wsdm2022-sec:system-design-subsec:skeleton-subsec:cost-estimation-eq:sample-table-acl-rru}
\end{equation}
\myequations{Sample estimation of the RRUs cost for the use of the access control list table by the \skeleton component.}

\begin{equation}
\begin{split}
\textrm{Data Storage} & = \$0.25 * \textrm{(Message Number * Avg. Payload Size (GB))} \\
              & = \$0.25 * (200 * (800/1024/1024/1024))) \\
              & = \$0.25 * (200 * 0.00000075) \\
              & = \$0.25 * 0.00014901\>\textrm{GB/Month} = \$0.00003725\quad(\$0 \iff \textrm{free tier})\\
              & \quad\>\>\textrm{threshold: 1 GB/Month} \\ 
              & \quad\>\>\textrm{note: on-demand capacity mode, standard table class} \\
              & \quad\>\>\textrm{free tier: 25 GB/Month } 
\end{split}
\label{cap:paper_wsdm2022-sec:system-design-subsec:skeleton-subsec:cost-estimation-eq:sample-table-acl-data-storage}
\end{equation}
\myequations{Sample estimation of the data storage cost for the use of the access control list table by the \skeleton component of \crowdframe.}

\begin{equation}
\begin{split}
\textrm{Table ACL} & = \textrm{WCUs} + \textrm{RCUs} + \textrm{Data Storage}\\
              & = \$0.045 + \$0.00045 + \$0.00003725 = \$0.04548725
\end{split}
\label{cap:paper_wsdm2022-sec:system-design-subsec:skeleton-subsec:cost-estimation-eq:sample-table-acl}
\end{equation}
\myequations{Sample overall estimation of the cost for the use of the data table by the \skeleton component of \crowdframe.}

Finally, Equation~\ref{cap:paper_wsdm2022-sec:system-design-subsec:skeleton-subsec:cost-estimation-eq:skeleton-total-cost} summarizes the cost of the use of the \skeleton component of \crowdframe for the task considered. It must be noted that data are written to the data table only during the task and read afterwards when downloading results. The cost of the RRUs component for the data table can therefore be charged at a later time. Furthermore, the cost must be intended on a per-month basis only for the storage components of both \index{Amazon!Web Services!S3} \index{Amazon!Web Services!DynamoDB}S3 and DynamoDB.

\begin{equation}
\begin{split}
\textrm{Skeleton Component} & = \textrm{S3} + \textrm{DynamoDB} \\
              & = \textrm{S3} + \textrm{Table Data} + \textrm{Table ACL} \\
              & = (\textrm{Storage} + \textrm{Data Retrieval} + \textrm{Data Transfer}) + \\
              &\quad\>(\textrm{WRUs} + \textrm{RRUs} + \textrm{Data Storage}) + \\
              &\quad\>(\textrm{WRUs} + \textrm{RRUs} + \textrm{Data Storage}) \\
              & = (\$0.00000014 + \$0.000856 + \$0.09) + (\$1.07 + \$0.03 + \$0.00005698) + \\
              &\quad\>(\$0.045 + \$0.00045 + \$0.00003725) \\
              & = \$1.23640037
\end{split}
\label{cap:paper_wsdm2022-sec:system-design-subsec:skeleton-subsec:cost-estimation-eq:skeleton-total-cost}
\end{equation}
\myequations{Sample overall estimation of the cost for the use of the data table by the \skeleton component of \crowdframe.}
\subsection{Search Engine}

\label{cap:paper_wsdm2022-sec:system-design-subsec:search-engine}

\crowdframe implements a component that allows the integration of a customizable search engine within the task body, as shown in Figure~\ref{cap:paper_wsdm2022-sec:design-fig:architecture}.

\subsubsection{Use Cases}

The \searchengine component imitates the standard approach followed by the most popular search engines, such as Google. This approach involves showing the search results retrieved for a given query provided by a user.

\begin{figure}[tpb]
  \centering
    \includegraphics[width=.9\linewidth]{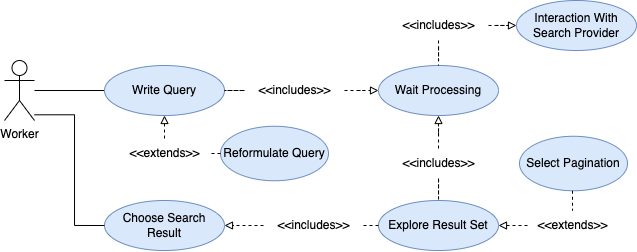}
\caption{Use case diagram of a worker performing a task deployed using \crowdframe.}
\label{cap:paper_wsdm2022-sec:system-design-subsec:search-engine-subsec:use-case-fig:worker}
\end{figure}

The diagram shown in Figure~\ref{cap:paper_wsdm2022-sec:system-design-subsec:search-engine-subsec:use-case-fig:worker} provides a high-level description of the interaction between a worker and the \searchengine component of \crowdframe. The worker writes a query in a simple text box shown by the user interface. The query is then sent to a search provider, and the worker waits for its processing. The search provider processes the query and returns a result set. The results are shown below the query box in a tabular format. Each search result is displayed to the worker with its URL, page name, and snippet. The result set is paginated by default. The worker can choose how many results to display per page. The worker can reformulate the query and obtain a new result set at any time.

The approach followed by the component deviates from the standard one when exploring the result set. Integrating a custom search engine allows workers to provide a URL for certain evaluation dimensions. The underlying goal can vary. The requester, for instance, may want the worker to provide some kind of evidence, as done by \citet{roitero2020crowd}. The user interface therefore shows a button to the right of each search result displayed. The worker explores each result and then finalizes the choice by clicking the corresponding button.

Figure~\ref{cap:paper_wsdm2022-sec:system-design-subsec:search-engine-subsec:use-case-fig:worker} shows a sample of the user interface of the \searchengine component. In more detail, the figure shows a result set made of 48 elements, split into pages of 5 elements each, retrieved for the query with text \spverb|Barack Obama|.

\begin{figure}[tpb]
  \centering
    \includegraphics[width=\linewidth]{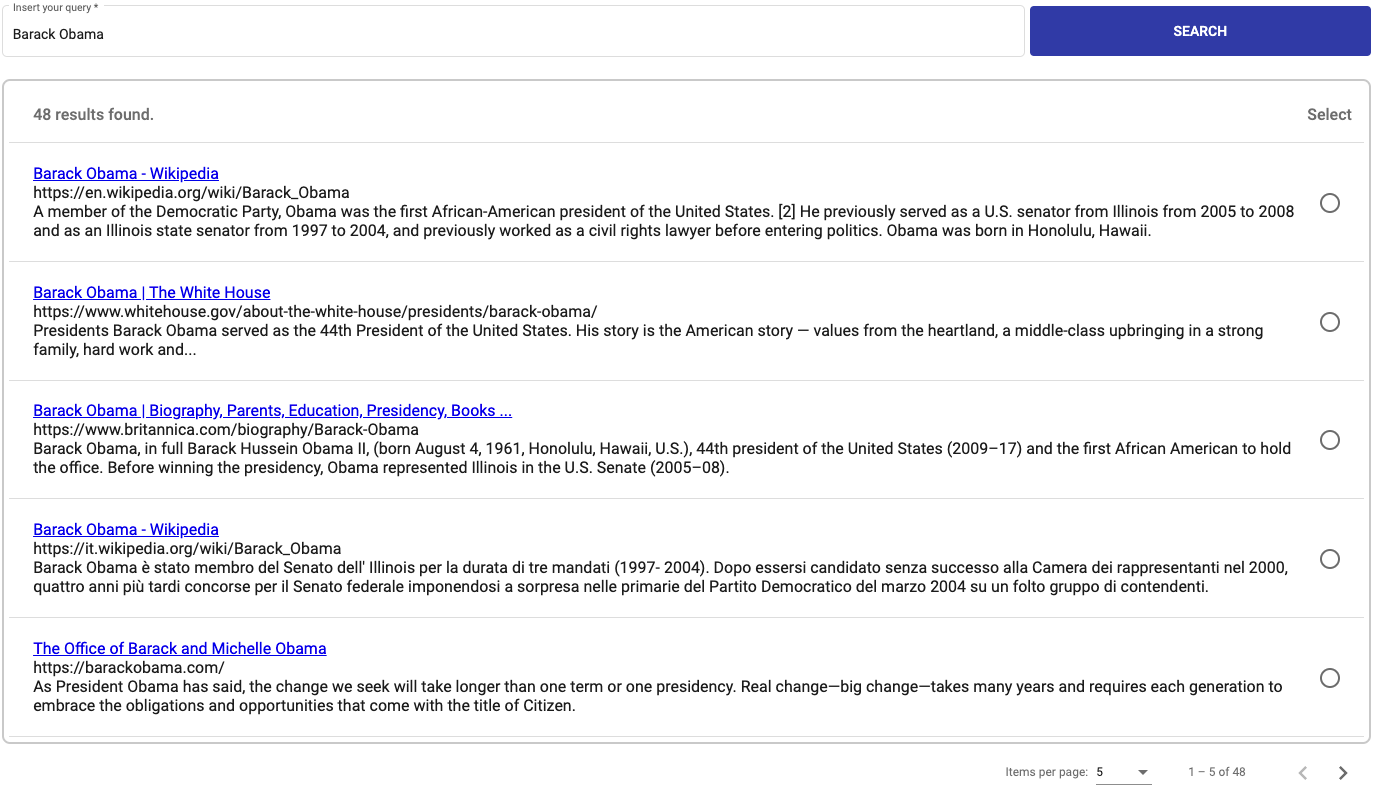}
\caption{Interface of the \searchengine component in \crowdframe.}
\label{cap:paper_wsdm2022-sec:system-design-subsec:search-engine-subsec:use-case-fig:interface-sample}
\end{figure}

\subsubsection{Architecture}

Figure~\ref{cap:paper_wsdm2022-sec:skeleton-fig:skeleton} details the interaction between the \searchengine component and the APIs of the search providers selected by the requester during task configuration. The query written by the worker is passed to a static method that encapsulates it in an HTTP message, along with the API key of the chosen search provider. The component implements the APIs of three different search providers.

The search provider processes the query and sends a second HTTP message in response. This raw response is parsed using the corresponding model and then decoded through an interface that abstracts away the underlying data structure. The result set is composed of multiple base responses, depending on how many search results the provider retrieves. This set is then passed to the user interface, which paginates and displays it to the worker.

The worker is required to select one of the results by clicking the corresponding radio button. The \searchengine component currently supports three search providers, implemented in sub-modules named \spverb|BingWebSearch|, \spverb|PubmedSearch|, and \spverb|FakerWebSearch|.

\begin{figure}[tpb]
  \centering
    \includegraphics[width=.9\linewidth]{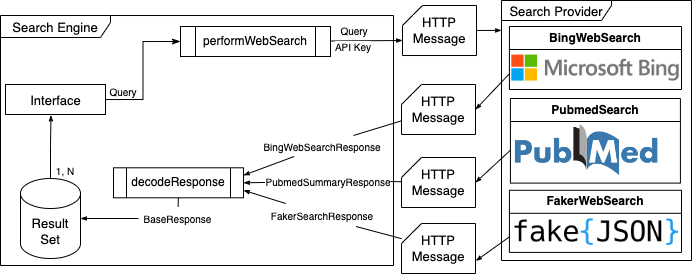}
\caption{Architecture of the \searchengine component of \crowdframe.}
\label{cap:paper_wsdm2022-sec:system-design-subsec:search-engine-subsec:architecture-fig:diagram}
\end{figure}

\subsubsection{Microsoft Search API}

\label{cap:paper_wsdm2022-sec:system-design-subsec:search-engine-subsec:bing}

The \spverb|BingWebSearch| \index{Bing!Web Search} provider implements the Bing Web Search API,\footnote{\url{https://www.microsoft.com/en-us/bing/apis/bing-web-search-api}} part of the \index{Bing!Web Search!API} Microsoft Search API. It is mainly designed to be used as a result of a direct user query or search, or as a result of an action within a system that can be logically interpreted as a user's search request. Acceptable search or search-like scenarios thus include a user who provides a query directly into a search box, a user who requests \lq\lq more information\rq\rq{} about a text or image using some kind of user interface control, and so on. The API returns and ranks implicitly whatever content is relevant to a query. It is possible to filter and control, to some extent, the results retrieved. For instance, it is possible to include or exclude specific types of results, return pages discovered within the last week, and so on.

\crowdframe relies on the \spverb|v7.0| version of the Bing Web Search API. To use the API, a developer must obtain a subscription key. Then, they can send HTTP GET messages to the API's endpoint,\footnote{\url{https://api.bing.microsoft.com/v7.0/search}} thereby retrieving results. The subscription key must be included in the \index{Ocp-Apim-Subscription-Key} \spverb|Ocp-Apim-Subscription-Key| header of the HTTP request. The GET parameter \spverb|q| is used to attach the user's query and must be URL-encoded. The component also provides four headers to improve the search experience for the worker. Table~\ref{cap:paper_wsdm2022-sec:skeleton-subsec:cost-estimation-tab:header-required} describes the headers sent to the API.

Furthermore, the component captures three headers returned in each response. These include, for instance, \spverb|X-MSEdge-ClientID|, which the component stores in a cookie after the first request, since it allows recognition of a user's device and should be attached to subsequent ones. Table~\ref{cap:paper_wsdm2022-sec:skeleton-subsec:cost-estimation-tab:header-capture} describes these headers and explains why they are captured. There are various GET parameters\footnote{\url{https://learn.microsoft.com/en-us/bing/search-apis/bing-web-search/reference/query-parameters}} that can be used to customize the results retrieved. The component attaches two additional parameters to the query text when sending a request to the API. Table~\ref{cap:paper_wsdm2022-sec:skeleton-subsec:cost-estimation-tab:query-parameters} describes the GET parameters added when sending each query.

\begin{table}[tpb]
\centering
\caption{Headers sent when issuing requests to the Bing Web Search API.}
\begin{tabular}{lp{4.8cm}p{4.8cm}}
\toprule
\textbf{Header} & \textbf{Value} & \textbf{Description} \\
\midrule
\spverb|User-Agent| & The user agent originating the request & Bing uses the user agent to provide mobile users with an optimized experience. \\
\midrule
\spverb|X-MSEdge-ClientID| & Alphanumeric string & Bing uses the client ID to provide users with consistent behavior across queries. \\
\midrule
\spverb|X-MSEdge-ClientIP| & The IPv4 or IPv6 address of the client device & The IP address is used to determine the user's location. Bing uses the location to set safe search behavior. \\
\midrule
\spverb|X-Search-Location| & A semicolon-delimited list of key/value pairs that describe the client's geographical location & Bing uses the location to set safe search behavior and return relevant local content. \\
\bottomrule
\end{tabular}
\label{cap:paper_wsdm2022-sec:skeleton-subsec:cost-estimation-tab:header-required}
\end{table}

\begin{table}[tpb]
\centering
\caption{Headers captured while receiving responses from the Bing Web Search API.}
\begin{tabular}{lp{4.8cm}p{4.8cm}}
  \toprule
\textbf{Header} & \textbf{Value} & \textbf{Description} \\
    \midrule
\spverb|X-MSEdge-ClientID| & Alphanumeric string returned after the first request & Bing uses the client id to provide users with consistent behavior across queries \\
    \midrule
    \spverb|BingAPIs-TraceId| & The ID of the log entry that contains the details of the request & Identifier of the request to provide to the support team in case of errors. \\
    \midrule
    \spverb|BingAPIs-Market| & The form is \spverb|<languageCode>-<countryCode>| & The market used by Bing to process the query \\
         \bottomrule
  \end{tabular}
\label{cap:paper_wsdm2022-sec:skeleton-subsec:cost-estimation-tab:header-capture}
\end{table}

\begin{table}[tpb]
\centering
\caption{Query parameters sent when issuing queries to the Bing Web Search API.}
\begin{tabular}{lp{11cm}}
  \toprule
   \textbf{Parameters} & \textbf{Description} \\
    \midrule
    \spverb|count| & The number of search results to return in a response. \\
    \midrule
    \spverb|offset| & The number of search results to skip. \\
    \midrule
    \spverb|mkt| & The market where the results come from. Typically, the country where the worker is making the query from. \\
         \bottomrule
  \end{tabular}
\label{cap:paper_wsdm2022-sec:skeleton-subsec:cost-estimation-tab:query-parameters}
\end{table}

Listing~\ref{cap:paper_wsdm2022-sec:system-design-subsec:search-engine-subsec:architecture-lst:bing-sample-request} shows a sample request sent to retrieve results for a query with the text \spverb|microsoft devices|, including the suggested headers and GET parameters. The request is issued using the \index{cURL}cURL software. The response is a JSON \index{JSON} object that contains various metadata fields and an array of retrieved web pages. This object is decoded by the component, as illustrated in Figure~\ref{cap:paper_wsdm2022-sec:skeleton-fig:skeleton}. The final result set for the query entered by the worker and retrieved via the Bing Web Search API is decoded using the static method shown in Figure~\ref{cap:paper_wsdm2022-sec:system-design-subsec:search-engine-subsec:architecture-fig:diagram}. The resulting array of base responses is returned to the user interface of the \searchengine component of \crowdframe.

\begin{lstlisting}[style=curlstyle, caption={Sample request including headers and query parameters sent to the Bing Web Search API.}, label={cap:paper_wsdm2022-sec:system-design-subsec:search-engine-subsec:architecture-lst:bing-sample-request}]
curl -H "Ocp-Apim-Subscription-Key: <yourkeygoeshere>" -H "X-MSEdge-ClientID: 00B4230B74496E7A13CC2C1475056FF4" -H "X-MSEdge-ClientIP: 11.22.33.44" -H "X-Search-Location: lat:55;long:-111;re:22" -A "Mozilla/5.0 (X11; Linux x86_64) AppleWebKit/537.36 (KHTML, like Gecko) Chrome/29.0.1547.65 Safari/537.36" https://api.bing.microsoft.com/v7.0/search?q=microsoft+devices&mkt=en-us&count=10&offset=0
\end{lstlisting}

\subsubsection{Entrez Programming Utilities}

The \index{Entrez Programming Utilities} \textsf{Entrez Programming Utilities}\footnote{\url{https://www.ncbi.nlm.nih.gov/books/NBK25501/}} (E-utilities) is a set of nine server-side programs that provide a stable interface to the Entrez query and database system at the National Center for Biotechnology Information (NCBI). These programs use fixed endpoints whose URL-based syntax translates a set of input parameters into the values required by the underlying software components to search for and retrieve the requested data. In other words, the \textsf{E-utilities} constitute the structured interface to the \textsf{Entrez} system, which includes 38 databases covering a variety of biomedical data. Table~\ref{cap:paper_wsdm2022-sec:search-engine-subsec:entrez-utilities-tab:utilities-summary} provides a brief description of each E-utility.

\textsf{PubMed}\footnote{\url{https://pubmed.ncbi.nlm.nih.gov/}} \index{PubMed} is a free search engine that primarily accesses the \index{MEDLINE} \textsf{MEDLINE}\footnote{\url{https://www.nlm.nih.gov/medline/medline_overview.html}} database, which contains references to the biomedical literature. The United States National Library of Medicine (NLM) at the National Institutes of Health (NIH) maintains the database as part of the \textsf{Entrez} system.

\begin{table}[tpb]
\centering
\caption{General description of the nine \textsf{E-utilities} provided by the \textsf{Entrez} system.}
\begin{tabular}{llp{8cm}}
  \toprule
   \textbf{E-utility} & \textbf{Goal} & \textbf{Description} \\
    \midrule
     \spverb|EInfo| & Database statistics & Provides the number of records indexed in each field of a given database, the date of the last update, and the available links to other Entrez databases. \\
     \midrule
     \spverb|ESearch| & Text searches & Responds to a text query with the list of matching UIDs (identifiers) in a given database for later use in other E-utilities, along with the term translations of the query. \\
     \midrule
     \spverb|EPost| & UID uploads & Accepts a list of UIDs from a given database and responds with the web environment of the uploaded dataset and its query key. \\
     \midrule
     \spverb|ESummary| & Document summary download & Responds to a list of UIDs in a given database with the corresponding document summaries. \\
     \midrule
     \spverb|EFetch| & Data record download & Responds to a list of UIDs in a given database with the corresponding data records in a specified format. \\
     \midrule
     \spverb|ELink| & Entrez links & Responds to a list of UIDs in a given database with either a list of related UIDs (and relevancy scores) in the same database or a list of linked UIDs in another Entrez database. \\
     \midrule
     \spverb|EGQuery| & Global query & Responds to a text query by providing the number of records matching the query in each Entrez database. \\
     \midrule
     \spverb|ESpell| & Spelling suggestions & Retrieves spelling suggestions for a text query in a given database. \\
     \midrule
     \spverb|ECitMatch| & PubMed batch citation search & Retrieves PubMed IDs (PMIDs) corresponding to a set of input citation strings. \\
    \bottomrule
  \end{tabular}
\label{cap:paper_wsdm2022-sec:search-engine-subsec:entrez-utilities-tab:utilities-summary}
\end{table}

\index{Entrez Programming Utilities!CitMatch} \index{Entrez Programming Utilities!Spell} \index{Entrez Programming Utilities!GQuery} \index{Entrez Programming Utilities!Link} \index{Entrez Programming Utilities!Summary} \index{Entrez Programming Utilities!Post} \index{Entrez Programming Utilities!Search} \index{Entrez Programming Utilities!Info} The NCBI recommends using an API key to access the E-utilities. Its usage helps to avoid overloading the underlying systems. Any IP address that sends more than 3 requests per second to the E-utilities without an API key receives an error message, while IP addresses with a valid key are allowed to send up to 10 requests per second. The API key can be obtained from the NCBI account page.\footnote{\url{https://www.ncbi.nlm.nih.gov/account/}} Each request must be sent to the base endpoint \url{https://eutils.ncbi.nlm.nih.gov/entrez/eutils/}. The API key is added by appending its value to the \spverb|api_key| GET parameter. The desired E-utility is specified by appending its lowercase name with the \spverb|.fcgi| suffix. The only exception is the ECitMatch utility, which uses the \spverb|.cgi| suffix. For example, using the ESummary utility involves sending requests to the endpoint \url{https://eutils.ncbi.nlm.nih.gov/entrez/eutils/esummary.fcgi}.

The \spverb|PubMedSearch| \index{PubMed!Search} provider interacts with the ESummary and ESearch utilities of the Entrez system. The E-utilities must be combined to create an effective and useful data pipeline. The system provides the Entrez History server, which simplifies transferring context and data between successive requests. 

The ESearch utility is the first block of the pipeline. The query provided by the worker is added as a parameter to the request. The request is sent to the PubMed search engine, which returns a list containing the identifiers of the items relevant to the query. A total of six GET parameters are used to customize the results returned by the utility, as shown in Table~\ref{cap:paper_wsdm2022-sec:search-engine-subsec:entrez-utilities-tab:esearch-params}.

\begin{table}[tpb]
\centering
\caption{Query parameters provided when sending requests to the ESearch utility of the Entrez system.}
\begin{tabular}{lp{11cm}}
  \toprule
   \textbf{Parameter} & \textbf{Usage} \\
    \midrule
    \spverb|db| & Database from which to retrieve results, which is \spverb|pubmed|. \\
    \midrule
    \spverb|term| & Text query for which to retrieve results. All special characters must be URL-encoded. Spaces may be replaced by the \spverb|+| character. \\
    \midrule
    \spverb|usehistory| & If set to \spverb|y|, ESearch will post the UIDs resulting from the search operation to the History server for use in subsequent calls. \\
    \midrule
    \spverb|retmode| & Format of the returned output. The JSON format is used by the search provider. \\
    \midrule
    \spverb|WebEnv| & Web environment returned by a previous ESearch, EPost, or ELink request. ESearch appends the result set retrieved to the one contained in the existing environment. The use of \spverb|usehistory| is required. \\
    \midrule
    \spverb|query_key| & Integer query key returned by a previous ESearch, EPost, or ELink request. ESearch finds the intersection of the result set identified by the key and the one retrieved for the current \spverb|term|. The use of \spverb|WebEnv| is required. \\
  \bottomrule
\end{tabular}
\label{cap:paper_wsdm2022-sec:search-engine-subsec:entrez-utilities-tab:esearch-params}
\end{table}

Listing~\ref{cap:paper_wsdm2022-sec:search-engine-subsec:entrez-utilities-list:esearch-request} shows a sample initial request sent to the ESearch utility to retrieve a JSON set of identifiers for a query having text \spverb|vaccines|. Subsequent requests can use the \spverb|WebEnv| and \spverb|query_key| parameters if needed. The ESummary \index{Entrez Programming Utilities!Summary} utility is the second and final block of the pipeline implemented by the search provider. The identifiers of each result item retrieved for the query written by the worker using the ESearch \index{Entrez Programming Utilities!Search} utility are used to fetch the details of the corresponding data records.

The GET parameters used to customize the results returned are those shown in Table~\ref{cap:paper_wsdm2022-sec:search-engine-subsec:entrez-utilities-tab:esearch-params}. The only difference is that the \spverb|term| parameter is replaced with the \spverb|id| parameter, which accepts either a single UID or a comma-delimited list of UIDs. The search provider uses all UIDs retrieved by the query sent to the ESearch utility. For instance, let us assume that one of the UIDs retrieved from the request shown in Listing~\ref{cap:paper_wsdm2022-sec:search-engine-subsec:entrez-utilities-list:esearch-request} is \spverb|36511263|. Listing~\ref{cap:paper_wsdm2022-sec:search-engine-subsec:entrez-utilities-list:esummary-request} shows the request sent to the ESummary utility to fetch the details of the data record.

The final result set for the query written by the worker and refined using the pipeline implemented by the \spverb|PubMedSearch| provider is then decoded using the static method shown in Figure~\ref{cap:paper_wsdm2022-sec:system-design-subsec:search-engine-subsec:architecture-fig:diagram}, and the array of base responses is returned to the user interface of the \searchengine component of \crowdframe.

\begin{lstlisting}[style=curlstyle, caption={Sample initial request sent to the ESearch utility of the Entrez system.}, label={cap:paper_wsdm2022-sec:search-engine-subsec:entrez-utilities-list:esearch-request}]
curl https://eutils.ncbi.nlm.nih.gov/entrez/eutils/esearch.fcgi?api_key=your_api_key&db=pubmed&term=vaccines&usehistory=y&retmode=json
\end{lstlisting}

\begin{lstlisting}[style=curlstyle, caption={Sample follow-up request sent to the ESummary utility of the Entrez system.}, label={cap:paper_wsdm2022-sec:search-engine-subsec:entrez-utilities-list:esummary-request}]
curl https://eutils.ncbi.nlm.nih.gov/entrez/eutils/esummary.fcgi?api_key=your_api_key&db=pubmed&id=36511263&retmode=json
\end{lstlisting}

\subsubsection{fakeJSON}

The \textsf{fakeJSON}\footnote{\url{https://fakejson.com/}} \index{fakeJSON} service provides a simple API for building mock data and supports frontend development, end-to-end testing, and data generation.\footnote{As of June 2025, the service is no longer available.} The underlying idea is that the developer sends a request directly to a single endpoint. In the request body, they specify the desired format of the response. The service then creates a JSON object populated with the requested data, initializing each field with random values. Requests can be sent using any of the standard HTTP methods such as \index{POST} \index{PUT} \index{DELETE} \index{PATCH} \texttt{POST}, \texttt{PUT}, \texttt{DELETE}, and \texttt{PATCH}. The service supports cross-origin resource sharing, allowing it to receive asynchronous requests from anywhere. Each request must be sent to the endpoint \url{https://app.fakejson.com/q}. 

The \texttt{FakerWebSearch} \index{FakerWebSearch} provider interacts with the service to retrieve fake search result data, enabling testing of the search engine component of \texttt{crowdframe} during task design. Every request sent to the service to generate fake data must comply with the same payload format. The payload must be a JSON object that includes a unique token, an optional set of parameters, and the \texttt{data} field. The unique token can be obtained from the account page.\footnote{\url{https://app.fakejson.com/member/token}} The parameters allow customization of the generated response's behavior depending on specific needs. For instance, they support delaying the response by a fixed number of seconds or adding a set of custom headers. Table~\ref{cap:paper_wsdm2022-sec:search-engine-subsec:fake-json-tab:parameters} shows the four parameters that can be included in the request payload.

\begin{table}[t]
\centering
\caption{Payload parameters used to customize the behavior of the responses generated by the \textsf{fakeJSON} service.}
\begin{tabular}{lp{10cm}}
  \toprule
   \textbf{Parameter} & \textbf{Usage} \\
    \midrule
    \spverb|code| & Specifies the HTTP status code to return in the fake response. \\
    \midrule
    \spverb|delay| & Delays the response by a fixed number of seconds. \\
    \midrule
    \spverb|headers| & Adds custom headers to the generated response. \\
    \midrule
    \spverb|consistent| & Prevents the service from returning a cached fake dataset. The only accepted value is \spverb|false|. \\
     \bottomrule
  \end{tabular}
\label{cap:paper_wsdm2022-sec:search-engine-subsec:fake-json-tab:parameters}
\end{table}

The \spverb|data| field of the request payload defines the format of the responses to be returned. It consists of fields, objects, and arrays, like any valid JSON structure, and follows an attribute–value pair format. The name of each key can be any string; however, the value must strictly follow the syntax of the corresponding data type defined by \textsf{fakeJSON}. The \spverb|_repeat| attribute can be added to an object to specify how many instances of that object the service should generate.

Listing~\ref{cap:paper_wsdm2022-sec:search-engine-subsec:fake-json-lst:sample-payload} shows a sample request sent to the service using the cURL software. The same listing also shows the payload sent by the search provider to generate fake data and enable the requester to test the search engine. The service is instructed to return a response with code 200, consisting of eight JSON objects. Each object contains a \spverb|text| field with a long string, a \spverb|name| field with a shorter string, and a \spverb|url| field with a randomly generated URL. The intent is to mimic search engine results, which typically include a page address, a page name, and a snippet describing the page. Various formats\footnote{\url{https://fakejson.com/documentation#request_data}} can be used to mock different types of data. The fake search result data generated by the \textsf{fakeJSON} \index{fakeJSON} service and encapsulated in the \spverb|FakerWebSearch| provider is then decoded using the static method shown in Figure~\ref{cap:paper_wsdm2022-sec:system-design-subsec:search-engine-subsec:architecture-fig:diagram}, and the resulting array of base responses is returned to the user interface of the \searchengine component of \crowdframe.

\begin{lstlisting}[style=curlstyle,
  caption={Sample request sent to the \textsf{fakeJSON} service using the \index{cURL}cURL command-line tool.},
  label={cap:paper_wsdm2022-sec:search-engine-subsec:fake-json-lst:sample-request}]
curl --request POST --url https://app.fakejson.com/q \
     --header 'content-type: application/json' \
     --data 'your_data_object'
\end{lstlisting}

\begin{lstlisting}[style=jsonstyle,
  caption={Example payload of a request sent to the \textsf{fakeJSON} service to generate fake search-result data.},
  label={cap:paper_wsdm2022-sec:search-engine-subsec:fake-json-lst:sample-payload}]
{
  "token": "...",
  "parameters": {
    "code": 200
  },
  "data": {
    "url": "internetUrl",
    "name": "stringShort",
    "text": "stringLong",
    "_repeat": 8
  }
}
\end{lstlisting}

Platforms that provide search APIs typically adopt a pricing model based on the number of queries issued and their rate per second. In light of this, a more accurate estimation of the usage cost of the \searchengine component of \crowdframe can be derived using data collected from previously deployed tasks, as done for the \skeleton component in Section~\ref{cap:paper_wsdm2022-sec:skeleton-subsec:cost-estimation}. The most recent variant of the task deployed by \citet{SOPRANO2021102710} (described in Chapter~\ref{cap:paper_ipm2021}) and \citet{draws2022bias} (described in Chapter~\ref{cap:paper_facct2022}) is therefore used to estimate this cost as well.

The variant of the task considered publishes 200 \index{HIT}HITs. Workers are required to provide a URL using the search engine for one of the evaluation dimensions. A total of 237 workers access the deployed task in an attempt to complete their assigned \index{HIT}HIT. In total, they issue 3520 different queries. The minimum number of queries issued by a worker is 2; this worker abandons the task shortly after starting. The maximum is 58, issued by a worker who either reformulates multiple queries for the same element or retries multiple times. On average, each worker issues roughly 15 queries.

The last query is issued 2 days, 8 hours, 2 minutes, and 17 seconds after the first one. These queries can be grouped into 2498 unique transactions-per-second \index{TPS}(TPS) blocks, where each block consists of queries sent within the same second. This indicates that queries were sent during 2498 distinct seconds between the first and last query. During each of these seconds, one or more workers may have issued one or more queries. These 2498 seconds (approximately 42 minutes in total) are scattered within the overall task duration. The smallest TPS block size is 1, meaning a single query was issued in a given second. The largest block contains 5 queries. The average TPS block size is 1.40. Note that these TPS blocks are not necessarily consecutive.

This analysis allows us to understand the query throughput to the search API platform. The task variant considered uses the \index{Bing!Web Search} \spverb|BingWebSearch| provider, which relies on the Microsoft Bing Search API. The API’s pricing plans\footnote{\url{https://www.microsoft.com/en-us/bing/apis/pricing}} vary based on three parameters: the maximum allowed TPS, the type of search service, and the cost per 1000 transactions. The cost per 1000 transactions differs depending on whether the Japanese market is targeted. \crowdframe uses the Bing Web Search subset for the United States market under the S3 pricing plan, which allows a maximum TPS of 100. This means that the API can receive up to 100 queries per second; exceeding this limit causes the service to throttle requests, resulting in slower response times.

Under this plan, the cost is \$4 per 1000 transactions. A transaction corresponds to a successful API request. Advanced features such as auto-completion may increase the number of transactions per query, but the implementation used by \crowdframe does not include such features. Since 3520 queries were issued, the fee is applied four times. The maximum TPS during the task was 5, well below the 100 TPS threshold, so no queries were throttled. Equation~\ref{cap:paper_wsdm2022-sec:search-engine-subsec:cost-estimation-eq:bing} shows the cost computation for the custom search engine in this task.

\begin{equation}
\begin{split}
\textrm{Bing Web Search} & = \$4 * \lceil\textrm{Query Number / Billing Threshold}\rceil \\
                         & = \$4 * \lceil 3520 / 1000\rceil = \$4 * \lceil 3.52\rceil = \$16  \\
                         & \quad\>\>\textrm{threshold: 1000 transactions}
\end{split}
\label{cap:paper_wsdm2022-sec:search-engine-subsec:cost-estimation-eq:bing}
\end{equation}
\myequations{Sample estimation of the cost for the usage of the \textsf{Bing Web Search} API.}

The usage of the \index{PubMed!Search} \spverb|PubmedSearch| provider does not require any form of payment. The Entrez \index{Entrez Programming Utilities} system and its E-utilities are publicly available for free, provided that users comply with the guidelines suggested by the NCBI. Similarly, the \spverb|FakerWebSearch| \index{FakerWebSearch} provider does not enforce any payment. 

The pricing model\footnote{\url{https://fakejson.com/pricing}} of the \textsf{fakeJSON} \index{fakeJSON} service includes a free plan limited to 1000 requests per day, which is typically sufficient for testing the task interface during its design. The service stops responding to requests once this threshold is reached. If needed, it is possible to upgrade to a higher-tier plan. The most affordable option is the \textsf{Developer} plan, which costs \$12 per month and allows up to \num{50,000} requests per day, along with access to additional features.

\subsection{Logger}

\label{cap:paper_wsdm2022-sec:system-design-subsec:logger}

\crowdframe implements a logging component that allows capturing worker behavior during the task, as shown in Figure~\ref{cap:paper_wsdm2022-sec:design-fig:architecture}.

\subsubsection{Architecture}

\label{cap:paper_wsdm2022-sec:system-design-subsec:logger-subsubsec:architecture}

\begin{figure}[tpb]
  \centering
    \includegraphics[width=0.7\linewidth]{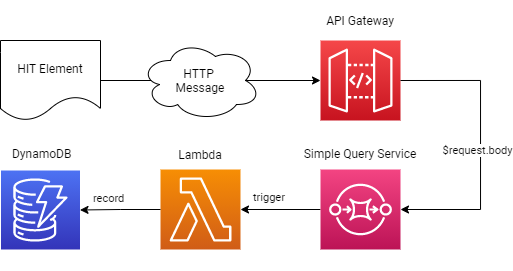}
\caption{Infrastructure of the \logger component in \texttt{crowdframe}.}
\label{cap:paper_wsdm2022-sec:logger-fig:pipeline}
\end{figure}

The log messages produced when capturing user behavior events are stored using a cloud-based logging server that relies on the infrastructure provided by Amazon Web Services. Figure~\ref{cap:paper_wsdm2022-sec:logger-fig:pipeline} shows an overview of the entire logging pipeline.

API Gateway\footnote{\url{https://aws.amazon.com/api-gateway/}} \index{Amazon!Web Services!API Gateway} is a service used to implement APIs for web applications or other AWS services. It addresses traffic management, cross-origin resource sharing (CORS) support, authorization and access control, request throttling, and monitoring of an API layer. In more detail, \crowdframe uses an HTTP API layer. Each user action is captured by the Logger component and sent to the layer using an HTTP message. CORS is an HTTP-header-based mechanism that allows a server to indicate any origins (in terms of port, domain, or scheme) other than its own from which a browser should permit loading resources. The API layer's CORS is configured to allow receiving POST messages only. The layer receives the messages through a single endpoint provided by the application. The body of each message is redirected to a queue upon reception.

Simple Queue Service\footnote{\url{https://aws.amazon.com/sqs/}} (SQS) \index{Amazon!Web Services!Simple Queue Service} is a service used to create and manage queues of messages. It allows the creation of two types of queues, namely Standard and FIFO. The first type aims to ensure the best delivery sequence, and each message may be delivered more than once if its processing fails or does not complete on time. The second type, on the other hand, ensures a first-in-first-out message delivery sequence where each message is processed exactly once. The main drawback is its higher cost. The use of a message queue allows decoupling the reception of log requests from their processing. The payload of each log request has a sequence number. Each request is independent, and the whole stream of messages can be reordered at a later time. \crowdframe thus uses a standard queue. The attached access policy allows only the gateway to send messages to the queue.

Lambda\footnote{\url{https://aws.amazon.com/lambda/}} \index{Amazon!Web Services!Lambda} is a service that provides serverless computing, allowing the execution of source code without explicitly provisioning or maintaining servers. One of the core features of the service is its ability to automatically scale depending on the size of the workload that the code must handle. In other words, Lambda allows running an algorithm only when needed, using a self-activating approach.

A serverless function deployed using Lambda polls the queue for new messages and collects batches of up to 100 messages. The function parses each log message, and the user action is stored as a record in the DynamoDB log table shown in Figure~\ref{cap:paper_wsdm2022-sec:design-fig:architecture}. The entire logging back-end can also be deployed on a private server. In that case, the requester can configure a custom endpoint to which each log message will be sent.
There is a trigger between the function and the queue, activated whenever the queue receives a new message. More specifically, the Lambda function runs a JavaScript \index{JavaScript} algorithm. It initially polls a batch of log requests, parses the JSON payload of each message, and adds the server time. The enriched payload is then stored in the table.

The function is configured to poll for new log requests every 20 seconds, using five parallel long polling connections by default. The service can dynamically reduce the interval between polling operations and increase the number of instances assigned to the function. This feature helps empty the queue when it starts filling up, thus avoiding overload. Otherwise, messages would need to be re-enqueued. A separate DynamoDB table is created for each batch deployed for each task, and the Logger component targets each table accordingly.

\subsubsection{Event Listening}

\label{cap:paper_wsdm2022-sec:system-design-subsec:logger-subsubsec:event-listening}

\textsf{Angular} \index{Angular} uses custom markup elements defined through an extension of HTML, allowing tags whose content is created and managed by client-side scripting in JavaScript or TypeScript. These elements are especially useful in dynamic environments, where flexibility is essential to support continuously evolving hypermedia experiences. To enable interaction with such elements, Angular provides a mechanism to bind event listeners directly within custom element tags. This approach is suitable when only a few elements are involved. However, in real-world scenarios where tens or hundreds of tags are dynamically generated, a more scalable solution is needed.

Angular addresses this need through \index{Directive} \spverb|Directives|.\footnote{\url{https://angular.io/guide/attribute-directives}} A \spverb|Directive| is a class that dynamically adds behavior to custom elements. A decorator specifies a CSS selector to target the elements, and the constructor passes a function that implements the desired behavior and attaches an event listener. The idea is to determine the set of markup elements to target based on the type of event to log.

Multiple types of events are monitored. Listing~\ref{cap:paper_wsdm2022-sec:system-design-subsec:logger-subsubsec:event-listening-lst:log-payload} shows the generic JSON payload of a log request triggered by an event. The \spverb|detail| object contains event-specific data. Each event is wrapped by the \logger component in a base payload, which is enriched with custom fields according to the event type, and then sent as an HTTP POST request to the AWS infrastructure. The server-side timestamp is added upon receipt. Section~\ref{appendix:log-events} describes the structure of the payloads for each monitored event, based on the HTML structure of \crowdframe.

\begin{lstlisting}[style=jsonstyle, caption={Payload structure of a log request sent by the \logger component of \crowdframe.}, label={cap:paper_wsdm2022-sec:system-design-subsec:logger-subsubsec:event-listening-lst:log-payload}]
{
    "bucket": "string",
    "worker": "string",
    "task": "string",
    "batch": "string",
    "unitId": "string",
    "try_current": "integer",
    "type": "string",
    "sequence": "integer",
    "client_time": "string",
    "server_time": "string", 
    "details": {
        ...
    }
}
\end{lstlisting}

There is a directive available for each event, and every directive shares the same base template. Each event is monitored relative to a custom element selected using its CSS selector. A dedicated function to handle each event is defined within the \logger component. Both the element and the logging function are provided to the directive constructor, which then adds the custom behavior to the targeted element. This event listening behavior is attached to the corresponding markup element only if the Logger component is enabled by the requester and configured to log that specific event. The requester can enable or disable each event through the Generator component.

To reduce the processing load caused by spammable events such as mouse clicks, movements, and scrolls, the event listeners apply debouncing and function call optimization. The \index{RxJS} \spverb|RxJS|\footnote{\url{https://rxjs.dev/}} library is used to further enhance logging capabilities. It enables the addition of event listeners while supporting a pipeline of operations to extract data and perform various manipulations. The use of this library allows for event composition. For instance, a text selection event begins only when the mouse button is pressed and held, and ends when the button is released.

\subsubsection{Performance Evaluation}

\label{cap:paper_wsdm2022-sec:system-design-subsec:logger-subsubsec:test-infrastucture}

Locust\footnote{\url{https://locust.io/}} \index{Locust} is a Python-based performance testing tool that allows defining custom user behavior and generating millions of simultaneous requests to evaluate system performance. The main script models a worker node that sends POST requests to the specified host address (i.e., the endpoint exposed by the API Gateway). In cluster mode, Locust launches a master node that coordinates the worker nodes, specifying the number of required nodes, the request spawn rate, and the test duration. When the test ends, the tool produces a file containing various performance statistics.

In the test setup, the master node is run on a local machine. Each cluster instance spawns two processes (one per core), each responsible for generating log requests. The initial configuration consists of 200 processes in total. Each process sends a message every 10 milliseconds. The test is executed in five rounds, with each round lasting five minutes. The number of worker nodes doubles at the start of each round, reaching a maximum of 1600 nodes.

\begin{figure}[tpb]
  \centering
    \includegraphics[width=0.9\linewidth]{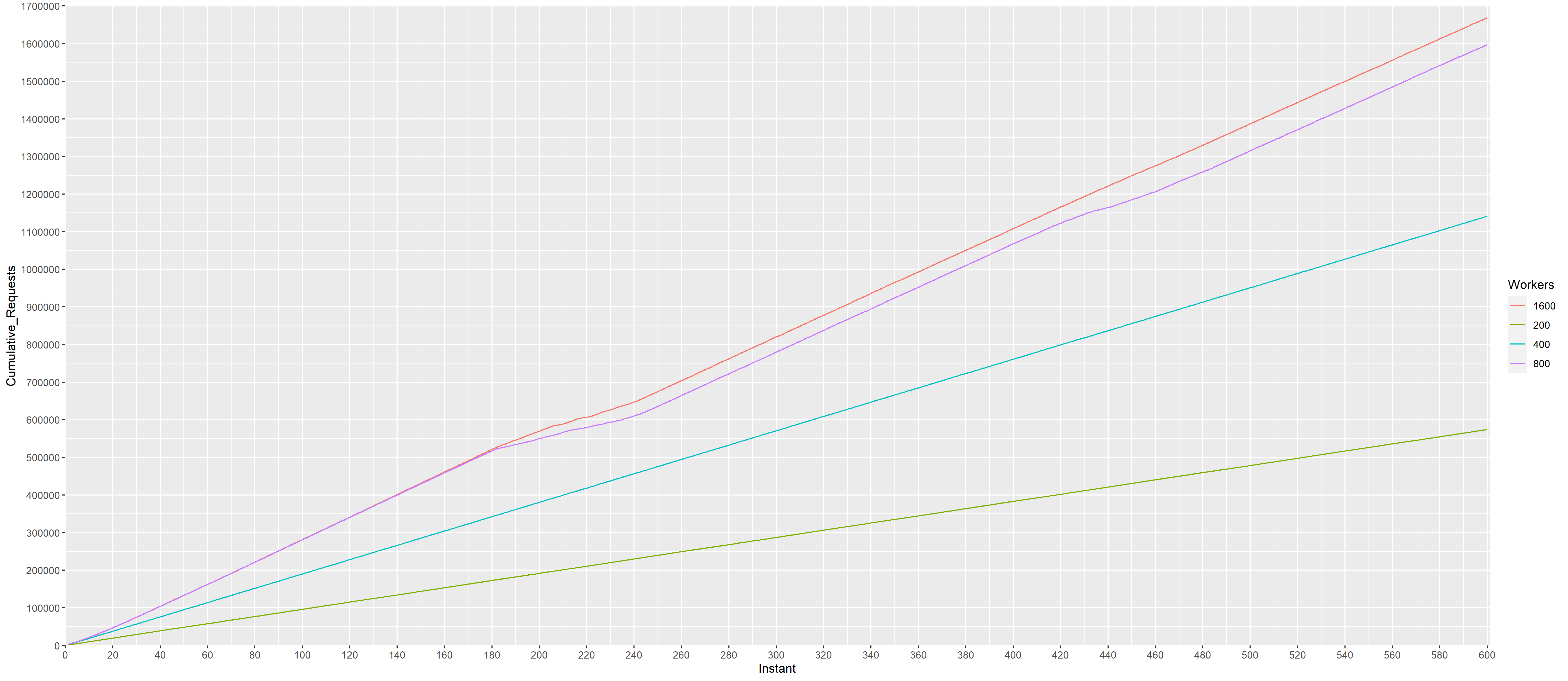}
\caption{Cumulative requests sent during each round of the performance evaluation tests.}
\label{cap:paper_wsdm2022-sec:system-design-subsec:logger-subsubsec:test-infrastucture-fig:requests-cumulative}
\end{figure}

\begin{figure}[tpb]
  \centering
    \includegraphics[width=\linewidth]{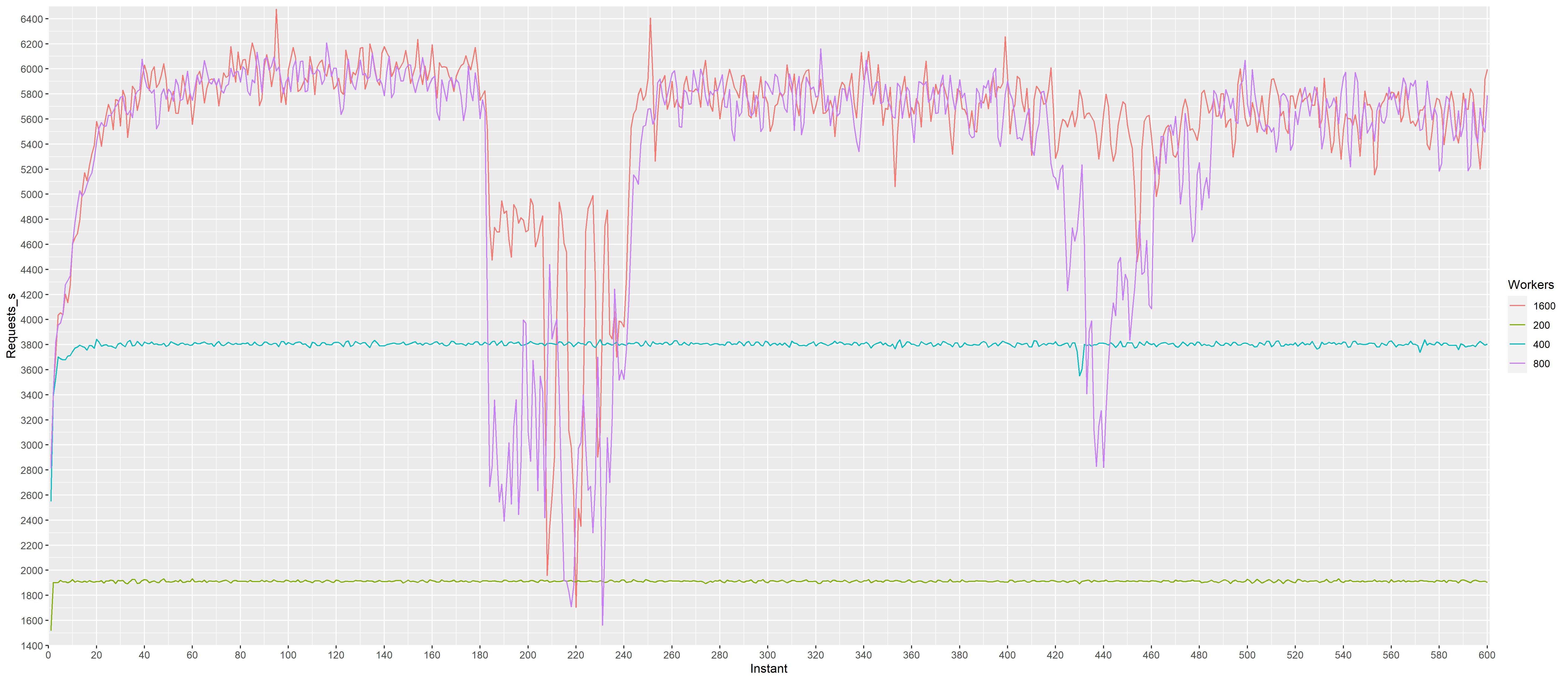}
\caption{Number of requests managed per second by the dedicated server during each round of the test.}
\label{cap:paper_wsdm2022-sec:system-design-subsec:logger-subsubsec:test-infrastucture-fig:requests-second}
\end{figure}

\begin{figure}[tpb]
  \centering
    \includegraphics[width=\linewidth]{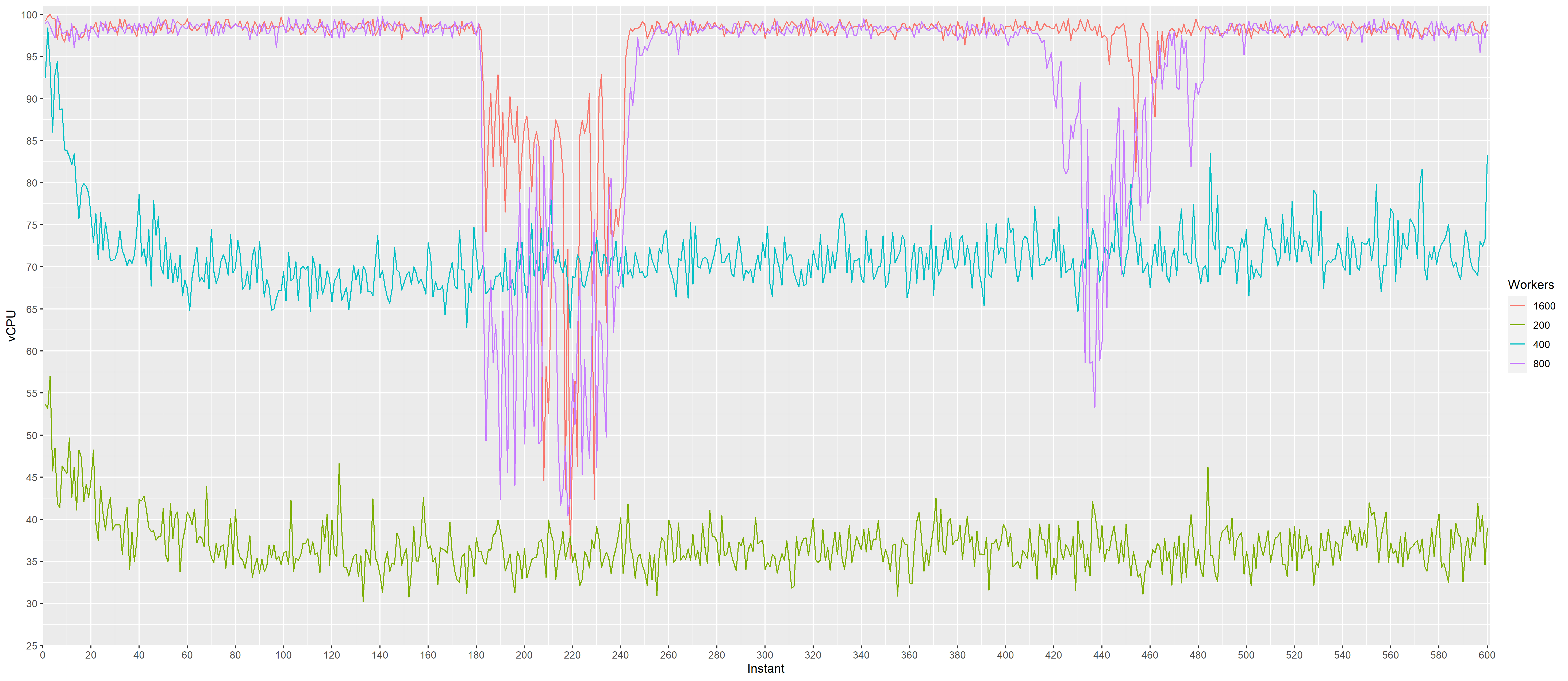}
\caption{CPU usage of the dedicated server during each round of the performance evaluation tests.}
\label{cap:paper_wsdm2022-sec:system-design-subsec:logger-subsubsec:test-infrastucture-fig:cpu-usage}
\end{figure}

\begin{figure}[tpb]
  \centering
    \includegraphics[width=\linewidth]{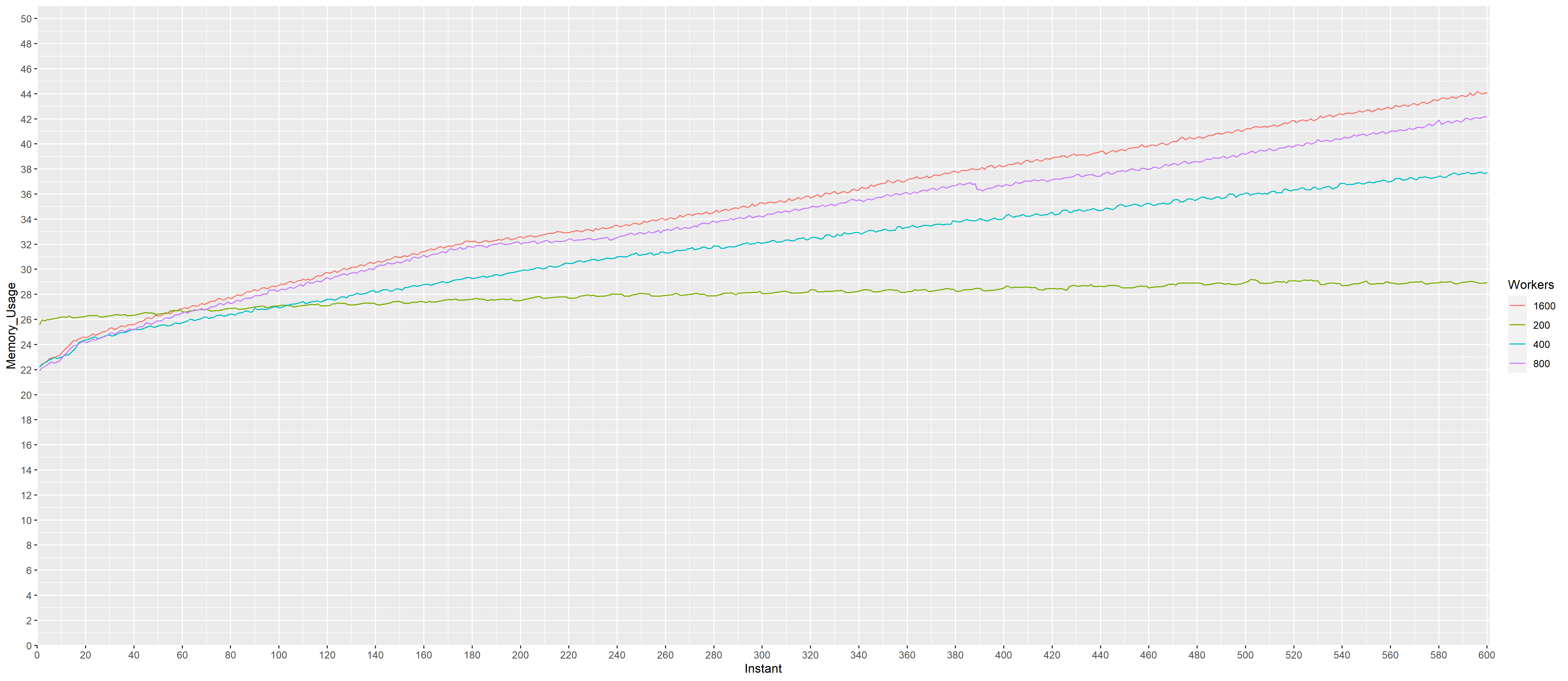}
\caption{Memory usage of the dedicated server during each round of the performance evaluation tests.}
\label{cap:paper_wsdm2022-sec:system-design-subsec:logger-subsubsec:test-infrastucture-fig:memory-usage}
\end{figure}

The pipeline has been initially tested on a dedicated server using an Amazon Elastic Compute Cloud\footnote{\url{https://aws.amazon.com/ec2/}} \index{Amazon!Web Services!Elastic Compute Cloud} instance based on a 2nd generation AMD EPYC \index{AMD EPYC} processor with 4 cores and 8 threads, running at frequencies up to 3.3 GHz and equipped with 16 GB of RAM. Figure~\ref{cap:paper_wsdm2022-sec:system-design-subsec:logger-subsubsec:test-infrastucture-fig:requests-cumulative} shows that the server is able to process up to 5000 requests per second without message loss. Beyond this limit, the server becomes overloaded and progressively fails to accept new requests. The mean (median) time estimated between each log message at this rate is 2.76 (1) seconds. Therefore, at least 5000 workers operating simultaneously are needed to overload the server. This behavior is confirmed by the number of requests managed per second, shown in Figure~\ref{cap:paper_wsdm2022-sec:system-design-subsec:logger-subsubsec:test-infrastucture-fig:requests-second}. CPU and memory usage decrease when more than 800 worker nodes are active, as shown in Figure~\ref{cap:paper_wsdm2022-sec:system-design-subsec:logger-subsubsec:test-infrastucture-fig:cpu-usage} and Figure~\ref{cap:paper_wsdm2022-sec:system-design-subsec:logger-subsubsec:test-infrastucture-fig:memory-usage}. 

The precise cause of the overload cannot be determined, as EC2 does not fully disclose the architectural details. However, a hypothesis can be made based on the data related to requests handled per second and CPU load. Specifically, the CPU is observed to operate at nearly 100\% usage for an entire minute before performance drops abruptly. This may be due to thermal throttling. In such a case, the machine reaches the maximum number of manageable concurrent connections after one minute, then stops accepting new requests.

The AWS-based pipeline has also been evaluated using a cluster of 100 EC2 instances. Each instance is of type \texttt{t3.micro}, which features two cores from an \index{Intel Xeon Platinum 8000} Intel Xeon Platinum 8000 series processor with a clock speed of up to 3.1 GHz, 1 GB of RAM, and up to 10 Gb/s of burst network bandwidth. Each instance runs an image retrieved from \index{Docker!Hub} Docker Hub,\footnote{\url{https://hub.docker.com/}} based on Ubuntu \index{Ubuntu} and containing a Locust \index{Locust} script used to coordinate interactions across the cluster. The cluster is managed using the Elastic Container Service.\footnote{\url{https://aws.amazon.com/ecs/}}

The first round of tests (i.e., with 200 worker nodes only) led to \num{16,538} Lambda\index{Amazon!Web Services!Lambda} invocations to manage the entire set of \num{1653801} requests sent by the cluster, compared to \num{575,000} requests processed by the dedicated server solution. The estimated mean number of requests processed per second by the pipeline is \num{5,512}. A negligible number of messages are lost due to network communication. This confirms that the API layer of the infrastructure can rapidly handle each request and forward it to the message queue, preventing network congestion. The maximum default limit of concurrent Lambda instances is \num{1,000}. This limit was not reached during the test. The maximum number of concurrent invocations observed was 206. In other words, the pipeline can scale effectively.

\subsubsection{Pilot Experiment}

\label{cap:paper_wsdm2022-sec:system-design-subsec:logger-subsubsec:real-world}

A variant of the task proposed by \citet{roitero2021crowd}, in which workers are asked to evaluate the truthfulness of information items, was published on \mturk to gather real log data. The data is used to estimate the impact of the \logger component on the overall cost of a crowdsourcing task deployed using \crowdframe. The component was configured to monitor every log event described in Appendix~\ref{appendix:log-events}.

\begin{figure}[tpb]
  \centering
    \includegraphics[width=0.85\linewidth]{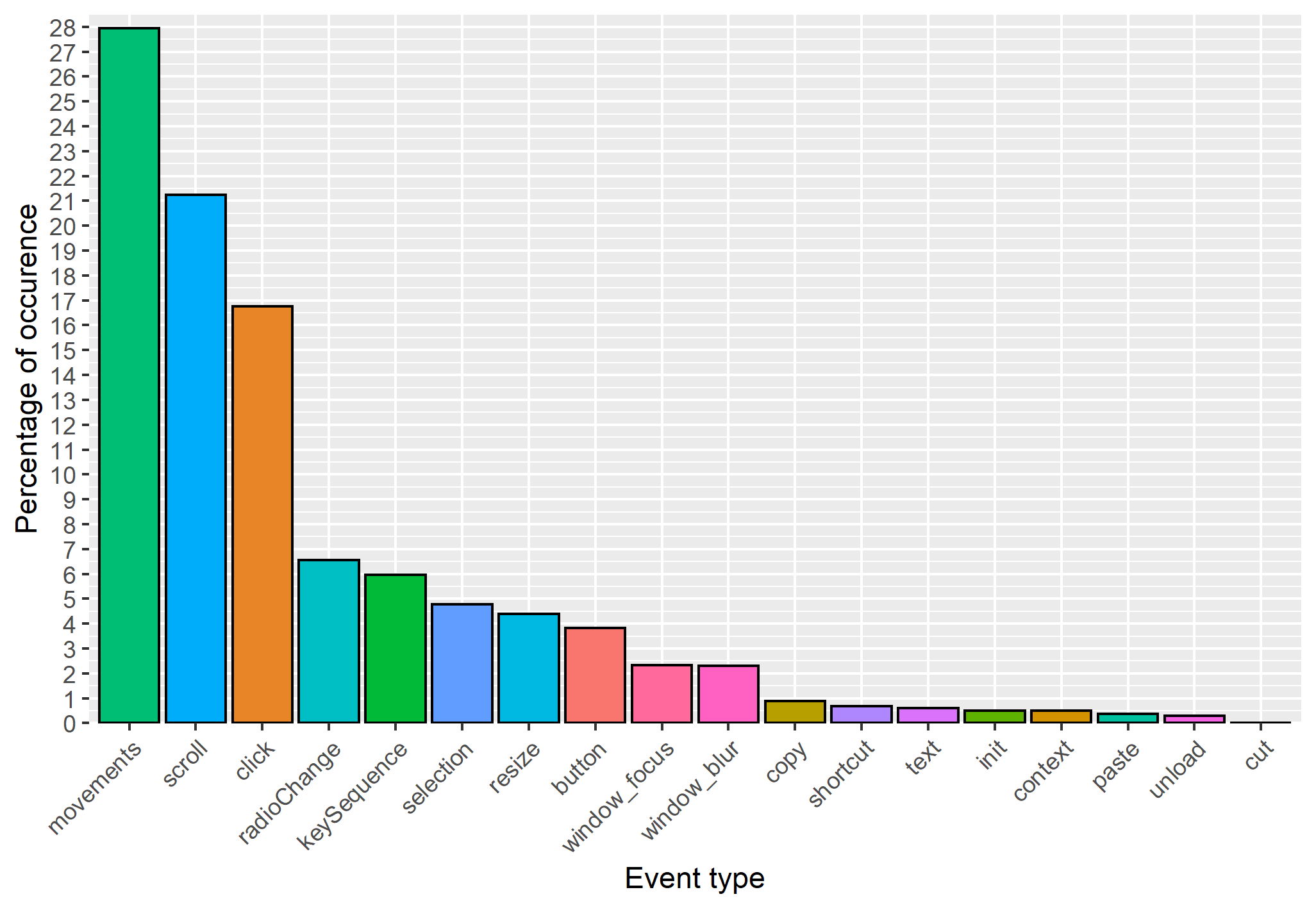}
\caption{Distribution of event types recorded during the pilot test using the \logger component.}
\label{cap:paper_wsdm2022-sec:logger-fig:event-dist}
\end{figure}

The task involved 46 workers, who produced \num{12,051} log requests. Each worker generated a mean of 262 requests and a median of 155. The minimum number of requests sent by a worker was 11, while the maximum was\num{1,192}. The mean task duration was 447 seconds and the median was 230 seconds. The shortest try performed by a worker lasted 2 seconds, while the longest lasted \num{4,588} seconds. The most frequent log event types collected were mouse movements, scrolls, and clicks. Figure~\ref{cap:paper_wsdm2022-sec:logger-fig:event-dist} shows the distribution of event types detected by the component during the task. As expected, the most frequent event detected was mouse movement.

\begin{figure}[tpb]
  \centering
    \includegraphics[width=0.85\linewidth]{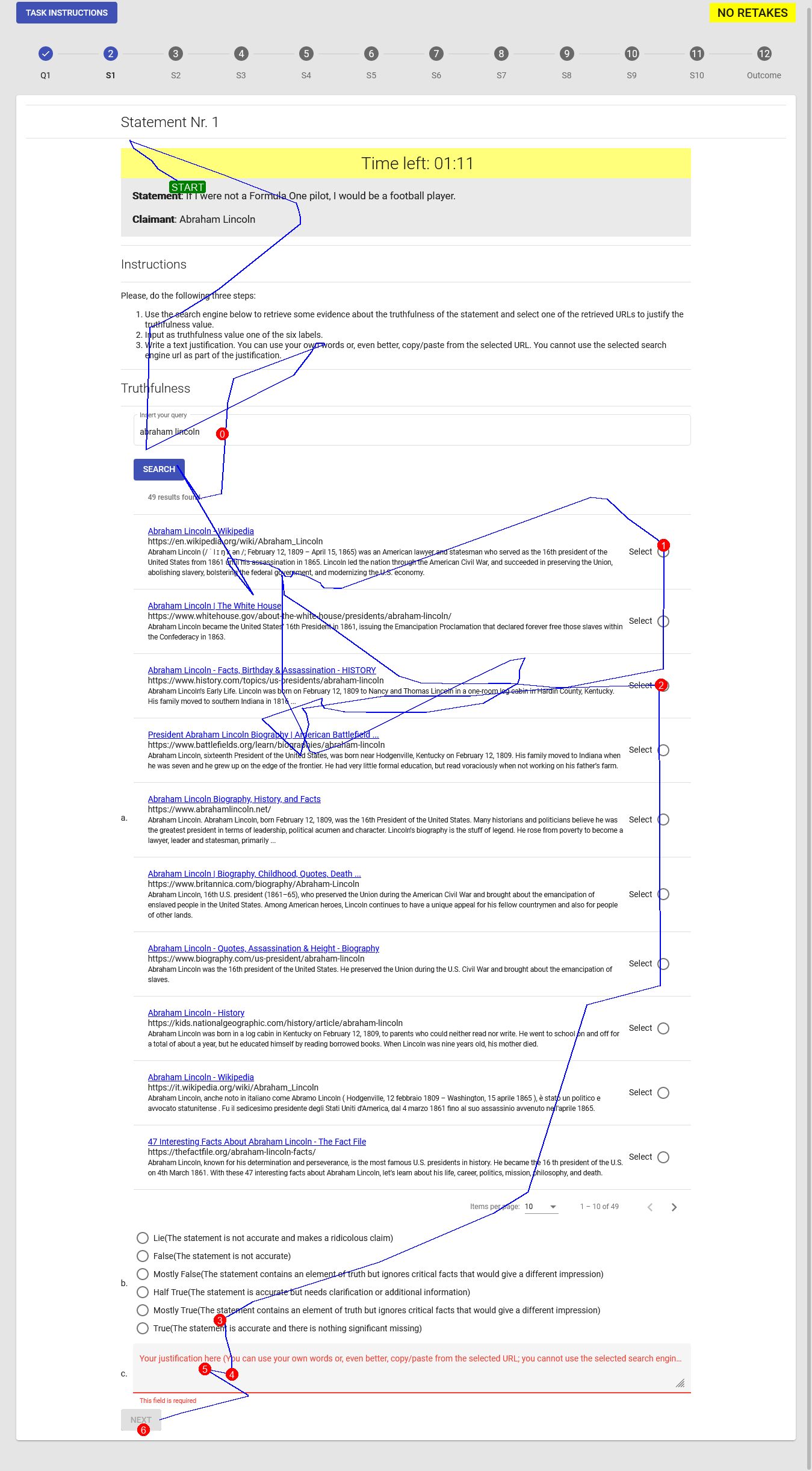}
\caption{Reconstruction of a worker's behavior while evaluating a \index{HIT}HIT element using log data collected by the \logger component.}
\label{cap:paper_wsdm2022-sec:logger-fig:log-trace-document}
\end{figure}

\begin{figure}[tpb]
  \centering
    \includegraphics[width=0.85\linewidth]{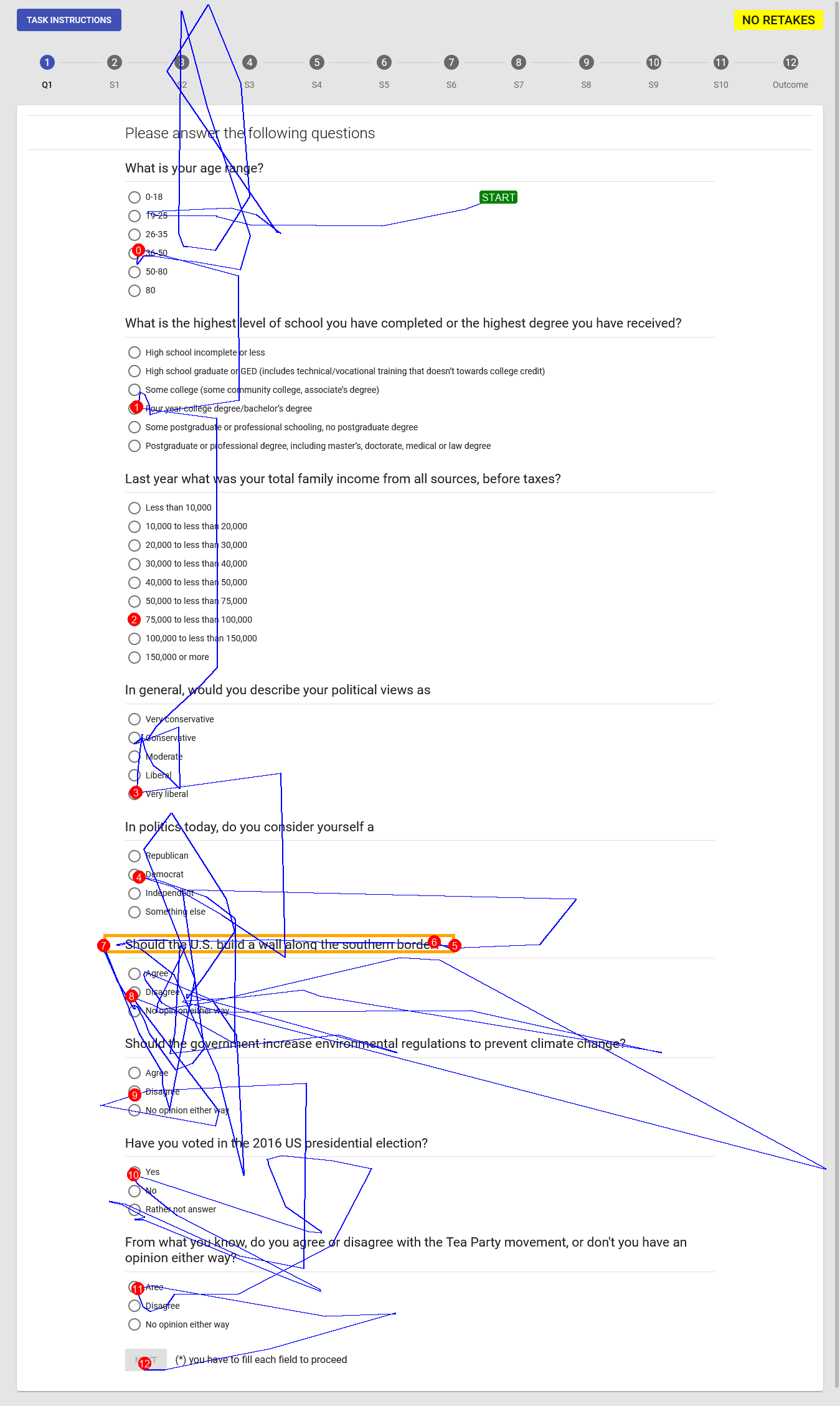}
\caption{Reconstruction of a worker's behavior while answering a questionnaire using log data collected by the \logger component.}
\label{cap:paper_wsdm2022-sec:logger-fig:log-trace-quest}
\end{figure}

Figure~\ref{cap:paper_wsdm2022-sec:logger-fig:log-trace-quest} reconstructs the behavior of a worker answering one of the questionnaires in the deployed task. Numbered points represent clicks, the blue line represents the movement trace, and the orange rectangles define sections of selected text. Figure~\ref{cap:paper_wsdm2022-sec:logger-fig:log-trace-document} reconstructs the behavior of a worker evaluating the first element of the \index{HIT}HIT assigned.

\subsubsection{Cost Estimation}

\label{cap:paper_wsdm2022-sec:system-design-subsec:logger-subsubsec:cost-estimation}

Let us hypothesize a monthly stream of one million log messages, where the average size of each message is based on the pilot test described in Section~\ref{cap:paper_wsdm2022-sec:system-design-subsec:logger-subsubsec:real-world}. Each message thus contains, on average, a header of 550 bytes and a body of 529 bytes, for a total size of 1079 bytes. The logging pipeline described in Figure~\ref{cap:paper_wsdm2022-sec:logger-fig:pipeline} is considered.

The first AWS service to consider is \index{Amazon!Web Services!API Gateway} API Gateway. For an HTTP API, the service applies a fee\footnote{\url{https://aws.amazon.com/api-gateway/pricing/}} of \$1 per million request units up to 300 million requests per month, and \$0.90 per million thereafter. The first million requests per month are free. Request units are defined as 512 KB of data; a request with 513 KB counts as two units. Given the average size of the requests considered, the cost for one million requests is expected to be approximately \$1. Equation~\ref{cap:paper_wsdm2022-sec:system-design-subsec:logger-subsubsec:cost-est-eq:est-api-gateway} shows the detailed computation of this cost component.

\begin{equation}
\begin{split}
\textrm{API Gateway} & = \$1 * (\textrm{Message Number} * \lceil\textrm{Average Message Size (KB) / Billing Factor}\rceil) \\
                     & = \$1 * (1000000 * \lceil\textrm{(1079/1024) / 512}\rceil \\
                     & = \$1 * 1\>\textrm{Billable Requests} = \$1\quad(\$0 \iff \textrm{free tier}) \\
                     & \quad\>\>\textrm{threshold: 300 million requests} \\
                     & \quad\>\>\textrm{free tier: first million requests} 
\end{split}
\label{cap:paper_wsdm2022-sec:system-design-subsec:logger-subsubsec:cost-est-eq:est-api-gateway}
\end{equation}
\myequations{Sample estimation of the API Gateway cost for the use of the \logger component.}

The second service to consider is \index{Amazon!Web Services!Simple Queue Service} Simple Queue Service. The service applies a fee\footnote{\url{https://aws.amazon.com/sqs/pricing/}} of \$0.40 per million request units per month, up to 100 billion requests. A message with a payload size of up to 64 KB is counted as a single request unit. The first million requests per month are free. The service also applies fees for data transfer; however, inbound transfers are free of charge. The same applies to outbound data sent to a Lambda function in the same region. Therefore, the expected cost for data transfer is \$0. Equation~\ref{cap:paper_wsdm2022-sec:system-design-subsec:logger-subsubsec:cost-est-eq:est-sqs} shows the detailed computation of this cost component.

\begin{equation}
\begin{split}
\textrm{Simple Queue Service} & = \$0.0000004000 * (\textrm{Multiplier} * \textrm{Requests Number (M)} ) \\
                    & = \$0.0000004000 * (\textrm{Multiplier} * \textrm{(Message Number} * \\
                    & \quad\> \textrm{(Average Message Size (KB) / Max Payload Size)))} \\
                     & = \$0.0000004000 * (1000000 * ((1079/1024) / 64)) \\
                     & = \$0.0000004000 * (1000000 * (0.016464)) \\
                     & = \$0.0000004000 * 16464\>\textrm{Requests}\\
                     & = \$0.0000004000 * 0\>\textrm{Billable Requests} = \$0 \\
                     & \quad\>\>\textrm{threshold: 1 million requests} \\
                     & \quad\>\>\textrm{free tier: first million requests} 
\end{split}
\label{cap:paper_wsdm2022-sec:system-design-subsec:logger-subsubsec:cost-est-eq:est-sqs}
\end{equation}
\myequations{Sample estimation of SQS cost for the use of the \logger component.}

The third service to consider is \index{Amazon!Web Services!Lambda} Lambda. The pricing of a function depends on several factors, such as the number of executions requested, the duration of each request, and the amount of memory allocated for processing. The performance estimation described in Section~\ref{cap:paper_wsdm2022-sec:system-design-subsec:logger-subsubsec:test-infrastucture} indicates that in an ideal setting, each invocation of a Lambda function retrieves 100 requests and processes them in at most 5 seconds. In a less optimal setting, each invocation retrieves up to 10 requests and executes them in around 500 milliseconds. 

The cost components\footnote{\url{https://aws.amazon.com/lambda/pricing/}} to consider are threefold. The service charges \$0.20 for every million requests. Then, it charges \$0.0000166667 for every GB-second of execution on an x86-based architecture, for the first 6 billion GB-seconds. Lastly, it charges \$0.0000021 for every second of usage of 128 MB of memory. 

In the ideal setting, \num{10,000} function invocations lasting 5 milliseconds each are needed. In the less optimal setting, \num{10000000} invocations lasting 50 milliseconds each are needed. A final average estimation can thus be \num{5000000} function invocations lasting 250 milliseconds each, processing a total of 0.54 GB. Equation~\ref{cap:paper_wsdm2022-sec:system-design-subsec:logger-subsubsec:cost-est-eq:est-lambda} shows the detailed computation of each cost component.

The last service to consider is \index{Amazon!Web Services!DynamoDB} DynamoDB. Section~\ref{cap:paper_wsdm2022-sec:skeleton-subsec:cost-estimation} describes the role of each cost component of DynamoDB. Equation~\ref{cap:paper_wsdm2022-sec:system-design-subsec:logger-subsubsec:cost-est-eq:est-dynamodb} shows the detailed computation of each cost component for the example considered.
Equation~\ref{cap:paper_wsdm2022-sec:system-design-subsec:logger-subsubsec:cost-est-eq:estimation} estimates the cost required to maintain the infrastructure of the \textrm{Logger} component of \crowdframe, assuming one million log requests are processed each month. After a year of usage, the total cost of the \logger component would be roughly \$51.

\begin{equation}
\begin{split}
\textrm{Lambda} & = \$0.0000166667 * \textrm{Max(Invocations Number} * \textrm{Avg. Proc. Time (s)} * \\
                & \quad\>\textrm{(Mem. Allocation (GB)), 0)} + \$0.0000002 * \textrm{Invocations Number} \\
                 & = \$0.0000166667 * \textrm{Max(500000} * (250/1000 * 0.125), 0) + \$0.0000002 * 500000 \\
                & = \$0.0000166667 * \textrm{Max(125000} * \textrm{0.125), 0)} + \$0.10 \\
                & = \$0.0000166667 * \textrm{Max(15625 GB/s, 0)} + \$0.10 \\
                & = \$0.0000166667 * 15625\>\textrm{Billable GB/s} + \$0.10 = \$0.3603\quad(\$0 \iff \textrm{free tier}) \\
                & \quad\>\>\textrm{thresholds: 1 million invocations, first 6 billions GB-Sec/Month} \\
                & \quad\>\>\textrm{free tier: subtract 40000 GB/s from Max() left hand side} 
\end{split}
\label{cap:paper_wsdm2022-sec:system-design-subsec:logger-subsubsec:cost-est-eq:est-lambda}
\end{equation}
\myequations{Sample estimation of the Lambda cost for the use of the \logger component.}

\begin{equation}
\begin{split}
\textrm{DynamoDB} & = \textrm{WRUs} + \textrm{RRUs} + \textrm{Data Storage}\\
                  & = \$1.25 * \textrm{(Message Number} * \lceil\textrm{Avg. Payload Size (KB) / Unit Amount}\rceil) + \\
                  & \quad\> \$0.25 * (0.5 * (\textrm{Message Number } * \lceil\textrm{Avg. Payload Size (KB) / Unit Amount}\rceil)) + \\
                  & \quad\> \$0.25 * \textrm{(Message Number * Avg. Payload Size (GB))} \\
                  & = \$0.00000125 * (1000000 * \lceil(1079/1024)/1\rceil) + \\
                  & \quad\> \$0.00000025 * (0.5 * (1000000 * \lceil(1079/1024)/4\rceil)) + \\
                  & \quad\> \$0.25 * (1000000 * ((1079/1024)/1024)/1024) \\
                  & = \$0.00000125 * (1000000 * 2\>\textrm{Billable WRUs}) + \\
                  & \quad\>\$0.00000025 * (1000000 * 0.5\>\textrm{Billable RRUs}) + \\
                  & \quad\>\$0.25 * (1\>\textrm{GB}) \\
                  & = \$0.00000125 * 2000000 + \$0.00000025 * 500000 + \$0.25 * 1 \\
                  & = \$2.50 + \$0.125 + \$0.25 = \$2.875\quad(\$2.67 \iff \textrm{free tier}) \\
                  & \quad\>\>\textrm{threshold: 1 million WRUs, 1 million RRUs} \\ 
                  & \quad\>\>\textrm{note: on-demand capacity mode, standard table class} \\
                  & \quad\>\>\textrm{free tier: 25 GB/Month data storage} 
\end{split}
\label{cap:paper_wsdm2022-sec:system-design-subsec:logger-subsubsec:cost-est-eq:est-dynamodb}
\end{equation}
\myequations{Sample estimation of the DynamoDB cost for the use of the \logger component.}

\begin{equation}
\begin{split}
\textrm{Logger} & = \textrm{API Gateway} + \textrm{SQS} + \textrm{Lambda} + \textrm{DynamoDB} \\
              & = \$1 + \$0 + \$2.875 + \$0.3603 \\ 
              & = \$4.2353
\end{split}
\label{cap:paper_wsdm2022-sec:system-design-subsec:logger-subsubsec:cost-est-eq:estimation}
\end{equation}
\myequations{Sample estimation of the overall cost for the use of the \logger component.}

\subsubsection{Log Events}

\label{appendix:log-events}

The following lists the events monitored by the \logger component of \crowdframe, as described in Section~\ref{cap:paper_wsdm2022-sec:system-design-subsec:logger}. The payloads presented are encapsulated within the \spverb|detail| object of the base payload shown in Listing~\ref{cap:paper_wsdm2022-sec:system-design-subsec:logger-subsubsec:event-listening-lst:log-payload}.

\myparagraph{Context} The log message with \lstinline[style=jsonstyle]!{"type": "context"}! contains information about the user agent and the IP address of the worker. It is the only message not containing the \spverb|section| field.

\begin{lstlisting}[style=jsonstyle]
{
  "ua": "string",
  "ip": "string"
}
\end{lstlisting}

\myparagraph{Mouse Movements} The log message with \lstinline[style=jsonstyle]!{"type": "movements"}!describes a mouse movement performed by the worker. The event is detected every 100 ms. The timestamp and (x, y) coordinates are mapped to a dictionary and buffered. The dictionaries in the buffer are pushed into an array within the \spverb|details| field after a dwell time of 500 ms. Each dictionary is added to the \spverb|points| array.

\begin{lstlisting}[style=jsonstyle]
{
  "section": "string",
  "points": [
    {
      "timeStamp": "string",
      "x": "integer",
      "y": "integer"
    },
    {
      "timeStamp": "string",
      "x": "integer",
      "y": "integer"
    },
    ...
  ]
}
\end{lstlisting}\myparagraph{Mouse Clicks} The log message with  \lstinline[style=jsonstyle]!{"type": "click"}! describes a mouse click performed by the worker. The event is detected when the left or right mouse button is pressed. Detection occurs after a fixed debounce time to prevent spamming excessive log requests. The timestamps of the first and last clicks in each click sequence are stored. Furthermore, the \spverb|(x, y)| coordinates, the targeted DOM element, and the number of clicks in the sequence are logged.

\begin{lstlisting}[style=jsonstyle]
{
  "section": "string",
  "mouseButton": "right || left",
  "startTime": "string",
  "endTime": "string",
  "x": "integer",
  "y": "integer",
  "target": "string",
  "clicks": "integer"
}
\end{lstlisting}

\myparagraph{Button Click} The log message with \lstinline[style=jsonstyle]!{"type": "button"}! describes a mouse click performed by the worker on a \spverb|button| DOM element. This event has a debounce time similar to the base click event, but different information is extracted. The information includes the targeted DOM button, the timestamp of the click, and the \spverb|(x, y)| coordinates.

\begin{lstlisting}[style=jsonstyle]
{
  "section": "string",
  "timestamp": "string",
  "button": "string",
  "x": "integer",
  "y": "integer"
}
\end{lstlisting}

\myparagraph{Shortcuts} The log message with \lstinline[style=jsonstyle]!{"type": "shortcut"}! records a keystroke combination used by the worker. Only combinations involving the keys \spverb|CTRL| or \spverb|ALT| are monitored.
Key combinations corresponding to shortcuts are tracked. From the event key pressed for the shortcut, the following fields are extracted:
\begin{itemize}[label=--]
    \item \lstinline[style=jsonstyle]!{"ctrl": "boolean"}! set to \spverb|true| if the \spverb|CTRL| or \spverb|CMD| key is pressed
    \item \lstinline[style=jsonstyle]!{"alt": "boolean"}! set to \spverb|true| if the \spverb|ALT| key is pressed
    \item \lstinline[style=jsonstyle]!{"key": "string"}! contains the value of the key pressed within the detected shortcut
\end{itemize}

\begin{lstlisting}[style=jsonstyle]
{
  "section": "string",
  "timestamp": "string",
  "ctrl": "boolean",
  "alt": "boolean",
  "key": "string"
}
\end{lstlisting}

\myparagraph{Keypress} The log message with \lstinline[style=jsonstyle]!{"type": "keySequence"}! describes a sequence of keystrokes performed by the worker. Each keypress is stored as a dictionary in a buffer array. Each dictionary contains the timestamp and the key pressed. The full sentence is reconstructed. The event handling completes after a dwell time of 1 second.

\begin{lstlisting}[style=jsonstyle]
{
  "section": "string",
  "keySequence": [
    {
      "timeStamp": "string",
      "key": "string"
    },
    {
      ...
    }
    ...
  ],
  "sentence": "string"
}
\end{lstlisting}

\myparagraph{Selection} The log message with \lstinline[style=jsonstyle]!{"type": "selection"}! describes a selection operation performed by the worker. The logged information includes the start and end timestamps and the content of the selection.

\begin{lstlisting}[style=jsonstyle]
{
  "section": "string",
  "startTime": "string",
  "endTime": "string",
  "selected": "string"
}
\end{lstlisting}

\myparagraph{Before Unload, Focus, and Blur} The log message \lstinline[style=jsonstyle]!{"type": "unload || window_focus || window_blur"}! describes the last log requests produced when the worker closes the task page or when the window gains or loses focus. The information includes only the corresponding timestamp.

\begin{lstlisting}[style=jsonstyle]
{
  "section": "string",
  "timestamp": "string"
}
\end{lstlisting}

\myparagraph{Scroll} The log message with \lstinline[style=jsonstyle]!{"type": "scroll"}! describes a scroll operation performed by the worker. Like mouse clicks and movements, a debounce time of 300 ms is applied to prevent spamming excessive log requests. The logged information includes the start and end timestamps and the (x, y) coordinates of the top-left corner from which the scroll started.

\begin{lstlisting}[style=jsonstyle]
{
  "section": "string",
  "startTimestamp": "string",
  "endTimestamp": "string",
  "x": "integer",
  "y": "integer"
}
\end{lstlisting}

\myparagraph{Resize} The log message with \lstinline[style=jsonstyle]!{"type": "resize"}! is detected when the worker resizes the task page window. The logged information includes the start and end timestamps and the updated window size.

\begin{lstlisting}[style=jsonstyle]
{
  "section": "string",
  "width": "integer",
  "height": "integer",
  "scrollWidth": "integer",
  "scrollHeight": "integer",
  "timestamp": "string"
}
\end{lstlisting}

\myparagraph{Copy, Cut, and Paste} The log message with \lstinline[style=jsonstyle]!{"type": "copy || cut || paste"}! is detected when the worker cuts, copies, or pastes content during the task. The information includes the timestamp of the event. The attribute \spverb|target| refers to the DOM element targeted by a \spverb|copy| or \spverb|cut| event. For a \spverb|paste| event, the \spverb|target| attribute is replaced by the \spverb|text| attribute, which contains the pasted text.

\begin{lstlisting}[style=jsonstyle]
 {
   "section": "string",
   "timestamp": "string",
   "target": "string"
 }
\end{lstlisting}

\myparagraph{Text Input Backspace and Blur} The log message with \lstinline[style=jsonstyle]!{"type": "text"}! is detected when the backspace key is pressed inside a text input. The event is also detected when the input loses focus (blur). The information includes the timestamp and the text contained in the input.

\begin{lstlisting}[style=jsonstyle]
{
  "section": "string",
  "timestamp": "string",
  "text": "string"
}
\end{lstlisting}

\myparagraph{Radio Group Input} The event with \lstinline[style=jsonstyle]!{"type": "radioChange"}! is detected when the worker chooses a new value within a radio button control. The information includes the timestamp and the new value of the radio button.

\begin{lstlisting}[style=jsonstyle]
{
  "section": "string",
  "timestamp": "string",
  "group": "string",
  "value": "string"
}
\end{lstlisting}

\myparagraph{Search Engine Queries and Results} The event with \lstinline[style=jsonstyle]!{"type": "query || queryResults"}! is detected when the worker queries the search engine and when results are retrieved. The information includes the query text in the former case and the list of URLs retrieved in the latter.

\begin{lstlisting}[style=jsonstyle]
{
  "section": "string",
  "query": "string"
}
\end{lstlisting}

\begin{lstlisting}[style=jsonstyle]
{
  "section": "string",
  "urlArray": [
    ...
  ]
}
\end{lstlisting}

\section{System Usage}

\label{cap:paper_wsdm2022-sec:usage}

Section~\ref{cap:paper_wsdm2022-sec:usage-sec:getting-started} describes the prerequisites and initial setup required to begin using \crowdframe. Section~\ref{cap:paper_wsdm2022-sec:usage-sec:env-var} provides the complete set of environment variables available to customize the system's behavior. Section~\ref{cap:paper_wsdm2022-sec:usage-sec:build-output} explains the output produced by a successful execution of the initialization script. Section~\ref{cap:paper_wsdm2022-sec:usage-sec:task-config} outlines how to access the task configuration interface, and Section~\ref{cap:paper_wsdm2022-sec:usage-sec:hits-format} details the format required to define the \index{HIT}HITs of a task, along with the procedure for allocating them. Section~\ref{cap:paper_wsdm2022-sec:usage-sec:quality-checks} describes how to implement a custom quality check on the answers provided by workers. Finally, Section~\ref{cap:paper_wsdm2022-sec:usage-sec:local-dev} illustrates how to enable the local development modality.

\subsection{Getting Started}

\label{cap:paper_wsdm2022-sec:usage-sec:getting-started}

The GitHub \index{GitHub} repository contains detailed and updated instructions.\footnote{\url{https://github.com/Miccighel/Crowd_Frame#readme}} Four (4) main prerequisites are required to start using \crowdframe. In more detail, these prerequisites are the \index{Amazon!Web Services!Command Line Interface}\textsf{AWS Command Line Interface}\footnote{\url{https://docs.aws.amazon.com/cli/latest/userguide/getting-started-install.html}} and distributions of \index{Node.js} \textsf{Node.js}\footnote{\url{https://nodejs.org/it/download/}} and \index{Python} \textsf{Python}.\footnote{\url{https://www.python.org/downloads/}} \index{Docker} \textsf{Docker}\footnote{\url{https://docs.docker.com/get-docker/}} may be optionally needed. \textsf{Yarn}\footnote{\url{https://yarnpkg.com/}} \index{Yarn} is used to manage the software dependencies. There are 15 steps to follow to successfully initialize \crowdframe and the overall infrastructure. The first 13 steps must be performed only once. Then, the task requester can vary the configuration and repeat the steps \#14 and \#15.
\begin{enumerate}
    \item Create a new \textsf{Amazon AWS} account.
    \item Create a new  \index{Amazon!Web Services!Identity Access Management}\textsf{IAM User}\footnote{\url{https://docs.aws.amazon.com/IAM/latest/UserGuide/id_users.html}} using a custom name such as \spverb|your_iam_user|.
    \item Attach the \spverb|AdministratorAccess| policy (cfr. Listing~\ref{cap:paper_wsdm2022-sec:usage-subsec:getting-started:adm-access-policy}).
    \item Generate a new access key pair.
    \item Store the access key in the \spverb|credentials.json| file (cfr. Listing~\ref{cap:paper_wsdm2022-sec:usage-subsec:getting-started:credentials}).
    \item Clone the repository in the local filesystem. 
    \item Enable the Yarn global library using the command: \spverb|corepack enable|.
    \item Move to the repository's folder: \spverb|cd ~/path/to/project|.
    \item Install the dependencies using the command: \spverb|yarn install|.
    \item Move to the data folder using the command: \spverb|cd data|.
    \item Create the environment file \spverb|.env| at path \spverb|your_repo_folder/data/.env|.
    \item Add to the environment file the subset of required variables (cfr. Listing \ref{cap:paper_wsdm2022-sec:usage-subsec:getting-started:env-var}).
    \item Install the Python packages required (cfr. Listing \ref{cap:paper_wsdm2022-sec:usage-subsec:getting-started:python-packages})
    \item Run the Python script \spverb|init.py|. The script will:
    \begin{itemize}[label=--]
        \item read the environment variables;
        \item setup the whole AWS infrastructure;
        \item generate a sample task configuration;
        \item deploy the sample source files on the public bucket.
    \end{itemize}
    \item Visit the task deployed using the link in the format:
    \begin{itemize}[label=--]
        \item  \url{https://your_deploy_bucket.s3.your_aws_region.amazonaws.com/your_task_name/your_batch_name/index.html}
    \end{itemize}
\end{enumerate}

\crowdframe interacts with various Amazon Web Services to deploy crowdsourcing tasks, store the resulting data, and perform other operations, as described in Section~\ref{cap:paper_wsdm2022-sec:system-design}. Each service used falls within the \index{Amazon!Web Services!Free Tier} AWS Free Tier program.\footnote{\url{https://aws.amazon.com/free/}} The task requester can set a budget limit using the \spverb|budget_limit| environment variable. Service usage will be blocked automatically if or when this limit is exceeded.

\begin{lstlisting}[style=jsonstyle, caption={Administrator access policy to be attached to the IAM user.}, label={cap:paper_wsdm2022-sec:usage-subsec:getting-started:adm-access-policy}]
{
  "Version": "2012-10-17",
  "Statement": [
    {
      "Effect": "Allow",
      "Action": "*",
      "Resource": "*"
    }
  ]
}
\end{lstlisting}

\begin{lstlisting}[style=textstyle, caption={Sample \protect\UseVerb{credentials.json} file used to store the IAM user's access key.}, label={cap:paper_wsdm2022-sec:usage-subsec:getting-started:credentials}]
[your_iam_user]
region = your_region
aws_access_key_id = your_key
aws_secret_access_key = your_secret
\end{lstlisting}

\begin{lstlisting}[style=textstyle,
  caption={Example subset of environment variables required to configure \crowdframe.},
  label={cap:paper_wsdm2022-sec:usage-subsec:getting-started:env-var}]
mail_contact=your_email_address
budget_limit=your_usd_budget_limit
task_name=your_task_name
batch_name=your_batch_name
admin_user=your_admin_username
admin_password=your_admin_password
server_config=none
aws_region=your_aws_region
aws_private_bucket=your_private_bucket_name
aws_deploy_bucket=your_deploy_bucket_name
\end{lstlisting}

\begin{lstlisting}[style=textstyle,
  caption={Python packages required for initializing \crowdframe. See the repository for the full and updated list.},
  label={cap:paper_wsdm2022-sec:usage-subsec:getting-started:python-packages}]
aiohttp==3.9.5
boto3==1.34.143
botocore==1.34.143
chardet==5.2.0
datefinder==0.7.3
docker==7.1.0
ipinfo==4.4.2
mako==1.3.5
numpy==1.26.4
pandas==2.2.2
pycountry==24.6.1
python-dotenv==1.0.1
python_dateutil==2.9.0
python_on_whales==0.71.0
pytz==2024.1
requests==2.32.3
rich==13.7.1
setuptools==70.3.0
toloka-kit==1.2.3
tqdm==4.66.4
\end{lstlisting}

\subsection{Environment Variables}

\label{cap:paper_wsdm2022-sec:usage-sec:env-var}

Table~\ref{cap:paper_wsdm2022-sec:usage-sec:env-var-tab:variables} describes the environment variables that can be set in the environment file to customize the behavior of \crowdframe.

\begin{longtable}{lp{6cm}cp{4cm}}
\caption{Environment variables used to customize \crowdframe.}
\label{cap:paper_wsdm2022-sec:usage-sec:env-var-tab:variables} \\
\toprule
\textbf{Variable} & \textbf{Description} & \textbf{Required} & \textbf{Value} \\
\midrule
\endfirsthead
\toprule
\textbf{Variable} & \textbf{Description} & \textbf{Required} & \textbf{Value} \\
\midrule
\endhead
\multicolumn{4}{r}{\footnotesize\itshape Continues on the next page} \\
\endfoot
\bottomrule
\endlastfoot
\spverb|profile_name| & Name of the IAM profile created during step 2 (cfr.\ Section~\ref{cap:paper_wsdm2022-sec:usage-sec:getting-started}). If unspecified, the default value is \spverb|default|. & No & \spverb|your_iam_user| \\
\midrule
\spverb|mail_contact| & Contact email to receive AWS budgeting-related communications & Yes & Valid email address \\ 
\midrule
\spverb|platform| & Platform used to publish the crowdsourcing task. Set to \spverb|none| to recruit workers manually. & Yes & \spverb|none|, \spverb|mturk|, \spverb|prolific|, or \spverb|toloka| \\ 
\midrule
\spverb|budget_limit| & Maximum monthly budget allowed in USD, e.g., 5.0 & Yes & Positive float number \\ 
\midrule
\spverb|task_name| & Identifier of the crowdsourcing task & Yes & Any string \\ 
\midrule
\spverb|batch_name| & Identifier of the current batch & Yes & Any string \\
\midrule
\spverb|batch_prefix| & Prefix for one or more task batch identifiers. Used to filter the final result set & No & Any string \\ 
\midrule
\spverb|admin_user| & Username of the admin user. Allows unlocking the \generator & Yes & Any string \\ 
\midrule
\spverb|admin_password| & Password of the admin user. Allows unlocking the \generator & Yes & Any string \\ 
\midrule
\spverb|aws_region| & AWS account region & Yes & Valid AWS region identifier, e.g., \spverb|us-east-1|\footnote{\url{https://docs.aws.amazon.com/AmazonRDS/latest/UserGuide/Concepts.RegionsAndAvailabilityZones.html}} \\
\midrule
\spverb|aws_private_bucket| & Name of the private S3 bucket to store task configuration and data & Yes & Unique string across AWS \\ 
\midrule
\spverb|aws_deploy_bucket| & Name of the public S3 bucket to deploy source code & Yes & Unique string across AWS \\ 
\midrule
\spverb|server_config| & Location of the worker behavior logging interface. Set to \spverb|aws| to deploy AWS infrastructure, \spverb|custom| to provide a custom endpoint, or \spverb|none| to disable logging & Yes & \spverb|aws|, \spverb|custom|, or \spverb|none| \\ 
\midrule
\spverb|enable_crawling| & Enables crawling of results retrieved by the search engine & No & \spverb|true| or \spverb|false| \\ 
\midrule
\spverb|enable_solver| & Allows deploying the \index{HIT}HITs solver locally. Requires Docker. This feature is \emph{experimental} & No & \spverb|true| or \spverb|false| \\ 
\midrule
\spverb|prolific_completion_code| & \prolific study completion code. Required if \spverb|prolific| is chosen as the platform & No & Valid \prolific completion code \\ 
\midrule
\spverb|toloka_oauth_token| & Token to access \toloka API. Required if \spverb|toloka| is chosen as the platform & No & Valid Toloka OAuth token \\ 
\midrule
\spverb|ip_info_token| & API key for \spverb|ipinfo.com| tracking functionalities & No & Valid API key \\ 
\midrule
\spverb|ip_geolocation_api_key| & API key for \spverb|ipgeolocation.io| tracking functionalities & No & Valid API key \\ 
\midrule
\spverb|ipapi_api_key| & API key for \spverb|ipapi.com| tracking functionalities & No & Valid API key \\ 
\midrule
\spverb|user_stack_token| & API key for \spverb|userstack.com| tracking functionalities & No & Valid API key \\  
\midrule
\spverb|bing_api_key| & API key for \spverb|BingWebSearch| search provider & No & Valid Bing Web Search API key \\ 
\midrule
\spverb|fake_json_token| & API key for \spverb|FakerWebSearch| search provider, returns dummy responses for testing & No & Valid \spverb|fakeJSON.com| API key \\ 
\end{longtable}

\subsection{Build Output}

\label{cap:paper_wsdm2022-sec:usage-sec:build-output}

Each execution of the initialization script of \crowdframe, described in Section~\ref{cap:paper_wsdm2022-sec:usage-sec:getting-started}, populates a build folder on the local filesystem at the path \spverb|cd ~/path/to/project/build/|. The folder may contain up to six subfolders, depending on the crowdsourcing platforms used. Table~\ref{cap:paper_wsdm2022-sec:usage-sec:build-output-tab:subfolders-summary} provides a general description of the contents of each subfolder.

\begin{table}[tpb]
\centering
\caption{Folder structure of the output of a \crowdframe build.}
\begin{tabular}{lp{7cm}}
  \toprule
   \textbf{Subfolder} & \textbf{Description}  \\
    \midrule
	\spverb|build/task/| & Contains the configuration of the task to deploy. \\
	\midrule
	\spverb|build/config/| & Contains the encrypted credentials used to unlock the \spverb|Generator| component. \\
	\midrule
	\spverb|build/environments/| & Contains the development and production environments. \\
	\midrule
	\spverb|build/mturk/| & Contains three files needed to publish the task using \mturk. \\
	\midrule
	\spverb|build/toloka/| & Contains six files needed to publish the task using Toloka. \\
	\midrule
	\spverb|build/skeleton/| & Contains an interface between the \index{HIT}HITs and the application and a file used to implement quality checks. \\
	\midrule
	\spverb|build/deploy/| & Contains the source files of the task to deploy. \\
	\bottomrule
  \end{tabular}
\label{cap:paper_wsdm2022-sec:usage-sec:build-output-tab:subfolders-summary}
\end{table}

\subsubsection{\protect\UseVerb{build/task}}

The folder contains the 8 configuration files of the deployed task. The \generator component described in Section~\ref{cap:paper_wsdm2022-sec:system-design-subsec:generator} populates each of these files along the setup steps. In other words, every task deployed using \crowdframe is configured using 8 special JSON \index{JSON} files. Table~\ref{cap:paper_wsdm2022-sec:usage-sec:build-output-tab:task-config} provides a general description of the content of each configuration file.

\begin{table}[tpb]
\centering
\caption{Configuration files of a task deployed using \crowdframe.}
\begin{tabular}{lp{7cm}}
  \toprule
   \textbf{File} & \textbf{Description}  \\
   \midrule
	\spverb|hits.json| & Contains the whole set of \index{HIT}HITs of the task. \\
    \midrule
	\spverb|questionnaires.json| & Contains the definition of each questionnaire of the task. \\
	\midrule
	\spverb|dimensions.json| & Contains the definitions of each evaluation dimension of the task. \\
	\midrule
	\spverb|instructions_general.json| & Contains the general instructions of the task. \\
	\midrule
	\spverb|instructions_evaluation.json| & Contains the evaluation instructions of the task. \\
	\midrule
	\spverb|search_engine.json| & Contains the configuration of the custom search engine. \\
	\midrule
	\spverb|task.json| & Contains several general settings of the task. \\
	\midrule
	\spverb|workers.json| & Contains settings concerning worker access to the task. \\
	\bottomrule
  \end{tabular}
\label{cap:paper_wsdm2022-sec:usage-sec:build-output-tab:task-config}
\end{table}

\subsubsection{\protect\UseVerb{build/environments}}

The folder contains the development and production environments of \crowdframe. Each environment includes the variables listed in Table~\ref{cap:paper_wsdm2022-sec:usage-sec:env-var-tab:variables} along with additional data. These environment files are overwritten each time the initialization script is executed to reflect any changes in the environment variables.

\subsubsection{\protect\UseVerb{build/config}}

This folder contains the encrypted credentials used to unlock access to the \generator component. It includes a single file named \spverb|admin.json|, which stores a hash generated using the \index{HMAC} HMAC \cite{bellare96hmac} scheme. The hash is derived from the values of the \spverb|admin_user| and \spverb|admin_password| variables listed in Table~\ref{cap:paper_wsdm2022-sec:usage-sec:env-var-tab:variables}.

\subsubsection{\protect\UseVerb{build/skeleton}}

The folder contains an interface called \spverb|Document| in the file \spverb|document.ts|. This interface acts as a bridge between the Angular application and the configured \index{HIT}HITs. It is generated during the execution of the initialization script if the attributes of the \index{HIT}HITs change. The folder also contains a file named \spverb|goldChecker.ts|, which provides a static method that developers can implement to perform custom quality checks when enabled for one or more evaluation dimensions.

\subsubsection{\protect\UseVerb{build/deploy}}

This folder contains the three source files built by \textsf{Angular} \index{Angular} that the initialization script deploys to the public S3 bucket. Table~\ref{cap:paper_wsdm2022-sec:usage-sec:build-output-tab:source-files} describes each of these source files. The role of the public bucket is simply to make them publicly accessible on the Internet. The entire client-side codebase is included, which means a developer can deploy these files on a private server without issues initializing the task.

\begin{table}[tpb]
\centering
\caption{Source files of a task deployed using \crowdframe.}
\begin{tabular}{lp{6cm}}
  \toprule
   \textbf{File} & \textbf{Description}  \\
   \midrule
	\spverb|index.html| & Markup of the task deployed. \\
    \midrule
	\spverb|styles.css| & Styling of the task deployed. \\
	\midrule
	\spverb|scripts.js| & Client side code of the task deployed. \\
	\bottomrule
  \end{tabular}
\label{cap:paper_wsdm2022-sec:usage-sec:build-output-tab:source-files}
\end{table}

\subsubsection{\protect\UseVerb{build/mturk}}

This folder contains the wrapper initialized for deployment on \mturk, saved in a file named \spverb|index.html|. The wrapper is initialized for the current task based on a general model stored in the file \spverb|model.html|. The initialization script uses the \index{Mako} template engine called \spverb|Mako|\footnote{\url{https://www.makotemplates.org/}} to generate the wrapper. 

The folder also contains a file named \spverb|tokens.csv|, which includes the input and output tokens required by the platform. These tokens correspond to those defined in the file that contains the complete set of \index{HIT}HIT definitions.

\subsubsection{\protect\UseVerb{build/toloka}}

The folder contains the wrapper initialized for deployment on \toloka, along with the input and output data specifications and the tokens to be provided. Table~\ref{cap:paper_wsdm2022-sec:usage-sec:build-output-tab:toloka-files} describes each file. The wrapper is generated using Mako from a model, similarly to \mturk. However, the final output is split into three different files due to \toloka's requester interface. The files \spverb|interface.html|, \spverb|interface.css|, and \spverb|interface.js| contain, respectively, the markup, the styling, and the client-side code of the wrapper. 

The two JSON \index{JSON} files provide the input and output data specifications to be used for the task deployment. The TSV \index{TSV} file contains the input and output tokens to be provided to the platform.

\begin{table}[tpb]
\centering
\caption{Content of the build folder to deploy a task on \toloka \index{Toloka}.}
\begin{tabular}{lp{8cm}}
  \toprule
   \textbf{File} & \textbf{Description}  \\
   \midrule
	\spverb|index.html| & Markup of the wrapper. \\
    \midrule
	\spverb|styles.css| & Styling of the wrapper. \\
	\midrule
	\spverb|scripts.js| & Client-side code of the wrapper. \\
	\midrule
	\spverb|input_specification.json| & Input data specification file. \\
	\midrule
	\spverb|output_specification.json| & Output data specification file. \\
	\midrule
	\spverb|tokens.tsv| & Token file to be provided to the platform. \\
	\bottomrule
  \end{tabular}
\label{cap:paper_wsdm2022-sec:usage-sec:build-output-tab:toloka-files}
\end{table}

\subsection{Task Configuration}

\label{cap:paper_wsdm2022-sec:usage-sec:task-config}

The \generator component must be accessed to configure the crowdsourcing task deployed. This involves four steps:
\begin{enumerate}
    \item Open the administrator panel by appending the suffix \spverb|?admin=true| to the task's URL (see step \#15, Section~\ref{cap:paper_wsdm2022-sec:usage-sec:getting-started}). 
    \item Click the \spverb|Generate| button to open the login form
    \item Input the admin credentials set in the corresponding environment variables (see Table~\ref{cap:paper_wsdm2022-sec:usage-sec:env-var-tab:variables})
    \item Proceed through each configuration step and upload the final configuration
\end{enumerate}

The final configuration can be uploaded using the \spverb|Upload| button. Table~\ref{cap:paper_wsdm2022-sec:system-design-sec:generator-tab:steps} provides a summary of each task configuration step. The initialization script synchronizes the local and remote configurations bidirectionally by downloading or uploading the most recent one.

\subsection{HITs Allocation}

\label{cap:paper_wsdm2022-sec:usage-sec:hits-format}

The \index{HIT}HITs for a crowdsourcing task designed and deployed using \crowdframe must be stored in a specific \index{JSON} \textsf{JSON} file. This file can be manually uploaded during the configuration of the crowdsourcing task, as described in Section~\ref{cap:paper_wsdm2022-sec:system-design-subsec:generator}. The file must comply with a specific format satisfying five requirements:
\begin{enumerate}
    \item It must contain an array of \index{HIT}HITs (also called \emph{units}).
    \item Each \index{HIT}HIT must have a unique input/output token attribute pair.
    \item The number of elements to assess must be specified for each \index{HIT}HIT.
    \item Each element must have an attribute named \spverb|id|.
    \item Each element can have an arbitrary number of additional attributes.
\end{enumerate}

Listing~\ref{cap:paper_wsdm2022-sec:system-design-subsec:hits-lst:hit-fragment} shows a valid example consisting of a single \index{HIT}HIT to be assessed within a crowdsourcing task configured using \crowdframe.

\begin{lstlisting}[style=jsonstyle,
  caption={Example of a valid \index{HIT}HIT set containing one task unit, designed and deployed using \crowdframe.},
  label={cap:paper_wsdm2022-sec:system-design-subsec:hits-lst:hit-fragment}]
[
  {
    "unit_id": "unit_0",
    "token_input": "ABCDEFGHILM",
    "token_output": "MNOPQRSTUVZ",
    "documents_number": 1,
    "documents": [
      {
        "id": "identifier_1",
        "text": "Lorem ipsum dolor sit amet"
      }
    ]
  }
]
\end{lstlisting}

\subsubsection{Manual Approach}

The requester can manually build the set of \index{HIT}HITs in the format required by \crowdframe. First, an attribute is selected whose values partition the dataset into different classes. The core idea is to create element pools, one for each class. Four parameters must be defined:
\begin{itemize}[label=--]
    \item the total number of elements to allocate across all \index{HIT}HITs
    \item the number of elements per \index{HIT}HIT
    \item the number of elements to allocate for each class
    \item the number of repetitions required for each element
\end{itemize}

Each class-specific pool must be updated to include all required repetitions. The set of \index{HIT}HITs is then built iteratively using a loop. A helper function is recommended to sample elements from each class until a sample without duplicates is obtained. If the condition is met, the sampled elements are removed from their respective pools.
The total number of \index{HIT}HITs depends on the chosen parameters. The resulting allocation matrix, i.e., the list of elements assigned to each \index{HIT}HIT, can be serialized for later use. Finally, this matrix can be used to generate the set of \index{HIT}HITs in the required format.

Algorithm~\ref{cap:paper_wsdm2022-sec:system-design-subsec:hits-alg:hit-allocation} provides a pseudocode representation of the allocation procedure. Suppose a requester wants to determine the number $m$ of \index{HIT}HITs to be published on the crowdsourcing platform. The sub-procedure $\Call{singleHIT}{\textit{...}}$, detailed in Algorithm~\ref{cap:paper_wsdm2022-sec:system-design-subsec:hits-alg:single-hit}, is used by the main algorithm to sample a set of elements without duplicates. Each sampled set is used to build a single \index{HIT}HIT within the entire allocation.

Let us hypothesize a requester aiming to allocate $n$ elements into \index{HIT}HITs containing $k$ positions each. If each element must appear in \index{$p$}$p$ different \index{HIT}HITs, then the total number of \index{HIT}HITs required is given by $m = \frac{n \cdot p}{k}$.

\begin{algorithm}
\caption{Procedure to allocate a dataset into HITs using the format required.}
\label{cap:paper_wsdm2022-sec:system-design-subsec:hits-alg:hit-allocation}
\begin{algorithmic}[1]
\State $\textit{elementsFiltered} \gets \Call{filterElements}{\textit{attribute},\textit{valuesChosen}}$
\State $\textit{classes} \gets \textit{valuesChosen}$
\State $\textit{pools} \gets \Call{List}$
\ForEach {$\textit{class} \in \textit{classes}$}
\State $\textit{elementsClass} \gets \Call{findElements}{\textit{elementsFiltered}, \textit{class}}$
\State $\textit{pool} \gets \Call{unique}{\textit{elementsClass}}$
\State $\textit{pools}.\Call{Append}{\textit{pool}}$
\EndFor
\State $\textit{totalElements} \gets \Call{len}{\textit{elementsFiltered}}$
\State $\textit{classElementsNumber} \gets \Call{len}{\textit{classes}}$
\State $\textit{hitElementsNumber} \gets \textit{k}$ 
\State $\textit{repetitionsElement} \gets \textit{p}$ 
\ForEach {$\textit{pool} \in \textit{pools}$}
\State $\textit{pool} \gets \Call{extendPool}{\textit{repetitionsElement}}$
\EndFor
\State $\textit{poolsDict} \gets \Call{mergePools}{\textit{pools},\textit{classes}}$
\State $\textit{hits} \gets \Call{List}$
\ForEach {$\textit{index} \in \Call{range}{(\textit{totalElements}*\textit{repetitionsElement})/\textit{hitElementsNumber}}$}
\State $\textit{hitSample} \gets \Call{singleHit}{\textit{poolsMerged}}$
\State $\textit{hitSample} \gets \Call{shuffle}{\textit{hitSample}}$
\State $\textit{hits}.\Call{append}{\textit{hitSample}}$
\EndFor
\State $\textit{hits}.\Call{serialize}{\textit{pathAssignments}}$
\State $\textit{hitsFinal} \gets \Call{List}$
\ForEach {$\textit{hit} \in \textit{hits}$}
\State $\textit{index} \gets \Call{index}{\textit{hit}}$
\State $\textit{unitId} \gets \Call{concat}{"unit\_",\textit{index}}$
\State $\textit{tokenInput} \gets \Call{randomString}{11}$
\State $\textit{tokenOutput} \gets \Call{randomString}{11}$
\State $\textit{hitObject} \gets \Call{BuildJSON}{\textit{unitId}, \textit{tokenInput}, \textit{tokenOutput}, \textit{hitElementsNumber}}$
\ForEach {$\textit{indexElem} \in \Call{range}{\textit{hitElementsNumber}}$}
\State $\textit{hitObject["documents"]} \gets \textit{hits[indexElem]}$
\EndFor
\State $\textit{hitsFinal}.\Call{append}{\textit{hitObject}}$
\EndFor
\State $\textit{hitsFinal}.\Call{serialize}{\textit{pathHits}}$
\end{algorithmic}
\end{algorithm}

\begin{algorithm}
\caption{Procedure to sample elements without duplicates for a single HIT.}
\label{cap:paper_wsdm2022-sec:system-design-subsec:hits-alg:single-hit}
\begin{algorithmic}[1]
\State $\textit{containsDuplicates} \gets \textit{True}$
\While {$\textit{containsDuplicates}$}
\State $\textit{sample} \gets \Call{List}$
\ForEach {$\textit{class} \in \textit{classes}$}
\ForEach {$\textit{indexClass} \in \Call{range}{\textit{classElementsNumber}}$}
\State $\textit{element} \gets \Call{random}{\textit{poolsDict[class]}}$
\State $\textit{sample}.\Call{append}{\textit{element}}$
\EndFor
\EndFor
\If{\Call{checkDuplicates}{\textit{sample}}==\textit{False}}
\State $\textit{containsDuplicates} \gets \textit{False}$
\EndIf
\EndWhile
\ForEach {$\textit{s} \in \textit{sample}$}
\ForEach {$\textit{c} \in \textit{classes}$}
\If{$\textit{s} \in \textit{pool[c]}$}
\State $\textit{pool[c]}.\Call{remove}{\textit{s}}$
\EndIf
\EndFor
\EndFor
\State \Return $\textit{sample}$
\end{algorithmic}
\end{algorithm}

\subsubsection{Automatic Approach}

\crowdframe provides an experimental solution to automatically allocate the elements to be evaluated into a set of \index{HIT}HITs. This feature currently works only when \crowdframe is used on the local filesystem. Future versions of the software will extend and stabilize this functionality. 

\citet{CESCHIA2022105995} propose a formal definition of the HIT allocation problem and solve it using a local search method. \crowdframe supports deploying an implementation of their solver and provides a communication interface to it. Docker must be installed on the local system, as a container is required to enable communication between the software and the solver.

\begin{figure}[tpb]
  \centering
    \includegraphics[width=.85\linewidth]{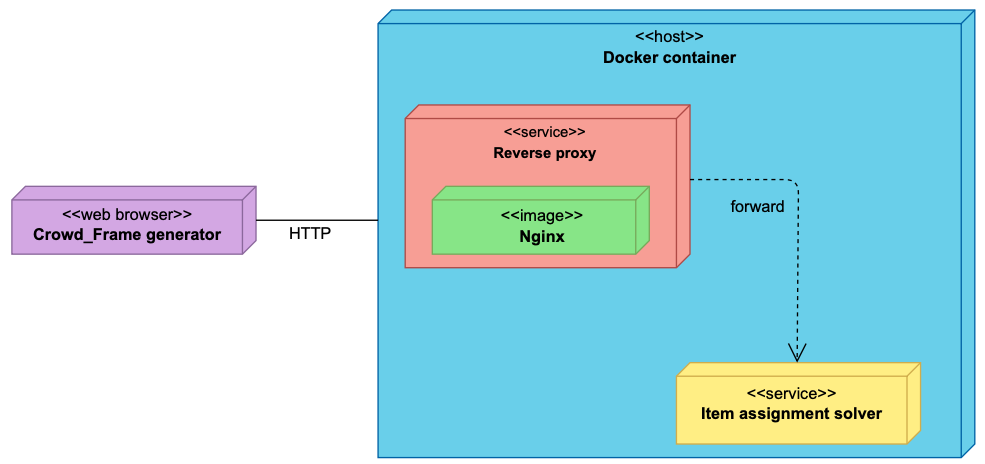}
\caption{Deployment diagram of the infrastructure enabling communication between \crowdframe and the solver.}
\label{cap:paper_wsdm2022-sec:system-design-subsec:hits-format-fig:solver-deploy}
\end{figure}

The Docker container includes two services: the first provides the solver implementation, while the second runs a reverse proxy based on the \index{Nginx} \textsf{Nginx}\footnote{\url{https://www.nginx.com/}} web server. The reverse proxy forwards HTTP messages to the solver, which processes them and sends back responses. Figure~\ref{cap:paper_wsdm2022-sec:system-design-subsec:hits-format-fig:solver-deploy} shows a deployment diagram illustrating the interaction among \crowdframe, the solver, and the reverse proxy.

The requester can enable this feature by setting the \spverb|enable_solver| environment variable, as shown in Table~\ref{cap:paper_wsdm2022-sec:usage-sec:env-var-tab:variables}. The solver can be used during the sixth step of the task configuration via the \generator component (cfr. Table~\ref{cap:paper_wsdm2022-sec:system-design-sec:generator-tab:steps}). The first step required to generate the solver’s input involves uploading the elements to be allocated across the \index{HIT}HITs. Each element must share the same set of attributes, and the dataset must be formatted as a \index{JSON} JSON array. In other words, the requester uploads the value of the \spverb|documents| object from Listing~\ref{cap:paper_wsdm2022-sec:system-design-subsec:hits-lst:hit-fragment}, omitting any \spverb|token_input|, \spverb|token_output|, or \spverb|unit_id| fields.

Next, the requester configures three parameters related to the allocation process: the number of workers that will evaluate each element, the total number of workers among whom the elements must be distributed, and the attributes used to categorize the elements into different \index{HIT}HITs. For each selected attribute (category), the requester also specifies how many elements should be assigned to each worker per category value. The system verifies whether each category–element count pair is feasible. If the validation is successful, the minimum number of workers required to cover the dataset is computed. The requester may increase this number at their discretion.

\begin{figure}[tbp]
  \centering
    \includegraphics[width=\linewidth]{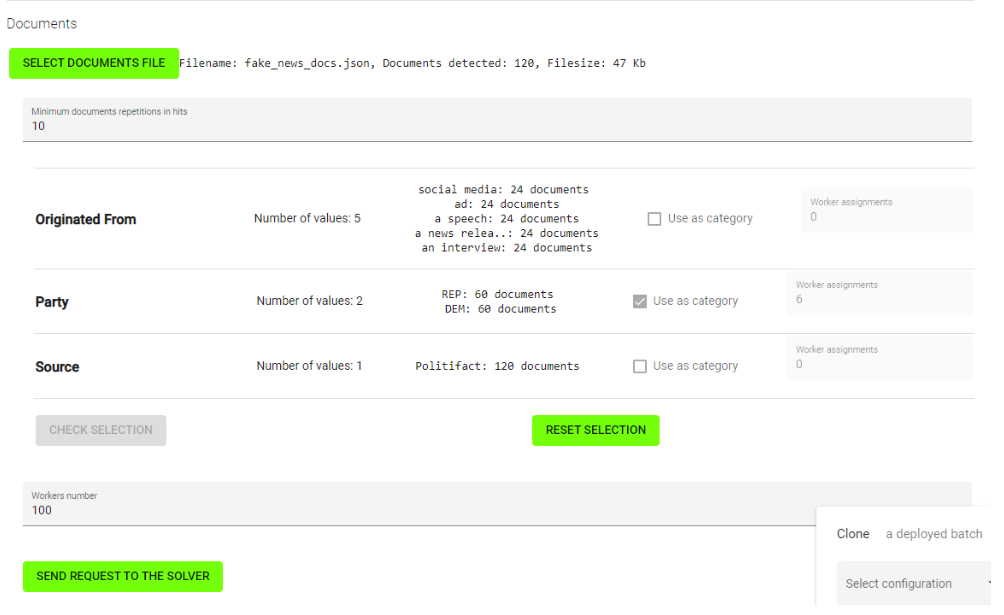}
\caption{Example configuration for the solver used to automatically allocate elements into a set of \index{HIT}HITs.}
\label{cap:paper_wsdm2022-sec:system-design-subsec:hits-format-fig:solver-screen}
\end{figure}

To illustrate when the verification may fail, consider a requester who selects an attribute named \spverb|A1| as a category. Suppose \spverb|A1| has two distinct values across the dataset, and the requester requires that each worker evaluate two elements for each value. Then, a second attribute named \spverb|A2| is also selected, which has three distinct values. The requester requires that each worker evaluate one element per value. This means that according to \spverb|A1|, each worker must evaluate four elements, whereas according to \spverb|A2|, they must evaluate three elements. Such an inconsistency renders the configuration invalid.

Figure~\ref{cap:paper_wsdm2022-sec:system-design-subsec:hits-format-fig:solver-screen} shows an example configuration for a subset of 120 statements sampled from \politifact, evaluated in the task by \citet{roitero2020crowd}. Listing~\ref{cap:paper_wsdm2022-sec:system-design-subsec:hits-lst:solver-elements} shows a fragment of these elements. In this example, the requester uploads a JSON file containing 120 elements to allocate and specifies that each element should be assigned to 10 different workers. The attribute \spverb|party| is chosen as the category, and each worker must evaluate 6 elements for each of its two values. This means each worker evaluates 12 elements in total. The system verifies that, given these constraints, at least 100 workers must be recruited.

The \generator component restricts category selection to attributes that are balanced, i.e., those that occur uniformly across the dataset. This design choice ensures that the input to the solver complies with the formal requirements defined by \citet{CESCHIA2022105995}. Once the configuration is validated, the requester can send the request to the solver, which computes and returns the allocation. Listing~\ref{cap:paper_wsdm2022-sec:system-design-subsec:hits-lst:solver-solution} shows an example of such a solution. The allocation is then used by \crowdframe to build the final set of \index{HIT}HITs, formatted according to the schema shown in Listing~\ref{cap:paper_wsdm2022-sec:system-design-subsec:hits-lst:hit-fragment}.

\begin{lstlisting}[style=jsonstyle,
  caption={Fragment of elements to be evaluated in the task published by~\citet{roitero2020crowd}.},
  label={cap:paper_wsdm2022-sec:system-design-subsec:hits-lst:solver-elements}]
[
  {
    "name": "REP_HALFTRUE_doc5",
    "statement": "Houston now has more debt per capita than California.",
    "claimant": "Rick Perry",
    "date": "2010",
    "originated_from": "ad",
    "id_par": "1796.json",
    "job": "Governor",
    "party": "REP",
    "source": "Politifact"
  },
  ...
]
\end{lstlisting}

\begin{lstlisting}[style=jsonstyle,
  caption={Fragment of the allocation built automatically using the solver integrated with \crowdframe.},
  label={cap:paper_wsdm2022-sec:system-design-subsec:hits-lst:solver-solution}]
{
  "finished": true,
  "runner": "BSA",
  "solution": {
    "Instance_id": "1658421484781",
    "Used_workers": 100,
    "Workers": [
      {
        "Assignments": [
          "REP_FALSE_doc8", "DEM_TRUE_doc5", "REP_BARELYTRUE_doc6",
          "DEM_MOSTLYTRUE_doc5", "REP_TRUE_doc7", "DEM_LIE_doc9"
        ],
        "Id": "W0"
      },
      {
        "Assignments": [
          "REP_TRUE_doc5", "DEM_TRUE_doc7", "REP_LIE_doc1",
          "DEM_TRUE_doc1", "REP_HALFTRUE_doc2", "DEM_FALSE_doc6"
        ],
        "Id": "W1"
      },
      ...
    ]
  },
  "completed": "2022-07-21 16:37:56",
  "started": "2022-07-21 16:37:56",
  "submitted": "2022-07-21 16:37:56",
  "task_id": "5447375273499815497"
}
\end{lstlisting}

\subsection{Quality Checks}

\label{cap:paper_wsdm2022-sec:usage-sec:quality-checks}

\crowdframe provides a mechanism to manually define custom quality checks, which are triggered for each evaluation dimension when the corresponding setting is enabled in the configuration. A custom quality check is implemented by defining the static method \spverb|performGoldCheck|, as described in Section~\ref{cap:paper_wsdm2022-sec:usage-sec:build-output}. Such checks are applied only to specific elements within a \index{HIT}HIT and only for a subset of the evaluation dimensions. An element is marked for quality checking by prepending the string \spverb|GOLD| to its \spverb|id| attribute.

Listing~\ref{cap:paper_wsdm2022-sec:system-design-subsec:local_dev-lst:hits-gold} shows a \index{HIT}HIT in which the second element is marked for quality checking. Listing~\ref{cap:paper_wsdm2022-sec:system-design-subsec:quality-checks-listing:default-method} presents the default implementation of the method generated by the initialization script. The \spverb|document| array contains the elements marked for the check, while the \spverb|answers| array includes the corresponding worker responses for the relevant evaluation dimensions. The logic of the quality check should be implemented between the two comment lines provided in the method template.

\begin{lstlisting}[style=jsonstyle,
  caption={Example of a valid \index{HIT}HIT containing two elements, one of which is used for a custom quality check.},
  label={cap:paper_wsdm2022-sec:system-design-subsec:local_dev-lst:hits-gold}]
[
  {
    "unit_id": "unit_0",
    "token_input": "ABCDEFGHILM",
    "token_output": "MNOPQRSTUVZ",
    "documents_number": 1,
    "documents": [
      { "id": "identifier_1", "text": "Lorem ipsum dolor sit amet" },
      { "id": "GOLD-identifier", "text": "Lorem ipsum dolor sit amet" }
    ]
  }
]
\end{lstlisting}

\begin{lstlisting}[style=customjs,
  caption={Default implementation of the static method used to perform custom quality checks in \crowdframe.},
  label={cap:paper_wsdm2022-sec:system-design-subsec:quality-checks-listing:default-method}]
export class GoldChecker {
  static performGoldCheck(goldConfiguration : Array<Object>) {
    let goldChecks = new Array<boolean>()
    /* If there are no gold elements there is nothing to be checked */
    if (goldConfiguration.length <= 0) {
      goldChecks.push(true)
      return goldChecks
    }
    for (let goldElement of goldConfiguration) {
      /* Element attributes */
      let document = goldElement["document"]
      /* Worker's answers for each gold dimension */
      let answers = goldElement["answers"]
      /* Worker's notes */
      let notes = goldElement["notes"]
      let goldCheck = true
      /* CONTROL IMPLEMENTATION STARTS HERE */
      /* The check for the current element holds if goldCheck remains true */
      ...
      /* CONTROL IMPLEMENTATION ENDS HERE */
      /* Push goldCheck inside goldChecks array for the current gold element */
      goldChecks.push(goldCheck)
    }
    return goldChecks
  }
}
\end{lstlisting}

\subsection{Local Development}

\label{cap:paper_wsdm2022-sec:usage-sec:local-dev}

\crowdframe provides a way to manually edit and test the task configuration locally, without deploying the full infrastructure. Enabling the local development mode involves the following steps:
\begin{enumerate}
    \item Navigate to the environments folder (see Section~\ref{cap:paper_wsdm2022-sec:usage-sec:build-output}):
    \begin{itemize}[label=--]
        \item \spverb|cd your_repo_folder/data/build/environments/|
    \end{itemize}
    \item Open the development environment file:
    \begin{itemize}[label=--]
        \item \spverb|environment.ts|
    \end{itemize}
    \item Set the variable \spverb|configuration_local| to \spverb|true|.
    \item Run the command \spverb|ng serve|.
\end{enumerate}

Alternatively, the requester may skip the first three steps and directly run \spverb|ng serve|. In this case, the local source code is used while the application is initialized with the remote configuration. Listing~\ref{cap:paper_wsdm2022-sec:system-design-subsec:local_dev-lst:environment} shows a valid development environment file that enables local testing of the task configuration.

Note that each execution of the \spverb|init.py| script overwrites the environment files. Therefore, local development mode must be re-enabled after running the script.

\begin{lstlisting}[style=customjs,
caption={Example of a development environment file enabling local development.},
label={cap:paper_wsdm2022-sec:system-design-subsec:local_dev-lst:environment}]
export const environment = {
    production: false,
    configuration_local: true,
    platform: "mturk",
    taskName: "your_task_name",
    batchName: "your_batch_name",
    region: "your_aws_region",
    bucket: "your_private_bucket",
    aws_id_key: "your_aws_key_id",
    aws_secret_key: "your_aws_key_secret",
    prolific_completion_code: false,
    bing_api_key: "your_bing_api_key",
    fake_json_key: "your_fake_json_key",
    log_on_console: false,
    log_server_config: "none",
    table_acl_name: "Crowd_Frame-your_task_name_your_batch_name_ACL",
    table_data_name: "Crowd_Frame-your_task_name_your_batch_name_Data",
    table_log_name: "Crowd_Frame-your_task_name_your_batch_name_Logger",
    hit_solver_endpoint: "None",
};
\end{lstlisting}

\section{Task Performing}

\label{cap:paper_wsdm2022-sec:usage-sec:task-performing}

Section~\ref{cap:paper_wsdm2022-sec:usage-sec:task-performing-sec:getting-started} provides some initial considerations regarding worker recruitment for a crowdsourcing task deployed using \crowdframe. Section~\ref{cap:paper_wsdm2022-sec:usage-sec:task-performing-subsec:manual} describes how to recruit workers manually. Section~\ref{cap:paper_wsdm2022-sec:usage-sec:task-performing-subsec:mturk} explains how to recruit workers using \mturk, while Section~\ref{cap:paper_wsdm2022-sec:usage-sec:task-performing-subsec:toloka} addresses recruitment via \toloka. Finally, Section~\ref{cap:paper_wsdm2022-sec:usage-sec:task-performing-subsec:prolific} focuses on recruitment through \prolific.

\subsection{Getting Started}

\label{cap:paper_wsdm2022-sec:usage-sec:task-performing-sec:getting-started}

Publishing a crowdsourcing task configured using \crowdframe involves selecting the platform to recruit the human workforce, although the requester can also manually recruit each worker. The process for publishing and starting the deployed task varies slightly depending on this choice.

\subsection{Manual Recruitment}

\label{cap:paper_wsdm2022-sec:usage-sec:task-performing-subsec:manual}

A task requester who intends to manually recruit each worker must:
\begin{enumerate}
    \item Set the environment variable \spverb|platform| (see Table~\ref{cap:paper_wsdm2022-sec:usage-sec:env-var-tab:variables}) to \spverb|none|.
    \item Generate and assign each worker an alphanumeric identifier, such as \spverb|randomWorkerId|.
    \item Append the identifier as a GET parameter to the task deployment link:
    \begin{itemize}[label=--]
        \item \spverb|?workerId=randomWorkerId|
    \end{itemize}
    \item Provide each worker with the complete deployment link:
    \begin{itemize}[label=--]
        \item \url{https://your_deploy_bucket.s3.your_aws_region.amazonaws.com/your_task_name/your_batch_name/index.html?workerId=randomWorkerId}
    \end{itemize}
    \item Wait for task completion.
\end{enumerate}

Steps \#2 and \#3 can be skipped, as the task URL can also be shared without a manual identifier. In this case, \crowdframe will automatically generate one.

\subsection{Amazon Mechanical Turk}

\label{cap:paper_wsdm2022-sec:usage-sec:task-performing-subsec:mturk}

A task requester who intends to recruit workers using \mturk must:
\begin{enumerate}
    \item Create the task on \mturk and set its general parameters and criteria (see Figure~\ref{cap:paper_wsdm2022-sec:crowdsourcing-platforms-fig:mturk-step-1}).
    \item Navigate to the build output folder for the platform:
    \begin{itemize}[label=--]
        \item \spverb|data/build/mturk/|
    \end{itemize}
    \item Copy the content of the wrapper file:
    \begin{itemize}[label=--]
        \item \spverb|data/build/mturk/index.html|
    \end{itemize}
    \item Paste the content into the \spverb|Design Layout| box on \mturk.
    \item Preview and save the task project.
    \item Publish the task and recruit a batch of workers by uploading the file containing the input/output tokens:
    \begin{itemize}[label=--]
        \item \spverb|data/build/mturk/tokens.csv|
    \end{itemize}
    \item Review the status of each submission using the \spverb|Manage| tab.
\end{enumerate}

\subsection{Toloka}

\label{cap:paper_wsdm2022-sec:usage-sec:task-performing-subsec:toloka}

A task requester who intends to recruit workers using \toloka must:
\begin{enumerate}
    \item Create the project and set its general parameters (see Figure~\ref{cap:paper_wsdm2022-sec:crowdsourcing-platforms-fig:toloka-step-1}).
    \item Navigate to the build output folder for the platform:
    \begin{itemize}[label=--]
        \item \spverb|data/build/toloka/|
    \end{itemize}
    \item Copy the markup, JavaScript code, and CSS styles of the wrapper:
    \begin{itemize}[label=--]
        \item \spverb|data/build/toloka/interface.html|
        \item \spverb|data/build/toloka/interface.js|
        \item \spverb|data/build/toloka/interface.css|
    \end{itemize}
    \item Paste each source file into the corresponding section of the \spverb|Task Interface| box in the \spverb|HTML/JS/CSS| editor (see Figure~\ref{cap:paper_wsdm2022-sec:crowdsourcing-platforms-fig:toloka-step-2}).
    \item Copy the input and output data specifications (see Listings~\ref{cap:paper_wsdm2022-sec:crowdsourcing-platforms-list:toloka-input-spec} and~\ref{cap:paper_wsdm2022-sec:crowdsourcing-platforms-list:toloka-output-spec}):
    \begin{itemize}[label=--]
        \item \spverb|data/build/toloka/input_specification.json|
        \item \spverb|data/build/toloka/output_specification.json|
    \end{itemize}
    \item Paste each specification into the appropriate section of the \spverb|Data Specification| box (see Figure~\ref{cap:paper_wsdm2022-sec:crowdsourcing-platforms-fig:toloka-step-3}).
    \item Copy the general instructions of the task:
    \begin{itemize}[label=--]
        \item \spverb|data/build/task/instructions_general.json|
    \end{itemize}
    \item Paste the text into the \spverb|Instructions for Tolokers| box using the source code editing mode.
    \item Create a new pool of workers by setting the audience and reward parameters (see Figure~\ref{cap:paper_wsdm2022-sec:crowdsourcing-platforms-fig:toloka-step-4}).
    \item Publish the task and recruit the workers for each pool by uploading the file containing the input/output tokens (see Listing~\ref{cap:paper_wsdm2022-sec:crowdsourcing-platforms-list:toloka-xlsx-file}):
    \begin{itemize}[label=--]
        \item \spverb|data/build/toloka/tokens.tsv|
    \end{itemize}
    \item Review the status of each submission on each pool's overview page.
\end{enumerate}

\subsection{Prolific}

\label{cap:paper_wsdm2022-sec:usage-sec:task-performing-subsec:prolific}

A task requester who intends to recruit workers using \prolific \index{Prolific} must:
\begin{enumerate}
    \item Create the study and set its general parameters (see Figure~\ref{cap:paper_wsdm2022-sec:crowdsourcing-platforms-fig:prolific-step-1}).
    \item Set the required data collection modality (see Figure~\ref{cap:paper_wsdm2022-sec:crowdsourcing-platforms-fig:prolific-step-2}):
    \begin{itemize}[label=--]
        \item Choose \spverb|External study link| as the data collection method.
        \item Provide the URL of the deployed task.
        \item Select the option to use URL parameters to record \prolific IDs.
        \item Rename the \spverb|PROLIFIC_PID| parameter to \spverb|workerId|.
        \item Choose to redirect participants upon completion using a custom URL.
        \item Copy the completion code from the URL (i.e., the value of the \spverb|cc| parameter).
        \item Set the \spverb|prolific_completion_code| environment variable to the copied value.
    \end{itemize}
    \item Define the audience of workers by setting the recruitment criteria (see Figure~\ref{cap:paper_wsdm2022-sec:crowdsourcing-platforms-fig:prolific-step-3}).
    \item Specify the overall cost of the study (see Figure~\ref{cap:paper_wsdm2022-sec:crowdsourcing-platforms-fig:prolific-step-4}).
    \item Monitor the status of submissions via the study's main page.
\end{enumerate}

\section{Task Results}

\label{cap:paper_wsdm2022-sec:usage-sec:task-result}

Section~\ref{cap:paper_wsdm2022-sec:usage-sec:task-result-sec:getting-started} describes the prerequisites and initial setup required to download the results of a task deployed using \crowdframe. Section~\ref{cap:paper_wsdm2022-sec:usage-sec:task-result-subsec:folder-task} details the contents of the \spverb|Task| folder, while Section~\ref{cap:paper_wsdm2022-sec:usage-sec:task-result-subsec:folder-data} covers those of the \spverb|Data| folder. Section~\ref{cap:paper_wsdm2022-sec:usage-sec:task-result-subsec:folder-resources} then focuses on the \spverb|Resources| folder, and Section~\ref{cap:paper_wsdm2022-sec:usage-sec:task-result-subsec:folder-crawling} on the \spverb|Crawling| folder. Finally, Section~\ref{cap:paper_wsdm2022-sec:usage-sec:task-result-subsec:folder-dataframe} describes the \spverb|Dataframe| folder.

\subsection{Getting Started}

\label{cap:paper_wsdm2022-sec:usage-sec:task-result-sec:getting-started}

The requester can download the final results of a crowdsourcing task deployed using \crowdframe by running the provided download script. This process involves four steps:
\begin{enumerate}
    \item Navigate to the main project folder: \spverb|cd ~/path/to/project/|
    \item Navigate to the data folder: \spverb|cd data|
    \item Run the \spverb|download.py| script. This script will:
    \begin{itemize}[label=--]
        \item Download and store snapshots of the raw data produced by each worker
        \item Refine the raw data into a tabular format
        \item Download and store the configuration of the deployed task
        \item Build and store support files containing worker and user agent attributes
    \end{itemize}
\end{enumerate}

The complete set of output data is stored in the results folder: \spverb|data/result/task_name/|, where \spverb|task_name| is the value of the corresponding environment variable listed in Table~\ref{cap:paper_wsdm2022-sec:usage-sec:env-var-tab:variables}. This folder is created by the download script if it does not already exist. It contains five subfolders, each corresponding to a different type of output data. Table~\ref{cap:paper_wsdm2022-sec:task-results-tab:results-subfolder} describes the contents of each subfolder.

\begin{table}[tpb]
\centering
\caption{Structure of the results folder of \crowdframe.}
\begin{tabular}{lp{11cm}}
  \toprule
  \textbf{Folder} & \textbf{Description} \\
  \midrule
  \spverb|Data| & Snapshots of the raw data produced by each worker. \\
  \midrule
  \spverb|Dataframe| & Refined, tabular versions of the raw data. \\
  \midrule
  \spverb|Resources| & Support files for each worker, including attributes about the worker and their user agent. \\
  \midrule
  \spverb|Task| & Backup of the task configuration. \\
  \midrule
  \spverb|Crawling| & Crawl of the pages retrieved via the search engine. \\
  \bottomrule
\end{tabular}
\label{cap:paper_wsdm2022-sec:task-results-tab:results-subfolder}
\end{table}

\subsection{\protect\UseVerb{result/Task}}

\label{cap:paper_wsdm2022-sec:usage-sec:task-result-subsec:folder-task}

The \spverb|Task| folder contains a backup of the task configuration. Table~\ref{cap:paper_wsdm2022-sec:usage-sec:build-output-tab:task-config} (Section~\ref{cap:paper_wsdm2022-sec:usage-sec:build-output}) describes its contents.

\subsection{\protect\UseVerb{result/Data}}

\label{cap:paper_wsdm2022-sec:usage-sec:task-result-subsec:folder-data}

The \spverb|Data| folder contains a snapshot of the data produced by each worker. Each snapshot is a JSON dictionary whose top-level object is an array. The download script creates one object for each batch of workers recruited within a crowdsourcing task. Listing~\ref{cap:paper_wsdm2022-sec:task-subsec:data-results-lst:snapshot} shows the snapshot for a worker with identifier \spverb|ABEFLAGYVQ7IN4| who participated in the batch \spverb|Your_Batch| of the task \spverb|Your_Task|. In this case, the snapshot contains an array with a single object. The \spverb|source_*| attributes represent the DynamoDB tables and the corresponding paths on the local filesystem.

\begin{lstlisting}[style=jsonstyle,
  caption={Snapshot of a worker who participated in a task with a single batch deployed using \crowdframe.},
  label={cap:paper_wsdm2022-sec:task-subsec:data-results-lst:snapshot}]
[
  {
    "source_data": "Crowd_Frame-Your-Task_Your-Batch_Data",
    "source_acl": "Crowd_Frame-Your-Task_Your-Batch_ACL",
    "source_log": "Crowd_Frame-Your-Task_Your-Batch_Logger",
    "source_path": "result/Your_Task/Data/ABEFLAGYVQ7IN4.json",
    "data_items": 1,
    "task": { ... },
    "worker": { ... },
    "ip": { ... },
    "uag": { ... },
    "checks": [ ... ],
    "questionnaires_answers": [ ... ],
    "documents_answers": [ ... ],
    "comments": [ ... ],
    "logs": [ ],
    "questionnaires": { ... },
    "documents": { ... },
    "dimensions": { ... }
  }
]
\end{lstlisting}

\subsection{\protect\UseVerb{result/Resources}}

\label{cap:paper_wsdm2022-sec:usage-sec:task-result-subsec:folder-resources}

The \spverb|Resources| folder includes two \index{JSON} JSON files for each worker. For a worker with ID \texttt{ABEFLAGYVQ7IN4}, the files are named with suffixes \texttt{_ip.json} and \texttt{_uag.json}, such as \texttt{ABEFLAGYVQ7IN4_ip.json}. The first contains attributes from reverse IP lookups; the second, from user agent strings. Since a worker may access the task from multiple devices or locations, each file can list several IP addresses or user agent strings.

Listing~\ref{cap:paper_wsdm2022-sec:task-subsec:resources-results-lst:ip_file} and Listing~\ref{cap:paper_wsdm2022-sec:task-subsec:resources-results-lst:uag_file} show a subset of the information provided by these two support files.

\begin{lstlisting}[style=jsonstyle,
  caption={Subset of the information obtained by performing the reverse lookup of a worker's IP address.},
  label={cap:paper_wsdm2022-sec:task-subsec:resources-results-lst:ip_file}]
{
  "<ip_address_1>": {
    "continent_code": "NA",
    "continent_name": "North America",
    "country_capital": "Washington D.C.",
    "country_code_iso2": "US",
    "country_code_iso3": "USA",
    "country_currency_code_iso3": "USD",
    "country_currency_name": "US Dollar",
    "country_currency_numeric": "840",
    "country_currency_symbol": "$",
    "country_flag_emoji": "...",
    "country_flag_emoji_unicode": "...",
    "country_flag_url": "...",
    "country_is_eu": false,
    "country_name": "United States",
    "country_name_official": "United States of America",
    "country_numeric": "840",
    "hostname": "...",
    "ip": "...",
    "ip_address_type": "...",
    "latitude": "...",
    "location_calling_code": "...",
    "location_coordinates": "...",
    "location_geoname_id": "...",
    "location_is_eu": false,
    "location_languages": [...],
    "location_name": "...",
    "location_postal_code": "...",
    "longitude": "...",
    "provider_name": "...",
    "region_code": "...",
    "region_code_full": "...",
    "region_name": "Louisiana",
    "region_type": "State",
    "timezone_name": "America/Chicago",
    ...
  },
  ...
}
\end{lstlisting}

\begin{lstlisting}[style=jsonstyle,
  caption={Subset of the information obtained by analyzing a worker's user-agent string.},
  label={cap:paper_wsdm2022-sec:task-subsec:resources-results-lst:uag_file}]
{
  "Mozilla/5.0 (Windows NT 10.0; Win64; x64) AppleWebKit/537.36 (KHTML, like Gecko) Chrome/105.0.5195.102 Safari/537.36": {
    "browser_engine": "WebKit/Blink",
    "browser_name": "Chrome",
    "browser_version": "105.0.5195.102",
    "browser_version_major": "105",
    "device_is_crawler": false,
    "device_is_mobile_device": false,
    "device_type": "desktop",
    "os_code": "windows_10",
    "os_family": "Windows",
    "os_family_code": "windows",
    "os_family_vendor": "Microsoft Corporation.",
    "os_icon": "https://assets.userstack.com/icon/os/windows10.png",
    "os_icon_large": "...",
    "os_name": "Windows 10",
    "os_url": "https://en.wikipedia.org/wiki/Windows_10",
    "ua": "...",
    "ua_type": "browser",
    "ua_url": "https://about.google/",
    ...
  },
  ...
}
\end{lstlisting}

\subsection{\protect\UseVerb{result/Crawling}}

\label{cap:paper_wsdm2022-sec:usage-sec:task-result-subsec:folder-crawling}

The \spverb|Crawling| folder contains a crawl of the web pages retrieved by the search engine when queried by a worker. A task requester who deploys a crowdsourcing task that uses the search engine within one or more evaluation dimensions can enable crawling by setting the \spverb|enable_crawling| variable shown in Table~\ref{cap:paper_wsdm2022-sec:usage-sec:env-var-tab:variables}. If this variable is enabled, the download script attempts to crawl each retrieved web page.

Initially, the script creates two subfolders named \spverb|Metadata/| and \spverb|Source/|. Each web page is assigned a \index{UUID} UUID (Universally Unique Identifier). For example, consider a page with the UUID \spverb|59c0f70f-c5a6-45ec-ac90-b609e2cc66d7|. The script attempts to download its source code and, if successful, saves it in the \spverb|Source| folder using a file named after the UUID with the suffix \texttt{_source}; the file extension depends on the page's source format. Then, metadata about the crawling operation is saved in the \spverb|Metadata| folder, in a file named after the UUID with the suffix \texttt{_metadata.json}. This metadata file enables the requester to check whether the operation succeeded, understand any failure reasons (such as the HTTP response code), and inspect all HTTP header values.

Listing~\ref{cap:paper_wsdm2022-sec:task-subsec:resources-results-lst:crawl_metadata} shows an example of metadata produced by the download script while attempting to crawl one of the retrieved pages.

\begin{lstlisting}[style=jsonstyle,
  caption={Metadata produced by the download script while attempting to crawl a web page retrieved by the \crowdframe\ search engine.},
  label={cap:paper_wsdm2022-sec:task-subsec:resources-results-lst:crawl_metadata}]
{
  "attributes": {
    "response_uuid": "59c0f70f-c5a6-45ec-ac90-b609e2cc66d7",
    "response_url": "...",
    "response_timestamp": "...",
    "response_error_code": null,
    "response_source_path": "...",
    "response_metadata_path": "...",
    "response_status_code": 200,
    "response_encoding": "utf-8",
    "response_content_length": 125965,
    "response_content_type": "text/html; charset=utf-8"
  },
  "data": {
    "date": "Wed, 08 Jun 2022 22:33:12 GMT",
    "content_type": "text/html; charset=utf-8",
    "content_length": "125965",
    "..."
  }
}
\end{lstlisting}

\subsection{\protect\UseVerb{result/Dataframe}}

\label{cap:paper_wsdm2022-sec:usage-sec:task-result-subsec:folder-dataframe}

The \spverb|Dataframe| folder contains a refined version of the data stored in each worker snapshot. The download script inserts the raw data into structures called \index{DataFrame} \lq\lq DataFrame\rq\rq{}. A \spverb|DataFrame|\footnote{\url{https://pandas.pydata.org/docs/reference/api/pandas.DataFrame.html}} is a two-dimensional data structure with labeled axes that can contain heterogeneous data. Such structures can be implemented as two-dimensional arrays or tables with rows and columns. The download script refines the raw data into up to ten tabular dataframes, serialized as \index{CSV} CSV files.

The final number of dataframes stored in the \spverb|Dataframe| folder depends on the environment variables configured by the task requester.

\begin{table}[tpb]
\centering
\caption{DataFrames produced when downloading the final results of a task.}
\begin{tabular}{lp{8.6cm}}
  \toprule
   \textbf{DataFrame} & \textbf{Description}  \\
    \midrule
	\spverb|workers_acl.csv| & Access control list of the workers. \\
	\midrule
	\spverb|workers_ip_addresses.csv| & Information about the workers’ IP addresses. \\
	\midrule
	\spverb|workers_user_agents.csv| & Information about the workers’ user agents. \\
	\midrule
	\spverb|workers_answers.csv| & Answers provided by workers for each evaluation dimension. \\
	\midrule
	\spverb|workers_questionnaire.csv| & Answers provided by workers for each questionnaire. \\
	\midrule
	\spverb|workers_dimensions_selection.csv| & Temporal order in which each worker selects values for evaluation dimensions. \\
	\midrule
	\spverb|workers_notes.csv| & Textual annotations provided by workers. \\
	\midrule
	\spverb|workers_urls.csv| & Search engine queries issued by workers and the retrieved results. \\
	\midrule
	\spverb|workers_crawling.csv| & Data about the crawling of web pages retrieved by the search engine. \\
	\midrule
	\spverb|workers_logs.csv| & Log data produced by the \spverb|Logger| component while workers perform the task. \\
	\midrule
	\spverb|workers_comments.csv| & Final comments provided by workers to the requester at the end of the task. \\
	\midrule
	\spverb|workers_mturk_data.csv| & Data about workers produced by \mturk. \\
	\midrule
	\spverb|workers_toloka_data.csv| & Data about workers produced by \toloka. \\
	\midrule
	\spverb|workers_prolific_study_data.csv| & Data about the study deployed on \prolific and its submissions. \\
	\midrule
	\spverb|workers_prolific_demographic_data.csv| & Demographic data of workers participating in a study published on \prolific. \\
	\bottomrule
  \end{tabular}
\label{cap:paper_wsdm2022-sec:task-subsec:resources-results-tab:dataframes}
\end{table}

Each \spverb|DataFrame| contains a variable number of rows and columns. Some \spverb|DataFrame|s provide general information and thus have one row per worker, such as \spverb|workers_info| and \spverb|workers_acl|. Others have higher granularity. For example, each row in the \spverb|workers_urls| DataFrame corresponds to a single result retrieved for a query submitted to the search engine while analyzing an element of a \index{HIT}HIT during a given attempt by a worker. Similarly, each row in the \spverb|workers_answers| DataFrame contains the answers provided for the evaluation dimensions related to a single \index{HIT}HIT element during a given attempt by a worker. The requester should therefore carefully examine each \spverb|DataFrame| to understand the nature of the data it contains before analysis.

Listing~\ref{cap:paper_wsdm2022-sec:task-subsec:resources-results-lst:dataframe-acl} provides an example of the access control list for a task with a single recruited worker. Listing~\ref{cap:paper_wsdm2022-sec:task-subsec:resources-results-lst:dataframe-answers} shows an example containing the answers provided by a single worker for two elements of the assigned \index{HIT}HIT.

\begin{lstlisting}[style=csvstyle,
  caption={Example of the \protect\UseVerb{workers-acl} DataFrame produced by \crowdframe.},
  label={cap:paper_wsdm2022-sec:task-subsec:resources-results-lst:dataframe-acl}]
worker_id,generated,in_progress,paid,platform,task_name,batch_name,unit_id,token_input,token_output,try_current,try_last,try_left,tries_amount,status_code,access_counter,time_arrival,time_arrival_parsed,time_submit,time_submit_parsed,time_completion,time_completion_parsed,time_expiration_nearest,time_expiration_nearest_parsed,time_expiration,time_expiration_parsed,time_expired,time_removal,time_removal_parsed,questionnaire_amount,questionnaire_amount_start,questionnaire_amount_end,dimensions_amount,documents_amount,ip_address,ip_source,user_agent,user_agent_source,folder,source_path,source_acl,source_data,source_log,study_id,session_id,n_submissions
ABEFLAGYVQ7IN4,False,False,False,mturk,Task-Sample,Batch-Sample,unit_0,KDSCKUOHINM,VQULVJHRTOZ,1,1,10,10,,1,"Thu, 14 Apr 2022 12:37:47 GMT",2022-04-14 12:37:47 00:00,"Thu, 14 Apr 2022 12:38:44 GMT",2022-04-14 12:38:44 00:00,,,,,,,True,"Thu, 14 Apr 2022 15:32:39 GMT",2022-04-14 15:32:39 00:00,3,3,0,0,0,75.41.166.226,cf,"Mozilla/5.0 (Windows NT 10.0; Win64; x64) AppleWebKit/537.36 (KHTML, like Gecko) Chrome/100.0.4896.75 Safari/537.36",cf,Task-Sample/Batch-Sample/Data/A3CGQOJC28OVGN/,result/Task-Sample/Data/A3CGQOJC28OVGN.json,Crowd_Frame-Task-Sample_Batch-Sample_ACL,Crowd_Frame-Task-Sample_Batch-Sample_Data,...,,,
\end{lstlisting}

\begin{lstlisting}[style=csvstyle,
  caption={Example of the \protect\UseVerb{workers-answers} DataFrame produced by \crowdframe.},
  label={cap:paper_wsdm2022-sec:task-subsec:resources-results-lst:dataframe-answers}]
worker_id,paid,task_id,batch_name,unit_id,try_last,try_current,action,time_submit,time_submit_parsed,doc_index,doc_id,doc_fact_check_ground_truth_label,doc_fact_check_ground_truth_value,doc_fact_check_source,doc_speaker_name,doc_speaker_party,doc_statement_date,doc_statement_description,doc_statement_text,doc_truthfulness_value,doc_accesses,doc_time_elapsed,doc_time_start,doc_time_end,global_outcome,global_form_validity,gold_checks,time_spent_check,time_check_amount
ABEFLAGYVQ7IN4,False,Task-Sample,Batch-Sample,unit_1,1,1,Next,"Wed, 09 Nov 2022 10:19:16 GMT",2022-11-09 10:19:16 00:00,0.0,conservative-activist-steve-lonegan-claims-social-,false,1,Politifact,Steve Lonegan,REP,2022-07-12,"stated on October 1, 2011 in an interview on News 12 New Jersey's Power & Politics show:","Today, the Social Security system is broke.",10,1,2.1,1667989144,1667989146.1,False,False,False,False,False
ABEFLAGYVQ7IN4,False,Task-Sample,Batch-Sample,unit_1,1,1,Next,"Wed, 09 Nov 2022 10:19:25 GMT",2022-11-09 10:19:25 00:00,1,yes-tax-break-ron-johnson-pushed-2017-has-benefite,true,5,Politifact,Democratic Party of Wisconsin,DEM,2022-04-29,"stated on April 29, 2022 in News release:","The tax carve out (Ron) Johnson spearheaded overwhelmingly benefited the wealthiest, over small businesses.",100,1,10.27,1667989146.1,1667989156.37,False,False,False,False,False
\end{lstlisting}

Each dataframe has its own characteristics and peculiarities. However, there are several rules of thumb that a requester should remember and consider when performing any kind of analysis:
\begin{itemize}[label=--]
    \item The attribute \spverb|paid| is present in all dataframes. It can be used to distinguish between workers who completed the task and those who did not. The requester may want to explore the data of workers who failed the task.
    \item The attribute \spverb|batch_name| is present in a subset of dataframes. It can be used to separate data among different batches of recruited workers. The requester may want to analyze each batch separately.
    \item The attributes \spverb|try_current| and \spverb|try_last| are present in a subset of dataframes. They can be used to distinguish each attempt made by a worker. The latter indicates the most recent try. The requester should consider the possible presence of multiple tries for each worker during analysis.
    \item The attribute \spverb|access_counter| is present in a subset of dataframes. It can be used to distinguish subsequent accesses made by each worker. The requester should note that a worker may close the browser tab and return later. Moreover, a given access recorded in the dataframe can span multiple tries if a worker retries immediately after failing without leaving.
    \item The attribute \spverb|action| is present in a subset of dataframes. It can be used to understand whether a worker moved to the previous or next \index{HIT}HIT element. Possible values are \spverb|Back|, \spverb|Next|, and \spverb|Finish|. The \spverb|Finish| value indicates the last element evaluated before completing a try. Only rows with \spverb|Next| or \spverb|Finish| describe the most recent answers for each element.
    \item The attribute \spverb|index_selected| is present in the \spverb|workers_urls| dataframe. It can be used to filter the search engine results. Results with a value different from $-1$ indicate those selected by the worker in the user interface. For example, a value of $4$ means three other results were selected previously, a value of $7$ means six others, and so on. The requester may want to analyze only the results with which the worker interacted.
    \item The attribute \spverb|type| is present in the \spverb|workers_logs| dataframe. It specifies the type of log record described by each row. Log records are generally sorted by global timestamp. The requester can use this attribute to split log records into subsets of the same type.
    \item The \spverb|workers_acl| dataframe contains useful information about each worker. The requester may want to merge it with other dataframes using the \spverb|worker_id| attribute as the key.
    \item The \spverb|workers_urls| dataframe contains all results retrieved by the search engine, while the \spverb|workers_crawling| dataframe contains information about the crawling of each result. The requester may want to merge the two dataframes using the \spverb|response_uuid| attribute as the key.
    \item The \spverb|workers_dimensions_selection| dataframe shows the temporal ordering of workers’ choices for evaluation dimensions. It is ordered by global timestamp alongside each worker’s selection. Rows belonging to the same worker may appear in different positions because multiple workers can perform the task concurrently. The requester should consider this when exploring the dataframe.
    \item The \spverb|workers_comments| dataframe provides the final comments from workers. Since providing a final comment is optional, this dataframe may be empty.
\end{itemize}

\section{Conclusions}

\label{cap:paper_wsdm2022-sec:conclusions}

\crowdframe is a relatively young software that still has a long way to go before becoming fully effective and accessible to the entire population of task requesters. There is significant potential to implement new features and strengthen its overall implementation. Despite this, the software has already been used to deploy several tasks, as outlined in Section~\ref{cap:paper_wsdm2022-sec:research_question} and described throughout this thesis.

\citet{roitero2020crowd} investigate whether crowdsourcing can be reliably used to assess the truthfulness of information items and to create large-scale labeled collections for information credibility systems. They deploy a crowdsourcing task to collect thousands of truthfulness judgments over two sets of statements using judgment scales with various granularity levels (Chapter~\ref{cap:paper_sigir2020}). 

\citet{roitero2021crowd} examine whether crowdsourcing is an effective and reliable method to assess truthfulness during a pandemic, focusing on statements related to \covid. Their study addresses (mis)information that is both sensitive and recent compared to the time of judgment. They deploy a crowdsourcing task where workers judge the truthfulness of a set of \covid-related statements and provide evidence for their judgments (Chapter~\ref{cap:paper_pauc2021}).

\citet{SOPRANO2021102710} hypothesize that a unidimensional truthfulness scale is insufficient to capture the subtle differences among publicly available statements for fact-checking. They thus propose a multidimensional notion of truthfulness and deploy a crowdsourcing task where workers judge statements across seven different truthfulness dimensions (Chapter~\ref{cap:paper_ipm2021}). 

\citet{soprano2023loyalty} design a questionnaire to understand how longitudinal studies are conducted by crowd workers and which factors influence participation across different crowdsourcing platforms. This questionnaire is included within a crowdsourcing task deployed on \mturk, \toloka, and \prolific to build a large-scale dataset of responses (Chapter~\ref{cap:paper_tsc2024}).

\citet{draws2022bias} perform a systematic exploratory analysis of publicly available crowdsourced data to identify potential systematic biases occurring when crowd workers perform tasks. They then deploy a crowdsourcing task to collect a novel set of truthfulness judgments to validate their hypotheses (Chapter~\ref{cap:paper_facct2022}).

\citet{brand2022neural} generate truthfulness predictions for claims using machine learning models and provide human-readable explanations. They deploy a crowdsourcing task to gather human evaluations on the impact of these explanations (Chapter~\ref{cap:paper_jdiq2022}). 

\citet{CEOLIN2022102107} introduce a rating system to identify review arguments and define weighted semantics using formal argumentation theory. They build an argumentation graph and deploy crowdsourcing tasks to evaluate argument contributions by comparing the argumentation dataset results with review upvotes (Section~\ref{cap:paper_is2022}).

These crowdsourcing tasks deployed by various researchers and practitioners demonstrate the effectiveness of \crowdframe in gathering crowd-powered data across multiple platforms and marketplaces.

\section{Future Directions}

\label{cap:paper_wsdm2022-sec:future_work}

\crowdframe still has some limitations that need to be addressed in future versions. Currently, the requester must manually build the overall set of \index{HIT}HITs to allocate the elements to be evaluated. This set must comply with the format described in Section~\ref{cap:paper_wsdm2022-sec:usage-sec:hits-format}. The feature for automatic allocation of elements is still experimental, and stabilizing it is a primary development goal. Even with this feature, the requester must upload the set of elements manually. Therefore, the user interface of the \generator component could be enhanced to facilitate this operation. Additionally, the interface between \crowdframe and the solver could be extended to specify attributes that characterize and differentiate each worker, such as their level of expertise.

The \generator component is currently bundled and shipped together with the \skeleton component. The use case diagrams in Figure~\ref{cap:paper_wsdm2022-sec:system-design-subsec:generator-subsec:use-case-fig:requester} and Figure~\ref{cap:paper_wsdm2022-sec:system-design-subsec:search-engine-subsec:use-case-fig:worker} show that different actors interact with these two components, which serve distinct procedures. Moreover, the user interface of the generator needs improvement to become truly effective. In the future, these two components should have separate implementations. The generator could be published as a standalone tool and be able to interact with the requester’s data buckets.

Initially, \crowdframe was designed to be completely client-side, and no backend is currently initialized. The client-side codebase communicates directly with Amazon Web Services. While the various building blocks of \crowdframe can be combined to create complex tasks, inexperienced requesters may find it difficult to design effective crowdsourcing tasks. The system should provide a set of ready-to-use templates to design and customize the most common types of crowdsourcing tasks supported by \mturk (see Figure~\ref{cap:paper_wsdm2022-sec:crowdsourcing-platforms-fig:mturk-step-1}).

Furthermore, \crowdframe currently lacks an interface to monitor the progress of deployed tasks. Requesters must use the user interface of the chosen crowdsourcing platform to review the status of each \index{HIT}HIT. A monitoring interface will be developed in the future to interact with crowdsourcing platforms via their software development kits (SDKs) and provide feedback within \crowdframe. These SDKs also allow automatic task publication and worker submission approval, which will reduce the manual steps described in Section~\ref{cap:paper_wsdm2022-sec:usage-sec:task-performing} and speed up the workflow. However, \crowdframe requires a more stable implementation before offering these features safely to task requesters.

Currently, \crowdframe relies on Amazon Web Services for backend infrastructure deployment. However, task requesters may prefer other cloud providers for various reasons. Future versions of the software will support alternative providers such as Microsoft Azure\footnote{\url{https://azure.microsoft.com/}} \index{Microsoft Azure} and Google Cloud Platform.\footnote{\url{https://cloud.google.com/}} \index{Google!Cloud Platform} Additionally, deploying the entire infrastructure on private servers will be facilitated.

\chapter{Chapter~\ref{cap:paper_sigir2020}: Questionnaires And Statement List}

\label{cap:paper_sigir2020-appendix:quest-stat}

The appendix reports the demographic questionnaire and the \index{Cognitive!Reflection Test}CRT tests used for the task described in Section~\ref{cap:paper_sigir2020-sec:exp-setup-subsec:crowdsourcing-task}. Such questionnaires are also employed for the tasks described in Chapter~\ref{cap:paper_pauc2021}, Chapter~\ref{cap:paper_ipm2021}, Chapter~\ref{cap:paper_facct2022} and Section~\ref{cap:paper_is2022}.

\section{Demographic Questionnaire}

 \label{cap:paper_sigir2020-appendix:quest-crt-sec:initial}

 \begin{enumerate}[leftmargin=*, align=left, label=Q\arabic*:]
     \item What is your age range?
     \begin{enumerate}[label=A\arabic*:]
         \item 0--18
        \item 19--25
         \item 26--35
         \item 36--50
         \item 50--80
         \item 80+
     \end{enumerate}
   
     \item What is the highest level of school you have completed or the highest degree you ave received?
     \begin{enumerate}[label=A\arabic*:]
       \item High school incomplete or less,
         \item High school graduate or GED (includes technical/vocational training that doesn't towards college credit)
           \item Some college (some community college, associate’s degree)
           \item Four year college degree/bachelor’s degree
           \item Some postgraduate or professional schooling, no postgraduate degree
           \item Postgraduate or professional degree, including master’s, doctorate, medical or law degree
     \end{enumerate}
    
     \item Last year what was your total family income from all sources, before taxes?
     \begin{enumerate}[label=A\arabic*:]
         \item Less than 10,000
         \item 10,000 to less than 20,000
         \item 20,000 to less than 30,000
         \item 30,000 to less than 40,000
         \item 40,000 to less than 50,000
         \item 50,000 to less than 75,000
         \item 75,000 to less than 100,000
         \item 100,000 to less than 150,000
         \item 150,000 or more
     \end{enumerate}
    
      \item In general, would you describe your political views as
     \begin{enumerate}[label=A\arabic*:]
         \item Very conservative
           \item Conservative
           \item Moderate
           \item Liberal
           \item Very liberal
     \end{enumerate}
    
      \item In politics today, do you consider yourself a
     \begin{enumerate}[label=A\arabic*:]
       \item Republican
           \item Democrat
           \item Independent
           \item Something else
     \end{enumerate}
    
      \item Should the U.S. build a wall along the southern border?
     \begin{enumerate}[label=A\arabic*:]
         \item Agree
           \item Disagree
           \item No opinion either way
     \end{enumerate}
    
      \item Should the government increase environmental regulations to prevent climate change?
     \begin{enumerate}[label=A\arabic*:]
         \item Agree
           \item Disagree
           \item No opinion either way
     \end{enumerate}
 \end{enumerate}

\section{Cognitive Reflection Test (CRT)}
 
 \label{cap:paper_sigir2020-appendix:quest-crt-sec:crt}
 
 \begin{enumerate}[leftmargin=*, align=left, label= CRT\arabic*:]
     \item If three farmers can plant three trees in three hours, how long would it take nine farmers to plant nine trees?
     \begin{itemize}
         \item Correct Answer: 3 hours
         \item Intuitive Answer: 9 hours
     \end{itemize}
    
     \item Sean received both the 5th highest and the 5th lowest mark in the class. How many students are there in the class?
     \begin{itemize}
         \item Correct Answer: 9 students
         \item Intuitive Answer: 10 students
     \end{itemize}
    
     \item In an athletics team, females are four times more likely to win a medal than males. This year the team has won 20 medals so far. How many of these have been won by males?
     \begin{itemize}
         \item Correct Answer: 4 medals
         \item Intuitive Answer: 5 medals
     \end{itemize}
 \end{enumerate}

\chapter{Chapter~\ref{cap:paper_pauc2021}: Statements List}

\label{cap:paper_pauc2021:-appendix:statements}

This appendix provides the list of statements used to perform the crowdsourcing experiment and the longitudinal study described in Chapter~\ref{cap:paper_pauc2021}. The statements are sampled from the \politifact \cite{wang2017liar} dataset.

\footnotesize
\begin{longtable}{rp{2.5cm}llp{9.5cm}}
\toprule
\textbf{ID} &   \textbf{Speaker} &  \textbf{Date} & \textbf{Ground Truth} & \textbf{Statement} \\
\midrule
\endhead
\midrule
\multicolumn{5}{r}{{Continues on the next page}} \\
\midrule
\endfoot
\bottomrule
\endlastfoot
\statement{1} &  Facebook User &  2020-25-03 &  \politifactzero &  If your child gets this virus their going to hospital alone in a van with people they don’t know... to be with people they don’t know you will be at home without them in their time of need. \\
 \statement{2} &  Donald Trump &  2020-30-03 &  \politifactzero &  We inherited a \lq\lq broken test\rq\rq{} for COVID-19. \\
 \statement{3} &  Facebook User &  2020-19-03 &  \politifactzero &  Says \lq\lq there is no\rq\rq{} COVID-19 virus. \\
 \statement{4} &  Facebook User &  2020-25-03 &  \politifactzero &  COVID literally stands for Chinese Originated Viral Infectious Disease. \\
 \statement{5} &  Bloggers &  2020-26-02 &  \politifactzero &  A post say \lq\lq hair weave and lace fronts manufactured in China may contain the coronavirus.\rq\rq{} \\
 \statement{6} &  Youtube Video &  2020-29-02 &  \politifactzero &  A video says that the Vatican confirmed that Pope Francis and two aides tested positive for coronavirus. \\
\statement{7} &  Ron Desantis &  2020-09-04 &  \politifactzero &  This particular pandemic is one where I don’t think nationwide, there’s been a single fatality under 25. \\
 \statement{8} &  Facebook User &  2020-16-03 &  \politifactzero &  The government is closing businesses to stop the spread of coronavirus even though \lq\lq the numbers are nothing compared to H1N1 or Ebola. Everyone needs to realize our government is up to something \ldots\rq\rq{} \\
 \statement{9} &  Facebook User &  2020-16-03 &  \politifactzero &  the U.S. is developing an \lq\lq antivirus\rq\rq{} that includes a chip to track your movement. \\
 \statement{10} &  Bloggers &  2020-31-03 &  \politifactzero &  Italy arrested a doctor \lq\lq for intentionally killing over 3,000 coronavirus patients.\rq\rq{} \\
 \addlinespace
 \statement{11} &  Facebook User &  2020-28-03 &  \politifactone &  Says COVID-19 remains in the air for eight hours and that everyone is now required to wear masks \lq\lq everywhere.\rq\rq{} \\
 \statement{12} &  Facebook User &  2020-27-03 &  \politifactone &  Says to leave objects in the sun to avoid contracting the coronavirus. \\
 \statement{13} &  Facebook User &  2020-23-03 &  \politifactone &  \lq\lq Slices of lemon in a cup of hot water can save your life. The hot lemon can kill the proliferation of\rq\rq{} the novel coronavirus. \\
 \statement{14} &  Facebook User &  2020-23-03 &  \politifactone &  Says the CDC now says that the coronavirus can survive on surfaces for up to 17 days. \\
 \statement{15} &  Viral Image &  2020-13-03 &  \politifactone &  Drinking \lq\lq water a lot and gargling with warm water \& salt or vinegar eliminates\rq\rq{} the coronavirus. \\
 \statement{16} &  Snapchat &  2020-23-03 &  \politifactone &  Says \lq\lq special military helicopters will spray pesticide against the Corona virus in the skies all over the country.\rq\rq{} \\
 \statement{17} &  Bloggers &  2020-20-03 &  \politifactone &  Says COVID-19 came to the United States in 2019. \\
 \statement{18} &  Bloggers &  2020-09-04 &  \politifactone &  Church services can’t resume until we’re all vaccinated, says Bill Gates. \\
 \statement{19} &  Facebook User &  2020-10-04 &  \politifactone &  Mass vaccination for COVID-19 in Senegal was started yesterday (4/8) and the first 7 CHILDREN who received it \lq\lq DIED on the spot.\rq\rq{} \\
 \statement{20} &  Facebook User &  2020-02-04 &  \politifactone &  Says video shows \lq\lq the Chinese are destroying the 5G poles as they are aware that it is the thing triggering the corona symptoms.\rq\rq{} \\
\addlinespace
 \statement{21} &  Turning Point Usa &  2020-25-03 &  \politifacttwo &  Says Nevada Governor Steve Sisolak \lq\lq has banned the use of an anti-malaria drug that might help cure coronavirus.\rq\rq{} \\
 \statement{22} &  Marco Rubio &  2020-19-03 &  \politifacttwo &  For coronavirus cases \lq\lq in the U.S. 38\% of those hospitalized are under 35.\rq\rq{} \\
 \statement{23} &  Image &  2020-15-03 &  \politifacttwo &  COVID-19 started because we eat animals. \\
 \statement{24} &  Facebook User &  2020-14-03 & \politifacttwo &  Italy has decided not to treat their elderly for this virus. \\
 \statement{25} &  Viral Image &  2020-13-03 & \politifacttwo &  President Trump, COVID-19 coronavirus: U.S. cases 1,329; U.S. deaths, 38; panic level: mass hysteria. President Obama, H1N1 virus: U.S. cases, 60.8 million; U.S. deaths, 12,469; panic level: totally chill. Do you all see how the media can manipulate your life. \\
 \statement{26} &  Facebook User &  2020-27-02 &  \politifacttwo &  Post says \lq\lq the blood test for coronavirus costs \$3,200.\rq\rq{} \\
 \statement{27} &  Deanna Lorraine &  2020-12-04 &  \politifacttwo &  Says of COVID-19 that Dr. Anthony Fauci "was telling people on February 29th that there was nothing to worry about and it posed no threat to the US public at large. \\
 \statement{28} &  Instagram Post &  2020-18-03 &  \politifacttwo &  Bill Gates and other globalists, in collaboration with pharmaceutical companies, are reportedly working to push tracking bracelets and ‘invisible tattoos’ to monitor Americans during an impending lockdown. \\
 \statement{29} &  Facebook User &  2020-28-03 &  \politifacttwo &  Says a \lq\lq 5G LAW PASSED while everyone was distracted\rq\rq{} with the coronavirus pandemic and lists 20 symptoms associated with 5G exposure. \\
 \statement{30} &  Facebook User &  2020-28-03 &  \politifacttwo &  Says for otherwise healthy people \lq\lq experiencing mild to moderate respiratory symptoms with or without a COVID-19 diagnosis\ldots only high temperatures kill a virus, so let your fever run high,\rq\rq{} but not over 103 or 104 degrees. \\
\addlinespace
  \statement{31} &  Facebook User &  2020-31-03 &  \politifactthree &  Ron Johnson said Americans should go back to work, because \lq\lq death is an unavoidable part of life.\rq\rq{} \\
 \statement{32} &  Jeff Jackson &  2020-19-03 &  \politifactthree &  North Carolina \lq\lq hospital beds are typically 85\% full across the state.\rq\rq{} \\
 \statement{33} &  Facebook User &  2020-15-03 &  \politifactthree &  So Oscar Health, the company tapped by Trump to profit from covid tests, is a Kushner company. Imagine that, profits over national safety. \\
 \statement{34} &  Brian Fitzpatrick &  2020-23-03 &  \politifactthree &  We've got to give the American public a rough estimate of how long we think this is going to take, based mostly on the South Korean model, which seems to be the trajectory that we are on, thankfully, and not the Italian model. \\
 \statement{35} &  Facebook User &  2020-10-03 &  \politifactthree &  Harvard scientists say the coronavirus is “spreading so fast that it will infect 70\% of humanity this year.” \\
 \statement{36} &  Drew Pinsky &  2020-03-03 &  \politifactthree &  You’re more likely to die of influenza right now” than the 2019 coronavirus. \\
 \statement{37} &  Michael Bloomberg &  2020-26-02 &  \politifactthree &  Says of President Donald Trump's actions on the coronavirus: \lq\lq No. 1, he fired the pandemic team two years ago. No. 2, hes been defunding the Centers for Disease Control.\rq\rq{} \\
 \statement{38} &  Joe Biden &  2020-05-04 &  \politifactthree &  45 nations had already moved \lq\lq to enforce travel restrictions with China\rq\rq{} before the president moved. \\
 \statement{39} &  Facebook User &  2020-07-04 &  \politifactthree &  Says Donald Trump \lq\lq himself has a financial stake in the French company that makes the brand-name version of hydroxychloroquine.\rq\rq{} \\
 \statement{40} &  Facebook User &  2020-01-04 &  \politifactthree &  \lq\lq Non-essential people get to file for unemployment and make two to three times more than normal,\rq\rq{} but essential workers still on the job get no pay raise. \\
\addlinespace
  \statement{41} &  Facebook User &  2020-29-03 &  \politifactfour &  Says a study projects Wisconsin’s coronavirus cases will peak on April 26, 2020. \\
 \statement{42} &  Facebook User &  2020-20-03 &  \politifactfour &  Says truck drivers are being turned away from fast-food restaurants during the COVID-19 pandemic. \\
 \statement{43} &  Facebook User &  2020-18-03 &  \politifactfour &  2019 coronavirus can live for \lq\lq up to 3 hours in the air, up to 4 hours on copper, up to 24 hours on cardboard up to 3 days on plastic and stainless steel.\rq\rq{} \\
 \statement{44} &  Facebook User &  2020-15-03 &  \politifactfour &  Bill Gates told us about the coronavirus in 2015. \\
 \statement{45} &  Chart &  2020-09-03 &  \politifactfour &  Says 80\% of novel coronavirus cases are \lq\lq mild.\rq\rq{} \\
 \statement{46} &  Lou Dobbs &  2020-02-03 &  \politifactfour &  The United States is \lq\lq actually screening fewer people (for the coronavirus than other countries) because we don’t have appropriate testing.\rq\rq{} \\
 \statement{47} &  Charlie Kirk &  2020-24-02 &  \politifactfour &  Three Chinese nationals were apprehended trying to cross our Southern border illegally. Each had flu-like symptoms. Border Patrol quickly quarantined them and assessed any threat of coronavirus. \\
 \statement{48} &  Bernie Sanders &  2020-08-04 &  \politifactfour &  It has been estimated that only 12\% of workers in businesses that are likely to stay open during this crisis are receiving paid sick leave benefits as a result of the second coronavirus relief package. \\
 \statement{49} &  Viral Image &  2020-08-04 &  \politifactfour &  Says a California surfer was \lq\lq alone, in the ocean,\rq\rq{} when he was arrested for violating the state’s stay-at-home order. \\
 \statement{50} &  Dan Patrick &  2020-31-03 &  \politifactfour &  Says for the coronavirus, \lq\lq the death rate in Texas, per capita of 29 million people, we're one of the lowest in the country.\rq\rq{} \\
\addlinespace
  \statement{51} &  Facebook User &  2020-02-04 &  \politifactfive &  On February 7, the WHO warned about the limited stock of PPE. That same day, the Trump administration announced it was sending 18 tons of masks, gowns and respirators to China. \\
 \statement{52} &  Pat Toomey &  2020-28-03 &  \politifactfive &  My mask will keep someone else safe and their mask will keep me safe. \\
 \statement{53} &  Andrew Cuomo &  2020-17-03 &  \politifactfive &  No city in the state can quarantine itself without state approval. \\
 \statement{54} &  Kelly Alexander &  2020-14-03 &  \politifactfive &  Says \lq\lq most\rq\rq{} NC legislators are in the \lq\lq high risk age group\rq\rq{} for coronavirus \\
 \statement{55} &  Viral Image &  2020-13-03 &  \politifactfive &  Says Spectrum will provide free internet to students during coronavirus school closures. \\
 \statement{56} &  Michael Dougherty &  2020-12-03 &  \politifactfive &  Some states are only getting 50 tests per day, and the Utah Jazz got 58. \\
 \statement{57} &  Blog Post &  2020-10-03 &  \politifactfive &  Whole of Italy goes into quarantine \\
 \statement{58} &  Dan Crenshaw &  2020-13-03 &  \politifactfive &  Says longstanding Food and Drug Administration regulations \lq\lq created barriers to the private industry creating a test quickly\rq\rq{} for the coronavirus. \\
 \statement{59} &  Viral Image &  2020-02-04 &  \politifactfive &  Photo shows a crowded New York City subway train during stay-at-home order. \\
 \statement{60} &  John Bel Edwards &  2020-05-04 &  \politifactfive &  Says of the coronavirus threat, \lq\lq there was not a single suggestion by anyone, a doctor, a scientist, a political figure, that we needed to cancel Mardi Gras.\rq\rq{} \\
\end{longtable}
\normalsize

\chapter{Chapter~\ref{cap:paper_tsc2024}: Survey Questions}

\label{cap:paper_tsc2024-appendix:survey-questions}

This appendix provides each question of the survey employed to investigate the barriers to running longitudinal tasks on crowdsourcing platforms. Section~\ref{cap:paper_tsc2024-sec:exp-setup-subsec:survey-design} and Section~\ref{cap:paper_tsc2024-sec:exp-setup-subsec:task} provide the details concerning the overall design of the survey and the crowdsourcing task. 

The questions are shown in order, as they were presented to the recruited workers. Each question is labeled with the corresponding survey part and index. When a question is labeled using sub-indexes, it means that it is nested in the survey. The text of each question is reported in \emph{italics}, together with the expected answer type using normal font and additional details written using \texttt{monospaced} font. 

In Appendix~\ref{cap:paper_tsc2024-appendix:survey-questions-pone}, several questions are labeled with the letter $X$. This labeling is a result of an explicit design choice in the crowdsourcing task, as described in Section~\ref{cap:paper_tsc2024-sec:exp-setup-subsec:task}.

\section{\pone: Current Perception Of Longitudinal Studies}

\label{cap:paper_tsc2024-appendix:survey-questions-pone}

\begin{myEnumerate}[label=1:, leftmargin=*, itemindent=0cm]
    \item \emph{Have you ever participated in a longitudinal study in the past, even if on other platforms?}
    \begin{myEnumerate}[label=1.1:, leftmargin=*, itemindent=0cm]
        \item \emph{How many?}
        \begin{itemize}
            \item[--] Integer number (X) in the interval $[0,3]$, such as $0\leq X\leq3$
        \end{itemize}
        \item[] \texttt{\small{numerical field, free text not allowed}}
        \begin{myEnumerate}[label=1.1.X:]
            \item \emph{Describe your experience with the longitudinal study nr. $X$}
            \begin{myEnumerate}[label=1.1.X.1:, leftmargin=*, itemindent=0cm]
                \item \emph{When was the study performed?}
                \begin{itemize}
                    \item[--] 1 month ago
                    \item[--] 2 months ago
                    \item[--] 3 to 5 months ago
                    \item[--] 6 to 12 months ago
                    \item[--] More than 1 year ago
                \end{itemize}
                \item[] \texttt{\small{closed-ended, radio button, free text not allowed}}
            \end{myEnumerate}
            \begin{myEnumerate}[label=1.1.X.2:, leftmargin=*, itemindent=0cm]
                \item \emph{How many sessions did the longitudinal study have?}
                \begin{itemize}
                    \item[--] Positive integer number
                \end{itemize}
                \item[] \texttt{\small{numerical field, free text not allowed}}
            \end{myEnumerate}
            \begin{myEnumerate}[label=1.1.X.3:, leftmargin=*, itemindent=0cm]
                \item \emph{Which was the time interval between each session?}
                \begin{itemize}
                    \item[--] 1 day
                    \item[--] 2 to 4 days
                    \item[--] 5 to 9 days
                    \item[--] 10 to 14 days
                    \item[--] 15 to 20 days
                    \item[--] 20 to 24 days
                    \item[--] 25 to 1 month
                    \item[--] 2 months
                    \item[--] 3 months
                    \item[--] 4 months
                    \item[--]  5 to 6 months
                    \item[--] 7 to 12 months
                    \item[--] More than 1 year
                    \item[--] Other (please, specify)
                \end{itemize}
                \item[] \texttt{\small{closed-ended, radio button, free text allowed}}
            \end{myEnumerate}
            \begin{myEnumerate}[label=1.1.X.4:, leftmargin=*, itemindent=0cm]
                \item \emph{What was the duration of each session?}
                \begin{itemize}
                    \item[--] 15 minutes
                    \item[--] 30 minutes
                    \item[--] 45 minutes
                    \item[--] 60 minutes
                    \item[--] 1 hour
                    \item[--] 2 hours
                    \item[--] 3 hours
                    \item[--] More than 3 hours
                    \item[--] Other (please, specify)
                \end{itemize}
                \item[] \texttt{\small{closed-ended, radio button, free text allowed}}
            \end{myEnumerate}
            \begin{myEnumerate}[label=1.1.X.5:, leftmargin=*, itemindent=0cm]
             \item \emph{Which was the crowdsourcing platform?}
                \begin{itemize}
                    \item[--] \mturk
                    \item[--] \prolific
                    \item[--] \toloka
                    \item[--] Other (please, specify)
                \end{itemize}
                \item[] \texttt{\small{closed-ended, radio button, free text allowed}}
            \end{myEnumerate}
            \begin{myEnumerate}[label=1.1.X.6:, leftmargin=*, itemindent=0cm]
             \item \emph{Which was the payment model?}
                \begin{itemize}
                    \item[--] Payment after each session
                    \item[--] Single final reward
                     \item[--] Other (please, specify)
                \end{itemize}
                \item[] \texttt{\small{closed-ended, checkbox, free text allowed}}
            \end{myEnumerate}
            \begin{myEnumerate}[label=1.1.X.7:, leftmargin=*, itemindent=0cm]
             \item \emph{How was your general satisfaction:}
                \begin{myEnumerate}[label=1.1.X.7.1:, leftmargin=*, itemindent=0cm]
                 \item \emph{Would you participate in the same study again?}
                    \begin{itemize}
                        \item[--] Yes
                        \item[--] No
                    \end{itemize}
                    \item[] \texttt{\small{closed-ended, radio button, free text not allowed}}
                \end{myEnumerate}
                \begin{myEnumerate}[label=1.1.X.7.2:, leftmargin=*, itemindent=0cm]
                 \item \emph{Please, tell us why}
                    \begin{itemize}
                        \item[--] Non-empty text
                    \end{itemize}
                    \item[] \texttt{\small{textual field}}
                \end{myEnumerate}
                \item[] \texttt{\small{question group}}
            \end{myEnumerate}
            \begin{myEnumerate}[label=1.1.X.8:, leftmargin=*, itemindent=0cm]
             \item \emph{What was the main incentives that convince you into participating in the longitudinal study?}
                \begin{itemize}
                    \item[--] Bonus
                    \item[--] Reward
                    \item[--] Interest on task
                    \item[--] Altruism (to help the research)
                    \item[--] Because the task was educative
                    \item[--] Other (please, specify)
                \end{itemize}
                \item[] \texttt{\small{closed-ended, radio-button, free text allowed}}
            \end{myEnumerate}
            \begin{myEnumerate}[label=1.1.X.9:, leftmargin=*, itemindent=0cm]
             \item \emph{Did you complete the task?}
                \begin{itemize}
                  \item[--]\emph{Yes}
                    \begin{myEnumerate}[label=1.1.X.9.1:, leftmargin=*, itemindent=0cm]
                    \item \emph{What were the main incentives that convinced you in completing the longitudinal study?}
                        \begin{itemize}
                            \item[--] Bonus
                            \item[--] Reward
                            \item[--] Interest on task
                            \item[--] Altruism (to help the research)
                            \item[--] Because the task was educative
                        \end{itemize}
                         \item[] \texttt{\small{closed-ended, radio-button, free text not allowed}}
                    \end{myEnumerate}
                \end{itemize}
                \begin{itemize}
                  \item[--]\emph{No}
                    \begin{myEnumerate}[label=1.1.X.9.2:, leftmargin=*, itemindent=0cm]
                    \item \emph{What are the reasons that made you dropout?}
                        \begin{itemize}
                            \item[--] Non-empty text
                        \end{itemize}
                        \item[] \texttt{\small{textual field}}
                    \end{myEnumerate}
                \end{itemize}
                \item[] \texttt{\small{closed-ended, radio-button, free text allowed}}
            \end{myEnumerate}
        \end{myEnumerate}
        \item[] \texttt{\small{question group, repeated $X$ times}}
    \end{myEnumerate}
    \item[] \texttt{\small{question group}}
\end{myEnumerate}

\begin{myEnumerate}[label=2:, leftmargin=*, itemindent=0cm]
\item \emph{Do you think this crowdsourcing platform is suitable to carry out longitudinal studies? Please, elaborate your answer}
    \begin{itemize}
        \item[--] Non-empty text
    \end{itemize}
    \item[] \texttt{\small{textual field}}
\end{myEnumerate}

\begin{myEnumerate}[label=3:, leftmargin=*, itemindent=0cm]
\item \emph{Longitudinal studies are not very common in crowdsourcing yet. Which of these statements do you agree with?}
    \begin{itemize}
        \item[--] Longitudinal studies are not optimally supported by current popular crowdsourcing platforms
        \item[--] Workers do not like to commit on daily effort
        \item[--] Reward and incentives are insufficient
        \item[--] Requesters do not need longitudinal participation since most of the tasks work with static data to annotate
        \item[--] Other (please, specify)
    \end{itemize}
    \item[] \texttt{\small{closed-ended, checkbox, free text allowed}}
\end{myEnumerate}

\newpage

\section{\ptwo: Possible Participation And Commitment To Longitudinal Studies}

\label{cap:paper_tsc2024-appendix:survey-questions-ptwo}

\begin{myEnumerate}[label=1:, leftmargin=*, itemindent=0cm]
    \item \emph{How many days would you be happy to commit to a longitudinal study (imagine a session of about 15 min per day)}
    \begin{itemize}
        \item[--] Positive integer number
    \end{itemize}
    \item[] \texttt{\small{numerical field, free text not allowed}}
\end{myEnumerate}

\begin{myEnumerate}[label=2:, leftmargin=*, itemindent=0cm]
    \item \emph{Which of the following would make you refuse participation in a longitudinal study?}
    \begin{itemize}
        \item[--] Too frequent
        \item[--] Too long
        \item[--] Other (please, specify)
    \end{itemize}
    \item[] \texttt{\small{closed-ended, checkbox, free text allowed}}
\end{myEnumerate}
\begin{myEnumerate}[label=3:, leftmargin=*, itemindent=0cm]
    \item \emph{What's your preferred frequency of participation in a longitudinal study?}
    \begin{itemize}
        \item[--] Daily
        \item[--] Every other day
        \item[--] Weekly
        \item[--] Biweekly
        \item[--] Monthly
        \item[--] Every six months
        \item[--] Yearly
    \end{itemize}
    \item[] \texttt{\small{closed-ended, radio button, free text not allowed}}
\end{myEnumerate}
\begin{myEnumerate}[label=4:, leftmargin=*, itemindent=0cm]
    \item \emph{What is your preferred session duration (in hours)?}
    \begin{itemize}
        \item[--] Positive integer number
    \end{itemize}
    \item[] \texttt{\small{numerical field, free text not allowed}}
\end{myEnumerate}
\begin{myEnumerate}[label=5:, leftmargin=*, itemindent=0cm]
    \item \emph{What do you consider an acceptable hourly payment for your work on this platform (in USD\$ dollars)?}
    \begin{itemize}
        \item[--] Positive integer number
    \end{itemize}
    \item[] \texttt{\small{numerical field, free text not allowed}}
\end{myEnumerate}
\begin{myEnumerate}[label=6:, leftmargin=*, itemindent=0cm]
    \item \emph{How much time would you be happy to allocate per day to work on longitudinal studies (in hours)?}
    \begin{itemize}
        \item[--] Positive integer number
    \end{itemize}
    \item[] \texttt{\small{numerical field, free text not allowed}}
\end{myEnumerate}
\begin{myEnumerate}[label=7:, leftmargin=*, itemindent=0cm]
    \item \emph{Which incentives would most motivate you to participate and engage in longitudinal studies?}
    \begin{itemize}
        \item[--] Final bonus to be awarded after the last contribution
        \item[--] Payment after each session
        \item[--] Progressive increment of payment
        \item[--] Progressive decrement of payment
        \item[--] Being penalized when skipping working sessions
        \item[--] Work on different tasks type to increase engagement diversity
        \item[--] Experimental variants of the same tasks to reduce repeatability
        \item[--] Other (please, specify)
    \end{itemize}
    \item[] \texttt{\small{closed-ended, checkbox, free text allowed}}
\end{myEnumerate}
\begin{myEnumerate}[label=8:, leftmargin=*, itemindent=0cm]
    \item \emph{What types of tasks would you like to perform in a longitudinal study?}
    \begin{itemize}
        \item[--] Information Finding - Such tasks delegate the process of searching to satisfy one’s information need to the workers in the crowd. For example, \lq\lq Find information about a company in the UK\rq\rq{}.
        \item[--] Verification and Validation - These are tasks that require workers in the crowd to either verify certain aspects as per the given instructions, or confirm he validity of various kinds of content. For example, \lq\lq Match the names of personal computers and verify corresponding information\rq\rq{}.
        \item[--] Interpretation and Analysis - Such tasks rely on the wisdom of the crowd to use their interpretation skills during task completion. For example, \lq\lq Choose the most suitable category for each URL\rq\rq{}.
        \item[--] Content Creation - Such tasks usually require the workers to generate new content for a document or website. They include authoring product descriptions or producing question-answer pair. For example, \lq\lq Suggest names for a new product\rq\rq{}.
        \item[--] Surveys - Surveys about a multitude of aspects ranging from demographics to customer satisfaction are crowdsourced. For example, \lq\lq Mother’s Day and Father’s Day Survey (18-29 year olds only!)\rq\rq{}.
        \item[--] Content Access - These tasks require the workers to simply access some content. For example, \lq\lq Click on the link and watch the video\rq\rq{}.
        \item[--] Other (please, specify)
    \end{itemize}
    \item[] \texttt{\small{closed-ended, checkbox, free text allowed}}
\end{myEnumerate}
\begin{myEnumerate}[label=9:, leftmargin=*, itemindent=0cm]
    \item \emph{What do you think are the benefits of being involved in longitudinal studies?}
    \begin{itemize}
        \item[--] No need to spend time regularly searching for new tasks to perform
        \item[--] No need to learn how to do the job (Learning curve)
        \item[--] Better productivity (more operationale)
        \item[--] Intermediate payments would increase trust on requester
        \item[--] Other (please, specify)
    \end{itemize}
    \item[] \texttt{\small{closed-ended, checkbox, free text allowed}}
\end{myEnumerate}
\begin{myEnumerate}[label=10:, leftmargin=*, itemindent=0cm]
    \item \emph{What do you think are the downsides that limit your interest in participating in longitudinal studies?}
    \begin{itemize}
        \item[--] Lack of flexibility
        \item[--] Long term commitment
        \item[--] Reward assigned at the end
        \item[--] Lack of diversity
        \item[--] Other (please, specify)
    \end{itemize}
    \item[] \texttt{\small{closed-ended, checkbox, free text allowed}}
\end{myEnumerate}
\begin{myEnumerate}[label=11:, leftmargin=*, itemindent=0cm]
    \item \emph{Do you have any additional suggestions for a requester who plans to design an attractive longitudinal study?}
    \begin{itemize}
        \item[--] Non-empty text
    \end{itemize}
    \item[] \texttt{\small{textual field}}
\end{myEnumerate}

\chapter{Chapter~\ref{cap:paper_ipm2021}: Task Instructions}

\label{cap:paper_ipm2021:-appendix:instructions}

This appendix reports the instruction text provided to each worker before starting the task described in Chapter~\ref{cap:paper_ipm2021}. The instructions contain descriptions of each truthfulness dimension as presented to the workers.

\begin{tcolorbox}[breakable,title=Task Instructions]
In this task, you will be asked to assess the truthfulness of eight statements by means of seven specific quality dimensions.

First, you will be asked to fill in one questionnaire and to answer three questions. Then, we will show you \emph{8 statements} made by popular people (for example, political figures) together with the information of who made the statement and on which date. For each statement, we ask you to search for evidence using our custom search engine and to tell \emph{how much do you agree with considering the statement true in general (as opposed to false)}; that is, its overall truthfulness. We also ask you to mark the evidence found in terms of an URL as well as your self-confidence about the topic, i.e., if you consider \emph{yourself expert / knowledgeable about its topic (as opposed to novice/beginner)}.

Then, we ask you to assess seven specific \emph{quality dimensions} by stating your level of agreement with them. All your answers are given on a 5 level scale, i.e., they must be selected among 5 different labels: (-2) Completely Disagree, (-1) Disagree, (0) Neither Agree Nor Disagree, (+1) Agree, (+2) Completely Agree. Each quality dimension is detailed in the following list. We provide a sample statement for each dimension so you can familiarize yourself with the seven dimensions. Please, note that there are some “positive” examples (i.e., statements that completely agree with the current dimension), and “negative” examples (i.e., statements that completely disagree with the current dimension). (Keep in mind that the examples are illustrative only, and it is likely that you may also need to use the rest of the labels in your answers). The seven dimensions we consider are the following:

\begin{itemize}
    \item \index{Correctness}\emph{Correctness}: the statement is expressed in an accurate way, as opposed to being incorrect and/or reporting mistaken information
    \begin{itemize}
        \item Example (which label is: +2 Completely Agree): ``It’s illegal to treat a minor without parental consent in the U.S. Even as hospitals are limiting visitors, minors will always be allowed to have one guardian present.''
    \end{itemize}
    \item \index{Neutrality}\emph{Neutrality}: the statement is expressed in a neutral / objective way, as opposed to subjective / biased
    \begin{itemize}
        \item Example (which label is: -2 Completely Disagree): ``The Labor Party has repeatedly claimed the Coalition needs to make cuts of \$70 billion to vital services to balance the budget.''
    \end{itemize}
    \item \index{Comprehensibility}\emph{Comprehensibility}: the statement is comprehensible / understandable / readable as opposed to difficult to understand
    \begin{itemize}
        \item Example (which label is: +2 Completely Agree): ``Florida ranks first among the nations for access to free prekindergarten.''
    \end{itemize}
    \item \index{Precision}\emph{Precision}: the information provided in the statement is precise / specific, as opposed to vague
    \begin{itemize}
        \item Example (which label is: -2 Completely Disagree): ``There were more deaths after the gun bans from guns than there were in the three years before Port Arthur.''
    \end{itemize}
    \item \index{Completeness}\emph{Completeness}: the information reported in the statement is complete as opposed to telling only a part of the story
    \begin{itemize}
        \item Example (which label is: -2 Completely Disagree): ``We inherited a broken test for COVID-19.''
    \end{itemize}
    \item \index{Speaker's Trustworthiness}\emph{Speaker's Trustworthiness}: the speaker is generally trustworthy / reliable as opposed to untrustworthy / unreliable / malicious
    \begin{itemize}
        \item Example (which label is: -2 Completely Disagree): ``Says video shows `the Chinese are destroying the 5G poles as they are aware that it is the thing triggering the corona symptoms.'''
    \end{itemize}
    \item \index{Informativeness}\emph{Informativeness}: the statement allows us to derive useful information as opposed to simply stating well known facts and/or tautologies
    \begin{itemize}
        \item Example (which label is: +2 Completely Agree): ``2019 coronavirus can live for up to 3 hours in the air, up to 4 hours on copper, up to 24 hours on cardboard, up to 3 days on plastic and stainless steel.''
    \end{itemize}
\end{itemize}

If you wish to change a previously given judgment, you can use the Back and Next buttons to navigate the task and revisit your answers. Please note that the statements are not presented in any particular order. You might see many good statements, many bad ones, or any combination. Try not to anticipate, and simply rate each statement after reading it. Note that you’ll need to answer \emph{all questions and fill in every field} in order to proceed in the task, otherwise you will not be able to proceed to the following steps. Note that there are some quality checks throughout the task, and if you do not perform these correctly you will not be able to terminate the task and get paid. The data from this task is being gathered for research purposes only. Participation is entirely voluntary, and you are free to leave the task at any point.
\end{tcolorbox}

\chapter{Chapter~\ref{cap:paper_ipm2023_bias}: \prisma Checklists}

\label{cap:paper_ipm2023_bias-appendix:prisma}

This appendix presents the complete checklists of the \lq\lq PRISMA 2020 Statement\rq\rq{} used for conducting the review on cognitive biases in a fact-checking context, as described in Section~\ref{cap:paper_ipm2023_bias-sec:methodology-subsec:prisma}. Table~\ref{cap:paper_ipm2023_bias-sec:methodology-subsec:prisma-tab:prisma-checklist} provides a summary of the \lq\lq Item \#\rq\rq{} and \lq\lq Section and Topic\rq\rq{} columns from the main checklist. Due to spacing issues, the checklists start on the following page.



# F.1   PRISMA 2020 Checklist For Abstracts

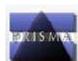 **PRISMA 2020 for Abstracts Checklist**

| Section and Topic | Item # | Checklist item | Reported (Yes/No) |
|---|---|---|---|
| **TITLE** | | | |
| Title | 1 | Identify the report as a systematic review. | Yes |
| **BACKGROUND** | | | |
| Objectives | 2 | Provide an explicit statement of the main objective(s) or question(s) the review addresses. | Yes |
| **METHODS** | | | |
| Eligibility criteria | 3 | Specify the inclusion and exclusion criteria for the review. | Yes |
| Information sources | 4 | Specify the information sources (e.g. databases, registers) used to identify studies and the date when each was last searched. | Not Relevant |
| Risk of bias | 5 | Specify the methods used to assess risk of bias in the included studies. | Not Relevant |
| Synthesis of results | 6 | Specify the methods used to present and synthesise results. | Yes |
| **RESULTS** | | | |
| Included studies | 7 | Give the total number of included studies and participants and summarise relevant characteristics of studies. | Yes |
| Synthesis of results | 8 | Present results for main outcomes, preferably indicating the number of included studies and participants for each. If meta-analysis was done, report the summary estimate and confidence/credible interval. If comparing groups, indicate the direction of the effect (i.e. which group is favoured). | Yes |
| **DISCUSSION** | | | |
| Limitations of evidence | 9 | Provide a brief summary of the limitations of the evidence included in the review (e.g. study risk of bias, inconsistency and imprecision). | Not Relevant |
| Interpretation | 10 | Provide a general interpretation of the results and important implications. | Yes |
| **OTHER** | | | |
| Funding | 11 | Specify the primary source of funding for the review. | No (specified at the end of the paper) |
| Registration | 12 | Provide the register name and registration number. | Not Relevant |

*From:* Page MJ, McKenzie JE, Bossuyt PM, Boutron I, Hoffmann TC, Mulrow CD, et al. The PRISMA 2020 statement: an updated guideline for reporting systematic reviews. BMJ 2021;372:n71. doi: 10.1136/bmj.n71

For more information, visit: http://www.prisma-statement.org/

# F.2   PRISMA 2020 Checklist

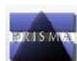 **PRISMA 2020 Checklist**

| Section and Topic | Item # | Checklist item | Location where item is reported |
|---|---|---|---|
| **TITLE** | | | |
| Title | 1 | Identify the report as a systematic review. | Title |
| **ABSTRACT** | | | |
| Abstract | 2 | See the PRISMA 2020 for Abstracts checklist. | Abstract, see PRISMA for Abstracts checklist |
| **INTRODUCTION** | | | |
| Rationale | 3 | Describe the rationale for the review in the context of existing knowledge. | Section 1.3 |
| Objectives | 4 | Provide an explicit statement of the objective(s) or question(s) the review addresses. | Section 8.1 |
| **METHODS** | | | |
| Eligibility criteria | 5 | Specify the inclusion and exclusion criteria for the review and how studies were grouped for the syntheses. | Section 8.2.2 |
| Information sources | 6 | Specify all databases, registers, websites, organisations, reference lists and other sources searched or consulted to identify studies. Specify the date when each source was last searched or consulted. | Section 8.2.2 |
| Search strategy | 7 | Present the full search strategies for all databases, registers and websites, including any filters and limits used. | Section 8.2.3 |
| Selection process | 8 | Specify the methods used to decide whether a study met the inclusion criteria of the review, including how many reviewers screened each record and each report retrieved, whether they worked independently, and if applicable, details of automation tools used in the process. | Section 8.2.3 |
| Data collection process | 9 | Specify the methods used to collect data from reports, including how many reviewers collected data from each report, whether they worked independently, any processes for obtaining or confirming data from study investigators, and if applicable, details of automation tools used in the process. | Section 8.2.3 |
| Data items | 10a | List and define all outcomes for which data were sought. Specify whether all results that were compatible with each outcome domain in each study were sought (e.g. for all measures, time points, analyses), and if not, the methods used to decide which results to collect. | Section 8.3.1 |
| | 10b | List and define all other variables for which data were sought (e.g. participant and intervention characteristics, funding sources). Describe any assumptions made about any missing or unclear information. | Section 8.3.1 |
| Study risk of bias assessment | 11 | Specify the methods used to assess risk of bias in the included studies, including details of the tool(s) used, how many reviewers assessed each study and whether they worked independently, and if applicable, details of automation tools used in the process. | Section 8.2.3 |
| Effect measures | 12 | Specify for each outcome the effect measure(s) (e.g. risk ratio, mean difference) used in the synthesis or presentation of results. | Not Relevant |
| Synthesis methods | 13a | Describe the processes used to decide which studies were eligible for each synthesis (e.g. tabulating the study intervention characteristics and comparing against the planned groups for each synthesis (item #5)). | Section 8.3.1 |
| | 13b | Describe any methods required to prepare the data for presentation or synthesis, such as handling of missing summary statistics, or data conversions. | Not Relevant |
| | 13c | Describe any methods used to tabulate or visually display results of individual studies and syntheses. | Section 8.3.2 |
| | 13d | Describe any methods used to synthesize results and provide a rationale for the choice(s). If meta-analysis was performed, describe the model(s), method(s) to identify the presence and extent of statistical heterogeneity, and software package(s) used. | Section 8.3.2 |
| | 13e | Describe any methods used to explore possible causes of heterogeneity among study results (e.g. subgroup analysis, meta-regression). | Not Relevant |
| | 13f | Describe any sensitivity analyses conducted to assess robustness of the synthesized results. | Not Relevant |
| Reporting bias assessment | 14 | Describe any methods used to assess risk of bias due to missing results in a synthesis (arising from reporting biases). | Section 8.2.3 |
| Certainty assessment | 15 | Describe any methods used to assess certainty (or confidence) in the body of evidence for an outcome. | Section 8.2.3 |
| **RESULTS** | | | |
| Study selection | 16a | Describe the results of the search and selection process, from the number of records identified in the search to the number of studies included in the review, ideally using a flow diagram. | Section 8.2.3 and Figure 8.2 |

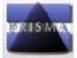 **PRISMA 2020 Checklist**

| Section and Topic | Item # | Checklist item | Location where item is reported |
|---|---|---|---|
| | 16b | Cite studies that might appear to meet the inclusion criteria, but which were excluded, and explain why they were excluded. | Section 8.2.2 |
| Study characteristics | 17 | Cite each included study and present its characteristics. | Section 8.3.1 |
| Risk of bias in studies | 18 | Present assessments of risk of bias for each included study. | Not Relevant |
| Results of individual studies | 19 | For all outcomes, present, for each study: (a) summary statistics for each group (where appropriate) and (b) an effect estimate and its precision (e.g. confidence/credible interval), ideally using structured tables or plots. | Section 8.3.2 |
| Results of syntheses | 20a | For each synthesis, briefly summarise the characteristics and risk of bias among contributing studies. | Section 8.3.2 |
| | 20b | Present results of all statistical syntheses conducted. If meta-analysis was done, present for each the summary estimate and its precision (e.g. confidence/credible interval) and measures of statistical heterogeneity. If comparing groups, describe the direction of the effect. | Not Relevant |
| | 20c | Present results of all investigations of possible causes of heterogeneity among study results. | Table 8.4 |
| | 20d | Present results of all sensitivity analyses conducted to assess the robustness of the synthesized results. | Section 8.3.2 |
| Reporting biases | 21 | Present assessments of risk of bias due to missing results (arising from reporting biases) for each synthesis assessed. | Not Relevant |
| Certainty of evidence | 22 | Present assessments of certainty (or confidence) in the body of evidence for each outcome assessed. | Not Relevant |
| **DISCUSSION** | | | |
| Discussion | 23a | Provide a general interpretation of the results in the context of other evidence. | Section 8.3.3 and Section 8.3.4 |
| | 23b | Discuss any limitations of the evidence included in the review. | Section 11.3.2 |
| | 23c | Discuss any limitations of the review processes used. | Section 11.3.2 |
| | 23d | Discuss implications of the results for practice, policy, and future research. | Section 11.2.2 |
| **OTHER INFORMATION** | | | |
| Registration and protocol | 24a | Provide registration information for the review, including register name and registration number, or state that the review was not registered. | Not Relevant |
| | 24b | Indicate where the review protocol can be accessed, or state that a protocol was not prepared. | Not Relevant |
| | 24c | Describe and explain any amendments to information provided at registration or in the protocol. | Not Relevant |
| Support | 25 | Describe sources of financial or non-financial support for the review, and the role of the funders or sponsors in the review. | End of the thesis, before references |
| Competing interests | 26 | Declare any competing interests of review authors. | End of the thesis, before references |
| Availability of data, code and other materials | 27 | Report which of the following are publicly available and where they can be found: template data collection forms; data extracted from included studies; data used for all analyses; analytic code; any other materials used in the review. | The thesis is self contained and all the material is included either in its text or in the appendix. |

*From:* Page MJ, McKenzie JE, Bossuyt PM, Boutron I, Hoffmann TC, Mulrow CD, et al. The PRISMA 2020 statement: an updated guideline for reporting systematic reviews. BMJ 2021;372:n71. doi: 10.1136/bmj.n71

For more information, visit: http://www.prisma-statement.org/



\chapter{Chapter~\ref{cap:paper_ipm2023_bias}: The \numbias Cognitive Biases}

\label{cap:paper_ipm2023_bias-appendix:biases}

This appendix reports the full list \numbias of cognitive biases found in the literature using the methodology described in Section~\ref{cap:paper_ipm2023_bias-sec:methodology}. The \numbiasused cognitive biases that might manifest in a fact-checking context, described in Section~\ref{cap:paper_ipm2023_bias-sec:methodology-subsec:selection-process}, have been derived from this list.

\begin{multicols}{2}
\begin{enumerate}[font=\sffamily]
\itemsep0em 
\item  \textsf{Action Bias}
\item  \textsf{Actor-observer Bias}
\item  \textsf{Additive Bias}
\item  \textsf{Agent Detection Bias}
\item  \textsf{Affect Heuristic}
\item  \textsf{Ambiguity Effect}
\item  \textsf{Anchoring Effect}
\item  \textsf{Anthropocentric Thinking}
\item  \textsf{Anthropomorphism}
\item  \textsf{Apophenia}
\item  \textsf{Association Fallacy}
\item  \textsf{Assumed Similarity Bias}
\item  \textsf{Attentional Bias}
\item  \textsf{Attribute Substitution}
\item  \textsf{Attribution Bias}
\item  \textsf{Authority Bias}
\item  \textsf{Automation Bias}
\item  \textsf{Availability Bias}
\item  \textsf{Availability Cascade}
\item  \textsf{Availability Heuristic}
\item  \textsf{Backfire Effect}
\item  \textsf{Bandwagon Effect}
\item  \textsf{Barnum Effect} (or \textsf{Forer Effect})
\item  \textsf{Base Rate Fallacy}
\item  \textsf{Belief Bias}
\item  \textsf{Ben Franklin Effect}
\item  \textsf{Berkson’s Paradox}
\item  \textsf{Bias blind Spot}
\item  \textsf{Bizarreness Effect}
\item  \textsf{Boundary Extension}
\item  \textsf{Cheerleader Effect}
\item  \textsf{Childhood Amnesia}
\item  \textsf{Choice-supportive Bias}
\item  \textsf{Cognitive Dissonance}
\item  \textsf{Commission Bias}
\item  \textsf{Compassion Fade}
\item  \textsf{Confirmation Bias}
\item  \textsf{Conformity}
\item  \textsf{Congruence Bias}
\item  \textsf{Conjunction Fallacy} (or \textsf{Linda Problem})
\item  \textsf{Conservatism Bias} (or \textsf{Regressive Bias})
\item  \textsf{Consistency Bias}
\item  \textsf{Context Effect}
\item  \textsf{Continued Influence Effect}
\item  \textsf{Contrast Effect}
\item  \textsf{Courtesy Bias}
\item  \textsf{Cross-race Effect}
\item  \textsf{Cryptomnesia}
\item  \textsf{Curse of Knowledge}
\item  \textsf{Declinism}
\item  \textsf{Decoy Effect}
\item  \textsf{Default Effect}
\item  \textsf{Defensive Attribution Hypothesis}
\item  \textsf{Denomination Effect}
\item  \textsf{Disposition Effect}
\item  \textsf{Distinction Bias}
\item  \textsf{Dread Aversion}
\item  \textsf{Dunning–Kruger Effect}
\item  \textsf{Duration Neglect}
\item  \textsf{Effort Justification}
\item  \textsf{Egocentric Bias}
\item  \textsf{End-of-history Illusion}
\item  \textsf{Endowment Effect}
\item  \textsf{Escalation of Commitment} (or \textsf{Irrational Escalation}, or \textsf{Sunk Cost Fallacy})
\item  \textsf{Euphoric Recall}
\item  \textsf{Exaggerated Expectation}
\item  \textsf{Experimenter’s Bias} (or \textsf{Expectation Bias})
\item  \textsf{Extension Neglect}
\item  \textsf{Extrinsic Incentives Bias}
\item  \textsf{Fading Affect Bias}
\item  \textsf{Fallacy of Composition}
\item  \textsf{Fallacy of Division}
\item  \textsf{False Consensus Effect}
\item  \textsf{False Memory}
\item  \textsf{False Uniqueness Bias}
\item  \textsf{Form Function Attribution Bias}
\item  \textsf{Framing Effect} (or \textsf{Frequency Illusion}, or \textsf{Baader–Meinhof Phenomenon})
\item  \textsf{Fundamental Attribution Error}
\item  \textsf{Gambler’s Fallacy}
\item  \textsf{Gender Bias}
\item  \textsf{Generation Effect} (or \textsf{Self-generation Effect})
\item  \textsf{Google Effect}
\item  \textsf{Group Attribution Error}
\item  \textsf{Groupshift}
\item  \textsf{Groupthink}
\item  \textsf{Halo Effect}
\item  \textsf{Hard–easy Effect}
\item  \textsf{Hindsight Bias}
\item  \textsf{Hostile Attribution Bias}
\item  \textsf{Hot-cold Empathy Gap}
\item  \textsf{Hot-hand Fallacy}
\item  \textsf{Humor Effect}
\item  \textsf{Hyperbolic Discounting}
\item  \textsf{IKEA Effect}
\item  \textsf{Illicit Transference}
\item  \textsf{Illusion of Asymmetric Insight}
\item  \textsf{Illusion of Control}
\item  \textsf{Illusion of Explanatory Depth}
\item  \textsf{Illusion of Transparency}
\item  \textsf{Illusion of Validity}
\item  \textsf{Illusory Correlation}
\item  \textsf{Illusory Superiority}
\item  \textsf{Illusory Truth Effect}
\item  \textsf{Impact Bias}
\item  \textsf{Implicit Bias}
\item  \textsf{Information Bias}
\item  \textsf{Ingroup Bias}
\item  \textsf{Insensitivity To Sample Size}
\item  \textsf{Intentionality Bias}
\item  \textsf{Interoceptive Bias} (or \textsf{Hungry Judge Effect})
\item  \textsf{Just-world Hypothesis}
\item  \textsf{Lag Effect}
\item  \textsf{Less-is-better Effect}
\item  \textsf{Leveling And Sharpening}
\item  \textsf{Levels-of-processing Effect}
\item  \textsf{List-length Effect}
\item  \textsf{Logical Fallacy}
\item  \textsf{Loss Aversion}
\item  \textsf{Memory Inhibition}
\item  \textsf{Mere exposure Effect} (or \textsf{Familiarity Principle})
\item  \textsf{Misattribution}
\item  \textsf{Modality Effect}
\item  \textsf{Money Illusion}
\item  \textsf{Mood-congruent Memory Bias}
\item  \textsf{Moral Credential Effect}
\item  \textsf{Moral Luck}
\item  \textsf{Naïve Cynicism}
\item  \textsf{Naïve Realism}
\item  \textsf{Negativity Bias}
\item  \textsf{Neglect of Probability}
\item  \textsf{Next-in-line Effect}
\item  \textsf{Non-adaptive Choice Switching}
\item  \textsf{Normalcy Bias}
\item  \textsf{Not Invented Here Syndrome}
\item  \textsf{Objectivity Illusion}
\item  \textsf{Observer-expectancy Effect}
\item  \textsf{Omission Bias}
\item  \textsf{Optimism Bias}
\item  \textsf{Ostrich Effect} (or \textsf{Ostrich Problem})
\item  \textsf{Outcome Bias}
\item  \textsf{Outgroup Homogeneity Bias}
\item  \textsf{Overconfidence Effect}
\item  \textsf{Parkinson’s Law of Triviality}
\item  \textsf{Part-list Cueing Effect}
\item  \textsf{Peak–end Rule}
\item  \textsf{Perky Effect}
\item  \textsf{Pessimism Bias}
\item  \textsf{Picture Superiority Effect}
\item  \textsf{Placement Bias}
\item  \textsf{Plan Continuation Bias}
\item  \textsf{Planning Fallacy}
\item  \textsf{Plant Blindness}
\item  \textsf{Positivity Effect} (or \textsf{Socioemotional Electivity Theory})
\item  \textsf{Present Bias}
\item  \textsf{Prevention Bias}
\item  \textsf{Primacy Effect}
\item  \textsf{Probability Matching}
\item  \textsf{Processing Difficulty Effect}
\item  \textsf{Pro-innovation Bias}
\item  \textsf{Projection Bias}
\item  \textsf{Proportionality Bias}
\item  \textsf{Prospect Theory}
\item  \textsf{Pseudocertainty Effect}
\item  \textsf{Puritanical Bias}
\item  \textsf{Pygmalion Effect}
\item  \textsf{Reactance Theory}
\item  \textsf{Reactive Devaluation}
\item  \textsf{Recency Effect}
\item  \textsf{Recency Illusion}
\item  \textsf{Reminiscence Bump}
\item  \textsf{Repetition Blindness}
\item  \textsf{Restraint Bias}
\item  \textsf{Rhyme As Reason Effect}
\item  \textsf{Risk Compensation} (or \textsf{Peltzman Effect})
\item  \textsf{Rosy Retrospection}
\item  \textsf{Salience Bias}
\item  \textsf{Saying Is Believing Effect}
\item  \textsf{Scope Neglect}
\item  \textsf{Selection Bias}
\item  \textsf{Self-relevance Effect}
\item  \textsf{Self-serving Bias}
\item  \textsf{Semmelweis Reflex}
\item  \textsf{Serial Position Effect}
\item  \textsf{Sexual Overperception Bias}
\item  \textsf{Shared Information Bias}
\item  \textsf{Social Comparison Bias}
\item  \textsf{Social Cryptomnesia}
\item  \textsf{Social Desirability Bias}
\item  \textsf{Source Confusion}
\item  \textsf{Spacing Effect}
\item  \textsf{Spotlight Effect}
\item  \textsf{Status Quo Bias}
\item  \textsf{Stereotypical Bias} (or \textsf{Stereotype Bias})
\item  \textsf{Stereotyping Subadditivity Effect}
\item  \textsf{Spacing Effect}
\item  \textsf{Subjective Validation}
\item  \textsf{Suffix Effect}
\item  \textsf{Surrogation}
\item  \textsf{Survivorship Bias}
\item  \textsf{System Justification}
\item  \textsf{Systematic Bias}
\item  \textsf{Tachypsychia}
\item  \textsf{Telescoping Effect}
\item  \textsf{Testing Effect}
\item  \textsf{Third-person Effect}
\item  \textsf{Time-saving Bias}
\item  \textsf{Tip of The Tongue Phenomenon}
\item  \textsf{Trait Ascription Bias}
\item  \textsf{Travis Syndrome}
\item  \textsf{Truth Bias}
\item  \textsf{Ultimate Attribution Error}
\item  \textsf{Unconscious Bias} (or \textsf{Implicit Bias})
\item  \textsf{Unit Bias}
\item  \textsf{Verbatim Effect}
\item  \textsf{Von Restorff Effect}
\item  \textsf{Weber–Fechner Law}
\item  \textsf{Well Travelled Road Effect}
\item  \textsf{Women Are Wonderful Effect}
\item  \textsf{Worse-than-average Effect}
\item  \textsf{Zero-risk Bias}
\item  \textsf{Zero-sum Bias}
\end{enumerate}
\end{multicols}

\chapter{Chapter~\ref{cap:paper_facct2022}: Questionnaires}

\label{cap:paper_facct2022-appendix:quest-crt}

This appendix reports the additional questionnaires used for the task described in Section~\ref{cap:paper_facct2022-sec:exp-setup-subsec:crowdsourcing-task}, the \lq\lq Belief in Science Scale\rq\rq{} (\index{BISS}BISS) questionnaire \cite{dagnall2019evaluation}, and the generalized version of the \lq\lq Citizen Trust in Government Organizations\rq\rq{} (\index{CTGO}CTGO) questionnaire \cite{grimmelikhuijsen2017validating}. The \index{BISS}BISS and \index{CTGO}CTGO questionnaires require an answer provided using a 5-level \index{Likert}Likert scale, ranging from \completelydisagree to \completelyagree. The task design includes also the questionnaires reported in Appendix~\ref{cap:paper_sigir2020-appendix:quest-crt-sec:initial} and Appendix~\ref{cap:paper_sigir2020-appendix:quest-crt-sec:crt}.

 \section{Citizen Trust in Government Organizations (CTGO)}
 
  \label{cap:paper_facct2022-appendix:quest-crt-sec:trust}

 \begin{enumerate}[align=left, leftmargin=*, label=CTGO\arabic*:]
 \item Politicians in general are capable.
 \item Politicians in general are effective.
 \item Politicians in general are skillful.
 \item Politicians in general are expert.
 \item Politicians in general carry out their duty very well.
 \item If citizens need help, the politicians will do their best to help them.
 \item Politicians in general act in the interest of citizens.
 \item Politicians in general are genuinely interested in the well-being of citizens.
 \item Politicians in general approach citizens in a sincere way.
 \item Politicians in general approach are sincere.
 \item Politicians in general keep their commitment.
 \item Politicians in general are honest.
 \end{enumerate}
 
 \section{Belief in Science Scale (BISS)}
 
 \label{cap:paper_facct2022-appendix:quest-crt-sec:belief}

 \begin{enumerate}[align=left, leftmargin=*, label=BIS\arabic*:]
 \item Science provides us with a better understanding of the universe than does religion.
 \item \lq\lq In a demon-haunted world, science is a candle in the dark.\rq\rq{} (Carl Sagan)
 \item We can only rationally believe in what is scientifically provable.
 \item Science tells us everything there is to know about what reality consists of.
 \item All the tasks human beings face are soluble by science.
 \item The scientific method is the only reliable path to knowledge.
 \item The only real kind of knowledge we can have is scientific knowledge.
 \item Science is the most valuable part of human culture.
 \item Science is the most efficient means of attaining truth.
 \item Scientists and science should be given more respect in modern society.
 \end{enumerate}
 
 \chapter{Section~\ref{cap:paper_is2022}: Multidimensional Scale For Reviews Quality Judgment}
 
 \label{cap:paper_is2022-appendix:mult-scale-rev}
 
 This appendix reports the adapted version of the multiple dimensions of truthfulness (Section~\ref{cap:paper_ipm2021-sec:exp-setup-subsec:seven-dim}) used to judge product review quality, as presented to workers during the crowdsourcing task. This list should be compared with the instructions in Appendix~\ref{cap:paper_ipm2021:-appendix:instructions}.

\begin{tcolorbox}[breakable,title=Review Quality Dimensions]
\begin{itemize}
    \item \index{Overall Truthfulness}\emph{Overall Truthfulness}: measures the overall truthfulness and trustworthiness of the review
    \begin{itemize}
        \item Example (label: +2 Completely Agree): ``They fit great, look great, are quite comfortable and are just what I was looking for!''
    \end{itemize}
    
    \item \index{Reliability}\emph{Reliability}: the review is considered reliable, as opposed to reporting unreliable information
    \begin{itemize}
        \item Example (label: +2 Completely Agree): ``They fit great, look great, are quite comfortable and are just what I was looking for!''
    \end{itemize}

    \item \index{Neutrality}\emph{Neutrality}: the review is expressed in objective terms, as opposed to being subjective or biased
    \begin{itemize}
        \item Example (label: -2 Completely Disagree): ``Love them!!''
    \end{itemize}

    \item \index{Comprehensibility}\emph{Comprehensibility}: the review is comprehensible / understandable / readable as opposed to difficult to understand
    \begin{itemize}
        \item Example (label: +2 Completely Agree): ``They run big. Order a full size smaller.''
    \end{itemize}

    \item \index{Precision}\emph{Precision}: the review is precise / specific, as opposed to vague
    \begin{itemize}
        \item Example (label: +2 Completely Agree): ``They run big. Order a full size smaller.''
    \end{itemize}

    \item \index{Completeness}\emph{Completeness}: the review is complete as opposed to partial
    \begin{itemize}
        \item Example (label: +2 Completely Agree): ``I actually have 3 pairs of these trainers. They are very comfortable, there is a neoprene sleeve that goes around your ankle that makes them the most comfortable for me compared to normal athletic shoes. They run a little narrow – for me this is perfect, but you may want to round up on the size or try on in the store first if your feet are on the wider side.''
    \end{itemize}

    \item \index{Informativeness}\emph{Informativeness}: the review allows deriving useful information as opposed to well-known facts and/or tautologies
    \begin{itemize}
        \item Example (label: +1 Agree): ``Love these shoes! Needed new running shoes and these are perfect. Light weight and fit great!''
    \end{itemize}
\end{itemize}
\end{tcolorbox}

\backmatter

\bibliographystyle{plainnat}
\bibliography{bibliography}

\printindex

\label{LastPages}

\end{document}